\documentclass[11pt,twoside,a4paper]{cernyrep}

\usepackage{graphicx}%
\usepackage{multirow}%
\usepackage{amsmath,amssymb,amsfonts}%
\usepackage{amsthm}%
\usepackage{mathrsfs}%
\usepackage[title]{appendix}%
\usepackage[table,xcdraw]{xcolor}
\usepackage{textcomp}%
\usepackage{manyfoot}%
\usepackage{booktabs}%
\usepackage{algorithm}%
\usepackage{algorithmicx}%
\usepackage{algpseudocode}%
\usepackage{listings}%

\usepackage{placeins}

\usepackage{subfiles}%
\usepackage{textgreek}

\usepackage{comment}
\usepackage{siunitx}
\usepackage{wrapfig}
\usepackage{tikz} 
\usepackage{physics}

\usepackage{tabularx}      
\usepackage{makecell}     

\usepackage{pdflscape}
\usepackage{booktabs} 
\usepackage{amsmath} 
\usepackage{makecell}
\usepackage{siunitx}
\usepackage{multirow} 

\usepackage{tcolorbox}

\usepackage[normalem]{ulem}

\usepackage[normalem]{ulem}
\usepackage{cancel}

\usepackage{hyperref}
\hypersetup{%
        colorlinks,
        breaklinks=true,
        plainpages=false,%
        citecolor=blue,
        linkcolor=blue,
        urlcolor=blue,
        bookmarksopen=true,%
        bookmarksnumbered=false,%
        bookmarksdepth=5%
}
\def\figureautorefname~#1\null{Fig.~#1\null}

\usepackage{wrapfig}
\usepackage{rotating}
\usepackage{wasysym}

\usepackage{tocloft}

\newcommand{\axion}{\ensuremath{\textsf{a}}\xspace}
\newcommand{\pt}{\ensuremath{p_\mathrm{T}}\xspace}
\newcommand{\kt}{\ensuremath{k_\mathrm{T}}\xspace}

\newcommand{\micron}{\ensuremath{\upmu \mathrm{m}}}
\newcommand{\lumi}{\ensuremath{\mathcal{L}}}
\newcommand{\keV}{\si{\kilo\electronvolt}}
\newcommand{\MeV}{\si{\mega\electronvolt}}

\newcommand{\Pgt}{\ensuremath{\mathrm{\tau}}}

\newcommand{\mlab}[1]%
    {\mbox{}\marginpar{\raggedright\hspace{0pt}\tiny {\color{red}{#1}}}}

\newcommand{\as}{\alpha_{\rm S}}

\newcommand{\epem}{\ensuremath{\Pep\Pem}}
\newcommand{\sqrts}{\sqrt{s}}

\providecommand{\qqbar}{\PQq\PAQq}
\providecommand{\uubar}{\PQu\PAQu}
\providecommand{\ddbar}{\PQd\PAQd}
\providecommand{\ssbar}{\PQs\PAQs}
\providecommand{\ccbar}{\PQc\PAQc}
\providecommand{\bbbar}{\PQb\PAQb}
\providecommand{\ttbar}{\PQt\PAQt}

\newcommand{\hz}{\PH\PZ}

\newcommand{\LumiInt}{\mathcal{L}_\text{int}}

\providecommand{\pythia}{\textsc{pythia}}

\begin{document}

\title{Future Circular Collider}

\pagenumbering{roman}
\setcounter{page}{1}

\newcommand{\main}{.}
\def\biblio{}

\begin{titlepage}
		\begin{center}
		\textbf{\Large Future Circular Collider}\\
        \vspace{0.5cm}
        \textbf{\LARGE Feasibility Study Report}\\
        \vspace{8 cm}
         \textbf{\Huge Volume 1}\\
        \vspace{1cm}
        \textbf{\Huge Physics, Experiments, Detectors}\\
        \vspace{3 cm}
        \textbf{ \today } \\
        \vspace*{\fill}
Submitted to the European Physics Journal ST, a joint publication of EDP Sciences,\\Springer Science+Business Media, and the Società Italiana di Fisica.
        \end{center}
\end{titlepage}

\section*{Note from the Editors}

\noindent One of the recommendations of the 2020 update of the European Strategy for Particle Physics was that “Europe, together with its international partners, should investigate the technical and financial feasibility of a future hadron collider at CERN with a centre-of-mass energy of at least 100 TeV and with an electron-positron Higgs and electroweak factory as a possible first stage.

\noindent In June 2021, the CERN Council launched the FCC Feasibility Study to be completed by 2025, in time for the next update of the European Strategy for Particle Physics. The study results are made publicly available through this FCC Feasibility Study Report, as input to the European Particle Physics Strategy update process, initiated by the CERN Council in March 2024. The studies presented in this FCC Feasibility Study Report do not imply any commitment by the CERN Member or Associate Member States to build the Future Circular Collider.

\noindent This report and the assumptions contained in it do not prejudge further territorial feasibility analysis by the Host States, France and Switzerland, as well as the outcome of their respective public debate and concertation processes, and future decisions of their relevant authorities.

\newpage

\section*{Acknowledgements}

We would like to thank the \textbf{International Steering Committee members:}\\

\vspace*{-3mm}

\begin{table}[!h]
	\small
	\centering
	\begin{tabular}{l}
		\toprule
		F.~Gianotti (Chair), CERN \\ 
		R.~Bello, CERN \\ 
		P.~Chomaz, CEA, France  \\ 
            M.~Cobal, INFN and University of Udine, Italy \\
		B.~Heinemann, DESY, Germany \\ 
		T.~Koseki, KEK, Japan \\ 
		M.~Lamont, CERN \\ 
		L.~Merminga, FNAL, United States \\ 
		J.~Mnich, CERN \\ 
		M.~Seidel, PSI and EPFL, Switzerland \\
		C.~Warakaulle, CERN \\
		\bottomrule
	\end{tabular}
\end{table}

\noindent and the \textbf{Scientific Advisory Committee members:}\\
\vspace*{-2mm}

\begin{table}[!h]
	\centering
	\small
	\vspace*{-3mm}
	\begin{tabular}{l}
		\toprule
		A.~Parker (Chair), Cambridge University, UK \\ 
		R.~Bartolini, DESY, Germany \\  
		A.~Chabert, SFTRF, France \\ 
		H.~Ehrbar, Heinz Ehrbar Partners LLC, Switzerland \\ 
		B.~Gavela Legazpi, UAM Madrid, Spain \\
		G.~Hiller, TU Dortmund, Germany \\
        S.~Krishnagopal, FNAL, U.S. \\		
        P.~Križan, University of Ljubljana, Slovenia\\
        P.~Lebrun, ESI, France\\
        P.~McIntosh, STFC, ASTeC,  UKRI, UK\\ 
        M.~Minty, BNL, U.S. \\  
        R.~Tenchini, INFN Sezione di Pisa, Italy  \\  
\bottomrule 
	\end{tabular}
\end{table}

\noindent for their continued guidance and careful reviewing that helped to complete this report successfully.\\

\newpage

\begin{wrapfigure}[4]{l}[0cm]{2cm}
  \begin{center}
  \vspace{-0.5cm}
    \includegraphics[width=2cm]{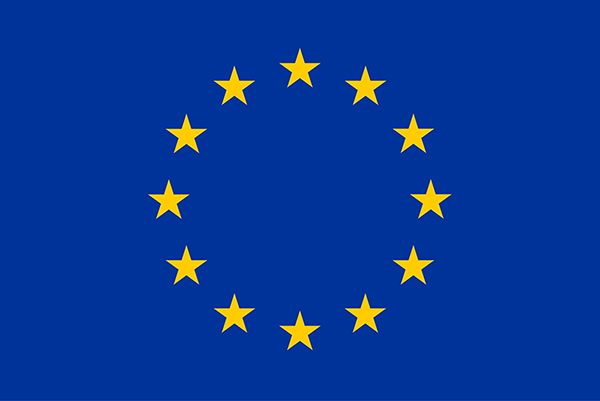}
  \end{center}
\end{wrapfigure}

\noindent
The research carried out by the international FCC collaboration hosted by CERN, which led to this publication, has received funding from the European Union's Horizon 2020 research and innovation programme under the grant numbers 951754 (FCCIS), 654305 (EuroCirCol), 764879 (EASITrain), 730871 (ARIES), 777563 (RI-Paths), 101086276 (EAJADE), 101004730 (iFAST), 101131435 (iSAS), 101131850 (RF2.0) and from FP7 under grant number 312453 (EuCARD-2).

\vspace{1em}

\begin{wrapfigure}[4]{l}[0cm]{2cm}
  \begin{center}
  \vspace{-0.85cm}
    \includegraphics[width=2cm]{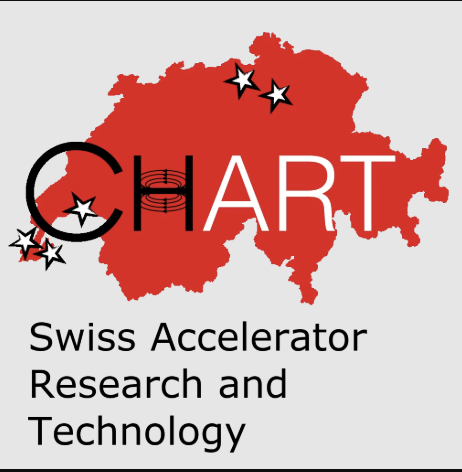}
  \end{center}
\end{wrapfigure}

\noindent This work has also benefited from the support of CHART (Swiss Accelerator Research and Technology, founded in 2016 as an umbrella collaboration for accelerator research and technology activities. Present partners in CHART are CERN, PSI, EPFL, ETH-Zurich and the University of Geneva.

\vspace*{\fill}
\noindent\textbf{Trademark notice:} All trademarks appearing in this report are acknowledged as such.\\

\newpage

\noindent This report was edited with the Overleaf.com collaborative writing and publishing system. Typesetting and final print preparation was performed using pdf\TeX 3.14159265-2.6-1.40.17\\
\\
\noindent Copyright CERN for the benefit of the FCC collaboration 2025\\
Creative Commons Attribution 4.0\\
\\
Knowledge transfer is an integral part of CERN's mission.\\
CERN publishes this volume Open Access under the Creative Commons Attribution 4.0 licence.\\ (\mbox{\url{http://creativecommons.org/licenses/by/4.0/}}) in order to permit its wide dissemination and use.
The submission of a contribution to the CERN document server shall be deemed to constitute the contributor's agreement to this copyright and license statement. Contributors are requested to obtain any clearances that may be necessary for this purpose.

\bigskip
\noindent This volume is indexed in: CERN Document Server (CDS):

\smallskip
\noindent CERN-FCC-PHYS-2025-0002\\
\noindent 10.17181/CERN.9DKX.TDH9\\
\noindent \url{https://cds.cern.ch/record/2928193}\\

\bigskip
\noindent This report edition should be cited as:

\smallskip
\noindent Future Circular Collider Feasibility Study Report Volume 1: Physics, Experiments, Detector, preprint edition edited by \mbox{M.~Benedikt et al.}, CERN physics reports,\\
\mbox{CERN-FCC-PHYS-2025-0002},\mbox{DOI 10.17181/CERN.9DKX.TDH9}, Geneva, 2025.\\
\noindent Available online: \url{https://cds.cern.ch/record/2928193}
\clearpage


\begingroup
\raggedright %

\section*{List of Editors at 31 March 2025}
M.~Benedikt$^{{1}}$~(Study Leader),
F.~Zimmermann$^{{1}}$~(Deputy Study Leader),
B.~Auchmann$^{{1, 2}}$,
W.~Bartmann$^{{1}}$,
J.P.~Burnet$^{{1}}$,
C.~Carli$^{{1}}$,
A.~Chanc{\'e}$^{{3}}$,
P.~Craievich$^{{2}}$,
M.~Giovannozzi$^{{1}}$,
C.~Grojean$^{{4, 5}}$,
J.~Gutleber$^{{1}}$,
K.~Hanke$^{{1}}$,
A.~Henriques$^{{1}}$,
P.~Janot$^{{1}}$,
C.~Louren{\c c}o$^{{1}}$,
M.~Mangano$^{{1}}$,
T.~Otto$^{{1}}$,
J.~Poole$^{{1}}$,
S.~Rajagopalan$^{{6}}$,
T.~Raubenheimer$^{{7}}$,
E.~Todesco$^{{1}}$,
L.~Ulrici$^{{1}}$,
T.~Watson$^{{1}}$,
G.~Wilkinson$^{{1,8}}$.
\section*{List of Chief Editors of Volume 1 Chapters at 31 March 2025}
P. ~Azzi$^{{9}}$,
G.~Bernardi$^{{10, 11, 12}}$,
A.~Blondel$^{{10, 13, 14}}$,
M.~Boscolo$^{{15}}$,
D.~{d'E}nterria$^{{1}}$,
M.~Dam$^{{16}}$,
J.~de~Blas$^{{17}}$,
B.~Francois$^{{1}}$,
A.~Freitas$^{{18}}$,
G.~Ganis$^{{1}}$,
J.~Keintzel$^{{1}}$,
M.~Klute$^{{19}}$,
M.~McCullough$^{{1}}$,
P.F.~Monni$^{{1}}$,
F.~Palla$^{{20}}$,
E.~Perez$^{{1}}$,
M.-A.~Pleier$^{{6}}$,
W.~Riegler$^{{1}}$,
F.~Sefkow$^{{4}}$,
M.~Selvaggi$^{{1}}$.
\section*{List of Contributors at 31 March 2025}
A.~Abada$^{{10, 21, 22}}$,
M.~Abbrescia$^{{23, 24}}$,
H.~Abdolmaleki$^{{25, 26}}$,
S.H.~Abidi$^{{6}}$,
A.~Abramov$^{{1}}$,
C.~Adam$^{{10, 27, 28}}$,
M.~Ady$^{{1}}$,
P.R.~Ad{\u z}i{\'c}$^{{29}}$,
I.~Agapov$^{{4}}$,
D.~Aguglia$^{{1}}$
% ...
\begin{itemize}
\item[${}^\dag$] deceased 
\item[$^{1}$] Switzerland - CERN, European Organization for Nuclear Research
\item[$^{2}$] Switzerland - PSI, Paul Scherrer Institute
\item[$^{3}$] France - CEA/Irfu, Commissariat {\`a} l'Energie Atomique et aux Energies Alternatives, Institut de recherche sur les lois fondamentales de l'Univers
\item[$^{4}$] Germany - DESY, Deutsches Elektronen-Synchrotron 
\item[$^{5}$] Germany - Humboldt-Universit\"at zu Berlin
\item[$^{6}$] United States - BNL, Brookhaven National Laboratory
\item[$^{7}$] United States - SLAC National Accelerator Laboratory
\item[$^{8}$] United Kingdom - University of Oxford
\item[$^{9}$] Italy - INFN, Istituto Nazionale di Fisica Nucleare, Sezione di Padova
\item[$^{10}$] France - CNRS/IN2P3, Centre National de la Recherche Scientifique, Institut National de Physique Nucl{\'e}aire et de Physique des Particules
\item[$^{11}$] France - APC, Laboratoire AstroParticule et Cosmologie
\item[$^{12}$] France - Universit{\'e} Paris Cit{\'e}
\item[$^{13}$] France - LPNHE, Laboratoire de Physique Nucl{\'e}aire et de Hautes {\'E}nergies
\item[$^{14}$] Switzerland - UNIGE, Universit{\'e} de Gen{\`e}ve
\item[$^{15}$] Italy - INFN, Istituto Nazionale di Fisica Nucleare, Laboratori Nazionali di Frascati
\item[$^{16}$] Denmark - NBI, Niels Bohr Institute
\item[$^{17}$] Spain - Universidad de Granada
\item[$^{18}$] United States - University of Pittsburgh
\item[$^{19}$] Germany - KIT, Karlsruher Institut f{\"u}r Technologie
\item[$^{20}$] Italy - INFN, Istituto Nazionale di Fisica Nucleare, Sezione di Pisa
\item[$^{21}$] France - IJCLab, Laboratoire de Physique des 2 Infinis Ir{\`e}ne Joliot Curie
\item[$^{22}$] France - Universit{\'e} Paris-Saclay et Universit{\'e} Paris-Cit{\'e}
\item[$^{23}$] Italy - INFN, Istituto Nazionale di Fisica Nucleare, Sezione di Bari
\item[$^{24}$] Italy - Universit{\`a} di Bari
\item[$^{25}$] Iran - IPM, Institute for Research in Fundamental Science
\item[$^{26}$] Iran - Malayer University
\item[$^{27}$] France - LAPP, Laboratoire d'Annecy de Physique des Particules
\item[$^{28}$] France - Universit{\'e} Savoie Mont Blanc
\item[$^{29}$] Serbia - University of Belgrade
\item[$^{30}$] Spain - ICMAB/CISC, Institut de Ci{\`e}ncia de Materials de Barcelona, Consejo Superior de Investigaciones Cient{\'\i}ificas
\item[$^{31}$] Italy - INFN, Istituto Nazionale di Fisica Nucleare, Sezione di Roma Tor Vergata
\item[$^{32}$] Italy - Universit{\`a} Roma Tor Vergata
\item[$^{33}$] T{\"u}rkiye - Yozgat Bozok {\"U}niversitesi
\item[$^{34}$] United States - Texas Tech University
\item[$^{35}$] T{\"u}rkiye - TOBB ETU, TOBB Ekonomi ve Teknoloji {\"U}niversitesi
\item[$^{36}$] Italy - INFN, Istituto Nazionale di Fisica Nucleare, Sezione di Bologna
\item[$^{37}$] Italy - Universit{\`a} di Bologna
\item[$^{38}$] Spain - CIEMAT, Centro de Investigaciones Energ{\'e}ticas, Medioambientales y Tecnol{\'o}gicas
\item[$^{39}$] Saudi Arabia - KACST, King Abdulaziz City for Science and Technology
\item[$^{40}$] Italy - Universit{\`a} Roma Tre
\item[$^{41}$] Italy - INFN, Istituto Nazionale di Fisica Nucleare, Sezione di Milano-Bicocca
\item[$^{42}$] Italy - Universit{\`a} di Milano-Bicocca
\item[$^{43}$] United Kingdom - University of Cambridge
\item[$^{44}$] T{\"u}rkiye - {\.{I}}stanbul {\"U}niversitesi
\item[$^{45}$] T{\"u}rkiye - Eski{\c s}ehir Teknik {\"U}niversitesi
\item[$^{46}$] Italy - INFN, Istituto Nazionale di Fisica Nucleare, Sezione di Napoli
\item[$^{47}$] Italy - Universit{\`a} di Napoli Federico II
\item[$^{48}$] United States - FNAL, Fermi National Accelerator Laboratory
\item[$^{49}$] Iran - University of Tehran
\item[$^{50}$] Iran- FUM, Ferdowsi University of Mashhad
\item[$^{51}$] United States - University of Texas Austin
\item[$^{52}$] France - IPHC, Institut Pluridisciplinaire Hubert Curien
\item[$^{53}$] France - Universit{\'e} de Strasbourg
\item[$^{54}$] Italy - INFN, Istituto Nazionale di Fisica Nucleare, Sezione di Milano
\item[$^{55}$] Italy - Universit{\`a} di Milano
\item[$^{56}$] Switzerland - PIBG, P{\^o}le Invert{\'e}br{\'e}s du Basin Genevois
\item[$^{57}$] Brazil - UFRN, Universidade Federal do Rio Grande do Norte
\item[$^{58}$] Switzerland - UNIBAS, University of Basel
\item[$^{59}$] Italy - Politecnico di Bari
\item[$^{60}$] Portugal - LIP, Laborat{\'o}rio de Instrumenta{\c c}{\~a}o e F{\'\i}sica Experimental de Part{\'\i}culas
\item[$^{61}$] United Kingdom - University of Manchester
\item[$^{62}$] United Kingdom - CI, Cockcroft Institute
\item[$^{63}$] United States - Brandeis University
\item[$^{64}$] Armenia - A. Alikhanyan National Laboratory
\item[$^{65}$] France - IP2I, Institut de Physique des 2 Infinis de Lyon
\item[$^{66}$] France - Universit{\'e} Claude Bernard Lyon 1
\item[$^{67}$] T{\"u}rkiye - Ankara {\"U}niversitesi
\item[$^{68}$] Sweden - Lund University
\item[$^{69}$] Italy - INFN, Istituto Nazionale di Fisica Nucleare, Sezione di Torino
\item[$^{70}$] United Arab Emirates - New York University Abu Dhabi
\item[$^{71}$] United States - Stony Brook University
\item[$^{72}$] Italy - INFN, Istituto Nazionale di Fisica Nucleare, Sezione di Perugia
\item[$^{73}$] Italy - Universit{\`a} di Perugia
\item[$^{74}$] Germany - Technische Universit{\"a}t M{\"u}nchen
\item[$^{75}$] T{\"u}rkiye - Hatay Mustafa Kemal {\"U}niversitesi
\item[$^{76}$] T{\"u}rkiye - Do{\u g}u{\c s} {\"U}niversitesi
\item[$^{77}$] People's Republic of China - Harbin Institute of Technology
\item[$^{78}$] United States - University of Wisconsin-Madison
\item[$^{79}$] United Kingdom - RAL, Rutherford Appleton Laboratory, Science and Technology Facilities Council
\item[$^{80}$] India - IMSc, Institute of Mathematical Sciences, Chennai
\item[$^{81}$] Switzerland - Universit{\"a}t Z{\"u}rich 
\item[$^{82}$] United States - University of New Mexico
\item[$^{83}$] France - CPPM, Centre de Physique des Particules de Marseille
\item[$^{84}$] France - Aix-Marseille Universit{\'e}
\item[$^{85}$] Italy - Universit{\`a} di Pisa
\item[$^{86}$] Hungary - HUN-REN Wigner Research Centre for Physics
\item[$^{87}$] United States - Catholic University of America
\item[$^{88}$] United States - Duke University
\item[$^{89}$] France - LLR, Laboratoire Leprince-Ringuet
\item[$^{90}$] France - {\'E}cole Polytechnique, Institut Polytechnique de Paris
\item[$^{91}$] Canada - TRIUMF, Canada's National Laboratory for Particle and Nuclear Physics
\item[$^{92}$] Canada - Simon Fraser University
\item[$^{93}$] Italy - FEEM, Fondazione Ente Nazionale Idrocarburi (ENI) Enrico Mattei
\item[$^{94}$] France - BRGM, Bureau de Recherches G{\'e}ologiques et Mini{\`e}res
\item[$^{95}$] T{\"u}rkiye - I{\c s}{\i}k {\"U}niversitesi
\item[$^{96}$] T{\"u}rkiye - {\.{I}}stanbul Teknik {\"U}niversitesi
\item[$^{97}$] Switzerland - UNIBE, University of Bern
\item[$^{98}$] Italy - Universit{\`a} di Roma la Sapienza
\item[$^{99}$] Italy - CNR-SPIN, Consiglio Nazionale delle Ricerche
\item[$^{100}$] France - Expert naturaliste et entomologiste
\item[$^{101}$] United States - ORNL, Oak Ridge National Laboratory
\item[$^{102}$] Austria - HEPHY, Institut f{\"u}r Hochenergiephysik
\item[$^{103}$] Switzerland - Geos, Bureau d'ing{\'e}nieurs conseils en g{\'e}otechnique, g{\'e}nie civil, hydraulique et environnement
\item[$^{104}$] Austria - TUWIEN, Technische Universit{\"a}t Wien
\item[$^{105}$] Switzerland - HEPIA, Haute {\'E}cole du Paysage, d'Ing{\'e}nierie et d'Architecture de Gen{\`e}ve
\item[$^{106}$] Switzerland - HES-SO University of Applied Sciences and Arts Western Switzerland
\item[$^{107}$] France - SETEC ALS, Soci{\'e}t{\'e} d'ing{\'e}nierie conseil en infrastructures de transport, g{\'e}nie civil et environnement
\item[$^{108}$] United States - East Texas A\&M University
\item[$^{109}$] Switzerland - Amberg Engineering Ltd
\item[$^{110}$] Switzerland - ECOTEC Environnement SA, Bureau d'{\'e}tudes et de conseil en environnement
\item[$^{111}$] United States - Southern Methodist University
\item[$^{112}$] Poland - IFJ PAN, Institute of Nuclear Physics, Polish Academy of Sciences
\item[$^{113}$] France - Cerema, {\'e}tablissement public pour l'{\'e}laboration, le d{\'e}ploiement et l'{\'e}valuation de politiques publiques d'am{\'e}nagement et de transport
\item[$^{114}$] United Kingdom - Rendel Ltd, Engineering design consultancy firm
\item[$^{115}$] Italy - INFN, Istituto Nazionale di Fisica Nucleare, Sezione di Roma Tre
\item[$^{116}$] T{\"u}rkiye - {\.{I}}stanbul Beykent {\"U}niversitesi
\item[$^{117}$] United States - University of Iowa
\item[$^{118}$] India - Indian Institute of Technology Kanpur
\item[$^{119}$] Switzerland - EPFL, {\'E}cole Polytechnique F{\'e}d{\'e}rale de Lausanne
\item[$^{120}$] Germany - Universit{\"a}t Hamburg, Fakult{\"a}t f{\"u}r Mathematik, Informatik und Naturwissenschaften
\item[$^{121}$] Belgium - VUB, Vrije Universiteit Brussel 
\item[$^{122}$] Italy - Scuola Superiore Meridionale
\item[$^{123}$] Affiliated with an institute formerly covered by a cooperation agreement with CERN
\item[$^{124}$] Canada - University of Saskatchewan and the Canadian Light Source
\item[$^{125}$] United Kingdom - University of Bristol
\item[$^{126}$] United States - Northwestern University
\item[$^{127}$] United States - University of Florida
\item[$^{128}$] Canada - York University
\item[$^{129}$] Italy - INFN, Istituto Nazionale di Fisica Nucleare, Sezione di Cagliari
\item[$^{130}$] Italy - Universit{\`a} di Cagliari
\item[$^{131}$] Italy - INFN, Istituto Nazionale di Fisica Nucleare, Sezione di Pavia
\item[$^{132}$] Canada - Queen's University
\item[$^{133}$] Portugal - CFTP-IST, Centro de F{\'\i}sica T{\'e}orica de Part{\'\i}culas, Instituto Superior Tecnico, Universidade de Lisboa
\item[$^{134}$] Sweden - Uppsala University
\item[$^{135}$] Germany - MPP, Max-Planck-Institut f{\"u}r Physik Garching
\item[$^{136}$] Ukraine - NSC KIPT, National Science Center Kharkiv Institute of Physics and Technology
\item[$^{137}$] Italy - Universit{\`a} degli Studi dell'Insubria
\item[$^{138}$] Germany - Albert-Ludwigs-Universit{\"a}t Freiburg
\item[$^{139}$] United Kingdom - JAI, John Adams Institute for Accelerator Science, University of Oxford
\item[$^{140}$] France - CNRS/INP, Centre National de la Recherche Scientifique, Institut de Physique
\item[$^{141}$] France - LPTHE, Laboratoire de Physique Th{\'e}orique et Hautes Energies
\item[$^{142}$] France - Sorbonne Universit{\'e}
\item[$^{143}$] Italy - INFN, Istituto Nazionale di Fisica Nucleare, Sezione di Genova
\item[$^{144}$] United States - LBNL, Lawrence Berkeley National Laboratory
\item[$^{145}$] Italy - IIT, Instituto Italiano di Tecnologia
\item[$^{146}$] T{\"u}rkiye - G{\"u}m{\"u}{\c s}hane {\"U}niversitesi
\item[$^{147}$] Switzerland - WSP Ing{\'e}nieurs Conseils SA
\item[$^{148}$] Mexico - UADY, Autonomous University of Yucatan
\item[$^{149}$] Italy - INFN, Istituto Nazionale di Fisica Nucleare, Sezione di Ferrara
\item[$^{150}$] Italy - Universit{\`a} di Ferrara
\item[$^{151}$] France - MARCELEON, Cabinet d'ing{\'e}nierie juridique et fonci{\`e}re
\item[$^{152}$] Italy - CSIL (Economic Research Institute)
\item[$^{153}$] Switzerland - ETHZ, Swiss Federal Institute of Technology Zurich
\item[$^{154}$] Spain - UAH, Universidad de Alcal\'a Madrid
\item[$^{155}$] France - CIA, Conseil Ing{\'e}nierie Acoustique
\item[$^{156}$] France - Evinerude, Bureau d'{\'e}tudes environnementales
\item[$^{157}$] France - LPCA, Laboratoire de Physique de Clermont Auvergne
\item[$^{158}$] France - Universit{\'e} Clermont Auvergne
\item[$^{159}$] Australia - ANSTO, Australian Synchrotron
\item[$^{160}$] India - Brahmananda Keshab Chandra College
\item[$^{161}$] United States - ANL, Argonne National Laboratory
\item[$^{162}$] Italy - INFN, Istituto Nazionale di Fisica Nucleare, Sezione di Lecce
\item[$^{163}$] Italy - Universit{\`a} di Palermo
\item[$^{164}$] Poland - Wroc{\l}aw University of Science and Technology
\item[$^{165}$] United States - Rutgers University
\item[$^{166}$] United States - Princeton University
\item[$^{167}$] Italy - Universit{\`a} di Padova
\item[$^{168}$] T{\"u}rkiye - IUE, {\.{I}}zmir Ekonomi {\"U}niversitesi
\item[$^{169}$] T{\"u}rkiye - Ege {\"U}niversitesi
\item[$^{170}$] Italy - INFN, Istituto Nazionale di Fisica Nucleare, Gruppo Collegato di Udine
\item[$^{171}$] Italy - Universit{\`a} di Udine
\item[$^{172}$] United States - University of Massachusetts Amherst
\item[$^{173}$] Italy - INFN, Istituto Nazionale di Fisica Nucleare, Sezione di Roma
\item[$^{174}$] Spain - CELLS/ALBA, Consortium for the Construction, Equipment and Exploitation of the Synchrotron Light Laboratory
\item[$^{175}$] France - LPSC, Laboratoire de Physique Subatomique et de Cosmologie
\item[$^{176}$] France - Universit{\'e} Grenoble Alpes
\item[$^{177}$] Italy - Universit{\`a} degli Studi di Napoli Parthenope
\item[$^{178}$] United States - National High Magnetic Field Laboratory
\item[$^{179}$] United States - Florida State University
\item[$^{180}$] South Africa - University of Johannesburg
\item[$^{181}$] Spain - IGFAE, Instituto Galego de Fisica de Altas Enerx{\'\i}as, Universidade de Santiago de Compostela
\item[$^{182}$] United Kingdom - LSE, London School of Economics
\item[$^{183}$] Spain - Universidade de Santiago de Compostela
\item[$^{184}$] United Kingdom - Imperial College London
\item[$^{185}$] United States - MIT, Massachusetts Institute of Technology
\item[$^{186}$] France - CETU, Centre d'Etude des Tunnels
\item[$^{187}$] United Kingdom - University of Liverpool
\item[$^{188}$] Switzerland - Linde Kryotechnik AG
\item[$^{189}$] Switzerland - ILF Consulting Engineers
\item[$^{190}$] Japan - Hokkaido University
\item[$^{191}$] Czech Republic - CUNI, Charles University
\item[$^{192}$] Italy - INFN, Istituto Nazionale di Fisica Nucleare, Sezione di Firenze
\item[$^{193}$] Malta - University of Malta
\item[$^{194}$] United States - BU, Boston University
\item[$^{195}$] T{\"u}rkiye - IBU, Bolu Abant {\.{I}}zzet Baysal {\"U}niversitesi
\item[$^{196}$] Germany - Julius-Maximilians-Universit{\"a}t W{\"u}rzburg
\item[$^{197}$] Australia - University of Adelaide
\item[$^{198}$] Finland - HIP, Helsinki Institute of Physics, University of Helsinki
\item[$^{199}$] Belgium - ULB, Universit{\'e} Libre de Bruxelles
\item[$^{200}$] Germany - IMA, Institut f{\"u}r Maschinenelemente, Universit{\"a}t Stuttgart
\item[$^{201}$] Belgium - CP3, Centre de Cosmologie, de Physique des Particules et de Ph{\'e}nom{\'e}nologie, Universit{\'e} Catholique de Louvain
\item[$^{202}$] Netherlands - NIKHEF, Nationaal instituut voor subatomaire fysica
\item[$^{203}$] People's Republic of China - IHEP, Chinese Academy of Sciences
\item[$^{204}$] Mexico - UAS, Universidad Aut{\'o}noma de Sinaloa
\item[$^{205}$] Austria - BOKU, Universit{\"a}t f{\"u}r Bodenkultur Wien
\item[$^{206}$] United States - Purdue University
\item[$^{207}$] India - University of Delhi
\item[$^{208}$] France - Ginger BURGEAP, bureau d'{\'e}tudes en environnement
\item[$^{209}$] United Kingdom - King's College London
\item[$^{210}$] United States - University of Maryland
\item[$^{211}$] Japan - KEK, High Energy Accelerator Research Organization
\item[$^{212}$] Switzerland - Geoenergy, Reservoir Geology and Basin Analysis Group
\item[$^{213}$] France - SETEC International, Soci{\'e}t{\'e} d'ing{\'e}nierie en charge des transports et des infrastructures
\item[$^{214}$] T{\"u}rkiye - KKU, K{\i}r{\i}kkale {\"U}niversitesi
\item[$^{215}$] France - SETEC LERM, Soci{\'e}t{\'e} d'ing{\'e}nierie conseil en mat{\'e}riaux de construction
\item[$^{216}$] Mexico - BUAP, Benem{\'e}rita Universidad Aut{\'o}noma de Puebla
\item[$^{217}$] United States - Stanford University
\item[$^{218}$] Austria - MUL, Montanuniversit{\"a}t Leoben, Lehrstuhl f{\"u}r Subsurface Engineering, Geotechnik und unterirdisches Bauen
\item[$^{219}$] Austria - MUL-ZaB, Underground Research Center, Zentrum am Berg
\item[$^{220}$] Spain - IFAE, Institut de F{\'\i}sica d'Altes Energies
\item[$^{221}$] Switzerland - FHNW, University of Applied Sciences Northwestern Switzerland
\item[$^{222}$] India - UPES, University of Petroleum and Energy Studies
\item[$^{223}$] France - GANIL, Grand Acc{\'e}l{\'e}rateur National d'Ions Lourds
\item[$^{224}$] France - Universit{\'e} Caen Normandie
\item[$^{225}$] Brazil - UFRGS, Universidade Federal do Rio Grande do Sul
\item[$^{226}$] Germany - Technische Universit{\"a}t Darmstadt
\item[$^{227}$] Poland - University of Silesia in Katowice
\item[$^{228}$] Portugal - Universidade de Coimbra
\item[$^{229}$] Brazil - UFPel, Universidade Federal de Pelotas
\item[$^{230}$] United States - University of California Santa Cruz
\item[$^{231}$] Italy - Universit{\`a} del Salento
\item[$^{232}$] United States - Brown University
\item[$^{233}$] United States - University of California Berkeley
\item[$^{234}$] Italy - Universit{\`a} di Torino
\item[$^{235}$] United States - Johns Hopkins University
\item[$^{236}$] People's Republic of China - Fudan University
\item[$^{237}$] France - LAPTh, Laboratoire d'Annecy-le-Vieux de Physique Th{\'e}orique
\item[$^{238}$] People's Republic of China - Dongguan University of Technology
\item[$^{239}$] T{\"u}rkiye - Kahramanmara{\c s} S{\"u}t{\c c}{\"u} {\.{I}}mam {\"U}niversitesi
\item[$^{240}$] Mexico - UNACH, Universidad Aut{\'o}noma de Chiapas
\item[$^{241}$] Mexico - MCTP, Mesoamerican Centre for Theoretical Physics
\item[$^{242}$] Mexico - UAZ, Universidad Aut\'onoma de Zacatecas
\item[$^{243}$] Finland - University of Jyv{\"a}skyl{\"a}
\item[$^{244}$] Germany - Technische Universit{\"a}t Dortmund
\item[$^{245}$] United Kingdom - Overleaf
\item[$^{246}$] United States - University of Virginia
\item[$^{247}$] United States - Cornell University
\item[$^{248}$] United States - FIT, Florida Institute of Technology
\item[$^{249}$] Switzerland - Shirokuma GmbH
\item[$^{250}$] Pakistan - National Centre for Physics
\item[$^{251}$] Germany - Johannes Gutenberg Universit{\"a}t Mainz
\item[$^{252}$] Pakistan - PAEC, Pakistan Atomic Energy Commission
\item[$^{253}$] India - Mathabhanga College
\item[$^{254}$] India - Harish-Chandra Research Institute
\item[$^{255}$] Germany - MPIK, Max-Planck-Institut f{\"u}r Kernphysik Heidelberg
\item[$^{256}$] Sweden - European Spallation Source ERIC
\item[$^{257}$] France - Centre de calcul de l'IN2P3
\item[$^{258}$] Germany - GSI, Helmholtzzentrum f{\"u}r Schwerionenforschung GmbH
\item[$^{259}$] Republic of Korea - IBS, Institute for Basic Science, Center for Theoretical Physics of the Universe
\item[$^{260}$] Poland - University of Warsaw
\item[$^{261}$] Slovenia - University of Ljubljana
\item[$^{262}$] Slovenia - Jozef Stefan Institute
\item[$^{263}$] France - Microhumus, Bureau d'{\'e}tude et d'ing{\'e}nierie sp{\'e}cialis{\'e} dans la gestion des sols d{\'e}grad{\'e}s
\item[$^{264}$] Germany - Fakult{\"a}t f{\"u}r Physik und Astronomie, Universit{\"a}t Heidelberg
\item[$^{265}$] T{\"u}rkiye - Ni{\u g}de {\"O}mer Halisdemir {\"U}niversitesi
\item[$^{266}$] T{\"u}rkiye - Giresun {\"U}niversitesi
\item[$^{267}$] Hungary - University of Miskolc
\item[$^{268}$] Greece - NTUA, National Technical University of Athens
\item[$^{269}$] Switzerland - Transmutex SA
\item[$^{270}$] Ireland - DIAS, Dublin Institute for Advanced Studies, School of Theoretical Physics
\item[$^{271}$] Poland - AGH, University of Science and Technology
\item[$^{272}$] Iran - University of Science and Technology of Mazandaran
\item[$^{273}$] United Kingdom - IPPP, Institute for Particle Physics Phenomenology, Durham University
\item[$^{274}$] T{\"u}rkiye - Bursa Uluda{\u g} {\"U}niversitesi
\item[$^{275}$] Republic of Korea - YU, Yonsei University
\item[$^{276}$] Affiliated with an international laboratory covered by a cooperation agreement with CERN
\item[$^{277}$] France - CPT, Centre de Physique Th\'eorique
\item[$^{278}$] France - Aix-Marseille Universit{\'e} et Universit{\'e} du Sud Toulon Var
\item[$^{279}$] Republic of Korea - KIAS, Korea Institute for Advanced Study
\item[$^{280}$] Canada - Carleton University
\item[$^{281}$] Greece - FEAC~Engineering~P.C.
\item[$^{282}$] Greece - UPATRAS, University of Patras
\item[$^{283}$] United States - University of Kansas
\item[$^{284}$] Greece - AUTH, Aristotle University of Thessaloniki
\item[$^{285}$] Germany - IML, Fraunhofer-Institut f{\"u}r Materialfluss und Logistik 
\item[$^{286}$] Germany - RWTH~Aachen, Rheinisch-Westf{\"a}lische Technische Hochschule Aachen
\item[$^{287}$] Switzerland - ZHAW, Zurich University of Applied Sciences
\item[$^{288}$] Gernany - Universit{\"a}t M{\"u}nster
\item[$^{289}$] South Africa - University of the Witwatersrand
\item[$^{290}$] Austria - Fachhochschule Technikum Wien
\item[$^{291}$] United States - University of California Irvine
\item[$^{292}$] United States - Northeastern University
\item[$^{293}$] France - ESI, European Scientific Institute
\item[$^{294}$] Republic of Korea - UOS, University of Seoul
\item[$^{295}$] Republic of Korea - KNU Kyungpook National University
\item[$^{296}$] Republic of Korea - KU, Korea University
\item[$^{297}$] Finland - Tampere University
\item[$^{298}$] United Kingdom - University of Edinburgh
\item[$^{299}$] Switzerland - BG Ing{\'e}nieurs Conseils
\item[$^{300}$] People's Republic of China - T.-D.~Lee Institute
\item[$^{301}$] People's Republic of China - Shanghai Jiao Tong University
\item[$^{302}$] Italy - CNR, Consiglio Nazionale delle Ricerche
\item[$^{303}$] United States - University of Michigan
\item[$^{304}$] United States - University of Pennsylvania
\item[$^{305}$] United States - University of Minnesota
\item[$^{306}$] France - ESRF, European Synchrotron Radiation Facility
\item[$^{307}$] United Kingdom - SUSSEX, University of Sussex
\item[$^{308}$] Italy - Universit{\`a} di Bari Aldo Moro
\item[$^{309}$] United Kingdom - University of Bath
\item[$^{310}$] Italy - Scuola Normale Superiore di Pisa
\item[$^{311}$] Brazil - Universidade de S{\~a}o Paulo
\item[$^{312}$] Austria - Universit{\"a}t Graz
\item[$^{313}$] Egypt - Center for High Energy Physics, Fayoum University
\item[$^{314}$] Egypt - Center of theoretical physics, British University in Egypt
\item[$^{315}$] Egypt - Cairo University
\item[$^{316}$] France - Sorbonne Universit{\'e} et Universit{\'e} Paris Cit\'e
\item[$^{317}$] Italy - Universit{\`a} di Genova
\item[$^{318}$] Spain - IFIC-CSIC/UV, Instituto de F\'{i}sica Corpuscular, Consejo Superior de Investigaciones Cient{\'\i}ficas/Universidad de Valencia
\item[$^{319}$] Switzerland - Service de g{\'e}ologie, sols et d{\'e}chets du canton de Gen{\`e}ve
\item[$^{320}$] Estonia - NICPB, National Institute for Chemical Physics and Biophysics
\item[$^{321}$] Estonia - UT, University of Tartu
\item[$^{322}$] Spain - Universidad de Salamanca
\item[$^{323}$] Mexico - UGTO, Universidad de Guanajuato
\item[$^{324}$] Switzerland - Edaphos engineering
\item[$^{325}$] Germany - Helmholtz-Zentrum Dresden-Rossendorf
\item[$^{326}$] India - Tata Institute of Fundamental Research Mumbai
\item[$^{327}$] Belgium - Universiteit Gent
\item[$^{328}$] Italy - CNR-IOM, Consiglio Nazionale delle Ricerche
\item[$^{329}$] Austria - JKU, Johannes Kepler Universit{\"a}t Linz
\item[$^{330}$] Norway - University of Stavanger
\item[$^{331}$] Colombia - Universidad Nacional de Colombia
\item[$^{332}$] Italy - Trento Institute for Fundamental Physics and Applications
\item[$^{333}$] United States - Caltech, California Institute of Technology
\item[$^{334}$] Italy - Universit{\`a} dalla Calabria
\item[$^{335}$] Spain - IFT, Instituto de F{\'\i}sica Te{\'o}rica, Universidad Aut{\'o}noma de Madrid
\item[$^{336}$] Spain - UPC, Universitat Polit\`{e}cnica de Catalunya
\item[$^{337}$] Portugal - Departamento de F{\'\i}sica, Universidade do Minho
\item[$^{338}$] Portugal - Centro de F{\'\i}sica das Universidades do Minho e do Porto
\item[$^{339}$] Portugal - LaPMET, Laboratory of Physics for Materials and Emergent Technologies
\item[$^{340}$] Norway - University of Bergen
\item[$^{341}$] Chile - SAPHIR, Instituto Milenio de F{\'\i}sica Subat{\'o}mica en la Frontera de Altas Energ{\'\i}as
\item[$^{342}$] Chile - Universidad Andres Bello
\item[$^{343}$] Brazil - IIP, International Institute of Physics
\item[$^{344}$] Japan - Tokyo International University
\item[$^{345}$] T{\"u}rkiye - {\.{I}}zmir Bak{\i }r{\c c}ay {\"U}niversitesi
\item[$^{346}$] Italy - INFN, Istituto Nazionale di Fisica Nucleare, Laboratori Nazionali del Gran Sasso
\item[$^{347}$] Italy - Universit{\'a} degli Studi di Cassino e del Lazio Meridionale
\item[$^{348}$] Serbia - Vin{\u{c}}a Institute of Nuclear Sciences
\item[$^{349}$] United States - Kennesaw State University
\item[$^{350}$] Italy - Universit{\`a} di Pavia
\item[$^{351}$] United States - Columbia University
\item[$^{352}$] United States - DOE, Department of Energy of the United States of America
\item[$^{353}$] France - IPSA, Institut Polytechnique des Sciences Avanc{\'e}es 
\item[$^{354}$] Italy - Universit{\`a} degli Studi del Sannio
\item[$^{355}$] Poland - UJ, Jagiellonian University
\item[$^{356}$] Austria - Universit{\"a}t Wien
\item[$^{357}$] Germany - Goethe-Universit{\"a}t Frankfurt, Institut f{\"u}r Angewandte Physik
\item[$^{358}$] Germany - HFFH, Helmholtz Forschungsakademie Hessen f{\"u}r FAIR
\item[$^{359}$] Romania - INCDTIM, National Institute for Research and Development of Isotopic and Molecular Technologies
\item[$^{360}$] United States - University of Arizona
\item[$^{361}$] Germany - Technische Universit{\"a}t Dresden
\item[$^{362}$] Latvia - RTU, Riga Technical University
\item[$^{363}$] United Kingdom - Lancaster University
\item[$^{364}$] Cyprus - University of Cyprus
\item[$^{365}$] Cyprus - Cosmos Open University
\item[$^{366}$] Brazil - CBPF, Centro Brasileiro de Pesquisas F{\'\i}sicas
\item[$^{367}$] Germany - Universit{\"a}t Bonn
\item[$^{368}$] Thailand - CMU, Chiang Mai University
\item[$^{369}$] United States - JLAB, Thomas Jefferson National Accelerator Facility
\item[$^{370}$] Ecuador - ESPOL, Escuela Superior Polit{\'e}cnica del Litoral
\item[$^{371}$] Croatia - IRB, Rudjer Boskovic Institute
\item[$^{372}$] France - Air Liquide Advanced Technologies
\item[$^{373}$] Netherlands - VU Amsterdam
\item[$^{374}$] France - ING{\'E}ROP ,Groupe d'ing{\'e}nierie et de conseil en mobilit{\'e} durable, transition {\'e}nerg{\'e}tique et cadre de vie
\item[$^{375}$] Poland - Warsaw University of Technology
\item[$^{376}$] Spain - IFCA, Instituto de F{\'\i}sica de Cantabria
\item[$^{377}$] Germany - Institut f{\"u}r Beschleunigerphysik und Technologie
\item[$^{378}$] T{\"u}rkiye - U{\c s}ak {\"U}niversitesi
\item[$^{379}$] Japan - ICEPP, International Center for Elementary Particle Physics, University of Tokyo
\item[$^{380}$] United Kingdom - Rudolf Peierls Centre for Theoretical Physics, University of Oxford
\item[$^{381}$] United Kingdom - All Souls College, University of Oxford
\item[$^{382}$] T{\"u}rkiye - Akdeniz {\"U}niversitesi
\item[$^{383}$] Switzerland - Latitude Durable SARL
\item[$^{384}$] Spain - USAL, Universidad de Salamanca
\item[$^{385}$] Germany - {PRISMA+} Cluster of Excellence
\item[$^{386}$] United States - Michigan State University
\item[$^{387}$] Spain - Universidad Complutense Madrid
\item[$^{388}$] Switzerland - scMetrology SARL
\item[$^{389}$] Portugal - IST, Instituto Superior Tecnico, Universidade de Lisboa
\item[$^{390}$] Portugal - CeFEMA, Center of Physics and Engineering of Advanced Materials
\item[$^{391}$] India - Indian Institute of Science Education and Research Mohali
\item[$^{392}$] United States - NIU, Northern Illinois University
\item[$^{393}$] India - Banaras Hindu University
\item[$^{394}$] Slovakia - Comenius University
\item[$^{395}$] Australia - Monash University
\item[$^{396}$] Slovakia - Slovak Academy of Sciences
\item[$^{397}$] Republic of Korea - KAIST, Korea Advanced Institute of Science and Technology
\item[$^{398}$] France - Amberg Engineering Chamb{\'e}ry
\item[$^{399}$] United Arab Emirates - Khalifa University of Science and Technology
\item[$^{400}$] United States - University of Tennessee
\item[$^{401}$] Austria - WIFO, {\"O}sterreichisches Institut f{\"u}r Wirtschaftsforschung
\item[$^{402}$] Brazil - Universidade do Estado do Rio de Janeiro
\item[$^{403}$] France - M{\'e}lica, NATURA SCOP, {\'E}tudes et expertises environnementales
\item[$^{404}$] Italy - Universit{\`a} LUM, Casamassima
\item[$^{405}$] Netherlands - University of Twente
\item[$^{406}$] Iran - Arak University
\item[$^{407}$] Italy - Universit{\`a} di Trieste
\item[$^{408}$] France - ForestAllia, Cabinet de gestion et d'expertise foresti{\`e}res
\item[$^{409}$] Germany - Universit{\"a}t Siegen
\item[$^{410}$] United States - University of Oregon
\item[$^{411}$] Germany - Universit{\"a}t Rostock
\item[$^{412}$] Switzerland - CEGELEC SA
\item[$^{413}$] Sweden - KTH, Royal Institute of Technology, Stockholm
\item[$^{414}$] Sweden - OKC, Oskar Klein Centre for Cosmoparticle Physics
\item[$^{415}$] United States - Brigham Young University
\item[$^{416}$] France - Expert foncier et agricole
\item[$^{417}$] Germany - ITSM, Institut f{\"u}r Thermische Str{\"o}mungsmaschinen und Maschinenlaboratorium, Universit{\"a}t Stuttgart
\item[$^{418}$] France - {\'E}cole Normale Sup{\'e}rieure de Lyon
\item[$^{419}$] Czech Republic - CTU, Czech Technical University
\item[$^{420}$] United States - University of Chicago
\item[$^{421}$] United States - Baylor University
\item[$^{422}$] United Kingdom - University of Birmingham
\item[$^{423}$] United Kingdom - University of Southampton
\item[$^{424}$] Switzerland - Swisstopo, Federal Office of Topography
\item[$^{425}$] United Kingdom - Daresbury Laboratory, Science and Technology Facilities Council
\item[$^{426}$] T{\"u}rkiye - IZTECH, {\.{I}}zmir Y{\"u}ksek Teknoloji Enstit{\"u}s{\"u}
\item[$^{427}$] Hong Kong - City University of Hong Kong
\end{itemize}

\endgroup

\clearpage

\section*{Abstract}
Volume 1 of the FCC Feasibility Report presents an overview of the physics case, experimental programme, and detector concepts for the Future Circular Collider (FCC). This volume outlines how  FCC would address some of the most profound open questions in particle physics, from precision studies of the Higgs and EW bosons and of the top quark, to the exploration of physics beyond the Standard Model. The report reviews the experimental opportunities offered by the staged implementation of  FCC, beginning with an electron-positron collider (FCC-ee), operating at several centre-of-mass energies, followed by a hadron collider (FCC-hh). Benchmark examples are given of the expected physics performance, in terms of precision and sensitivity to new phenomena, of each collider stage. Detector requirements and conceptual designs for FCC-ee experiments are discussed, as are the specific demands that the physics programme imposes on the accelerator in the domains of the calibration of the collision energy, and the interface region between the accelerator and the detector. The report also highlights advances in detector, software and computing technologies, as well as the theoretical tools /reconstruction techniques that will enable the precision measurements and discovery potential of the FCC experimental programme. The content and structure of this report are guided by the scope and priorities defined in the mandate of the FCC Feasibility Study. It is therefore not intended to serve as an exhaustive review of the full physics potential of FCC. Several topics, already covered in earlier reports such as the FCC CDR, are not reiterated here or are addressed only briefly, in alignment with the study’s focus. This volume reflects the outcome of a global collaborative effort involving hundreds of scientists and institutions, aided by a dedicated community-building coordination, and provides a targeted assessment of the scientific opportunities and experimental foundations of the FCC programme.

\clearpage

\section*{Preface from CERN's Director-General}

In 2021, in response to the 2020 update of the European Strategy for Particle Physics, the CERN Council initiated the Future Circular Collider (FCC) Feasibility Study. 

This report summarises an immense amount of work carried out by the international FCC collaboration over several years. It covers, inter alia, physics objectives and potential, geology, civil engineering, technical infrastructure, territorial implementation, environmental aspects, R\&D needs for the accelerators and detectors, socio-economic benefits and cost. It constitutes important input for the ongoing update of the European Strategy for Particle Physics.

The Feasibility Study required engagement with a broad range of stakeholders. In particular, throughout the Study, CERN has been accompanied by its two Host States, France and Switzerland, and has been working with entities at local, regional and national level. I am very grateful to the Host State authorities and teams for their invaluable help. Furthermore, significant sections of the Study were supported by the European Union under the Horizon 2020 and Horizon Europe framework programmes. The Study also greatly benefited from contributions from accelerator laboratories and universities from across Europe, such as the Swiss Accelerator Research and Technology (CHART) initiative, and from the Americas, Asia, Africa and Australia. 

The proposed FCC integrated programme consists of two possible stages: an electron–positron collider serving as a Higgs-boson, electroweak and top-quark factory running at different centre-of-mass energies, followed at a later stage by a proton–proton collider operating at an unprecedented collision energy of around 100 TeV. The complementary physics programmes of each stage match the physics priorities expressed in the 2020 update of the European Strategy for Particle Physics. 

A major achievement of the Feasibility Study is the choice of placement of the collider ring and the entire infrastructure, including the surface sites and the access shafts, which was developed and optimised over several years following the principle “avoid, reduce, compensate”. Sustainability studies have assessed energy efficiency, land use, water and resource management, and socio-economic impact, ensuring that the FCC is designed in accordance with the latest environmental and societal standards. 

I would like to thank all contributors to this report for their hard work and commitment, which allowed the outstanding results presented here to be achieved.

\begin{flushright}
\textbf{Fabiola Gianotti}\\
CERN, Director-General
\end{flushright}

\clearpage

\section*{Preface from the FCC Collaboration Board Chair}

Building on the earlier Future Circular Collider (FCC) Conceptual Design Study conducted between 2014 and 2018, the FCC Feasibility Study (2021–2025) has been undertaken by a robust international collaboration, now comprising over 160 institutes worldwide.  The FCC “integrated programme”, developed in the framework of the Feasibility Study, consists of an initial electron-positron collider, the FCC-ee, which could be followed by a proton-proton collider, the FCC-hh. This staging takes into account the physics priorities as formulated in the updates of the European Strategy for Particle Physics of 2012 and 2020, as well as the relative technology readiness and costs of the FCC-ee and FCC-hh.

Over the years, I have closely followed the steady progress of the study, representing the FCC collaboration at the international steering committee and participating in annual FCC Week meetings, which include sessions of the International Collaboration Board. The commitment and enthusiasm of the members of the collaboration has always been impressive. The collective effort is clearly visible. Participation by students and early-career researchers is increasing. There is a shared determination and momentum to move forward.

The strong international collaboration around the FCC and its global network provide a solid foundation for the future of this project. The FCC community continues to grow, with increasing engagement from new institutes and partners worldwide. This broad support will be essential as the project enters its next phase.

The FCC Feasibility Study demonstrates not only the technical viability of the project, but also the strength of the international community that supports it. As we move towards the next step in the decision-making phase, this collective effort is key to showing a possible path forward. The FCC promises far-reaching scientific opportunities and long-term benefits for innovation, training, and global collaboration in science and technology.

\begin{flushright}
\textbf{Philippe Chomaz}\\
CEA, Chair of the FCC International Collaboration Board
\end{flushright}

\clearpage


\tableofcontents

\cleardoublepage
\pagenumbering{arabic}
\setcounter{page}{1}

\section{Overview}
\label{sec:overview}
The particle physics landscape has been profoundly influenced by the discovery of a Higgs boson with a mass around 125\,GeV at the LHC~\cite{CMS:2012qbp,ATLAS:2012yve}.
The long-predicted matrix of particles and interactions of the Standard Model (SM) is now complete, and this consistent and predictive theory has so far been successful at  describing all phenomena accessible to collider experiments.
After almost 15 years of LHC operation, the remarkable precision measurements and the many exploratory searches have demonstrated its validity and excluded signs of new physics across an order of magnitude of energies around the TeV scale.
This notwithstanding, several fundamental experimental facts remain unexplained in the current framework, such as the abundance of matter over antimatter, the evidence for dark matter, or the non-zero neutrino masses, and many theoretical issues definitively also require physics beyond the present Standard Model (BSM), altogether calling for intensified collider exploration. For the first time since the Fermi theory, however, the field of possible explanations to these unanswered questions provides little guidance for what form this exploration may take.

The confirmation by the LHC of the SM predictions up to the TeV range requires a different approach to addressing these open questions. 
Solutions could exist at even higher energy, at the price of either an \emph{unnatural} value of the weak scale or an ingenious but still elusive structure. 
Instead, radically new physics scenarios have recently been devised, which  often include light and feebly coupled structures~\cite{Graham:2015cka, Espinosa:2015eda, Arkani-Hamed:2020yna}.
Neither the mass scale (from meV to ZeV) of this new physics nor the intensity of the couplings to the SM (from 1 to $10^{-12}$ or less) are known, thus calling for a new, broad, and powerful tool of exploration. 
To experimentally push the limits of the unknown as far as possible with a real chance of discovery, this tool must be able to address the following goals in the broadest and deepest possible way.

\begin{itemize}
    \item Map the properties of the Higgs and electroweak (EW) gauge bosons, pinning down their interactions with accuracies order(s) of magnitude better than today, and acquiring
    sensitivity to, e.g., the processes that led to the formation of today's Higgs vacuum field during the time span between $10^{-12}$ and $10^{-10}$\,s after the Big Bang.
    \item Sharpen our knowledge of already identified particle physics phenomena 
    with a comprehensive and accurate campaign of precision electroweak, QCD, flavour, Higgs, and top measurements, sensitive to tiny deviations from the predicted Standard Model behaviour and probing energy scales far beyond the direct kinematic reach. 
    Such a campaign requires running at the intensity frontier with large collected event samples, exquisitely precise experimental conditions and theoretical calculations, as well as a maximal amount of synergies within the programme. 
    \item Improve by orders of magnitude the sensitivity to rare and elusive phenomena at low energies, including the possible discovery of light particles with very small couplings (e.g., massive neutrinos, and/or axion-like particles). 
    In particular, the search for dark matter should seek to reveal, or conclusively exclude, dark sector  candidates belonging to broad classes of models.
    \item Improve, by at least an order of magnitude, the direct discovery reach for new particles at the energy frontier.
\end{itemize}

The entirely new context that led to these ambitious objectives calls for the highest integrated luminosities at the electroweak and Higgs scales, and for parton-parton collision centre-of-mass energies an order of magnitude above the TeV scale, already extensively probed by the LHC (and further so by the HL-LHC) direct searches.
The 2021 update of the European Strategy for Particle Physics (ESPPU)~\cite{esppu} and the U.S.\ Community Study on the Future of Particle Physics (Snowmass~'21)~\cite{Butler:2023glv} now 
require an $\Pep\Pem$ Higgs factory with the highest priority, to complete and deepen the trailblazing Higgs boson measurements performed at the LHC and HL-LHC. 
In addition, the 2021 ESPPU and the 2023 U.S.\ P5 report have both highlighted the long-term vision~\cite{esppu,P5:2023wyd} to operate a 10\,TeV parton centre-of-mass energy (pCM) collider following this $\Pep\Pem$ programme. 
Specifically, Europe's longer-term ambition~\cite{esppu} is to operate a proton-proton collider at the highest achievable energy, sensitive to energy scales an order of magnitude higher than those reached at the LHC. 
By providing considerable advances in sensitivity, precision, and energy far above the TeV scale, the Future Circular Collider (FCC) matches the present landscape and the above requirements to near perfection. 
Indeed, the FCC starts with a luminosity-frontier $\Pep\Pem$ machine spanning centre-of-mass energies from below the \PZ pole to beyond the top-pair production threshold (FCC-ee), later evolving to an energy-frontier hadron collider (FCC-hh). 
Both machines are strongly motivated in their own right, as shown in the rest of this section.
In particular, the full programme will provide unique sensitivity to new physics in the Higgs sector: 
with almost three million Higgs bosons, FCC-ee will offer in a few years model-independent measurements of 
the Higgs boson mass, width, and couplings to \PZ, \PW, \Pgt, \PQb, \PQc, 
order-of-magnitude more precise than today, to probe the possible BSM origin of Electroweak Symmetry Breaking. 
Ultimately, FCC-hh will produce 20 billion Higgs bosons and, in combination with FCC-ee, provide incomparable measurements of the Higgs self-coupling, top Yukawa coupling and of other rare or invisible modes probing in new directions the processes that led to the formation of today’s Higgs vacuum field.
Efforts to understand, analyse, and document the broad physics potential of such an ambitious programme started in 2012~\cite{Gomez-Ceballos:2013zzn} and culminated in the FCC Conceptual Design Report (CDR) in 2018~\cite{fcc-phys-cdr, fcc-ee-cdr, FCC-hhCDR}. 

From the scale of Fermi interactions to that of the \PW and \PZ bosons, from the EW precision measurements to the observation of the top quark and of the Higgs boson, precision has always provided the route to new discoveries; the FCC will be no exception in this respect. 
Any deviation from the SM predictions, interpreted as the manifestation of higher-dimensional operators, will point to a new energy scale that will be explored directly later on. 
Furthermore, correlations among several deviations will be instrumental in characterising the structure of new physics and to reveal its exact or approximate global symmetries. 
In that regard, the fact that the SM features some a priori `accidental' selection rules (lepton and baryon number conservation, custodial symmetry, suppression of flavour-changing neutral currents, smallness of the mixing among the different quarks, collective suppression of CP violation effects) that generic new physics scenarios do not share, is a welcome virtue and a promise for future discoveries or deeper understanding. 
The expected FCC precision is such that much will be learned, regardless of the outcome: theories beyond the SM will be very much constrained, even with a null result, and will further guide new models from these precision measurements. 
When a new physics signal is eventually observed, these precision measurements (whether they agree with or deviate from the SM) will be precious in establishing the nature of the signal.

The presently proposed schedule of the FCC programme has 15 years of FCC-ee operation followed by 25 years of FCC-hh operation, interleaved with a shutdown of 10 years to dismantle the lepton collider and install the hadron collider in the tunnel, for a grand-total of 50 years, a similar duration to that of the current LEP + LHC programme (1989--2041). 
This sequence optimises the overall investment and its science value for several fundamental reasons: 
{\it (i)}~the powerful physics complementarity of the two machines, leading to a uniquely broad exploration potential; 
{\it (ii)}~the synergy of the infrastructure, which leads to a considerable cost saving and reduces the financial burden; 
{\it (iii)}~the implementation schedule, which opens a time window of at least 25 years for the development of  the critical technology of high-field magnets for the hadron collider, reducing the financial and technological risks; and 
{\it (iv)}~the duration of the programme and its strategic importance, which makes it conceivable to obtain the upfront funding for the common infrastructure. 
The recyclable infrastructure and the large integrated luminosities also significantly improve the global sustainability of the particle physics worldwide endeavour over the 21$^\text{st}$ century, by minimising the operation time, the electricity consumption, the cost, and the carbon emissions for a given scientific outcome~\cite{Janot:2022jtn,Blondel:2024mry}.

\subsection{FCC-ee: A great Higgs factory, and so much more}
\label{sec:overview_ee}

\begin{figure}[ht]
\centering
\includegraphics[width=0.77\textwidth]{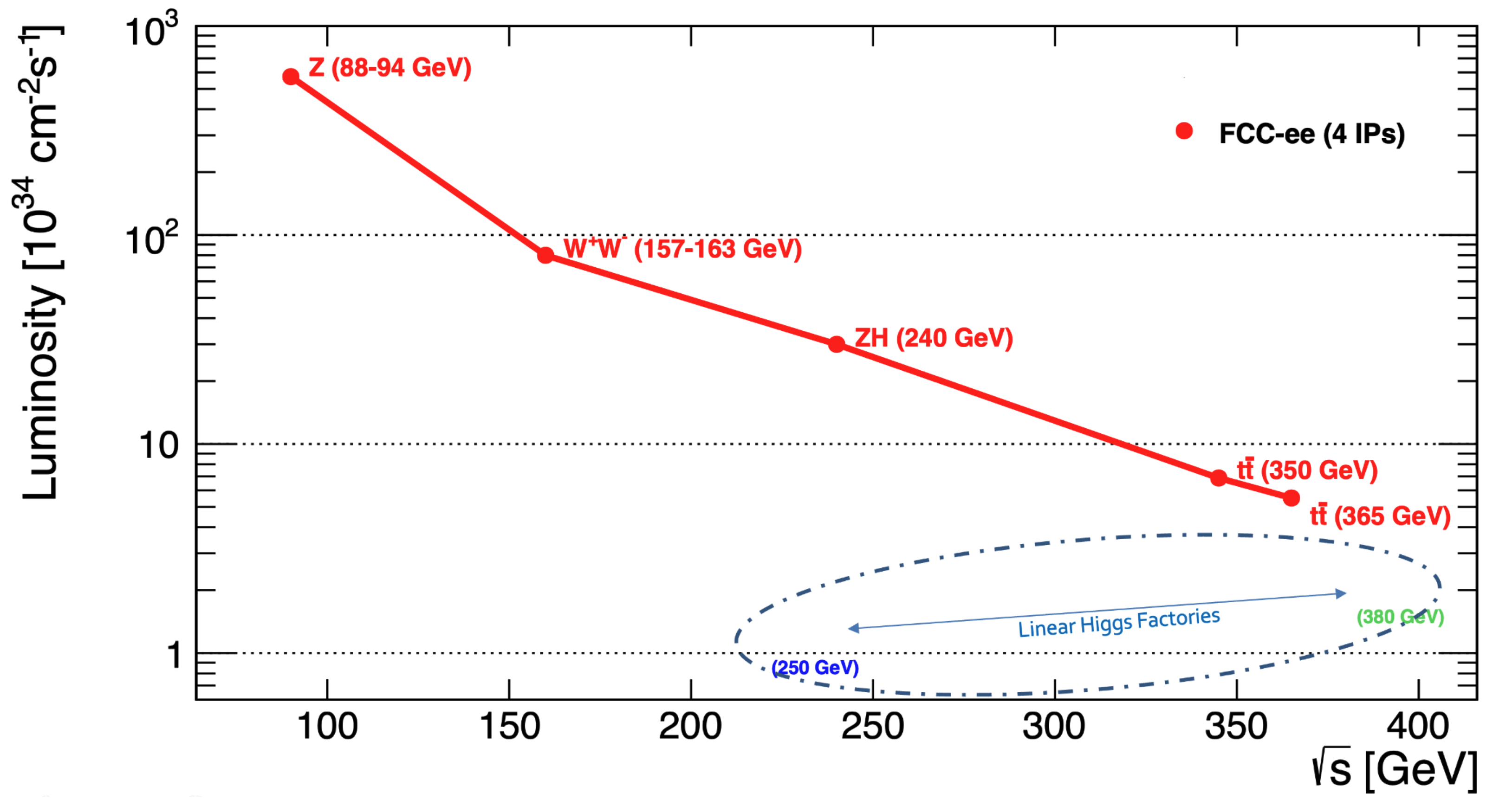}
\caption{The FCC-ee baseline design luminosity, summed over 4~IPs, displayed as a function of the centre-of-mass energy, 
from the \PZ pole to the $\PQt\PAQt$ threshold and beyond (red curve).
The luminosity typically achievable by linear $\Pep\Pem$ Higgs factories with a single IP in their baseline design between 250 and 380\,GeV is also indicated (dash-dotted oval in the lower-right corner of the figure).}
\label{fig:FCCeeLumis} 
\end{figure}

With its high luminosity, its clean experimental conditions, its multiple interaction regions, and a range of energies that cover the four heaviest elementary particles known today, FCC-ee offers a uniquely broad and powerful physics exploration programme as a Higgs, electroweak, QCD, flavour, and top factory, with high potential for discoveries. 
The baseline plan, confirmed by the feasibility study mid-term review recommendations, considers operating detectors at four interaction points (IPs), spanning the $\Pep\Pem$ centre-of-mass energies around the \PZ pole, the $\PW\PW$ threshold, the $\PZ\PH$ production maximum, up to the $\PQt\PAQt$ threshold and just above. 
The current values for the luminosities expected at these energies are displayed in Fig.~\ref{fig:FCCeeLumis} and the envisioned 15-year experimental programme is summarised in Table~\ref{tab:seqbaseline}, together with the numbers of events expected at each energy. 

\begin{table}[ht]
\renewcommand{\arraystretch}{0.95}
\centering
\caption{The baseline FCC-ee operation model with four interaction points, showing the centre-of-mass energies, design instantaneous luminosities for each IP, and integrated luminosity per year summed over 4~IPs.  
The integrated luminosity values correspond to 185 days of physics per year and a 75\% operational efficiency (i.e., $1.2 \times 10^7$ seconds per year)~\cite{Bordry:2645151}, in the \PZ, $\PW\PW$, $\PZ\PH$, and $\PQt\PAQt$ baseline sequence. 
The last two rows indicate the total integrated luminosity and number of events expected to be produced in the four detectors. 
The number of $\PW\PW$ events includes all $\sqrt{s}$ values from 157.5\,GeV up.
\label{tab:seqbaseline}}
\begin{tabular}{lccccccc}
\hline 
Working point & \PZ pole & $\PW\PW$ thresh.\ & $\PZ\PH$ & \multicolumn{2}{c}{$\PQt\PAQt$} \\ \hline
$\sqrt{s}$ {(GeV)} & 88, 91, 94 & 157, 163 & 240 & 340--350 & 365 \\ 
Lumi/IP {($10^{34}$\,cm$^{-2}$s$^{-1}$)} & 140 & 20 & 7.5 & 1.8 & 1.4 \\ 
Lumi/year {(ab$^{-1}$)} & 68 & 9.6 & 3.6 & 0.83 & 0.67 \\ 
Run time {(year)} & 4 & 2 & 3 & 1 & 4 \\ 
Integrated lumi.\ {(ab$^{-1}$)} & 205 & 19.2 & 10.8 & 0.42 & 2.70 \\ \hline
 &  &  & $2.2 \times 10^6$ $\PZ\PH$ & \multicolumn{2}{c}{$2 \times 10^6$ $\PQt\PAQt$} \\
Number of events &  $6 \times 10^{12}$ \PZ & $2.4 \times 10^8$ $\PW\PW$ & $+$ & \multicolumn{2}{c}{$+\,370$k $\PZ\PH$} \\
 &  &  & 65k $\PW\PW \to \PH$ & \multicolumn{2}{c}{$+\,92$k $\PW\PW \to \PH$} \\ \hline
\end{tabular} 
\end{table}

\begin{figure}[ht]
\centering
\includegraphics[width=0.85\textwidth]{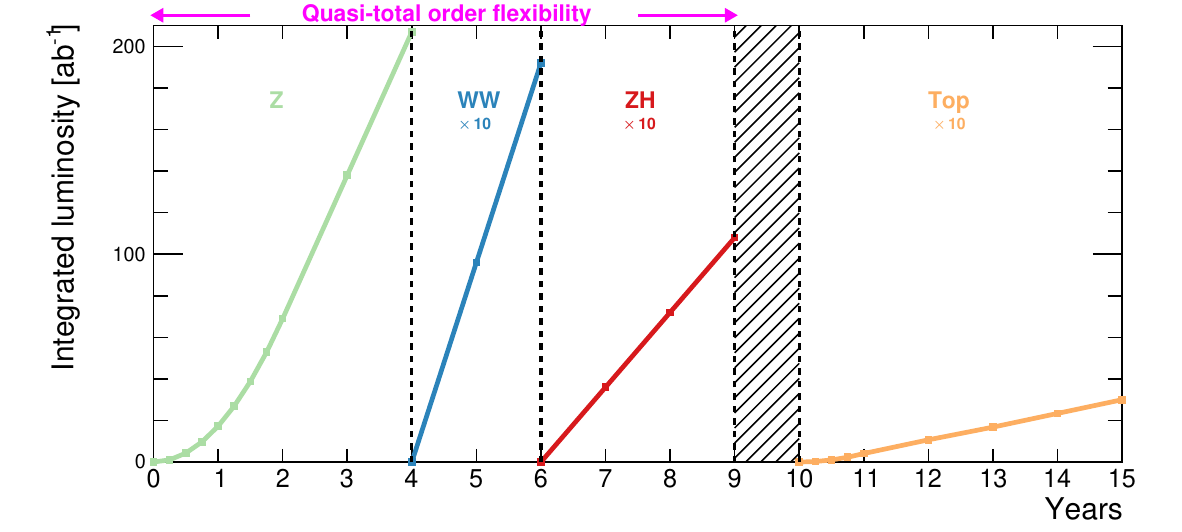}
\caption{\small Baseline operation model for FCC-ee with four interaction points, showing the integrated luminosity at the \PZ pole (green), the $\PW\PW$ threshold (blue), the Higgs factory (red), and the $\PQt\PAQt$ threshold (orange) as a function of time. 
In this baseline model, the sequence of events follows the increase in collision energy, but there is quasi-total flexibility in the sequence all the way to 240\,GeV (see inset). 
The integrated luminosities delivered during the first two years at the \PZ pole and the first year at the $\PQt\PAQt$ threshold are half of the annual design value. 
The hatched area indicates the shutdown time needed to prepare for the higher energy runs at the $\PQt\PAQt$ threshold and beyond.}
\label{fig:seqbaseline} 
\end{figure}

The currently envisioned working hypotheses for the operation model~\cite{janot_2024_nfs96-89q08}, i.e.\ for the overall sequence and the duration of each step, will be continuously optimised in the coming years. 
An example of a baseline sequence is displayed in Fig.~\ref{fig:seqbaseline}, 
with the \PZ, $\PW\PW$, and $\PZ\PH$ runs assumed to happen in this chronological order. 
In reality, however, there will be quasi-total flexibility in the choice of the running sequence. 
(See inset below: `A quasi total flexibility'.) 
This flexibility will accommodate the likely requests of the user community and will also fit the possible need for runs at different energies in view of complementary measurements.

\begin{tcolorbox}[colback=red!5!white,
                  colframe=red!75!black,
                  title=A quasi-total flexibility
                 ]
\footnotesize In Ref.~\cite{note-FCCeeSequence} it was deemed essential to establish the technical and financial feasibility of scheduling a \PZ pole run after the $\PZ\PH$ run or the $\PW\PW$ threshold run, ideally with a first \PZ pole run during the initial period of FCC-ee operation. 
Indeed, the \PZ pole run, while extremely fertile in physics opportunities, is undoubtedly the most ambitious and demanding part of the programme from all perspectives (accelerator, energy calibration, detectors systematic biases, theory calculations).
It will be extremely challenging to achieve all the goals of the \PZ pole run during the first four years of the collider operation. 
\tcblower
\footnotesize This requirement from physics was compounded by a recommendation from the mid-term review committees to 
{\it consolidate (and ideally simplify) the design of the RF system to allow efficient energy-staging, as well as to reduce complexity, risk, and cost; and to study options to avoid the 1-cell/2-cell RF cavity reconfiguration between \PZ and $\PZ\PH$/$\PW\PW$ running, in order to simplify the SRF system implementation and to improve flexibility in the physics programme.} 
The new versatile RF system designed accordingly (see Ref.~\cite{FCC-PED-FSR-Vol2})
enables a quasi-total flexibility to choose the running sequence. 
For example, it would allow for short (few weeks) initial \PZ pole and $\PW\PW$ threshold runs, to commission the collider and the detectors, to establish the resonant depolarisation procedures for centre-of-mass energy calibration, etc. 
The $\PZ\PH$ run could then proceed early, before going back to the \PZ pole and the $\PW\PW$ threshold, both now at full luminosity, with fully functional resonant depolarisation and a complete understanding of the collider.
\end{tcolorbox}

It will ultimately be up to the experimental collaborations and the relevant scientific committees to optimise the time flexibility 
and to tailor the FCC-ee operation scenario according to a number of factors that cannot be mastered today. 
For example, external events such as the FCC-hh magnet readiness or the CERN financial situation may call for a change in the overall duration of the FCC-ee running time. 
Meanwhile, new avenues will be explored during the next phase of the FCC study, between April 2025 and the end of 2027, 
towards increasing the luminosities at all energies, 
as it would be highly beneficial across the whole programme and would make a qualitative difference in a number of cases. 

The original motivation for an $\Pep\Pem$ circular collider was to create a high-luminosity Higgs Factory, operating at $\sqrt{s}=240$\,GeV in the LEP/LHC tunnel~\cite{Blondel:2011fua,Blondel:2012ey,Azzi:2012yn}. 
Choosing to build it in the 80--100\,km tunnel that would ultimately be hosting a 100\,TeV hadron collider was cardinal in making this machine unique on the Higgs factory market. 
To start with, such a tunnel enables a highly versatile Higgs Factory that extends much beyond the study of the Higgs boson alone, for three main reasons:

\begin{enumerate}
\item A circumference of 80--100\,km is required for an $\Pep\Pem$ circular collider to reach (for the first time in $\Pep\Pem$ collisions) the top-pair production threshold, making the FCC-ee energy range optimal to sharpen and challenge our current physics knowledge by studying all SM particles. 
In particular, enabling precise top quark measurements is essential for the overall FCC electroweak and Higgs precision physics programme.

\vspace{0.4cm}

\item For a given power and centre-of-mass energy, the luminosity of a circular collider is roughly proportional to the ring circumference. 
A circumference of 80--100\,km makes FCC-ee the $\Pep\Pem$ electroweak, Higgs, and top factory project with the highest luminosity proposed to date, able to produce, when summing all the data collected at the 4~IPs, $6\times 10^{12}$ \PZ bosons, $2.4\times 10^{8}$ \PW pairs, almost $3\times 10^6$ Higgs bosons, and $2\times 10^6$ top pairs, in as little as 15 years. 
(See also inset below: `Rationale for the operation model').
\end{enumerate}

\begin{tcolorbox}[colback=blue!5!white,
                  colframe=blue!75!black,
                  title=Rationale for the operation model 
                 ]
\footnotesize (From the mid-term review) 
{\it It would also be interesting to add more information in the final Feasibility report about the optimisation of the durations of the various energy runs (\PZ, $\PW\PW$, $\PZ\PH$, $\PQt\PAQt$) as well as a more detailed and quantitative discussion demonstrating the importance of the $\PQt\PAQt$ running.}
\tcblower
\footnotesize The baseline set of centre-of-mass energies and integrated luminosities is found to be the most sensible way to distribute luminosity within the 15 years overall running constraint. 
It is deemed sufficient to establish a minimal, yet remarkable, physics outcome for such a collider, with the smallest possible set of centre-of-mass energies that enables a study of all particles of the SM with a real chance of discovery. 

\begin{itemize}
\item At the \PZ pole ($\sqrt{s} = 91.2$\,GeV), at least $5\times 10^{12}$ \PZ (in two years) enables otherwise unreachable flavour (\PQb, \PQc, $\Pgt$) physics, studies of QCD and hadronisation, the search for rare or forbidden decays, and the exploration of the dark sector. 
Together with the runs at $\sqrt{s} \simeq 88$ and $94$\,GeV, the \PZ pole data yield 50 to 1000-fold improved measurements of the \PZ line-shape and many electroweak precision observables (EWPO), including the \PZ invisible width. 
An integrated luminosity of at least 40\,ab$^{-1} $ at the two energies, just below and just above the \PZ pole (one year each), was deemed appropriate for a direct, unique, and statistically limited determination of $\alpha_\text{QED}(m_{\PZ})$ and $\alpha_\mathrm{S}(m_{\PZ})$, 
which otherwise would greatly limit the new physics interpretation of all measurements of $\sin^2\theta_{\PW}^\text{eff}$. 
One of the findings of the study has been the realisation of how rich the run at and around the \PZ pole is for the FCC-ee physics outcome.
\item An integrated luminosity of at least 20\,ab$^{-1}$ in two years around the $\PW^+\PW^-$ production threshold, evenly shared between $\sqrt{s} = 157.5$ and 162.5\,GeV, is needed for the measurement of the \PW mass (and decay width) with a statistical precision commensurate with the expected precision of the centre-of-mass energy determination. 
These data are also important for, in particular, the determination of the number of neutrino species and for an independent measurement of the strong coupling constant.
\item An integrated luminosity of at least 10\,ab$^{-1}$ in three years at $\sqrt{s} = 240$\,GeV provides model-independent measurements of the Higgs boson couplings from a combination of its  branching fractions and of the total $\PZ\PH$ production cross section. 
In particular, a per-mil precision can be reached on the coupling of the Higgs boson to the \PZ, which greatly constrains new physics coupled to the Higgs boson.  
\item A short run to scan the $\PQt\PAQt$ threshold ($\sqrt{s} = 340$--350\,GeV) allows the measurement of the top-quark mass, a fundamental parameter of the Standard Model, with a precision of $\cal O$(10\,MeV), for which hadron colliders cannot compete. 
Because the prediction of EWPO's is, in various ways, sensitive to $m_\text{top}$, the discovery power of this EWPO exploration is limited by the uncertainty on $m_\text{top}$. 
For example, matching the precision of the SM predictions from the EWPO measurements to the 180\,keV (resp.\ 4\,keV) statistical uncertainty on the \PW mass (resp.\ \PZ width) requires a 20\,MeV (resp.\ 15\,MeV) knowledge of $m_\text{top}$.  
It is only when the mass of the top quark mass is measured with this precision that the FCC-ee EWPO measurements give their best sensitivity, typically increasing the reach in new physics energy scale by 60\%. 
\item 
The $\PZ\PH$ cross section dependence on $\sqrt{s}$ provides sensitivity to the Higgs boson self-coupling when data at 365\,GeV are available, allowing a 5\,$\sigma$ discovery of this coupling to be contemplated. 
More importantly, the per-cent measurement of the top EW couplings 
{\it(i)}~matches the EWPO ppm precision at the \PZ pole and the $\PW\PW$ threshold; 
and {\it (ii)}~keeps the theoretical uncertainties on the top Yukawa coupling determination 
at FCC-hh at the per-cent level, a pre-requisite for the model-independent determination of the Higgs boson self-coupling.
\end{itemize}
\end{tcolorbox}

\begin{itemize}
\item[3.] A circumference of 80--100\,km is also required for FCC-ee to offer unparalleled control of the centre-of-mass energy, not only at the \PZ pole but also at the $\PW\PW$ threshold, with the use of resonant depolarisation~\cite{Blondel:2019jmp,Blondel:2021zix}. 
At the \PZ pole, the centre-of-mass energy scale should be known to 1\,ppm or better, 
with a point-to-point residual uncertainty of 28\,keV and a per-mil level spread. 
These parameters are key inputs to the electroweak precision programme~\cite{Maestre:2021zvq}. 
Details are given in Chapter~\ref{sec:epol}. 
The resulting precisions on the \PZ mass and width, the forward-backward asymmetry of muon pairs around the \PZ pole, the electromagnetic coupling constant $\alpha_\text{QED}(m_{\PZ})$, as well as on the \PW mass, are listed in the first rows of Table~\ref{tab:EWPO}.  
\end{itemize}

\noindent Several other fundamental aspects add to the uniqueness of FCC-ee.

\begin{enumerate}

\item One of the findings of the FCC feasibility study has been the realisation of how rich the intensity-frontier multi-Tera-Z run is for the FCC-ee physics outcome. 
It promises comprehensive measurements of the \PZ lineshape and many EWPOs with at least fifty-fold improved precision, as well as direct and uniquely precise determinations of the
$\alpha_\text{QED}(m_{\PZ})$~\cite{Janot:2015gjr,Riembau:2025ppc} and $\alpha_\mathrm{S}(m_{\PZ})$ interaction couplings (Table~\ref{tab:EWPO}), which would otherwise be dominant parametric uncertainties for virtually all precision SM calculations. 
The comparison of these data with commensurately accurate SM predictions is a way to reveal the existence of new physics (or to severely constrain its properties) through virtual loops or mixing: a factor of 50 in precision corresponds to a factor of 7 in energy scale, representing a step towards discovery 
(Section~\ref{sec:PhysicsCaseBSM} and Refs.~\cite{Allwicher:2024sso,Gargalionis:2024jaw}) 
similar to that from LHC to FCC-hh.  
This \PZ pole run offers more than `just higher precision'. 
It also enables otherwise unreachable flavour (\PQb, \PQc, \Pgt) physics, studies of QCD and hadronisation, the search for rare or forbidden decays, the exploration of the dark sector, and significantly increases sensitivity to new feebly interacting particles, altogether further increasing the prospects for discovery. 

\item A unique feature of circular colliders is the ability to serve several IPs. Another finding of the FCC Feasibility Study has been the demonstration of the importance of operating four detectors (instead of only two in the 2014--2018 Conceptual Design Study), which led to an optimisation of the ring layout with a new four-fold periodicity 
(further increasing the synergies with FCC-hh). 
In a configuration with 4~IPs, the FCC science value for the investment is maximised in multiple ways. 
(See inset below: `The benefits of four interaction points'.) 

\item Finally, FCC-ee stands out among the Higgs Factory projects for its unique opportunity to access the Higgs boson coupling to electrons~\cite{Ghosh:2015gpa,Dery:2017axi,Altmannshofer:2015qra}, through the resonant production process \mbox{$\Pep\Pem \to \PH$} at $\sqrt{s} = 125$\,GeV~\cite{dEnterria:2021xij}. 
This measurement relies on the combination of the high luminosity, the possibility of operating four detectors, the continuous ppm centre-of-mass energy control, and the ability for centre-of-mass monochromatisation~\cite{FausGolfe2021}. 
This is a unique opportunity for FCC-ee, and one of its toughest challenges. 

\end{enumerate}

As a Higgs factory, FCC-ee has all the advantages of an $\Pep\Pem$ collider running in the vicinity of the $\PZ\PH$ cross section maximum: the measurement of this cross section by counting events with an identified \PZ boson 
(and for which the mass recoiling against the \PZ gathers around the Higgs boson mass~\cite{Azzurri:2021nmy}, independently of the Higgs boson properties details) 
provides a precise and model-independent determination of $\kappa_{\PZ}$, the Higgs boson coupling to the \PZ. 
This absolute measurement can then be used as a `standard candle' by all other measurements, including those made at hadron or muon colliders. 
The position of the \PZ recoil mass peak also provides an accurate measurement of the Higgs boson mass from the precise knowledge of the centre-of-mass energy. 
In combination with the measurement of the rate of $\PZ\PH$ events with a $\PH \to \PZ \PZ^\ast$ decay, proportional to $\kappa_{\PZ}^4/\Gamma_{\PH}$, a model-independent determination of the total width $\Gamma_{\PH}$ can be obtained. 
The analysis of the other decays provides a set of model-independent partial width and coupling measurements. 

\begin{table}[p]
\centering
\caption{Experimental (statistical and systematic) precision expected for a selection of measurements accessible at FCC-ee, 
compared with the present world-average precision~\cite{ParticleDataGroup:2024cfk}. 
Some of the FCC-ee experimental systematic uncertainties (4$^\text{th}$ column) are initial estimates from early 2021~\cite{Blondel:2021ema} and others have been improved and consolidated during the Feasibility Study. 
A goal of further studies will be to improve them down to the level of the statistical uncertainties (3$^\text{rd}$ column) with new ideas and innovative methods.
This set of measurements, together with those of the Higgs boson properties, 
achieves indirect sensitivity to new physics up to a scale $\Lambda$ of 100\,TeV 
in an Effective Field Theory (EFT) description with dimension-6 operators (Chapter~\ref{sec:physics}) 
and possibly much higher in specific new physics (non-decoupling) models.  
\label{tab:EWPO}}
\setlength{\tabcolsep}{1.5pt}
\renewcommand{\arraystretch}{0.80}
{\small
\begin{tabular}{lrclccr}
\toprule
Observable  &  & present &          &  FCC-ee  &  FCC-ee  &  Comment and   \\
            & value  &$\pm$& uncertainty  &  {\bf Stat.} &   Syst.\     & leading  uncertainty \\
\midrule
$m_{\PZ}$ (keV)  &  91\,187\,600   & $\pm$ &  2000    & {\bf 4} & 100  & From \PZ line shape scan  \\
$  $  &  & &    &   &  & Beam energy calibration \\
\midrule
$\Gamma_{\PZ}$ (keV)  & 2\,495\,500   & $\pm$ &  2300    & {\bf 4}  &  12  & From \PZ line shape scan  \\
$  $  &  & &    &   &   &  Beam energy calibration  \\
\midrule
$\sin^2{\theta_{\PW}^\text{eff}} ~ (\times 10^6)$  & 231,480   & $\pm$ &  160   & {\bf 1.2}   &  1.2  & 
From $A_\text{FB}^{\PGm\PGm}$ at \PZ peak\\
$  $  &  & &    &   &   &  Beam energy calibration  \\
\midrule
$1/\alpha_\text{QED} (m_{\PZ}^2) ~ (\times 10^3) $  & 128\,952  & $\pm$ &  14   & {\bf 3.9}   &  small  &
From $A_\text{FB}^{\PGm\PGm}$ off peak\\
& & & & {\bf 0.8} & tbc & From $A_\text{FB}^{\PGm\PGm}$ on peak\\
$  $  &  & &    &   &  &  QED\&EW uncert.\ dominate  \\
\midrule
$ R_{\ell}^{\PZ} ~(\times 10^3)$  & 20\,767 & $\pm$ &  25   & {\bf 0.05}   & 0.05   &  Ratio of hadrons to leptons \\
$  $  &  & &  &  &  &    Acceptance for leptons  \\
\midrule
$ \alpha_\mathrm{S} (m_{\PZ}^2) ~(\times 10^4) $  & 1\,196 & $\pm$ &  30  &  {\bf 0.1}  &  1  &
Combined ${R_{\ell}^{\PZ}},\,\Gamma^{\PZ}_\text{tot},\,\sigma^{0}_\text{had}$ fit\\
\midrule
$ \sigma^{0}_\text{had}~(\times 10^3)$ (nb) & 41\,480.2 & $\pm$ &  32.5    & {\bf 0.03}  &  0.8  &  Peak hadronic cross section  \\
$  $  &  & &    &  &   &  Luminosity measurement  \\
\midrule 
$ {N_{\PGn}}  (\times 10^3) $  & 2\,996.3  & $\pm$ &  7.4   & {\bf 0.09}   &  0.12 &  \PZ peak cross sections \\
$  $  &   &  &    &   &  &   Luminosity measurement \\
\midrule
$ R_{\PQb} ~(\times 10^6) $  & 216\,290 & $\pm$ &  660   & {\bf 0.25}    &  0.3  &  Ratio of $\PQb\PAQb$ to hadrons  \\
\midrule
$ A_\text{FB}^{\PQb,0} ~(\times 10^4) $  & 992 & $\pm$ &  16   & {\bf 0.04}   &  0.04  &  \PQb-quark asymmetry at \PZ pole  \\
$  $  &  & &  &   &  &    From jet charge \\
\midrule
$ A_\text{FB}^{\text{pol},\Pgt} ~(\times 10^4) $  & 1\,498 & $\pm$ &  49  & {\bf 0.07}   &  0.2  &  \Pgt\ polarisation asymmetry  \\
$  $  &  & &    &   &  &  \Pgt\ decay physics \\
\midrule
\Pgt\ lifetime (fs)  &  290.3  & $\pm$ &  0.5   &  {\bf 0.001}   & 0.005   &  ISR, \Pgt\ mass  \\
\midrule 
\Pgt\ mass (MeV)  & 1\,776.93 & $\pm$ & 0.09  & {\bf 0.002} & 0.02   & estimator bias, ISR, FSR \\
\midrule 
\Pgt\ leptonic ($\PGm \PGnGm \PGnGt$) BR (\%) & 17.38 & $\pm$ & 0.04  & {\bf 0.00007} & 0.003   &  PID, $\pi^0$ efficiency \\ 
\midrule 
$ m_{\PW}$ (MeV)  &  80\,360.2 & $\pm$ &  9.9    & {\bf 0.18}  & 0.16  & From $\PW\PW$ threshold scan \\
$  $  &  & &    &   &  &  Beam energy calibration  \\
\midrule
$\Gamma_{\PW}$ (MeV)  & 2\,085   & $\pm$ &  42    & {\bf 0.27}  & 0.2  & From $\PW\PW$ threshold scan \\
$  $  &  & &    &  &   &  Beam energy calibration  \\
\midrule
$ \alpha_\mathrm{S} (m_{\PW}^2) ~ (\times 10^4)$  & 1\,010   & $\pm$ & 270 &  {\bf 2}  & 2  &
  Combined $R_{\ell}^{\PW},\,\Gamma^{\PW}_\text{tot}$ fit\\
\midrule
$ N_{\PGn} ~ (\times 10^3) $  & 2\,920 & $\pm$ &  50   & {\bf 0.5}   & small   &   Ratio of invis.\ to leptonic \\
$  $  &  & &    &   &  & in radiative \PZ returns  \\
\midrule
$m_\text{top}$ (MeV) &  172\,570 & $\pm$ &  290    & {\bf 4.2}  & 4.9 
& From $\PQt\PAQt$ threshold scan \\
$  $  &  & &    &   & &  QCD uncert.\ dominate \\
\midrule
$\Gamma_\text{top}$ (MeV) &  1\,420   & $\pm$ &  190    &{\bf 10}  & 
6 
& From $\PQt\PAQt$ threshold scan \\
$  $  &  & &    & & &  QCD uncert.\ dominate \\
\midrule
$\lambda_\text{top}/\lambda_\text{top}^\text{SM}$  &   1.2   & $\pm$ &  0.3    & {\bf 0.015} & 
0.015 
& From $\PQt\PAQt$ threshold scan \\
$  $  &  & &    &   &  &  QCD uncert.\ dominate \\
\midrule
$\PQt\PQt\PZ$ couplings &
   & $\pm$ &  30\%   & {\bf 0.5--1.5} \%  & small  & From $\sqrt{s}=365$\,GeV run  \\
\bottomrule
\end{tabular}
}
\end{table}

\begin{tcolorbox}[colback=green!5!white,
                  colframe=green!75!black,
                  title=The benefits of four interaction points
                 ]
\footnotesize (From the mid-term review) 
{\it The Scientific Advisory Committee recommends to construct 4~IPs for FCC-ee from the beginning. 
The Scientific Policy Committee finds the proposal to include more than two IPs an attractive option. 
However, we expect, for the final Feasibility report, arguments focusing not only on the na{\"\i}ive luminosity gain but also on the overall physics optimisation.}
\tcblower
\footnotesize

{\bf Detector Diversity}
The diverse set of demanding physics measurements and searches at FCC-ee leads to many different challenging detector requirements, which cannot be simultaneously satisfied by only one or even two detectors. 
Instead, four FCC-ee interaction points allow for a broader diversity of detector solutions to be explored and, therefore, a wider community with diverse interests and skills to join the FCC-ee project, with a better perspective of optimally covering all FCC-ee physics opportunities. 
Detector diversity also allows cross-checks of results across experiments: a single experiment is quite vulnerable to unforeseen effects, and consistency between two experiments can still happen by chance. 
Such bad luck is much less likely with four diverse detectors. 
(See remark about redundancy below.)  

{\bf Improved sustainability} 
Operating FCC-ee with 4~IPs provides a net gain of luminosity (typically up to a factor 1.7 with respect to the 2~IP option) for a given electricity consumption. 
The physics outcome of 15 years operation with 4~IPs would require 25 years with just 2~IPs. 
Running with four experiments would decrease the FCC-ee operation carbon emissions in the same proportions. Folding in the construction carbon budget of the tunnel the additional two IPs and two detectors, it is found that operating FCC-ee with 4~IPs would reduce the total carbon footprint by $\sim$\,15\% for the same physics outcome as with 2~IPs, both with today's figures and with those expected in the construction and operation times of FCC-ee.

{\bf Reduced and more robust systematic uncertainties} 
In order to fully benefit from the considerable FCC-ee event samples, especially at the \PZ pole, most of the work will focus on reducing the systematic uncertainties to match the statistical precision. 
As a matter of fact, many key measurements are potentially affected by detector-related systematic biases. 
These biases are uncorrelated between experiments, so that the related systematic uncertainties scale down as $1/\sqrt{N_\text{experiments}}$. 
Precision is also about redundancy: measuring the same quantity in several experiments can reveal sources of errors that would have been overlooked otherwise. 
The illustrative example closest to FCC-ee comes from the LEP measurement of the \PZ mass with the 1991 data, when a large discrepancy of $\sim 20 \pm 5$\,MeV was noticed between the measurements from L3 and OPAL, on the one hand, and the measurements from ALEPH and DELPHI, on the other. 
The investigation that followed identified and solved the origin of the issue, but it could have remained unnoticed for a long time (or forever) had there been only ALEPH and DELPHI (or only L3 and OPAL) around the ring.

{\bf Collider monitoring} 
With the large collision rate at the \PZ pole, each detector will act as sophisticated beam instrumentation by allowing quick and high-quality measurements of many beam parameters: positions and sizes of the luminous regions; longitudinal and transverse boosts of the collision at the four IPs (beam energy difference, centre-of-mass energy spread, crossing angle); the dependence of these parameters on the exact position of the collision within each bunch; and other information we have not thought of yet. 
These measurements at four different points of the collider provide a huge amount of information on the beam properties, which is useful for collider performance optimisation and, more specifically, on the `energy model', which is the cornerstone of the precision programme at the \PZ pole, at the $\PW\PW$ threshold, and even more so at the Higgs boson resonance ($\sqrt{s} = 125$\,GeV).   

{\bf Time flexibility} 
The additional flexibility offered by two additional interaction points could be a game changer for the FCC-ee scientific outcome. 
For example, reaching the $5\sigma$ discovery level for the Higgs self-coupling, contemplating a first measurement of the electron Yukawa coupling, or securing a thorough search for sterile neutrinos, would simply be missed opportunities with only two IPs (a little over 2\,$\sigma$ significance expected).
The larger luminosity expected with four IPs allows for a different time allocation to the different centre-of-mass energies, to be optimised when the time comes, by the FCC-ee user community.

{\bf Community building} 
The FCC-hh will require the participation of an even larger community of users than the LHC. 
The large collaborations operating the four detectors at FCC-ee would raise the overall profile of the worldwide high-energy frontier community to a level more appropriate to support the ultimate high-energy proton collider.

{\bf Integrated luminosity gain} 
Many FCC-ee measurements are statistically limited, either directly (e.g., the direct determination of $\alpha_{\rm QED}(m_{\rm Z}^2)$ from the muon forward-backward asymmetry measurements at $\sqrt{s}=88$ and 94\,GeV) or because their experimental systematic uncertainties would continue to improve with additional data, as is the case for many critical observables (e.g., the \PZ width, the effective mixing angle, etc.). 
Any increase of integrated luminosity is also welcome for searches for feebly interacting particles or for rare/forbidden processes. 
Therefore, these measurements and searches immediately benefit from four interaction points. 
\end{tcolorbox}

The FCC-ee best-measured Higgs boson couplings, $\kappa_{\PZ}$ and $\kappa_{\PW}$, with a precision of 0.1\% and 0.23\%, after eight years of operation at $\sqrt{s} \geq 240$\,GeV, respectively, would require half a century to be reached with linear $\Pep\Pem$ Higgs factories running either at 250 and 500\,GeV or at 380 and 1500\,GeV, making FCC-ee the most time- and cost-effective, as well as the most sustainable, Higgs factory of all proposed options~\cite{Janot:2022jtn,Blondel:2024mry}. 
More details on the physics programme are given in Chapter~\ref{sec:physics}. 

The many new opportunities of physics measurements and searches at FCC-ee create at least as many challenges. 
Reaching experimental and theoretical systematic uncertainties commensurate with the statistical precision of the many measurements feasible at FCC-ee requires a careful study of the detector concepts, possibly of the mode of operation, and of theoretical developments. 

The FCC-ee physics programme presents a number of key theoretical challenges~\cite{Heinemeyer:2021rgq}. 
Generally speaking, the aim is either to provide the tools to compare experimental observations to theoretical predictions at a level of precision similar or better than the (statistical) experimental uncertainties (Table~\ref{tab:EWPO}), or to identify the additional calculations, tools, observables, or experimental inputs that are required to achieve this level of precision.
Another essential line of research to be followed jointly by theorists and experimenters is to identify observables, or ratios of observables, for which experimental and/or theoretical uncertainties can be reduced. 
Finally, both for motivational purposes and for prioritisation, the relative impact of the various measurements on the search for new physics should be evaluated. The theoretical work motivated by the FCC programme can be organised as follows.

\begin{itemize}
\item  Calculation of QED (mostly), EW, and QCD corrections to (differential) cross sections, needed to convert experimental measurements to so-called `pseudo-observables': couplings, masses, partial widths, asymmetries, etc., without altering significantly the possible new-physics contributions in the original measurements.
Appropriately accurate event generators are essential for the implementation of these effects in the experimental procedures.
\item  Calculation of the pseudo-observables with the precision required in the framework of the Standard Model so as to take full advantage of the experimental precision.
\item 
Identification of the limiting issues, such as the questions related to the definition of parameters, in particular the treatment of quark masses and, more generally, QCD objects.
\item Investigation of the sensitivity of the proposed (or new) experimental observables to the effect of new physics in a number of important, specific scenarios. 
This essential work must be done at an early stage, before the project is fully designed, since it potentially affects the priorities for detector concepts and for the running plan.
\end{itemize}

A community of theorists has already risen to the precision challenges, especially at the \PZ peak~\cite{Blondel:2018mad}. 
An evaluation of the options for the path ahead can be found in Refs.~\cite{Blondel:2019qlh,Freitas:2019bre,Heinemeyer:2021rgq}, and is developed in Chapter~\ref{sec:theory}.

A complete set of detector specifications and their impact on the physics measurements is another main deliverable of the feasibility study. 
The performance of a number of options for the various detector components of possible future detectors: calorimeters~\cite{Aleksa:2021ztd}, tracking and vertex detectors~\cite{Barchetta:2021ibt}, muon detectors~\cite{Braibant:2021wts}, luminometers~\cite{Dam:2021sdj}, and particle identification devices~\cite{Wilkinson:2021ehf} are being studied to understand how they could meet the requirements. 
More details are given in Chapter~\ref{sec:concepts}. 
The following gives a brief idea of the type of requirements that are under study. 

\begin{itemize}
\item {\bf Higgs and top physics}
A set of requirements have been established by past linear collider studies for Higgs and top physics at 240 
and 340--365\,GeV, arising in particular from the desired performance of particle-flow reconstruction, flavour tagging, or lepton momentum measurement. 
They need to be adapted to take into account the cleaner FCC-ee experimental environment, the smaller beam-pipe radius, and the smaller maximum operating energy~\cite{Azzi:2021gwg}. 
Additional requirements arise, related to the need for a more accurate Higgs boson mass determination~\cite{Azzurri:2021nmy} prior to the $s$-channel Higgs boson production run, and for a more accurate $\PZ\PH$ cross section measurement, to enable a determination of the  Higgs self-coupling. 
The $s$-channel Higgs boson production also leads to demanding requirements on the centre-of-mass energy monochromatisation, while keeping a high luminosity and calling for the most sensitive analysis to separate Higgs boson decays from the huge background of $\Pep\Pem$ annihilation events~\cite{dEnterria:2021xij}. 

\item {\bf Tera-Z challenges}
The high luminosity and large event rate at the \PZ pole (over 100\,kHz), to which simulated data need to be added, turn into considerable challenges for data taking, storage and processing. 
The \PZ lineshape determination, which is based on cross section measurements as a function of the centre-of-mass energy for hadronic and leptonic decays of the \PZ, requires accurate mechanical construction and in-situ alignment of the luminometer and detector end-caps, in view of acquiring precise knowledge of the central tracker and calorimeter acceptance, for the dilepton and diphoton events (and, to a lesser extent, for hadronic events). 
The point-to-point centre-of-mass-energy uncertainties of the resonance scan, which can be verified in particular by means of the muon pair mass and boost reconstruction, are most relevant for the \PZ width and the forward-backward asymmetry measurements. 
The accuracy of the mass reconstruction needed sets stringent constraints on the stability of the momentum reconstruction over time and scan points~\cite{Maestre:2021zvq}. 
With a statistical precision of 4\,\keV for both the \PZ mass and the \PZ width, the centre-of-mass determination is critical for the physics output of the Tera-Z run~\cite{Blondel:2019jmp,note_EPOL}.

\item {\bf Physics with \PW pairs} 
The sub-MeV precision on the \PW mass expected from a scan of the $\PW\PW$ production threshold seems, a priori, more demanding on the centre-of-mass energy calibration, and on the theoretical understanding of the cross section, than on the detector itself. 
At higher energies, the requirements on jet angular calibrations and the possible dependence of the precise knowledge of the composition of hadrons in these jets, however, should be revisited. 
Similarly, requirements on lepton angle, energy reconstruction, and energy scale, should be established~\cite{Azzurri:2021yvl}, although they are probably covered by the requirements from the \PZ run.

\item {\bf Flavour challenges}
The formidable progress of the performance of vertexing devices in the past two decades, paired with the five times smaller beam-pipe diameter with respect to LEP, leads to unprecedented \PQb- and \PQc-tagging performance. 
The resulting expected statistical uncertainties on the flavour EWPOs, such as $R_{\PQb,\PQc}$ or $A_\text{FB}^{\PQb,\PQc}$, 
allow for much larger improvements (by up to a factor of 2000) with respect to the LEP measurements of other EWPOs. 
In addition, the rich FCC-ee flavour programme can be fully exploited only if the detector is equipped with hadron identification covering the effective momentum range at the \PZ resonance~\cite{WilkinsonFCCPhys3} and electromagnetic calorimetry with a superb energy resolution~\cite{AleksanFCCPhys4}.

\item {\bf Tau physics}
The $\Pgt$ measurements listed in Table~\ref{tab:EWPO} (lifetime, mass, and branching fractions) have the potential to achieve a determination of the Fermi constant at the level of a few $10^{-5}$. 
These measurements provide some of the most demanding detector requirements on momentum resolution (for the mass resolution), on the knowledge of the vertex detector dimensions (for the tau lifetime), and on $\Pe/\PGm/\pi$ separation over the whole momentum range (for the leptonic branching ratios). 
The \Pgt-based EWPOs, $R_{\Pgt}^{\PZ}$, $A_\text{FB}^\text{pol,\Pgt}$, and $P_{\Pgt}$, as well as the detailed study of the hadronic spectral functions, require fine granularity and high efficiency in the tracker and electromagnetic calorimeter~\cite{Dam:2021ibi}.

\item {\bf Feebly interacting particles} 
The current heavy neutral lepton (HNL) search strategy is based on a rather conservative signal selection, requiring in particular an HNL decay less than 1\,m away from the interaction point. 
With four IPs, the discovery region extends to the physical regions favoured by the see-saw models of neutrino masses.
It would be possible to extend it by detecting HNL decays further away from the IP, e.g., by making use of the large amount of cavern space surrounding the detector~\cite{Chrzaszcz:2020emg}. 
Other feebly interacting particles, decaying into non-pointing or delayed photons, pose a different set of challenges, which could possibly benefit from a high-precision timing detector. 
See Section~\ref{sec:HNL} for details. 
\end{itemize}

Integrating all these (initial) detector requirements will be a considerable challenge, commensurate with the unique set of measurement opportunities and with the discovery potential offered by FCC-ee.  
A refined list of detector requirements is being derived from a number of case studies inspired by the set of physics benchmark measurements at FCC-ee (encompassing those presented in Table~\ref{tab:EWPO}), as described in Refs.~\cite{SnowmassBenchmarkProcesses,Azzi:2021ylt}. 
The four FCC interaction points will not be too many to cover the resulting variety of detector requirements and physics opportunities. 
More details are given in Chapter~\ref{sec:requirements}.

The experimental environment at FCC-ee, where the incoming beams cross typically a few million times before interacting, is gentle and clean compared to hadron colliders, muon colliders and even linear $\Pep\Pem$ colliders. 
The main challenges have been extensively studied by the Machine Detector Interface group~\cite{Boscolo:2019awb}:
\begin{itemize}
    \item The design of the $\Pep$ and $\Pem$ rings is asymmetric around each interaction region, in order to eliminate the need for strong bending magnets upstream of the collision points and thus minimise the synchrotron radiation background in the detectors.
    \item The large number of bunches and the low level of beamstrahlung radiation result in a low rate of pile-up events, typically $2\times 10^{-3}$ at the \PZ pole and less than $10^{-2}$ at the top energies.
    \item The strong focusing optics requires a short free space (2.2\,m) between the last beam elements. 
    A compensating solenoid and quadrupole assembly that fits in a forward dead cone of no more than 100\,mrad, has been designed and a complete mock-up is being prepared.   
    \item The luminosity monitors situated in front of this assembly could be adapted to the geometry with two beam pipes crossing at an angle of 30\,mrad.
    \item The small beam transverse dimensions allow for a beam pipe of 10\,mm radius and for the first layer of the vertex detector to be installed very close to the vacuum chamber.
    \item In the current design with compensating solenoids, the detector solenoid magnetic field must be limited to 2\,T when operating at the \PZ pole, to avoid a blow up of the vertical beam emittance and a resulting loss of luminosity. 
    Alternative designs with non-local compensation are being studied.  
\end{itemize}

These conditions are favourable on several accounts. 
The 100\,mrad low angle dead cone offers the possibility to extend the detector coverage down to a polar angle of $\cos{\theta}\simeq 0.99$, while the 10\,mm radius beam pipe should be a prime starting point for high efficiency \PQb- and \PQc-flavour tagging against light quarks and gluons. 
The 2\,T magnetic field limit is not a significant handicap since the momentum of the produced partons is typically distributed around 50\,GeV and does not exceed 183\,GeV. 
If needed, an increase of the magnetic field to 3\,T or more can be envisioned at higher energies, without loss of luminosity. 
More details about the challenges pertaining to the machine-detector interface and beam backgrounds are given in Chapter~\ref{sec:mdi}.

The work for the particle physicists is now clearly cut out: design the experimental setup and prepare the theoretical tools that can, demonstrably, fully exploit the capabilities of FCC-ee. 
Experimentation at FCC-ee is both relatively easy and extremely demanding. 
The experimental conditions are clean, with essentially no pile-up, a well-defined and controllable centre-of-mass energy, and benign beamstrahlung and synchrotron radiation effects. 
The sophistication arises from the very richness of the programme. 
Matching the experimental and theoretical accuracy to the statistical precision, and the detector configuration to the variety of channels and discovery cases, will be the real challenge.
\subsection{FCC-hh: The energy-frontier collider with the broadest exploration potential}

The physics landscape emerging from the FCC-ee, complemented by the HL-LHC exploration, will significantly raise the bar for the performance requirement of subsequent colliders. 
Several challenges remain open.

\begin{itemize}
\item The extension of Higgs boson property measurements to several processes driven by smaller effective couplings 
(e.g., rare decays such as $\PH \to \PGg\PGg$, $\PGmp\PGmm$, $\PZ\PGg$), 
and to processes requiring higher energy (e.g., $\PQt\PAQt\PH$ and $\PH\PH$ production).
\item The extension of studies of EW dynamics to a regime well beyond the scale of the EW symmetry breaking.
\item The direct exploration of the multi-TeV energy scale, 
to identify and directly study the sources of possible deviations found in the FCC-ee precision measurements, 
and to systematically extend the mass reach for a broad spectrum of BSM theories, 
particularly those whose impact on low-energy EW and Higgs observables is suppressed and those whose primary production processes require initial-state gluons.
\end{itemize}

The 100\,TeV FCC-hh provides the most complete and effective facility to tackle these challenges. 
The study of the FCC-hh physics potential is very mature; it occupied the first phase of the efforts towards the FCC CDR, thanks to a world-wide effort whose results have been documented in extensive reports~\cite{Arkani-Hamed:2015vfh,Mangano:2017tke}, partly summarised in Volumes~1 and~3 of the CDR~\cite{fcc-phys-cdr,FCC-hhCDR}. 
Some of the key results are recalled here. 
Recent preliminary studies of physics performance at different collider energies, in the \mbox{80--120}\,TeV range, have been done in the context of this feasibility study, and are reported in Section~\ref{sec:PhysicsCaseHHscenarios}. 
Section~\ref{sec:PhysicsCaseHHFPF} outlines the opportunities offered by a forward-physics facility, which would expand the landscape of FCC-hh physics measurements presented in the CDR.

On the Higgs boson side, the data sample produced by FCC-hh of over 20~billion Higgs bosons will bring the absolute determination of the couplings to muons, to photons, to the top quark, and to $\PZ\PGg$ below the per-cent level. 
The large production yield of Higgs bosons at large transverse momentum allows measurements to be performed in kinematic regions with optimal signal-to-background ratios and reduced experimental systematic uncertainties. 
The possibility to accurately measure the ratio of those couplings to couplings precisely measured at the FCC-ee 
(e.g., the $\PH\PZ\PZ^{*}$ and $\PQt\PAQt\PZ$ couplings), 
enables the FCC-hh to deliver sub-percent absolute measurements of those couplings, which are beyond the reach of lepton colliders. 
The large $\PH\PH$ production rate, furthermore, will bring the uncertainty on the Higgs self-coupling below the 5\% level, even with conservative estimates of the experimental systematic uncertainties. 
The studies of Higgs boson production in very-high $Q^2$ processes, whether in direct, associated, or vector boson fusion/scattering production, 
test the existence of higher-dimensional effective operators in ways that are complementary to what is accessible even at the highest-energy lepton colliders. 

The direct search for new particles extends the mass reach to the several tens of TeV range, in particular around 40\,TeV for $s$-channel produced EW or coloured resonances, and in the 10--20\,TeV range for pair-produced strongly-interacting particles, such as squarks, stops, vector-like top partners, and gluinos. 
Weakly interacting particles can be probed up to 1.5--5\,TeV, depending on their weak multiplet assignment, allowing in particular the theoretically limited spectrum of possible WIMP dark-matter candidates to be covered.

Likewise, the search for partners of the Higgs boson extends to the multi-TeV region, a sensitivity that, together with the precise determination of the Higgs boson self-coupling, constrains, in a unique way, extensions of the Higgs sector leading to a strong first-order EW phase transition in the early universe.

Last but not least, the FCC-hh is a unique facility to extend the study of the less known manifestation of the SM dynamics, namely the behaviour of high-density and high-temperature strongly interacting plasmas, 
through the analysis of data collected in central lead-lead collisions at nucleon-nucleon collision energies up to $\sqrt{s} \approx 40$\,TeV~\cite{Dainese:2016gch}. 
Despite the significant progress made over the last decades at the SPS, at RHIC, and at the LHC regarding our understanding of the properties of the hot QCD medium created in high-energy heavy-ion collisions, grounded on increasingly precise and diverse experimental measurements, much remains unknown about the physics behind this most exotic area of QCD.
Details of the concrete physics potential of heavy-ion running at the FCC-hh are provided in Refs.~\cite{Dainese:2016gch,Mangano:2017tke}. 
Aside from its intrinsic and undisputed scientific value, this part of the physics programme, exclusively accessible at hadron colliders, offers the opportunity to enlarge the community of scientists attracted to the FCC, providing them the sole facility that can continue and extend the study of the high-density and high-temperature QCD medium produced in high-energy heavy ion collisions and the associated collective QCD phenomena, presently being actively pursued at the LHC.

The precision programme of the known elements of the SM should be complemented by a direct exploration of the unknown. Exploration requires breadth. 
To understand the properties of the visible Universe, large-scale structure surveys are performed, gravitational waves with pulsar timing and much more are sought. 
In the same vein, to understand the microverse this must be explored with as many different microscopes as possible. 
The FCC-ee will directly explore the evasive and weakly-coupled light new-physics frontier, 
while the FCC-hh will offer a direct way to scan the new energy frontier, above the TeV scale. 
Together, the FCC-ee and the FCC-hh complement each other in this exploration of the unknown in a more comprehensive way than any other energy frontier projects presently under consideration.

\cleardoublepage
\section{Specificities of the FCC physics case}
\label{sec:physics}
\subsection{The impact of FCC in particle physics}
\label{sec:PhysicsCaseIntro}

Particle physics is the science of studying fundamental processes on the smallest accessible scales. 
The last sixty years of particle physics have led to a radical overhaul of the understanding of the Universe, 
culminating in the landmark discovery of the Higgs boson in 2012. 
In this context, FCC may be viewed as a general-purpose particle physics facility, aimed at completing the extensive testing of the SM, 
exploring remaining open questions, and finding leads towards understanding the fundamental origins of the Universe. 
The combination of the most powerful $\epem$ Higgs and electroweak factory with the ultimate 100\,TeV-class hadron collider 
is unique in its capacity to carry out a new phase of exploration in new regimes of precision and energy, 
and in a controlled and repeatable experimental environment. 
More specifically, FCC will shed light on the many fundamental open questions about how the Universe came to be.

\textbf{What is the origin of the Higgs boson?}  
The Higgs boson discovered at LHC, with a mass of $125.20 \pm 0.11$\,GeV, 
is, so far, consistent with the SM version of Electroweak Symmetry Breaking (EWSB), 
which predicts the full Higgs phenomenology with no free parameter. 
These predictions must be tested with the highest possible precision with FCC-ee and FCC-hh. 
Indeed, there are many reasons to believe that the underlying microscopic physics, from which the Higgs boson emerges, 
is associated with some of the deepest mysteries in particle physics.  
One such mystery concerns the origin of the Higgs boson mass and the naturalness puzzle, 
which FCC is uniquely placed to address by a combination of precision Higgs coupling measurements at the per-mil level 
and electroweak precision measurements improved by a factor of 50 or much more with respect to the current precision,
with direct high-energy exploration to comprehensively probe symmetry-based explanations for the electroweak hierarchy.

\textbf{What is the origin of the presence of matter in the Universe?} 
The predominance of matter over antimatter, a.k.a.\ the baryon asymmetry of the universe (BAU), calls for both C and CP violation, 
as well as an out-of-equilibrium epoch in cosmological history. 
The SM EWSB does not suffice to predict a significant BAU, but particle physics can propose two hypotheses to accommodate this observation. 
A first possibility is that the Higgs potential is modified via a change of the Higgs self-coupling, leading to a first-order electroweak phase transition (EWPT). 
The FCC measurement of the Higgs boson self-coupling will reach a level of precision deemed adequate to clarify 
if the EWPT played a role in setting an out-of-equilibrium phase, 
a necessary condition for creating the observed matter-antimatter asymmetry. 
Another possibility is that the Higgs boson couples to neutrinos, 
which would generate a fermion-number-violating Majorana mass term, 
leading in turn to leptogenesis and, possibly, to the right amount of BAU. 
With its very high luminosity, FCC is best suited for the direct observation of the resulting heavy neutral leptons (HNL).

\textbf{What is the origin of mass and flavour?} 
The Higgs mechanism is responsible for generating mass, but the reason for the pattern of Yukawa couplings is yet to be understood. 
The FCC unique exploration potential of flavour physics and of new symmetries and forces 
would lead to a deeper understanding of approximate conservation laws,
such as baryon and lepton number conservation 
(or the absence thereof in case of Majorana neutrinos);
would probe the limits of lepton flavour universality and violation; and 
could reveal new selection rules governing the fundamental laws of nature. 
A precise comparison of the  Yukawa couplings for the three fermion families will also be crucial in this endeavour. 
The measurements of the muon and top Yukawa couplings at FCC-hh can be compared to the tau and charm Yukawa coupling measurements at FCC-ee with sub-per-cent precision. 
A first measurement of the strange Yukawa coupling is also possible at FCC-ee. 
Finally, an FCC-ee run at the $s$-channel Higgs resonance 
uniquely offers sensitivity to the electron Yukawa coupling, tantalisingly close to its SM value.  

\textbf{What is the nature of dark matter?} 
The unprecedented luminosity and detector environment of FCC-ee make it uniquely sensitive to exploring weakly-coupled dark sectors. 
Dark matter candidates, such as heavy axions, dark photons, etc., provide long-lived particle signatures.  
If dark matter is a doublet or triplet weakly interacting massive particle (WIMP), 
FCC-hh will cover the entire parameter space up to the upper mass limit for a thermal relic. 
A range of complementary detector facilities could also be envisioned to extend the FCC capabilities for neutrino physics, long-lived particles, and forward physics.

\textbf{What lies beyond the Standard Model?} 
An impressive number of electroweak precision observables (EWPOs) will be measured by FCC-ee, 
with accuracies ranging from a factor 50 (e.g., for the \PW mass) to a factor 1000 (e.g., for $R_{\PQb}$) better than today (Table~\ref{tab:EWPO}). 
This new precision regime will allow the effect of further weakly-coupled heavy particles to be detected up to masses of several tens of TeV 
(and much more in the non-decoupling configuration). 
Alternatively, this precision will bring sensitivity to particles mixing with the SM fermions in one part in 100\,000, 
up to very high mass scales (more than 1000\,TeV for HNL). 
The SM, possibly extended with additional light degrees of freedom, 
might be a low-energy effective field theory (EFT) approximation of a microscopic, ultraviolet (UV) theory from which it originates. 
The FCC-ee will improve the sensitivity to all EFT Wilson coefficients by one to several order(s) of magnitude, 
while FCC-hh can further extend the reach of direct and indirect exploration in a complementary way. 
The combined FCC programme is the most powerful general survey of this uncharted EFT territory 
and of the existence of new particles beyond the known fermions and bosons.

\textbf{Relation with exotic astrophysical and cosmological signals} 
Stochastic gravitational waves from cosmological phase transitions or astrophysical signatures of high-energy gamma rays 
are examples of phenomena that can arise due to exotic new physics which is, by contrast, 
directly accessible only to a facility offering the highest possible energy, such as FCC. 
Examples include confining new physics in a dark sector, decaying or co-annihilating TeV-scale WIMPs, or a modified EWPT. 
More precise top quark and Higgs boson mass measurements would also have important implications in the understanding of the SM vacuum metastability.

As an all-in-one-facility, FCC provides the versatility, redundancy, and cross-correlations ne\-cessary to identify the fundamental origins of new phenomena. 
In general, its varied range of precise measurements of particle properties and interaction dynamics 
promises a set of guaranteed results with an unrivalled breadth with respect to other potential future facilities.

This chapter presents the unique role of FCC to address these fundamental questions, with a special focus on: 
the combined characterisation of the Higgs boson at FCC-ee and FCC-hh (Section~\ref{sec:PhysicsCaseHiggsEW}); 
the direct discovery potential resulting from the FCC-ee clean environment, high integrated luminosity, and large acceptance rates of the detectors (Section~\ref{sec:PhysicsCaseBSM}); 
and the excellent opportunities in flavour physics (Section~\ref{sec:PhysicsCaseFlavour}). 
The advantages of FCC-hh over multi-TeV lepton colliders to complement FCC-ee are presented in Section~\ref{sec:PhysicsCaseHHvsHEL}. 
Finally, the physics reach of FCC-hh as a function of its centre-of-mass energy, 
and the possibility of a forward-physics facility at FCC-hh, 
are briefly discussed in Sections~\ref{sec:PhysicsCaseHHscenarios} and~\ref{sec:PhysicsCaseHHFPF}.

\subsection{Characterisation of the Higgs boson: role of EW measurements and of FCC-hh}
\label{sec:PhysicsCaseHiggsEW}

The Higgs boson is certainly a central piece of the current understanding of the fundamental structure of matter and physics laws. 
Its discovery in 2012, and the rich measurement programme over the past thirteen years, 
have been successful milestones in high-energy physics that, maybe more importantly, 
enabled the formulation of relevant questions about the nature of new physics beyond the weak scale, 
in very concrete and often quantitative terms. 
As already shown in various studies~\cite{deBlas:2019rxi, DeBlas:2019qco, deBlas:2022ofj}, 
FCC-ee offers fantastic opportunities for learning more about the Higgs boson, 
beyond what can be achieved at HL-LHC (Table~\ref{tab:HiggsKappa3}) or at other proposed Higgs factories~\cite{Blondel:2024mry}.

\begin{table}[ht]
\centering
\caption{Expected 68\%~CL relative precision of the $\kappa$ parameters (Higgs couplings relative to the SM) 
and of the Higgs boson total decay width $\Gamma_{\PH}$, 
together with the corresponding 95\%~CL upper limits on the untagged (undetected events), $\cal{B}_{\text{unt}}$, 
and invisible, $\cal{B}_{\text{inv}}$, branching ratios at HL-LHC, FCC-ee (combined with HL-LHC), and the FCC integrated programme.
For the HL-LHC numbers, a $|\kappa_V| \leq 1$ constraint is applied (denoted with an asterisk), 
since no direct access to $\Gamma_{\PH}$
is possible at hadron colliders;
this restriction is lifted in the combination with FCC-ee.
The `--' indicates that a particular parameter has been fixed to the SM value, due to lack of sensitivity.
From Ref.~\cite{deBlas:2019rxi}, updated with 4~IPs, the baseline luminosities of Table~\ref{tab:seqbaseline}, and the most recent versions of the data analysis. 
For some of the entries, the $\kappa$ precision starts being limited by the projected SM parametric uncertainties, e.g. in $m_{\PQb}$~\cite{Freitas:2019bre}. For these entries, the precision obtained by neglecting such parametric uncertainties is also reported (separated by a $/$).
}
\label{tab:HiggsKappa3}
\begin{tabular}{ c   c    c     c }
\toprule
Coupling & HL-LHC 
& FCC-ee 
& FCC-ee $+$ FCC-hh\\ 
\midrule 
$\kappa_{\PZ}$ (\%) &  1.3$^\ast$ 
& 0.10  
& 0.10 \\ 
$\kappa_{\PW}$ (\%)  &   1.5$^\ast$  
& 0.29 
& 0.25 \\
$\kappa_{\PQb}$ (\%)  & 2.5$^\ast$  
& 0.38 / 0.49 
& 0.33 / 0.45 \\ 
$\kappa_{\Pg}$ (\%)    &   2$^\ast$ 
& 0.49 / 0.54 
& 0.41 / 0.44 \\
$\kappa_{\PGt}$ (\%)   & 1.6$^\ast$   
& 0.46 
& 0.40 \\
$\kappa_{\PQc}$ (\%)   & --  
&0.70 / 0.87 
& 0.68 / 0.85 \\
$\kappa_{\PGg}$ (\%)  &  1.6$^\ast$  
& 1.1 
& 0.30 \\
$\kappa_{\PZ\PGg}$ (\%)    & 10$^\ast$  
& 4.3
& 0.67\\
$\kappa_{\PQt}$ (\%)   & 3.2$^\ast$ 
& 3.1 
& 0.75 \\
$\kappa_{\PGm}$ (\%)     & 4.4$^\ast$  
& 3.3 
& 0.42 \\
$\vert \kappa_{\PQs}\vert$ (\%) & -- & $_{-67}^{+29}$ & $_{-67}^{+29}$\\
$\Gamma_{\PH}$ (\%) & -- 
& 0.78
& 0.69 \\ 
$\cal{B}_{\text{inv}}$ ($<$, 95\% CL) 
& $1.9 \times 10^{-2}$ $^\ast$   
& $5 \times 10^{-4}$ 
& $2.3 \times 10^{-4}$ \\
$\cal{B}_{\text{unt}}$ ($<$, 95\% CL) & $4 \times 10^{-2}$ $^\ast$ 
& $6.8\times 10^{-3}$ 
& $6.7\times 10^{-3}$\\
\bottomrule
\end{tabular}
\end{table}

In a record time, FCC-ee will bring the Higgs programme into a sub-percent precision area, 
allowing the classical as well as the quantum mechanical effects of new physics on the Higgs couplings to be probed and, 
from the pattern of the deviations, new selection rules distinguishing different models to be discovered. 
Concrete examples proving the relevance of the sub-percent precision target can be found in Table~5 of Ref.~\cite{Barklow:2017suo}, 
where several BSM models beyond the reach for direct discovery at HL-LHC are considered, 
and for which per-cent level deviations in the Higgs couplings are predicted. 
A sub-per-cent precision is a clear necessary target to expose such deviations.
As mentioned in Section~\ref{sec:overview_ee}, the Higgs precision programme at 240\,GeV (365\,GeV) 
can be achieved within three (eight) years of operation with FCC-ee, 
while other colliders considered at CERN would need half a century to reach a similar precision~\cite{Blondel:2024mry}.

These phenomenological projections are now being confirmed by independent experimental studies, 
with different detector set-ups~\cite{HiggsNoteRecoil, note_Higgs_to_hadrons, HiggsInvisibleNote}. 
Further directions in the Higgs precision programme also need to be more systematically investigated beyond what was done so far, 
in particular in the context of specific flavour scenarios or considering BSM sources of CP violation. 
This document, instead, emphasises the benefit of the interplay between Higgs and electroweak measurements, 
a specificity of FCC-ee that was not discussed in detail in the FCC CDR~\cite{fcc-phys-cdr, fcc-ee-cdr} 
and has been studied afterwards~\cite{DeBlas:2019qco, deBlas:2022ofj}.

The interpretation of current Higgs boson measurements at LHC is so far not hindered by the limited precision of the electroweak measurements at LEP and SLC. 
With FCC-ee targeting an order-of-magnitude improvement in the precision of Higgs boson properties in the main channels, 
the current (experimental and theoretical) precision on electroweak quantities would become a limitation. 
The \PZ-pole run of FCC-ee is instrumental in avoiding contamination from electroweak coupling uncertainties in the Higgs boson characterisation. 
If the electroweak symmetry is linearly realised in the SM fields, 
the interplay between the Higgs and electroweak sectors is even deeper~\cite{Gupta:2014rxa}.
Indeed, $\epem \to \PWp\PWm$ production is then sensitive to some of the same new-physics effects as Higgs boson production and decay processes, 
making both types of measurements complementary.

\begin{figure}[t]
\centering
\includegraphics[width=0.95\textwidth]{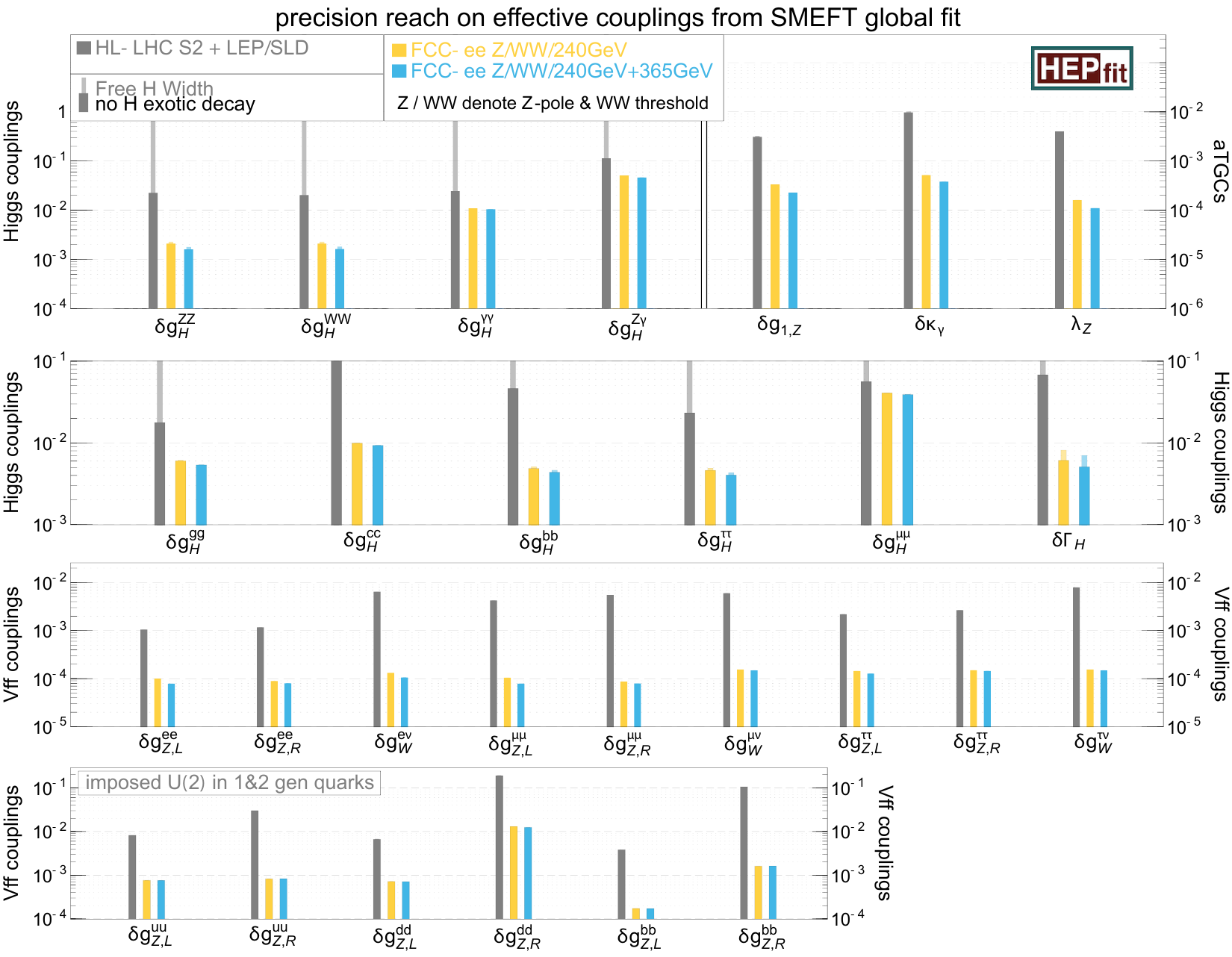}
\caption{Results of a global SMEFT fit to HL-LHC (grey) and FCC-ee (yellow, up to 240\,GeV, and blue, up to 365\,GeV) data, 
interpreted in terms of the 68\% probability sensitivity to Higgs and electroweak effective couplings. 
Adapted from Ref.~\cite{deBlas:2022ofj}.}
\label{fig:Global_Fit}
\end{figure}

The Standard Model Effective Field Theory (SMEFT) framework  is adopted, truncated to operators of dimension six~\cite{Buchmuller:1985jz, Grzadkowski:2010es}. 
The SMEFT is an appropriate framework for enumerating and quantifying the different possibilities in a relatively model-independent way. 
It assumes that new physics arises at a scale $\Lambda$, significantly above the electroweak scale, with the Higgs boson embedded in a $\text{SU}(2)_L$ doublet. 
The current status of the global SMEFT fit is shown in Fig.~\ref{fig:Global_Fit}, adapted from Ref.~\cite{deBlas:2022ofj}. 
In this figure, the results of the fit for the different dimension-six operators 
affecting at leading order either the electroweak processes 
(including anomalous triple gauge couplings, aTGCs, and boson-fermion couplings, Vff)  
or the Higgs processes, or both simultaneously, 
are projected on a more physically meaningful set of effective couplings capturing the effects of new physics.
More details can be found in Ref.~\cite{deBlas:2022ofj}.

\begin{figure}[ht]
\centering
\includegraphics[width=0.80\textwidth]{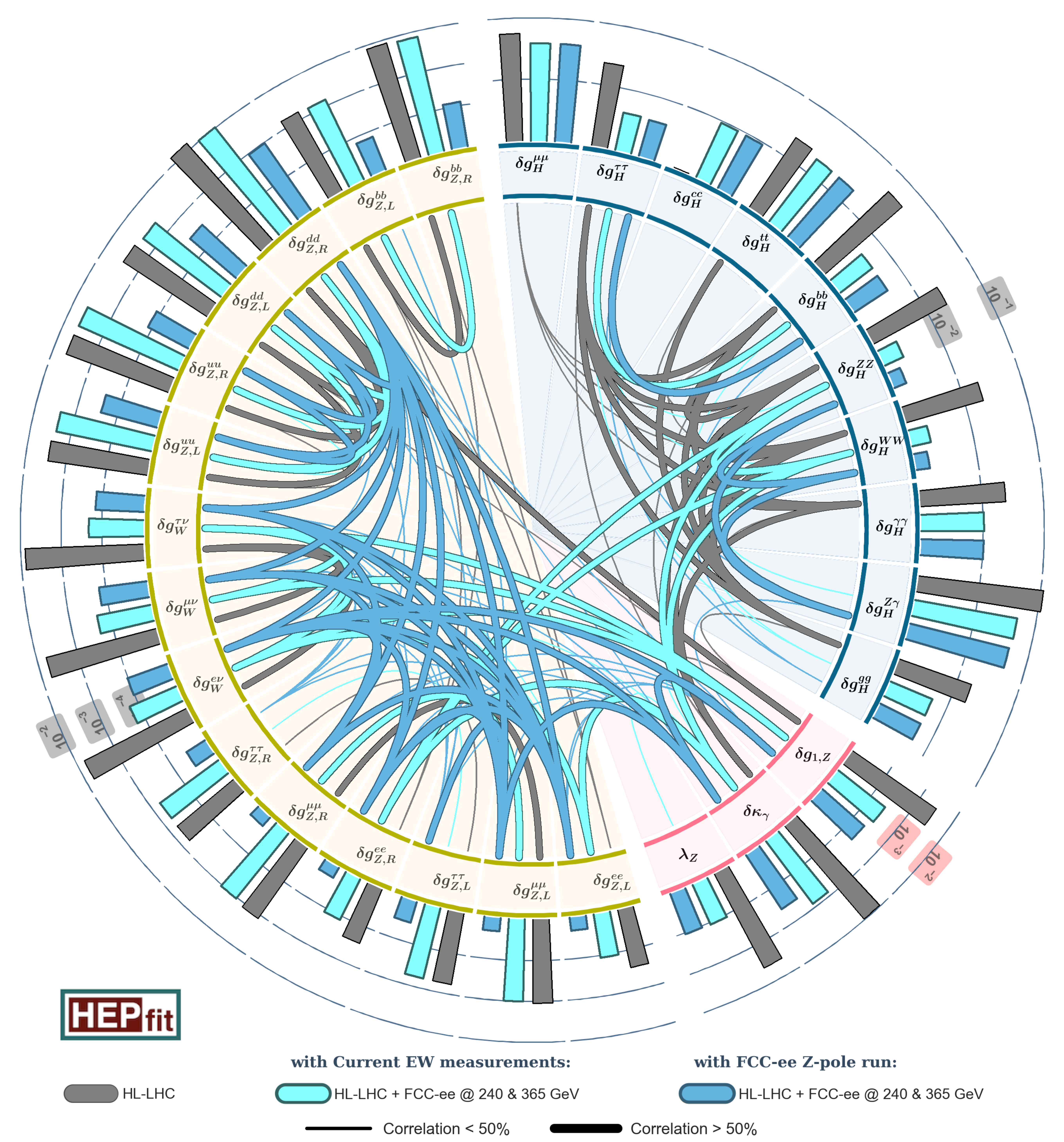}
\caption{Correlations between the determination of different EW/aTGC/Higgs interactions from a fit to future projections at HL-LHC and FCC-ee within the SMEFT framework. 
The bars on the exterior of the circle indicate the expected sensitivity in the determination of corresponding coupling (with different scales for the different sectors). 
The light/dark blue lines in the interior of the circle show the results excluding/including the FCC-ee programme of EW measurements at the \PZ pole. 
The \PZ-pole run is essential to isolate the Higgs measurements (there is no dark blue line connecting the Higgs and EW sectors) 
and to ensure that the extraction of the Higgs couplings is not hindered by the uncertainties on the EW coupling determination.
Adapted from Refs.~\cite{DeBlas:2019qco,deBlas:2022ofj}.}
\label{fig:Corr_Plot}
\end{figure}

The interplay between Higgs and electroweak measurements is illustrated in Fig.~\ref{fig:Corr_Plot}, 
which shows the expected precision in the effective coupling determination. 
The correlations are displayed as internal lines of variable thickness and are visibly reduced 
when including \PZ-pole data at FCC-ee (dark blue) on top of the current electroweak measurements (light blue). 
The importance of \PZ-pole measurements is summarised below, followed by a discussion of the importance of the diboson process for Higgs physics.

\begin{figure}[ht]
\centering
\includegraphics[width=0.80\textwidth]{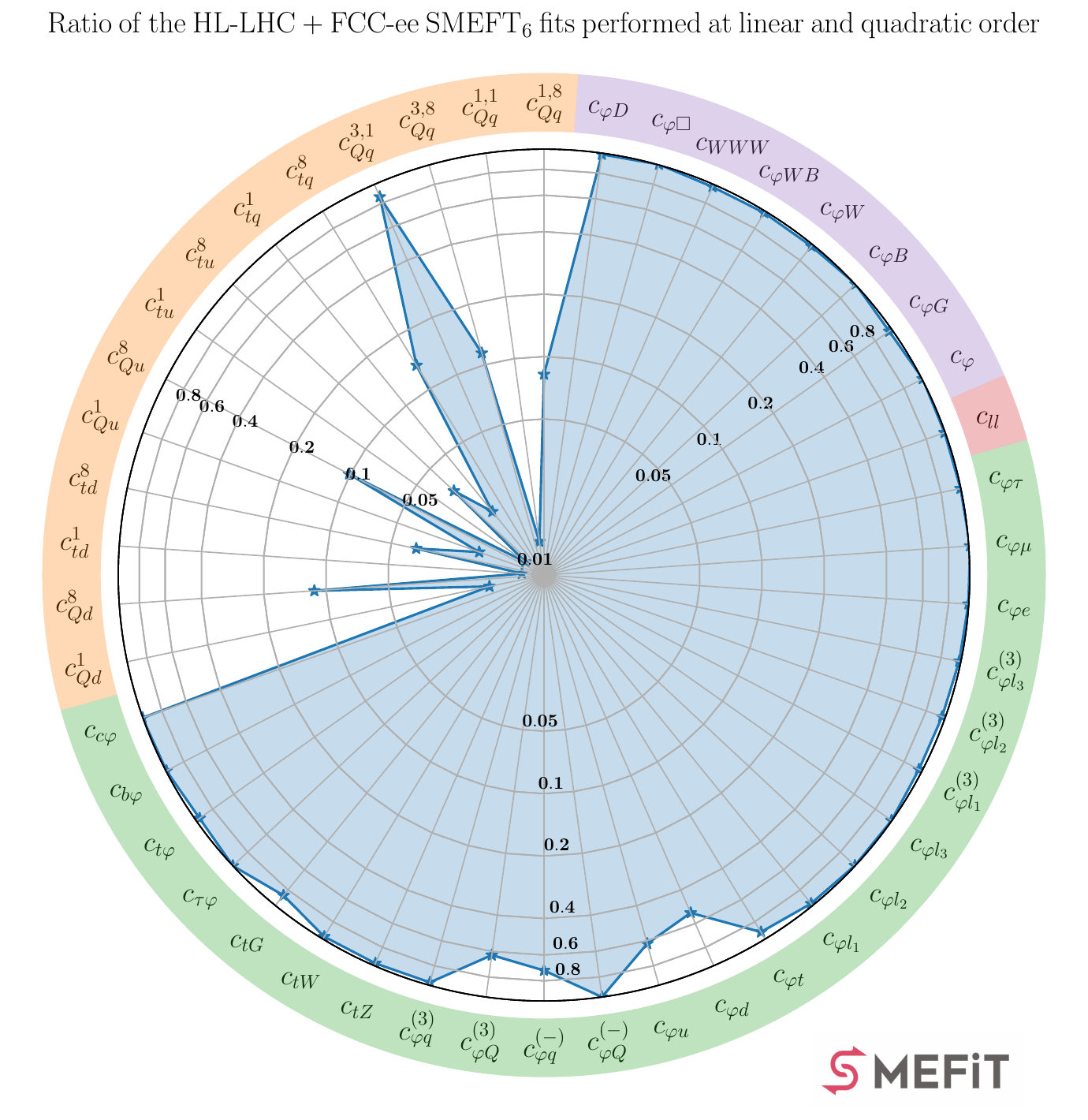}
\caption{Impact of quadratic contributions from dimension-6 operators, i.e., at ${\cal O}(\Lambda^{-4})$, in the SMEFT fit of Ref.~\cite{Celada:2024mcf}. 
A $U(2)_Q \times U(2)_u \times U(3)_d \times \left( U(1)_L \times U(1)_e \right)^3$ flavour symmetry is assumed. 
For each of the different Wilson coefficients considered, 
the blue star indicates the ratio of the bound obtained in the pure dimension-6 (linear) fit to that including quadratic dimension-six contributions. 
The difference between the two fits is minimal for most operators entering in FCC-ee observables at tree level, 
i.e., leaving aside the purely hadronic top-quark operators in the top-left quadrant. The FCC-ee fit includes EWPOs at the $\PZ$-pole, light fermion-pair production, Higgs boson production ($\PZ\PH +$ VBF), diboson production, and top pair production.}

\label{fig:ImpactQuad_Plot}
\end{figure}

The SMEFT results discussed in this section were obtained from fits with only linear effects in $\Lambda^{-2}$, consistently with the dimension-6 expansion. 
To estimate the theory uncertainty associated with the neglected higher-order terms in the EFT expansion, 
Fig.~\ref{fig:ImpactQuad_Plot} shows the ratio of the bounds on the Wilson coefficients obtained in the linear case 
to those including quadratic contributions from dimension-6 operators, 
which are formally of the same order as dimension-8 contributions. 
These results derive from the HL-LHC$+$FCC-ee SMEFT fit of Ref.~\cite{Celada:2024mcf}.
All the displayed operator coefficients can be accessed at FCC-ee, with the exception of the operators in the top-left quadrant, 
which enter in top-quark measurements at HL-LHC but do not contribute to FCC-ee observables at tree level. 
So, in almost all cases, the results of the linear fit are unchanged after the addition of quadratic effects. 
This observation can be used as an indication that the uncertainty associated to effects of ${\cal O}(\Lambda^{-4})$ is well controlled 
by FCC-ee precision measurements and is expected to be small. 
Another way to gauge the usefulness of the SMEFT analysis is to 
compare the bound on the scale of new physics inferred from the fit of the dimension-6 Wilson coefficients 
to the actual limits set in explicit BSM models, as will be reported in the next section. 
Of course, this does not exclude that (non-decoupling) new physics could still exist at much lower scales.

\subsubsection{Impact of \texorpdfstring{\PZ}{Z}-pole measurements in Higgs couplings determination}

The importance of having precise \PZ-pole measurements for the extraction of Higgs couplings from future $\epem$ Higgs data is clear, 
given that the $\PZ\Pe\Pe$ vertex also enters in $\PZ\PH$ production. 
A precise-enough extraction of the $\PZ\Pe\Pe$ vertex is therefore needed to avoid extra uncertainties from new physics effects in these interactions. The left-handed and right-handed $\PZ\Pe\Pe$ couplings can be extracted from the $\PZ \to \epem$ partial decay width and the left-right asymmetry parameter $A_{\rm e}$, determined at FCC-ee from the hadronic-to-leptonic cross-section ratios, the dilepton forward-backward asymmetries, and the tau polarisation forward-backward asymmetry, measured at the Z pole. This $\PZ\Pe\Pe$ coupling measurement also helps mitigating the (centre-of-mass-energy growing) new-physics effects in the $\PH\PZ\Pe\Pe$ interactions, 
directly correlated with the new-physics contributions to the $\PZ\Pe\Pe$ vertex in the SMEFT formalism.

\begin{figure}[ht]
\centering
\includegraphics[width=\textwidth]{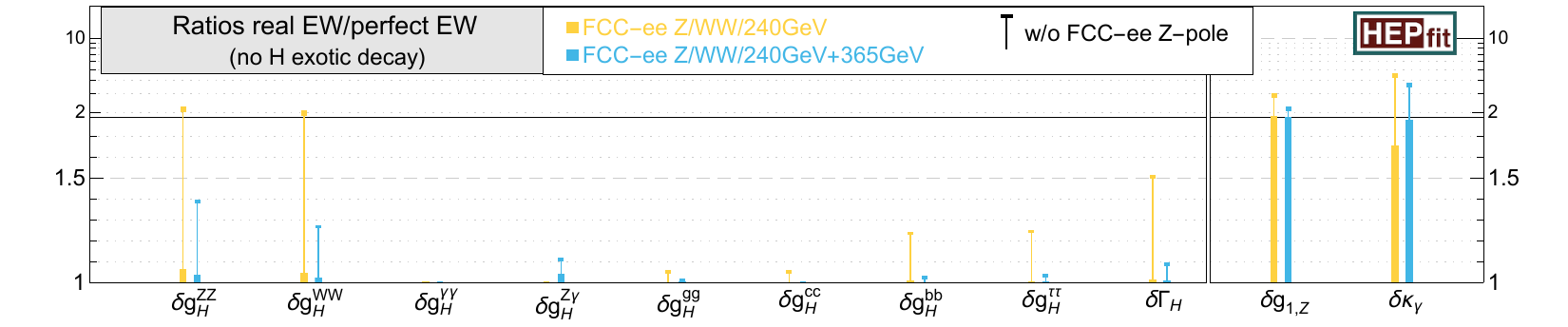}
\caption{Impact of \PZ-pole measurements on the Higgs coupling, Higgs decay width, and anomalous triple gauge-boson coupling ($\delta g_{1,\PZ}$ and $\delta \kappa_\gamma$) precision from the SMEFT fit interpretation. 
The ratio to the precision that would be obtained with infinitely precise \PZ-pole measurements is shown for fits 
that either exclude (`T' bars) or include (solid bars) the projected FCC-ee \PZ-pole run measurements, 
with the Higgs programme at 240\,GeV (yellow bars) and the additional data at 365\,GeV (blue bars). 
The colour code is the same as in Fig.~\ref{fig:Global_Fit}. Adapted from Ref.~\cite{deBlas:2022ofj}. 
}
\label{fig:ZpoleHiggsUnc}
\end{figure}

The impact of \PZ-pole measurements in Higgs coupling fits was studied in Ref.~\cite{DeBlas:2019qco} and, more recently, in Ref.~\cite{deBlas:2022ofj}. 
As a result of the reduced correlations between the $\PH\PZ\PZ$ and $\PZ\Pe\Pe$ couplings, displayed in Fig.~\ref{fig:Corr_Plot}, 
and of the corresponding reduction of the uncertainty in the $(\PH)\PZ\Pe\Pe$ interactions, 
the precision of the extraction of the Higgs couplings is improved, 
as illustrated in Fig.~\ref{fig:ZpoleHiggsUnc}.
In this figure, the deterioration in the precision of the Higgs coupling extraction 
(with respect to an extraction with perfectly known \PZ couplings, dubbed \emph{perfect EW} in the figure) 
is shown in two configurations: 
one with the future FCC-ee \PZ-pole measurements, for which the Higgs coupling precision is close to the perfect EW case; 
and the other, limited to the  precision of LEP/SLC \PZ-pole measurements, 
for which a deterioration of up to a factor of two in precision  would be observed 
(for the Higgs programme at 240\,GeV). 
Data at 365\,GeV already significantly alleviate the impact of \PZ-pole measurements on most couplings, 
though a 30--40\% precision improvement is still observed for the Higgs couplings to the \PZ and \PW bosons.

\begin{figure}[ht]
\centering
\includegraphics[width=0.48\textwidth]{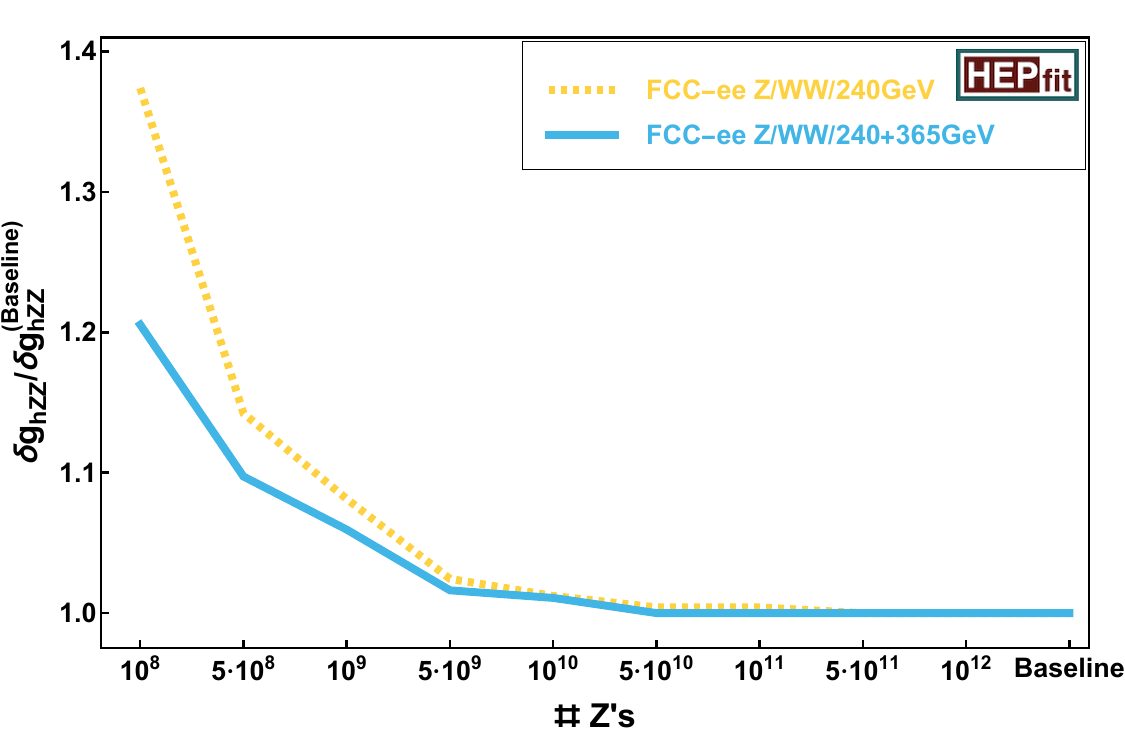}
\includegraphics[width=0.48\textwidth]{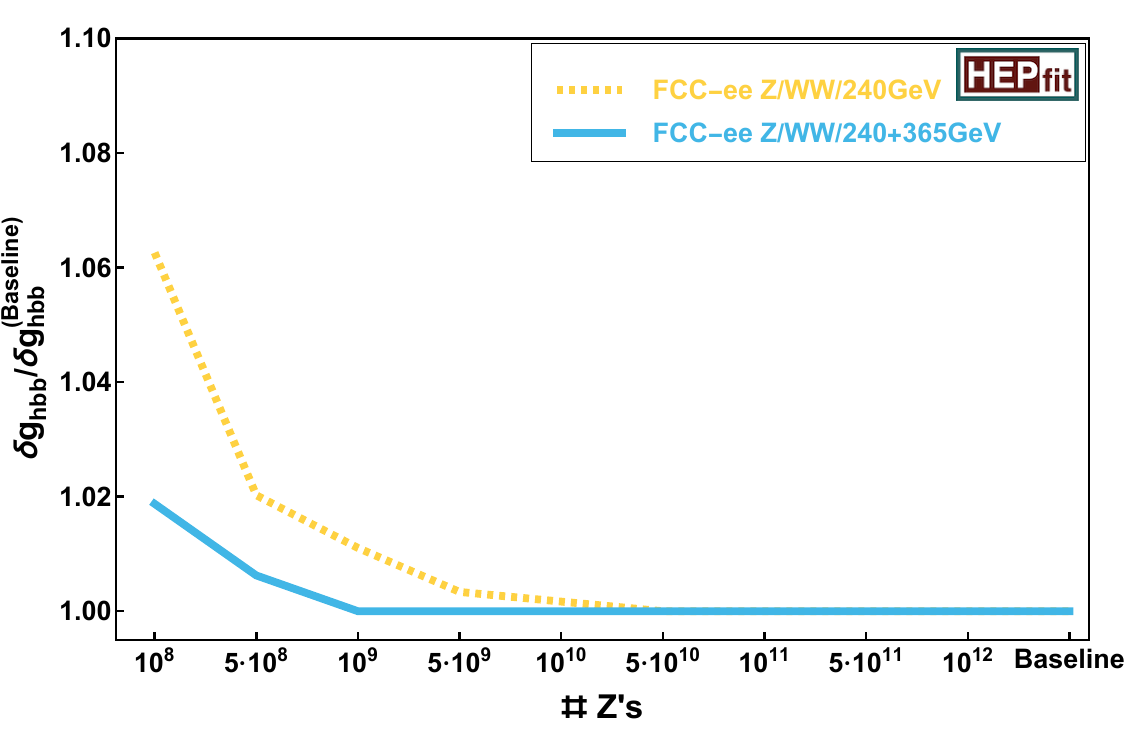}
\caption{Estimated deterioration in the precision of the $\PH\PZ\PZ$ (left) and $\PH\PQb\PQb$ (right) couplings 
as a function of the number of \PZ's produced at FCC-ee,  
with respect to the baseline of $6\times 10^{12}$ events, 
for the FCC-ee programme up to 240\,GeV (yellow) and up to 365\,GeV (blue).}
\label{fig:ZpoleHiggsUncVar}
\end{figure}

Not all the $6 \times 10^{12}$ \PZ bosons expected during the \PZ-pole run are needed 
to reach the precision on the $\PZ\Pe\Pe$ vertex required to fully mitigate the EW contamination in the Higgs couplings determination.
An estimate of the number of events needed to reduce such contamination to a level that does not hinder the Higgs couplings extraction 
is displayed in Fig.~\ref{fig:ZpoleHiggsUncVar}, 
which shows the deterioration in the precision of the most precise $\PH\PZ\PZ$ and $\PH\PQb\PQb$ couplings with respect to the full statistics of the FCC-ee \PZ-pole run, 
as a function of the number of events.
From the perspective of Higgs couplings, a few $10^9$ \PZ's, 
equivalent to less than a day of operation at the \PZ pole, 
are enough to saturate the Higgs coupling precision from the 240 and 365\,GeV runs.

\subsubsection{Impact of diboson measurements in Higgs couplings determination}

Together with gauge invariance, 
the assumption that electroweak symmetry is linearly realised in SMEFT implies a series of correlations 
between the new-physics effects on the electroweak properties of the SM particles, 
for example the aforementioned correlation between contributions to the $\PZ\Pe\Pe$ and $\PH\PZ\Pe\Pe$ interactions.
Another correlation relates modifications of the Higgs couplings to anomalous triple-gauge couplings (aTGCs): 
two out of the three aTGCs ($\delta g_{1,\PZ}$ and $\delta \kappa_\gamma$) are closely related to the Higgs couplings to vector bosons.

\begin{figure}[ht]
\centering
\includegraphics[width=\textwidth]{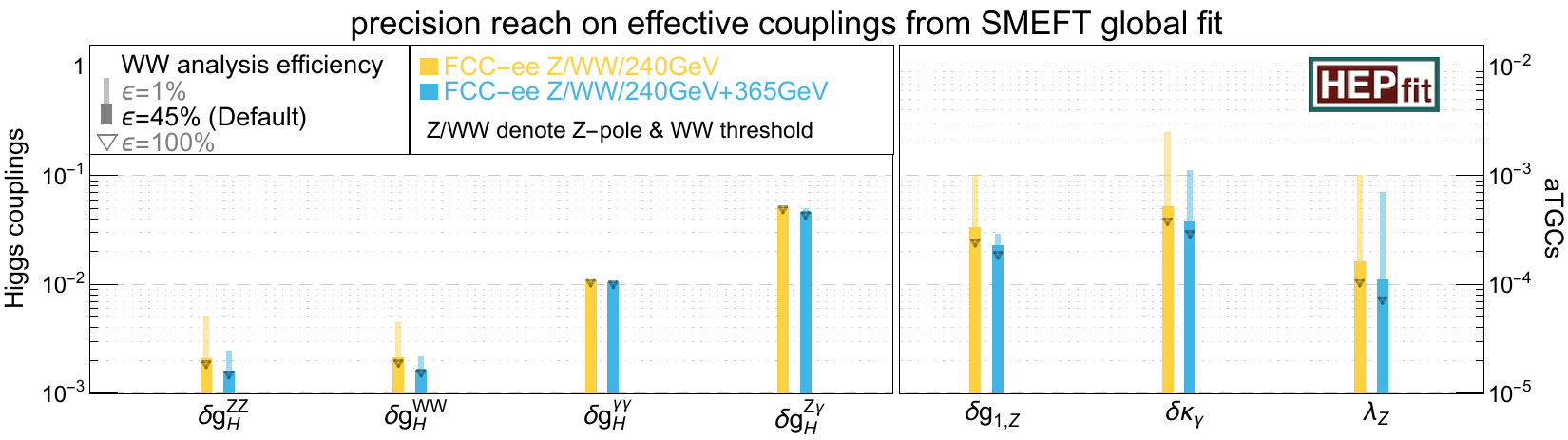}
\caption{Relative precision of the determination of the Higgs couplings and aTGCs in the SMFET fit, 
as a function of the selection efficiency assumed in the $\epem \to \PWp\PWm$ study: 1\% (light shades), 45\% (solid bars, chosen as default value), and 100\% (triangles). The sensitivity comes mainly from the high-energy runs and the dependency at the $\PW\PW$ threshold remains small.
From Ref.~\cite{DeBlas:2019qco}.}
\label{fig:WWefficiency}
\end{figure}

As shown in Ref.~\cite{Falkowski:2015jaa} for existing measurements and in Refs.~\cite{DeBlas:2019qco,deBlas:2022ofj} for FCC-ee projections, 
diboson and Higgs measurements are indeed sensitive to the same types of new-physics effects, 
so that diboson measurements can help constrain Higgs couplings (and vice-versa).
In the current SMEFT fits, all angular information available from the $\epem \to \PWp\PWm$ process is encoded in optimal observables, 
which bring a significant improvement~\cite{DeBlas:2019qco} with respect to the fits performed in the FCC CDR~\cite{fcc-phys-cdr,fcc-ee-cdr}, 
where only the (binned) \PW angular distribution was exploited to constrain the aTGCs. 
The inclusion of the full diboson kinematic information in the SMEFT fit also constrains all other possible interactions that can modify this process. 
The achieved relative precision on the Higgs couplings to gauge bosons and on the aTGCs is displayed in Fig.~\ref{fig:WWefficiency}, 
for three assumed values of the selection efficiency for the $\epem \to \PWp\PWm$ events: 
1\%, 45\% (default value in the following), and 100\%. 
A detailed experimental study of the projected precision is required to assess the ultimate sensitivity of the diboson measurements. 

\subsubsection{Complementarity between \texorpdfstring{\PZ}{Z}-pole and higher-energy runs}

Electroweak precision physics at the \PZ pole is typically thought to mainly constrain the $\hat{S}$ and $\hat{T}$ oblique parameters characterising the gauge 2-point functions, 
and not be as sensitive as higher energy runs for the $\hat{W}$ and $\hat{Y}$ parameters, 
the aTGCs, the Higgs self-energy and self-coupling, Higgs-vector-vector interactions, and a variety of four-fermion operators that, at tree-level, only enter off-pole observables~\cite{Barbieri:2004qk}. 
All these interactions, however, enter at higher-orders in the \PZ-pole observable theoretical predictions. 
The ultra-high precision provided by the \mbox{Tera-\PZ} event samples can overcome the extra loop-suppression factor of these operators 
and actually provide relevant constraints, competitive and complementary to higher-energy measurements 
where these interactions enter at leading order~\cite{Maura:2024zxz, Greljo:2024ytg}. 

\begin{figure}[t]
\centering
\includegraphics[width=\textwidth]{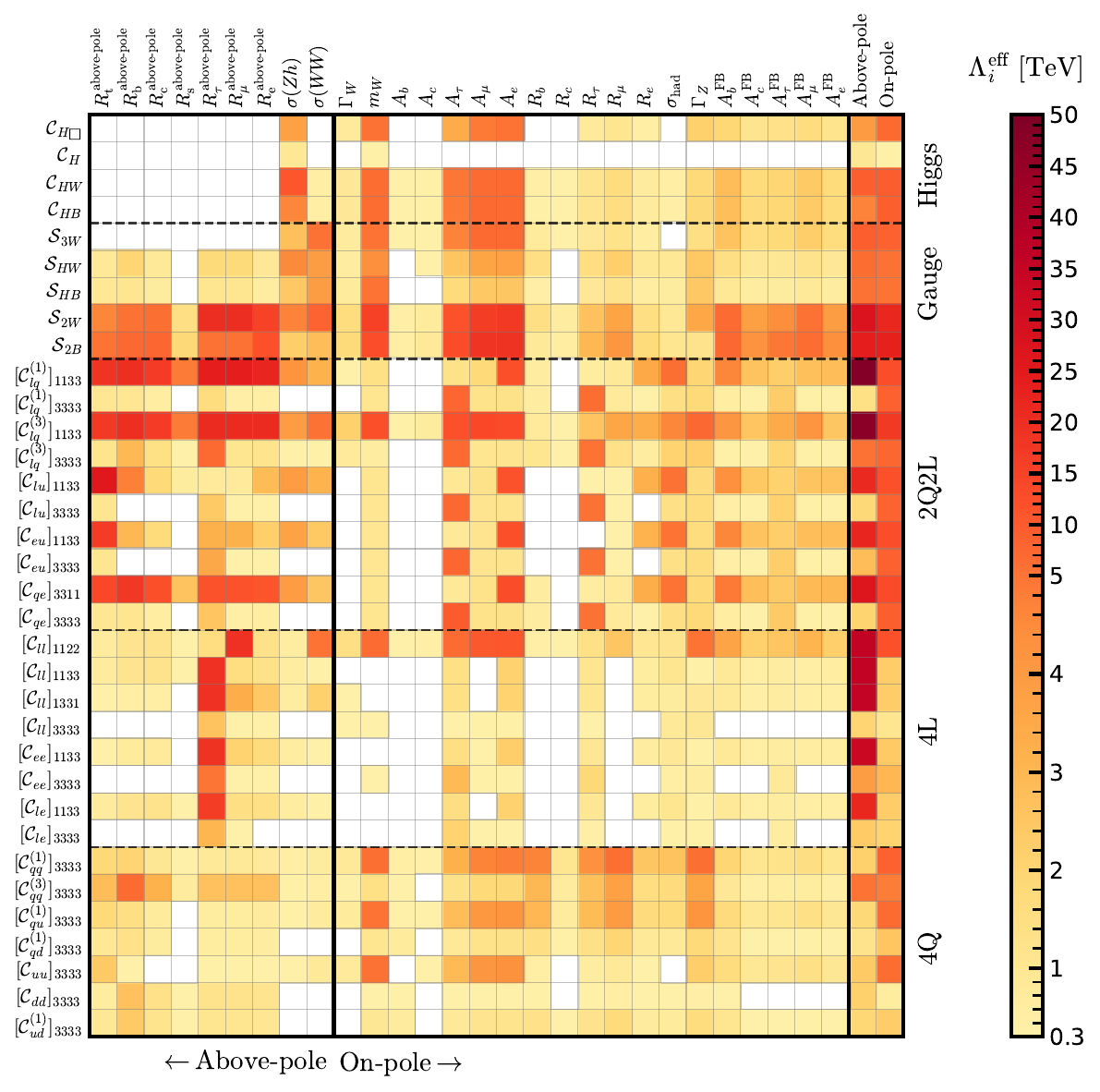}
\caption{The 95\% CL projected sensitivities at FCC-ee to four-fermion, gauge, and Higgs dimension-6 operators, 
in terms of the effective scale of new physics $\Lambda$ (in TeV), 
from measurements of \PZ-pole observables (right part of the table) 
and of higher-energies observables (left part of the table).  
The rightmost two columns show the above- and on-pole combinations of all observables.
Adapted from Ref.~\cite{Maura:2024zxz}.}
\label{fig:onpolevsoffpole}
\end{figure}

\begin{figure}[ht]
\centering
\includegraphics[width=0.46\textwidth]{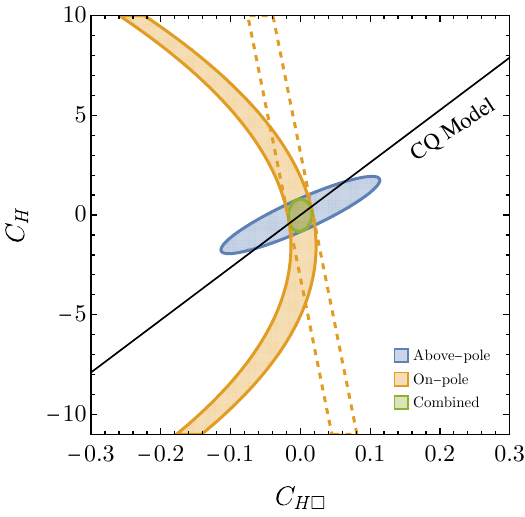}
\includegraphics[width=0.49\textwidth]{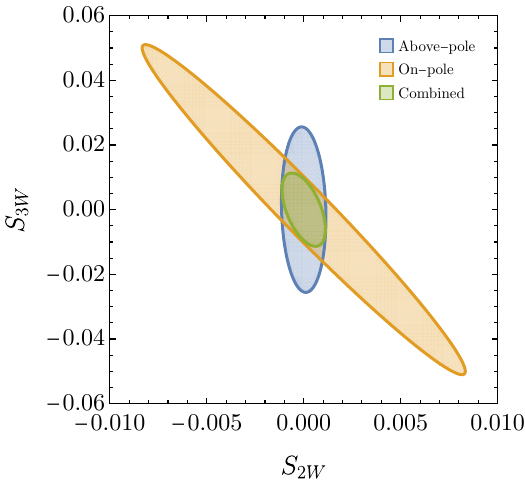}
\caption{Projected 68\% CL sensitivities at FCC-ee, from measurements at the \PZ pole (orange) and in higher-energy runs (blue), 
in the planes of Higgs self-energy vs.\ self-coupling (left) and of the $\hat{W}$ oblique parameter vs.\ an operator modifying anomalous triple-gauge couplings (right). 
The overall combined sensitivities (green) are much more constraining than the individual bounds. 
In the right plot, $S_{2W}$ and $S_{3W}$ are the coefficients of the dimension-6 operators 
$\mathcal{O}_{2W} = -1/2\, \left(D^\mu \, W_{\mu\nu}^I \right) \left(D_\rho \, W^{I\rho\nu}\right)$ and 
$\mathcal{O}_{3W} = \epsilon_{IJK} \, W_\mu^{I\nu} \, W_\nu^{J\rho} \, W_\rho^{K\mu}$,
both normalised with a scale of 1\,TeV. 
In the left plot, the region between the dashed orange lines corresponds to linearised on-pole constraints in the SMEFT operator coefficient
and the solid black line corresponds to a weak quadruplet scalar model. 
Adapted from Ref.~\cite{Maura:2024zxz}.}
\label{fig:onpoleoffpolecomplementarity}
\end{figure}

This statement is illustrated in Fig.~\ref{fig:onpolevsoffpole}, 
which shows the sensitivity (in terms of effective scale of new physics $\Lambda$) to dimension-6 SMEFT operators entering in the four-fermion, 
gauge, and Higgs sectors, from the corresponding constraints in \PZ-pole observables on the one hand, 
and in higher-energy (denoted `above-pole') observables on the other. 
Besides often being better, the on-pole bounds also constrain different directions in the SMEFT space, 
as illustrated in Fig.~\ref{fig:onpoleoffpolecomplementarity} for two specific examples. 
As a consequence, the combination of on-pole and above-pole data very substantially reduces the overall volume of parameter space 
and, accordingly, tightens the constraints on specific models. 
The scope of the \mbox{Tera-\PZ} electroweak precision programme of FCC-ee is, therefore, 
far wider and more general than a na\"{\i}ve extrapolation of the already amazingly accurate SM confirmation from LEP data would suggest.

\subsubsection{Complementarity and synergy between FCC-ee and FCC-hh}

\begin{figure}[ht]
\centering
\includegraphics[width=\textwidth]{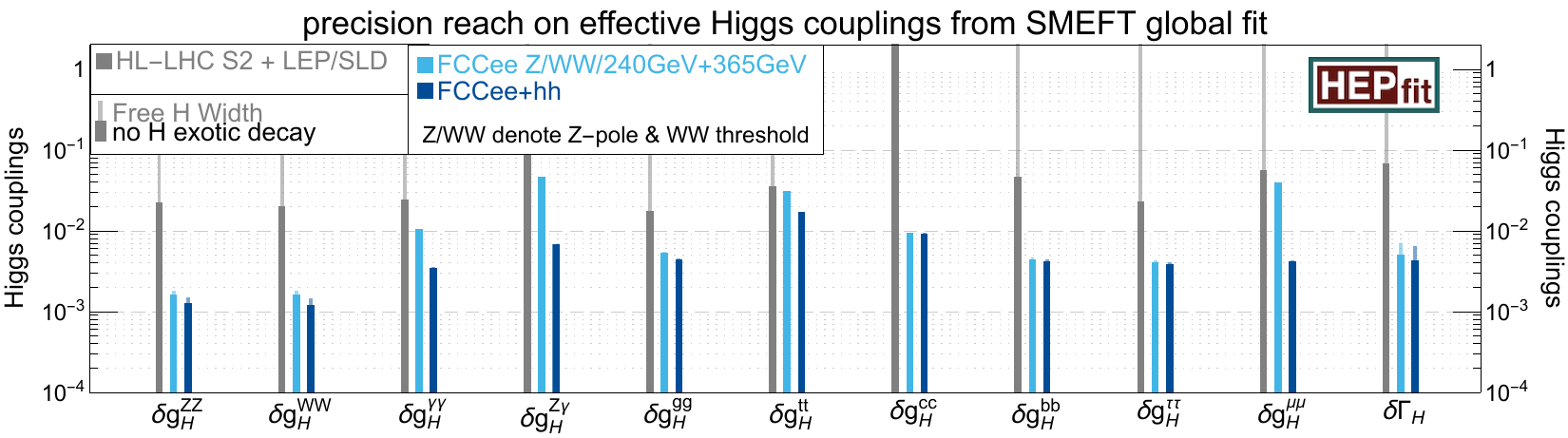}
\caption{Higgs coupling precision improvements with the global SMEFT fit from HL-LHC to FCC-hh, 
benefitting from the FCC-ee accurate and absolute coupling determination to optimally exploit the hadron collider measurements.
}
\label{fig:hhimproveRare}
\end{figure}

As shown in Table~\ref{tab:HiggsKappa3}, 
the precision of many Higgs boson couplings is expected to improve by about one order of magnitude with respect to the (partial and not assumption-free) knowledge 
that will be available at the end of the HL-LHC era. 
The precision of several couplings, such as $\PH\PGm\PGm$, $\PH\PQt\PQt$, $\PH\PGg\PGg$ or $\PH\PZ\PGg$, will however remain dominated by the HL-LHC knowledge,
even if they will only be made assumption-free by the FCC-ee absolute determination of the $\PH\PZ\PZ$ coupling. 
The few per-cent HL-LHC precision will be reduced by an order of magnitude at FCC-hh, by measuring ratios of branching ratios, 
free of systematic uncertainties and normalised by the FCC-ee precise coupling determination~\cite{fcc-phys-cdr}. 
The successive improvements of the Higgs boson coupling precision when going from HL-LHC to FCC-hh, 
benefitting on the way from the FCC-ee absolute and accurate coupling determination, are shown in Fig.~\ref{fig:hhimproveRare}.

\begin{figure}[t]
\centering
\includegraphics[width=0.49\textwidth]{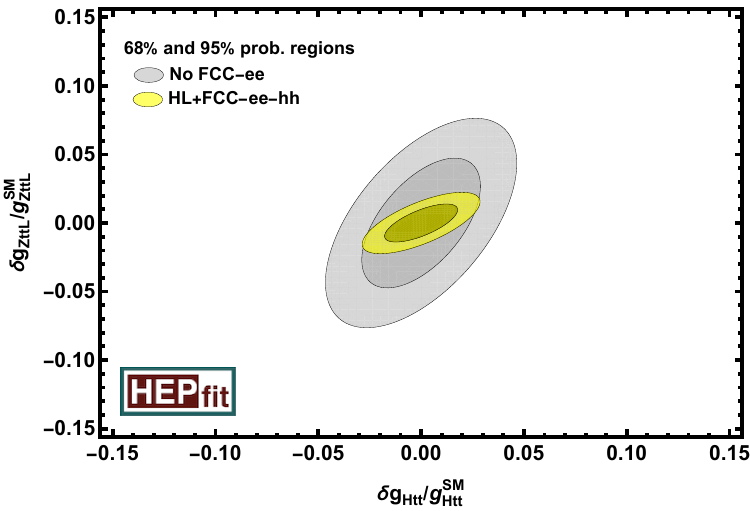}
\includegraphics[width=0.49\textwidth]{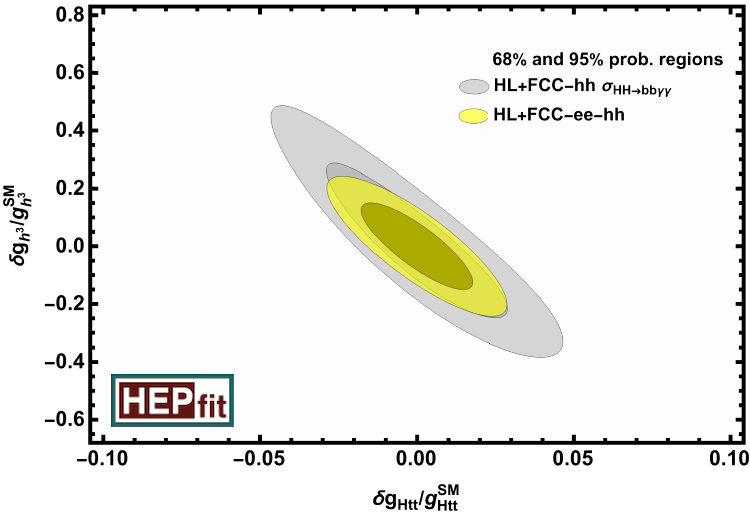}
\caption{Results from ``toy'' fits to simulated data to illustrate the benefit of FCC-ee measurements 
in the FCC-hh determinations of the top Yukawa coupling (left) and of the Higgs self-coupling (right). 
In these fits, the total $\PH \PH \to \PQb\PAQb \PGg\PGg$ cross section is used as a proxy 
to discuss the interplay between the two couplings, following the parameterisation of Ref.~\cite{Azatov:2015oxa}. 
Simplifying assumptions are made on how the top EW couplings and the Higgs couplings to bosons and fermions are measured with a standalone FCC-hh.
The yellow (grey) areas show the 68\% and 95\% probability contours obtained assuming that the FCC-ee measurements are (are not) 
used in the coupling extraction from the FCC-hh data.}
\label{fig:Interplay}
\end{figure}

The FCC-ee role in the top Yukawa coupling determination is twofold. 
On the one hand, and as for all other Higgs couplings, 
the $\PH\PZ\PZ$ absolute determination at 240\,GeV enables a model-independent determination of the top Yukawa coupling from the HL-LHC data. 
On the other hand, the determination of the $\PZ\PQt\PAQt$ couplings from $\epem \to \PQt\PAQt$ at 365\,GeV~\cite{Janot:2015yza,defranchis_2024_cqd16-xhk71} 
can be used to normalise the measurement of the $\PQt\PAQt\PH$ cross section at FCC-hh to that of the $\PQt\PAQt\PZ$ cross section, 
and to reduce the uncertainty in the top Yukawa coupling down to the per-cent level~\cite{Mangano:2015aow}, as illustrated in the left panel of Fig.~\ref{fig:Interplay}. 

In this discussion, it is assumed that contributions from, e.g., dipole or four-fermion interactions in $\epem \to \PQt\PAQt$ or $\Pp\Pp \to \PQt\PAQt {\rm X}$ are negligible. 
The contributions of these four-fermion interactions to the $\PQt\PAQt\PH$ cross section at FCC-hh could be constrained from $\ttbar$ measurements. 
To constrain the contributions of the $\Pe\Pe\PQt\PQt$ four-fermion interactions to the $\epem \to \PQt\PAQt$ cross section, 
extra handles might need to be identified, during the next phase of the study, 
to lift the approximate flat directions that could appear in a global top-quark analysis.

\begin{figure}[t]
\centering
\vskip -0.5cm
\includegraphics[width=0.9\textwidth]{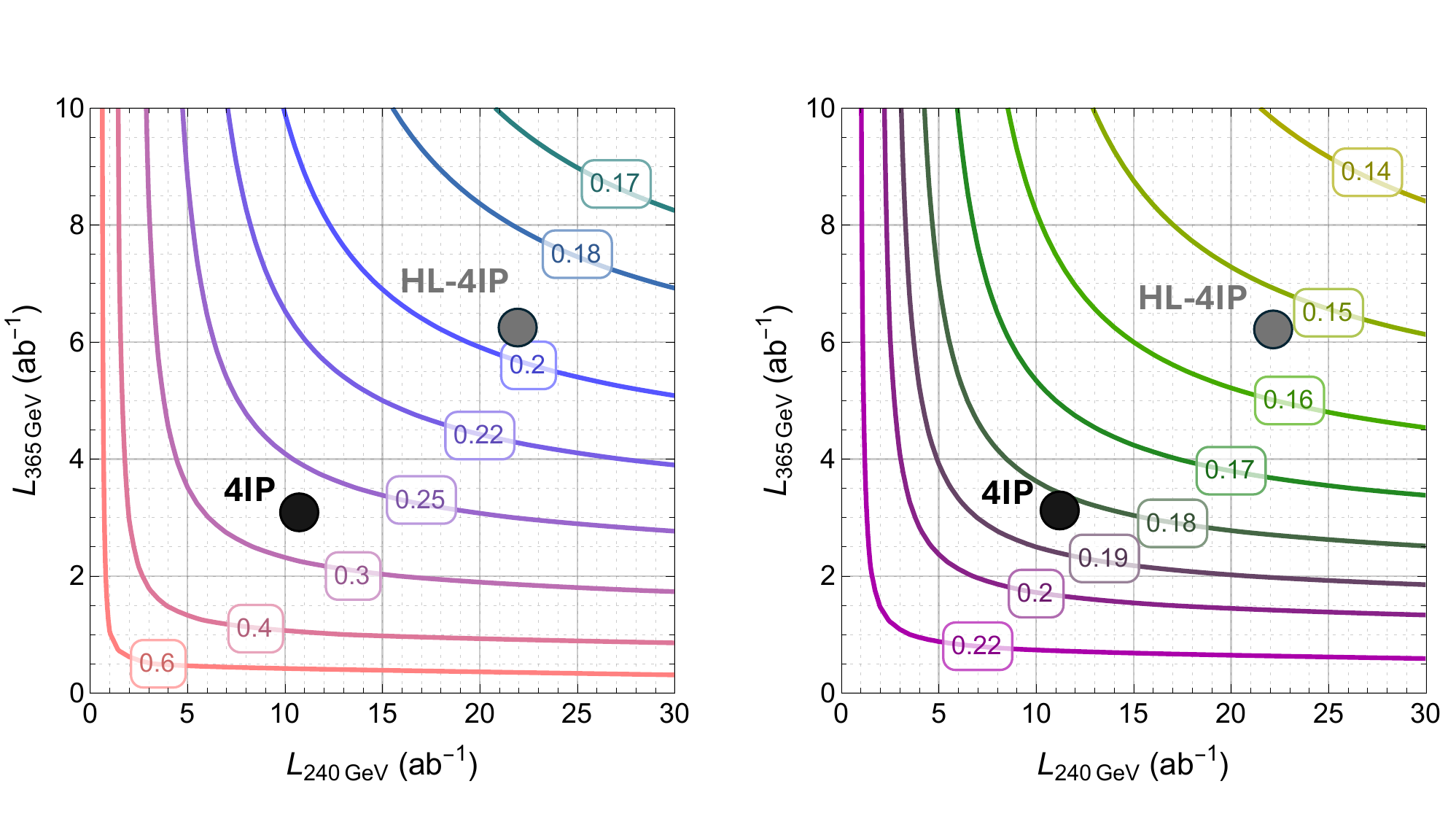}
\caption{The projected Higgs self-coupling relative precision from the SMEFT global fit, 
as a function of the integrated luminosities of the 240 and 365\,GeV runs, 
with FCC-ee alone (left) and FCC-ee combined with HL-LHC (right). 
The 4\,IP and HL-4\,IP dots represent the FCC-ee baseline scenario 
and an hypothetical scenario (described in the text) with doubled integrated luminosities at 240 and 365\,GeV, respectively.}
\label{fig:Hselfcoupling1}
\end{figure}

The Higgs boson self-coupling, $\kappa_\lambda$, will start being probed at HL-LHC, 
with an uncertainty that was estimated at the time of the 2020 ESPPU to be about 50\%, 
in a fit where only deformations of $\kappa_\lambda$ are allowed. 
Since then, improved analysis techniques and the inclusion of additional decay channels have reduced this uncertainty to 26\%~\cite{HLLHCProjections}.
The larger acceptance of the upgraded HL-LHC trackers is likely to bring it well below 25\%. 
In $\epem$ collisions, one-loop radiative corrections, that involve the Higgs self-coupling, to the Higgs boson production cross sections, 
are proportional to $\kappa_\lambda$. 
These corrections amount to several per-cent in the SM at 240\,GeV and strongly depend on the centre-of-mass energy~\cite{McCullough:2013rea}.
With this energy dependence, the sub-percent precision of the FCC-ee Higgs cross-section measurements at 240 and 365\,GeV 
suffice to disentangle the effect of new physics in the $\PH\PZ\PZ$ coupling and in $\kappa_\lambda$, 
and allows a stand-alone determination of $\kappa_\lambda$ with a precision of 28\%, as shown in the left panel of Fig.~\ref{fig:Hselfcoupling1}. 
Besides being quantitatively competitive with HL-LHC projections, 
this precision is also qualitatively distinct, as it is achieved within an SMEFT framework that accounts for a broad variety of potential new physics effects.

The combination of FCC-ee with HL-LHC therefore lifts the assumption made for the sole HL-LHC extraction, 
and significantly improves the overall precision on $\kappa_\lambda$ to 18\%, as shown in the right panel of Fig.~\ref{fig:Hselfcoupling1}.
If deemed important, it would be possible to double the FCC-ee integrated luminosity at 240 and 365\,GeV, 
and reach a precision close to 15\% on $\kappa_\lambda$, in a high-luminosity scenario dubbed HL-4\,IP in Fig.~\ref{fig:Hselfcoupling1}. 
The doubling of the integrated luminosity can be achieved either by running twice as long at these centre-of-mass energies, 
or by doubling the instantaneous luminosity, with a larger number of colliding bunches and/or a smaller vertical $\beta^\ast$ 
(along with a smaller vertical emittance), but without any hardware modifications to the collider, as explained in Ref.~\cite{frankAnnecy}. 

Recent studies~\cite{Asteriadis:2024xuk, Asteriadis:2024xts}
have shown that other new physics interactions, 
entering at the same order in perturbation theory, but absent from the SMEFT framework considered in this report, 
could marginally affect the $\kappa_\lambda$ extraction from single-Higgs processes.
In rare and extreme cases, the value of $\kappa_\lambda$ extracted from single production at FCC-ee data could differ from that with pair production at HL-LHC, 
a difference that would allow this new physics to be identified and constrained. 
More pragmatically, a global fit is needed to bound these other interactions and to mitigate their impact in the determination of $\kappa_\lambda$. 
The more operators are considered, the more observables are needed in the global fit. 
Some operators currently weakly bounded by experimental data, e.g., the four-fermion $\Pe\Pe\PQt\PQt$ operator, 
will require further investigation in the next phase of the study.

\begin{figure}[ht]
\centering
\includegraphics[width=0.4\textwidth]{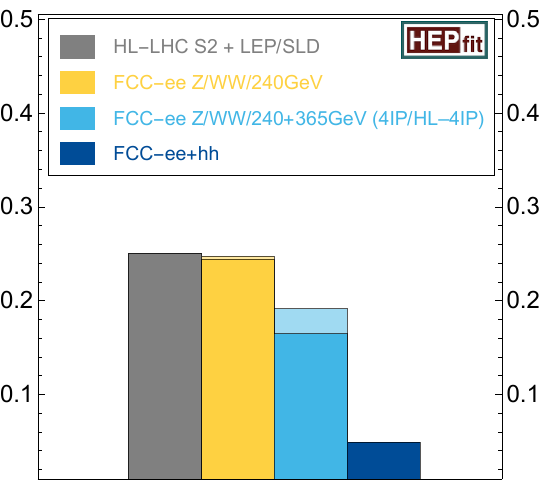}
\caption{Improvement in the determination of $\kappa_\lambda$ at FCC-ee 
(the darker colours are for the HL-4\,IP optimised scenario) 
and, subsequently, at FCC-hh. 
}
\label{fig:Hselfcoupling2}
\end{figure}

This subtlety is anecdotal in the grand vision of the FCC integrated project. 
Indeed, the Higgs self-coupling will be uniquely and unambiguously probed at FCC-hh via Higgs boson pair ($\PH\PH$) production. 
Current estimates, combining the $\PQb\PAQb \PGg\PGg$, $\PQb\PAQb \PGtp\PGtm$, $\PQb\PAQb\PZ\PZ$, and 4\,\PQb decay channels, 
suggest that a precise determination, with an uncertainty as small as 3.4\%, 
would be within the reach of a 100\,TeV $\Pp\Pp$ collider~\cite{Mangano:2020sao}, 
and probably better by a factor of two with progress similar to those made in the HL-LHC projections~\cite{HLLHCProjections} and in recent FCC-hh studies~\cite{Gallo:2024lin,mangano_2025_bzhc2-mem17}. 
The role of the different FCC stages in the determination of the Higgs self-coupling is summarised in Fig.~\ref{fig:Hselfcoupling2}.

As for HL-LHC, this FCC-hh precision refers to an exclusive determination of the Higgs self-coupling, 
i.e., only considering deformations of $\kappa_\lambda$. 
Other new physics interactions, however, could modify the $\PH\PH$ production or decay rates and it is important to keep their uncertainties under control. 
While an inclusive analysis taking into account all these effects has not yet been undertaken, 
an illustration of the impact of such effects is displayed in the right panel of Fig.~\ref{fig:Interplay}. 
This figure shows how the uncertainty in the top Yukawa coupling modifies the $\kappa_\lambda$ precision 
from the measurement of the total $\Pg\Pg \to \PH\PH \to \PQb\PAQb \PGg\PGg$ cross section. 
The top Yukawa coupling enters with several powers in the different diagrams, contributing to $\sigma(\Pg\Pg \to \PH\PH)$. 
As noted above, the precise determination of the top Yukawa coupling at FCC-hh relies 
on the measurement of the $\sigma(\Pp\Pp \to \ttbar\PH)/\sigma(\Pp\Pp \to \ttbar\PZ)$ ratio, 
where the $\ttbar\PZ$ coupling itself should be allowed to vary within its experimental uncertainty, 
rather than being fixed to its assumed SM value. 
The plots in Fig.~\ref{fig:Interplay} show the impact of using, or not, the direct FCC-ee measurement of the $\ttbar\PZ$ coupling 
in the extraction of the top Yukawa coupling at FCC-hh (left panel) and the consequent impact on the $\kappa_\lambda$ determination (right panel). 

Finally, a concern was expressed 
about the sensitivity of the $\kappa_\lambda$ determination at HL-LHC if its true value were about twice the SM value ($\kappa_\lambda = 2$), 
for which the destructive interference between the box and triangle diagrams of the $\PH\PH$ production by gluon fusion in $\Pp\Pp$ collisions 
would significantly decrease the cross section. 
While the HL-LHC relative precision would actually not degrade for $\kappa_\lambda = 2$, the synergy with FCC-ee is, once more, remarkable here. 
Indeed, at FCC-ee the $\epem \to \PZ\PH$ production cross section is a linear function of the true value of $\kappa_\lambda$: 
$\sigma_{\PZ\PH} = \sigma_0 \times (1 + \alpha \, \kappa_\lambda)$, with $\alpha > 0$ (i.e., with a constructive interference). 
The FCC-ee relative precision on $\kappa_\lambda$, therefore, evolves as $1/\kappa_\lambda$ when the true value of $\kappa_\lambda$ increases. 
For $\kappa_\lambda = 2$, a stand-alone precision of 14\% would therefore be achieved at FCC-ee 
with an explicit EFT fit similar to that of Figs.~\ref{fig:Hselfcoupling1} and~\ref{fig:Hselfcoupling2}, 
improved to less than 10\% in the high-luminosity scenario.

\subsection{Discovery landscape}
\label{sec:PhysicsCaseBSM}

The discovery landscape for BSM physics at FCC-hh has been well documented in Ref.~\cite{Mangano:2017tke} 
and in the FCC CDR~\cite{fcc-phys-cdr}.  
Here, instead, the FCC-ee discovery landscape is discussed. 
While several results were already presented in the CDR, 
continuous work has taken place since, to refine or extend the scope of BSM searches, 
a notable example being the search for long-lived particles (LLPs) and heavy neutral leptons (HNLs).

\subsubsection{BSM exploration potential}

The capability and versatility of FCC-ee to explore open questions about the origins and nature of the Universe is immense. 
In addition to probing the known elementary particles and fundamental forces with the highest precision, 
it can survey uncharted territory both directly towards ultra-weak couplings and indirectly at very high energies, 
up to 100\,TeV, for so-far unknown particles and interactions. 

At the centre of many mysteries lies the Higgs boson. 
Besides enabling a much sharper picture of the Higgs boson and testing its (non-)SM nature, 
the $\PZ\PH$ run can also provide crucial model-independent sensitivity to its invisible decay mode(s), 
which probe the Higgs portal to dark sectors. 
Ultimately, an explanation for the origin of the Higgs mechanism itself is sought. 
From what underlying theory does the Higgs sector emerge? 
While this outstanding question is sufficient motivation in itself, 
a more fundamental description of the nature of electroweak symmetry breaking is, moreover, 
expected to be associated with new theoretical principles addressing the hierarchy problem, 
a naturalness strategy that has proven useful in the past 
and whose success or failure now would be of profound importance~\cite{Giudice:2008bi, Craig:2022uua}. 
These theories typically extend the symmetries of the SM, for example in supersymmetric, composite, 
or extra-dimensional frameworks~\cite{Fox:2022tzz}. 
More recent proposals involving light new physics include novel types of cosmological dynamics~\cite{Asadi:2022njl}. 
Alternatively, something radically new altogether may manifest in direct searches, 
indirectly by measuring higher-dimensional operator coefficients 
or spectacularly in unforeseen types of exotic signatures. 
It is crucial to fully explore the Higgs boson as much as possible above the TeV scale.

\begin{figure}[t]
\centering
\includegraphics[width=0.7\linewidth]{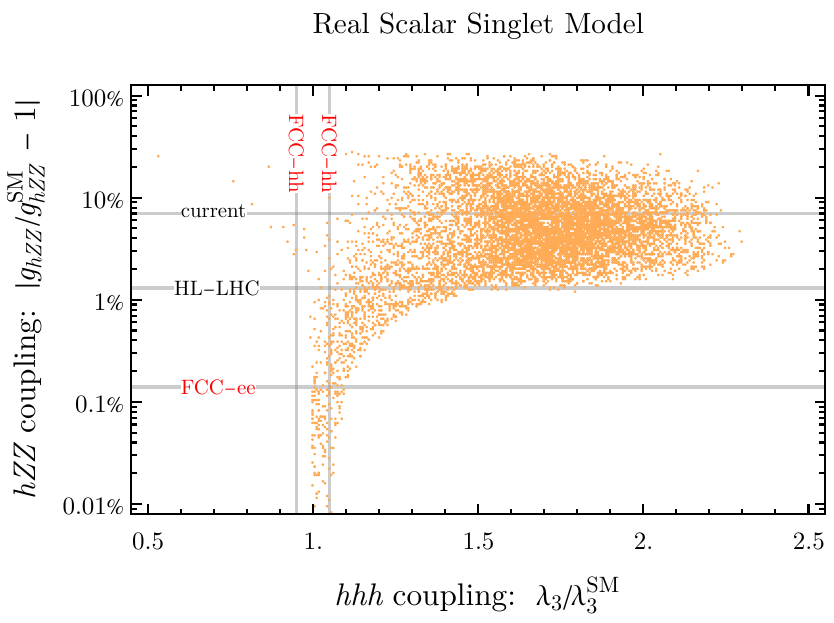}
\caption{Scan of the parameter space for a real scalar singlet model, 
where all points shown have a first-order electroweak phase transition, 
plotted in the plane of the fractional change in the Higgs coupling to a pair of \PZ bosons relative to its SM value (`hZZ coupling') 
vs.\ the triple-Higgs coupling normalised to its SM value (`hhh coupling').
The space outside the two FCC-hh lines and above the FCC-ee line can be covered at those facilities. 
Adapted from Ref.~\cite{Huang:2016cjm} and scaled to baseline luminosities.  
}
\label{fig:realscalarsinglet}
\end{figure}

A deeper understanding of the scalar sector of the SM can also show 
whether the early Universe underwent a first-order electroweak phase transition (FOEWPT) or not, 
knowledge that is crucial for understanding the matter-antimatter asymmetry. 
As an example, in the real scalar singlet model, 
a FOEWPT is correlated with a modification of the Higgs coupling to \PZ bosons 
in a way that can be explored almost entirely by FCC-ee (Fig.~\ref{fig:realscalarsinglet}). 
Furthermore, a FOEWPT can also lead to gravitational wave (GW) signatures, 
which could be detectable by future GW observatories, such as LISA. 
Since it will be difficult to disentangle these signals from other astrophysical phenomena, 
precisely measuring the Higgs properties at colliders could have an important role to play 
in settling this important question~\cite{Friedrich:2022cak}. 
The synergy between cosmology and particle physics has been fruitful in the past 
and may well lead to a more profound understanding of the nature of the electroweak phase transition, 
exploring whether it had a role to play in the origin of matter.

The power of FCC-ee to probe the Higgs boson at much higher resolution than accessible today 
is that it enables peering further into the cloud of quantum fluctuations surrounding it. 
The envisioned precision indirectly opens a window onto physics at multi-TeV energies, 
around the level of current electroweak precision observables from LEP but uniquely sensitive to the Higgs boson. 
Given the number of possible BSM Feynman diagrams contributing virtually to these quantum fluctuations, 
there is a tremendous variety of possible new physics scenarios solving shortcomings of the SM that cannot be tested by other means. 
Similar to the past Fermi theory, the SMEFT is an EFT that maps out the path for continued exploration of what lies far beyond.  
For this essential programme of indirect BSM exploration through the SMEFT, FCC-ee is just indispensable. 

Non-decoupling particles  get most of their mass from the Higgs mechanism,  
requiring a more general EFT framework than SMEFT, known as HEFT, 
to accurately capture their low-energy phenomenology~\cite{Banta:2021dek}. 
Their coupling to the Higgs boson has a lower bound and unitarity requires their mass to have an upper limit. 
Their parameter space is therefore finite and can be comprehensively covered by FCC-ee, as detailed in Ref.~\cite{Crawford:2024nun}.
An interesting open question --- 
is the Higgs responsible for most of the mass of any other particles beyond the SM ones? ---
could then be basically settled. 

A further qualitative leap in resolving power comes from studying the \PZ boson at the \PZ-pole run of FCC-ee. 
A \mbox{Tera-\PZ} factory, with five orders of magnitude more \PZ bosons than at LEP, 
is another truly exciting prospect in its breadth of SM physics and spectacular BSM sensitivity. 
While the measurements at a Higgs factory bring the knowledge of the Higgs sector 
up to the high-precision standards of LEP electroweak precision observables, 
a new era of ultra-high precision observables will begin with \mbox{Tera-\PZ}. 
This ultra-high precision gives FCC-ee an indirect sensitivity to scales of several tens of TeV for weakly-coupled new physics, 
while the sensitivity for strongly coupled physics can reach 100\,TeV. 
To consolidate this sensitivity, the statistical precision reached by the very large \mbox{Tera-\PZ} event samples 
must be matched by a new generation of challenging theoretical calculations at higher electroweak loop orders, 
the development of which is another benefit of the FCC-ee physics programme: 
pushing the boundaries of experiment also expands the frontiers of theory and the understanding of quantum field theory.

After the \PZ boson, the \PW boson can also be used as a precision tool at FCC-ee. 
Its mass is one of the most precisely measured parameters that can be calculated in the SM and is thus of utmost importance. 
The production of almost $5 \times 10^8$ \PW bosons during the planned runs at the $\PW\PW$ threshold and above 
will provide tests of the SM and of a plethora of BSM models with a precision at least one order of magnitude better than today.
The combination of multi-\PZ and \PW production will measure the gauge-boson sector with the sharpest cutting-edge precision, 
leading not only to an unprecedented overview at the microscopic scale of electroweak physics 
but also offering sensitivity to a far broader range of physics intricately linked across the SM, 
and SMEFT more generally, by precision quantum effects.

Going to its highest energy, FCC-ee can explore physics associated with the heaviest known particle, the top quark, 
with the $\ttbar$ run. 
Its mass plays a fundamental role in the prediction of SM processes 
and for the cosmological fate of the metastable vacuum if the SM were extended up to $10^{10}$\,GeV 
or higher~\cite{Degrassi:2012ry,Buttazzo:2013uya,Bednyakov:2015sca,Andreassen:2017rzq}. 
An improvement by more than an order of magnitude, achievable at FCC-ee~\cite{defranchis_2024_cqd16-xhk71}, 
is necessary to perform the above-mentioned precision tests at the quantum level. 
This improvement goes hand in hand with significant progress in the experimental determination 
of the strong coupling constant~\cite{dEnterria:2022hzv}, one of nature's fundamental parameters. 
The large top quark mass suggests that it plays a unique role, 
not only in Higgs and flavour physics, but also in relation with new, so far unobserved, phenomena. 
Indeed, proposed solutions to important open questions motivate couplings of BSM physics to the third generation, 
which also happens to be among the least constrained sectors of the SM. 
Therefore, the precise measurements of the top quark mass and other top quark characteristics 
are crucial for BSM precision exploration.

As a heavy flavour factory, FCC-ee can also investigate the other crucial third generation particles, 
the tau lepton and the bottom quark. 
At the \PZ-pole run, a record number of `clean' \PB mesons and \PGt leptons are expected to be produced, 
the rare decays of which offer exceptional insight into flavourful processes beyond several tens of TeV in energy. 
Exploring the flavour structure of the SM in new regimes can yield clues as to its origin 
and potentially reveal BSM symmetries and selection rules. 
In addition to testing BSM flavour models, 
it provides a valuable experimental probe of CP violation and lepton flavour violation/universality 
that can arise generically in BSM extensions.

Finally, following the proposals and approvals at LHC of `parasite' detectors that enhance its physics potential, 
it is possible to also envision additional experiments at FCC at different locations~\cite{Wang:2019xvx,Tian:2022rsi, Lu:2024fxs}. 
Notably, the civil engineering provides FCC-ee with much bigger detector caverns (Chapter~\ref{sec:caverns}) 
than needed for a lepton collider, in order to use them later for FCC-hh. 
It would then be possible to, e.g., 
install instrumentation in the cavern walls to search for new long-lived particles~\cite{Chrzaszcz:2020emg}. 

Even in the absence of no-lose theorems for the discovery of BSM physics, whether in particle physics or anywhere else, 
the case for a general-purpose particle observatory to carry out fundamental science at the smallest accessible scales 
remains as strong as ever in the face of unanswered questions. 
The purpose of FCC-ee is to improve the knowledge of the Universe and explore its fundamental origins. 
On both these fronts, it is guaranteed to make significant progress.

\subsubsection{Tera-\texorpdfstring{\PZ}{Z} sensitivity to heavy new physics}

The extensive physics programme and unprecedented precision of FCC-ee 
renders it uniquely suited to a general exploration of the zeptoscale. 
In particular, the understanding of the role of the \mbox{Tera-\PZ} programme at FCC-ee 
in discovering evidence for BSM physics has evolved significantly in recent years.  
The picture that has emerged is that, if there is heavy new BSM physics coupled to the SM fields that modifies SM observables, 
\mbox{Tera-\PZ} is well placed to find traces of it.  
This reasoning has been developed in the context of concrete models concerned with the flavour puzzle~\cite{Allwicher:2023shc} 
and the microscopic origins of the Higgs boson~\cite{Stefanek:2024kds, Knapen:2024bxw}. 
Even if a broad and agnostic view of the motivation for the presence of new physics is taken, 
\mbox{Tera-\PZ} offers discovery potential for almost any scenario 
that leads to tree-level modifications of the SM~\cite{Allwicher:2024sso, Gargalionis:2024jaw}.

\begin{figure}[ht]
\centering
\includegraphics[width=0.75\linewidth]{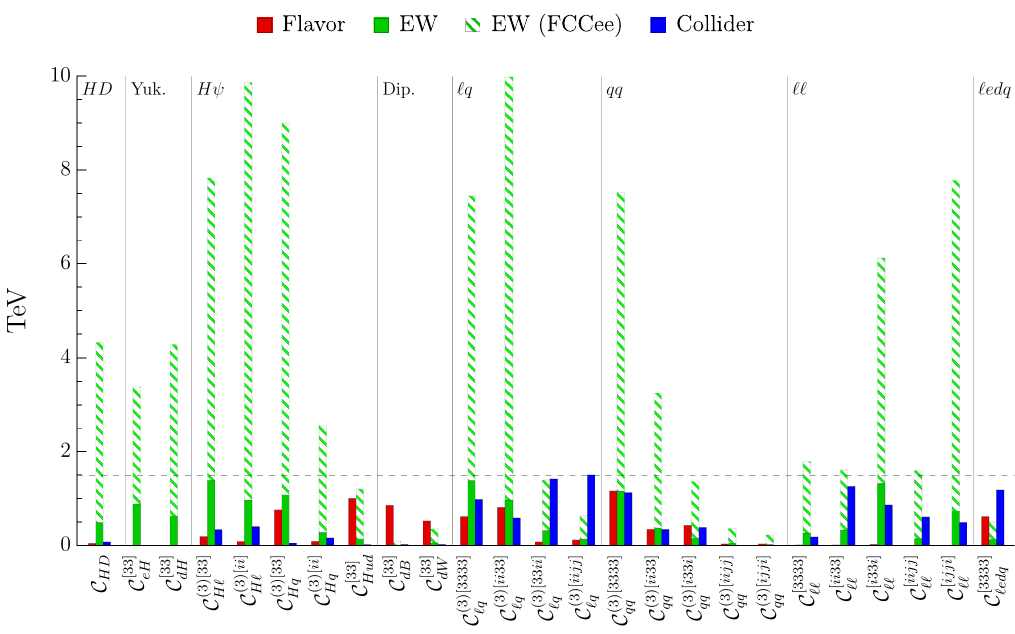}
\caption{The reach of FCC-ee precision electroweak observables (hatched) 
compared to present day flavour, electroweak, and high energy collider constraints, 
for all operators consistent with $\text{U}(2)^5$-symmetric flavour physics at the new physics matching scale. 
The bounds are 3\,$\sigma$ single-parameter fits, 
obtained running from a scale of 3\,TeV with full resummation of the logarithmic terms. 
Taken from Ref.~\cite{Allwicher:2023shc}. }
\label{fig:U(2)}
\end{figure}

When considering the possibility of new physics at the TeV scale, 
it is incumbent upon any model-builder to account for all aspects relating to flavour.  
Since precision flavour observables have, for decades, 
provided a powerful indirect window onto physics at the shortest accessible distances, 
it is common that putative new physics scenarios fall foul of flavour constraints, 
requiring the mass scale of new states to greatly exceed the TeV scale.  
It is important, therefore, to identify and classify new physics scenarios 
that can be consistent with the present suite of experimental constraints.  
In Ref.~\cite{Allwicher:2023shc} it was reported that new physics with an approximate $\text{U}(2)^5$ flavour symmetry 
could exist at very low scales whilst remaining consistent with constraints. 
The high energy states in such a scenario necessarily give rise to modifications of the SM, 
with effects captured by families of dimension-6 SMEFT operators.  
The coefficients of these operators are shown in Fig.~\ref{fig:U(2)}, 
alongside present-day constraints from flavour, precision EW, and high-energy collider probes.   

It is clear that new physics states as low as the TeV scale are presently compatible with observations.  
On the other hand, projections from a precision EW programme at FCC-ee, driven primarily by the \mbox{Tera-\PZ} programme, 
are shown to probe significantly beyond this scale, as far as 10\,TeV. 
Not only is the sensitivity of operators that enter electroweak precision at tree level enhanced to the tens of TeV scale, 
there are also many more third generation operators being constrained at loop level through RGE evolution. 
This observation would significantly impact the understanding of the new physics landscape at high energies 
and its interplay with flavour, providing opportunities to discover the effects of heavy new physics.

The previous discussion concerns the flavour structure of new physics scenarios.  
In contrast, Ref.~\cite{Stefanek:2024kds} focuses specifically on the question of Higgs compositeness.  
Composite Higgs models are a class of scenarios in which the Higgs boson is a composite bound state 
of a new strongly-coupled dynamics at the TeV scale and a potential precursor of extra spatial dimensions. 
These models offer compelling answers to the question of the microscopic origins of the Higgs sector of the SM, 
in analogy with the way in which QCD explains the existence and properties of the pions.  
However, a significant challenge to such a scenario follows from the magnitude of the Higgs boson Yukawa coupling to the top quark.  
This large coupling is difficult to accommodate in the most basic realisations, 
but it can arise if the left and/or right-handed top quarks are also partially composite.  
This observation thus brings flavour considerations to the fore when attempting to understand the origins of the Higgs sector.

\begin{figure}[ht]
\centering
\includegraphics[width=0.325\linewidth]{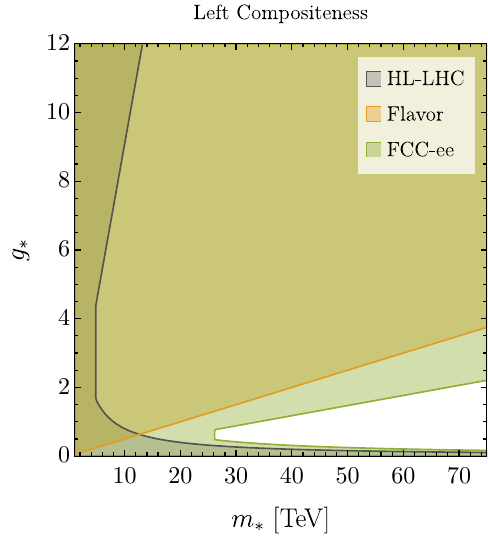}    
\includegraphics[width=0.325\linewidth]{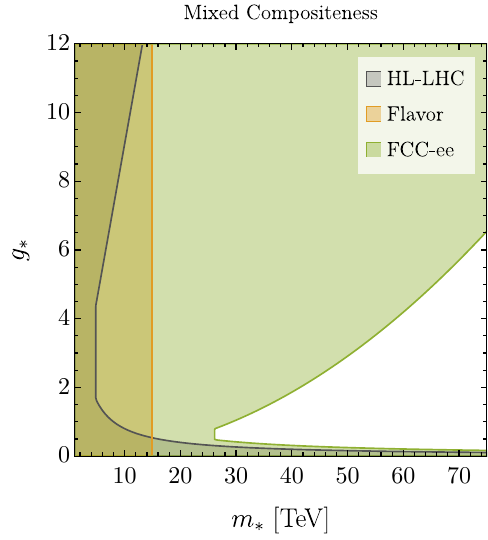}    
\includegraphics[width=0.325\linewidth]{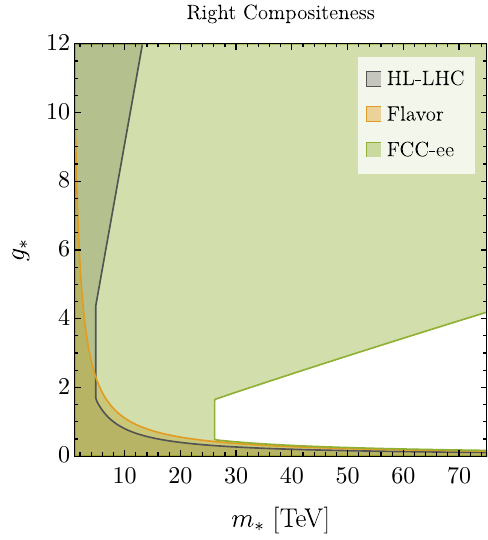}
\caption{The FCC-ee reach for probing Higgs and top compositeness through the precision EW programme, 
as compared to nearer-term HL-LHC and flavour measurements.  
Taken from Ref.~\cite{Stefanek:2024kds}.  
Left, mixed, and right top partial compositeness are considered (left to right panels).}
\label{fig:Stefanek}
\end{figure}

Reference~\cite{Stefanek:2024kds} considers the role that an FCC-ee precision EW programme 
would play in probing the possibility of partial top quark compositeness. 
A custodial symmetry is typically assumed of such models to evade current electroweak precision bounds from the $T$ parameter, 
equivalent to a custodially-violating dimension-6 operator. 
However, the SM violation of custodial symmetry itself leads to a next-to-leading-log RG running into the $T$ parameter 
that can no longer be so easily evaded at FCC-ee. 
This renders large swathes of parameter space accessible to FCC-ee, as shown in Fig.~\ref{fig:Stefanek}, 
where FCC-ee projections are compared to nearer-term HL-LHC and flavour projections. 
A huge increase in projected reach is observed, to compositeness scales of at least 25\,TeV. 

A final consideration, developed in Refs.~\cite{Allwicher:2024sso, Gargalionis:2024jaw}, 
could be considered an `informed agnostic' perspective on the nature of heavy new physics.  
The leading low energy effects of any heavy new physics scenarios with states whose mass does not depend significantly on the Higgs field 
can be captured in the dim-6 operators of the SMEFT.  
However, that does not conversely mean that any pattern of dim-6 SMEFT operators 
corresponds to a legitimate heavy new physics scenario.  
Thus, it is too simplistic to perform EFT fits in the context of future colliders without paying heed to concrete UV scenarios.

\begin{table}[ht]
\centering
\caption{Heavy scalars, fermions, and vectors that can contribute to SMEFT at dimension-6 
with corresponding representations under $\text{SU}(3)_C \times \text{SU}(2)_L \times \text{U}(1)_Y$.  
Taken from Ref.~\cite{deBlas:2017xtg}.}
\label{tab:Granadadict}
    {\small
      \begin{tabular}{l|cccccccc}
        Scalar &
        ${\cal S}$ &
        ${\cal S}_1$ &
        ${\cal S}_2$ &
        $\varphi$ &
        $\Xi$ &
        $\Xi_1$ &
        $\Theta_1$ &
        $\Theta_3$ \\
         &
        $\left(1,1\right)_0$ &
        $\left(1,1\right)_1$ &
        $\left(1,1\right)_2$ &
        $\left(1,2\right)_{\frac 12}$ &
        $\left(1,3\right)_0$ &
        $\left(1,3\right)_1$ &
        $\left(1,4\right)_{\frac 12}$ &
        $\left(1,4\right)_{\frac 32}$ \\[1.3mm]
        &&&&&&&\\[-0.4cm]
         &
        ${\omega}_{1}$ &
        ${\omega}_{2}$ &
        ${\omega}_{4}$ &
        $\Pi_1$ &
        $\Pi_7$ &
        $\zeta$ &
        & \\ 
         &
        $\left(3,1\right)_{-\frac 13}$ &
        $\left(3,1\right)_{\frac 23}$ &
        $\left(3,1\right)_{-\frac 43}$ &
        $\left(3,2\right)_{\frac 16}$ &
        $\left(3,2\right)_{\frac 76}$ &
        $\left(3,3\right)_{-\frac 13}$ \\[1.3mm]
        &&&&&&&\\[-0.4cm]
         &
        $\Omega_{1}$ &
        $\Omega_{2}$ &
        $\Omega_{4}$ &
        $\Upsilon$ &
        $\Phi$ &
        &
        & \\
         &
        $\left(6,1\right)_{\frac 13}$ &
        $\left(6,1\right)_{-\frac 23}$ &
        $\left(6,1\right)_{\frac 43}$ &
        $\left(6,3\right)_{\frac 13}$ &
        $\left(8,2\right)_{\frac 12}$ \\[1.3mm]
\hline
Fermion &
        $N$ & $E$ & $\Delta_1$ & $\Delta_3$ & $\Sigma$ & $\Sigma_1$ & & \\
         &
        $\left(1, 1\right)_0$ &
        $\left(1, 1\right)_{-1}$ &
        $\left(1, 2\right)_{-\frac{1}{2}}$ &
        $\left(1, 2\right)_{-\frac{3}{2}}$ &
        $\left(1, 3\right)_0$ &
        $\left(1, 3\right)_{-1}$ & & \\[1.3mm]
        &&&&&&&\\[-0.4cm]
         &
        $U$ & $D$ & $Q_1$ & $Q_5$ & $Q_7$ & $T_1$ & $T_2$ \\
         &
        $\left(3, 1\right)_{\frac{2}{3}}$ &
        $\left(3, 1\right)_{-\frac{1}{3}}$ &
        $\left(3, 2\right)_{\frac{1}{6}}$ &
        $\left(3, 2\right)_{-\frac{5}{6}}$ &
        $\left(3, 2\right)_{\frac{7}{6}}$ &
        $\left(3, 3\right)_{-\frac{1}{3}}$ &
        $\left(3, 3\right)_{\frac{2}{3}}$ & \\[1.3mm]
\hline
        Vector &
        ${\cal B}$ &
        ${\cal B}_1$ &
        ${\cal W}$ &
        ${\cal W}_1$ &
        ${\cal G}$ &
        ${\cal G}_1$ &
        ${\cal H}$ &
        ${\cal L}_1$ \\
         &
        $\left(1,1\right)_0$ &
        $\left(1,1\right)_1$ &
        $\left(1,3\right)_0$ &
        $\left(1,3\right)_1$ &
        $\left(8,1\right)_0$ &
        $\left(8,1\right)_1$ &
        $\left(8,3\right)_{0}$ &
        $\left(1,2\right)_{\frac 12}$ \\[1.3mm]
        &&&&&&&\\[-0.4cm]
         &
        ${\cal L}_3$ &
        ${\cal U}_2$ &
        ${\cal U}_5$ &
        ${\cal Q}_1$ &
        ${\cal Q}_5$ &
        ${\cal X}$ &
        ${\cal Y}_1$ &
        ${\cal Y}_5$ \\
         &
        $\left(1,2\right)_{-\frac 32}$ &
        $\left(3,1\right)_{\frac 23}$ &
        $\left(3,1\right)_{\frac 53}$ &
        $\left(3,2\right)_{\frac 16}$ &
        $\left(3,2\right)_{-\frac 56}$ &
        $\left(3,3\right)_{\frac 23}$ &
        $\left(\bar 6,2\right)_{\frac 16}$ &
        $\left(\bar 6,2\right)_{-\frac 56}$ \\[1.3mm]
      \end{tabular}
    }
\end{table}

\begin{figure}[ht]
\centering
\includegraphics[width=0.65\linewidth]{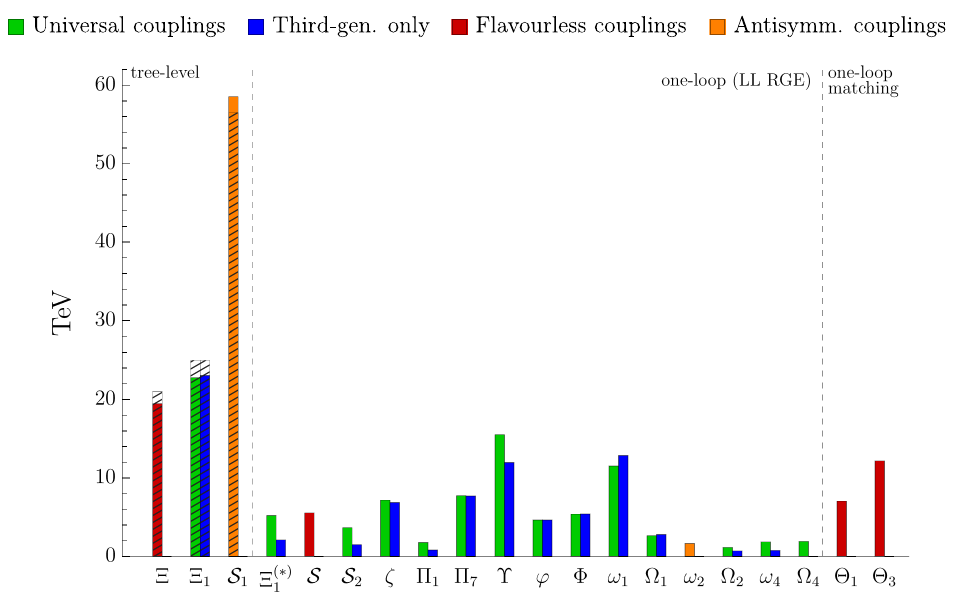}
\caption{Projected bounds on the masses of new scalar fields (the bounds are 2\,$\sigma$ fits, 
obtained running with the leading-logarithmic terms from a scale of 2\,TeV, 
at which the couplings to the SM particles are assumed to be unity, down to the mass of the \PZ).
`Flavourless couplings' correspond to models in which couplings of the new states to fermions are absent, 
`Universal couplings' correspond to equal couplings to all generations, and 
`Third-gen only' to new physics coupled only to the third generation fermions. 
The vertical dashed lines separate fields that contribute to EWPOs at tree level (left), 
via one-loop RG evolution (middle), and via one-loop matching (right). 
From Ref.~\cite{Allwicher:2024sso}, scaled to baseline luminosities.}
\label{fig:scalars6}
\end{figure}

\begin{figure}[t]
\centering
\includegraphics[width=0.65\linewidth]{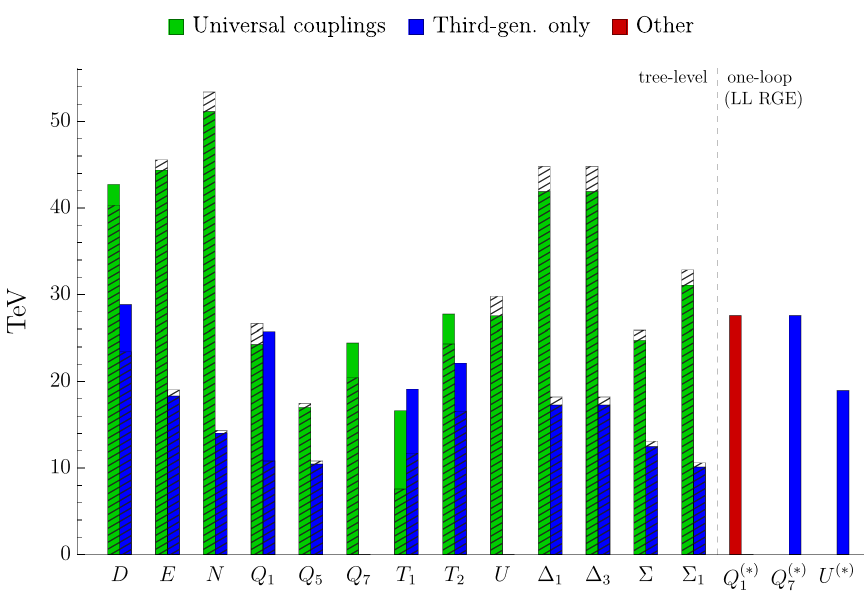}
\caption{Projected bounds on the masses of new vector fields. Same as for Fig.~\ref{fig:scalars6}.
}
\label{fig:fermions6}
\end{figure}

\begin{figure}[t]
\centering
\includegraphics[width=0.65\linewidth]{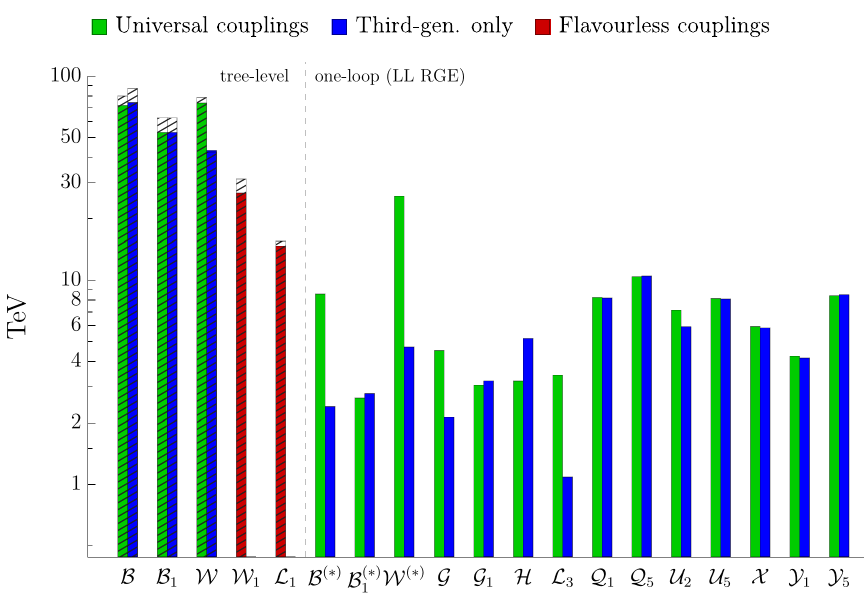}
\caption{Projected bounds on the masses of new fermion fields. Same as for Fig.~\ref{fig:scalars6}.
}
\label{fig:vectors6}
\end{figure}

As an informed middle ground, 
Refs.~\cite{Allwicher:2024sso, Gargalionis:2024jaw} considered all possible new physics scenarios 
where any dim-6 SMEFT operators are generated at tree-level.  
The number of possibilities is large, but finite, as detailed in Table~\ref{tab:Granadadict}.  
If any such scenario exists and modifies SM properties in any sector of the SM at dim-6, 
the precision EW programme at FCC-ee has sensitivity, 
as shown in Figs.~\ref{fig:scalars6}, \ref{fig:fermions6}, and~\ref{fig:vectors6}. 

As a matter of fact, FCC-ee can effectively probe virtually all heavy particles linearly coupled to the SM, 
with a sensitivity that reaches up to 100\,TeV at tree-level 
and $\mathcal{O}$(1--10)\,TeV at the one-loop level for $\mathcal{O}(1)$ couplings. 
This reach could extend even higher in the case of strongly-coupled new physics. 
Quantum RG effects are crucial.  
Any treatment that considers the SMEFT contributions of these states only at tree-level 
would overlook the majority of the sensitivity of the FCC-ee \mbox{Tera-\PZ} programme to their existence.

\begin{figure}[ht]
\centering
\includegraphics[width=0.65\linewidth]{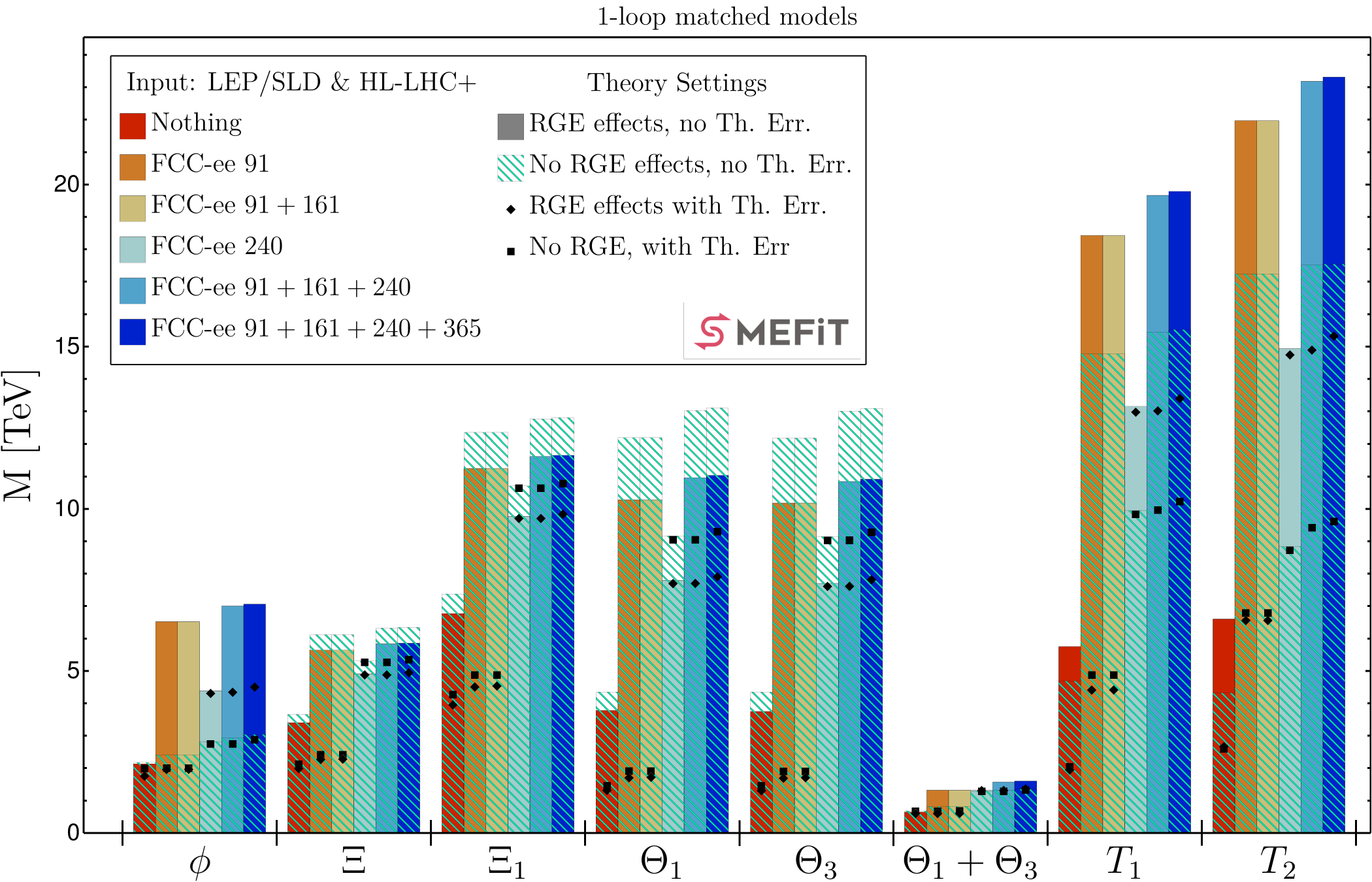}
\caption{Sensitivity (95\% CL) to a selection of models from Table~\ref{tab:Granadadict} 
with Wilson coefficients fully matched at one-loop 
(the bounds are obtained from a $\chi^2$ fit and the couplings of the new particle are set to unity at the matching scale, 
taken to be the mass of this heavy particle, 
the different operators are then run down to the \PZ boson mass with full numerical resummation of the logarithmic terms).   
In the case of new physics coupled to SM fermions, 
heavy scalar and fermions couple to the third generation only, 
while vectors couple to third-generation quarks and all generations of leptons.  
The starting scenario corresponds to all LEP/SLD and LHC constraints anticipated by the end of HL-LHC operation.
Additional combinations of FCC-ee runs are then shown. 
Black markers denote anticipated constraints were the theory uncertainties to remain as at present.  
The $\Theta_1+\Theta_3$ model corresponds to the custodial quadruplet model of Refs.~\cite{Logan:2015xpa,Chala:2018ari,Durieux:2022hbu}.}
\label{fig:SMEFITmatched}
\end{figure}

\begin{figure}[ht]
\centering
\includegraphics[width=0.65\linewidth]{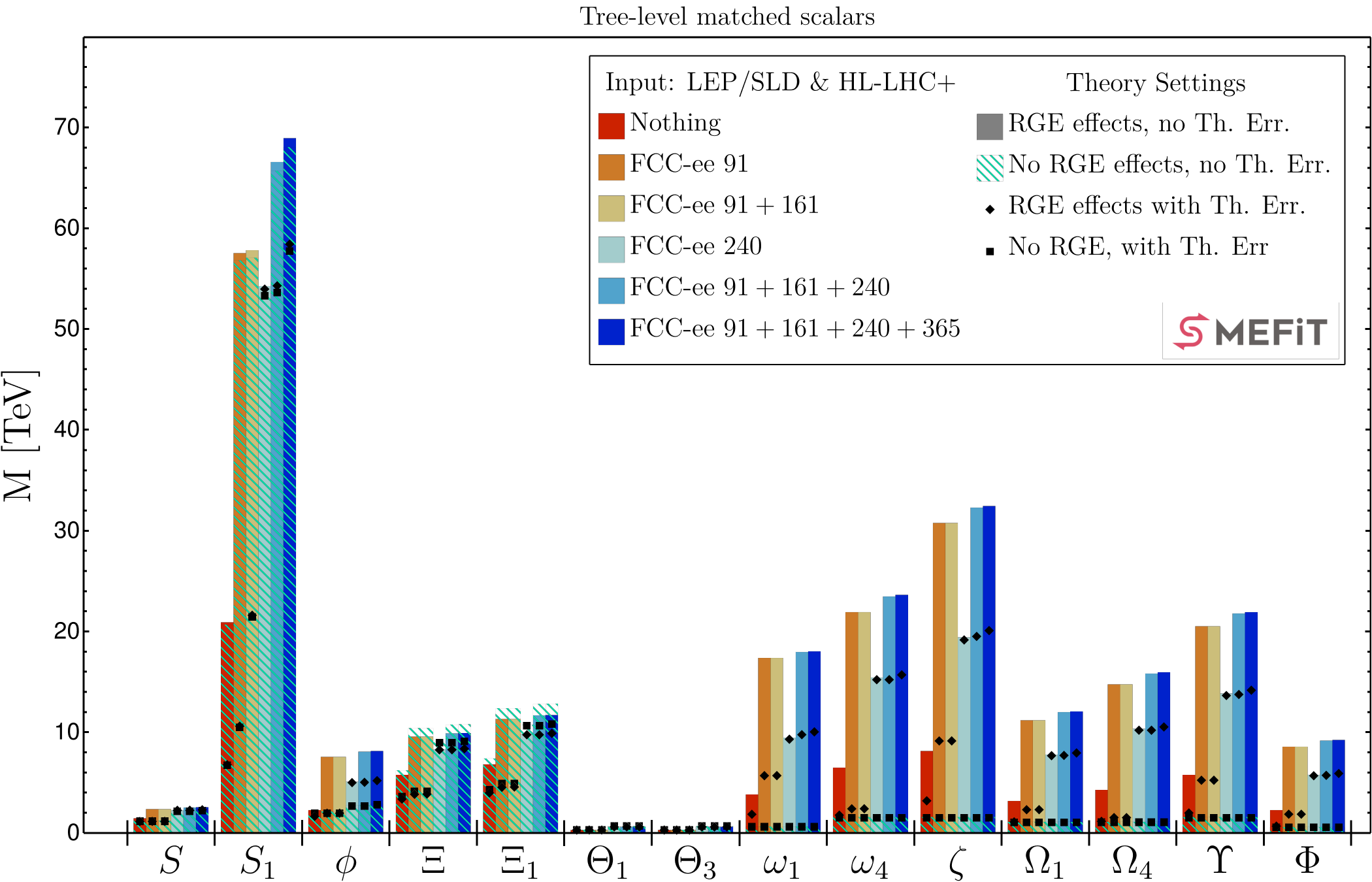}
\caption{As Fig.~\ref{fig:SMEFITmatched}, but without the one-loop matched terms, for scalars.}
\label{fig:SMEFITscalars}
\end{figure}

\begin{figure}[ht]
\centering
\includegraphics[width=0.65\linewidth]{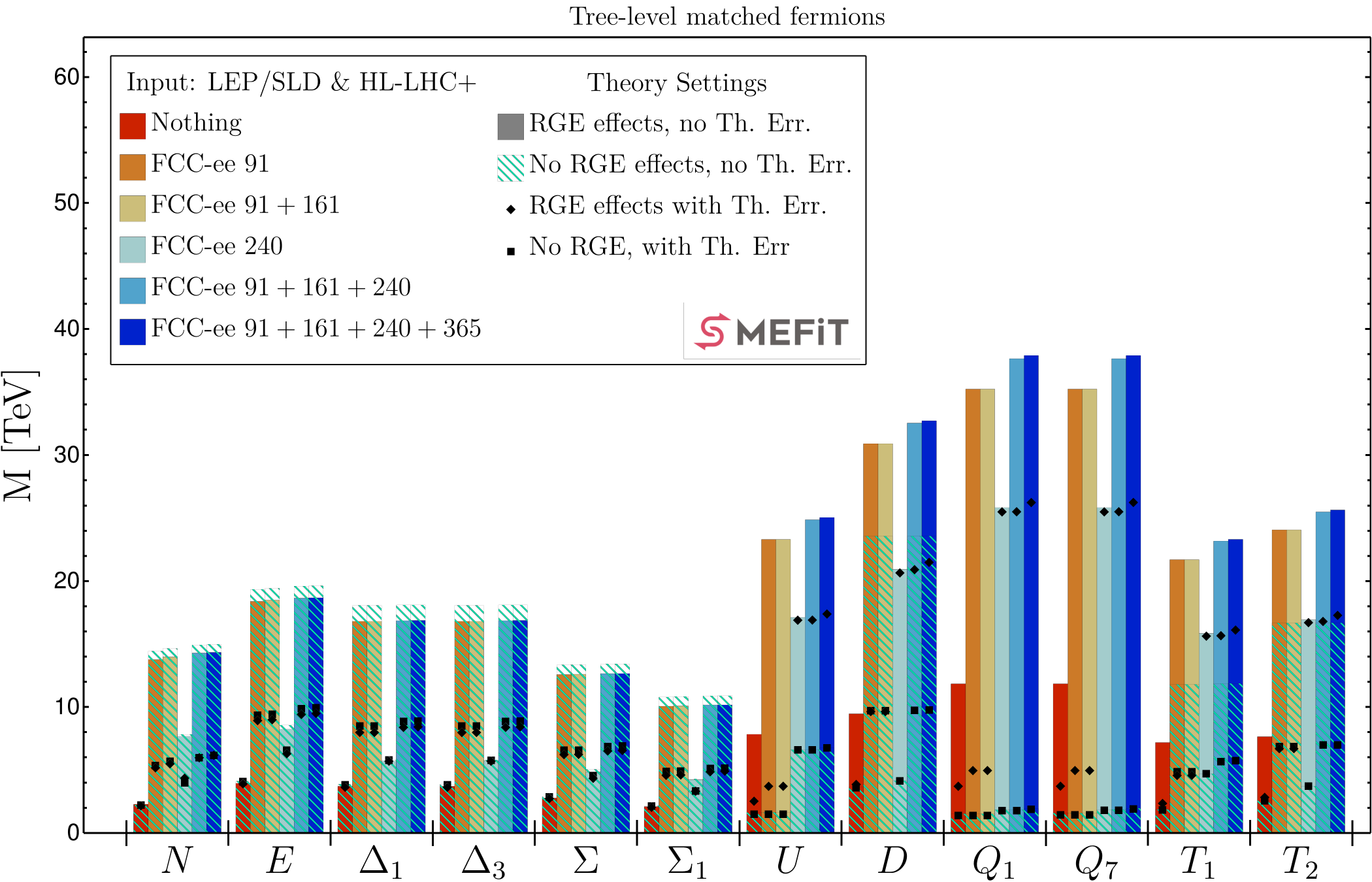}
\caption{As Fig.~\ref{fig:SMEFITmatched}, but without the one-loop matched terms, for fermions.}
\label{fig:SMEFITfermions}
\end{figure}

\begin{figure}[ht]
\centering
\includegraphics[width=0.85\linewidth]{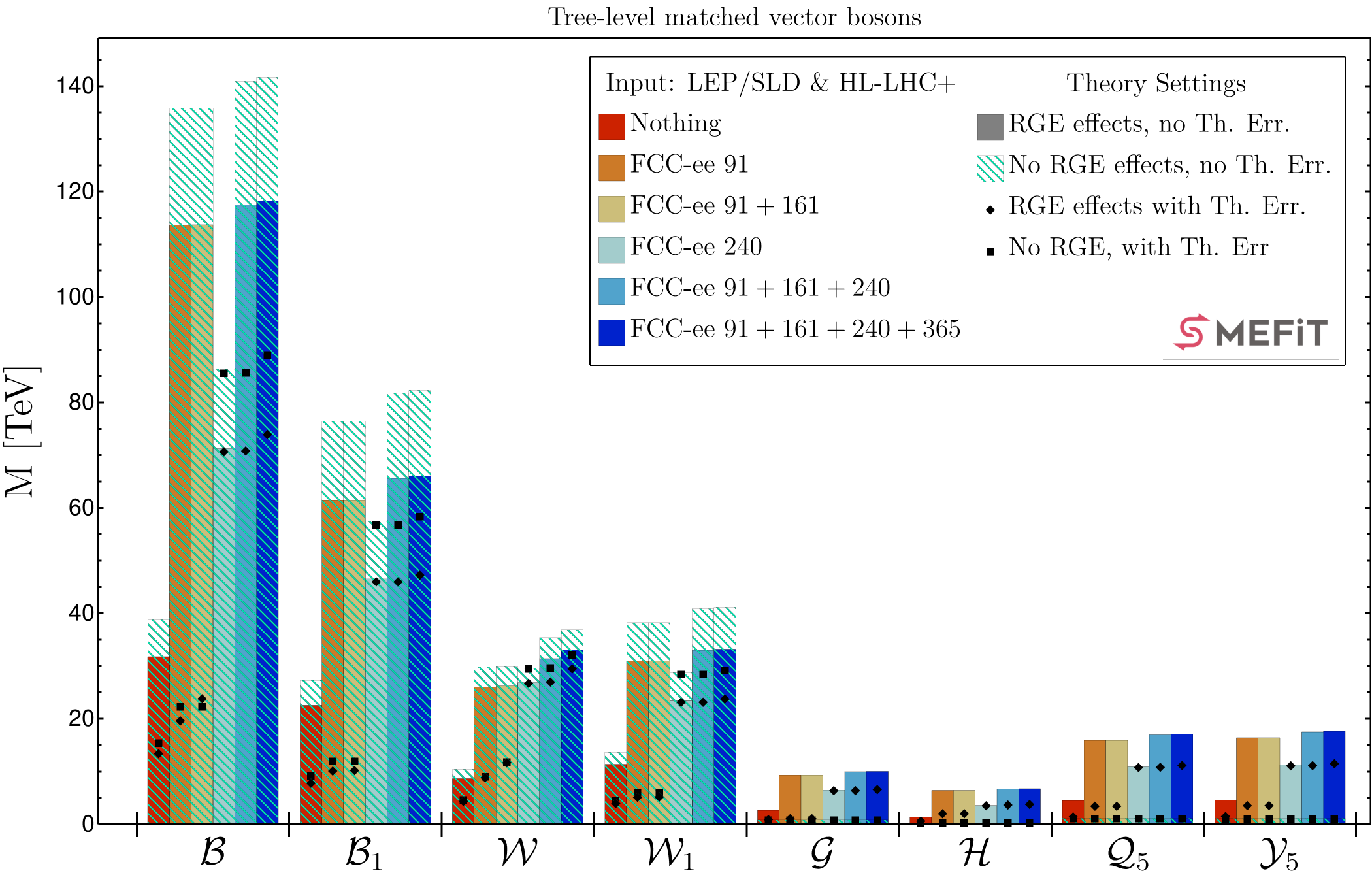}
\caption{As Fig.~\ref{fig:SMEFITmatched}, but without the one-loop matched terms, for vectors.  
The models $\mathcal{B}$ and $\mathcal{W}$ include flavour-universal couplings to leptons. 
See Ref.~\cite{terHoeve:2023pvs} for details.}
\label{fig:SMEFITvectors}
\end{figure}

Reference~\cite{Allwicher:2024sso} only considers the \mbox{Tera-\PZ} and $\PW\PW$ runs. 
Recently, the SMEFit Collaboration studied the impact of the various runs combined and individually. 
Details can be found in Refs.~\cite{terHoeve:2023pvs,Celada:2024mcf,smefit:2025}, 
including the slightly different EW-input scheme, 
which leads to some minor differences with respect to the results of Ref.~\cite{Allwicher:2024sso} for some models.  
Figure~\ref{fig:SMEFITmatched} demonstrates the exploratory power of the various runs 
to a variety of models in which the full one-loop matching is performed at the matching scale.  
Figures~\ref{fig:SMEFITscalars}, \ref{fig:SMEFITfermions}, and~\ref{fig:SMEFITvectors} 
consider selected scalar, fermion, and vector scenarios with RG-evolved Wilson coefficients 
but not one-loop matching, as in Figs.~\ref{fig:scalars6}, \ref{fig:fermions6}, and~\ref{fig:vectors6}.



\begin{figure}[ht]
\centering
\includegraphics[width=0.4\linewidth]{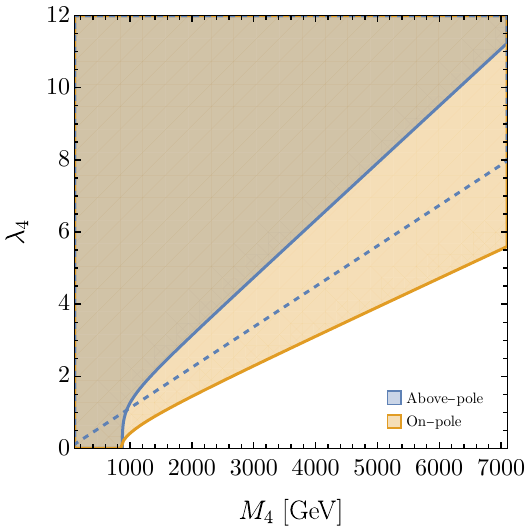}
\includegraphics[width=0.55\linewidth]{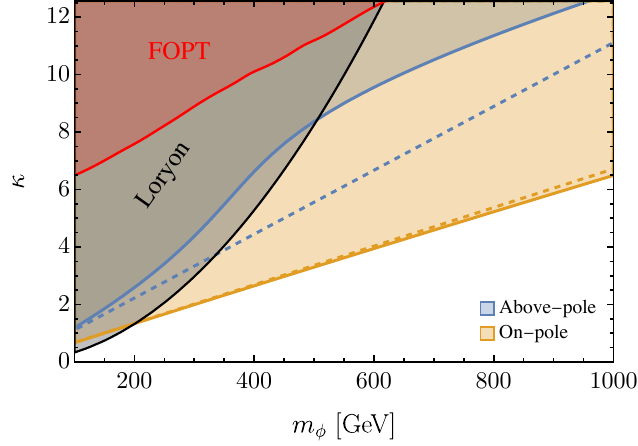}
\caption{FCC-ee sensitivity at 68\% CL on the weak custodial quadruplet scalar (left) and real singlet scalar (right) models in the mass vs.\ coupling plane when combining \PZ pole data (orange) with $\PZ\PH$ measurements (blue). 
For the real singlet case, the constraints exclude a first order phase transition (FOPT) for the mass region shown. See Ref.~\cite{Maura:2024zxz} for further information. }
\label{fig:custodialquadrupletandRSS}
\end{figure}

The picture that emerges from Figs.~\ref{fig:SMEFITmatched}, \ref{fig:SMEFITscalars}, \ref{fig:SMEFITfermions}, 
and~\ref{fig:SMEFITvectors} is that the \mbox{Tera-\PZ} programme of FCC-ee is, indeed, 
extremely powerful in searching for new physics, extending well beyond the scales probed at HL-LHC, 
and is also highly complementary to the Higgs run. 
The latter offers loop-induced sensitivity to the Higgs trilinear coupling~\cite{McCullough:2013rea}, 
which could probe custodially symmetric models, as shown for the $\Theta_1 + \Theta_3$ model in Fig.~\ref{fig:SMEFITmatched}. 
This weak custodial quadruplet scalar model and the Higgs trilinear coupling can, moreover, be probed on the \PZ pole, 
as shown on the left plot of Fig.~\ref{fig:custodialquadrupletandRSS}, 
with the combination significantly extending the coverage of the model parameter space~\cite{Maura:2024zxz}.

\begin{figure}[ht]
\centering
\includegraphics[width=0.49\linewidth]{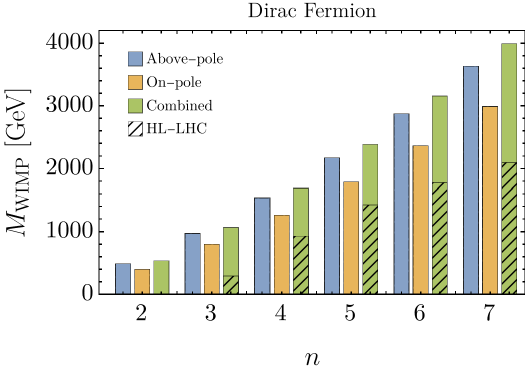}
\includegraphics[width=0.49\linewidth]{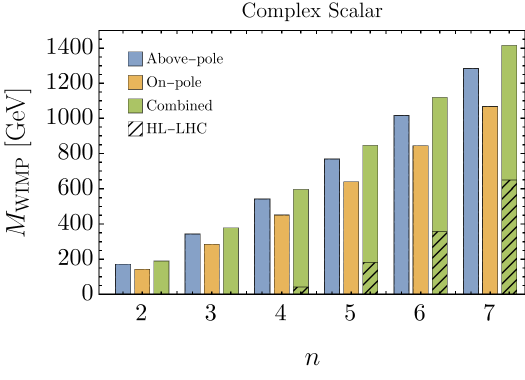}
\caption{Projected 95\%~CL for $n$-plet Dirac fermion and complex scalar weakly interacting massive particles with zero hypercharge. 
From Ref.~\cite{Maura:2024zxz}.}
\label{fig:wimps}
\end{figure}

The \mbox{Tera-\PZ} run is crucial to definitively answer the fundamental question of 
whether any particles other than the SM ones obtain most of their mass from the Higgs mechanism. 
Such so-called `loryon' candidates have a finite parameter space that can be almost entirely probed by current and future colliders, 
with a remaining open window for the elusive real singlet scalar model~\cite{Crawford:2024nun}. 
This window can potentially be closed by precision measurements at the \PZ pole, 
as shown on the right plot of Fig.~\ref{fig:custodialquadrupletandRSS}, from Ref.~\cite{Maura:2024zxz}, 
thus allowing this important question to be definitively settled. 
The high sensitivity of FCC-ee to the $\hat{W}$ and $\hat{Y}$ parameters, 
both at the \PZ pole and in higher energy runs, 
also enables constraining elusive BSM candidates, such as weakly interacting massive particles. 
The combination of on-pole and above-pole data, shown in Fig.~\ref{fig:wimps}, 
indeed gives a projected sensitivity beyond the indirect sensitivity reach of Drell--Yan searches at HL-LHC~\cite{Maura:2024zxz}. 

It will be critical to reduce theory uncertainties (Chapter~\ref{sec:theory}) 
in order to capitalise on the full potential of the FCC-ee data.  
Indeed, for a number of scenarios, 
the \mbox{Tera-\PZ} programme would probe the parameter space inaccessible to the runs at higher energies 
only in the context of improved theory uncertainties.  
An additional noteworthy aspect concerns the synergy with FCC-hh.  
Indirect probes at FCC-ee approach or exceed the 10\,TeV scale.
As a result, the only motivated successor to such a probe would be a collider capable of directly probing beyond that parameter space 
or one capable of directly producing the new states for which indirect evidence may have emerged at FCC-ee.  
In other words, a successor collider would necessarily have to be capable of directly probing beyond the 10\,TeV scale. 
The only such plausible facility is FCC-hh, strengthening the combined FCC physics case.

It should be noted that exceptions from the exploratory power of \mbox{Tera-\PZ} can exist.  
It was recently shown in Ref.~\cite{Davighi:2024syj} that a class of anomaly-free $\PZ^\prime$ models could exist 
in which the charge assignments lead to a suppression of electroweak-precision-sensitive operators at tree and one-loop level.  
Nonetheless, such a scenario would be strongly constrained above the \PZ-pole, 
both at FCC-ee and HL-LHC, demonstrating additional complementarity between runs and colliders.

The individual models considered are, themselves, not being proposed as well-motivated in any subjective sense.  
Rather, the collection of scenarios and the full family of Wilson coefficient patterns they populate 
can be taken as being qualitatively representative of the broad family of Wilson coefficients that could arise in UV scenarios.  
Thus the fact that the FCC-ee \mbox{Tera-\PZ} programme has sensitivity to this wide class of models, 
whether at tree-level or one-loop, is indicative of the fact that it will have sensitivity to generic UV scenarios.  
To argue otherwise would require the construction of an explicit counterexample model.  
Within the class of models considered in Ref.~\cite{Allwicher:2024sso}, 
the two counterexample states are $\Omega_4$ and a special case for $\mathcal{G}$, 
for which only the four top-right interaction is generated up to one loop. 
However, even BSM modifying four-top operators can be highly constrained through its next-to-leading-log running into the $T$ parameter 
at the \PZ pole~\cite{Allwicher:2023shc, Stefanek:2024kds}, 
as also seen in Figs.~\ref{fig:SMEFITscalars} and~\ref{fig:SMEFITvectors}.

\begin{figure}[ht]
\centering
\includegraphics[width=0.4\linewidth]{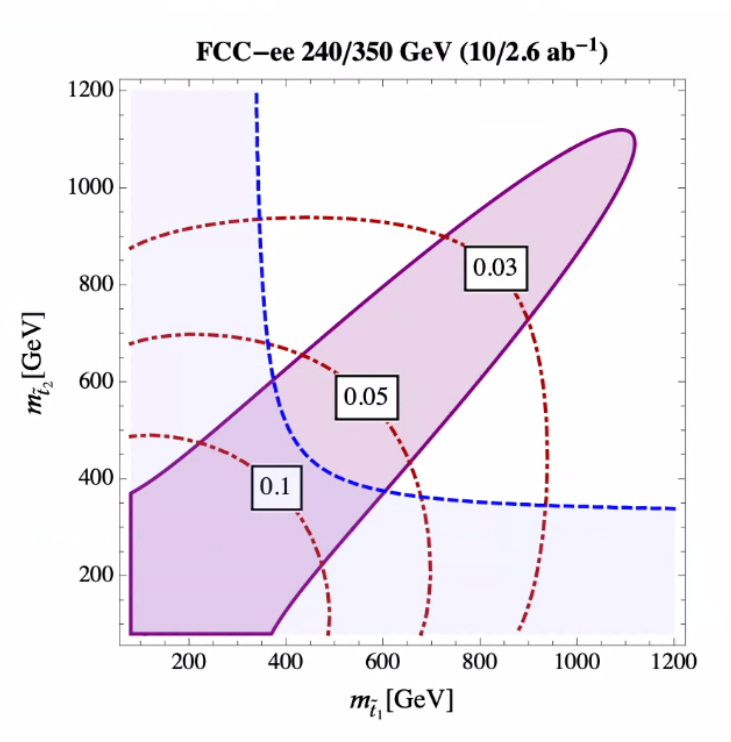}
\includegraphics[width=0.5\linewidth]{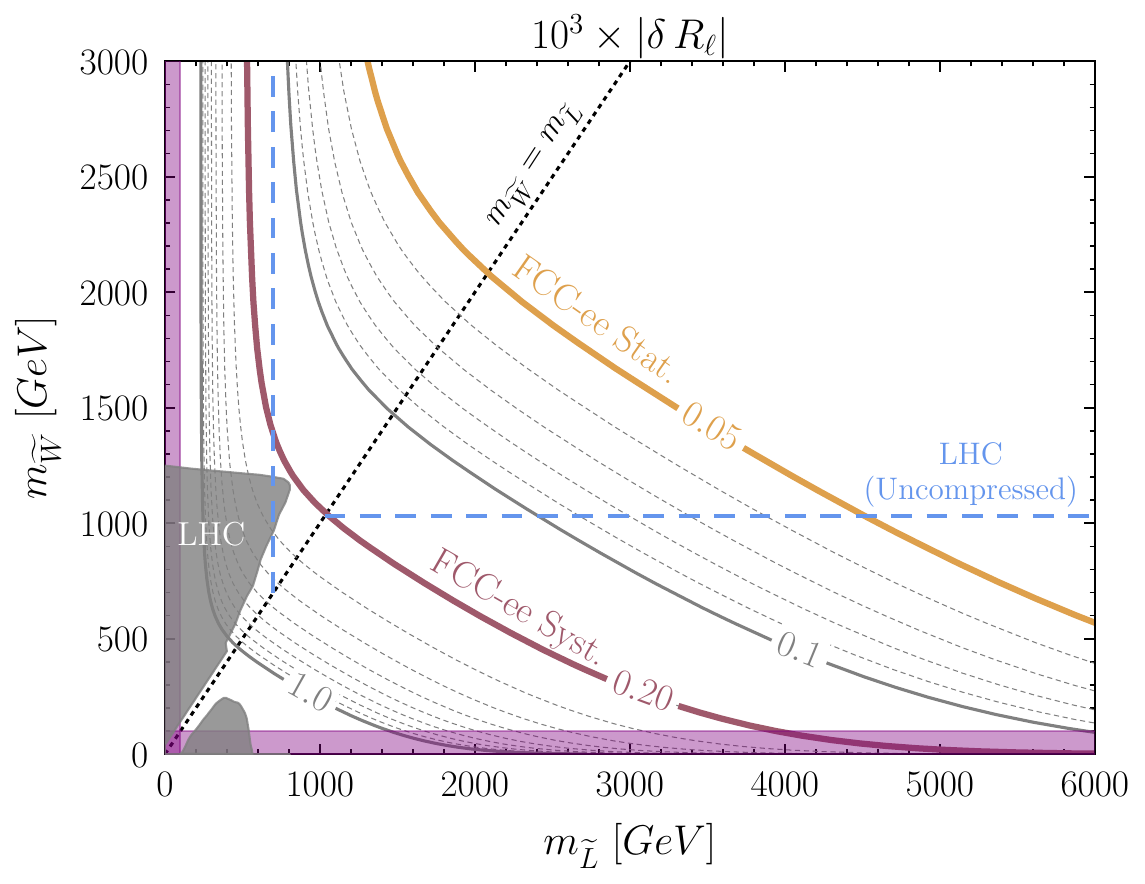}
\caption{Left: Projected 2\,$\sigma$ sensitivity of Higgs couplings to stops at FCC-ee 
in the parameter space of stop masses $m_{\tilde t_1}$ vs.\ $m_{\tilde t_2}$;
from Ref.~\cite{Fan:2014axa}. 
Right: FCC-ee \PZ branching ratio projected statistical and systematic 1\,$\sigma$ uncertainties 
in the left-handed selectron mass vs.\ pure wino mass plane;
from Ref.~\cite{Knapen:2024bxw} scaled to the baseline luminosities.}
\label{fig:susy}
\end{figure}

Finally, as another concrete UV scenario, 
supersymmetric models are a well-motivated approach to understanding the origin of the Higgs and addressing the hierarchy problem, 
where the BSM particles are, instead, typically expected to arise from new weakly-coupled dynamics. 
The Higgs couplings at one loop, in particular to gluons and photons, 
can probe the contributions of superpartners such as the stop, the coloured scalar partner of the top. 
Such indirect probes are complementary to more model-dependent direct searches and their sensitivity can reach around a TeV, 
as shown on the left plot of Fig.~\ref{fig:susy}, taken from Ref.~\cite{Fan:2014axa}. 
The right plot of Fig.~\ref{fig:susy}, from Ref.~\cite{Knapen:2024bxw}, 
shows the projected \PZ branching ratio sensitivity at FCC-ee to the selectron and pure wino, 
together with the current direct search constraints at LHC.

\subsubsection{Flavour deconstruction at FCC-ee}
\label{sec:otherBSM}

To give an example
of concrete scenarios explored through a combination of the unprecedented precision 
on \PZ-pole and $\PW\PW$-threshold observables with excellent capabilities as a flavour factory, 
a special class of theories based on `flavour deconstruction'~\cite{Davighi:2023iks} is considered, 
whereby part of the electroweak symmetry is resolved into three generation-specific copies $G_i$ 
(with the Higgs boson coupled to $G_3$). 
This symmetry, which arises, e.g., from gauge-flavour unification~\cite{Davighi:2022fer} 
or from extra-dimensions~\cite{Fuentes-Martin:2022xnb}, 
is broken to the SM in two steps that generate hierarchical fermion masses and mixings. 
The last breaking step, $G_{12} \times G_3 \to G$, 
should occur close to the TeV scale to avoid destabilising the Higgs potential, 
delivering TeV-mass gauge bosons in the adjoint of $G$ coupled mostly to the third generation. 

These electroweakly-charged and flavour non-universal gauge bosons give rise to a rich phenomenology 
across EWPOs, flavour observables, and high-energy LHC measurements,
which has been quantified by considering deconstructed hypercharge~\cite{Davighi:2023evx} 
and deconstructed $SU(2)_L$~\cite{Davighi:2023xqn} (Table~\ref{tab:pheno}). 
Firstly, EWPOs are shifted at tree-level because the heavy gauge bosons directly couple to the Higgs boson.
Constraints from LEP are already strong and the expected precision brought by FCC-ee would all but cover the `natural' regime 
in which these flavour models can be reconciled with the hierarchy problem. 
Secondly, being intrinsically non-universal theories of flavour, 
many rare \PB- and \PGt-decays receive tree-level shifts that will also be measured with excellent precision at \mbox{Tera-\PZ}.
While more detailed studies are needed to fully assess these prospects, 
key processes will be $\PB \to \PK^{(\ast)}\PGn\PAGn$, $\PQb \to \PQs\PGt\PGt$ transitions 
that are a unique opportunity at FCC-ee, and LFUV measurements in \PGt decays. 
In these models, the shift in $\PB \to \PK^{(\ast)}\PGn\PAGn$ 
is directly correlated to that in $\PBs \to \PGm\PGm$, 
which should be measured to few percent precision at HL-LHC~\cite{LHCb:2018roe}.
Lastly, these models are constrained by high-energy Drell--Yan measurements of $\Pp\Pp \to \ell\ell$, 
that should improve by nearly a factor of two after HL-LHC. 

Other flavour deconstruction models have been proposed where, for instance, 
a new force-carrying gauge boson field is associated with a TeV-scale $U(1)_{Y_3}$ $\PZ^\prime$, 
coupling primarily to the third family of fermionic fields~\cite{Allanach:2025wfi}.  
These models, currently favoured by measurements of \PB decays in tension with the SM predictions, 
can be thoroughly tested by FCC-ee.

To summarise, FCC-ee is an optimal machine for probing theories of flavour deconstruction 
because it achieves comparable sensitivity across diverse measurements 
traditionally separated into `electroweak' and `flavour' categories, 
but whose effects in such flavour models are inextricably tied. 
Viewing the electroweak and flavour programmes of FCC-ee together, 
and in combination with high-\pt and \mbox{low-\pt}  HL-LHC data, 
is therefore crucial to understanding its full power. 

\begin{table}[t]
\centering
\caption{Constraints from flavour, high \pt, and EWPOs 
on the mass of the gauge bosons predicted by flavour deconstruction.} 
\label{tab:pheno}
\begin{tabular}{c c c c}
\toprule
& Deconstructed $SU(2)_L$ & Deconstructed $U(1)_Y$ \\ \midrule
Electroweak: \PZ-pole \& $\PW\PW$-threshold & 9\,TeV (5\,TeV if exc.\ $m_{\PW}$) & 2\,TeV  \\
Flavour: $\PBs \to \PGm\PGm$ (up-alignment) & 7.5\,TeV & 2\,TeV \\
High \pt: Drell--Yan $\Pp\Pp \to \Pe\Pe, \PGm\PGm, \PGt\PGt$ & 4.5\,TeV & 3.5\,TeV \\ 
\midrule
EW projection FCC-ee & \multirow{2}{*}{30\,TeV} & \multirow{2}{*}{7\,TeV}\\
(on and above \PZ-pole \& \PW-pole)& & \\
\bottomrule
\end{tabular}
\end{table}

\subsubsection{Heavy Neutral Leptons}
\label{sec:HNL}

In its minimal version, the SM Lagrangian does not accommodate neutrino masses, 
but this historical accident can easily be overcome by minimally extending the SM. 
Nonzero neutrino masses offer many theoretical and phenomenological opportunities 
to address pending questions in the understanding of Nature.

Massive neutrinos can oscillate in flavour space, as experimentally observed since 1998. 
With three families, this phenomenon naturally leads to the possibility of breaking CP symmetry in the leptonic sector. 
Furthermore, fermion (or lepton) number violation might be observed. 
In the SM, fermion number conservation (FNC) stems from charge conservation for charged fermions
and from angular momentum conservation for massless neutrinos~\cite{Kayser:1989iu}. 
Thus, FNC is considered to be an accidental conservation law, which has no reason to hold if neutrinos are massive. 
The combination of the two possibilities opens the door to shedding light on the question of the existence of matter via leptogenesis.

\begin{figure}[ht]
\centering
\includegraphics[width=25em]{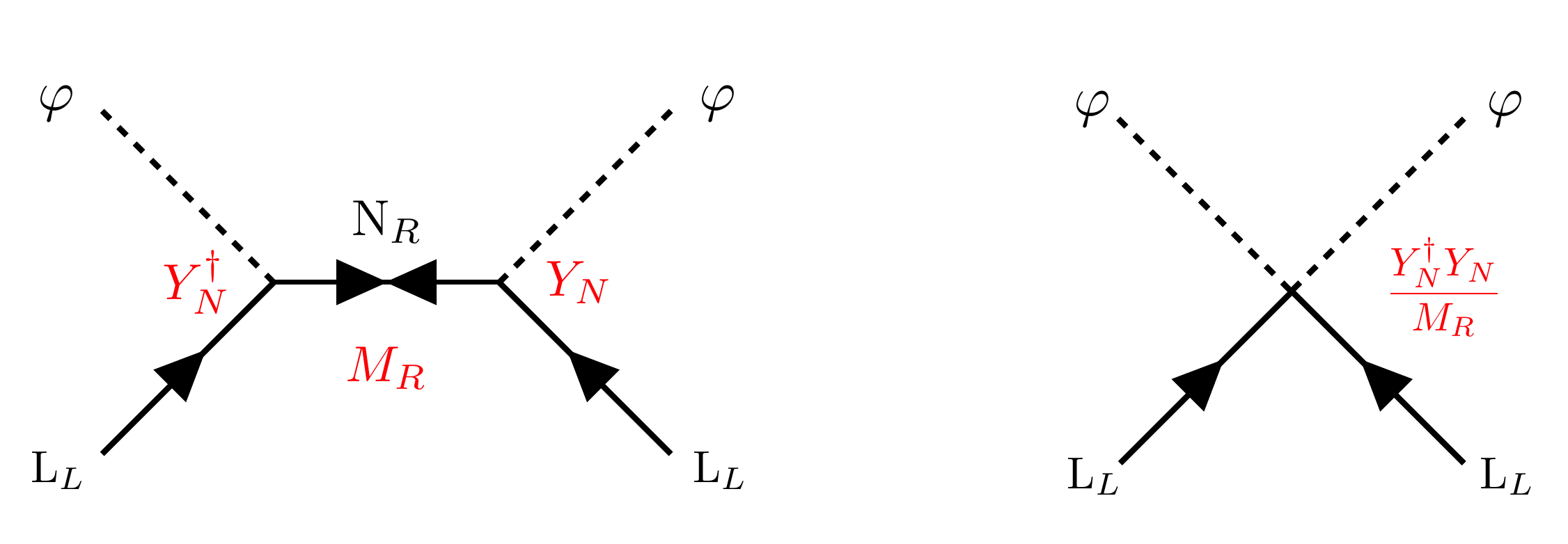} 
\caption{Schematic illustration of the type-I see-saw mechanism. 
Left: The left-handed lepton EW doublets $\mathrm{L}_L$ have Yukawa-like interactions $Y_N$, 
with the Higgs EW doublet $\varphi$, mediated by the heavy right-handed neutrinos $\mathrm{N}_R$. 
Right: At energies below the mass $M_R$ of these heavy neutrinos, 
an effective fermion-number-violating dimension-5 operator, 
involving two left-handed leptons and two Higgs fields, 
describes the low-energy degrees of freedom. 
After electroweak symmetry breaking, a Majorana mass is generated for the active left-handed neutrinos. 
This mass is proportional to the square of the original Yukawa coupling, 
to the squared Higgs vacuum expectation value $v$,
and to $1/M_R$. 
Heavy Neutral Leptons (HNLs), aligned with the right-handed neutrinos in the limit $M_R \gg v$, complete the spectrum.}
\label{fig:Seesaw}
\end{figure}

When the FNC rule is relaxed, active neutrinos can mix with right-handed neutrinos, 
via Yukawa interactions mediated by the Higgs field. 
This minimal scenario is a concrete realisation of the (type-I) see-saw mechanism (Fig.~\ref{fig:Seesaw}). 
The right-handed neutrinos, which do not carry electric charge, weak isospin, or colour, 
have no interaction that would distinguish the particle from the antiparticle: they are naturally Majorana particles. 
At energies much below the Majorana mass, 
the mixing is described by an effective dimension-5 Weinberg operator that, 
after electroweak symmetry breaking, generates an effective mass for the active neutrinos. 
Having two mass terms for each family, the neutrinos undergo level splitting thus the `see-saw', 
with typically light active neutrinos and heavy sterile neutrinos.  
Unlike for other SM fermions, 
the Yukawa coupling is not proportional to the mass of the light neutrinos, 
but could either be made proportional to the geometric mean of the masses of the light and heavy particles, 
or be enhanced with a symmetry that protects the masses of the light neutrinos~\cite{Agrawal:2021dbo}. 
The resulting heavy mass eigenstates are called Heavy Neutral Leptons (HNLs) and are nearly sterile particles. 
The ensuing rich phenomenology includes the possible observation of LLPs, 
for which, in the mass range 20--80\,GeV, 
a thorough and conclusive search can be performed at FCC-ee, during the \mbox{Tera-\PZ} run.

Generalising beyond this minimal scenario, FCC-ee can search for HNLs in the large sample of \PZ bosons produced resonantly. 
The \PZ decay to $\PGn_L + \text{HNL}$ followed by the decays of the heavy neutral lepton 
$\text{HNL} \to \ell + \PW^*$, 
$\text{HNL} \to \PGn + \PZ^*$, 
leads to a wealth of final state signatures that can be exploited at FCC-ee~\cite{Blondel:2014bra}. 
The \PZ branching ratio to HNLs is proportional to the squares of the mixing angles $U^2_{ij}$ between the SM neutrinos and the HNL, 
where $i,j$ run over the three neutrino flavours.

The very large number of \PZ boson decays allows the exploration of very low $U^2_{ij}$ values, for HNL masses below the \PZ mass. 
The explored values  may result in HNLs with a measurable path in the detectors, 
yielding spectacular signatures with jets or leptons produced far from the interaction vertex, with no SM backgrounds. 
In particular, for decay lengths in the detector between 1\,mm and 2\,m, 
a discovery would be possible for HNL masses up to 60--70\,GeV down to $U^2$ values of ${\cal O}(10^{-11})$. 
For larger HNL masses, up to the kinematic limit of the \PZ mass, a prompt analysis would be needed, 
which, because of significant SM backgrounds, would allow discovery for $U^2$ values of ${\cal O}(10^{-9})$.

\begin{figure}[ht]
\centering
\includegraphics[width=0.8\textwidth]{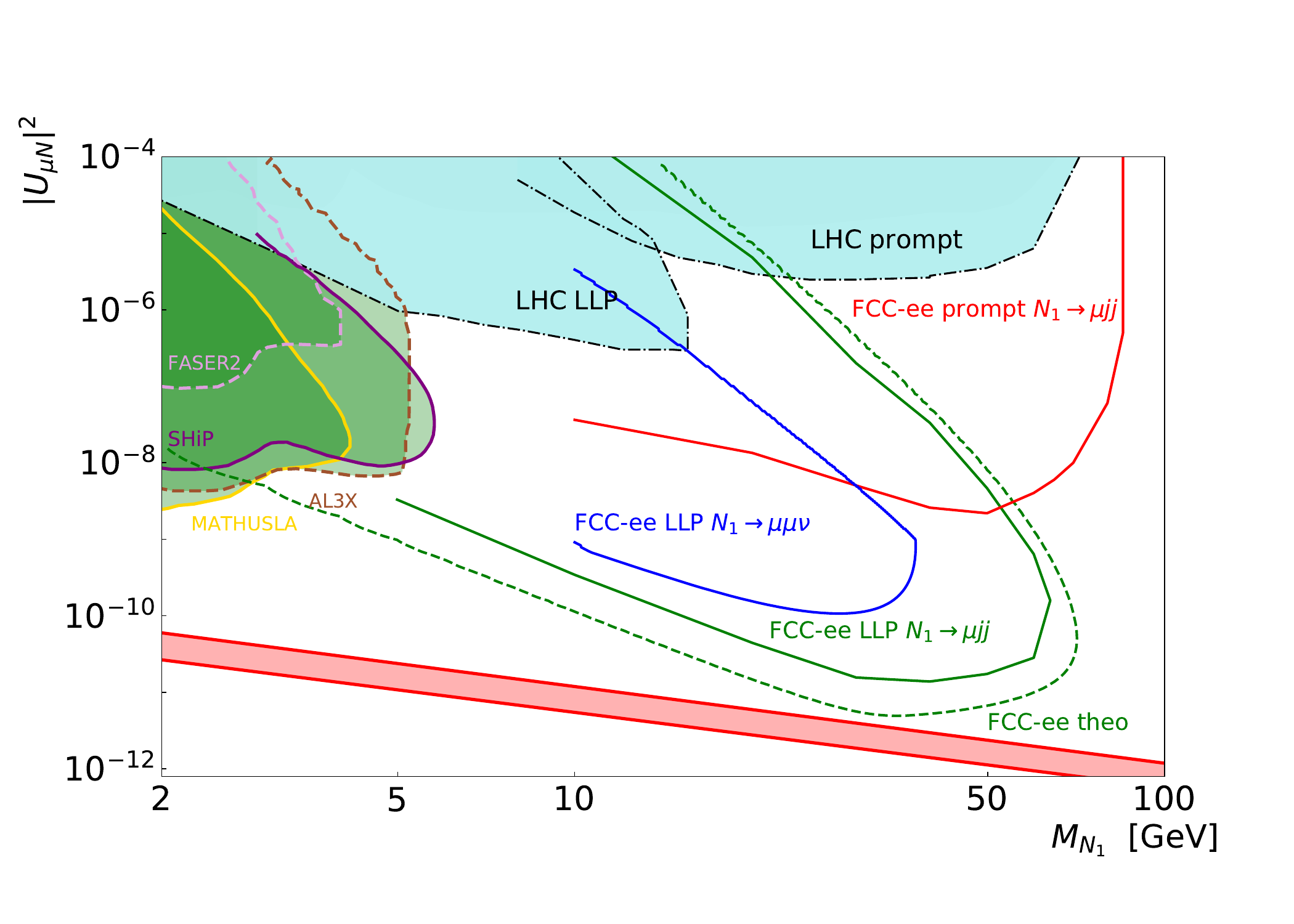}
\caption{Discovery potential in the $m_{N}-|U_{\mu N}|^2$ plane. 
The FCC-ee potential is based on the decay channel $\text{HNL} \to \ell \PGn j j$ 
and is shown as a red (green) line for the prompt (long-lived) analyses described in the text. 
The blue line shows the reach of a search for long-lived particles in the decay channel $\text{HNL} \to \PGmp\PGmm \PGn$.
The dashed green line bounds the area where, out of $6 \times 10^{12}$ \PZ bosons, 
three events are produced with visible HNL decays inside an FCC-ee detector, 
i.e., with a displacement smaller than 5\,m
and larger than 0.5\,mm (based on the analytical formulas in Ref.~\cite{Drewes:2022rsk}).
The requirement to explain the light neutrino masses imposes a lower bound, indicated as a pink band,  
on the total HNL mixing (summed over flavours). 
The width of this band indicates the uncertainty in this lower bound 
due to the current lack of knowledge about the absolute scale and the ordering of the light neutrino masses. 
Light neutrino oscillation data can be explained anywhere above this band, 
in particular in models where the neutrino masses are protected by a symmetry related to approximate lepton-number conservation. 
Furthermore, this region could also accommodate the observed matter-antimatter asymmetry 
via a leptogenesis mechanism~\cite{Drewes:2021nqr}. 
The existing limits from LHC searches are shown as turquoise areas. 
The expected discovery potential of projected experimental searches based on long baseline experiments 
are shown as green areas and are taken from the website accompanying Ref.~\cite{Antel:2023hkf}, 
where all the original works are cited.}
\label{fig:hnlreach}
\end{figure}

Detailed simulation studies were performed to verify the experimental feasibility of these analyses 
and to determine the corresponding requirements on detector design. 
The reach in parameter space was first studied in a toy model commonly used to compare different experimental approaches, 
rather than a complete model for neutrino masses. 
The model features a single Majorana HNL mixing with a single flavour of active neutrinos. 
The signal samples and all of the SM decays of the \PZ boson, 
as well as four-fermion processes yielding the same final state as the HNL decays, 
were passed through a parametrised simulation of the IDEA detector (Chapter~\ref{sec:concepts}) and then analysed. 
Both semi-leptonic and fully leptonic decays of the HNL were studied, 
for the cases of mixing with an electron neutrino and with a muon neutrino. 
The fully leptonic decay mode provides a clean experimental final state, 
with a good reach for long-lived decays, but with a significant price in branching fraction. 
The semileptonic decay $\text{HNL} \to \ell \PGn j j$ has a branching fraction $\sim$\,50\%, 
allows full kinematic reconstruction of the neutrino decay, 
and was studied both for long-lived and prompt decays. 
The results are shown in Fig.~\ref{fig:hnlreach} and documented in 
Refs.~\cite{MSc_Moulin, MSc_Critchley, note_HNL_Polesello_Valle_mujj,Blondel:2022qqo, 
note_HNL_Presilla_Giappichini,note_HNL_Williams}. 

The conclusion of these studies is that searches at FCC-ee enable the HNL discovery 
over a mass range beyond the reach of specialised detectors for LLP searches being developed for HL-LHC 
and for mixing values much smaller than those covered by future searches at HL-LHC, 
for both prompt and long-lived signatures. 
Besides the work based on parametrised detector simulations, 
the model is also being studied~\cite{note_HNL_CLD_Andrea} 
through a detailed \textsc{Geant4} simulation of the ILD detector (Chapter~\ref{sec:concepts}).

The models featuring a single HNL are useful for assessing, in a simplified way, the parameter space coverage of the experiments. 
In order to explain the observed neutrino oscillations, however, at least two HNLs are needed. 
A realistic model~\cite{Tastet:2021vwp} featuring two Majorana neutrinos with coupling to all three flavours of active neutrinos 
was studied in the fully leptonic final state featuring two electrons or muons in the final state. 
The patterns of couplings were chosen such as to be compatible with neutrino oscillation data 
based on the benchmarks proposed in Ref.~\cite{Abdullahi:2022jlv}. 
As for the single-neutrino analyses, 
the study was performed with the parametrised simulation of the IDEA detector and featured a full background analysis. 
An estimate of the reach was obtained for different coupling scenarios, 
corresponding to both normal and inverted neutrino mass hierarchy.

The phenomenology of the `Symmetry Protected Seesaw Scenario'~\cite{Shaposhnikov:2006nn, Kersten:2007vk} 
has been presented in Refs.~\cite{Antusch:2016ejd, Antusch:2022ceb}. 
This model adds two right-chiral neutrinos, required to explain the observed light neutrino mass squared differences, 
provided with a protective lepton number-like symmetry (LNLS) 
to ensure that the two heavy neutral leptons form an almost degenerate pair of pseudo-Dirac neutrinos.
The main phenomenological signature of this model is the production of lepton flavour violating final states, 
with a probability that oscillates with the length of the flight path of the HNL in the detector~\cite{Antusch:2023jsa}. 
The region in parameter space where this oscillation would be detectable at FCC-ee 
is mapped in detail in the phenomenological study documented in Ref.~\cite{Antusch:2024otj}. 
An experimental study was performed based on this work, 
with the same parametrised detector simulation and software as the studies described above~\cite{note_HNL_Polesello_Valle}. 
For HNL masses below 35\,GeV, for favourable ratios of the HNL width and of the mass difference between the two pseudo-Dirac states, 
a striking oscillation signal would be observed, allowing the measurement of the mass difference.

A large fraction of the model parameter space consistent with the see-saw mechanism, 
including those with symmetry-enhanced neutrino masses, 
such as the inverse~\cite{Mohapatra:1986bd} or linear~\cite{Akhmedov:1995ip,Akhmedov:1995vm} models, 
would be probed. 
For masses around 40\,GeV, the FCC-ee sensitivity would reach the smallest value of the mixing angle 
theoretically compatible with the mass range experimentally still allowed for the active neutrinos.

With HNLs (as well as with other light and feebly-coupled particles, such as ALPs), 
FCC-ee is simultaneously a discovery and a precision tool, in a single collider. 
In view of the cost and time scales associated with the construction of each new collider, 
such an advantage cannot be overstated.
Indeed, the number of HNLs that can be detected during the \PZ-pole run 
grows as $|U_{\ell N}|^4$ if their decay length in the laboratory exceeds the detector size, 
or as $|U_{\ell N}|^2$ if it is smaller than the effective detector dimensions
As a direct consequence, 
up to a million HNL decays can be observed during the \mbox{\PZ-pole} run
if $|U_{\ell N}|$ is near the current upper experimental limits~\cite{Drewes:2022rsk}. 
Such a large sample opens the way for the measurement of several quantities, 
from which important information about the underlying particle physics model can be extracted,
as the following.
\textit{i)}~The branching ratios of HNL decays into different SM generations 
can be measured with per-cent accuracy~\cite{Antusch:2017pkq}, 
which allows the consistency with neutrino oscillation data to be checked;
many parameters of the see-saw model to be constrained~\cite{Drewes:2024bla} 
(including CP-violating phases and the absolute neutrino mass scale); 
the leptogenesis hypothesis~\cite{Antusch:2017pkq} to be tested; 
and underlying flavour and CP symmetries~\cite{Drewes:2024pad} to be probed.
\textit{ii)}~The number of HNL events and the measurement of the HNL lifetime 
can reveal information about the violation of lepton number~\cite{Blondel:2022qqo,Drewes:2022rsk} 
as well as underlying flavour and CP symmetries~\cite{Drewes:2024pad}.
\textit{iii)}~Finally, the angular distribution and the energy spectrum of the HNL decay products 
can provide information regarding the lepton number violation~\cite{Blondel:2021mss} 
and the HNL mass splitting~\cite{Antusch:2023jsa,Antusch:2024otj,note_HNL_Polesello_Valle}.

The combination of these observables may provide important information about the HNL role in particle physics and cosmology, 
in particular in the context of neutrino masses and leptogenesis. 
It can also shed light on the properties of the underlying particle-physics theory 
within which the see-saw model is embedded (such as flavour and CP symmetries). 

\subsubsection{Dark matter and dark sectors}
\label{sec:dark}

Understanding the origin\,/\,nature of dark matter (DM) is a central question in contemporary physics, 
connecting particle physics and astrophysics. 
Despite decades of searches across experiments, the favoured DM candidates, the WIMPs, have not been found, 
and the lower bounds on their masses are progressively pushed beyond the reach of TeV-class $\epem$ colliders.

Mono-photon searches at FCC-ee can play a significant role in probing the so far unexplored parameter range 
allowed by the WIMP relic density constraints~\cite{Horigome:2021qof}.  
In particular, predictive models of leptophilic candidates with a thermal origin can be tested.  
Missing energy signatures at FCC-ee can probe much of the parameter space 
for which DM direct annihilation into a dilepton yields the observed relic density 
in Higgs-like models with mass-proportional couplings to charged leptons~\cite{Cesarotti:2024rbh}. 
Models of asymmetric dark matter can also be probed at FCC-ee, 
with a sensitivity to DM-lepton interactions improved by almost an order of magnitude~\cite{Roy:2024ear}. 
Additionally, it would be possible to detect spin 0, 1, and~1/2 DM at FCC-ee 
in the context of simple, consistent, and renormalisable field theories 
that provide the correct DM abundance and satisfy direct detection, indirect detection, 
and collider constraints~\cite{Grzadkowski:2020frj}. 

Hidden sectors, consisting of new, invisible particles that interact almost imperceptibly with the SM, 
are rapidly gaining attention as they could hold the answer, 
not only to the dark matter problem but also to a variety of other open questions in the field. 
The dark sector could contain a multitude of hidden particles. 
After all, the visible sector is non-minimal, 
so there is no good reason why the dark sector should contain only one type of dark matter particle and nothing else. 
Benefiting from a clean environment, high luminosity, and large acceptance, 
FCC-ee can directly scrutinise the $\mathcal{O}$(1--100)\,GeV mass range for 
so-far unknown particles, with interactions that would otherwise have been too feeble to detect. 

Dark sectors typically exhibit a stable particle, fundamental or composite, that could be a DM candidate, 
and one or more mediator particles coupled to the SM via a neutral portal. 
The spin of the mediator particle defines the portal: 
vector, e.g., dark photon; 
scalar or pseudoscalar, e.g., Higgs portal; 
or fermion, e.g., sterile neutrino. 
These well-motivated dark sectors can be thoroughly explored at FCC-ee 
in a mass-coupling range that is inaccessible by any other means with the same sensitivity. 
Furthermore, small couplings of dark sector particles to SM particles could give rise to LLPs 
and other unconventional collider signatures, such as dark showers.  

Dark photons may mix kinetically with the SM hypercharge gauge boson. 
This possibility can be studied in $\epem \to \mathrm{A}^\prime \gamma$ production, 
where the dark photon A$^\prime$ decays to $\PGmp\PGmm$. 
The sensitivity to small coupling values could be significant, 
reaching around $\sim$\,$2 \times 10^{-4}$ for $m_\mathrm{A^\prime}$ ranging from 10 to 80\,GeV at FCC-ee~\cite{Karliner:2015tga}. 
Alternatively, dark photons or dark $\PZ^\prime$ bosons can also be investigated 
via $\epem \to \mathrm{A}^\prime \PH$ or $\PZ^\prime \PH$ production at the $\PW\PW$ threshold or above~\cite{Giffin:2020jtl}.

Finally, the associated production of a neutrino and a dark sector fermion can also be searched for at FCC-ee 
in mono-photon signatures~\cite{Ge:2023wye}. 
This kind of search has sensitivity up to mass values in excess of 1\,TeV. 
Other searches, such as invisible decays of a dark Higgs boson~\cite{Haghighat:2022qyh}
or \PZ boson decays to an invisible dark photon~\cite{Cobal:2020hmk} can also be exploited.

\subsubsection{Axion-like particles}
\label{sec:alps}

Axion-like particles (ALPs) are generically expected in many BSM extensions, 
most famously in QCD axion models originally introduced to tackle the strong CP problem~\cite{Bauer:2018uxu}. 
In these models, ALPs can be heavier than the QCD axion and serve as mediators to the dark sector. 
Astrophysical and beam-dump experiments constrain the ALP-photon coupling at low masses, 
but are less stringent in the 0.1--100\,GeV range~\cite{Agrawal:2021dbo,dEnterria:2021ljz}, 
potentially accessible at FCC-ee via $\epem \to \axion \gamma$~\cite{Bauer:2018uxu} 
and $\epem \overset{\gamma\gamma}{\longrightarrow} (\epem) \axion$~\cite{RebelloTeles:2023uig}, 
with the ALP $\axion$ decaying as $\axion \to \gamma\gamma$. 
Various aspects of the ALP phenomenology at FCC-ee are addressed 
in Refs.~\cite{Cheung:2023nzg,Calibbi:2022izs,Liu:2022tqn,Zhang:2021sio}

The FCC-ee sensitivity to the $\epem \to \axion \gamma$ channel was studied in Ref.~\cite{note_ALP_Polesello} 
with the parametrised simulation of the IDEA detector.
Two final states were addressed: 
the `prompt' case, where the ALP decays near the interaction vertex, 
and the `monojet' case, where the ALP decays outside the detector, 
yielding the signature of a monochromatic photon recoiling against missing energy and mass. 
For both channels, a full kinematic analysis was developed to separate the signal from irreducible SM backgrounds.

Sensitivity to photon couplings down to $<10^{-3}$\,TeV$^{-1}$ can be obtained at FCC-ee 
in the $m_\axion$ range from 0.1 to 100\,GeV, 
extending current limits by more than two orders of magnitude in the \mbox{Tera-\PZ} run. 
In particular, the `monophoton' signature would cover a difficult region in the vicinity of 1\,GeV, 
which is not accessible to beam dump experiments. 
In an intermediate region in mass and at small coupling values, 
the ALP lifetime is large enough to give rise to measurable paths inside the detector, 
providing LLP signatures that could be accessible by exploiting the pointing capabilities of the calorimeters.

The ALP coupling to \PZ and Higgs bosons can also be probed via $\epem \to \PZ \axion$ or $\PH\axion$, 
with visible \PZ boson decays or $\PH \to \PQb\PAQb$, 
and either $\axion \to \gamma\gamma$ or $\axion \to \ell^+\ell^-$. 
Decays to SM particles other than photons are less constrained and provide an additional opportunity for ALP discovery at FCC-ee.
A preliminary study of the sensitivity of FCC-ee to ALPs decaying into two gluons is presented in Ref.~\cite{note_ALP_Andrea}.

\begin{figure}[ht]
\centering
\includegraphics[width=\linewidth]{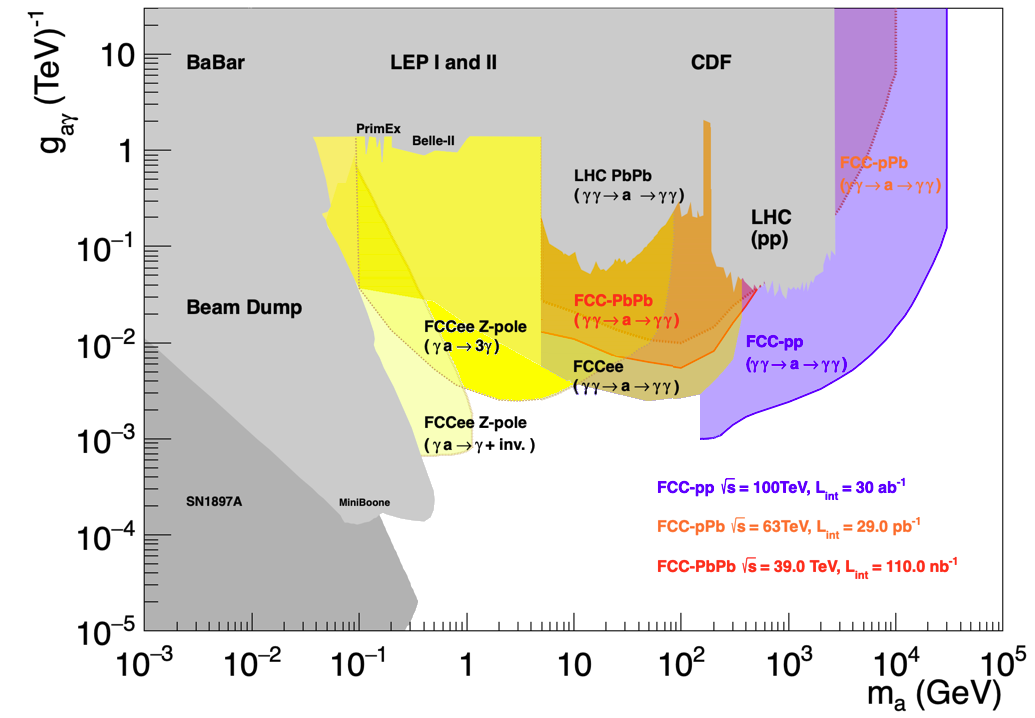}
\caption{Projected sensitivity for ALPs in the photon coupling vs.\ ALP mass plane 
from the three following processes at FCC-ee: 
$\epem \to \gamma \axion \to 3\gamma$ (yellow area),  
photon-fusion $\gamma\gamma \to  \axion \to 2\gamma$ (salmon area)~\cite{RebelloTeles:2023uig},
and $\epem \to \gamma \axion \to \gamma + \text{INV}$ (orange area). 
Existing limits (in grey) are adapted from Refs.~\cite{Agrawal:2021dbo,Antel:2023hkf}. Also shown are projected exclusion limits at 95\% C.L. on the ALP-photon coupling as a function of the ALP mass expected from searches for $\gamma\gamma\to \axion \to \gamma\gamma$ in pp (violet), pPb (dark pink), and PbPb (orange) collisions at FCC-hh \cite{PRebello_DdE25}.}
\label{fig:axion}
\end{figure}

The projected sensitivity of the ALP search is illustrated in Fig.~\ref{fig:axion}, 
where the importance of FCC-ee is conspicuous in probing ALPs coupling to photons 
with a lifetime below cosmological/astrophysical scales, i.e., with a mass above 1\,MeV. 

\subsubsection{Exotic decays of the Higgs and \texorpdfstring{\PZ}{Z} bosons}
\label{sec:exoHZ}

The large Higgs boson event sample at FCC-ee allows a direct search for exotic decays of the Higgs boson. 
Exotic decays are predicted by a variety of BSM theories~\cite{Curtin:2013fra, Cepeda:2021rql,Ma:2020mjz, Alipour-Fard:2018lsf}, 
including extended scalar sectors, SUSY, and dark sector models with Higgs or vector portals. 
Higgs branching ratios up to four orders of magnitude lower than the expected HL-LHC reach will be probed at FCC-ee,
giving access to entirely new channels~\cite{Liu:2016zki}. 
  
\begin{figure}[ht]
\centering
\includegraphics[width=0.7\linewidth]{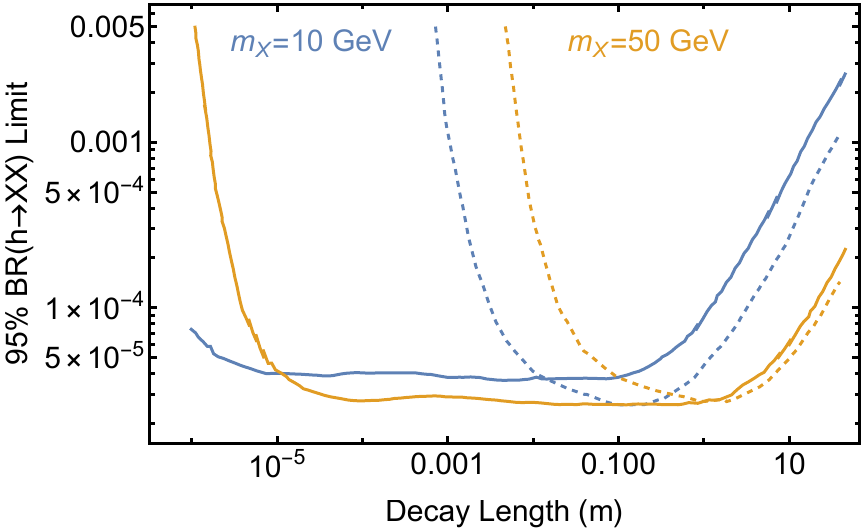}
\caption{Estimate of the sensitivity to LLPs in exotic Higgs portal decays; adapted from Ref.~\cite{Alipour-Fard:2018lsf}. 
Solid (dashed) lines correspond to dedicated light (heavy) LLP search strategies.}
\label{fig:HiggsLLP}
\end{figure}

Substantial work in preparing searches for Higgs boson decays into long-lived particles (LLPs) is ongoing,  
focusing on signatures with displaced vertices~\cite{note_LLPs_ExoticHiggsDecays}. 
The results suggest that FCC-ee is sensitive to long-lived scalars with decay lengths of order 1\,mm to 10\,m, 
with the peak sensitivity for decay lengths around 0.3\,m, as shown in Fig.~\ref{fig:HiggsLLP}. 
Additional studies, 
comparing different detector setups for exotic Higgs decays, 
have also been initiated~\cite{note_exoticHiggs_Larson}.

Searches for exotic \PZ decays to LLPs can also be carried out at FCC-ee during the \PZ pole run. 
In models of $R$-parity-violating supersymmetry, studied in Ref.~\cite{Wang:2019orr}, 
the FCC-ee sensitivity to long-lived light neutralinos in \PZ decays exceeds that of hadron colliders 
or of future dedicated LLP experiments by three orders of magnitude.
Comparable gains in sensitivity can be obtained in other searches, such as \PZ decays into hidden mesons~\cite{Cheng:2019yai}.

The available parameter space of new SM-neutral vector bosons (${\PZ}^\prime$) that couple exclusively to leptons 
could be significantly extended, 
offering the best performance of all future collider projects in the kinematically allowed mass range 
(from 5 to 360\,GeV)~\cite{GonzalezSuarez:2024dsp}.

In brief, by taking full advantage of the large \mbox{Tera-\PZ} event samples, 
the dark sector programme at FCC-ee extends the intensity-frontier discovery potential significantly. 
It provides access to very-weakly-interacting dark-sector particles 
with couplings so small that they are usually considered typical of beam-dump experiments, 
but in a mass range wholly inaccessible to any other intensity-frontier experiment.  
Given that there is no preferred mass range for such states, 
this mass-range extension by more than two orders of magnitude provides unique scientific opportunities in particle physics.

\subsubsection{Other new physics searches}

Other possible studies at FCC-ee include, for example, 
searches for lepton-flavour violation (LFV) in different 
manners~\cite{Hajahmad:2024arh,Munbodh:2024shg,Altmannshofer:2023tsa,Ho:2022ipo,Calibbi:2021pyh,Kamenik:2023hvi}, 
compositeness~\cite{Stefanek:2024kds}, leptoquarks~\cite{Crivellin:2020ukd}, 
new scalars~\cite{magnan2024}, 
and more.   
In this instance, the production of hidden-valley light quarks would perturb the QCD cascade, 
generating correlations among the final state particles that are absent in the SM.  
A detailed experimental study is in progress.  

\subsubsection{Complementarity and synergy between FCC-ee and FCC-hh}

\begin{figure}[ht]
\centering
\includegraphics[width=0.6\linewidth]{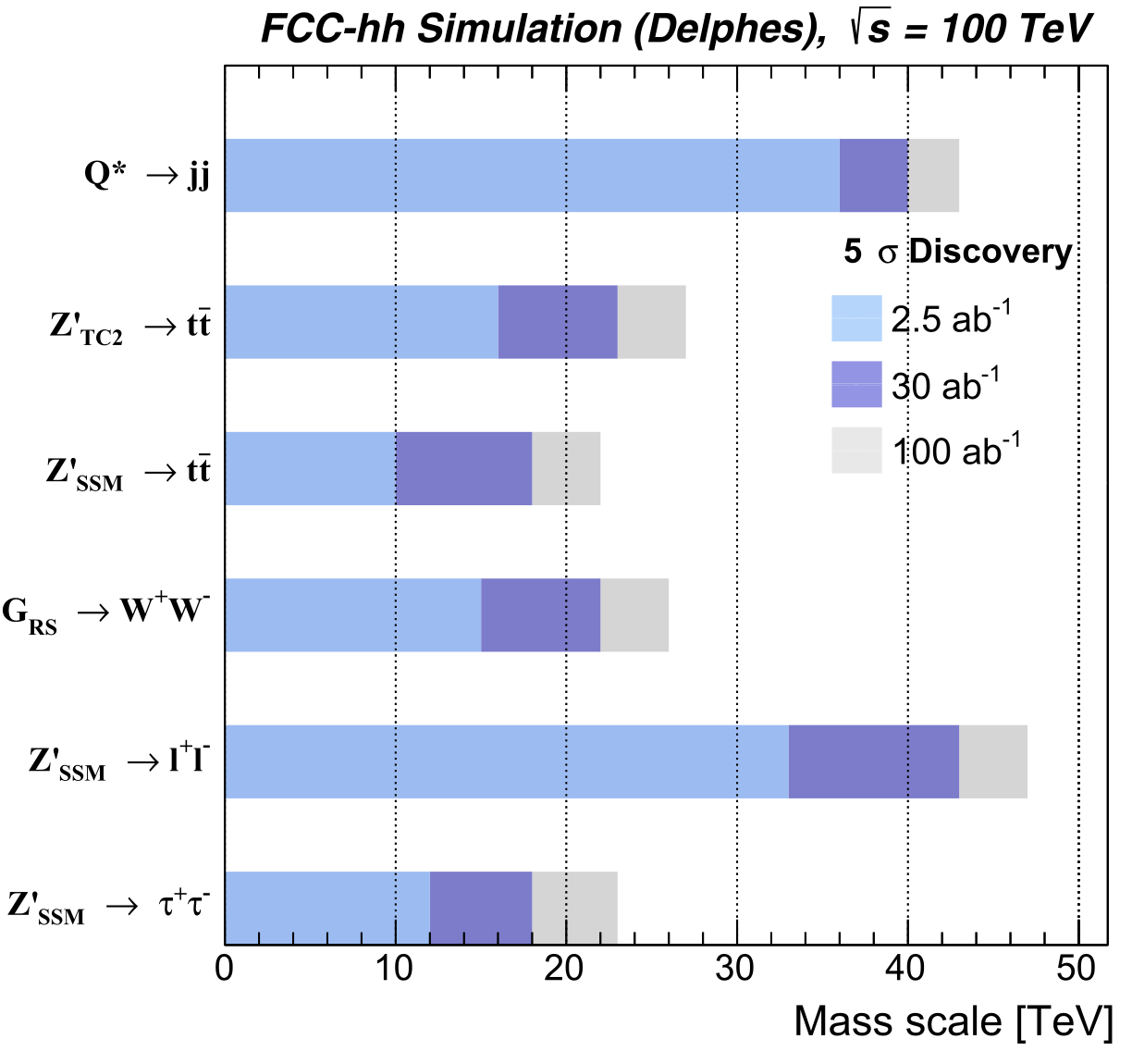}
\caption{Summary of the 5\,$\sigma$ discovery reach, 
as a function of the resonance mass, for different FCC-hh luminosity scenarios. 
From Ref.~\cite{Helsens:2019bfw}.}
\label{fig:schannel}
\end{figure}

The FCC-hh will complement and substantially extend the FCC-ee physics reach in nearly all possible directions. 
The seven-fold centre-of-mass energy increase with respect to LHC 
enhances the potential for observing new particles at mass scales up to 40\,TeV, as shown in Fig.~\ref{fig:schannel}. 
Indirectly, it will be sensitive to energies well above its kinematic reach of 100\,TeV, 
for example in the tails of Drell--Yan distributions. 
Should any deviations from SM expectations be observed at FCC-ee, FCC-hh has the potential to pinpoint its microscopic origin.  
Some specific synergies between FCC-ee and FCC-hh in this regard are highlighted in the next paragraphs.

As already alluded to in Section~\ref{sec:PhysicsCaseHiggsEW}, 
there are many synergies between FCC-ee and FCC-hh that come together 
to make the integrated FCC physics programme more than the sum of its parts. 
For example, the absolute determinations of Higgs couplings and its total width at FCC-ee 
is necessary to fully exploit the potential of FCC-hh that is only able to measure relative coupling ratios. 
The clean environment of FCC-ee enables the precise determination of SM parameters necessary to reduce uncertainties for FCC-hh. 
Moreover, certain types of BSM physics can only be accessed by FCC-ee and not FCC-hh, and vice versa.   
For instance, the nature of the electroweak phase transition will be probed by the precise measurement of the Higgs self-coupling 
and the search for additional Higgs partners at FCC-hh, together with potential deviations in the Higgs couplings at FCC-ee.

More concretely, if a deviation in $\kappa_{\PZ}$ were observed at the level of 0.5\%,  
this hint would be less than half of one sigma at HL-LHC (with some unavoidable assumptions), 
but would correspond to an unambiguous 5\,$\sigma$ discovery at FCC-ee.  
In Ref.~\cite{Abu-Ajamieh:2020yqi}, 
it was shown that for a BSM modification of this coupling 
new physics would have to show up at an energy scale of 
\begin{equation}
E_{\text{Max}} \simeq \frac{1.1 \text{TeV}}{\sqrt{|\kappa_Z-1|}} \, .
\end{equation}
Thus, a deviation of 5\,$\sigma$ at FCC-ee would correspond to an energy scale of 15\,TeV, 
essentially guaranteeing that FCC-hh would be able to discover the heavy states 
ultimately responsible for the low energy Higgs coupling deviation, 
illustrating a compelling, dynamic, interplay between FCC-ee and FCC-hh.

\begin{figure}[ht]
\centering
\includegraphics[width=1.0\linewidth]{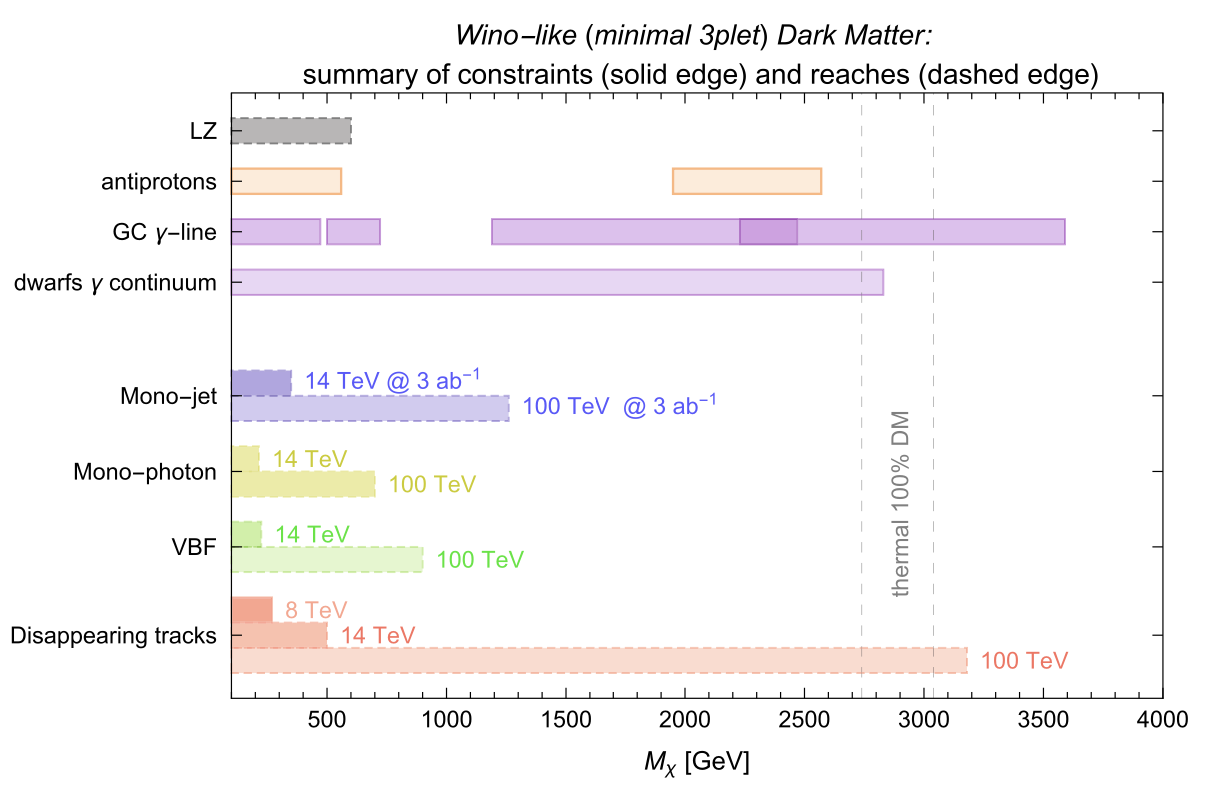}
\caption{The projected sensitivity of searches for a WIMP triplet DM candidate 
in final states with disappearing tracks at FCC-hh~\cite{fcc-phys-cdr}.
Adapted from Ref.~\cite{Cirelli:2014dsa}.}
\label{fig:DM}
\end{figure}

This synergy also extends to astrophysics and cosmology. 
Beyond-the-Standard-Model physics leading to a phase transition of some new sector, 
anywhere from the electroweak to the TeV scale, 
could yield gravitational wave signatures accessible to the next-generation gravitational wave observatories, 
such as LISA, the origin of which could then potentially be directly accessed at FCC-hh~\cite{Friedrich:2022cak}. 
The detection of high-energy gamma rays in astrophysical observatories such as CTA~\cite{Montanari:2022buj} 
may be due to a TeV-scale WIMP, 
but with a large degree of uncertainty inherent to such indirect observation methods. 
The FCC-hh will be directly sensitive to the entire upper mass range,
of around 1 and 3\,TeV for triplet and doublet thermal WIMP dark matter, respectively~\cite{fcc-phys-cdr}, 
far beyond the reach of any conventional dark matter direct detection experiment, as illustrated in Fig.~\ref{fig:DM}. 
Cosmic-ray physics will furthermore benefit from hadron collision data at unprecedented energy and luminosity. 

The multiplicity of partons in the proton makes FCC-hh the most versatile high-energy parton collider 
for the broadest possible exploration of elementary particle processes. 
Similarly to LHC, its capabilities may be augmented by hosting complementary detectors for 
neutrino physics, dark sectors, and long-lived particles that may otherwise escape undetected.   
Together, FCC-ee and FCC-hh can truly enter the tens of TeV scale, indirectly and directly, 
to fully explore the open questions that LHC has only begun to touch upon. 

\subsection{Selected topics in flavour physics}
\label{sec:PhysicsCaseFlavour}

The peculiar structure of quark and lepton masses, as well as the quark mixing angles, 
is ad~hoc within the SM and is likely the low-energy imprint of some new dynamics. 
This structure implies approximate flavour symmetries that give rise to the strong suppression 
of a series of flavour-changing processes within the SM,  
whose precise experimental study allows probing dynamics at scales well above the electroweak scale.  
It is necessary to push the precision flavour frontier further and the FCC-ee programme offers unique opportunities in this respect.
Despite the fact that many tests in this sector have been performed in the recent past, 
without reporting significant deviations from the SM, the discovery potential remains high.

One class of unique opportunities concerns \PQb-hadron and \PGt decays, and, in particular, $\PQb \to \PGt$ transitions. 
These processes directly test models where 
new physics is coupled mainly to the third generation 
--- a general feature of many explicit BSM scenarios. 
Moreover, the class of models that can be tested indirectly via such studies 
is also the one less constrained by direct searches (given the suppressed couplings to light quarks). 
Here, the FCC-ee programme offers a particularly large margin of improvement 
over the accuracy expected from currently approved flavour physics experiments, 
both at HL-LHC and at lower-energy $\epem$ colliders (in particular Belle~II at Super~KEKB).
The key features of the flavour-physics programme during the \PZ-pole run at FCC-ee can be summarised as follows.

\begin{itemize}
\item \emph{Clean environment} (as in \PB-factories), 
with momentum and tagging efficiency of the pair-produced \PQb's, \PQc's, and \PGt's from \PZ decays, 
with $\sim$\,10 times more $\bbbar$ and $\ccbar$ pairs than the total collected by Belle~II (Table~\ref{tab:flavouryields}).
\item \emph{Boosted \PQb's and \PGt's}, leading to a significantly higher efficiency (compared to \PB-factories) 
for modes with missing energy (especially multiple-\PGn modes) and inclusive modes, 
as well as smaller uncertainties in lepton ID efficiencies.  
\end{itemize}

\begin{table}[ht]
\centering
\caption{Yields of heavy-flavoured particles produced at FCC-ee for $6 \times 10^{12}$ \PZ decays~\cite{Monteil:2021ith}.} 
\label{tab:flavouryields}
\begin{tabular}{c ccccccc} 
\toprule
Particle species & \PBz & \PBp & \PBzs & \PGLb & \PBpc & $\PQc\PAQc$ & $\PGtm\PGtp$ \\
\midrule
Yield ($\times 10^9$)    & 370 &  370 &  90  & 80 & 2 & 720 & 200 \\
\bottomrule
\end{tabular} 
\end{table}

These features lead to to the identification of the following class of observables as particularly promising: 
i)~neutral-current rare \PQb- and \PQc-hadron decays with $\PGtp\PGtm$ and $\PGn\PAGn$ pairs in the final state; 
ii)~charged-current \PQb-hadron decays with a $\PGt\PGn$ pair in the final state; 
iii)~CP-violating observables in \PQb- and \PQc-hadron decays involving neutral particles in the final states 
(\PGpz, \PKS, \PGg, \ldots);  
iv)~lepton flavour violating \PGt decays; 
v)~precision tests of lepton universality in \PGt decays.

In addition to these \emph{classical} flavour studies via low-energy probes, 
a truly unique opportunity offered by FCC-ee is the possibility of 
determining flavour parameters from \PW decays,
as well as searching for flavour-violating decays of \PZ and Higgs bosons. 
Within the SM, these processes are possibly too rare to be detected, 
but their search is an effective way to constrain or discover BSM dynamics. 
Some further ideas uniquely relevant to the FCC-ee flavour programme 
are discussed in Refs.~\cite{Grossman:2021xfq,Chrzaszcz:2021nuk, Monteil:2021ith}

The wide-ranging FCC-ee flavour physics programme cannot be easily summarised in a few pages. 
The most interesting classes of observables are listed below, 
together with a discussion of a few examples that illustrate the physics reach and the diversity of this programme.

\subsubsection{Lepton universality tests in \texorpdfstring{\PGt}{tau} decays} 

\begin{figure}[ht]
\centering 
\includegraphics[width=0.4\textwidth]{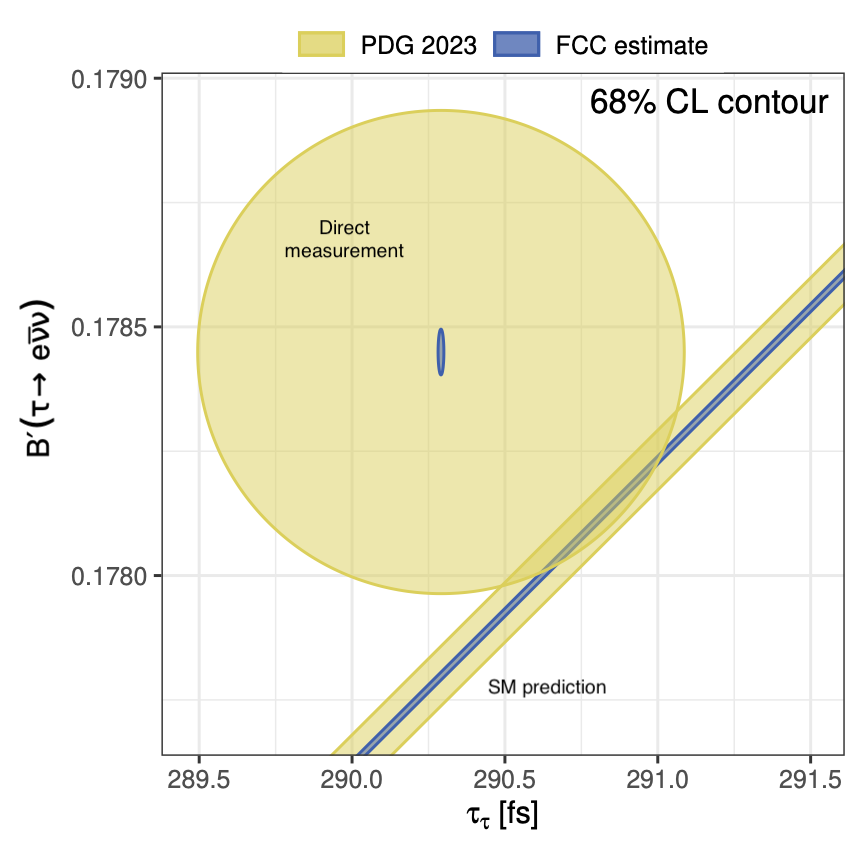}
\includegraphics[width=0.4\textwidth]{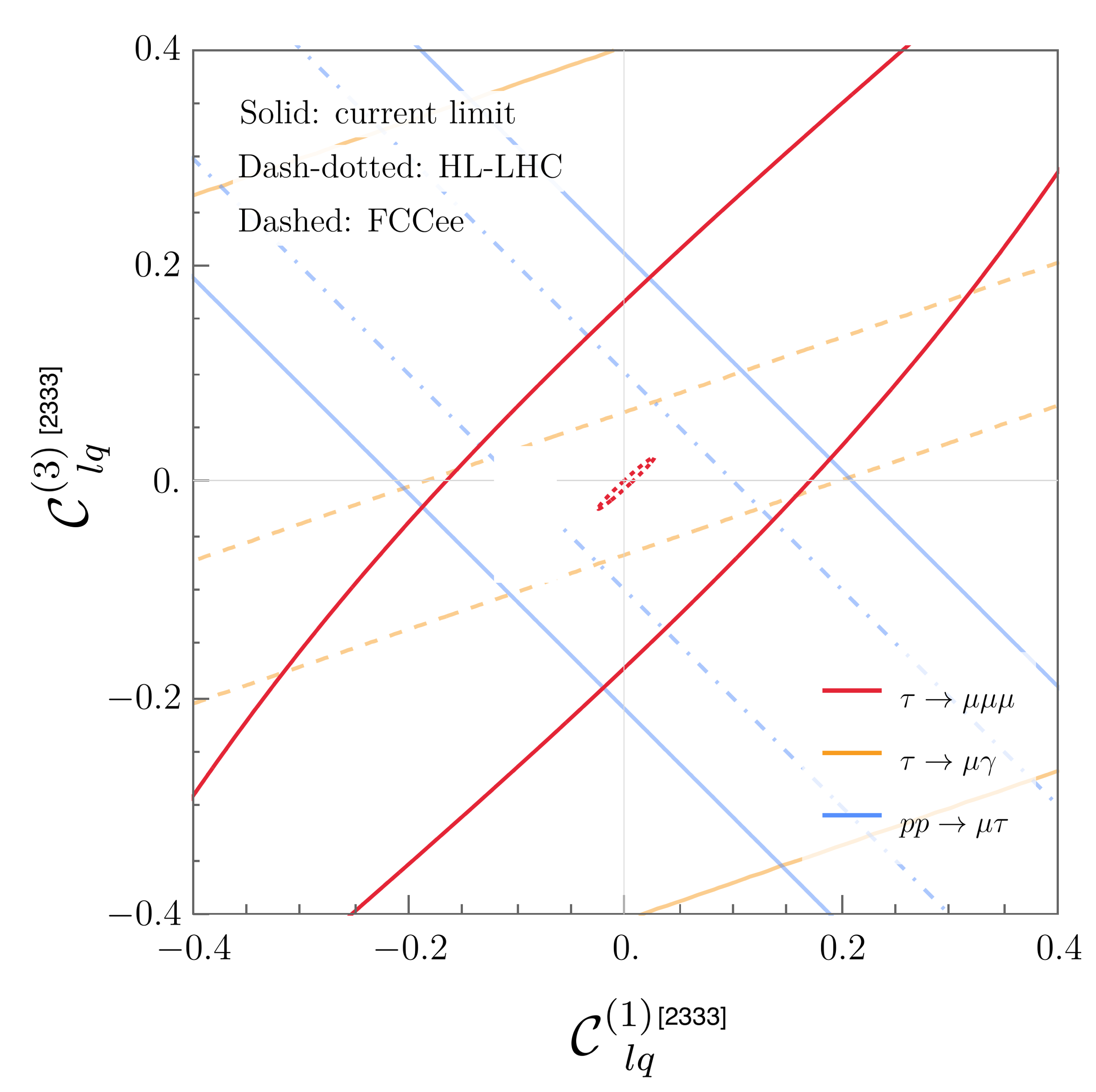}
\caption{Left: Direct measurements of $\mathcal{B}(\PGt \to \ell \PGn \PAGn)$ branching fraction and \PGt lifetime (ellipse) 
compared to the SM prediction (band); 
the width of the SM band is determined by the \PGt mass uncertainty. From Ref.~\cite{Lusiani:2024tau}.
Right: FCC-ee impact in constraining, at 95\%~CL, 
the coefficients of representative dimension-six LFV operators normalised to the 1\,TeV scale.
Adapted from Ref.~\cite{Plakias:2023esq}, scaled to the baseline FCC-ee luminosities.
\label{fig:tauLFU}}
\end{figure}

The major improvement expected at FCC-ee in the precision of measurements of 
the leptonic \PGt decay branching fraction, the \PGt mass, $m_\PGt$, and the \PGt lifetime, $\tau_{\PGt}$, 
corresponds to a jump of more than one order of magnitude (from $10^{-3}$ to $10^{-4}$) 
in tests of universality between third-generation and light leptons~\cite{Lusiani:2018zvr, Dam:2018rfz}.
An illustration of the FCC-ee reach on these measurements is shown in the left panel of Fig.~\ref{fig:tauLFU}~\cite{Lusiani:2024tau}. 
With this level of precision, 
models addressing the hint of non-universality observed in $\PQb \to \PQc \PGt \PGn$ decays~\cite{Allwicher:2021ndi} 
could either be unambiguously confirmed, leading to a major discovery, or ruled out. 
Theoretical work (on electroweak and radiative corrections), achievable with present knowledge but not yet available, 
is needed to reduce the theoretical systematic uncertainties on these measurements 
well below the projected experimental systematic uncertainties. 

\subsubsection{Lepton flavour violating \texorpdfstring{\PGt}{tau} decays} 

Searches for LFV \PGt decays are null tests of the SM and, thus, are pristine targets for indirect NP searches. 
A variety of explicit models, predicting rates not far from  current exclusion bounds, 
exist in the literature~\cite{Cornella:2021sby, Ardu:2021koz, Plakias:2023esq}, 
and FCC-ee will offer the ultimate experimental sensitivity on virtually 
all \PGt LFV decay modes~\cite{Li:2018cod, Dam:2018rfz, Banerjee:2022xuw}.
For example, the 90\%~CL projected limit on $\mathcal{B}(\PGt \to 3 \PGm)$ 
would be $2 \times 10^{-11}$ in absence of signal~\cite{Lusiani:2024tau}, 
which is three orders of magnitude below the current limit ($2.1 \times 10^{-8}$~\cite{ParticleDataGroup:2024cfk}) 
and more than an order of magnitude better than the final Belle~II projection (with 50\,ab$^{-1}$). 
The impact of such bounds is illustrated in the right panel of Fig.~\ref{fig:tauLFU}.
    
\subsubsection{Rare \texorpdfstring{\PQb}{b}-hadron decays with \texorpdfstring{$\PGtp\PGtm$}{taubar tau} pairs in the final state} 

In all these processes, there is an unexplored gap of about three orders of magnitude 
between SM predictions~\cite{Bobeth:2013uxa, Bouchard:2013mia, Kamenik:2017ghi} 
and experimental data~\cite{BaBar:2016wgb,LHCb:2017myy} 
--- a gap that represents a wide unexplored parameter range 
of motivated new-physics models~\cite{Buttazzo:2017ixm,Capdevila:2017iqn,Bauer:2021mvw}. 
At the SM predicted rates, FCC-ee is expected to be able to measure these decays, 
in particular $\PBz \to \PK^{*0} \PGtp \PGtm$~\cite{Li:2020bvr, miralles_2024_9t6w9-mmd12}.

\subsubsection{Charged-current \texorpdfstring{\PQb}{b}-hadron decays with a \texorpdfstring{$\tau\nu$}{tau nu} pair in the final state} 

The persisting anomalies in exclusive $\PQb \to \PQc\PGt\PGn$ decays~\cite{HFLAV:2022esi} 
give a strong phenomenological motivation for deeper studies 
of these processes~\cite{Fajfer:2012jt,Alonso:2015sja,Bernlochner:2017jka, Ho:2022ipo}.
A unique opportunity offered by the FCC-ee programme would be the experimental access 
to the clean inclusive~\cite{Ligeti:2014kia,Ligeti:2021six} 
and leptonic~\cite{Amhis:2021cfy, Zheng:2020ult} modes, 
as well as the suppressed $\PQb \to \PQu$ transitions~\cite{Aebischer:2022oqe,Fuentes-Martin:2019mun}.  
For example, the measurement of the \mbox{$\PB \to \PGt \PGn$} branching ratio at FCC-ee~\cite{Zuo:2023dzn} 
is expected to yield ultimate precision on $|V_{\PQu\PQb}|$, 
with an estimated relative uncertainty of only 1\%, 
which is a factor of three below current best estimates~\cite{HFLAV:2022esi} 
and more than a factor of two below the projected precision from the $\PB \to \PGt \PGn$ measurement at Belle~II~\cite{Belle-II:2018jsg}.

\subsubsection{Rare \texorpdfstring{\PQb}{b}- and \texorpdfstring{\PQc}{c}-hadron decays to di-neutrino final states} 

\begin{figure}[ht]
\centering 
\includegraphics[width=0.5\textwidth]{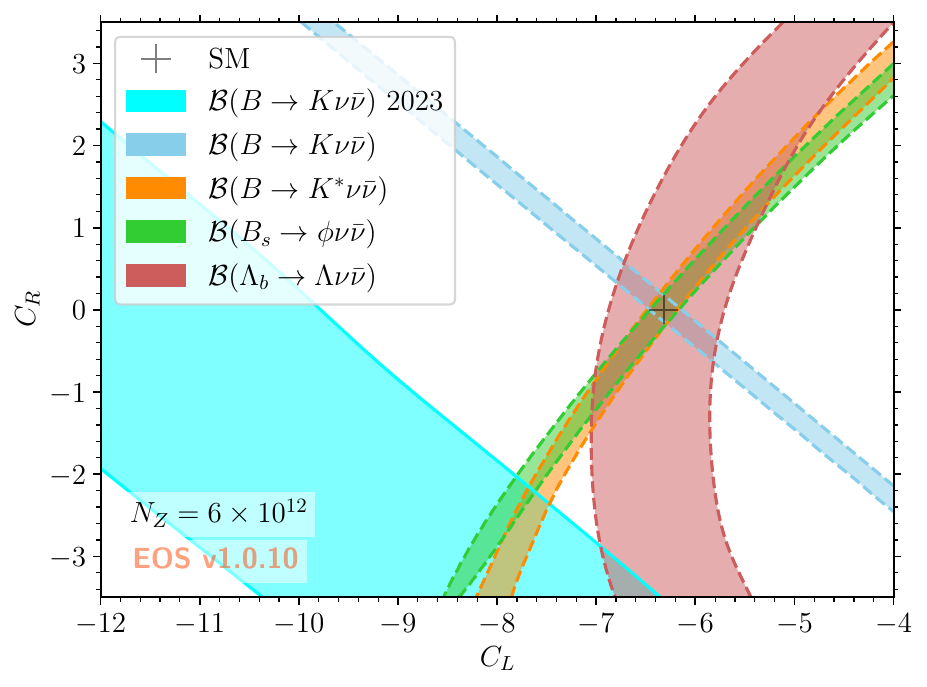}
\includegraphics[width=0.4\textwidth]{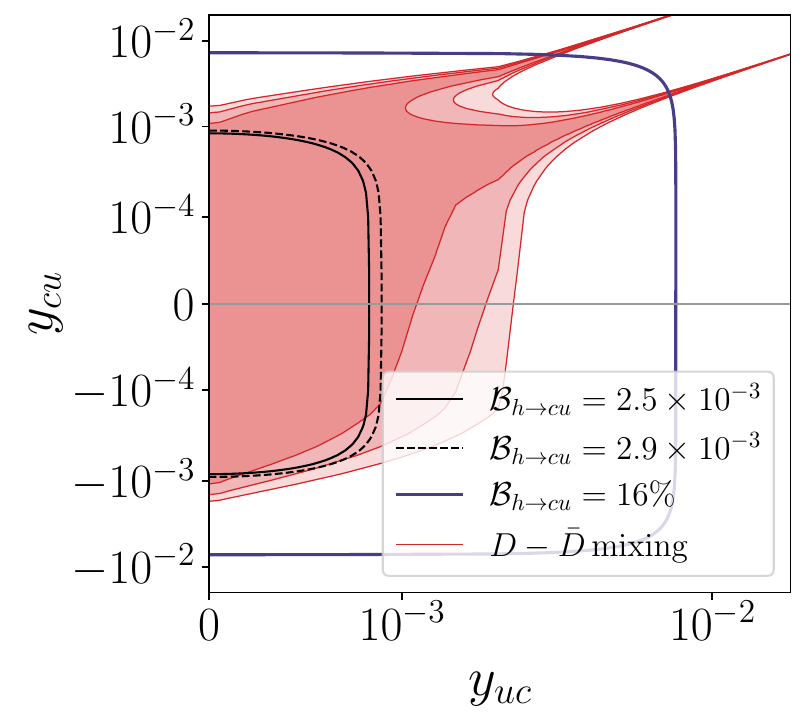}
\caption{Left: Comparison between the current constraint 
on the relevant effective $\PQb \to \PQs \PGn\PAGn$ transition operator coefficients ($C_{L,R}$) 
due to existing measurements~\cite{Belle-II:2023esi} (cyan band) 
and the (systematic-error dominated) sensitivities predicted at FCC-ee (blue, orange, green, and red bands)~\cite{Amhis:2023mpj}. 
The projected regions denote the 68\% probability of the marginal posterior density, assuming that all observables are SM-like. 
The Wilson coefficients are normalised to an effective scale of 6.5\,TeV, 
up to the CKM factors $V_{\PQt\PQs} \times V_{\PQt\PQb}$~\cite{Amhis:2023mpj}.
Right: Current (from \PD-\PAD mixing) and projected (from FCC-ee searches for $\mathrm{h} \to \PQc \PQu$) 
limits on effective charm-up flavour-changing Yukawa couplings of the Higgs boson~\cite{Kamenik:2023hvi}. 
The 68\%, 95\%, and 99.7\%~CL regions allowed by \PD-\PAD mixing data are depicted from darker to lighter red.}
\label{fig:bsnunu}
\end{figure}

A recent sensitivity study~\cite{Amhis:2023mpj} has shown that the branching ratios of the rare 
$\PBz \to \PK^{*0} \PGn\PAGn$, 
$\PBz \to \PKzS \PGn\PAGn$, 
$\PBs \to \PKzS \PGn\PAGn$, and 
$\PGLzb \to \PGL \PGn\PAGn$ decays 
could be determined at FCC-ee with precisions of 0.53\%, 1.20\%, 3.37\%, and 9.86\%, respectively, 
relative to the current SM estimates~\cite{Bause:2021cna}. 
In particular, the precision on the \PB decay modes would exceed that expected from Belle~II by an order of magnitude, 
while the \PBs and \PGLzb modes, accessible at FCC-ee, 
would uniquely constrain possible new physics in $\PQb \to \PQs \PGn\PAGn$ transitions, 
as illustrated in the left panel of Fig.~\ref{fig:bsnunu}.

Furthermore, the large number of expected events could uniquely allow more detailed differential studies of decay kinematics 
beyond simple branching ratios. 
Such measurements can also efficiently probe possible BSM effects 
in terms of the production of light invisible particles in the final state, 
mimicking the missing energy signature of the neutrino pair in the SM~\cite{Bolton:2024egx}.
 
In the charm sector, processes mediated by the $\PQc \to \PQu \PGn\PAGn$ transition are unique probes of up-quark FCNCs 
because they do not receive long-distance QCD contributions plaguing related non-leptonic or rare semileptonic decays~\cite{Artuso:2008vf}. 
In addition, SM gauge invariance and approximate $U(2)$ flavour symmetry 
relate these quark-flavour-changing processes to rare (semi)leptonic kaon decays~\cite{Bause:2020auq, Bause:2020xzj, Fajfer:2023nmz}, 
thus making them crucial in (over)constraining BSM effects in FCNCs involving the first two quark generations.  
Finally, the light dark sector coupling to $\PQc \to \PQu$ quark currents~\cite{Kamenik:2011vy, Li:2023sjf} 
can be probed with the missing energy signature, 
similar to the $\PQb \to \PQs$ transition in the \PQb-quark sector. 
Currently, the only experiment probing these processes is BES~III, which set the first upper limit, 
$\mathcal B(\PDz \to \PGpz \PGn \PAGn) < 2.1 \times 10^{-4}$ at 90\%~CL~\cite{BESIII:2021slf}. 
Based on na{\"\i}ve luminosity scaling, 
this bound could be improved by more than two orders of magnitude at FCC-ee 
and would thus make it relevant to probe flavour $U(2)$ invariant BSM scenarios~\cite{Fajfer:2023nmz}.  

\subsubsection{Rare decays and CPV studies with neutrals}

The discovery of CP violation in singly Cabibbo-suppressed \PD decays by LHCb~\cite{LHCb:2019hro} 
opened a new and largely unexplored venue for CPV studies. 
One of the outstanding puzzles is to understand whether the observed size of CPV 
is due to SM effects enhanced by long distance QCD dynamics, 
or a possible signal of NP~\cite{Chala:2019fdb, Dery:2019ysp}. 
Recent phenomenological studies using isospin~\cite{Grossman:2012eb, Gavrilova:2023fzy} 
and $U$-spin~\cite{Grossman:2019xcj, Bause:2022jes} expansions 
have shown that the measurements of decay rates and CP asymmetries in related \PD and \PDs decays 
to final states involving \PGpz and \PKzS, 
which are not possible at LHCb but would be accessible at FCC-ee, could yield a definite answer.
Important representative studies also include CP violation in 
radiative singly Cabibbo suppressed modes~\cite{Isidori:2012yx, Gisbert:2020vjx, Adolph:2022ujd}
and the doubly-radiative decay $\PDz \to \PGg \PGg$ 
required for the SM prediction of $\PDz \to \PGmp \PGmm$~\cite{Burdman:2001tf}.  
    
In the \PB sector, rates and CP asymmetries of rare radiative decays, mediated by the $\PQb \to \PQs \PGg$ transition, 
such as $\PB \to \PK^* \PGg$ and $\PB \to \mathrm{X_s} \PGg$, 
are most-sensitive probes of quark flavour violating dipole transitions~\cite{Fajfer:2023gie} 
in both minimally violating and $U(2)$ flavour models;
they are also crucial probes of flavour dynamics in composite Higgs models~\cite{Glioti:2024hye}. 
Improvements in the understanding of these decays~\cite{Benzke:2010tq, Paul:2016urs} 
are needed to fully leverage the prospective sizes of the FCC-ee event samples~\cite{BKgamma}. 
Recent steps in this direction are documented in Refs.~\cite{Gunawardana:2019gep,Misiak:2020vlo}.
Finally, time-dependent studies of rare decays, 
such as $\PBs \to \PGmp \PGmm$~\cite{Buras:2013uqa, Fleischer:2017yox}, 
$\PB \to \PKS \ell^+ \ell^-$~\cite{Descotes-Genon:2020tnz, Fleischer:2022klb}, $B_s \to \phi \mu^+\mu^-$~\cite{Kwok2024},
and $\PB \to \PKS \PGn\PAGn$~\cite{Descotes-Genon:2022gcp}, 
offer novel and complementary probes of CP violation beyond the SM, 
some only accessible at FCC-ee.     

\subsubsection{On-shell \texorpdfstring{\PZ}{Z}, \texorpdfstring{\PW}{W}, and Higgs flavour-changing decays}

Flavour-changing decays offer unique novel probes of SM and BSM flavour dynamics 
and can be searched for in inclusive final states 
(profiting from the unparalleled jet-flavour tagging capabilities of the detectors) 
or in exclusive final states~\cite{dEnterria:2023wjq}. 
In particular, $\PW \to \PQb \PQc$ decays could yield the ultimate precision on $|V_{\PQc\PQb}|$, 
with a relative uncertainty well below 1\%~\cite{WWVcb,WWVcb2}. 
Similarly, rare four-body \PZ decays could uniquely determine individual flavour neutrino couplings~\cite{Durieux:2015hsi}.
Furthermore, lepton-flavour-violating Higgs and \PZ decays involving \PGt leptons in the final state 
would improve existing LEP and LHC constraints by orders of magnitude~\cite{Abada:2014cca, Qin:2017aju, Dam:2021ibi}. 
Finally, FCC-ee could also directly probe FCNC decays of \PZ and Higgs bosons~\cite{Kamenik:2023hvi}, 
possibly even reaching SM expectations for the former (in the case of $\PZ \to \PQb\PQs$) 
and surpassing indirect constraints from neutral meson oscillation measurements for the latter, 
as shown in the right panel of Fig.~\ref{fig:bsnunu}. 
Such direct measurements are also particularly important to lift degeneracies in BSM fits 
and disentangle possible UV sources of low energy FCNC effects~\cite{Crivellin:2022rhw}. 

\subsubsection{Combined studies and synergy between the flavour and EW programmes}
\label{sect:flavour-synergy}

Besides the interest of the specific observables illustrated above, 
probably the most interesting aspect of the FCC flavour programme lies 
in the possibility of combining several precision flavour-violating and flavour-conserving measurements. 
This combination offers a unique opportunity to characterise a possible NP signal, 
when it emerges, as illustrated in the next two paragraphs. 

\subsubsubsection{New-physics in \texorpdfstring{\PB-\PAB}{B-Bbar} mixing}

\begin{figure}[ht]
\centering
\includegraphics[width=0.32\textwidth]{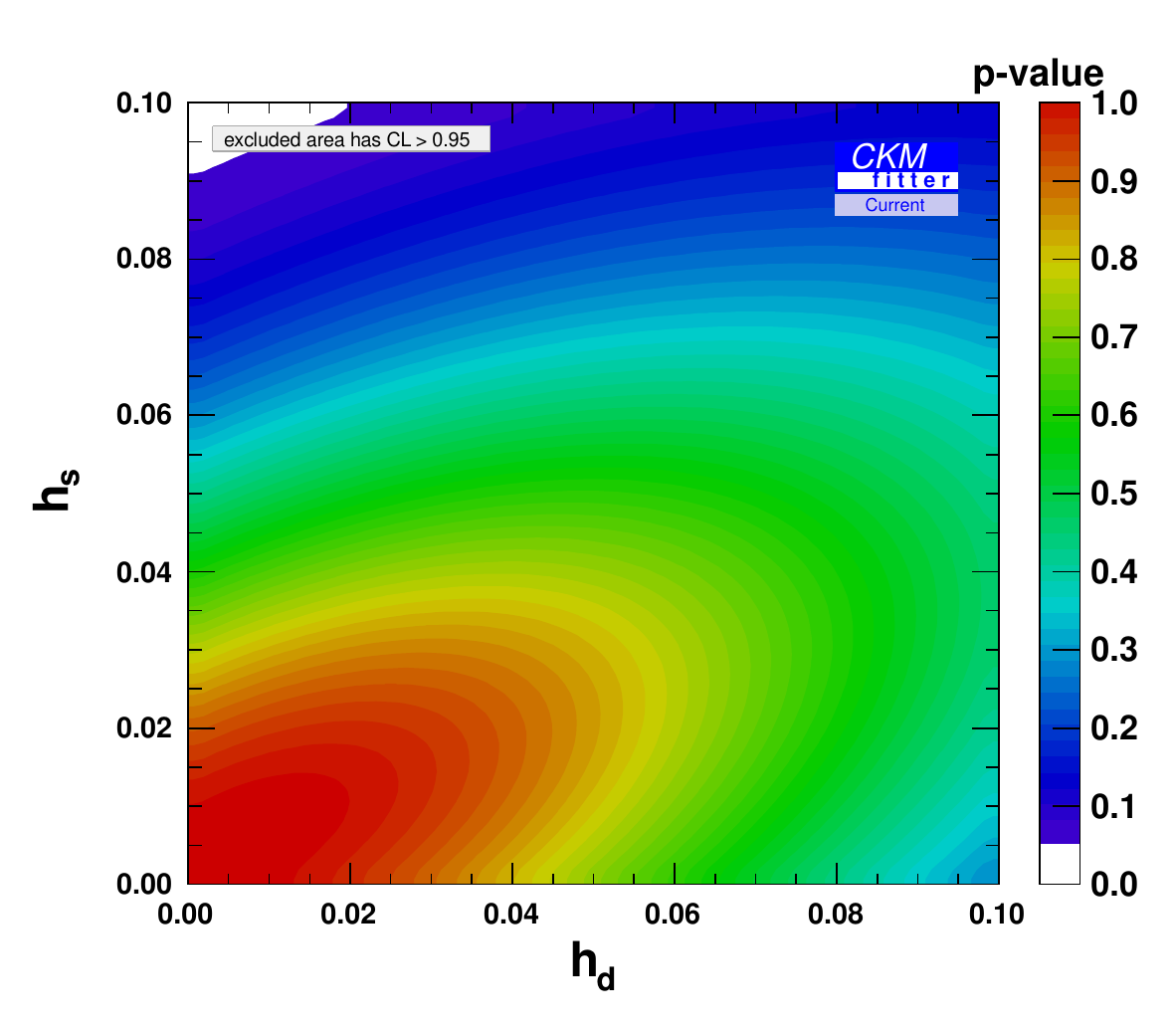}
\includegraphics[width=0.32\textwidth]{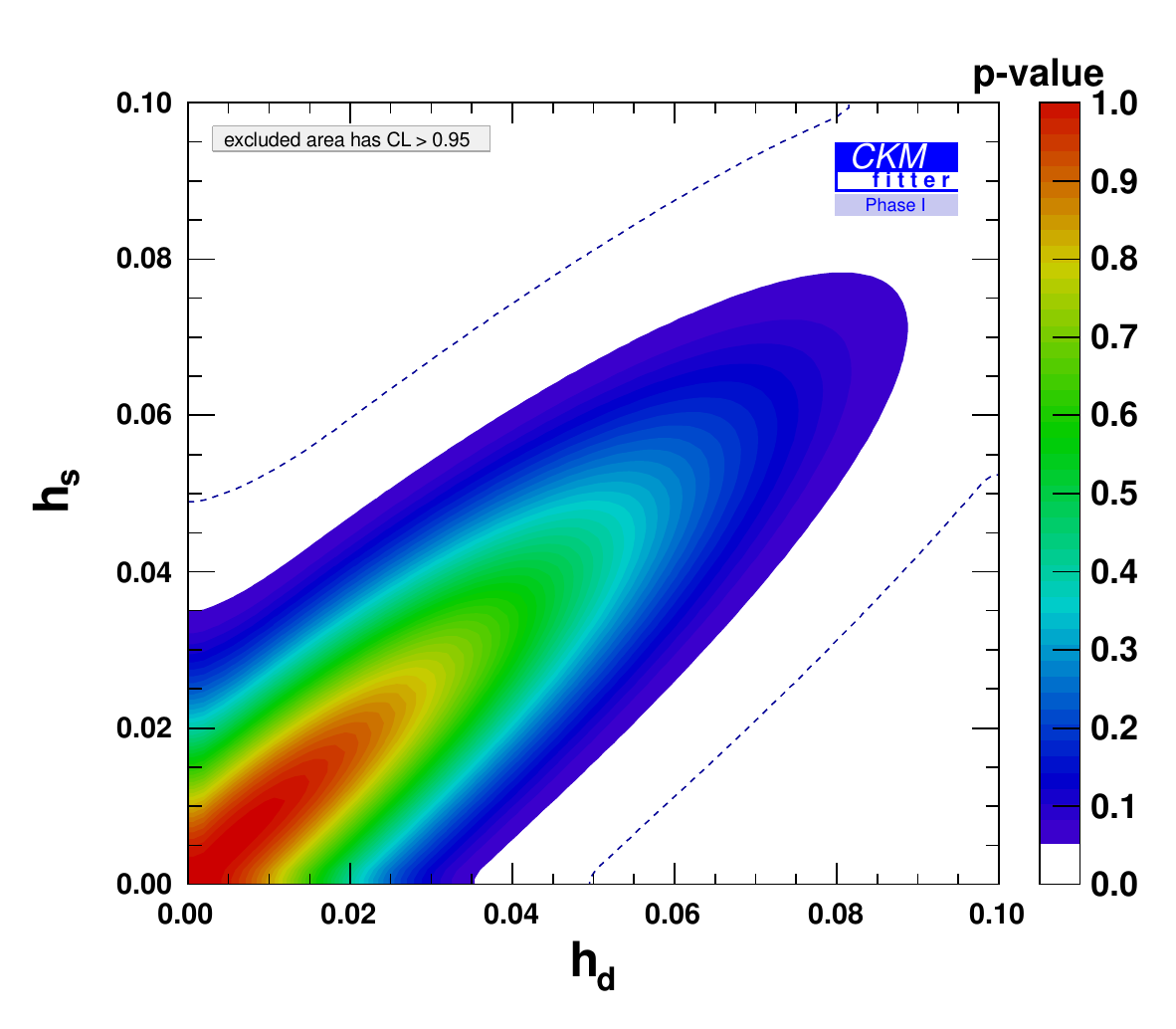}
\includegraphics[width=0.32\textwidth]{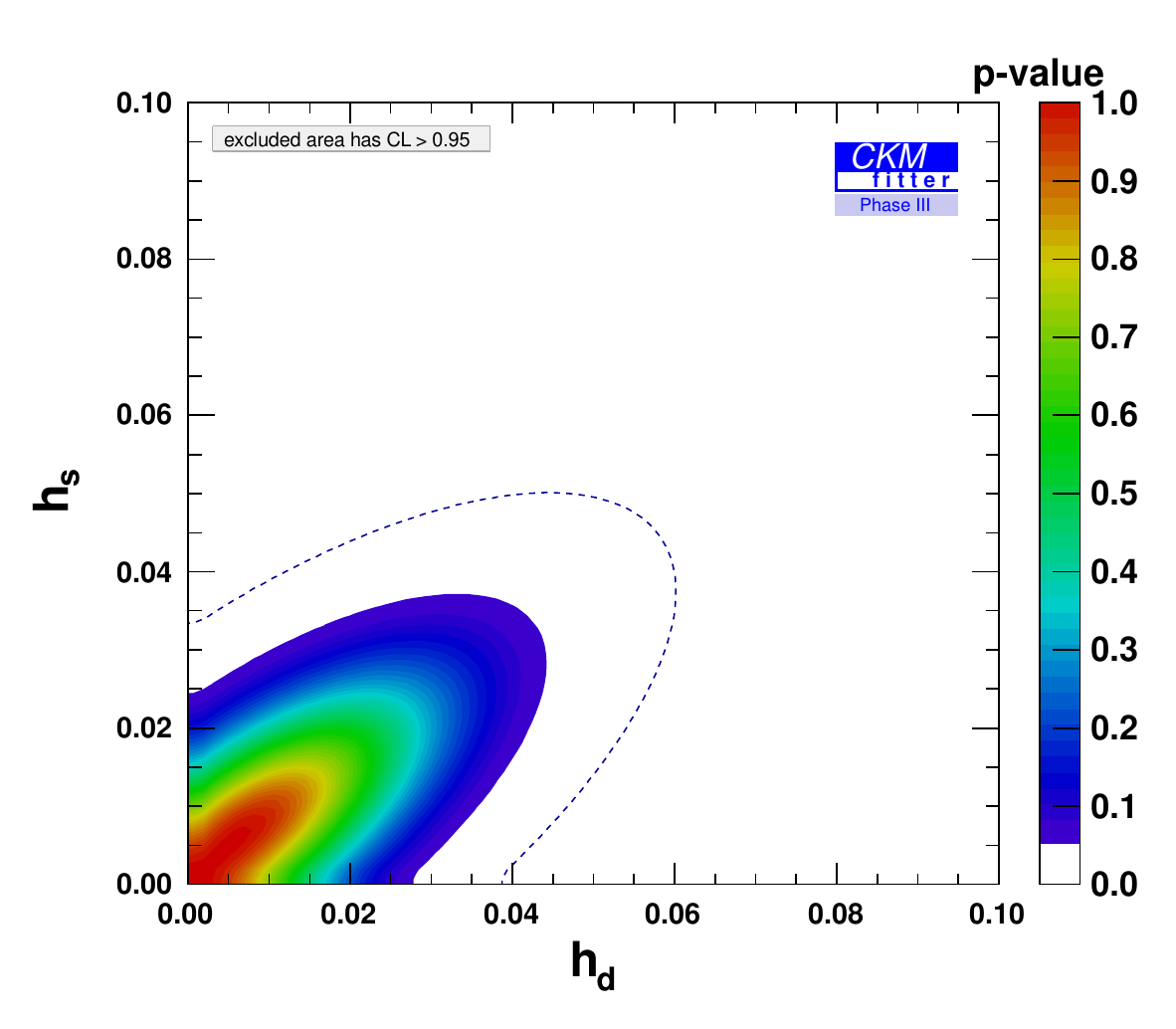}
\caption{Sensitivities to relative BSM amplitudes ($h_{\PQd}$ and $h_{\PQs}$) in \PBd and \PBs mixings:
current values (left), 
LHCb (50\,fb$^{-1}$) and Belle~II (50\,fb$^{-1}$) projections (centre),
and FCC-ee projections (right).
From Ref.~\cite{Charles:2020dfl}. 
A limiting factor of the FCC-ee sensitivity is the uncertainty on $V_{\PQc\PQb}$, 
expected to be reduced with the $\PW\PW$ threshold run and from improved flavour-tagging algorithms.
\label{fig:dF2}}
\end{figure}

Measurements of $\phi_{\PQs}$ from $\PBs \to \PJGy\, \PGf$ and $\PBs \to \PGf \PGf$ at FCC-ee 
could challenge present theory uncertainties~\cite{Aleksan:2022mtk, Gerlach:2022hoj}. 
Similarly, the full exploitation of experimental precision on $\Delta m_{\PQs}$ 
will hinge upon the determination of $|V_{\PQc\PQb}|$ at or below the sub-per-cent level~\cite{Lenz:2019lvd, Charles:2020dfl} 
from either (inclusive) semileptonic \PB decays, 
leptonic \PBc decays~\cite{Amhis:2021cfy, Zheng:2020ult}, 
or on-shell $\PWp \to \PAQb \PQc$ decays~\cite{WWVcb, WWVcb2}. 
As an illustration of the increased sensitivity achieved by combining all these measurements, 
the NP reach in the $\PB_{\PQs,\PQd}$ mixing amplitudes is shown in Fig.~\ref{fig:dF2}.
Mainly due to the increased precision on $|V_{\PQc\PQb}|$, 
the effective NP scale sensitivity would increase by a factor of three compared to current measurements 
and by a factor of 1.5 compared to the ultimate precision expected at Belle~II.   
It is worth stressing that FCC-ee will also deliver the ultimate determination of the CKM-matrix $\gamma$ angle 
from $\PB \to \PD \PK$ decays~\cite{Brod:2013sga}, 
allowing a per-mil test of CKM unitarity in the $\PQb \to \PQs$ triangle~\cite{Aleksan:2021fbx, Aleksan:2021gii, Aleksan:2024jwm} 
(one order of magnitude better than the current sensitivity), 
as well as the possibility to test the SM predictions for mixing-induced semileptonic CP asymmetries 
in both $\PBd$ and $\PBs$ decays~\cite{Artuso:2015swg,Jubb:2016mvq, Lenz:2020efu}.

\subsubsubsection{Synergy with the EW programme}

\begin{figure}[ht]
\centering 
\includegraphics[width=0.42\textwidth]{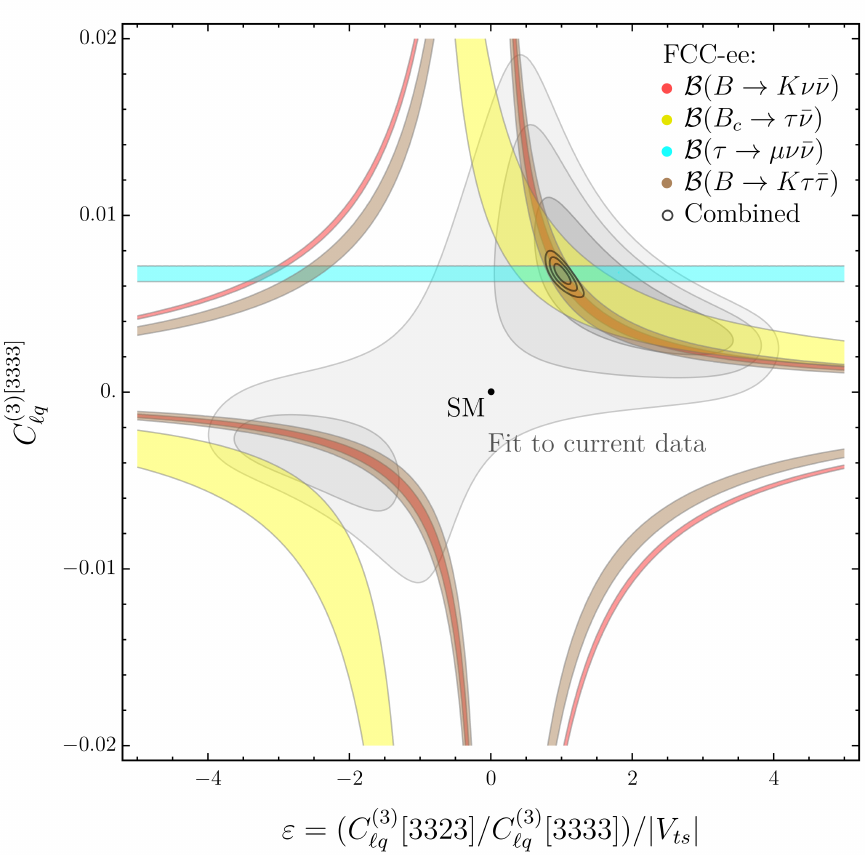} 
\includegraphics[width=0.42\textwidth]{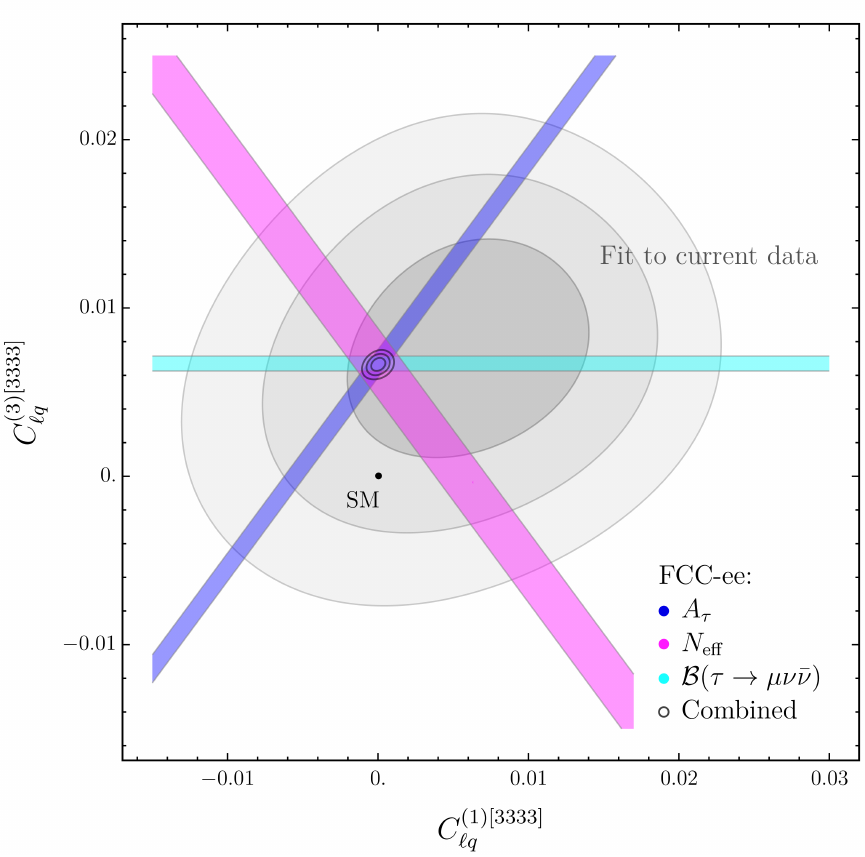}
\caption{Constraints on semileptonic effective operators describing new physics coupled mainly to the third generation 
with minimal $U(2)$ breaking~\cite{Allwicher:2023shc}.
The grey regions show the result of the current fit (68\%, 95\%, and 98\% CL), 
including flavour, electroweak, and Drell--Yan data 
(the slight tension with the SM is driven by $\PQb \to \PQc \PGt \PAGn$ data).
The small ellipses are the result of an hypothetical fit 
with FCC-ee projected uncertainties on flavour and electroweak observables, 
assuming a signal compatible with current data.
The coloured bands are the individual constraints of the observables in this scenario.
The FCC constraints follow from the projections of a precision of 3\%, 1.6\%, 0.01\%, and 20\% 
on the $\mathcal{B} (\PB \to \PK \PGn\PAGn)$, 
$\mathcal{B}(\PBc \to \PGt \PAGn)$, 
$\mathcal{B}(\PGt \to \PGm \PGn\PAGn)$, 
and $\mathcal{B}(\PB \to \PK \PGtp\PGtm)$ values, respectively. 
From Ref.~\cite{Allwicher:2025new}.}
\label{fig:FlavEWFCC}
\end{figure}

A particularly interesting synergy emerges when combining flavour-violating and EW observables. 
The discovery potential offered by this combination in a variety of NP models 
has only recently begun to be explored~\cite{Garosi:2023yxg, Allwicher:2023shc, Grunwald:2023nli}.
An illustration of this potential is presented in Fig.~\ref{fig:FlavEWFCC} 
in the context of models describing generic NP coupled mainly to the third generation, 
as defined in Ref.~\cite{Allwicher:2023shc}. 

As can be seen in this case, different observables probe the same parameter space in different directions, 
and their combination is essential to fully characterise the hypothetical NP framework, 
with, in particular, the synergy between rare \PB decays, rare \PGt decays, 
and the flavour-conserving effective couplings of the \PZ boson. 

\subsection{FCC-hh specificities compared to high-energy lepton colliders}
\label{sec:PhysicsCaseHHvsHEL}

With operation expected to start in the mid-2040's, 
the guaranteed deliverables and primary targets of FCC-ee are 
the precise and model-independent measurement of a broad range of key couplings of the Higgs boson, 
a fifty- to thousand-fold improvement in the precision of EW parameters, 
and a precise determination of the top quark mass and EW couplings.  
A 3\,TeV muon collider has also been proposed, for a timescale similar to that of FCC-ee~\cite{Black:2022cth}.
The best-measured Higgs couplings, $\kappa_{\PZ}$ and $\kappa_{\PW}$, 
will be measured at FCC-ee with expected uncertainties of 0.11\% and 0.29\%, respectively (Table~\ref{tab:HiggsKappa3}),  
much better than the 0.9\% and 0.4\% uncertainties foreseen at a 3\,TeV muon collider~\cite{MuonCollider:2022xlm}.  
This observation, combined with the \mbox{Tera-\PZ} programme and the $10^6$ top pairs at FCC-ee, 
as compared to orders of magnitude fewer at a 3\,TeV muon collider, 
renders FCC-ee the obvious priority for Higgs, electroweak, top, and flavour physics to follow HL-LHC.  
This section considers the question of what should follow FCC-ee as a high-energy exploration facility, 
comparing and contrasting frontier ($\gtrsim 10$'s\,TeV) lepton collider options with FCC-hh.

\subsubsection{Generalities}

Exploration in particle physics cannot be quantified by the volume of parameter space explored, nor by the scale of the energy reached.  
Indeed, LHC has shown that the microscopic world that can be explored by hadron colliders 
is so multidimensional and rich that no metric can fairly convey the depth 
with which the laws of nature are explored by colliding protons.  
Yet, looking to future high-energy facilities, this is precisely what must be attempted.

The core attribute of a proton collider is that it does not simply collide a single particle and an antiparticle but, instead, 
collides anything found in the proton; 
any of the lightest five quarks and antiquarks, the gluons, the photon, and the EW gauge bosons. 
Thus, a proton collider has many particle colliders in one, operating at the highest plausibly achievable energies, 
with an array of production channels for discovery and exploration that is factorially greater 
than what can be achieved by colliding single fundamental particles 
--- besides the additional possibility of colliding heavy ion in the same machine.  
The price to be paid for this great leap in exploratory power is the inherent messiness of the final state from colliding composite objects.  
However, this challenge has been amply met at LHC and there is no reason to expect something else at a future proton collider.

Were some hints for new physics at high energies to appear in the Run~3 of LHC or at HL-LHC, 
it is highly unlikely that identification of the microscopic production channel from which it emerged would be possible at HL-LHC. 
Depending on the new physics at stake, 
the FCC-ee precision data may exhibit interesting patterns of deviations that 
would refine the interpretation of these hints and give precious information about their quantum origin. 
At high energy, however, the only way to guarantee further exploration and characterisation of whatever may arise at LHC 
would be to collide protons once again, with greater energy and larger collected event samples, 
possibly activating additional production channels and dynamical regimes: 
this is the only way to cover all possible production channels.

Also, it is worth noting that the discovery of the Higgs boson opened windows beyond the gauge paradigm 
to a new class of fundamental forces and to the flavour puzzle.  
Modern particle physics exploration should be equipped to navigate this new world.  
The case, for example, of a new neutral scalar coupled to fermions with a strength proportional to their mass is enlightening. 
Clearly, the third generation quarks provide the most immediate access to such a sector, 
whether through production by $\bbbar$ annihilation or through gluon fusion via a top quark loop.  
This is, after all, how the Higgs boson was discovered.  
Such scenarios should not be blind spots for any future facility and, indeed, 
a future proton collider would provide excellent discovery opportunities for states that do not carry electroweak charges.
Some more concrete, yet still general, future discovery possibilities are considered below.

\subsubsection{Resonance searches}

A key exploratory task for any high-energy exploration facility is to search thoroughly for new high-mass resonances. 
There are two complementary approaches to probe the existence of high-mass states: 
the direct search for a mass peak 
and the indirect evidence emerging from deviations in precise measurements of processes accurately predicted within the SM. 
The indirect approach extends the sensitivity of a collider well beyond its kinematic reach. 
As clearly shown by LEP2, 
the superior precision of a lepton collider can therefore 
match the sensitivity of direct searches at hadron colliders of much higher centre-of-mass energies. 
The limitation of the indirect approach, however, 
is that the interpretation of such deviations is very much dependent on model assumptions. 
For example, the new state masses can vary significantly as a function of the assumptions on their couplings. 
Outside a well-defined theoretical BSM framework, for which there is no compelling prejudice today, 
this ambiguity significantly limits the ability to point to new measurements that would clarify the source of new physics. 
Direct search, when possible, therefore provides a preferable (yet synergistic) approach, 
when the indirect and direct sensitivity of lepton and hadron colliders are numerically comparable. 

For this reason, the focus here is on the comparison of FCC-hh and high-energy lepton colliders to the case of direct detection. 
Present theory priors do not favour any particular possibility, 
thus the breadth of couplings that can be explored is a paramount consideration in planning for the future.
For this purpose, a proton collider has a further unique virtue, sampling a wide array of initial states, 
from gluons to photons or EW gauge bosons and quarks of a variety of flavours, 
and from neutral initial states to charged or even coloured, 
any of which could lead to the formation of a new resonance.
For example, 15 $\qqbar^{(')}$ initial states can create exotic bosonic resonances, 
both in flavour-conserving and flavour-violating channels. 
Besides, 20 $\mathrm{V}+\PQq$ initial states could create excited quark states (with $\mathrm{V} = \PGg$, \Pg), 
or new heavy quarks through charged ($\mathrm{V} = \PW$) or neutral ($\mathrm{V} = \PZ$) current transitions. 
Furthermore,
8~$\mathrm{V}\mathrm{V}^\prime$ initial states can give rise to EW resonances 
or axion-like particles ($\mathrm{V},\mathrm{V}^\prime = \PGg, \PW, \PZ$), or to $\Pg\Pg$ resonances.

Another key component of exploration combines breadth with energy.  
As a figure of merit, proton colliders can be compared with lepton colliders to assess their relative reach in energy for resonances.  
Following Ref.~\cite{AlAli:2021let}, an optimistic scenario is considered, 
where a new high-energy resonance (serendipitously) resides at the kinematic limit of a lepton collider 
and can be produced by that initial 
state\,\footnote{For lower masses, radiative-return production would be required to enable discovery.}. 
The energy of proton collisions that would be necessary to produce the same number of resonant events, 
given the same integrated luminosity, is a natural figure of merit here. 

\begin{figure}[ht]
\centering
\includegraphics[width=0.55\textwidth]{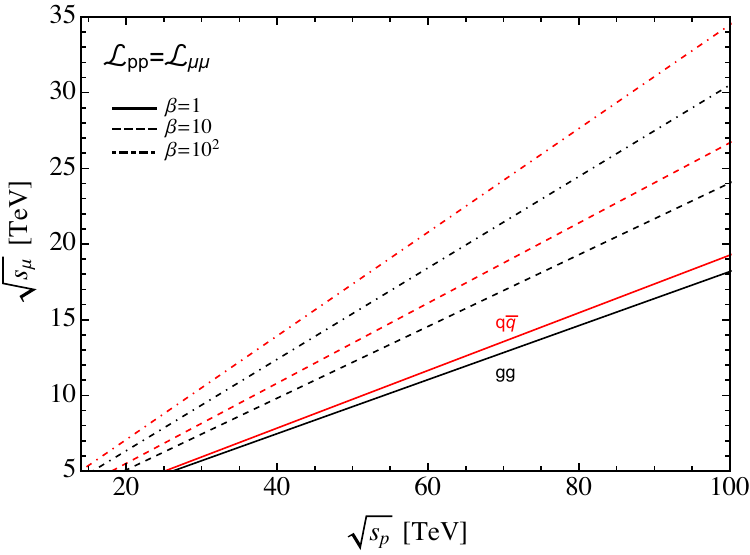} 
\caption{Energies of equivalent event numbers, 
assuming equal integrated luminosities, 
between a high-energy muon collider and a proton collider;
$\beta=1$ corresponds to the same partonic cross section between muons and proton partons while
larger values of $\beta$ correspond to enhanced partonic cross sections for quarks and gluons, 
as would be the case for resonances coupled primarily through QCD.  
Note that $\beta<1$ would require muons to be more strongly coupled to new resonances than quarks, 
requiring muon flavour to play a role.}
\label{fig:resonancecomp}
\end{figure}

The resulting equivalent CM energy is shown in Fig.~\ref{fig:resonancecomp}, 
under the assumption that the parton-level production cross sections are a factor $\beta = 1$, 10, 100 larger than 
the lepton-collider production cross section on resonance for $\qqbar$ and $\Pg\Pg$ initial states.  
It is noteworthy that, 
even under the significant assumption that the production cross section via leptons matches that of quarks or gluons ($\beta = 1$),  
a 100\,TeV proton collider significantly exceeds, for such a resonance, the reach of a 10\,TeV lepton collider.  
For larger parton production cross sections relative to leptons, 
the gap in coverage between the two classes of facilities further increases.  
Such a comparison, however, also conceals the richness of the physics programme for such $>10$\,TeV resonances at a proton collider.

To go further, the direct discovery prospects for narrow resonances of mass $M_\mathrm{R} > 10$\,TeV, 
which would be inaccessible to a 10\,TeV muon collider, is investigated.  
In the narrow width approximation, at a proton collider operating at proton CM energy $\sqrt{s}$, 
the production cross section for a resonance R of spin $S$ coupled to the initial state partons `yy' 
and decaying into the final state `xx' is
\begin{equation}
\sigma = r \frac{C_{\rm yy}}{s} \, ,
\end{equation}
where, for gluons and quarks, the factor $C_{\rm yy}$ is related to the parton luminosity as
\begin{equation}
C_{\Pg\Pg} =\frac{\pi^2}{8} \int_\tau^1 \frac{dx}{x} \, f_{\Pg}(x) f_{\Pg}(\tau x) ~,
\quad
C_{\qqbar} =\frac{4\pi^2}{9} \int_\tau^1 \frac{dx}{x} \, 
\left[ f_{\PQq}(x) f_{\PAQq}(\tau x) + f_{\PAQq}(x) f_{\PQq}(\tau x) \right] \, ,
\label{eqCi}
\end{equation}
where $\tau = M_\mathrm{R}^2/s$ and the parton distribution functions $f_{\Pg,\PQq,\PAQq}$ 
are evaluated at $Q^2 = M_\mathrm{R}^2$. 
The parameter $r$ encodes all the model-dependent factors,
\begin{equation}
r = (2 S+1) B_{\rm yy} B_{\rm xx} \frac{\Gamma_\mathrm{R}}{M_\mathrm{R} } \, .
\label{eq:r}
\end{equation}

The width-to-mass ratio essentially encodes the microscopic coupling and hence parametrises the nature of the underlying theory.  
For these purposes, whenever $\Gamma_\mathrm{R} / M_\mathrm{R} \lesssim 0.1$, 
the theory is approximately perturbative and the resonance effectively narrow.  
In Eq.~(\ref{eq:r}), $B_{\rm yy}$ is the branching ratio into the partons whose collision produced the resonance 
and $B_{\rm xx}$ is the branching ratio into whichever final state is under consideration.  
The latter could more prosaically be a pair of SM particles or something much more exotic, such as pairs of long-lived particles.

\begin{figure}[ht]
\centering
\includegraphics[width=0.55\textwidth]{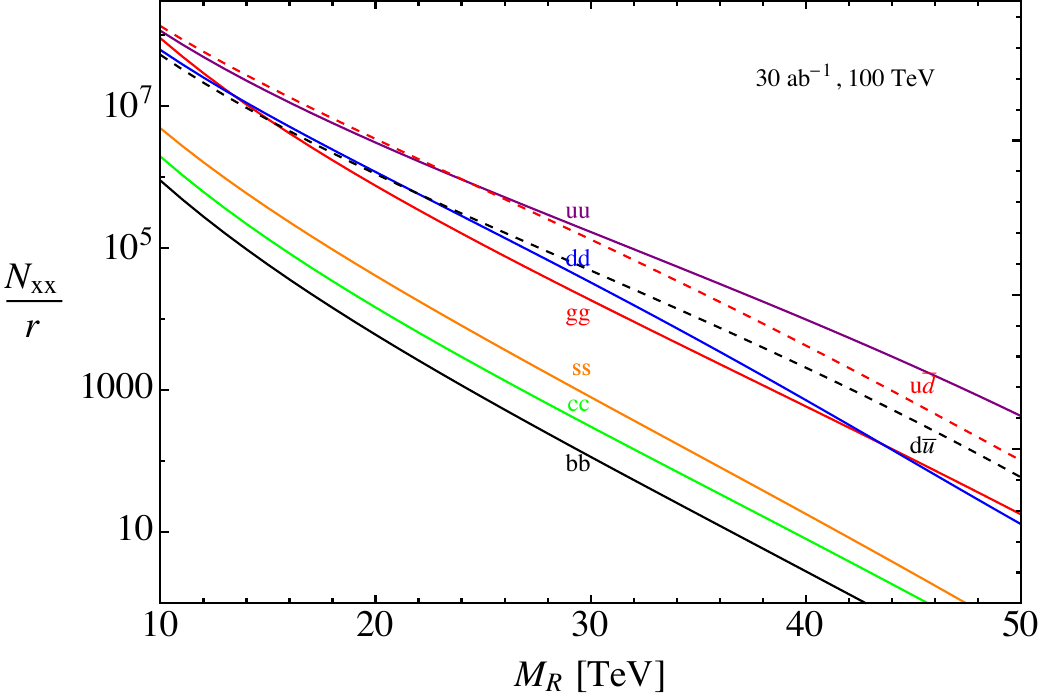} 
\caption{The number of resonance-production events at a 100\,TeV proton collider, 
relative to the model-dependent factor $r$ defined in Eq.~(\ref{eq:r}).  
A non-exhaustive variety of initial states yy is considered and the final-state decay product xx is unspecified.}
\label{fig:resonance}
\end{figure}

Figure~\ref{fig:resonance} shows the number of $\Pp\Pp \to R \to xx$ events 
for a variety of production channels at masses above 10\,TeV.  
This number of events is weighted relative to the model-dependent factor $r$.  
Simply as an illustrative example, a scalar resonance could plausibly have $r = 10^{-2}$.  
Thus, in this case, the total number of events in Fig.~\ref{fig:resonance} should be multiplied by $10^{-2}$.  
The number of events plausibly available, hence the breadth of the exploration, 
to a 100\,TeV proton collider for resonances above the 10\,TeV scale is vast.  
Smaller values of $r$ are also possible, 
with events occurring above energies of 10\,TeV even for extremely narrow resonances, down to $r \sim 10^{-8}$.  
For example, the SM Higgs boson has $r \approx 3 \times 10^{-5}$.

If the final states were rare collider objects, such as long-lived particles, 
then these many-orders of magnitude of events would be highly visible, 
offering a vast programme of not only discovery but also characterisation. 
Presumably, for such exotic decays, the discovery reach would essentially saturate kinematic and PDF limits, up to $\sim$\,50\,TeV.  
On the other hand, if the final states were not rare but visible, such as to photons or leptons, 
then such a resonance would also stand out above the background. 
This is confirmed, for example, for the Sequential Standard Model scenario reach for leptonic decays shown in Fig.~\ref{fig:schannel}, 
where the reach extends beyond 40\,TeV.  
These are the easiest discovery opportunities that would maximise the opportunities of the statistical reach of the large event samples.

Finally, even in scenarios where the final state has a significant background, 
the number of events is large enough that a discovery would still be expected.  
This point is confirmed by detailed studies for specific scenarios.  
For instance, Fig.~\ref{fig:schannel} also shows the discovery reach for a variety of $s$-channel resonances.  
Even for a dijet final state, for which there is overwhelming background, the discovery reach extends to the 40\,TeV scale.

\subsubsection{Pair production of new particles}

\begin{figure}[ht]
\centering
\includegraphics[width=0.55\textwidth]{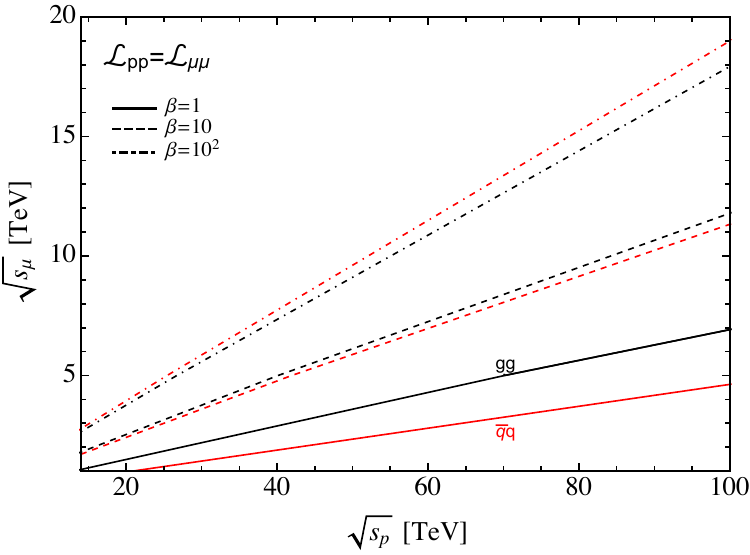} 
\caption{Same as Fig.~\ref{fig:resonancecomp}, but for pair-produced resonances.}
\label{fig:paircomp}
\end{figure}

Figure~\ref{fig:paircomp} shows the result of a similar exercise as for the resonances, 
again following Ref.~\cite{AlAli:2021let} and assuming the same integrated luminosity for each collider, 
and wondering what energy of muon collider would produce the same number of pair-production events 
as a proton collider operating at a CM energy $\sqrt{s_{\Pp}}$.  
To do this, it is assumed that the muon collider can produce states close to threshold and, 
considering partonic cross sections
\begin{equation}
\hat{\sigma}_{\rm ij\to xx}(s_\mu) = \beta \hat{\sigma}_{\mu^+ \mu^-\to \rm xx}(s_\mu)
\end{equation}
(where $\rm ij = \Pg\Pg, \qqbar$ are the initial state partons and xx are the pair-produced new states), 
it is also assumed that the energy dependence of the cross section is inversely proportional to the partonic energy-squared.  
As before, $\beta \approx 1$ qualitatively corresponds to new states that couple equally strongly to the muons and the partons.  
Such scenarios would involve the production of purely electroweak states, or states with flavour-independent couplings. 
Instead, $\beta \approx 10$--100 corresponds to states that may also carry colour.  
Going beyond this, states that only couple to gluons or couple proportionally to fermion masses would have $\beta \gg 100$.

\begin{figure}[ht]
\centering
\includegraphics[width=0.55\textwidth]{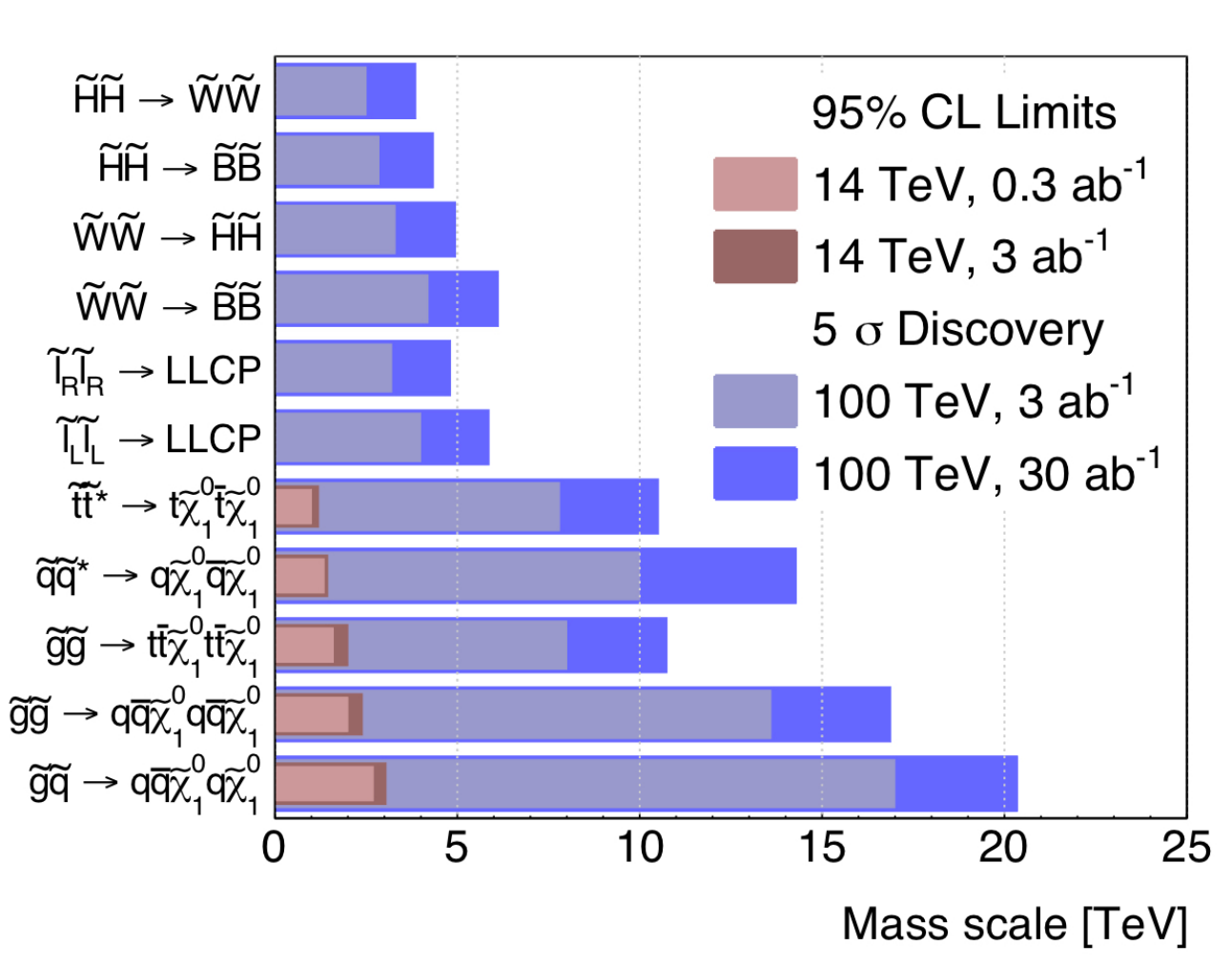}
\caption{Summary of the 5\,$\sigma$ discovery reach for different SUSY particles. 
From Ref.~\cite{Golling:2016gvc}.}
\label{fig:SUSY}
\end{figure}

It can be seen from Fig.~\ref{fig:paircomp} that only for the case of $\beta \approx 1$, 
hence essentially only purely EW states, 
does a 10\,TeV muon collider have an estimated reach extending beyond that of a proton collider.  
This is confirmed in the simple example of supersymmetry, 
where the estimated discovery reach for coloured particles shown in Fig.~\ref{fig:SUSY} 
for a 100\,TeV proton collider extends beyond 10\,TeV, 
hence beyond the kinematic limit of a 10\,TeV muon collider.  
For purely electroweak particles, the proton collider reach saturates at around the 5\,TeV scale, 
close to the kinematic limit of a 10\,TeV muon collider.

This treatment has, however, been relatively na{\"\i}ve in the sense that it is essentially rooted in the gauge paradigm.  
It can instead be assumed that, at these energy scales, the new physics couples only through dimension-6 contact interaction. 
In this case, the production cross section at the partonic level scales proportional to the partonic energy-squared.  
Such a possibility is plausible if the states that connect the two sectors are heavy.

\begin{figure}[ht]
\centering
\includegraphics[width=0.55\textwidth]{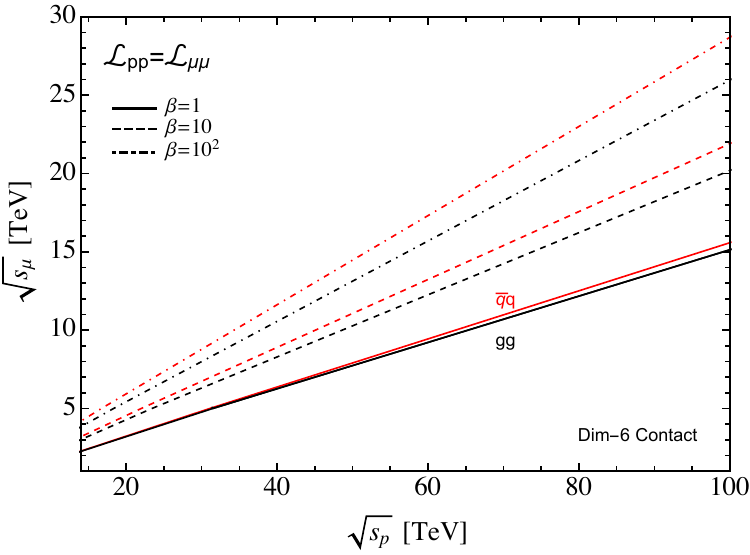} 
\caption{Same as Fig.~\ref{fig:paircomp}, but for production through a higher-dimension operator.}
\label{fig:paircomp6}
\end{figure}

Figure~\ref{fig:paircomp6} shows the approximately equivalent colliders under this assumption, 
where the outcome is very different.  
In this case, a 100\,TeV proton collider exceeds the reach of a 10\,TeV muon collider 
even assuming comparable partonic cross sections at $s_\mu$, which may not be the case.

\subsection{Physics reach of alternative FCC-hh \texorpdfstring{$\sqrt{s}$}{sqrt(s)} options}
\label{sec:PhysicsCaseHHscenarios}

The reduction of the FCC tunnel length, with respect to the CDR, from 100 to 90.7\,km, 
and the range of values for the magnetic field strength of dipoles that might be ready for mass production 
on the timescale of the final definition of the FCC-hh accelerator, 
motivate a study of beam-energy scenarios alternative to the canonical $\sqrt{s}=100$\,TeV used for the CDR physics studies. 

The first results of this study are presented here, considering the cases of $\sqrt{s}=80$ and 120\,TeV. 
The former is motivated by the combined effect of a reduction in ring size and a less aggressive assumption, 
with respect to the CDR, for the strength of Nb$_3$Sn dipoles finally available (14\,T instead of 16\,T). 
The higher energy value is relevant in scenarios where high-temperature superconducting (HTS) dipoles, 
in the range of 20\,T, become available. 
A more precise and detailed assessment of the full operational scenarios, 
including therefore the estimates of the collider luminosity, is underway. 
The definition of these scenarios will be modulated by assumptions about, and interplay between, 
key elements such as overall power consumption and experimental pileup. 
The results of these more complete studies, and their impact on the physics programme, will be documented at a later stage. 

For this report, the same luminosity setting is assumed for all three energies, 80, 100, and 120\,TeV, 
namely a total of 30\,ab$^{-1}$, obtained by combining the event samples collected by two general-purpose detectors. 
The goal of the study is not to provide an accurate update of the 100\,TeV results 
but rather to focus on the loss or gain in performance for some of the key deliverables highlighted in the FCC-hh CDR, 
namely the precision study of Higgs properties, the search for high-mass resonances, 
and the potential to discover or exclude scenarios of WIMP dark matter candidates. 
The chosen energies (80 and 120\,TeV) may not match precisely the values that will emerge from the more detailed accelerator studies. 
Preliminary indications suggest, however, that they provide a 
good reference for the physics performance with dipole field strengths ranging between 12 and 20\,T. 

\subsubsection{Higgs properties}

The CDR has shown that the baseline FCC-hh programme can improve to sub-percent level the precision of Higgs decays 
such as $\PGg\PGg$, $\PZ\PGg$, and $\PGm\PGm$, 
which are too rare to be detected with that kind of precision at FCC-ee or any other proposed $\epem$ Higgs factory.
This precision is achieved by considering the ratio of Higgs bosons produced above some \pt threshold 
and decaying to a given channel, relative to those decaying to four leptons. 
For the latter, the branching ratio is expected to be known from FCC-ee with a per-mil precision. 
The ideal value of the \pt cut is defined by the optimum balance between statistical and systematic uncertainties;
typical values are in the 100 to 200\,GeV range. 
In this \pt range, the Higgs production rate is reduced (increased) by about 30\% (35\%), 
when changing $\sqrt{s}$ from 100 to 80 (120)\,TeV. 
Folding this modified statistical uncertainty with the systematic uncertainties 
leads to the projections for the measurement precision of Higgs couplings shown in Table~\ref{tab:FCChh_sce_H}.  
The 1\% statistical precision expected at 100\,TeV for the determination of 
the $\PQt\PAQt\PH\,/\,\PQt\PAQt\PZ$ production cross-section ratios is modified to 1.2\% and 0.85\%, at 80 and 120\,TeV, respectively. 

\begin{table}[ht]
\centering
\caption{Projected uncertainties of Higgs coupling measurements at FCC-hh, 
for different $\sqrt{s}$ scenarios, with an integrated luminosity of 30\,ab$^{-1}$.}
\label{tab:FCChh_sce_H}
\begin{tabular}{ l   c c   c }
\toprule
Coupling & 100 TeV & 80 TeV & 120 TeV   \\
 & CDR baseline & & \\
\midrule 
$\delta g_{\PH\PGg\PGg}/g_{\PH\PGg\PGg}~(\%)$
& 0.4 & 0.4 & 0.4 \\
$\delta g_{\PH\PGm\PGm}/g_{\PH\PGm\PGm}~(\%)$
& 0.65 & 0.7 & 0.6 \\
$\delta g_{\PH\PZ\PGg}/g_{\PH\PZ\PGg}~(\%)$
& 0.9 & 1.0 & 0.8 \\
$\delta g_{\PH\PGm\PGm}/g_{\PH\PGm\PGm}~(\%)$
& 0.65 & 0.7 & 0.6 \\
\bottomrule
\end{tabular}
\end{table}
\begin{table}[ht]
\centering
\caption{Projected uncertainties of SM Higgs self-coupling measurements at FCC-hh, 
for different $\sqrt{s}$ scenarios, under different assumptions regarding the experimental systematic uncertainties, 
as defined in Ref.~\cite{Mangano:2020sao}.}
\label{tab:FCChh_sce_HH}
\begin{tabular}{ l   c c   c }
\toprule
Systematic uncertainties & 100\,TeV & 80\,TeV & 120\,TeV   \\
 scenario & CDR baseline & & \\
\midrule 
I (\%) & 3.4 & 3.8 & 3.1 \\
II (\%)
& 5.1 & 5.6 & 4.7 \\
III (\%)
& 7.8 & 8.4 & 7.3 \\
\bottomrule
\end{tabular}
\end{table}

The evolution of the Higgs self-coupling measurement is shown in Table~\ref{tab:FCChh_sce_HH}. 
Three different scenarios for the systematic uncertainties are considered~\cite{Mangano:2020sao}: 
(I)~corresponds to systematic uncertainties comparable to those of the ATLAS and CMS experiments during the Run~2 of LHC;
(III)~is a conservative estimate based on 2019 projections for the HL-LHC performance; 
and (II)~is an intermediate scenario. 
Significant progress has been made since 2019 and the results, even for the baseline 100\,TeV operations, 
will be updated by the time of the ESPP discussions.

\subsubsection{WIMP DM search}

\begin{figure}[ht]
\centering
\includegraphics[width=0.98\linewidth]{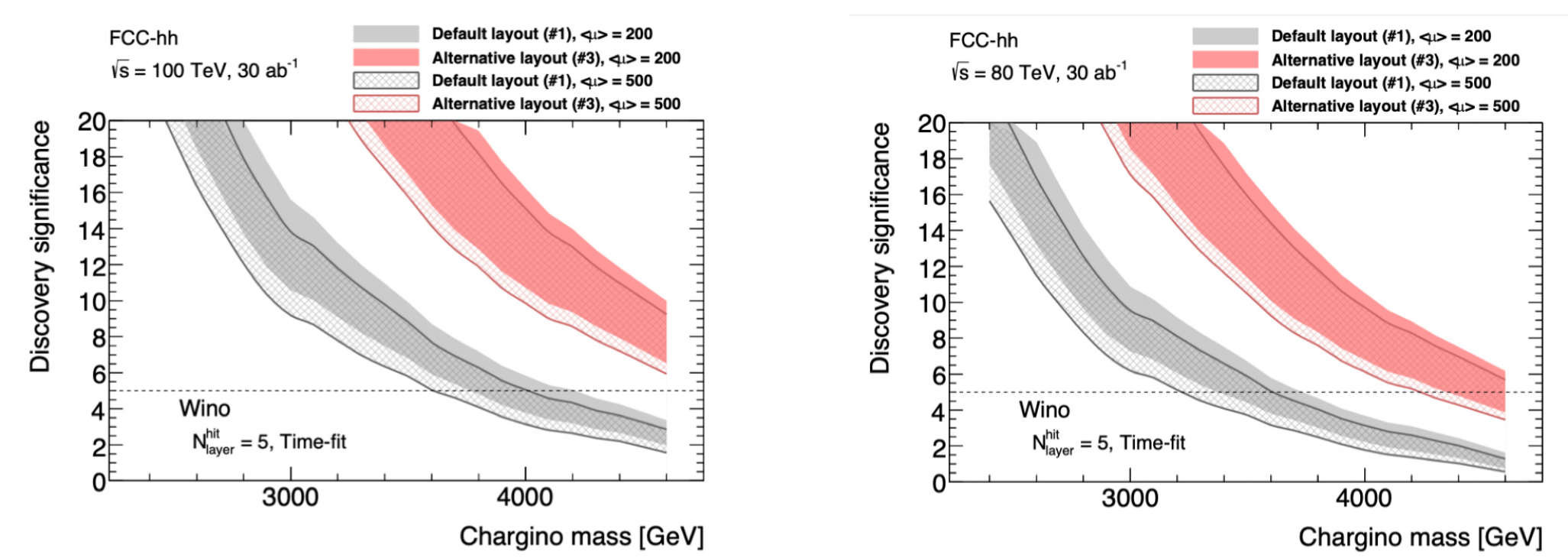}
\includegraphics[width=0.98\linewidth]{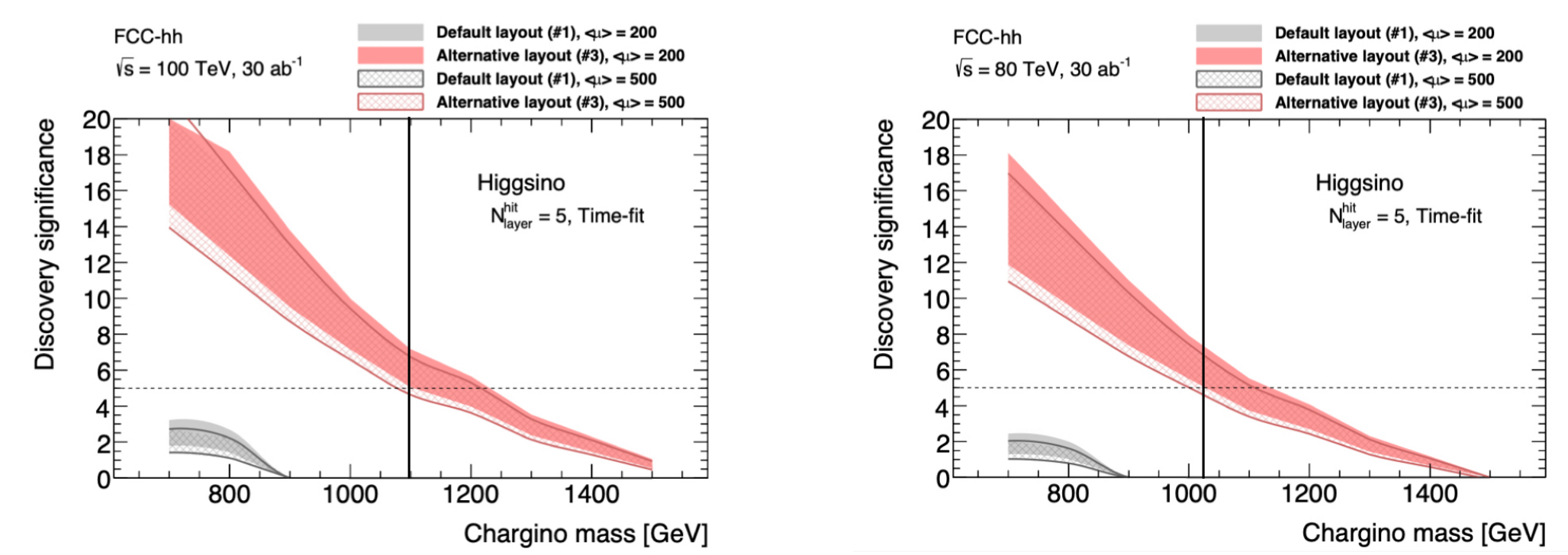} 
\caption{The 5\,$\sigma$ discovery reach for wino and higgsino dark matter candidates, 
via disappearing-track signals for $\sqrt{s} = 100$ (left) and 80\,TeV (right). 
The red band corresponds to the alternative pixel-detector layout proposed in Ref.~\cite{Saito:2019rtg}, 
where the details of the analysis are discussed.}
\label{fig:wimpdm}
\end{figure}

A key result obtained for the CDR is the proof that an FCC-hh experiment could conclusively discover, or exclude, 
broad classes of dark matter candidates. 
For example, weakly interacting neutral components of doublet or triplet weak isospin fermions, 
such as the higgsino and wino gauge fermions in supersymmetric theories, 
are required by cosmology to have masses smaller than about 1 and 3.5\,TeV, respectively. 
It was shown in Ref.~\cite{Saito:2019rtg} that a search for disappearing charged tracks 
in events with large missing transverse energy could fully cover these mass ranges 
(as shown by the red bands in the left-side plots of Fig.~\ref{fig:wimpdm}). 
The analysis has been repeated at 80\,TeV, rescaling the signal strength while keeping the same background level, 
leading therefore to a conservative extrapolation to lower energy. 
The results are shown in the right-side plots of Fig.~\ref{fig:wimpdm}. 
The $\cal{O}$(10\%) loss in reach still enables the  5\,$\sigma$ discovery up to the cosmology limit, 
even for the more critical case of the higgsino. 
Is is expected that a re-optimisation of the analysis should cover the relevant mass range 
down to slightly smaller collision energies, possibly above $\sqrt{s} = 70$\,TeV.

\subsubsection{High-mass reach}

The general potential of FCC-hh to directly explore the existence of new large-mass particles is discussed above. 
The beam energy is clearly the main driver for the mass reach. 
The detector performance and the pileup environment have a minor impact on the energy-dependence of this reach, 
a dependence that can be studied with a simple extrapolation based on the evolution with energy of the partonic luminosities, 
using, e.g., the Collider Reach tool~\cite{ColliderReach}.  
Table~\ref{tab:FCChh_sce_res} shows the extrapolation to 80 and 120\,TeV of the 5\,$\sigma$ reach 
for several possible $s$-channel resonances.
Details of the 100\,TeV analysis and the definition of the various models listed here
\begin{table}[ht]
\centering
\caption{Beam-energy dependence of the projected 5\,$\sigma$ discovery reach, 
with the new particle mass in TeV, for $s$-channel BSM resonances, in various decay modes. 
The definition of the various models can be found in Ref.~\cite{Helsens:2019bfw}.}
\label{tab:FCChh_sce_res}
\begin{tabular}{ l   c   c  c }
\toprule
Resonance & 100 TeV 
& 80 TeV & 120 TeV \\ 
\midrule 
Q$^*$ & 40 & 33 & 46 \\
$\PZ^\prime_\mathrm{TC2} \to \ttbar$ & 23 & 20 & 26 \\
$\PZ^\prime_\mathrm{SSM} \to \ttbar$ & 18 & 15 & 20 \\
G$_\mathrm{RS}\to \PW\PW$ & 22 & 19 & 25 \\
$\PZ^\prime_\mathrm{SSM}\to \ell\ell$ & 43 & 36 & 50 \\
$\PZ^\prime_\mathrm{SSM}\to \PGt\PGt$ & 18 & 15 & 20 \\
\bottomrule
\end{tabular}
\end{table}
can be found in Ref.~\cite{Helsens:2019bfw}. 
A reduction of the collision energy to 80\,TeV leads to a 15--20\% loss in mass reach, 
while running at 120\,TeV would increase the reach by 10--15\%. 
It goes without saying that, to this date, 
there is no specific model that precisely requires the existence of such new resonances in 
some of the mass ranges that would be (or not be) covered at the different energies. 
The impact of the beam-energy change is reduced for particles well below the kinematic endpoint, 
whose discovery reach is mostly limited by weak couplings, backgrounds or systematic uncertainties 
(as, for example, in the case of DM~WIMPS discussed above). 

\subsection{A forward-physics facility at FCC-hh}
\label{sec:PhysicsCaseHHFPF}

\begin{figure}[ht]
\centering
\includegraphics[width=0.8\textwidth]{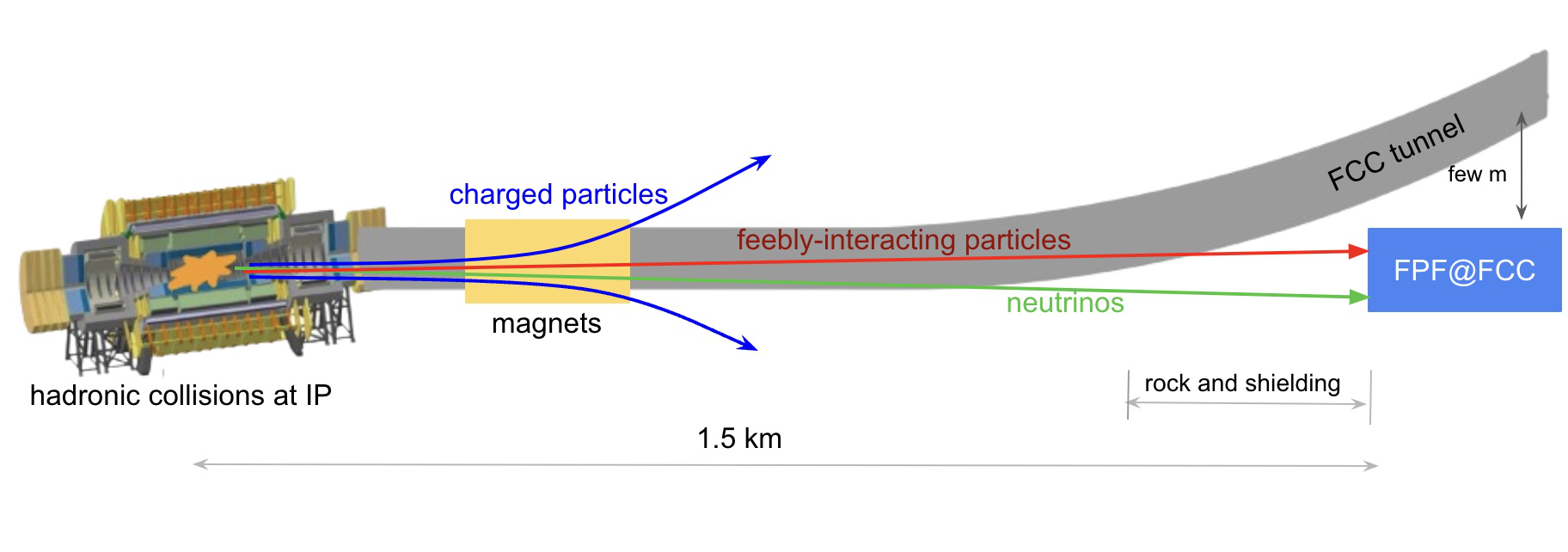} 
\includegraphics[width=0.8\textwidth]{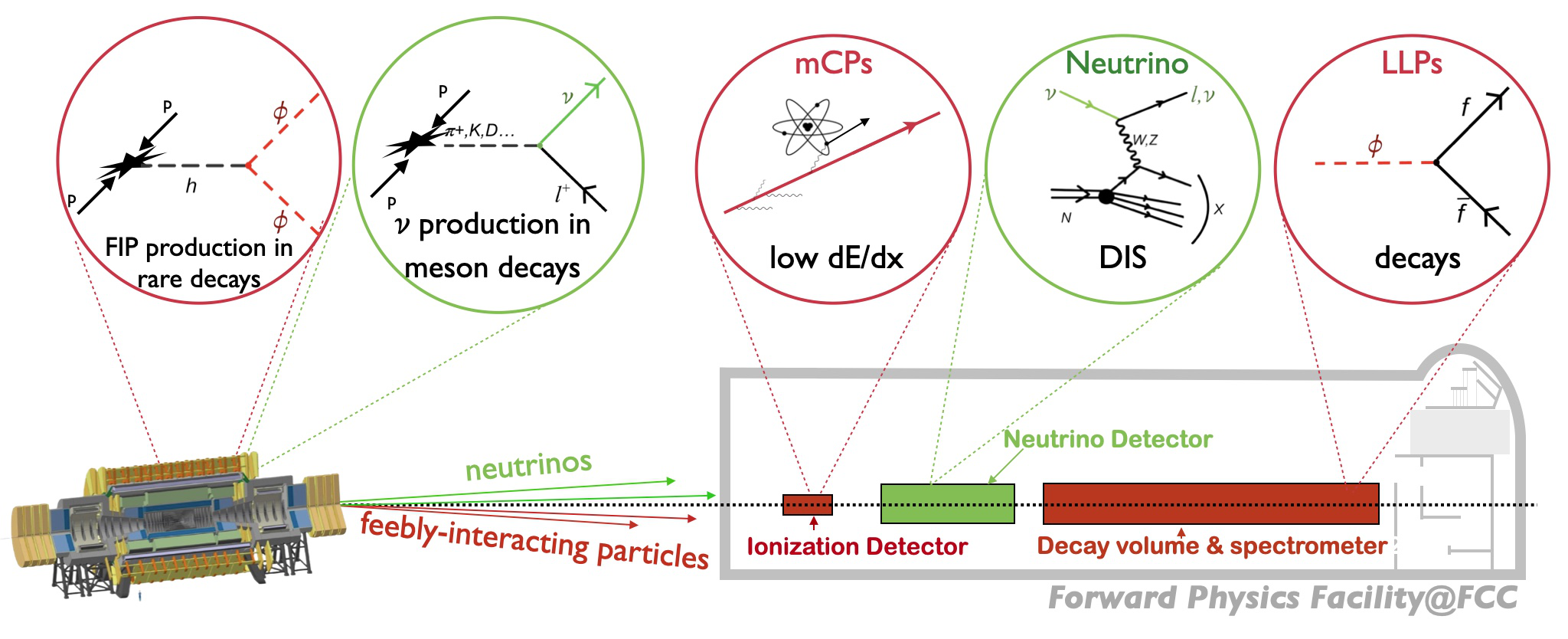}
\caption{High-energy light particles (neutrinos, LLPs/FIPs) produced in FCC-hh collisions could be detected at the FPF@FCC, 
located at around 1.5\,km away from the IP, enabling sensitivity to a variety of SM and BSM signatures.}
\label{fig:schematicphysics}
\end{figure}

The immense breadth of the physics programme available to a hadron-collider facility is well proven by LHC, 
with its programme of $\Pp\Pp$, pN, and NN collisions (where N stands for heavy ion). 
The four large LHC experiments are accompanied by a multitude of smaller, dedicated experiments, 
which further push the exploitation of the scientific opportunities. 
A first exploration of the opportunities offered by the programme of heavy ion collisions at FCC-hh 
was presented in Ref.~\cite{Mangano:2017tke} and in the CDR~\cite{fcc-phys-cdr}. 
The additional unique opportunities that could arise from the exploitation of the high intensity beam of energetic particles, 
including neutrinos, produced by $\Pp\Pp$ collisions in the forward region, is briefly discussed here.
At LHC, these particles are studied with the far-forward experiments 
FASER($\nu$)~\cite{FASER:2020gpr,FASER:2022hcn,FASER:2023tle,FASER:2024hoe} 
and SND@LHC~\cite{SNDLHC:2022ihg,SNDLHC:2023pun}, 
and new dedicated experiments have been proposed in the context of a 
forward physics facility (FPF)~\cite{Feng:2022inv,Anchordoqui:2021ghd} operating concurrently with HL-LHC.
A FPF-like suite of far-forward experiments integrated within FCC-hh 
would provide unique opportunities for neutrino physics, QCD studies, and BSM searches.
Such FPF@FCC could be located around 1.5~km away from the IP (Fig.~\ref{fig:schematicphysics}), where a dedicated cavern would be excavated aligned with the line-of-sight (LoS).
It would have a length between 100 and 500\,m, depending on the detector setup. 
Representative applications of the FPF@FCC experiments are summarised here. 
A more detailed discussion can be found in Ref.~\cite{MammenAbraham:2024gun}.

\subsubsection{Neutrino physics}

\begin{figure}[ht]
\centering
\includegraphics[width=0.9\linewidth]{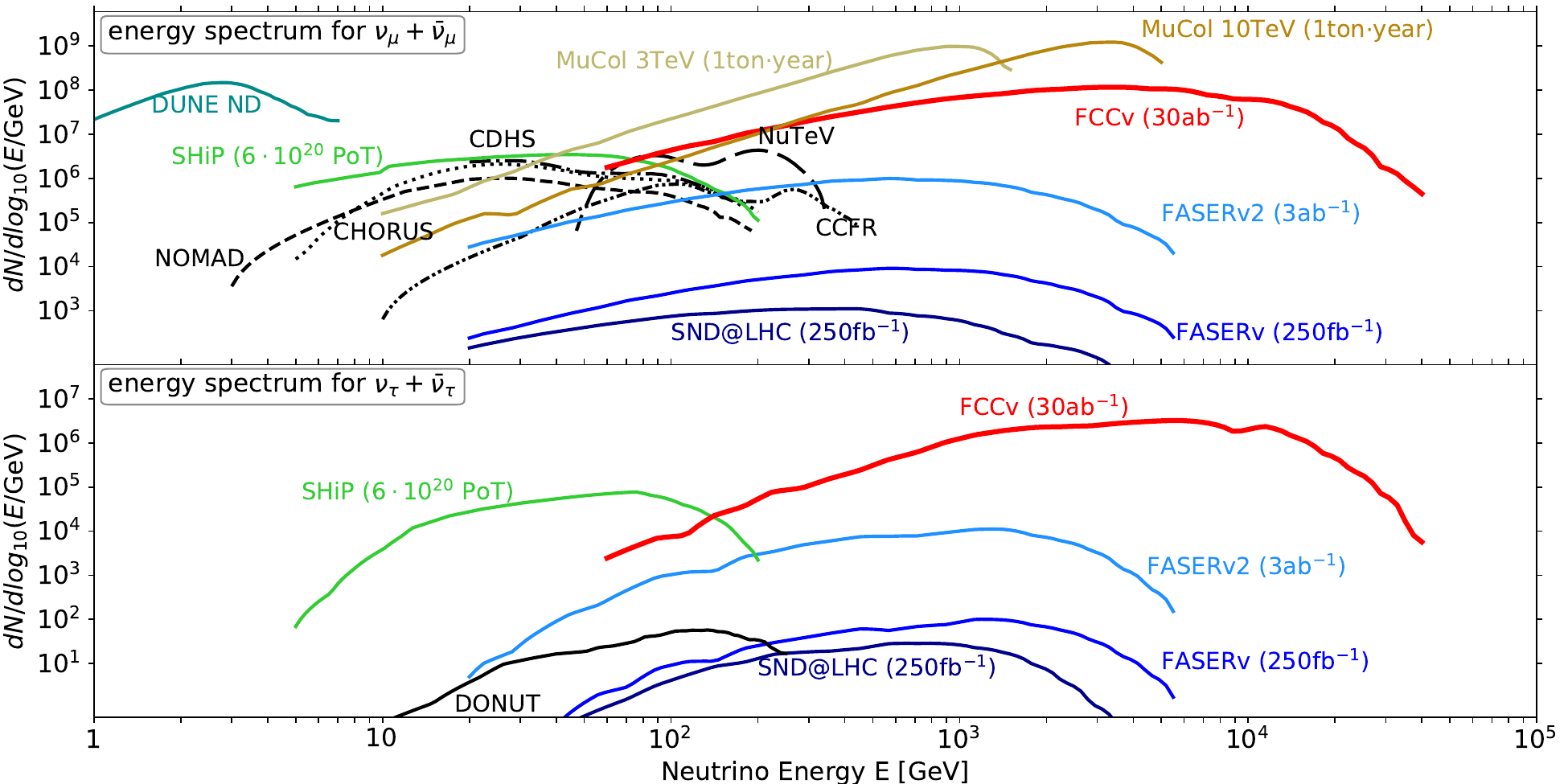}
\caption{The muon and tau neutrino scattering yields as a function of $E_{\PGn}$ at FCC$\PGn$ and other experiments.}
\label{fig:fluxes-global-overview}
\end{figure}

For neutrino detection, two options are considered:
a FASER$\PGn$2-like detector~\cite{Feng:2022inv}, dubbed FCC$\PGn$, with a length of 6.6\,m, 
and a deeper variant, FCC$\PGn$(d), with a length of 66\,m.
These detectors could collect~\cite{MammenAbraham:2024gun} up to $10^9$ electron/muon neutrinos and $10^7$ tau neutrinos 
for $\mathcal{L}_{\Pp\Pp} = 30$\,ab$^{-1}$, an increase of several orders of magnitude with respect to the (HL-)LHC yields.
These neutrinos would exhibit the highest energies ever achieved in a laboratory (up to 40\,TeV), 
overlapping with those from astrophysical sources.
The event yields forecast at the FPF@FCC would outperform all previous, ongoing, and future (proposed) neutrino experiments 
for all three generations, as indicated in Fig.~\ref{fig:fluxes-global-overview}.
These unprecedented event rates allow the precise fingerprinting of neutrino properties, 
such as their flavour universality and non-standard interactions. 

As a representative application, the electromagnetic properties of neutrinos are considered, 
long recognised as a window to new physics~\cite{Giunti:2014ixa}, 
in particular on the measurement of the neutrino charge radius $\langle r^2_{\PGn} \rangle$, 
which modifies the neutral-current DIS cross section for neutrinos.
It can be extracted~\cite{Ismail:2020yqc, MammenAbraham:2023psg} 
by searching for deviations in the ratio between NC and CC DIS events as compared to the 
$\langle r^2_{\PGn} \rangle = 0$ baseline. 
The upper left panel of Fig.~\ref{fig:BSM} shows the projected sensitivity to 
$\langle r^2_{\PGn} \rangle$ at FPF@FCC considering only statistical uncertainties.
World-leading bounds would be achieved, 
measuring the (SM) neutrino charge radius for \PGne and \PGnGm, 
and reaching down to five times the SM value for \PGnGt.
Realising this potential for precision neutrino physics 
demands an excellent modelling of CC and NC neutrino DIS interactions within the detector, 
combined with high-precision event generators for neutrino DIS~\cite{FerrarioRavasio:2024kem,Buonocore:2024pdv,vanBeekveld:2024ziz}.

\subsubsection{BSM sensitivity}

\begin{figure}[ht]
\centering
\includegraphics[width=0.4\textwidth]{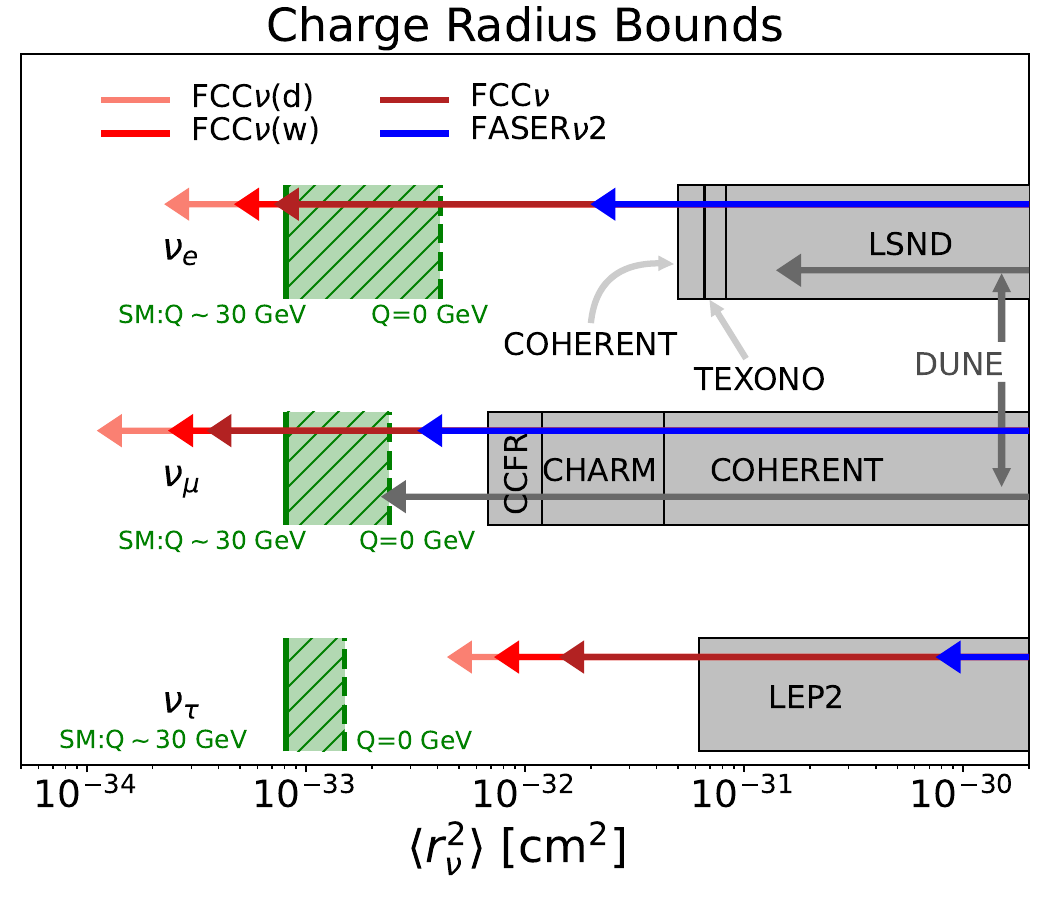} 
\includegraphics[width=0.45\textwidth]{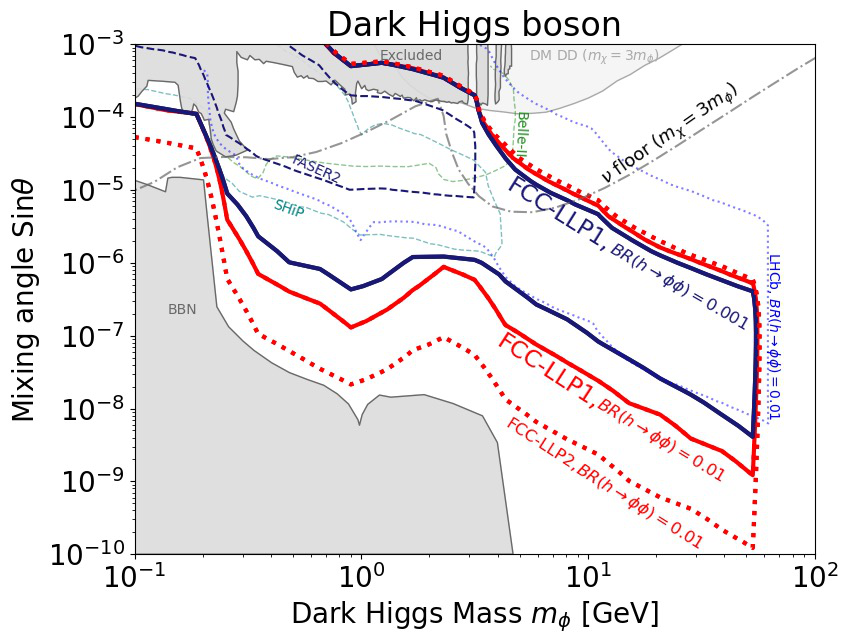}
\includegraphics[width=0.45\textwidth]{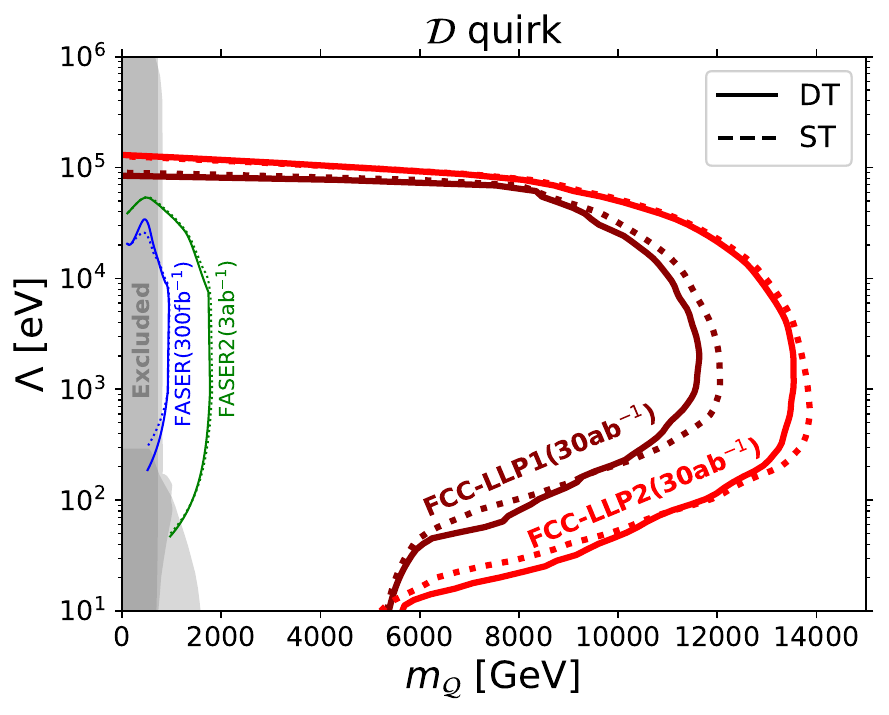}
\includegraphics[width=0.45\textwidth]{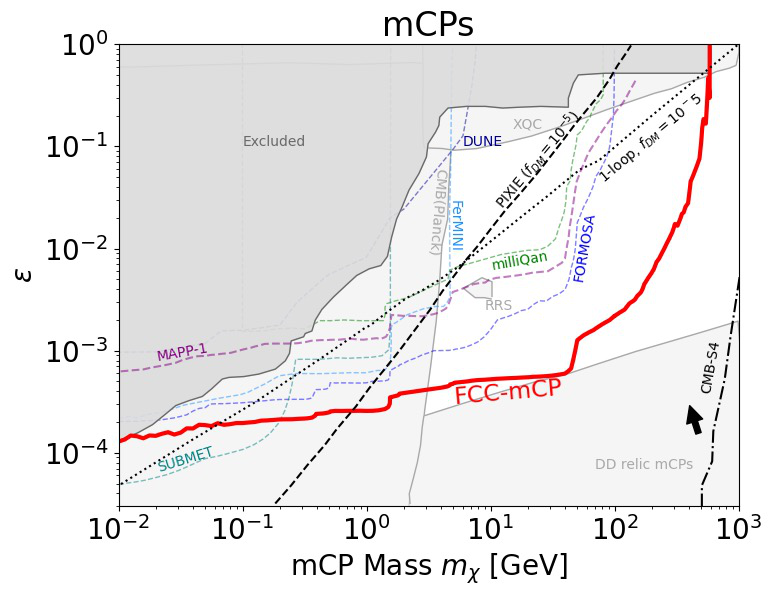}
\caption{Top left: Sensitivity of the FCC$\PGn$ detectors to the neutrino charge radius.
Top right: Reach for a dark Higgs boson $\phi$ at the FPF@FCC.
Bottom left: Discovery potential for $\mathcal{D}$-type quirks with mass $m_{\mathcal{Q}}$ and confinement scale $\Lambda$. 
Bottom right: Sensitivity to mCPs with mixing parameter $\epsilon$.} 
\label{fig:BSM}
\end{figure}

The FPF@FCC is sensitive to a variety of BSM models~\cite{MammenAbraham:2024gun}, 
including dark Higgs bosons~\cite{Kling:2021fwx}, 
relaxion-type scenarios, 
quirks~\cite{Batell:2022pzc,Li:2021tsy, Li:2023jrt, Feng:2024zgp}, 
and millicharged particles (mCPs).
These BSM signatures fall, in most cases, outside the coverage of the main FCC-hh detectors, 
and hence such a far-forward facility would markedly increase the discovery capabilities of FCC-hh.
The sensitivity to dark Higgs bosons, $\mathcal{D}$-type quirks, and mCPs of the FPF@FCC is summarised in Fig.~\ref{fig:BSM}.
The enormous rates of forward Higgs production at 100\,TeV enable the discovery of LLPs from Higgs decays 
with masses as large as 62\,GeV ($\simeq m_{\PH}/2$) 
and couplings as small as $\sin\theta \sim 10^{-8}$ (beyond the reach of FCC-ee); 
of quirks with masses up to 14\,TeV for a broad range of confinement scales $\Lambda$; 
and of mCPs with masses up to several hundreds of GeV and charge $q \sim 10^{-3} e$, 
closing the gap between (future) accelerator-based bounds and dark matter direct detection searches.

\subsubsection{QCD and hadronic structure}

\begin{figure}[ht]
\centering
\includegraphics[width=0.3\textwidth]{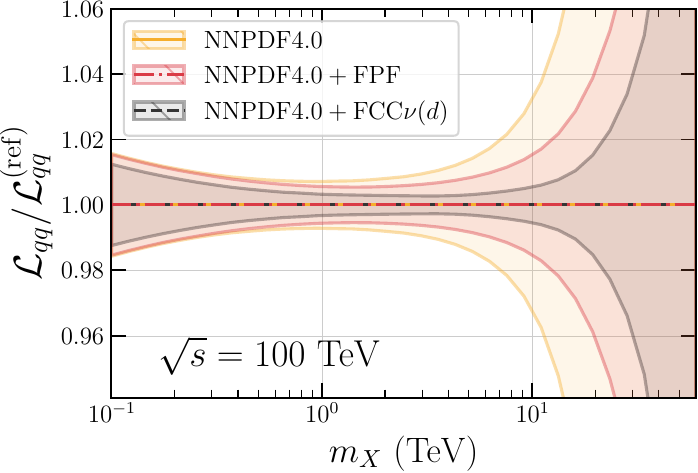} 
\includegraphics[width=0.3\textwidth]{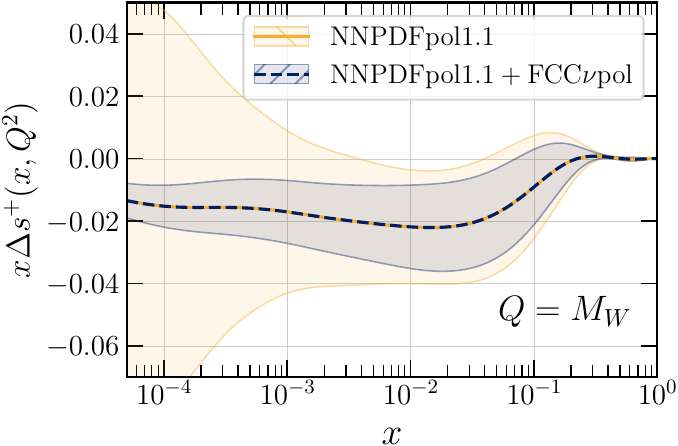}  
\includegraphics[width=0.3\textwidth]{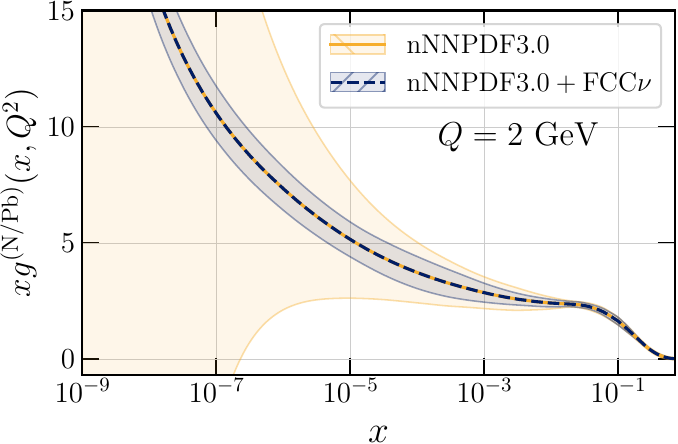}  
\caption{Projected constraints from the FPF@FCC experiments on 
the unpolarised (left) and polarised (middle) PDFs, 
and on the nuclear PDFs of lead nuclei (right).}
\label{fig:QCD}
\end{figure}

The enormous samples of multi-TeV neutrinos available at the FPF@FCC (Fig.~\ref{fig:fluxes-global-overview}) 
enable high-resolution probes of the unpolarised and polarised structure of protons and of heavy nuclei, 
as summarised by the representative applications of Fig.~\ref{fig:QCD}.

First, high-energy neutrino DIS structure functions~\cite{Cruz-Martinez:2023sdv,Candido:2023utz} 
provide a precise quark/antiquark flavour separation at large-$x$, 
which in turn reduces the theoretical uncertainties entering high-mass searches at FCC-hh.
Second, neutrino DIS on a polarised target~\cite{Forte:2001ph,Mangano:2001mj} 
resolves the spin structure of the proton~\cite{Nocera:2014gqa,Borsa:2024mss} 
with complementary information to related experiments, 
such as those at the Electron-Ion Collider~\cite{AbdulKhalek:2021gbh}.
Third, detecting neutrinos originating from proton-lead collisions provides 
information on nuclear structure~\cite{Klasen:2023uqj,Eskola:2021nhw,AbdulKhalek:2022fyi} down to $x \sim 10^{-9}$, 
where nuclear PDFs are unconstrained and novel dynamical regimes of QCD, 
such as the Colour Glass Condensate, are expected to dominate.
This kinematic region is also of prime importance for astroparticle physics experiments.

\subsubsection{Summary}

Integrating far-forward experiments into FCC-hh would extend its scientific potential 
in several synergetic directions from QCD and neutrino physics to BSM searches in a cost-efficient manner.
While at this early stage no attempt has been made to optimise the accelerator infrastructure 
or define the FPF@FCC detector technology more precisely, 
the results shown here motivate further studies of how to fully realise this unique physics potential.

\cleardoublepage
\section{Theoretical calculations}
\label{sec:theory}
To fully leverage the significantly improved experimental precision in \PZ-pole observables, \PW boson and top quark masses, \PQb and \PGt decays, 
as well as a broad array of Higgs observables, 
it is essential to obtain Standard Model predictions with an accuracy that matches the anticipated statistical uncertainties of FCC-ee. 
The expectation that most systematic experimental uncertainties can be reduced to the level of the statistical precision further underscores this requirement.
Such theoretical predictions are necessary for both inclusive and exclusive processes, 
with the latter requiring integration with Monte Carlo generators. 
They encompass a wide range of technical aspects, including fixed-order perturbative corrections, resummation calculations, 
improved parton showers, non-perturbative hadronization effects, and lattice QCD. 
Any discrepancies between experimental data and Standard Model predictions (anomalies) could provide crucial insights, potentially guiding the way towards new discoveries.
In addition, detailed precision analyses of BSM effects within concrete models and effective theories will open up a broad spectrum of new prospects.

On top of that, FCC-hh has unique capabilities for testing SM phenomena at ultra-high energies, 
particularly the mechanism of electroweak symmetry breaking and broad coverage for direct particle discovery. 
Theory calculations are needed to evaluate (often large) backgrounds, expected signal rates, and to optimise the experimental search strategies. 
The high precision reached by the LHC, due to further improve in the HL-LHC era, 
has been and remains a powerful stimulus for the theory community to enhance the calculation reliability. 
This progress is fully aligned with the future needs of FCC-hh.

A very active effort (in lattice and in perturbative calculations) is ongoing worldwide to improve the theoretical control of flavour observables, 
in view of the current and forthcoming experimental progress made by LHCb and Belle~II. 
This effort will continue alongside the experimental advances made by these experiments in the next decades, 
undertaking the required steps towards the ultimate precision goals required by FCC. 
Suitably refined precision calculations for flavour observables should also be identified and relevant actions must be planned accordingly. 
Present and future work focuses on 
(a)~calculating higher-order perturbative QCD and EW corrections (including, for the latter, their matching at the weak scale), 
and (b)~understanding an order-of-magnitude better than before the separation and interplay between 
perturbative and non-perturbative contributions with the goal of constraining the latter from data as much as possible.
Improved lattice calculations, from algorithmic progress and hardware developments, are a further essential ingredient. 

For all the measurement areas where theoretical improvements are needed to match the FCC precision goals --- EW, QCD, flavour, BSM --- 
the large interplay between the development of EW and QCD calculations, 
in terms of impact on the total theoretical systematic uncertainties and of computational challenges and tools, is remarkable.
Two-loop QCD corrections are often of comparable size to one-loop EW corrections, 
and they both play a critical role at the required level of precision. 
Their interplay is thus essential, and often it is the source of added complexity when both EW and QCD corrections are mixed 
(e.g., in the presence of off-shell decays of weak bosons to quarks). 
On the formal side, progress towards the understanding of multi-scale multi-loop diagrams is of common value to both EW and QCD, as well as to BSM studies. 
Techniques inspired by effective field theory (EFT) are becoming ubiquitous, 
and tools initially developed for flavour physics (soft-collinear effective theory, or SCET) 
are now relevant to describe QCD event shapes, resummations, jet-vetoed final states, and much more. 
This global synergy of developments in different specific areas opens a whole new dimension for future coordinated theoretical efforts. 

If any deviation from the SM expectations were observed at FCC-ee/hh, 
the interpretation would require calculations to the requisite precision in BSM models 
or models with higher-dimensional EFT operators to the requisite precision.
These higher-order corrections can be straightforwardly achieved by adapting 
the techniques developed for standard model (QED, EW, and QCD) calculations. 
Therefore, improving the accuracy of SM predictions at large currently remains the main priority of the community of experts on radiative corrections. 
This chapter focuses on a brief review of the status, future needs and prospects for theoretical improvements of relevance to EW, Higgs, jets, and top quark measurements, 
including the MC event-generator perspective, and the actions under way to support and coordinate such efforts.

\subsection{Electroweak corrections}
\label{sec:theocalc}

To meet the precision goals of FCC-ee, 
significant advances in calculations of higher-order radiative corrections and in MC generators will be needed~\cite{Freitas:2019bre,Blondel:2018mad,EWPrecisionWshop}. 
For instance, electroweak NNLO corrections for various pair production processes 
($\epem \to \Pf\bar{\Pf}$, $\epem \to \PGg\PGg$, $\epem \to \PWp\PWm$, $\epem \to \PZ\PH$) 
are needed, as well as MC tools for the simulation of multiple photon radiation beyond leading-logarithmic (LL) approximation. 
Even higher perturbative orders, including three-loop corrections in the full SM and leading four-loop corrections, 
are required to interpret precision measurements at the \PZ pole.
Table~\ref{tab:obs1} provides a few illustrative examples of precision quantities and the theory calculations required to extract them from data.

\setlength{\tabcolsep}{6pt}
\begin{table}[ht]
\caption{A few sample precision quantities of interest for the FCC-ee programme, 
their current and projected experimental uncertainties, 
and the required theory input for their \emph{extraction} from the data. 
The last two columns show the current state of the art for calculations of this theory input
and higher-order calculations needed to reach the FCC-ee precision target.
More details can be found in Ref.~\cite{Freitas:2019bre}.}
\label{tab:obs1}
\setlength{\tabcolsep}{5pt}
\begin{tabular}{p{1.8cm}p{1.6cm}p{2.8cm}p{2.7cm}p{2.5cm}p{2.5cm}@{}p{0cm}}
\toprule
Quantity & Current precision & \raggedright FCC-ee stat.\ (syst.)~precision 
& \raggedright Required\newline theory input & \raggedright Theory status\newline as of today & \raggedright Needed theory improvement$^\dagger$ & \\
\midrule
$m_{\PZ}$~(MeV) & 2.0 & 0.004 (0.1) 
& \multirow{3}{2.7cm}{non-resonant $\epem \to \Pf\bar{\Pf}$, initial-state radiation (ISR)} 
& \multirow{3}{2.5cm}{NLO,\newline ISR logarithms up to 6$^\text{th}$ order} & \multirow{3}{2.8cm}{NNLO for $\epem \to \Pf\bar{\Pf}$ } \\[.7ex]
\cmidrule(r){1-3}
$\Gamma_{\PZ}$~(MeV) & 2.3 & 0.004 (0.012) & & & \\[.7ex]
\cmidrule(r){1-3} 
$\sin^2\theta_\text{eff}^\ell$ & $1.6\times 10^{-4}$ & 1.2 $(1.2) \times 10^{-6}$ & & & \\[1ex]
\midrule
$m_{\PW}$~(MeV) & 9.9 & 0.18 (0.16) 
& \raggedright lineshape of $\epem \to \PW\PW$ near threshold 
& \raggedright NLO ($\epem \to 4\Pf$\newline or EFT framework) 
& \raggedright NNLO for $\epem \to \PW\PW$, $\PW \to \Pf\bar{\Pf}'$\newline in EFT setup & \\
\midrule
\raggedright $\PH\PZ\PZ$ coupling & \raggedright -- $^\ast$  & 0.1\% 
& \raggedright cross section for $\epem \to \PZ\PH$ & \raggedright NLO EW plus partial NNLO QCD/EW & \raggedright full NNLO EW & \\
\midrule
$m_\text{top}$~(MeV) & 290 & 4.2 (4.9) & \raggedright threshold scan $\epem \to \PQt \PAQt$ 
& \raggedright N$^3$LO QCD, NNLO EW, resummations up to NNLL, $\mathcal{O}$(30\,MeV) scale uncert.\ 
& \raggedright Matching fixed orders with resum\-mations, merging with MC, $\as$ (input) & \\
\bottomrule
\end{tabular}\\
\footnotesize{$^\dagger$\,The necessary theory calculations mentioned are a minimum baseline; 
additional partial higher-order contributions may also be required.}\\
\footnotesize{$^\ast$ No absolute value for the HZZ coupling can be extracted from the LHC data without additional assumptions.}
\end{table}

\setlength{\tabcolsep}{6pt}

\begin{table}[ht]
\centering
\caption{Required theory calculations for the \emph{prediction} of the listed precision quantities within the SM. 
These predictions are needed for comparison with the quantities extracted from data (Table~\ref{tab:obs1}), 
for the purpose of testing the validity of the SM and probing BSM physics.
More details in Ref.~\cite{Freitas:2019bre}.}
\label{tab:obs2}
\begin{tabular}{p{1.2cm}p{3.5cm}p{3cm}p{3.5cm}@{}p{0cm}}
\toprule
Quantity & \raggedright Required theory input & \raggedright Theory status\newline as of today & \raggedright Needed theory improvement$^\ddagger$ & \\
\midrule
$\Gamma_{\PZ}$ & \multirow{3}{3.5cm}{vertex corrections for $\PZ \to \Pf\bar{\Pf}$} & \multirow{3}{3cm}{NNLO + partial higher orders} & \multirow{3}{3.5cm}{N$^3$LO EW + partial higher orders} & \\[1ex]
\cmidrule(r){1-1} 
$\sin^2\theta_\text{eff}^\ell$ &&& \\[.7ex]
\midrule
$m_{\PW}$ & \raggedright SM corrections to the muon decay rate & \raggedright NNLO + partial higher orders & \raggedright N$^3$LO EW + partial higher orders & \\
\bottomrule
\end{tabular}\\
\footnotesize{$^\ddagger$\,The mentioned needed theory calculations are a minimum baseline; \\
additional partial higher-order contributions may also be required.}
\end{table}

Additional theory input is necessary for the interpretation of the experimentally determined values of these quantities, 
i.e., to test the validity of the SM and probe possible physics beyond the SM, as illustrated in Table~\ref{tab:obs2} for some of the same examples as above.
In this context, the three-loop $\as^3$ corrections to the semileptonic $\PQb \to \PQc$ decay~\cite{Fael:2020tow} 
and the three-loop QED corrections to the muon decay~\cite{Fael:2020tow,Czakon:2021ybq} were recently calculated in the Fermi approximation. 
These results are a milestone in perturbative calculations and an important step towards precision calculations for FCC-ee. 
In particular, the $\alpha^3$ QED contribution translates to a shift of the muon lifetime of $(-9 \pm 1) \times 10^{-8}$\,$\mu$s.
With the current measurement of $\tau_{\PGm} = 2.1969811 \pm 0.0000022$\,$\mu$s, 
the new correction terms are almost two orders of magnitude smaller than the experimental uncertainty. 
Thus, an updated value of $G_\mathrm{F}$ can only be extracted once the latter has been improved.

The Fermi constant $G_\mathrm{F}$ is one of several fundamental parameters that need to be precisely determined 
in order to make quantitative predictions for electroweak precision observables (\mbox{EWPOs}) 
within the SM or concrete BSM realisations through data-theory comparisons. 
Other examples of such quantities include the strong coupling $\as$ and the top-quark mass $m_\text{top}$. 
Many of these parameters can be obtained from FCC-ee data with much reduced uncertainty compared to today (Table~\ref{tab:EWPO}), 
but their extraction from the experimental measurements requires significant theory input 
(as shown in Section~\ref{sec:qcdcalc} for $\as$ and in the last row of Table~\ref{tab:obs1} for $m_\text{top}$).

Similarly, the interpretation of cross section measurements requires the precise determination of the luminosity 
through measurements of small-angle Bhabha scattering and/or large-angle photon-pair production (Section~\ref{sec:lumi}). 
For this purpose, the rates for these processes need to be computed at least with NNLO QED corrections, 
logarithmically enhanced higher-order effects, and a careful treatment of virtual hadronic corrections 
(which appear at NLO for Bhabha scattering and NNLO for two-photon production)~\cite{deBlas:2024bmz,Jadach:2018jjo}.

To achieve the precision goals outlined in Tables~\ref{tab:obs1} and~\ref{tab:obs2}, 
significant improvements in calculation techniques for higher-order radiative corrections are required, 
including full two-loop corrections for $2\to 2$ scattering processes, 
as well as decay processes with full three-loop corrections and approximate four-loop contributions in a large-mass expansion.

Recently, significant progress has been achieved in semi-numerical multi-loop calculation techniques based on dispersion relations~\cite{Song:2021vru,Freitas:2022hyp} 
and on series solutions of differential equations~\cite{Dubovyk:2022frj,Liu:2021wks,Liu:2022mfb,Liu:2022chg,Hidding:2022ycg,Armadillo:2022ugh}. 
These techniques have enabled the first calculations of electroweak two-loop corrections for the main Higgs boson production process at FCC-ee, 
$\epem \to \PZ\PH$~\cite{Freitas:2022hyp,Chen:2022mre}, 
and they are expected to be useful for calculations of NNLO corrections for other pair production processes as well.

The approach of Refs.~\cite{Song:2021vru,Freitas:2022hyp} exploits dispersion relations and Feynman parameters 
to express one subloop of a two-loop diagram in terms of integrals with up to three variables. 
Then the second subloop becomes a simple one-loop integral, with well-known analytical results. 
For divergent diagrams, the singularities can be removed with systematically constructed subtraction terms, which can also be evaluated analytically. 
These procedures lead to two- or three-dimensional finite numerical integrals that can be evaluated with good precision 
within minutes on a single CPU core for a single diagram class. 
The method does not use any reduction to master integrals, which saves computing resources.

Another powerful method for multi-loop integrals is the method of differential equations~\cite{Kotikov:1990kg,Kotikov:1991hm,Kotikov:1991pm,Remiddi:1997ny,Henn:2013pwa}. 
While it can be difficult to find analytical solutions to these equations, it was recently realised that solutions can be efficiently obtained 
to arbitrarily high precision, using deep series expansions~\cite{Dubovyk:2022frj,Liu:2021wks,Liu:2022chg,Hidding:2022ycg,Armadillo:2022ugh}. 
By matching series about different expansion points, solutions can be constructed that consistently extend across thresholds and other singular points. 
To fully determine the differential equation solutions, boundary values are also needed, and can be evaluated analytically or numerically for special kinematic points 
(such as zero momentum or unphysical Euclidean momentum). 
This step is made particularly simple with the auxiliary flow method, 
which uses boundary conditions at infinity of an unphysical (auxiliary) variable, 
where they can be evaluated in terms of algebraic recurrence relations~\cite{Liu:2022mfb}.

The series expansion approach is not only restricted to two-loop integrals, but is also a promising tool for three-loop SM corrections. 
It requires a reduction to master integrals, which is the computational bottleneck for this method. 
Recent developments in integral reduction methods~\cite{Klappert:2020nbg,Chestnov:2022alh,Fontana:2023amt} can help to push the envelope on this front. 
In addition, the integral reduction can be performed numerically with much lower computing resources, 
and the functional dependence on kinematic variables can be reconstructed with suitable expansions or interpolations. 
Progress was also made towards analytical solutions of differential equations for Feynman integrals, using functions beyond multiple polylogarithms~\cite{Bourjaily:2022bwx,Pogel:2022ken}.

For a reliable phenomenological description of $\epem \to \PZ\PH$, 
one of the next steps is to combine the NNLO result with a calculation of the process $\epem \to \PH \Pf\bar{\Pf}$, 
where the fermion pair may be off the \PZ resonance. 
These off-shell contributions are important because of the relatively large decay width of the \PZ boson. 
For the projected FCC-ee precision, it is sufficient to compute $\epem \to \PH \Pf\bar{\Pf}$ at NLO, 
which can be straightforwardly accomplished with existing automated tools.

\subsection{QCD precision calculations}
\label{sec:qcdcalc}

This section addresses some of the main theoretical challenges regarding Quantum Chromodynamics (QCD) calculations on the path towards FCC-ee.
The considerations below are inspired by previous reviews~\cite{Monni:2021rcz} 
and largely by discussions that took place at topical FCC-ee workshops~\cite{EWPrecisionWshop,QCDPrecisionWshop}.

\subsubsection{QCD studies in \texorpdfstring{$\PZ/\PGg^*\to \text{jets}$}{Z/gamma* to jets}}
\label{sec:QCDZdec}

The very large FCC-ee integrated luminosities at the \PZ boson pole and at higher energies 
provide an excellent opportunity to expand the
knowledge of strong interactions in QCD final states, 
both in the area of jet physics and for the extraction of the strong coupling constant, $\as(m^2_{\PZ})$, 
from hadronic observables in $\PZ/\PGg^* \to \text{jets}$ events.

\paragraph{Jet physics and shape observables}
\label{sec:evshp}

Fine details of QCD final states can be investigated through a range of jet observables 
such as event shapes (designed to describe the geometric properties of hadronic events), 
jet rates, and jet substructure observables.

Owing to their sensitivity to QCD radiation, 
these observables are widely used to extract $\as$~\cite{dEnterria:2022hzv}, 
calibrate non-perturbative hadronisation models~\cite{Dasgupta:2003iq},  
or study QCD dynamics within jets~\cite{Marzani:2019hun}. 
Fully differential calculations for the process $\epem \to \PZ/\PGg^*\to \PQq\PAQq + {\rm X}$ at N$^3$LO in QCD ($\as^{3}$)
for massless partons in the final state can be derived starting from the results of
Refs.~\cite{Gehrmann-DeRidder:2007vsv,Gehrmann-DeRidder:2008qsl,Weinzierl:2008iv,Weinzierl:2009ms,DelDuca:2016ily}
with the inclusive cross section at N$^3$LO~\cite{Gorishnii:1990vf}. 
Similarly, the production of heavy (notably bottom) quarks, 
$\epem \to \PZ/\PGg^* \to \mathrm{Q}\bar{\mathrm{Q}} + X$, can be described at NNLO in QCD 
using the predictions of Refs.~\cite{Bernreuther:1997jn,Nason:1997tz,Nason:1997nw,Chetyrkin:2000zk}.
For higher jet multiplicities, 
the computation of QCD radiative corrections with massless final-state partons has been extended to NLO
for $\epem \to \PZ/\PGg^* \to n ~ \text{jets}$ with $n=5$~\cite{Frederix:2010ne} and $n=6$, 7~\cite{Becker:2011vg}.

The essential NNLO QCD calculations for final states with four or five jets are beyond the current state of the art. With the recent progress in the calculation of the necessary two-loop scattering amplitudes (see, e.g., Refs.~\cite{Hartanto:2019uvl,Badger:2021nhg,Abreu:2021asb} for the five-point case), these NNLO predictions will arguably become available in the coming years.
Future developments in the computation of such multi-scale amplitudes may benefit from novel computational techniques 
discussed in Section~\ref{sec:theocalc} and in 
Refs.~\cite{Bierenbaum:2010cy,Badger:2013gxa,vonManteuffel:2014ixa,Ita:2015tya,Buchta:2015wna,Peraro:2016wsq,Liu:2017jxz,
Abreu:2017xsl,Capatti:2019ypt,Capatti:2020xjc,Liu:2021wks,Bobadilla:2021pvr}. 
A recent review of modern computational methods can be found in Refs.~\cite{Heinrich:2020ybq,EWPrecisionWshop}.

The perturbative description of kinematic regimes that require the all-order resummation of radiative corrections 
has also improved substantially in the past decade, 
and the state-of-the-art calculations for standard global event shapes and jet rates in
$\epem \to \PZ/\PGg^* \to \PQq\PAQq + X$ have reached NNLL order and
beyond~\cite{Becher:2008cf,Abbate:2010xh,Chien:2010kc,Monni:2011gb,
Becher:2012qc,Mateu:2013gya,Hoang:2014wka,deFlorian:2004mp,Banfi:2014sua,
Banfi:2016zlc,Tulipant:2017ybb,Moult:2018jzp,Banfi:2018mcq,Bell:2018gce,
Procura:2018zpn,Ebert:2020sfi,Bris:2020uyb,Moult:2022xzt,Duhr:2022yyp}. 
Resummations for multijet final states are desirable for QCD phenomenology at FCC-ee, 
while currently only a limited number of predictions is available beyond the NLL order for
$\epem \to \PZ/\PGg^* \to \PQq\PAQq \Pg + {\rm X}$ 
observables~\cite{Catani:1991hj,Larkoski:2018cke,Chen:2019bpb,Arpino:2019ozn}.
Further progress is therefore necessary in this area, 
requiring dedicated analytical and numerical resummation techniques.
Furthermore, the computation of non-global observables~\cite{Dasgupta:2001sh,Banfi:2002hw}, 
sensitive to the geometric pattern of soft QCD radiation, 
has recently been pushed to the NNLL order~\cite{Banfi:2021owj,Banfi:2021xzn,FerrarioRavasio:2023kyg,Becher:2023vrh}, 
and a number of dedicated phenomenological applications to $\epem$ collisions are becoming 
available~\cite{Dasgupta:2002bw,Banfi:2002hw,Weigert:2003mm,Hatta:2013iba,Becher:2015hka,
Larkoski:2015zka,Becher:2016mmh,Balsiger:2019tne,Banfi:2021xzn,Becher:2023vrh}.
Computational techniques for jet substructure observables have also witnessed outstanding progress in the last decade, 
resulting in an array of applications at lepton colliders to the study of observables measured 
on groomed jets~\cite{Frye:2016okc,Baron:2018nfz,Kardos:2020gty,Dasgupta:2022fim}
or the study of fragmentation and spin correlations 
in jet physics~\cite{Jain:2011xz,Chen:2020adz,Neill:2020bwv,Karlberg:2021kwr,Chen:2022muj,Chen:2023zlx,vanBeekveld:2023lsa}.

\paragraph{Measurements of the strong coupling constant}

Among the main challenges to match FCC-ee experimental accuracy is
the reduction of the uncertainties of the QCD parameters,
and notably the strong coupling constant, $\as(m^2_{\PZ})$~\cite{dEnterria:2022hzv}.
The current world average~\cite{ParticleDataGroup:2024cfk}, 
which reaches a $\sim$\,0.8\% uncertainty ($\as(m^2_{\PZ}) = 0.1180 \pm 0.0009$), 
is mainly constrained by the precision of lattice calculations~\cite{FlavourLatticeAveragingGroupFLAG:2021npn}, 
expected to be improved by a factor of two in the next decade~\cite{DallaBrida:2019wur,DelDebbio:2021ryq}.

At FCC-ee, a precision of 0.1\% on $\as(m^2_{\PZ})$ can be contemplated (Table~\ref{tab:EWPO}). 
The most precise determination comes from a combined fit of three EW pseudo-observables at the Z pole: 
the \PZ boson hadronic partial width,
the total hadronic cross section at the resonance peak, 
and the ratio of hadronic to leptonic branching fractions.
These inclusive quantities are particularly suitable to extract $\as$ 
given the small non-perturbative hadronisation corrections,
which scale with the centre-of-mass energy $Q$ as $(\Lambda_\text{QCD} / Q)^6$~\cite{Dokshitzer:1995qm}.
With the $6 \times 10^{12}$ \PZ bosons produced at FCC-ee, 
the total experimental uncertainty in $\as$ extractions from fits of the above quantities 
is of the order of 0.1\%~\cite{dEnterria:2020cpv}, 
hence requiring a substantial reduction of the corresponding theoretical uncertainties.
The status of theory computations for these observables is well-advanced:
QCD corrections are known up to N$^4$LO~\cite{Baikov:2008jh,Baikov:2012er,Herzog:2017dtz} 
and N$^3$LO corrections for massive bottom quarks are also known in a power series in $m_{\PQb}^2 / Q^2$~\cite{Chetyrkin:2000zk}.
On the other hand, EW and mixed QCD-EW corrections are available at least up to two loops 
(see, e.g., Refs.~\cite{Dubovyk:2018rlg,Dubovyk:2019szj,Chen:2020xot} and references therein), 
as discussed in Section~\ref{sec:theocalc} of this report.

Other very accurate extractions of $\as$ at FCC-ee can be derived from 
\PGt~\cite{Pich:2016bdg,Boito:2016oam} and \PW~\cite{dEnterria:2016rbf,dEnterria:2020cpv} hadronic and leptonic decays, 
using high-order perturbative QCD computations. 
Existing studies on the $\as$ extraction from the decay of EW bosons indicate that 
the calculation of higher-order corrections\footnote{State-of-the-art calculations can be found in, e.g., 
Refs.~\cite{Baikov:2008jh, Baikov:2012er, Freitas:2014hra, Freitas:2013dpa, Chen:2020xzx, Dubovyk:2019szj}.}
up to $\mathcal{O}(\as^5)$, $\mathcal{O}(\alpha^3)$, and $\mathcal{O}(\as,\alpha^3)$ or $\mathcal{O}(\as^2,\alpha^2)$ 
for QCD, EW, and QCD$\,\oplus\,$EW, respectively,
are needed to match the expected statistical experimental precision on inclusive \PW and \PZ hadronic observables~\cite{dEnterria:2020cpv}.
In the case of \PGt decays, 
open theoretical questions are related to the treatment of non-perturbative effects~\cite{Pich:2013lsa,Davier:2013sfa,Boito:2014sta}, 
as well as to the difference between extractions relying on contour-improved (CIPT)~\cite{Pivovarov:1991rh,LeDiberder:1992jjr} 
and fixed-order (FOPT) perturbative calculations adopted in the fits~\cite{Hoang:2021nlz,Benitez-Rathgeb:2021gvw}.
Recent investigations indicate that CIPT calculations might require a more robust estimate of 
non-perturbative corrections~\cite{Hoang:2020mkw,Hoang:2021nlz,Golterman:2023oml}, 
with potential ways forward discussed in Refs.~\cite{Benitez-Rathgeb:2022hfj,Benitez-Rathgeb:2022yqb}. 
Given these open questions, $\as$ fits based on CIPT have been excluded from the latest world average~\cite{ParticleDataGroup:2024cfk}.
A deeper understanding of these aspects is necessary for robust determinations of $\as$ from \PGt decays at FCC-ee.

The sensitivity to $\as$ of differential observables, e.g., in the final states discussed in the previous section, 
makes them also suitable for precise extractions of the strong coupling.
Currently, different $\as$ determinations from these observables are included 
in the world average~\cite{Jones:2003yv,Dissertori:2007xa,Bethke:2009ehn,Becher:2008cf,Davison:2009wzs,
Dissertori:2009ik,Gehrmann:2010uax,Chien:2010kc,Abbate:2010xh,OPAL:2011aa,Gehrmann:2012sc,Abbate:2012jh,
Hoang:2015hka,Kardos:2018kqj,Dissertori:2009qa,Schieck:2012mp,Verbytskyi:2019zhh,Kardos:2020igb},
and can differ from each other by up to a few standard deviations.
Besides the high-accuracy perturbative ingredients described above,
such fits require input on hadronisation effects. 
Non-perturbative radiation induces ${\cal O}(\Lambda_\text{QCD}^p / Q^p)$ changes in the observables 
(with typically $p = 1$) that must be estimated to achieve the desired precision.
Aside from the different observables considered in the fit, 
a central difference between these different $\as$ determinations 
lies in how hadronisation corrections are evaluated 
(with either Monte Carlo models or analytic techniques)~\cite{Korchemsky:1994is,Dokshitzer:1995qm,Dokshitzer:1997iz,Beneke:1997sr,Dokshitzer:1998pt,
Gardi:1999dq,Korchemsky:1999kt,Dasgupta:1999mb,Salam:2001bd,Gardi:2001di,Gardi:2003iv,Mateu:2012nk,Agarwal:2020uxi}. 
The yet suboptimal internal consistency of $\as$ determinations in $\epem$ jet observables 
hints at unsatisfactory modelling of non-perturbative QCD dynamics, which poses a major bottleneck for precision phenomenology at FCC-ee 
and requires substantial improvements in our understanding of this kinematic regime. 
First innovative steps towards this ambitious goal are being taken within a dispersive model of the QCD coupling, 
which has recently been used to extract the leading non-perturbative scaling in three-jet 
final states~\cite{Luisoni:2020efy,Caola:2021kzt,Caola:2022vea,Nason:2023asn}.

Data from FCC-ee will be very beneficial for deepening the understanding of hadronisation corrections. 
On the one hand, energies higher than those of previous lepton colliders would arguably justify 
the further development of analytic models based on a power expansion in $\Lambda_\text{QCD}/Q$.
On the other hand, 
the energy span and experimental accuracy of FCC-ee are instrumental in gaining better control of non-perturbative dynamics in MC generators, 
which will be beneficial in all measurements expected at FCC-ee, 
such as $\epem \to \PQt\PAQt$, $\epem \to \PWp\PWm$, and $\epem \to \PZ \PH$.
A complementary approach is given by the design of observables with reduced sensitivity to hadronisation, 
e.g., through jet-substructure techniques~\cite{Frye:2016okc,Baron:2018nfz,Kardos:2020gty,Dasgupta:2022fim},
which open promising avenues for complementary extractions of $\as$~\cite{Marzani:2019evv}.
An in-depth study of the effectiveness of these techniques at FCC-ee energies, 
as well as the estimate of the remaining hadronisation corrections, 
are highly desirable in the coming years~\cite{Marzani:2019evv,Hoang:2019ceu}.

\subsubsection{QCD aspects of Higgs physics}

Reaching theoretical uncertainties aligned with the projections of Table~\ref{tab:HiggsKappa3} for Higgs precision studies 
requires dedicated developments in different areas of both QCD and EW calculations. 
While EW corrections are discussed in Section~\ref{sec:theocalc}, 
this section focuses on QCD aspects.
The clean experimental conditions at FCC-ee allow a detailed and exclusive study of the hadronic decays of the Higgs boson. 
Partial widths are currently theoretically known at the per-cent level, 
with a  parametric uncertainty on $\as(m^2_{\PZ})$ that will be significantly reduced at FCC-ee with the expected 0.1\% precision achieved in this parameter. 
More specifically, in the case of $\PH \to \PQb\PAQb$, 
N$^4$LO QCD corrections are known in the limit of massless bottom quarks~\cite{Baikov:2005rw,Davies:2017xsp,Herzog:2017dtz}, 
and N$^4$LO QCD corrections to $\PH \to \Pg\Pg$ have been computed in the heavy-top-mass limit~\cite{Herzog:2017dtz}.
The simulation of Higgs decays is paramount both for correcting the fiducial acceptance of the experiments
and for studying kinematic distributions of the decay products and jet observables.
These distributions are sensitive to quark Yukawa couplings~\cite{Gao:2016jcm,Bi:2020frc} or 
to new physics states~\cite{Kaplan:2009qt,Curtin:2013fra,Liu:2016zki,Liu:2016ahc,Gao:2019ypl}.
The sensitivity to light-quark Yukawa couplings can be enhanced by means of modern quark and gluon tagging techniques~\cite{Bedeschi:2022rnj}.

A key application is the extraction of the strange-quark Yukawa coupling, 
for which preliminary studies have shown promising results (Section~\ref{sec:PhysPerf_strangeTagging}).
Among the challenges in the realisation of this measurement, 
a relevant theory bottleneck is the separation of the $\PH \to \PQs\PAQs$ decay from 
the Dalitz decay of a Higgs to a pair of either gluons (QCD mediated) or photons (EW mediated) 
followed by a gluon/photon splitting into strange quarks. 
Initial investigations~\cite{QCD4Higgs-SalamSoyez} indicate that 
this background can be drastically suppressed by a cut in the invariant mass of the pair of jets 
originating from the fragmentation of the two strange quarks, $m_{j_1 j_2} \gtrsim 100$\,GeV.
A robust theoretical control of the $\PH \to \PQs\PAQs$ signal in this region 
requires accurate perturbative calculations as well as the resummation of logarithmic corrections 
stemming from soft-gluon radiation off the final-state strange quarks near the Higgs mass threshold, $m_{j_1 j_2} \sim m_{\PH}$.
Together with advances in perturbative calculations, 
a second essential element in this endeavour is the development of improved hadronisation models 
to distinguish the non-perturbative fragmentation of (strange) quarks from that of gluons. 
Such models are instrumental for the reliable training of jet taggers,
and their calibration within future Monte Carlo generators will highly benefit from the precise QCD data collected at FCC-ee.

Recently, considerable steps have been taken in the calculation of theoretical predictions for Higgs decays, 
which are essential for the implementation of the above ideas. 
The predicted kinematic distributions of the $\PH \to \PQb\PAQb$ decay products are 
known up to N$^3$LO in the limit of 
massless \PQb quarks~\cite{Anastasiou:2011qx,DelDuca:2015zqa,Caola:2017xuq,Ferrera:2017zex,Mondini:2019gid,Gauld:2019yng}
and NNLO (and partially beyond) mass corrections are
available~\cite{Chetyrkin:1995pd,Larin:1995sq,Harlander:1997xa,Bernreuther:2018ynm,Primo:2018zby,Behring:2019oci,Mondini:2020uyy,Behring:2020uzq}. 
Similarly, differential QCD predictions for $\PH \to \Pg\Pg$ are now available up to N$^3$LO in the large-top-mass limit~\cite{Chen:2023fba}.
Finite quark mass corrections to $\PH \to \Pg\Pg$ are relevant at the level of precision foreseen at FCC-ee
and could be included up to NNLO in QCD in the near future with state-of-the-art 
calculations~\cite{Djouadi:1995gt,Spira:1995rr,Larin:1995sq,Schreck:2007um,Melnikov:2016qoc,Melnikov:2017pgf,Kudashkin:2017skd,Frellesvig:2019byn}.
Calculations of hadronic event shapes and jet resolutions at NLO~\cite{Gao:2019mlt,Luo:2019nig,Gao:2020vyx} and
NNLO~\cite{Mondini:2019vub} in QCD have also been performed in recent years, 
including the treatment of mass logarithms in the relevant virtual
amplitudes~\cite{Banfi:2013eda,Grazzini:2013mca,Melnikov:2016emg,Liu:2017vkm,Liu:2018czl,Liu:2020tzd,Anastasiou:2020vkr,Liu:2021chn,Liu:2022ajh}
and of Sudakov logarithms in event-shape distributions~\cite{Ju:2023dfa}.
Similar to \PZ boson decays, 
the relatively low-energy scale involved in the Higgs boson decays ($\sim m_{\PH}$) 
implies that non-perturbative QCD effects have a sizeable impact on most differential distributions of hadronic Higgs decay products.
Therefore, the considerations made in Section~\ref{sec:QCDZdec} apply here as well. 
The FCC-ee Higgs programme will benefit enormously from future developments in the modelling of hadronisation effects.

\subsubsection{QCD modelling of the top-quark threshold}
\label{sec:QCDtop}

The properties of top quark, 
such as its mass, width, and EW couplings (Table~\ref{tab:EWPO}) are scheduled to be measured at FCC-ee
with runs at centre-of-mass energies between 340 and 365\,GeV.
The top mass can be extracted with high precision at FCC-ee through a threshold scan, 
where the top-quark pair is non-relativistic.
Suitable definitions of short-distance top-mass schemes unaffected by ambiguities related to infrared physics 
are available~\cite{Bigi:1997fj,Hoang:1998hm,Hoang:1998ng,Beneke:1998rk,Pineda:2001zq},
and can be exploited for precise predictions of the  $\sigma_{\ttbar}$ vs.\ $\sqrt{s}$ lineshape.
Although NNLO and N$^3$LO fixed-order QCD calculations are available~\cite{Chen:2016zbz,Chen:2022vzo}, 
$\sigma_{\ttbar}$ receives a substantial contribution from Coulomb-type interactions in this non-relativistic kinematic regime.
These effects can be accurately described in the context of effective field theories 
derived from non-relativistic QCD~\cite{Thacker:1990bm,Lepage:1992tx,Bodwin:1994jh}, 
valid when the top quark velocity is of order $\as$. 
A lot of effort has been devoted to computing predictions in this framework, 
including QCD effects up to N$^3$LO~\cite{Beneke:2015kwa} 
(see also Refs.~\cite{Hoang:2000yr,Beneke:2013jia} and references therein),
approximate NNLL renormalisation-group improved corrections~\cite{Hoang:2013uda}, 
and the inclusion of EW effects within an analogous EFT framework~\cite{Beneke:2017rdn}.

The description of the final state $\PWp\PWm \PQb\PAQb + \mathrm{X}$
requires the inclusion of non-resonant channels that do not involve the creation of a pair of on-shell top quarks. 
These non-resonant channels critically demand embedding the aforementioned non-relativistic EFT
into the unstable particle EFT~\cite{Beneke:2003xh,Beneke:2004km},
where current predictions reach NNLO accuracy for the non-resonant part~\cite{Beneke:2017rdn}.
Projections for FCC-ee quote an expected theoretical accuracy for the top mass of ${\cal O}$(50)\,MeV~\cite{Beneke:2017rdn} 
in the potential-subtracted top-mass scheme~\cite{Beneke:1998rk} 
(see also Refs.~\cite{Vos:2016til,Abramowicz:2018rjq}), 
with ${\cal O}$(30)\,MeV scale uncertainties~\cite{Beneke:2024sfa,defranchis_2024_cqd16-xhk71}. 
Improving further on these results represents a formidable challenge for the field of precision calculations, 
far beyond the current state of the art. 
The main steps include the computation of N$^4$LO corrections in the non-relativistic EFT framework, 
as well as the description of QED effects at NNLL both in the collinear limit 
(e.g., ISR~\cite{Bertone:2019hks,Frixione:2019lga}) and in the soft limit 
(as discussed in Ref.~\cite{BenekeEWPrecisionWshop}).
Moreover, the optimal exploitation of the FCC-ee measurements might also require the N$^3$LO calculation for the non-resonant channels.

The theoretical description of differential distributions is less accurate than that of the inclusive quantities just discussed, 
and reaches either NLO or NNLO accuracy only for specific observables~\cite{Hoang:1999zc,Bach:2017ggt}.
Further progress is needed in these computations, which are central to controlling precisely the effect of experimental cuts. 
Some aspects of these calculations pose considerable theoretical challenges, 
for instance, concerning the differential calculations in the non-relativistic limit, 
or the assessment of non-factorisable radiative corrections to the decays of the two top quarks~\cite{Melnikov:1995fx}.

\subsection{Monte Carlo event generators}

Among the theoretical developments necessary for the FCC-ee physics programme, 
the area of Monte Carlo (MC) event generators 
(see, e.g., Ref.~\cite{Campbell:2022qmc} for a review of the current state of the art) 
plays a pivotal role.
These tools are instrumental not only for the accurate simulation of QCD and QED effects,
but also for the calibration of the detectors as well as the training of analysis tools.
The precision expected at FCC-ee requires a significant improvement in MC generators, with respect to their previous generation,
far exceeding the current state of the art.
The following paragraphs summarise some of the corresponding challenges, regarding QCD and EW (notably QED) aspects.

\subsubsection{QCD aspects}
\label{sec:QCDmc}

Monte Carlo generators simulate QCD corrections in three stages:
the hard scattering at high momentum transfer; 
the parton shower stage, in which the system evolves towards low momentum scales; 
and the non-perturbative stage, which implements the transition of final state partons into hadrons.
A first area where substantial improvement is necessary concerns the formulation of parton shower algorithms, 
given that current public tools are generally limited to leading logarithmic (LL) accuracy. 
The precision goals of FCC-ee arguably demand NNLL accuracy or beyond, 
which is currently the subject of widespread investigations within the community. 
Specifically, state-of-the-art algorithms achieving NLL accuracy for broad ranges of observables have recently been 
formulated~\cite{Dasgupta:2018nvj,Bewick:2019rbu,Dasgupta:2020fwr,Forshaw:2020wrq,Hamilton:2020rcu,Nagy:2020rmk,
Nagy:2020dvz,Hamilton:2021dyz,Karlberg:2021kwr,Herren:2022jej,vanBeekveld:2022zhl,vanBeekveld:2022ukn,vanBeekveld:2023lfu,Assi:2023rbu} 
with novel techniques that help bridge the field of parton showers with 
QCD resummations~\cite{Dasgupta:2018nvj,Bewick:2019rbu,Dasgupta:2020fwr,Forshaw:2020wrq,Nagy:2020rmk,Nagy:2020dvz,Herren:2022jej}.
Significant progress has also been made in the formulation of amplitude-level evolution~\cite{AngelesMartinez:2018cfz,Forshaw:2019ver}, 
which would ultimately allow a systematic treatment of soft quantum interference effects in parton showers.
Beyond NLL, several conceptual and technical problems become relevant, 
spanning from the inclusion of higher-order matrix elements~\cite{Jadach:2011kc,Li:2016yez,Hoche:2017hno,Dulat:2018vuy,Ravasio:2023anp,vanBeekveld:2024qxs}, 
to the formulation of consistent schemes for the treatment of virtual corrections in the soft and 
collinear limits~\cite{Catani:1990rr,Banfi:2018mcq,Catani:2019rvy,Dasgupta:2021hbh,vanBeekveld:2023lsa,vanBeekveld:2024qxs}, 
and to the realisation of matching schemes to NLO matrix element corrections that preserve the shower accuracy~\cite{Hamilton:2023dwb}.
The significant progress in the understanding of these aspects has recently resulted in the first class of NNLL parton-shower algorithms~\cite{vanBeekveld:2024wws} 
capable of achieving this perturbative accuracy for event-shape observables at lepton colliders. 
This progress suggests that fully general NNLL parton showers may become available for phenomenology applications in the coming years.

A second aspect with much scope for improvement is the matching of parton showers to higher-order calculations for the hard scattering. 
Presently, an array of techniques allows NLO~\cite{Frixione:2002ik,Nason:2004rx,Frixione:2007vw,Hamilton:2012rf,Jadach:2015mza,Nason:2021xke} 
or NNLO~\cite{Hamilton:2013fea,Alioli:2013hqa,Hoche:2014uhw,Monni:2019whf,Monni:2020nks} 
QCD calculations (and possibly beyond~\cite{Prestel:2021vww}) 
to be matched to LL parton showers.
These techniques have been applied to the differential simulation of Higgs boson decays~\cite{Bizon:2019tfo,Hu:2021rkt}, 
but will arguably need to be revisited in light of the recent progress in advancing the accuracy of parton showers~\cite{Hamilton:2023dwb}.

Further necessary developments concern the accurate production of particle pairs at threshold 
(e.g., $\ttbar$ or $\PWp\PWm$), 
for which significant challenges arise from the inclusion of effects related to non-relativistic dynamics 
and unstable particle decay (Section~\ref{sec:QCDtop}) within event generators.
Notable progress in the development of resonance-aware matching has been made for top-pair production in 
hadronic collisions~\cite{Jezo:2015aia,Frederix:2016rdc,Bach:2017ggt}. 
However, a full description of the hierarchy of scales involved in the process
(top-quark velocity $v$, $m_\text{top}$, and width $\Gamma_\text{top}$) 
is not available in existing generators and requires significant conceptual advances.

A final area that requires significant improvement in view of the precision required at FCC-ee is the modelling of non-perturbative corrections,
especially in the fragmentation of light partons and heavy quarks, 
central in the simulated performance of flavour tagging algorithms.
Possible advances may come from a combination of generators with higher perturbative accuracy, as discussed above, 
and more versatile parametrisations and tuning of non-perturbative effects, 
for which the use of machine learning technology offers a promising alternative to current models~\cite{Ilten:2022jfm,Ghosh:2022zdz,Chan:2023ume}.
The tuning of these models benefit from the ability to select large high-purity data samples enriched with specific flavours
(such as gluons and \PQb quarks), essential for the direct training of jet taggers on experimental data.
Valuable experimental data for the calibration and testing of non-perturbative corrections could be collected 
from measurements of observables performed with hadronic invariant masses below the \PZ-boson resonance. 
Such measurements are accessible either via dedicated runs at lower $\sqrt{s}$ 
or from hard initial-state radiation away from the \PZ-pole run. 
Preliminary feasibility studies~\cite{verbytskyi24,verbytskyi25} indicate that $\mathcal{O}(10^9)$ events 
can be collected at various energy points over the 20--80\,GeV hadronic mass range, 
enabling a variety of measurements useful for the development and testing of non-perturbative models.

Finally, the experimental accuracy expected in resonance production,
such as $\epem \to \PWp\PWm \to \PQq\PAQq^\prime \PQq^{\prime\prime}\PAQq^{\prime\prime\prime}$ or $\epem \to \ttbar$, 
will likely demand improved models of colour reconnection~\cite{Christiansen:2015yca}, 
which will be calibrated with the accurate data available for these processes.

\subsubsection{QED aspects}
\label{sec:EWmc}

The tools based on Yennie, Frautschi, and Suura (YFS) exponentiation~\cite{Yennie:1961ad} 
provide a promising framework for systematically improving the precision of QED (and QCD) radiative processes in MC generators. 
While the soft logarithms, arising from the real and virtual enhanced regions of phase space, 
are resummed to infinite order in the YFS formalism, the remaining collinear logarithms are not. 
These algorithms can be incorporated order by order, 
and the associated divergences can be regularised by including masses for all leptons. 
The YFS treatment is completely exclusive for multiple photon emission.

The \textsc{kkmc} event generator uses a variant of YFS exponentiation called Coherent Exclusive Exponentiation (CEEX). 
The most recent version, 5.00.2, of the \textsc{kkmc} package~\cite{Arbuzov:2020coe} includes improvements for the simulation of fermion pair production. 
Work is ongoing to include additional collinear contributions in the YFS framework~\cite{Jadach:2023pka}, 
beam spread parametrisations, and better descriptions of tau decays~\cite{Banerjee:2019mle}.

YFS resummation has been implemented in the {\textsc{Sherpa}}-3 series in an automatic framework~\cite{Krauss:2022ajk} 
for both the initial (ISR) and final (FSR) state QED radiation. 
Initial-final state interference (IFI) QED effects within the YFS framework are currently  implemented in \textsc{kkmc} but not in \textsc{Sherpa}, 
for which they remain an essential target for future improvements,
especially for the prediction of  forward-backward asymmetries at FCC-ee.
Using {\textsc{Sherpa}}'s automated matrix element generators, the ISR resummation can be applied to any $\epem$ process, 
while final state radiation is currently restricted to leptonic states only. 
The matching of this resummation to higher-order perturbative corrections is also automated within {\textsc{Sherpa}}, 
including the effects of collinear logarithms. 
The procedure renders finite the higher-order corrections, which by themselves can be infrared divergent. 
One-loop electroweak corrections are provided by automated tools such as \textsc{Recola}~\cite{Actis:2016mpe}, 
while {\textsc{Sherpa}} automatically creates the YFS subtraction term such that the end result is infrared finite. 
The real corrections can also be included in a similar fashion using internal automated tools. 
Since the subtraction has been automated within the \textsc{Sherpa} framework, 
the current limitation on the perturbative accuracy is missing higher-order corrections, in particular multiloop calculations. 
However, as these calculations become available, it is relatively simple to include them within the YFS resummation framework.

These higher-order corrections can be provided by external electroweak libraries, 
which contain process-dependent multi-loop matrix elements 
and that can be interfaced with MC tools. 
The \textsc{Dizet~6.45}~\cite{Arbuzov:2023afc} package contains most of the currently known higher-order corrections for \PZ-pole physics. 
For systematic extensions to higher levels of precision, however, a flexible object-oriented code library would be more suitable, 
which is the goal of the new \textsc{Griffin} project~\cite{Chen:2022dow}. 
Its object-oriented structure also simplifies the interfacing with MC generators, fitting tools, and other external programs.

The study of aspects related to ISR has been the subject of recent work~\cite{Frixione:2022ofv}.
An alternative approach to YFS factorisation for the description of ISR is based on collinear factorisation, 
as in Refs.~\cite{Frixione:2012wtz,Frixione:2021zdp,Bertone:2022ktl}. 
Unlike what happens in the YFS framework, a systematic resummation of collinear logarithms is implemented, 
while soft corrections are only approximately described. 
Recent work has expanded this formulation to the NLL order~\cite{Frixione:2019lga,Bertone:2019hks}, 
which is relevant for the precision targets of FCC-ee~\cite{Bertone:2022ktl}. 
Such a formalism also offers a promising avenue for an alternative and independent simulation of QED radiation in parton showers.

\subsection{Organisation and support of future activities to improve theoretical precision}

As discussed in Section~\ref{sec:theocalc}, the important targets for FCC precision physics have been defined. 
Theoretical progress is limited by several factors, 
ranging from the analytic control and classification of the integral structures due to appear at higher-loop order 
to the numerical challenges of integration or event generation. 
A rather diverse expertise is needed to address all the different aspects of the problem. 
Most calculations to improve the precision of EW and QCD observables at FCC-ee are therefore currently ongoing as part of efforts by individual groups, 
historically engaged in this activity. 
The physics groups of the FCC Feasibility Study are overseeing the coordination of the different efforts, via their regular meetings and dedicated workshops. 
The workshops are an opportunity to discuss and share information on the progress of the various groups, 
as well as to refine the definition of the precision targets and milestones. 
Recent examples of these workshops include 
`Precision EW and QCD Calculations for the FCC Studies: methods and tools'~\cite{Blondel:2018mad}, 
`Precision calculations for future $\epem$ colliders: targets and tools'~\cite{EWPrecisionWshop},
`Parton Showers for future $\epem$ colliders'~\cite{QCDPrecisionWshop}, and, in 2024,
`Frontiers in precision phenomenology: resummation, amplitudes, and subtraction'~\cite{QCDFrontiersWshop}.

The Future Colliders Unit at CERN has made resources available to host researchers engaged in FCC studies and, in particular, precision studies, 
either as long-term Scientific Associates (SASS) or as short-term visitors (STV). 
These visitors interact with the current CERN Theory Group staff and fellows, 
providing a constant presence of world experts on EW and QCD precision calculations. 

Activities related to HL-LHC and its precision goals are intimately connected with FCC studies: 
the LHC precision needs are pushing both the fields of QCD, EW, and mixed QCD$\,\oplus\,$EW calculations. 
On the QCD side, higher-order perturbative calculations develop technology that can be adapted to EW higher-loop cases, 
in  addition to pushing the precision of QCD predictions for FCC-ee at the \PZ peak. 
A more profound understanding of non-perturbative effects of interest to the LHC 
(e.g., the study of leading and subleading power corrections) 
directly impacts the leading theoretical uncertainties ultimately affecting the precision of, for example, the extraction of $\as$ at FCC-ee. 
The sensitivity of LHC measurements to EW effects, furthermore, is pushing the development of higher-order EW calculations, 
with direct benefit for the pure EW programme of FCC-ee. 
As a consequence, all the efforts of CERN and of the wider community, dedicated to precision physics at the LHC must be seen as 
milestones towards the achievement of the FCC-ee precision goals. 

In this context, a new initiative is being undertaken by the LHC Physics Centre at CERN (LPCC), 
with the creation of a global effort dedicated to coordinate and support studies in the domain of event generators and multi-loop calculations, 
also building on the success of the former MCnet network~\cite{mcnet}. 
This activity, in addition to continuing the coordinated Monte Carlo development work established by MCnet 
--- opening it to MC codes not covered by MCnet so far and preserving the student mentoring and educational activities --- 
will support research to improve the projected computing performance of event generators and of multi-loop codes, 
exploring the opportunities offered by new hardware high-performance computing architectures (e.g., GPUs). 
The numerical complexity and CPU cost are recognised as a limiting factor in implementing the highest-available theoretical precision in event generators 
that can be used for realistic studies of the experimental data. 
These improvements are, therefore, a critical step towards the realistic study of FCC-ee detector performance in the context of precision observables. 

To summarise: the theoretical field of precision studies for FCC-ee is multifaceted, 
touching on diverse challenges in both the formal and numerical fields and thus calling on diverse expertise. 
The structures that have been put in place during the FCC feasibility study allow these developments to be addressed and coordinated, 
engaging the most qualified and experienced researchers in the community. 
These steps towards the goal of matching theoretical and experimental uncertainties are important and promising, 
in order to optimally exploit the data from FCC-ee and, later, from FCC-hh. While it will be years before the necessary precision goals are attained, the efforts initiated during the period of the FCC feasibility study have began defining a clear roadmap. The approval of the FCC project will catalyse the engagement of the theory community. The past and ongoing successes in enhancing the precision of LHC predictions, beyond any previous expectation, give confidence in the feasibility of attaining the FCC theory precision targets.

\cleardoublepage
\section{Detector requirements}
\label{sec:requirements}
\subsection{Introduction}
\label{sec:PhysPerf_Introduction}

The detector performance specifications required by the physics programme of a future electron-positron Higgs factory 
operating at the $\PZ\PH$ production threshold and above, have been studied in the past for linear colliders. 
The different FCC-ee experimental environment, on the one hand, 
and the rich FCC-ee electroweak, QCD, and flavour physics programme offered by the very large event samples anticipated at the \PZ resonance 
(the so-called `Tera-\PZ' run), on the other, come with specific and entirely new challenges. 
The statistical uncertainties expected on key electroweak measurements,
both at the \PZ peak and at the $\PW\PW$ threshold, 
call for a superb control of the systematic uncertainties, 
and put commensurate demands on the acceptance, construction quality and stability of the detectors.
The specific discovery potential of feebly coupled particles in the huge FCC-ee event samples 
should also be kept in mind when designing the detectors. 

General considerations on the requirements for an FCC-ee detector were outlined in Refs.~\cite{Azzi:2021ylt, Blondel:2021ema}. 
In this section, a few benchmark analyses~\cite{SnowmassBenchmarkProcesses} are selected and their sensitivity dependence on various aspects of the detector performance are studied. 
In some cases, these studies result in a quantified performance requirement. 
It is understood that the extremely broad range of FCC-ee measurements has not yet been completely covered, and that some important requirements may be missing. 
The analyses performed for this report make use of the simulation of the performance of the detector concepts considered so far 
and described in Sections~\ref{sec:CLDDetCon} (CLD), \ref{sec:IDEADetCon} (IDEA), and~\ref{sec:ALLEGRODetCon} (ALLEGRO). 
For the reader's convenience, a brief account of these three concepts is given below, in Section~\ref{sec:PhysPerf_DetectorConcepts}; 
they are an important input to further optimisations and to the development of new concepts 
that might be better adapted to parts of the FCC-ee physics programme. 
As mentioned in Section~\ref{sec:overview}, FCC-ee will accommodate four interaction points. 
A single experiment may not be perfect in all aspects; 
a configuration with 4~IPs allows a range of detector solutions that will cover all physics opportunities.

Most of the studies reported here have been performed using computer-generated events processed through 
a fast detector simulation performed with the \textsc{Delphes} package. 
The chosen baseline is the IDEA detector~\cite{fcc-phys-cdr} described in Section~\ref{sec:IDEADetCon}. 
More details about the \textsc{Delphes} simulations used for the studies reported here can be found in Section~\ref{sec:delphes}.
To investigate the impact of the detector performance, variations around this detector baseline have been studied, 
either by smearing the reconstructed objects at the analysis level or by producing dedicated event samples. 
For some tracker-related studies, the performance expected with the CLD detector~\cite{Bacchetta:2019fmz} has also been looked at. 
The CLD main tracker consists of a set of layers of silicon sensors, in sharp contrast with the gaseous drift chamber of the IDEA detector. 
One study using a \textsc{Geant} simulation of the ALLEGRO noble-liquid calorimeter has also been made. 
A few preliminary performance studies or physics analyses have been performed using a full simulation of the CLD detector, 
as shown in Section~\ref{sec:PhysPerf_FullSim}.
After a brief overview of the detector benchmarks implemented in the simulations used for the studies reported here,
the following sections address, in turn, the requirements on the various subdetectors.

\subsection{Brief overview of the current detector concepts}
\label{sec:PhysPerf_DetectorConcepts}

Two detector concepts have been studied for FCC-ee at the time of the CDR: 
CLD, a consolidated option based on the detector design developed for CLIC, 
with a silicon tracker and a 3D-imaging highly-granular calorimeter; 
and IDEA, an innovative design developed specifically for FCC-ee, 
with a short-drift wire chamber and a dual-readout calorimeter. 
Since then, a third concept, ALLEGRO, has been proposed and is being developed, 
with an electromagnetic calorimeter based on a noble liquid. 
The three concepts considered so far have similar overall dimensions, 
with a length of 11--13\,m and a height of 10--12\,m. 
More details on these detector concepts are given in Section~\ref{sec:concepts}.

\subsubsection{The IDEA detector concept}

The tracking system of IDEA\footnote{Innovative Detector for an Electron-positron Accelerator.}
includes a silicon vertex detector (VXD) surrounded by a low mass drift chamber and an external layer of silicon micro-strip detectors. 
In the version used for the studies reported here, the VXD consists of 
a cylindrical `barrel' section made of five single layers of sensors, at radii between $R = 1.2$ and 31.5\,cm, 
and of two `endcap' sections (one on each side of the interaction point), each made of three disks of single layers of sensors. 
The drift chamber (DC) has a total length of 4\,m and consists of 112 layers of wires, placed at radii between 35 and 200\,cm. 
It allows `continuous tracking' and particle identification, as will be shown below. 
The total amount of DC material crossed by a particle emitted at 90$^{\circ}$ from the detector axis 
(defined by the bisector of the two beam axes) is 1.6\% of a radiation length. 
The DC is surrounded by a silicon wrapper that provides one precise spatial point at the end of the particle trajectory 
and measures the time-of-arrival of charged particles. 
The tracking system is immersed in a 2\,T magnetic field parallel to the detector axis, 
provided by a thin superconducting solenoid covering radii between 2.1 and 2.4\,m. 
Outside the tracker, a preshower made of $\mu$RWELL detectors and a dual-readout calorimeter of 2\,m depth and 7 interaction lengths, 
made of lead and fibres, measure the position and energy of electromagnetic and hadron showers. 
The dual-readout calorimeter (DRC) is sensitive to the independent signals from scintillation and Cerenkov light production, 
resulting in an excellent energy resolution for both electromagnetic and hadron showers.  
The simulations used here include a modification of the CDR calorimeter design, 
in which a 20\,cm deep (dual readout) electromagnetic calorimeter made of crystals is added upstream of the DRC, following Ref.~\cite{Lucchini:2020bac}. 
Finally, the muon system consists of layers of chambers based on $\mu$RWELL detectors embedded in the magnet return yoke. 

\subsubsection{The CLD detector concept}

The CLD\footnote{CLIC-Like Detector.}
detector has been adapted to the FCC-ee specificities from the CLIC detector model; 
it features a silicon pixel vertex detector and a silicon tracker, followed by a highly granular calorimeter. 
For the simulations used here, the barrel section of the VXD consists of three double-layers of sensors, at radii ranging between 1.2 and 5.8\,cm, 
and two endcap sections, consisting of three disks each, built as double-layer devices. 
The silicon tracker is made of six barrel layers, at radii from 12.7\,cm to 2.1\,m, and of two sets of eleven endcap disks. 
The material budget for the tracker modules is estimated to be 1.1--2.1\% of a radiation length per layer. 
The tracker is surrounded by a silicon-tungsten electromagnetic calorimeter (ECAL), which is 30\,cm or 22\,$X_0$ deep, 
and by a scintillator-steel hadron calorimeter (HCAL), that extends up to $R = 3.6$\,m; 
the combined ECAL plus HCAL thickness corresponds to 6.5 interaction lengths. 
The very high granularity of the calorimeter makes it particularly well suited for `particle flow' reconstruction techniques that, 
in CLD, are the key for reaching very good resolutions. 
A superconducting solenoid, delivering a 2\,T magnetic field, surrounds the calorimeter, at a radius of about 3.7\,m.  
The iron return yoke is instrumented with resistive plate chambers (RPC) that form the muon detection system. 
A detailed description of the CLD detector can be found in Ref.~\cite{Bacchetta:2019fmz}.

\subsubsection{The ALLEGRO detector concept}

The design of ALLEGRO\footnote{A Lepton-Lepton collider Experiment with Granular Read-Out.} 
is more recent than the IDEA and CLD designs. 
Its tracking system consists of a silicon vertex detector and a main tracker, which could either also be made of silicon sensors or be a gaseous detector. 
Surrounding the tracker is a high granularity noble-liquid ECAL, 
which could be made of lead (or tungsten) and liquid argon or, alternatively, of tungsten and liquid krypton. 
The magnet coil separates the ECAL from the high-granularity HCAL, 
which could be made of steel absorber plates interleaved with scintillator tiles, as envisaged for the CLD calorimeter. 
For the muon system, several options are under consideration.

\subsection{Measurement of the tracks of charged particles}
\label{sec:ch_part_tracking}

The detection of charged particles and the measurement of their momenta before they traverse a large amount of high density material 
is crucial for all physics analyses and for a successful particle-flow reconstruction~\cite{MLPF_NOTE}. 
This section deals with properties that are mostly related to the main tracker: 
the precise measurement of the track momentum and angles, and the reconstruction of highly-displaced vertices. 
The determination of the track impact parameters, which is largely provided by the vertex detector, is treated in Section~\ref{sec:PhysPerf:VXD}. 
The measurement of quantities related to the ionisation energy per unit length, useful to identify the nature of a charged particle, 
is covered in Section~\ref{sec:PhysPerf_PID}.

\subsubsection{Track momentum resolution: the measurement of the Higgs boson mass}
\label{sec:PhysPerf_HiggsRecoil}

The track momentum resolution is a key handle in many analyses. 
In particular, it directly impacts the reconstructed masses, which are often used to suppress backgrounds. 
Both IDEA and CLD tracker designs would offer an excellent track momentum resolution. 
For example, for 10\,GeV (50\,GeV) muons emitted at an angle of 90$^{\circ}$ with respect to the detector axis, 
the momentum resolution is about 0.5$\permil$ (1.5$\permil$) with the very light drift chamber of IDEA 
and about 2.5$\permil$ (3$\permil$) with the heavier full silicon tracker of CLD, 
the latter being dominated by the effect of multiple scattering. 
Some examples from B~physics, showing how the exceptional momentum resolution of the IDEA tracker separates the signals from the backgrounds, 
are presented in Sections~\ref{sec:PhysPerf_Bs2DsK} and~\ref{sec:PhysPerf_muons}.

In this section, the needs on the track momentum resolution are illustrated with the measurement of the Higgs boson mass, $m_{\PH}$,
which needs to be known to better than 4\,MeV (its intrinsic width), 
in view of a potential run at the Higgs resonance (Section~\ref{sec:eYukawa}).

\begin{figure}[ht]
\centering
\includegraphics[width=0.48\textwidth]{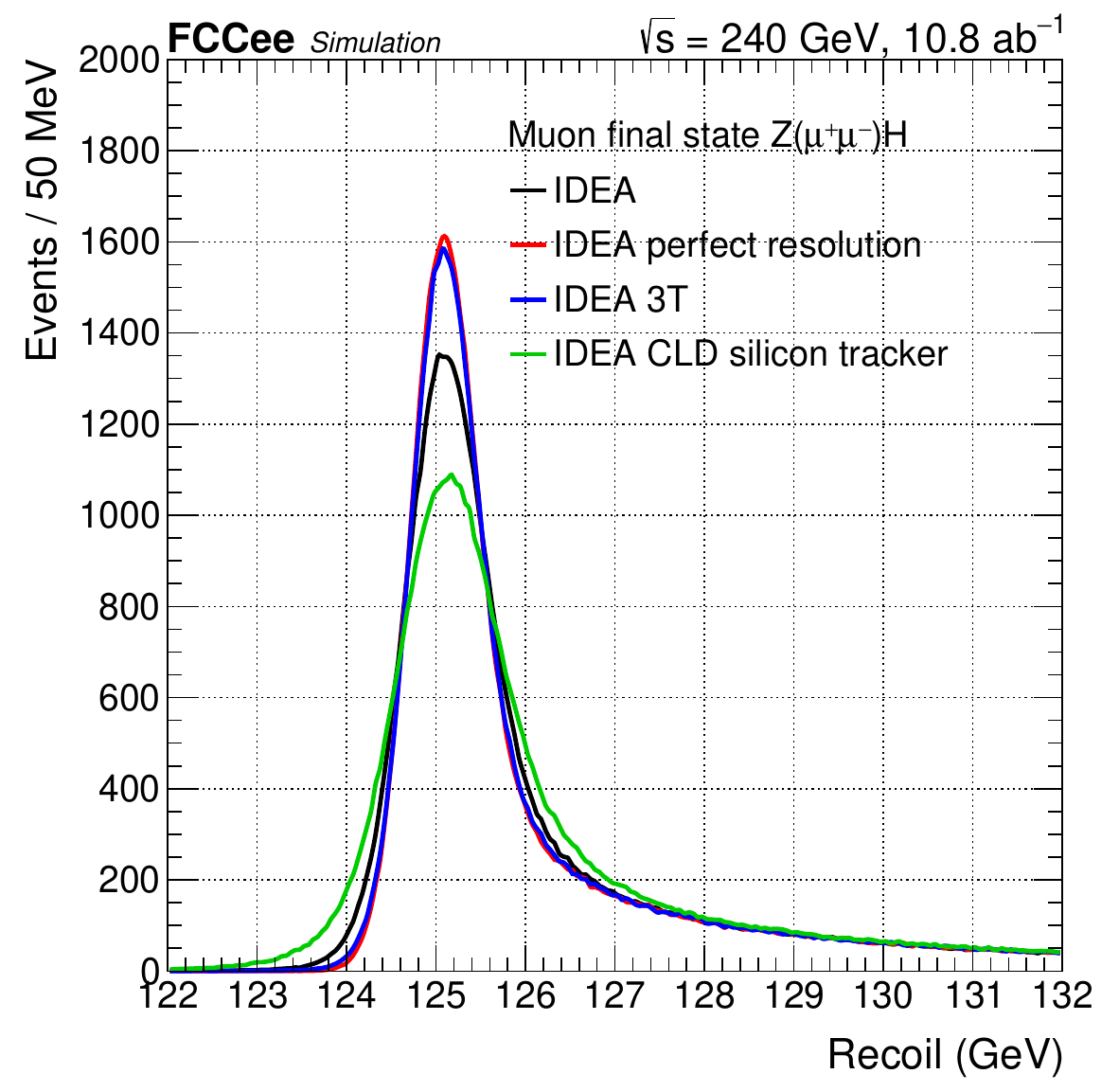}
\includegraphics[width=0.48\textwidth]{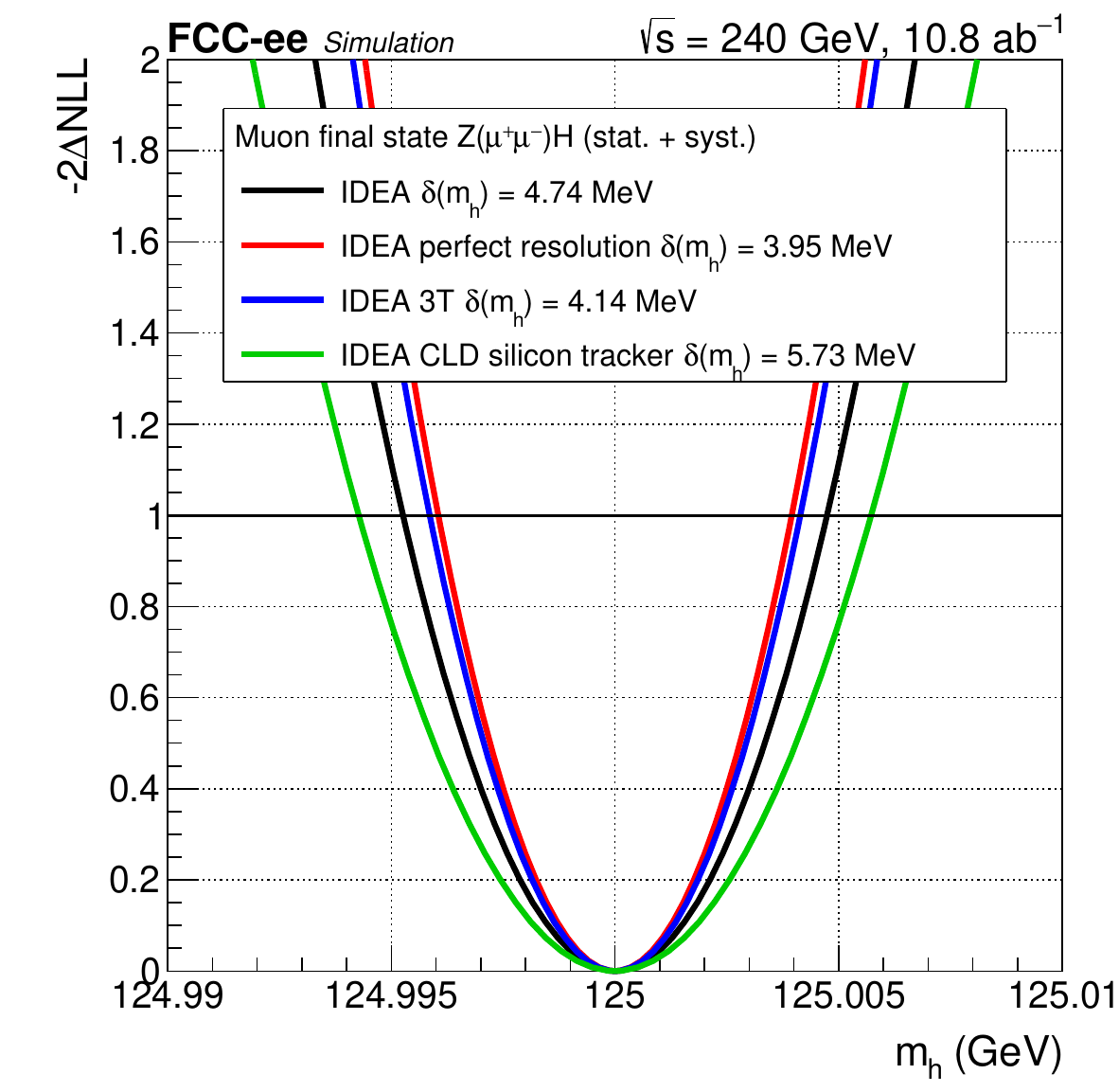}
\caption{Left: Fits to the distribution of the Higgs boson recoil mass in $\PZ\PH$ events, where the \PZ decays into muons, 
assuming an ideal momentum resolution (red), such that the resolution on the recoil mass is determined by the beam energy spread, 
or the momentum resolution of the IDEA (black and blue) or CLD (green) detectors. 
The blue curve corresponds to a 3\,T magnetic field, instead of the 2\,T field used for the other distributions. 
Right: Corresponding Higgs boson mass measurements.}
\label{fig:mH-recoil}
\end{figure}

This measurement, described in detail in Ref.~\cite{HiggsNoteRecoil}, exploits the Higgs-strahlung process, 
i.e., the associated production of a \PZ and a Higgs boson. 
Since, at a lepton collider, the total energy and momentum of the final state are well known, the mass of the system that recoils against the \PZ, 
called recoil mass and denoted $m_\text{recoil}$, 
can be reconstructed exclusively considering the \PZ decay products, irrespective of the Higgs boson decay. 
The $m_\text{recoil}$ distribution exhibits a sharp peak at $m_{\PH}$,
so that a fit to this distribution provides a precise measurement of the Higgs boson mass.  
Experimentally, the channel where the \PZ decays into a pair of muons offers the best resolution. 
A fit to the $m_\text{recoil}$ distribution for $\PZ(\PGm\PGm) \PH$ events at $\sqrt{s} = 240$\,GeV 
is displayed in the left panel of Fig.~\ref{fig:mH-recoil}, for various assumptions on the muon momentum resolution, 
while the right panel shows the result of a likelihood fit of the distribution of signal events in the presence of background, with the same assumptions. 
Even with a perfect measurement of the momentum of the muons, 
the width of the $m_\text{recoil}$ peak would be limited by the beam energy spread (BES), 
which amounts\footnote{The beam energy spread values used for the studies reported here are taken from Ref.~\cite{talk_IS_workshop_december}.} 
to 0.185\% of the beam energy at $\sqrt{s} = 240$\,GeV, 
so that $m_{\PH}$ would be determined with an uncertainty of 3.95\,MeV 
(including the contribution of other sub-dominant systematic uncertainties~\cite{HiggsNoteRecoil}), 
for an integrated luminosity of 10.8\,ab$^{-1}$.

The decay muons, with typical momentum of ${\cal{O}}(50)$\,GeV, should be measured with a momentum resolution better than the BES,
so that the mass resolution be not limited by the momentum 
measurement\footnote{Having a momentum resolution for ${\cal{O}}(50)$\,GeV muons better than or comparable to the BES is also an important requirement in the search for $\PZ \to \PGt\PGm$ lepton flavour violating decays, at $\sqrt{s} = 91$\,GeV. The analysis strategy requires a clear tau decay in one hemisphere and a beam-energy muon in the other, 
in order to suppress the $\PZ \to \PGt\PGt$ background~\cite{Dam:2018rfz}.}. 
This goal is achieved with the IDEA detector (black curves in Fig.~\ref{fig:mH-recoil}), only considering the muon decay channel. 
The full silicon tracker of CLD, in its current implementation, performs less well (green curves) because of the larger amount of material, 
which leads to increased multiple scattering, the factor that dominates the muon momentum resolution, even for the relatively high momentum range of interest. 
However, the goal of 4\,MeV on the measured Higgs boson mass would most likely be achieved by combining the muon and electron decay channels 
(Section~\ref{sec:PhysPerf_ECAL}). 
The blue curves in Fig.~\ref{fig:mH-recoil} show what would be expected, with the IDEA tracker, 
if the detector magnetic field were increased 
from 2 to 3\,T\footnote{For the highest luminosity runs at the \PZ peak, 
the magnetic field of the detector is constrained to not exceed about 2\,T, as explained in Section~\ref{sec:mdi}. 
At $\sqrt{s} = 240$\,GeV, the blow-up of the beam emittance that results from a higher magnetic field is negligible 
and running, for example, with a 3\,T field, which would provide a better track momentum resolution, is not ruled out.}; 
the result is a 33\% improvement of the momentum resolution and a 14\% improvement on the total mass uncertainty.

\subsubsection{Track momentum resolution: the \texorpdfstring{\PZ}{Z} width and the stability of the momentum scale}

\begin{figure}[t]
\centering
\includegraphics[width=0.4\textwidth]{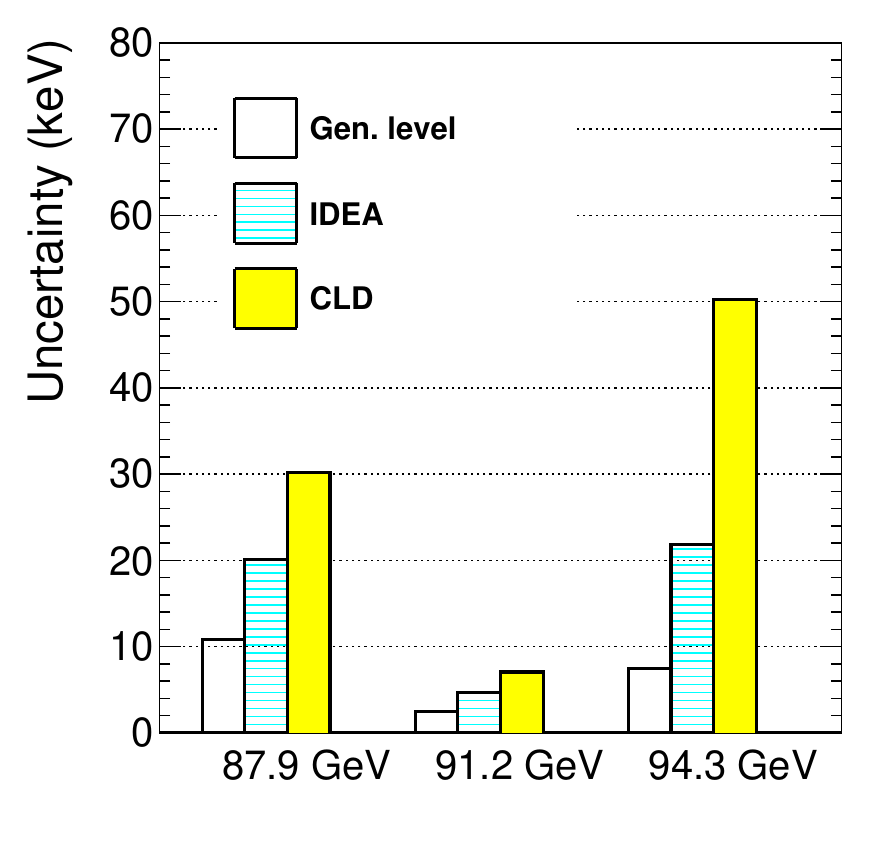} 
\caption{Statistical uncertainty in the peak position of the measured dimuon invariant mass distribution, 
expected with the full FCC-ee event sample of the \PZ lineshape scan 
(125\,ab$^{-1}$ at the \PZ peak and 40\,ab$^{-1}$ at each off-peak energy). 
This uncertainty is shown assuming an ideal detector resolution (leftmost bars), 
the resolution of the IDEA tracker (middle bars), 
and that of the CLD tracker (rightmost bars).}
\label{fig:ptp-sqrt-uncertainty}
\end{figure}

The point-to-point uncertainty on the centre-of-mass energy in the scan of the \PZ lineshape is, 
together with the knowledge of the BES and the point-to-point uncertainty on the integrated luminosity, 
a dominant source of systematic uncertainty in the \PZ width measurement. 
The reconstructed peak position of the dimuon invariant mass distribution in $\epem \to \PGmp\PGmm$ events provides a measurement of the centre-of-mass energy. 
As proposed in Ref.~\cite{Blondel:2019jmp} and developed in Ref.~\cite{perez_2024_gyqhp-m0480}, 
the aforementioned uncertainty can be assessed from the difference in this reconstructed peak position between the energy points used in the lineshape scan. 
More details about the method are given in Section~\ref{sec:EPOL_input_from_the_experiments}. 
Figure~\ref{fig:ptp-sqrt-uncertainty} shows the statistical uncertainty with which the peak position is expected to be determined 
at $\sqrt{s} = 87.9$, 91.2, and 94.3\,GeV, with the full FCC-ee event sample. 
With the track momentum resolution provided by the IDEA tracker, this precision reaches 20\,keV for the two off-peak energies, 
such that the difference in the collision energy at these two points can be measured with an uncertainty of about 28\,keV, 
leading to an 11\,keV uncertainty on the extracted \PZ width. 
With the resolution offered by the CLD tracker, this difference is measured with two times larger uncertainty, 
resulting in an uncertainty of 22\,keV in the \PZ width.

Such a precision, however, requires that the scale of the momentum measurements 
(and, in particular, of the magnetic field) 
be stable at the level of a few $10^{-7}$ (20\,keV over $\sqrt{s}$, to ensure a 20\,keV uncertainty on the peak position) 
or, at least, that its variations be monitored at that level. 
A monitoring at this level of precision may be difficult to achieve with magnetic NMR probes. 
However, the large samples of well-known resonances, 
in particular of \PKS decaying into $\PGpp\PGpm$, provide an in-situ monitoring at this exceptional precision. 
An efficient and high purity algorithm that reconstructs $\PKS \to \PGpp\PGpm$ decays is presented below. 
It allows a \PKS to be reconstructed in about every second \PZ event, when the \PZ decays hadronically. 
With a mass resolution better than 400\,keV with the IDEA detector, 
the position of the \PKS mass peak can be determined with a relative uncertainty of $2.3 \times 10^{-9}$ 
with an integrated luminosity of 40\,ab$^{-1}$ at $\sqrt{s} = 87.9$\,GeV 
(and even better at $\sqrt{s} = 94.3$\,GeV, given the larger cross section). 
This dataset could, hence, be split into, e.g., 100 subsamples in time, 
and the \PKS resonance be reconstructed in 100 bins, 
to ensure monitoring of the scale stability at the required level of $2.3 \times 10^{-7}$.

\subsubsection{Angular resolutions}
\label{sec:PhysPerf_TrackAngles}

The polar and azimuthal angles of the momentum of a prompt charged particle, at its production vertex, 
are given by the angles of the corresponding track at its distance of closest approach to the detector axis.
The angular resolutions in the two detector concepts considered so far vary between 
about 20\,$\mu$rad for high momentum particles produced in the central region of the detector and a few mrad for soft forward particles~\cite{Bacchetta:2019fmz, softwareNote}. 
The contribution of these angular resolutions to the uncertainty in the recoil mass shown above is negligible compared to that of the momentum resolution. 
For the reconstruction of the mass of heavy-flavoured hadrons, 
the momenta of the daughter particles must be taken at the decay vertex of the hadron 
and the resolution of the azimuthal angle crucially relies on the reconstruction of this decay vertex (see Section~\ref{sec:PhysPerf:VXD}). 
Nevertheless, for all examples considered involving \PB mesons 
(see, e.g.,\ Sections~\ref{sec:PhysPerf_Bs2DsK} and~\ref{sec:PhysPerf_muons}), 
the mass resolution remains completely dominated by the momentum resolution.

A precise determination of the polar and azimuthal angles is crucial for exploiting the over-constrained kinematics of many processes in $\epem$ collisions. 
That is the case, in particular, of muon pair production.
As shown in Ref.~\cite{Blondel:2019jmp}, for $\epem \to \PGmp\PGmm (\PGg)$ dimuon events, 
the crossing angle and the longitudinal momentum imbalance can be reconstructed event-by-event from the sole measurement of the polar and azimuthal angles of both muons. 
The BES, which is a crucial ingredient in the extraction of the \PZ width at Tera-\PZ, 
can then be derived from the width of the longitudinal momentum imbalance distribution. 
To ensure that the BES uncertainty has a negligible effect on the extracted \PZ width, 
muon tracks from \PZ decays must be measured with an  angular resolution of 0.1\,mrad or better, 
a requirement fulfilled~\cite{Bacchetta:2019fmz, softwareNote} by the detector concepts presented in the CDR. 
Moreover, the mean of the longitudinal momentum imbalance distribution provides the longitudinal boost. 
Its determination constrains the energy losses of the beams in their separate rings~\cite{note_EPOL} and, consequently, 
has a crucial influence on the precision of the \PZ and \PW mass measurements.

Another example that exploits the angles of dimuon events is the determination of the centre-of-mass energy well above the $\PW\PW$ threshold, 
where the resonant depolarisation method cannot be applied. 
In particular, at $\sqrt{s} = 240$\,GeV, the centre-of-mass energy needs to be determined to ${\cal{O}}$(1)\,MeV 
in order not to spoil the Higgs boson mass determination from the recoil mass. 
The over-constrained kinematics of radiative return events $\epem \to \PZ (\PGg)$ with $\PZ \to \PGmp\PGmm$ 
(accompanied by an initial-state radiated photon along the direction of one of the beams) 
allows the determination of $\sqrt{s}$ from the precise knowledge of the $\PZ$ mass and the muon angles alone. 
The target precision for the $\sqrt{s}$ measurement may set tighter requirements on the angular resolutions than those given above.

\subsubsection{Number of tracker layers and highly-displaced vertices}
\label{sec:PhysPerf_Kshorts}

The reconstruction of long-lived particles (LLPs) that decay into charged particles within the tracker volume 
after having crossed the innermost layers of the vertex detector requires that the main tracker be able to efficiently identify such highly-displaced vertices. 
Typical use cases are the searches for LLPs predicted in many models that extend the Standard Model 
(Section~\ref{sec:PhysicsCaseBSM} and Refs.~\cite{note_HNL_Polesello_Valle_mujj,note_LLPs_ExoticHiggsDecays})
or the reconstruction of decays that involve \PKS or \PGL hadrons.

As an example, the case of \PKS mesons produced in $\PBp \to \PKp \PDz$ decays, followed by the decay of the \PDz meson into a $\PKS \PGpz$ pair, is considered here. 
The reconstruction of the $\PKS \to \PGpp \PGpm$ decay is detailed in Ref.~\cite{Aleksan:2021fbx}. 
The \PKS candidates are built from pairs of opposite-charge particle tracks that do not come from the primary vertex and that can be fitted to a common vertex, 
with a reconstructed mass close to the nominal \PKS mass. 

\begin{figure}[ht]
\centering
\includegraphics[width=0.48\textwidth]{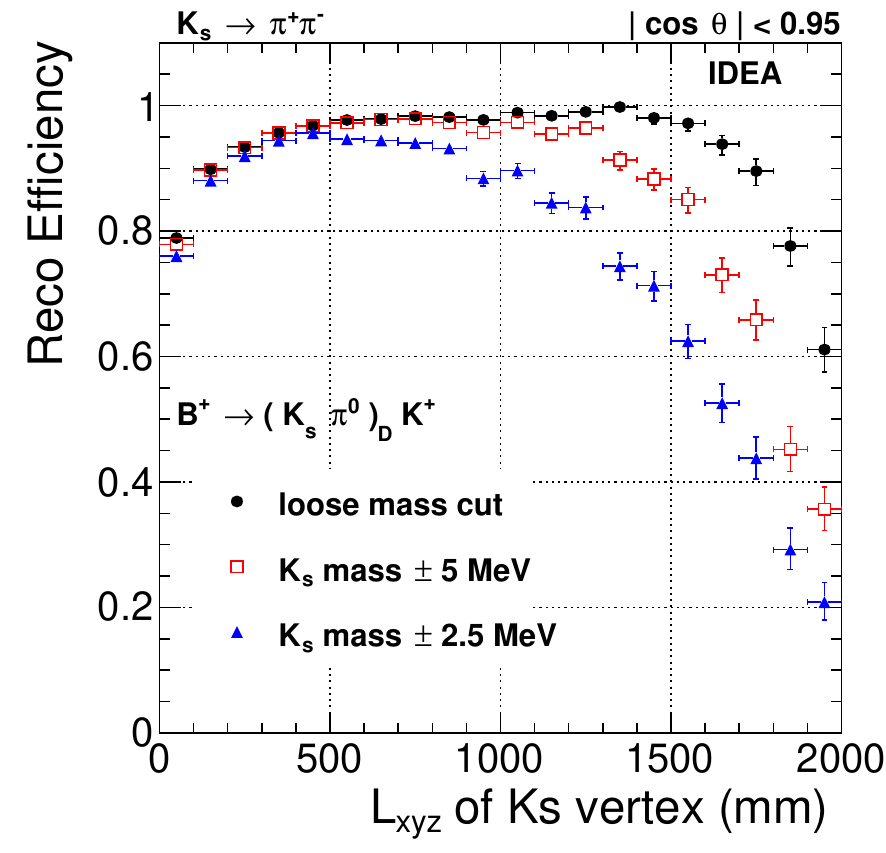}
\includegraphics[width=0.48\textwidth]{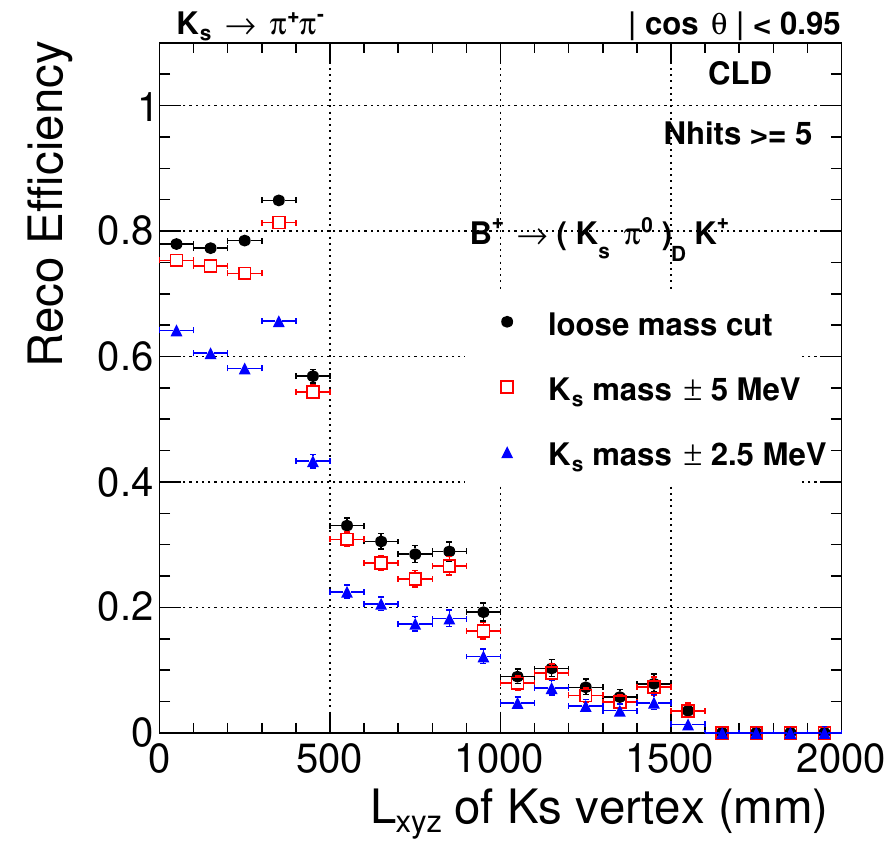}
\caption{The \PKS reconstruction efficiency for kaons emitted in $\PBp \to (\PKS \PGpz)_{\PD} \, \PKp$ decays,
as a function of the distance $L_{xyz}$ between the \PKS decay vertex and the interaction point, 
for several thresholds on the \PKS mass candidate, for the IDEA (left) and CLD (right) detector concepts.}
\label{fig:Ksreco_efficiency}
\end{figure}

The \PKS reconstruction efficiency obtained with the IDEA detector is shown in Fig.~\ref{fig:Ksreco_efficiency}~(left), 
as a function of the distance between the \PKS decay vertex and the interaction point, $L_{xyz}$. 
The efficiency is defined for the \PKS mesons that come from $\PBp \to (\PKS \PGpz)_{\PD} \, \PKp$ decays 
for which the daughter pions satisfy the acceptance requirement $|\cos \theta| < 0.95$. 
Within a mass window of $\pm 5$\,MeV, the efficiency remains higher than 80\%, as long as the \PKS meson decays within 1.5\,m from the interaction 
point\footnote{The drop in efficiency observed at small flight distances is due to the correlation of the flight distance with the \PKS momentum: 
the pion tracks from \PKS mesons that decay close to the IP are softer and more affected by multiple scattering or may lead to curling tracks (`loopers').}.
At larger distances, the efficiency drops because the pion tracks only traverse a small fraction of the tracker. 
The right panel of Fig.~\ref{fig:Ksreco_efficiency} shows the corresponding efficiency for events processed through a \textsc{Delphes} simulation of the CLD detector. 
As expected, the performance is much worse than that of IDEA. 
Steps corresponding to the positions of the tracker layers are clearly visible. 
For example, since there are only four barrel layers at a radial distance larger than 40\,cm, 
the efficiency for selecting 5-hit tracks displaced by more than 40\,cm vanishes in the central region, explaining the first step seen in the figure. 
Also, the \PKS mesons that decay within a few tens of centimetres from the IP have a lower detection efficiency than that of the IDEA drift chamber, 
given the larger amount of material of the full silicon tracker and the correspondingly larger multiple scattering and worse resolutions on the \PKS reconstructed vertex and mass. 

While some optimisations could be made to the full silicon tracker considered here, 
it is clear that an efficient reconstruction of $\PKS \to \PGpp \PGpm$ decays and, more generally, 
of long-lived particles that lead to late appearing tracks 
calls for a highly transparent tracker, a large tracking volume, and a considerable number of measurement layers. 

\subsubsection{Work ahead}

\begin{figure}[ht]
\centering
\includegraphics[width=0.465\textwidth]{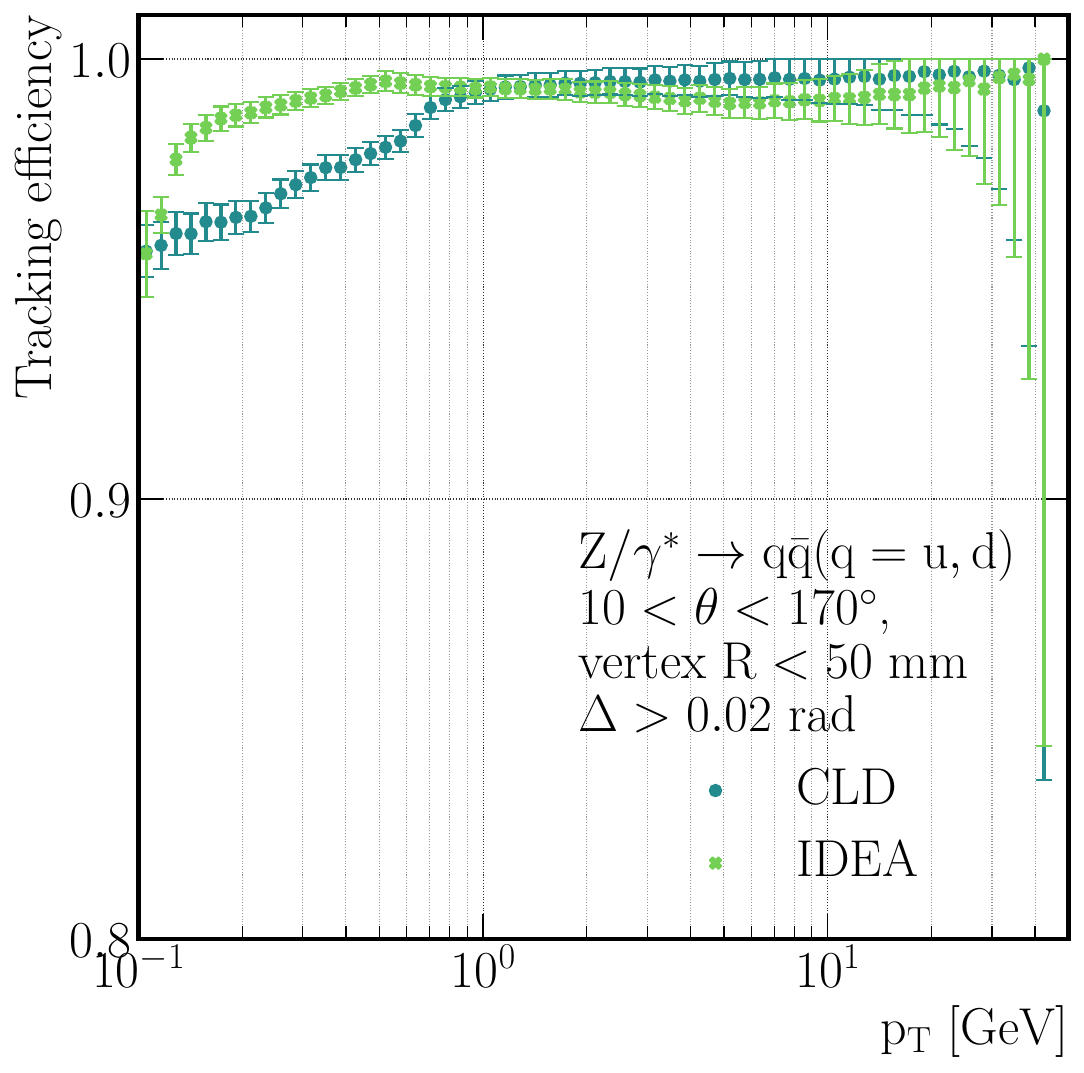} 
\includegraphics[width=0.475\textwidth]{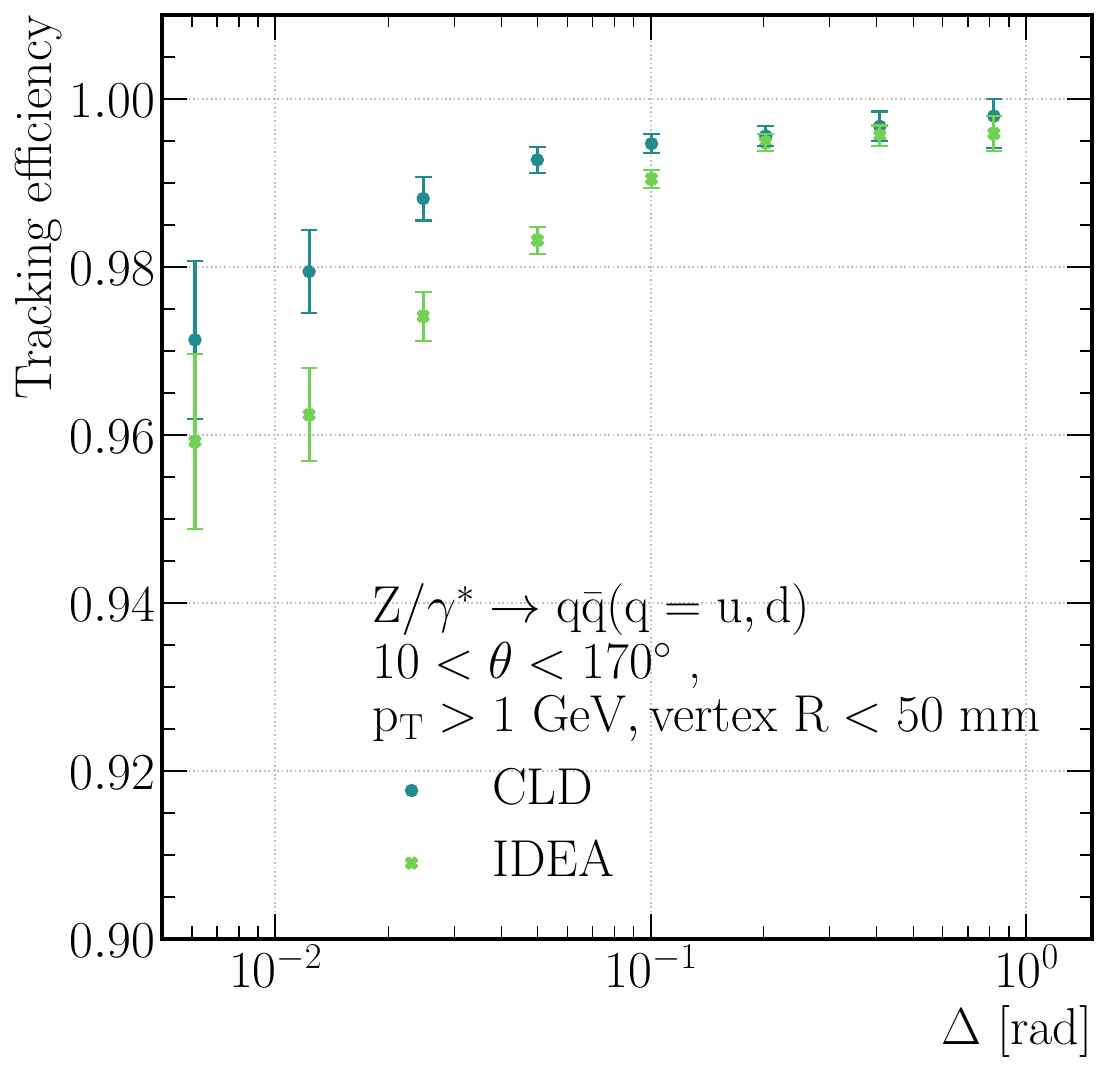}
\caption{Tracking efficiency evaluated with full simulation of the CLD and IDEA detectors using the ML-based approaches, 
as a function of \pt (left) and of the angle $\Delta_\text{MC}$ to the closest generated charged particle (right).}
\label{fig:mltracking}
\end{figure}

\paragraph{Tracker acceptance and efficiencies}
The search for the $\PB \to \PK^{*}\, \PGt \PGt$ rare decay, described in more detail in Section~\ref{sec:DetReq_vertex}, sets requirements on the track efficiencies. 
Demanding that both taus decay into three prongs is particularly useful for this analysis and leads to a final state with six soft pion tracks. 
A momentum acceptance down to 100--150\,MeV is necessary to maintain a good signal efficiency. 
With a magnetic field of 2\,T, the requirement of a minimum number of, e.g., 5~hits to reconstruct a track,
implies that there be at least five measurement layers at a radius smaller than 33\,cm (50\,cm) to reconstruct a track with a transverse momentum of 100\,MeV (150\,MeV). 
This requirement on the position of the layers of the vertex detector and of the innermost layers of the main tracker 
is fulfilled by both the IDEA and the CLD tracker designs~\cite{softwareNote, Bacchetta:2019fmz}. 
As an illustration, the tracking efficiency as a function of transverse momentum is shown in Fig.~\ref{fig:mltracking}~(left), 
for tracks reconstructed with a novel machine-learning algorithm, discussed in Section~\ref{sec:PhysPerf_FullSim}. 
In addition, the track reconstruction efficiency as a function of the distance to the closest charged hadron in a jet is shown in Fig.~\ref{fig:mltracking}~(right). 
It indicates that CLD features a slightly larger efficiency in dense hadronic environments, 
possibly due to a better separation capability of two close-by tracks, driven by the superior single-point spatial resolution provided by silicon sensors.

The measurement of the luminosity from $\epem \to \PGg \PGg$ events (Section~\ref{sec:PhysPerf_ECAL}), to a precision of about $10^{-5}$, 
requires a very precise knowledge of the large background from $\epem \to \epem$ events. 
The targeted precision will set a requirement on the $\Pe / \PGg$ separation that, in turn, 
will constrain the tracker inefficiency and the precision with which this inefficiency is known. 
In addition, the ratio of the partial width of the \PZ boson decay into hadrons to that into muons, $R_{\PGm}$, 
will be measured with a statistical precision of about $5 \times 10^{-6}$ at Tera-\PZ. 
Counting the number of $\PGmp\PGmm$ events with a similar level of systematic uncertainty 
sets a very challenging requirement on the knowledge of the tracking (and muon chamber) efficiency.   

\paragraph{Track angles}
The angular resolutions provided by the current tracker designs comply, with a good safety margin, with the requirements set by the studies done so far, as shown above. 
Further studies are needed to check if other measurements set tighter requirements. 
Beside the measurement of $\sqrt{s}$ from the radiative return events previously mentioned, 
methods that are being developed to precisely determine dilepton acceptance in-situ (Section~\ref{sec:PhysPerf_gammagamma}) may call for better angular resolutions. 

\subsubsection{Preliminary conclusions}

The performance of a gaseous tracker and of a full silicon tracker have been shown and quantified with several examples. 
\begin{itemize}
\item Having a large number of measurement points along the tracks, as offered by a gaseous tracker, 
is crucial for an efficient reconstruction of \PKS and $\Lambda$ hadrons or other long-lived particles that decay into charged particles, 
and is a clear bonus for an experiment with a strong focus on flavour or BSM physics.
\item The momentum resolution offered by both designs looks adequate for Higgs boson measurements. 
This statement probably holds as well for most electroweak measurements, with the notable exception of the \PZ width measurement. 
For flavour physics at the \PZ peak, where low momentum tracks are involved, 
a low mass gaseous tracker is advantageous since the momentum resolution is minimally affected by multiple scattering.
\item The tracker volume, extending to a radius of about 2\,m, 
may have to be reduced a little in order to free some space to accommodate a detector dedicated to charged-hadron particle identification 
(which may be needed in particular for the CLD option, see Section~\ref{sec:PhysPerf_PID}). 
A reduction by ${\cal{O}}$(20)\,cm would degrade the momentum resolution by about 15\%, 
as shown in Section~\ref{sec:Detectors_MainTracking}.
This effect might be partly compensated by reducing the amount of material in the CLD tracker layers. 
\end{itemize}

\subsection{Requirements for the vertex detector}
\label{sec:PhysPerf:VXD}

The measurement of the track impact parameters is driven by the performance of the vertex detector, 
as it provides very precise spatial points in the tracker layers closest to the beams. 
A precise measurement of these parameters is crucial for reconstructing vertices, 
for efficient identification of heavy quarks and of taus, 
and for accurate lifetime measurements. 
The resolution on these parameters typically scales as
\begin{equation}
\sigma( d_0 ) = a \oplus \frac{b}{p \sin^{3/2} \theta} \, ,
\label{eq:track_impact_parameter}
\end{equation}
where the asymptotic term, $a$, is driven by the single hit resolution, 
while the second term represents the contribution of multiple scattering and depends on the material of the vertex detector layers and of the beam pipe. 
The radial distance of the first layer of the vertex detector is also crucial~\cite{Bedeschi:2022rnj}. 
In the simulations used for the studies reported here, the radius of the beam pipe is 1\,cm, 
that of the innermost layer of the vertex detector is 1.2\,cm, 
and the material crossed by a particle emitted at a polar angle $\theta$ before the first VXD measurement corresponds to $0.67 \% \, / \, \sin \theta$ of a radiation length.

\subsubsection{Heavy-flavour tagging and the Higgs boson coupling to charm quarks}
\label{sec:PhysPerf_Hcc}

Previous flavour tagging studies made for linear colliders~\cite{Behnke:2013lya,Linssen:2012hp} colliders 
concluded that $a \approx 5$\,$\mu$m and $b \approx 15$\,$\mu$m/GeV 
(in Eq.~\ref{eq:track_impact_parameter}) 
are adequate for measuring the Higgs boson couplings to bottom and charm quarks. 
These requirements have been revisited in the context of the present study. 
In particular, major progress has been made in recent years in the field of flavour tagging algorithms. 
Machine-learning approaches, on one hand, and the ever increasing computing power, on the other, 
have significantly improved the performance of flavour tagging with respect to that of the algorithms that were the state-of-the-art a decade ago, 
so that more ambitious goals can be contemplated today. 
The algorithm developed for the studies reported here is based on {\textsc{ParticleNet}}~\cite{Qu:2019gqs} 
and uses state-of-the-art jet representations and an advanced graph neural network architecture. 
This algorithm is referred to as {\textsc{ParticleNetIDEA}} and is described in more detail in Ref.~\cite{Bedeschi:2022rnj}. 
With this new tool, an efficient and pure identification of jets induced by gluons or even strange quarks (see Section~\ref{sec:PhysPerf_strangeTagging}) 
is available and the tagging efficiency of bottom and charm quarks is much larger than what was achieved with the previous generation of algorithms, 
for the same purity.

This algorithm is a key ingredient in the measurement of the Higgs boson couplings to bottom, charm, and strange quarks, as well as to gluons, 
as described in detail in Ref.~\cite{note_Higgs_to_hadrons}. 
The analysis exploits the $\PZ\PH$ events at $\sqrt{s} = 240$\,GeV and considers the channels where the \PZ boson decays 
to pairs of muons or electrons, to jets, or to neutrinos, the latter channel currently providing the best sensitivity. 
The probabilities that each jet originates from a bottom, charm, or strange quark, or from a gluon, 
are determined by the \textsc{ParticleNetIDEA} algorithm and are used to categorise events into orthogonal categories, 
each enriched in one of the different Higgs boson decay modes and depleted in the others. 

The branching ratios of the Higgs boson decays to $\PQb\PAQb$, $\PQc\PAQc$, $\PQs\PAQs$, and $\Pg\Pg$ pairs
are measured in all categories with a simultaneous fit to the recoil mass distributions 
(and, for the $\PZ \to \PGn \PAGn$ channel, to the visible mass distributions).

The anticipated precision of the coupling measurements is summarised in Section~\ref{sec:PhysicsCaseHiggsEW}. 
With an integrated luminosity of 10.8\,ab$^{-1}$, the $\PH \to \PQb\PAQb$ and $\PH \to \PQc\PAQc$ event yields 
(cross section times branching ratio) 
can be measured with a precision of 0.21\% and 1.66\%, respectively.

The dependence of this precision on alternative beam pipe and vertex detector configurations 
has been studied and compared to the sensitivity obtained with the baseline IDEA detector design. 
Beam pipes with twice or half the nominal material budget have been studied. 
A vertex detector with the barrel layers shifted by 0.5\,cm towards larger radii 
and an option where the heavier beam pipe is complemented by a single-point resolution degraded by a factor of~2 
have also been considered. 
The measurement of the $\PH \to \PQc\PAQc$ and $\PH \to \PQb\PAQb$ branching fractions is marginally improved or degraded for the assumed better or worse scenarios. 
A transverse impact parameter resolution of \mbox{2--3}\,$\mu$m is achieved with the baseline IDEA vertex detector. 
A resolution worsened by a factor of~2 remains largely sufficient to identify displaced tracks 
originating from boosted \PD and \PB meson decays produced in $\PH \to \PQb\PAQb$ and $\PQc \PAQc$ decays 
and has, therefore, a small impact on the expected precision. 
A comprehensive discussion on the impact of vertex detector assumptions on the tagging performance can be found in Ref.\cite{sciandra_2024_gt813-z8602}.

\subsubsection{Reconstruction of vertices and the measurement of the 
\texorpdfstring{$\PB \to \PK^* \PGt\PGt$}{B to K* tau tau} branching fraction}
\label{sec:DetReq_vertex}

With the current version of the IDEA detector, the primary vertex, for example in events where a \PZ decays to hadrons, 
is typically reconstructed with a resolution of 2--3\,$\mu$m in the $x$ and $z$ coordinates, 
and of a few tens of nm in the $y$ direction 
(driven by the r.m.s.\ of the beam position in the vertical direction, 
as shown in Table~\ref{tab:IR_parameters}, Section~\ref{sec:mdi}), 
using a beamspot constraint in the vertex fit. 
For displaced vertices, the resolution depends very much on the track multiplicity of the vertex, on the track momenta and angles, and on the angular separation of the tracks. 
In the various \PB-physics processes that have been looked into so far, 
the resolution on the 3D distance between the interaction point and the displaced vertex typically varies between $\sim$\,10 and $\sim$\,80$\mu$m. 
For example, in the $\PBs \to \PDs \PK$ decay, where the \PDs decays to $\PK\PK\PGp$, 
the resolution of 15\,$\mu$m obtained on the \PBs decay vertex is sufficient for time-dependent CP violation measurements, 
since it is ${\cal{O}}$(40) times smaller than the distance travelled by the \PBs during one oscillation period. 

However, some processes of interest for flavour physics impose very strong demands on the reconstruction of vertices.
So far, the most stringent requirements on the performance of the vertex detector come from the measurement of the branching ratio of the, 
yet unobserved, $\PB \to \PK^* \PGt\PGt$ decay.  
This branching fraction is predicted to be very small in the Standard Model, at the level of ${\cal{O}}(10^{-7})$, 
and the current experimental upper limit is larger than this prediction by three orders of magnitude. 
New physics contributions can potentially result in a significant enhancement of this branching fraction. 
Consequently, its measurement is an important goal of the FCC-ee physics programme. 
A detailed analysis of the feasibility of this measurement at the \PZ pole is reported in Ref.~\cite{miralles_2024_9t6w9-mmd12}, and is summarised below. 

\begin{figure}[ht]
\centering
\includegraphics[width=0.5\textwidth]{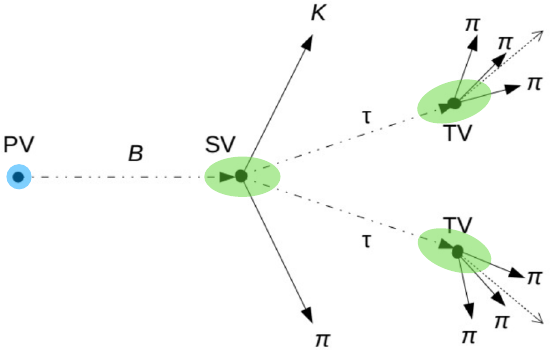} 
\caption{Decay chain under study for a measurement of the $\PB \to \PK^* \PGt\PGt$ branching fraction.}
\label{fig:KstarTauTau_decaychain}
\end{figure}
\begin{figure}[ht]
\centering
\includegraphics[width=0.67\textwidth]{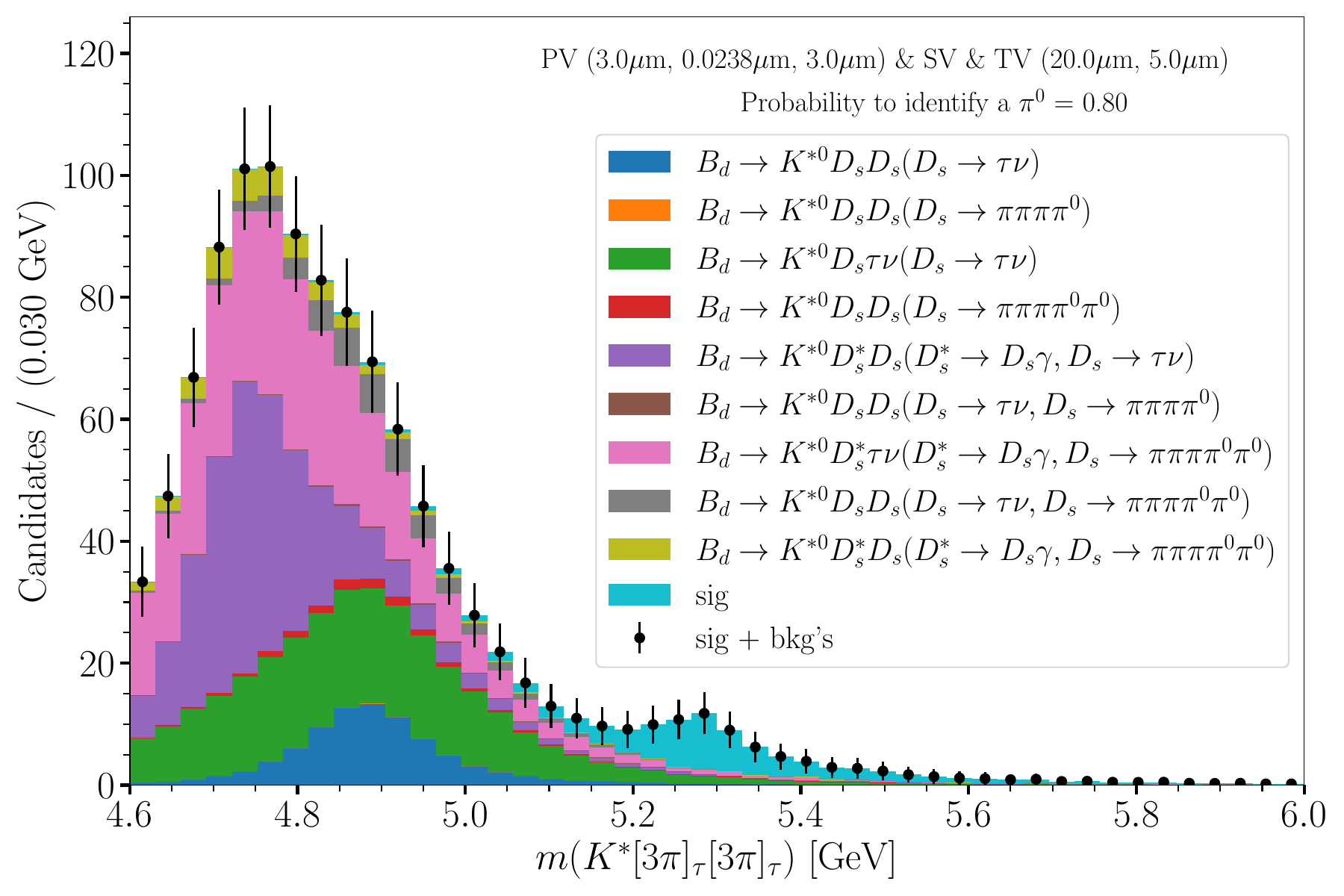}
\caption{Distribution of the mass of the \PB candidates after the full selection, 
assuming that the secondary and tertiary vertices can be reconstructed with 
a resolution of 5\,$\mu$m (20\,$\mu$m) in the transverse (longitudinal) direction. 
The normalisation corresponds to a total of $6 \times 10^{12}$ produced \PZ bosons.}
\label{fig:KstarTauTau_mass}
\end{figure}

When both \PGt leptons decay into three charged pions and a neutrino, 
there are enough kinematic constraints to completely reconstruct the decay, as illustrated in Fig.~\ref{fig:KstarTauTau_decaychain}. 
The \PGt decay vertices (TV) are obtained from their daughter pion tracks. 
The secondary vertex (SV) is reconstructed from the two tracks (the kaon and the pion) produced by the $\PK^*$ decay. 
The flight direction of each \PGt is then determined from the line that joins the SV to each corresponding TV. 
The component of the neutrino momentum that is transverse to this flight direction is obtained as the opposite 
of the transverse component of the momentum of the three charged pions. 
The \PGt mass constraint finally provides the remaining longitudinal component of the neutrino momentum.
Clearly, this reconstruction relies critically on the precise reconstruction of the secondary and tertiary vertices. 

Several $\PQb \to \PQc \PAQc \PQs$ and $\PQb \to \PQc \PGt \PGn$ transitions lead to final states that are similar to that of the signal, 
apart from the presence of additional neutrinos and neutral pions; 
a powerful \PGpz identification is needed to suppress these backgrounds. 
A \PGpz identification efficiency of 80\% is assumed here. 
A multi-variate analysis allows the background to be reduced to a manageable, albeit still large, level. 
As an example, Fig.~\ref{fig:KstarTauTau_mass} shows the distribution of the mass of the \PB candidates 
for a case where the resolution of the position of the secondary and tertiary vertices is 
20\,$\mu$m in the longitudinal direction and 5\,$\mu$m in the transverse 
direction\footnote{For each tertiary or secondary vertex, the longitudinal direction is defined by the flight direction of the decaying particle.}. 
The remaining background under the \PB mass peak is dominated by irreducible background processes.

\begin{figure}[t]
\centering
\includegraphics[width=0.7\textwidth]
{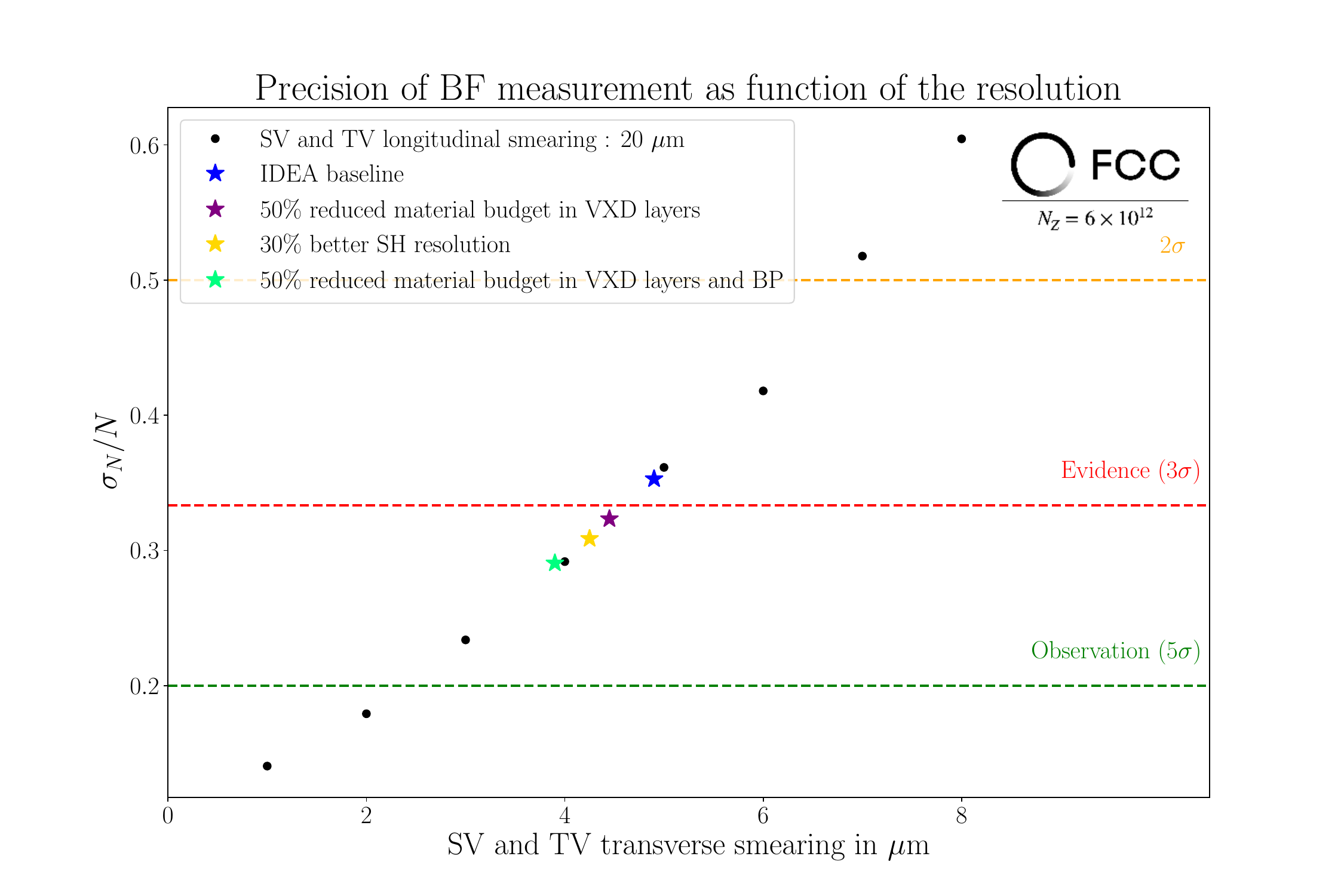}
\caption{Statistical uncertainty on the $\PB \to \PK^* \PGt \PGt$ branching fraction expected 
from an event sample corresponding to $6 \times 10^{12}$ \PZ bosons, 
as a function of the resolution of the position of the secondary and tertiary vertices, in the transverse direction. 
The blue star shows the precision expected with the version of the IDEA detector used in the nominal simulations. 
The additional star symbols show how the sensitivity improves in alternative simulations of the IDEA detector.}
\label{fig:KstarTauTau_vertexResolutions} 
\end{figure}

The numbers of signal ($N$) and background events are determined in the 5.0--6.0\,GeV range of invariant mass from a maximum likelihood fit. 
The significance of the signal in this range is used as a conservative figure of merit of the reconstruction performance. 
This significance has been determined for several assumptions on the vertex resolution. 
It shows a very strong dependence on the resolution of the position of the secondary and tertiary vertices in the transverse 
direction\footnote{The dependence on the resolution in the longitudinal direction is much milder.}, 
as can be seen from the black dots in Fig.~\ref{fig:KstarTauTau_vertexResolutions}. 
With an event sample corresponding to $6 \times 10^{12}$ produced \PZ bosons, 
a resolution better than $\sim$\,4.3\,$\mu$m is needed to see evidence (3\,$\sigma$) of the decay mode under study 
and it should be better than $\sim$\,2.2\,$\mu$m for the observation (5\,$\sigma$) of this decay. 
With the version of the IDEA detector that is used as a baseline for this report this resolution is only 5\,$\mu$m, 
as shown by the blue star symbol in Fig.~\ref{fig:KstarTauTau_vertexResolutions}. 
To reach the evidence and observation thresholds, 
the resolution on the track impact parameters needs to be improved by about 10\% and 40\%, respectively, 
with respect to the nominal IDEA 
resolutions\footnote{Restricting the invariant mass to the 5.2--5.6\,GeV range would yield a local significance of the signal of about 5\,$\sigma$, but this value is obtained assuming an accurate knowledge of the functional shapes of signal and background candidates, which requires a study that goes beyond the scope of this work.}.

In order to investigate if and how these improvements can be achieved, additional \textsc{Delphes} signal samples have been produced, 
with an alternative tracker description, with less tracker 
material\footnote{The decay of interest comprises eight charged particles plus two neutrinos and, hence, features low momentum charged particles to be reconstructed. 
The reduction of the material is therefore one natural improvement to test, in view of minimising multiple Coulomb scattering in the resolution sources.} 
and an improved single hit resolution. 
The results are shown in Fig.~\ref{fig:KstarTauTau_vertexResolutions} by the additional star symbols.
An improvement of the single hit resolution in the VXD layers by 30\% 
(i.e., from 3 to 2\,$\mu$m in the barrel layers) 
brings the sensitivity beyond the 3\,$\sigma$ evidence threshold, as shown by the yellow symbol. 
This threshold is also reached with a 50\% reduction of the VXD material, as shown by the magenta symbol. 
However, the impact of this sole improvement is limited by the material of the beam pipe, 
which causes charged particles to experience multiple scattering before they enter the detector. 
Indeed, in the nominal IDEA simulations used here, the material of the beam pipe is twice that of a VXD layer, 
so that to profit from the reduction in the material of the VXD layers the beam pipe transparency also needs to be increased. 
The green symbol in Fig.~\ref{fig:KstarTauTau_vertexResolutions} shows that, 
by reducing by 50\% the material of both the beam pipe and the VXD layers, 
the sensitivity increases to about 3.5\,$\sigma$, under the SM hypothesis for the considered branching fraction. 
Additional handles will be exploited to increase the sensitivity beyond the 5\,$\sigma$ discovery level.

\subsubsection{Vertex resolutions and heavy-flavour electroweak precision observables}
\label{sec:PhysPerf_HF_EWPO}

The FCC-ee operation at the \PZ-pole offers an immense potential for the measurement of the heavy-quark electroweak precision observables, 
the partial decay-width ratios $R_{\PQb (\PQc)} = \Gamma(\PZ \to \PQb \PAQb \, (\PQc \PAQc) ) \,/\, \Gamma(\PZ \to \text{hadrons})$, 
and the heavy-quark forward-backward production asymmetries $A_{\text{FB}}^{\PQb (\PQc)}$. 
A huge improvement of the statistical uncertainty is expected, with respect to the LEP measurements (by up to a factor of 2000), 
thanks not only to the large luminosity increase but also to the improvements in vertex detector technologies and tagging algorithms.
Experimental strategies that go beyond the current state-of-the-art methods also have to be developed, 
in order to reduce the systematic uncertainties in these measurements, down to a level commensurate with the expected statistical precision.
Earlier measurements have shown that the contamination from light quarks is a large source of systematic uncertainty. 
Taking advantage of the very large number of \PZ bosons that will be produced at FCC-ee, 
a novel tagging technique has been pioneered in Ref.~\cite{note_Lars} for measuring $R_{\PQb}$
and $A_{\text{FB}}^{\PQb}$, which relies on the exclusive reconstruction of a selected list of \PQb-hadron decay modes. 
It allows hemispheres containing a \PQb-hadron to be tagged with a purity larger than 99.8\% and, 
using selected decay modes of charged \PB mesons, the charge of the hemisphere is determined unambiguously. 
After removing the contamination from light quark processes, 
the remaining source of systematic uncertainty for the $R_{\PQb}$ measurement is the correlation of tagging efficiencies 
between the two hemispheres of an event. 
This uncertainty has been studied in detail in Ref.~\cite{note_Lars}, 
using generated events passed through a full simulation (including the reconstruction steps) of the CLD detector. 
In particular, an improved selection of the \PQb-hadron tracks based on their displacement with respect to the luminous region 
has been shown to minimise the impact on the  uncertainty of the exact knowledge of the correlation. 
The hemisphere correlation would need to be known with a relative precision of only 10\% 
to ensure that the corresponding systematic uncertainty on $R_{\PQb}$ is similar to the statistical uncertainty, at the level of 0.015\%. 
This results in a $R_{\PQb}$ determination that is 20 times more precise than the current value, with a lot of potential to do even better.
For the measurement of $A_{\text{FB}}^{\PQb}$, the remaining source of systematic uncertainty is the size of the QCD corrections. 
By exploiting acolinearity~\cite{AlcarazMaestre:2020fmp} or \PQb-hadron energy selection requirements, 
these corrections would need to be known with a relative uncertainty of 1\% for the statistical and systematic uncertainties 
to contribute equally to the total uncertainty. 
As a result, an improvement by a factor of 50 compared to the precision of the average of the LEP measurements would be obtained.

\begin{figure}[t]
\centering
\includegraphics[width=0.7\textwidth]{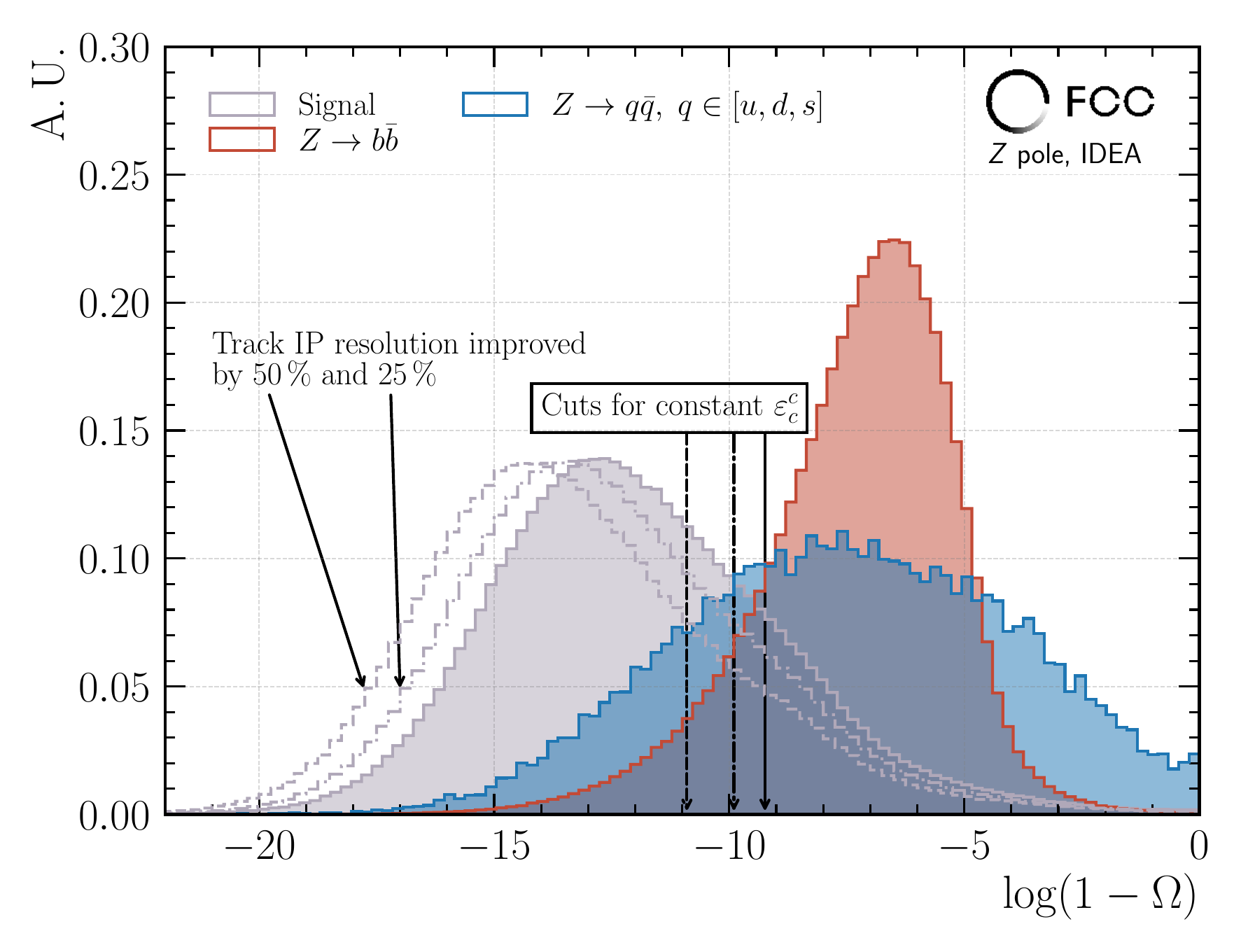}
\caption{Distribution of the pointing variable $\log (1 - \Omega)$ for signal and background events, 
where $\PADz \to \PKp \PGpm$ decays (and their charge-conjugates) are used to tag charmed hemispheres. 
The dashed and dash-dotted histograms show how the discrimination improves 
when the resolution of track impact parameters is improved by 25\% and 50\%, respectively, with respect to the nominal IDEA performance.}
\label{fig:Rc_Omega}
\end{figure}

These very small uncertainties motivate the use of exclusively reconstructed hadrons to measure $R_{\PQc}$ as well. 
A first study, described in Ref.~\cite{note_Lars_Rc}, exploits $\PADz \to \PKp \PGpm$ decays. 
The main contamination arises from $\PZ \to \PQb \PAQb$ events and needs to be reduced as much as possible 
to reach the expected statistical precision of $5.3 \times 10^{-5}$ 
(corresponding to an improvement by a factor of about 60 compared to the statistically most precise measurement 
from the SLD Collaboration~\cite{SLD:2005zyw}). 
This contamination can be reduced by demanding that the `pointing angle', 
i.e., the angle between the \PADz momentum and the line that joins the primary vertex and the decay vertex of the \PADz, be close to zero. 
Figure~\ref{fig:Rc_Omega} illustrates how a cut on the cosine of this angle, $\Omega$, separates the signal from the background. 
The measurement of $\Omega$ clearly depends on the reconstruction of the vertices, 
and the figure also shows how the discrimination power increases when the resolution on the longitudinal and transverse track impact parameters improves. 

\subsubsection{Alignment, overall scale of the detector, and the measurement of the tau lifetime}
\label{sec:PhysPerf_Taulifetime}

Precise measurements of the mass, the lifetime, and the leptonic branching fraction of the tau offer a crucial test of lepton flavour universality (LFU) since, 
up to small radiative and electroweak corrections, 
the following relation~\cite{Pich:2013lsa,CLEO:1996oro} holds between the muon and tau masses, 
their lifetimes $\tau_{\PGm}$ and $\tau_{\PGt}$, 
their charged-current coupling constants $g_{\PGm}$ and $g_{\PGt}$, 
and their branching fractions into electrons and neutrinos:
\begin{equation}
\left(\frac{g_{\PGt}}{g_{\PGm}}\right)^2 \simeq 
\frac{\tau_{\PGm}}{\tau_{\PGt}} \, \frac{{\cal{B}}(\PGt \to \Pe\PGnGt\PAGne) }{{\cal{B}}(\PGm \to \Pe\PGnGm\PAGne) }\, \left( \frac{m_{\PGm}}{m_{\PGt}} \right)^5 \, . 
\label{eq:tau_LFU}
\end{equation}
This test, which can be seen as a determination of the Fermi constant from the tau, 
is currently limited by the precision of the \PGt lifetime, $290.3 \pm 0.5$\,fs, 
and by that of the $\cal{B}(\PGt \to \Pe \PGnGt \PAGne)$ branching fraction, $17.82 \pm 0.04 \%$. 
The measurement of $\cal{B}(\PGt \to \PGm \PGnGt \PAGnGm)$ allows additional similar tests of $(g_{\PGt} / g_{\Pe})$ and $(g_{\PGm} / g_{\Pe})$.
With the $\sim$\,$2 \times 10^{11}$ tau pair events expected at Tera-\PZ, and a much improved determination of the three key 
ingredients\footnote{The knowledge of the \PGt lepton mass is also expected to be much improved from pair production threshold measurements at a next-generation tau-factory.} 
of Eq.~(\ref{eq:tau_LFU}), FCC-ee has the potential to test LFU at an unprecedented level. 
In particular, the statistical uncertainty in the tau lifetime is expected to be about 15~ppm~\cite{Lusiani:2024tau}. 
The current measurements, from the Belle and LEP experiments, are statistically limited and, 
for the two single most precise measurements (Belle~\cite{Belle:2013teo} and DELPHI~\cite{DELPHI:2003zcz}), 
the alignment of the vertex detector is the dominant source of systematic uncertainty. 
While this alignment could, at first sight, appear to be a concern in view of a lifetime measurement at the level of 15~ppm  
(which corresponds to a few tens of nm on the flight distance of \PGt leptons produced at 91\,GeV), 
a more careful investigation shows that this is not the case~\cite{Lusiani:2024tau}.
Indeed, the offset on the decay length caused by misalignment effects averages to zero, to first order, 
when integrating over the azimuthal angle, as was first noted in Ref.~\cite{Wasserbaech:1998ur}. 
The expected systematic uncertainty due to misalignment, scaled from the value quoted by DELPHI according to the luminosity increase, is below 4~ppm. 
Other sources of systematic uncertainty have been considered in Ref.~\cite{Lusiani:2024tau}. 
One of them reflects the knowledge of the overall length scale of the vertex detector. 
For this uncertainty to be smaller than one half of the statistical uncertainty on the \PGt lifetime, that scale should be known to better than 5~ppm. 
At LEP and at the \PB-factories this scale was only known to 100~ppm. 
However, recent developments made for the MUonE experiment indicate that a precision of 5~ppm is within reach, thanks to the use of optical techniques~\cite{9803636}. 
The leading systematic uncertainties are expected to be due to the modelling of initial state radiation and to the knowledge of the tau mass, and a total systematic uncertainty of 16~ppm on the tau lifetime is deemed within reach~\cite{Lusiani:2024tau}.

\subsubsection{Preliminary conclusions}

\begin{itemize}

\item One example has been shown where the physics outcome of FCC-ee would gain from having better vertex detector performance than the baseline detectors considered so far. 

\item Minimising the amount of material in front of the vertex detector is important. 
Indeed, the material of the beam pipe is a limiting factor in some cases, 
in particular for observing the rare $\PB \to \PK^{*} \PGt \PGt$ decay.

\item It should be noted that these requirements, tighter than the ones presented for a linear collider detector, 
will have to be reached despite the additional constraints set by the FCC-ee environment on the readout electronics of the detector: 
(i)~its power budget is tighter than for a detector operating at a linear collider 
(since power-pulsing the electronics is not possible with collisions occurring every $\sim$\,20\,ns), 
and (ii)~it should handle a hit rate of about 200\,MHz/cm$^2$ in the innermost vertex detector layer at the \PZ peak (Section~\ref{sec:MDI_backgrounds}).

\end{itemize}

\subsection{Requirements for charged hadron particle identification}
\label{sec:PhysPerf_PID}

Charged hadron particle identification (PID) is essential for the FCC-ee flavour physics programme and brings significant benefits for other areas. 
The momentum range over which good PID capabilities are necessary is broad, in sharp contrast with the PID needs of a \PB-factory. 
For flavour physics, for example, it extends from the lowest momentum of the reconstructed charged-particle tracks up to about 40\,GeV.  
In a gaseous tracker like the IDEA drift chamber, the determination of the number of ionisation clusters per unit length (d$N$/d$x$)
is a promising approach that should allow charged kaons to be efficiently separated from charged pions in a large momentum range, from about 1.5\,GeV to several tens of GeV. 
The expected resolution on d$N$/d$x$ 
is about 2\%\footnote{The promising performance of the d$N$/d$x$ approach as determined from calculations, 
in particular the upper momentum bound, is being checked with test beam data.}, 
significantly better than that of the well-known measurement of ionisation energy per unit length (d$E$/d$x$).

Time-of-flight (TOF) measurements, in a single layer at a distance of about 2\,m from the interaction point, 
fill the gap around 1\,GeV\footnote{This statement holds for a magnetic field of 2\,T; the case of a higher field value is briefly considered in Section~\ref{sec:PhysPerf_TOF}.}, 
where the average energy loss per unit length of kaons and pions is similar. 
However, they can only provide a 3\,$\sigma$ $\PGp / \PK$ separation at low momenta, up to about 3\,GeV with the 30\,ps resolution assumed for the studies reported 
here\footnote{Even with a 10\,ps resolution, the momentum range would only extend up to about 5\,GeV.}. 
Innovative solutions are being studied that would provide good PID capabilities in the absence of a specific energy loss measurement, 
over the large momentum range of interest and despite the tight spatial constraints. 
A conceptual design exists for a compact RICH detector, which looks promising and could comfortably cover the required momentum range~\cite{forty_2024_6g0gs-7kw30, Forty_ARC, Tat_ARC}. 
An overview of the motivations for PID and of the possible detector solutions can be found in Ref.~\cite{Wilkinson:2021ehf}. 
This section illustrates the demands on PID performance with the measurement of the Higgs boson coupling to strange quarks and with two \PB-physics processes.

\subsubsection{Strange tagging and the Higgs boson coupling to strange (and charmed) quarks}
\label{sec:PhysPerf_strangeTagging}

The progress in the development of flavour-tagging algorithms in recent years allows, for the first time, 
a relatively efficient and pure tagging of jets induced by strange quarks. 
Various analysis efforts are ongoing within the circular and linear colliders communities. 
They all rely heavily on the identification power of state-of-the-art machine learning approaches. 

\begin{figure}[ht]
\centering
\includegraphics[width=0.49\textwidth]{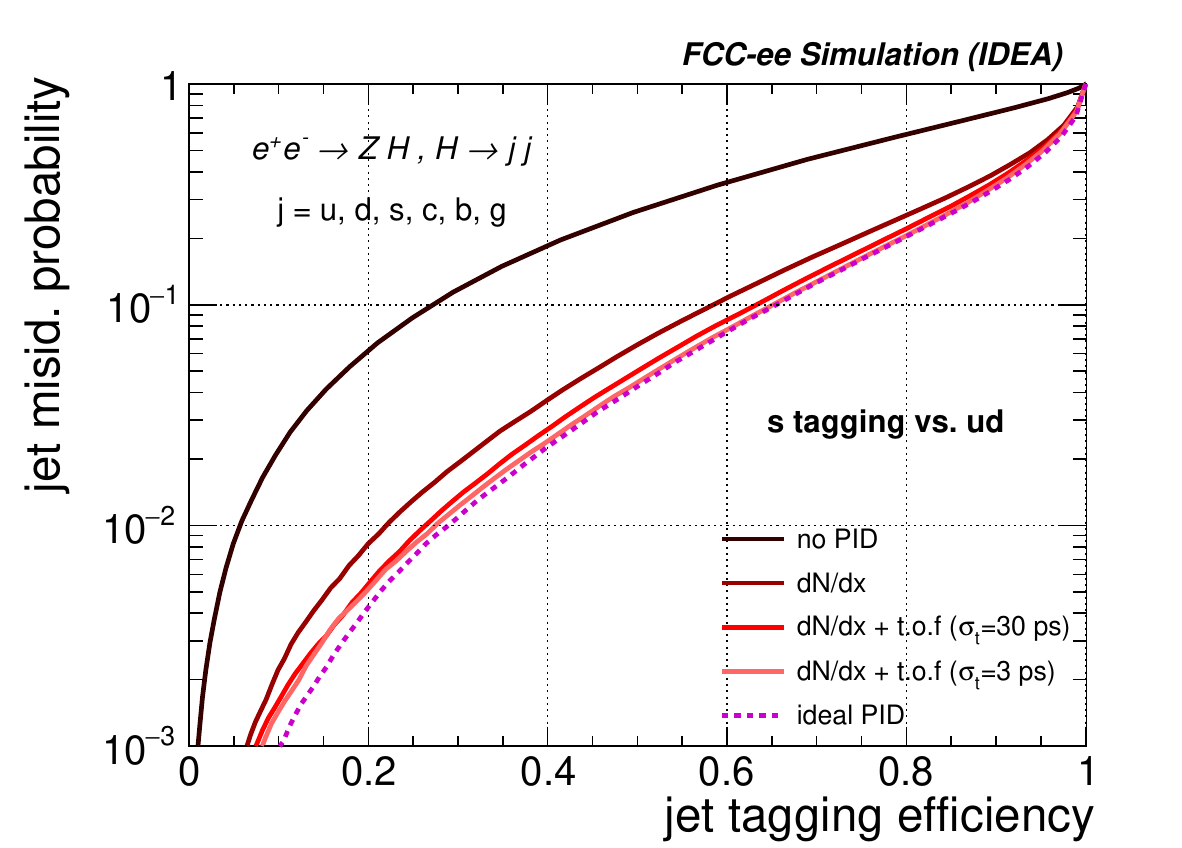}
\includegraphics[width=0.49\textwidth]{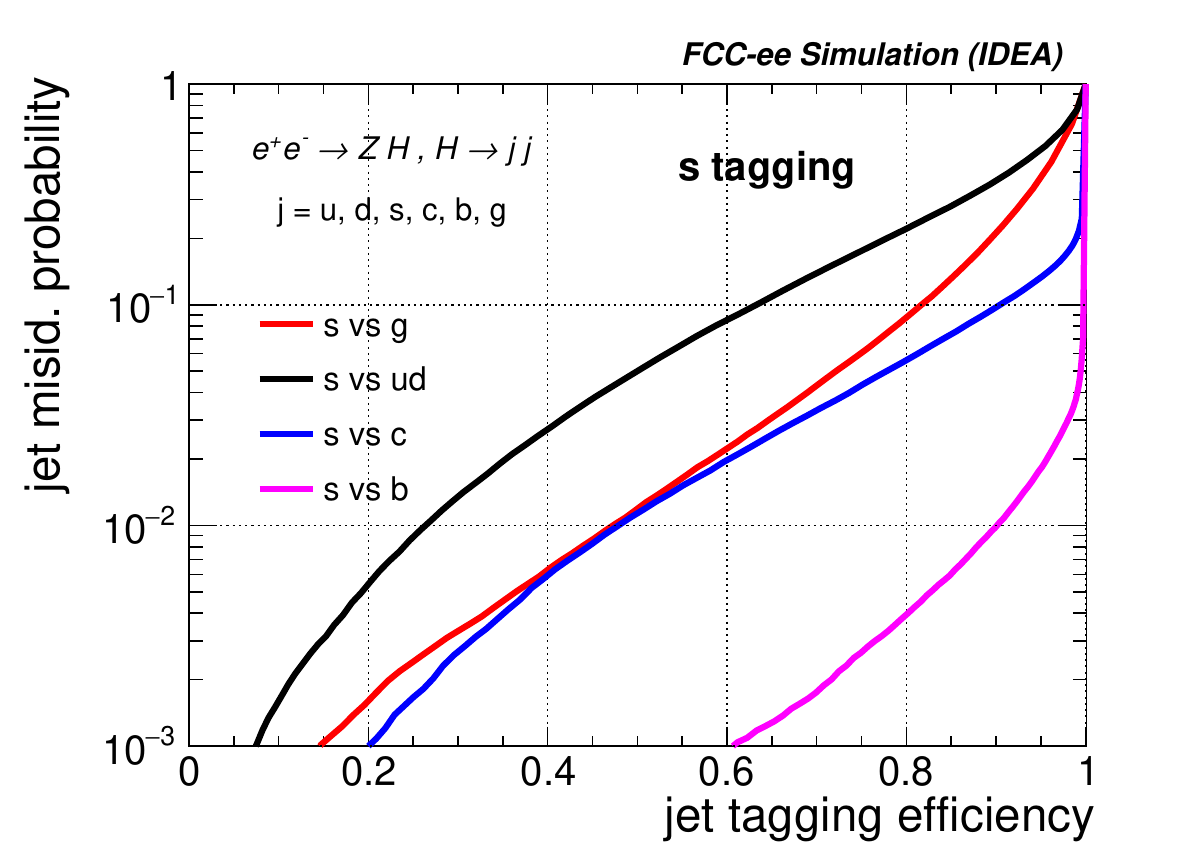}
\caption{Performance of strange-quark tagging in jet samples from $\PZ(\PGn \PAGn) \PH$ events at $\sqrt{s} = 240$\,GeV, 
where the Higgs boson decays into quarks of a well-defined flavour or into gluons. 
The left panel shows the impact of PID on the separation power between jets induced by strange quarks and those induced by \PQu or \PQd quarks. 
The right panel shows how the algorithm separates strange jets from all other flavours. 
The `\PQs vs.\ $\PQu\PQd$' (black) curve in the right panel is identical to the red curve in the left panel.}
\label{fig:strange_tagging}
\end{figure}

Figure~\ref{fig:strange_tagging} illustrates the performance of strange-quark tagging with the baseline IDEA detector used for this report, 
using the {\textsc{ParticleNetIDEA}} algorithm~\cite{Bedeschi:2022rnj}. 
As shown in the left panel, when no PID information is provided to the algorithm, 
the separation power between jets induced by strange quarks and those induced by \PQu or \PQd quarks is limited, 
as it mostly comes from the slightly harder energy spectrum of the constituents of a strange 
jet\footnote{The number of reconstructed $\PKS \to \PGpp \PGpm$ decays from displaced tracks associated with the jets (see Section~\ref{sec:PhysPerf_Kshorts}) 
is not yet explicitly included in the set of variables used by the {\textsc{ParticleNetIDEA}} algorithm; adding this variable may improve the performance shown here.}. 
When the specific energy loss measured along the tracks with the cluster counting technique is also used, 
the misidentification probability of jets induced by \PQu or \PQd quarks is reduced by about one order of magnitude, for the same \PQs-quark tagging efficiency. 
In the kinematic range considered here, spanned by quarks produced in $\PZ (\PGn \PAGn) \PH(\PQq \PAQq)$ events at $\sqrt{s} = 240$\,GeV, 
the time-of-flight information only brings a mild performance improvement. 
The right panel of Fig.~\ref{fig:strange_tagging} shows how the algorithm separates strange jets from all other flavours, 
the separation from $\PQu\PQd$ jets being, as expected, the most challenging. 
For a strange-quark tagging efficiency of 80\% (90\%), the mis-tag efficiency of $\PQu\PQd$ jets reaches 20\% (40\%).

This tagging algorithm has been used to categorise $\PZ(\ell \ell, j j) \PH(jj)$ events 
(where $\ell$ stands for \Pe, \PGm or \PGn) 
into mutually orthogonal categories enriched in one of the different Higgs boson decays (Section~\ref{sec:PhysPerf_Hcc}) 
and to extract the Higgs boson branching fractions to $\PQb \PAQb$, $\PQc \PAQc$, $\PQs \PAQs$, and $\Pg\Pg$.
With an integrated luminosity of $10.8$\,ab$^{-1}$, the Higgs boson branching fraction to strange quarks would be measured with a 105\% uncertainty. 
More details can be found in Ref.~\cite{note_Higgs_to_hadrons}.

\begin{figure}[ht]
\centering
\includegraphics[width=0.49\textwidth]{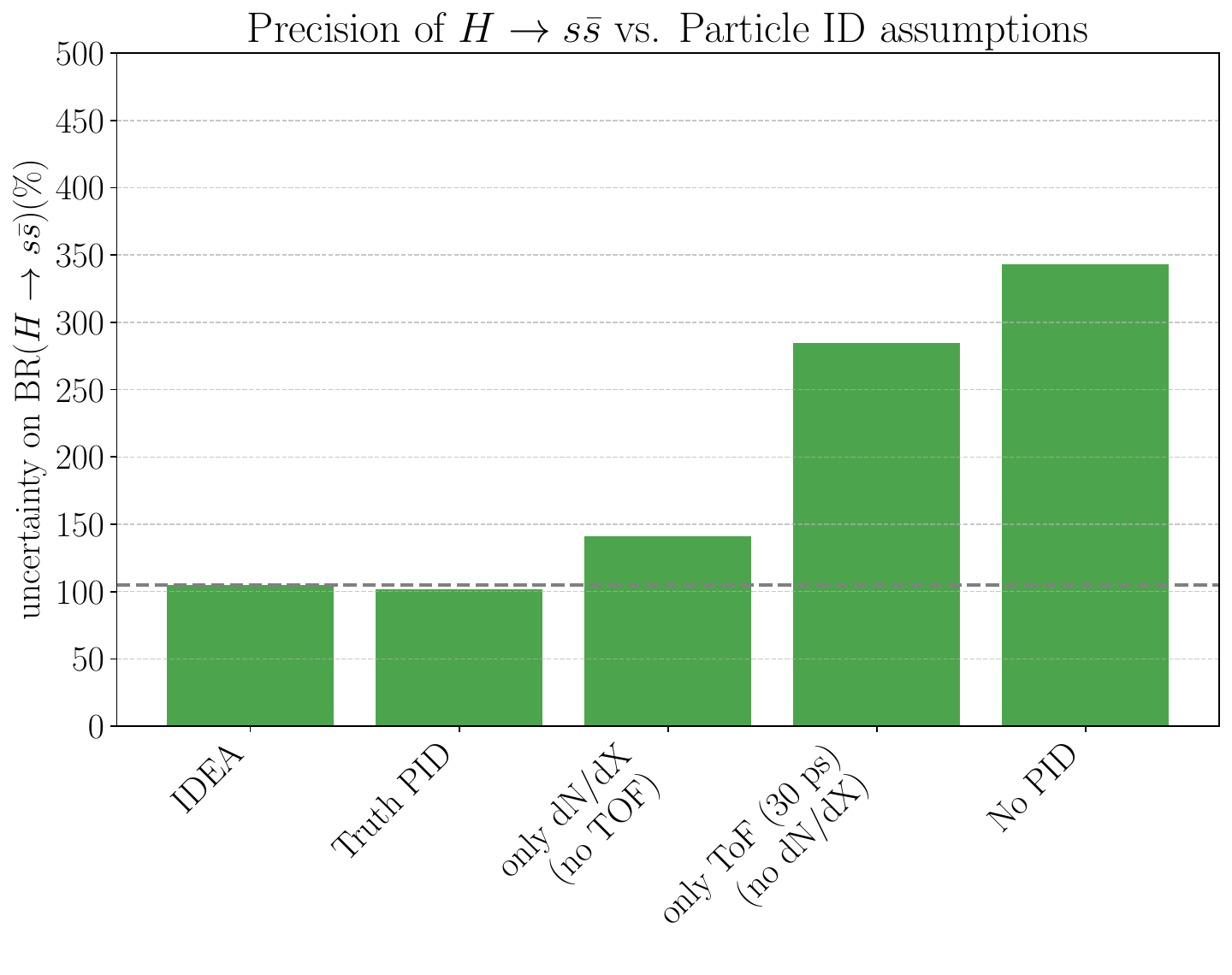}
\includegraphics[width=0.49\textwidth]{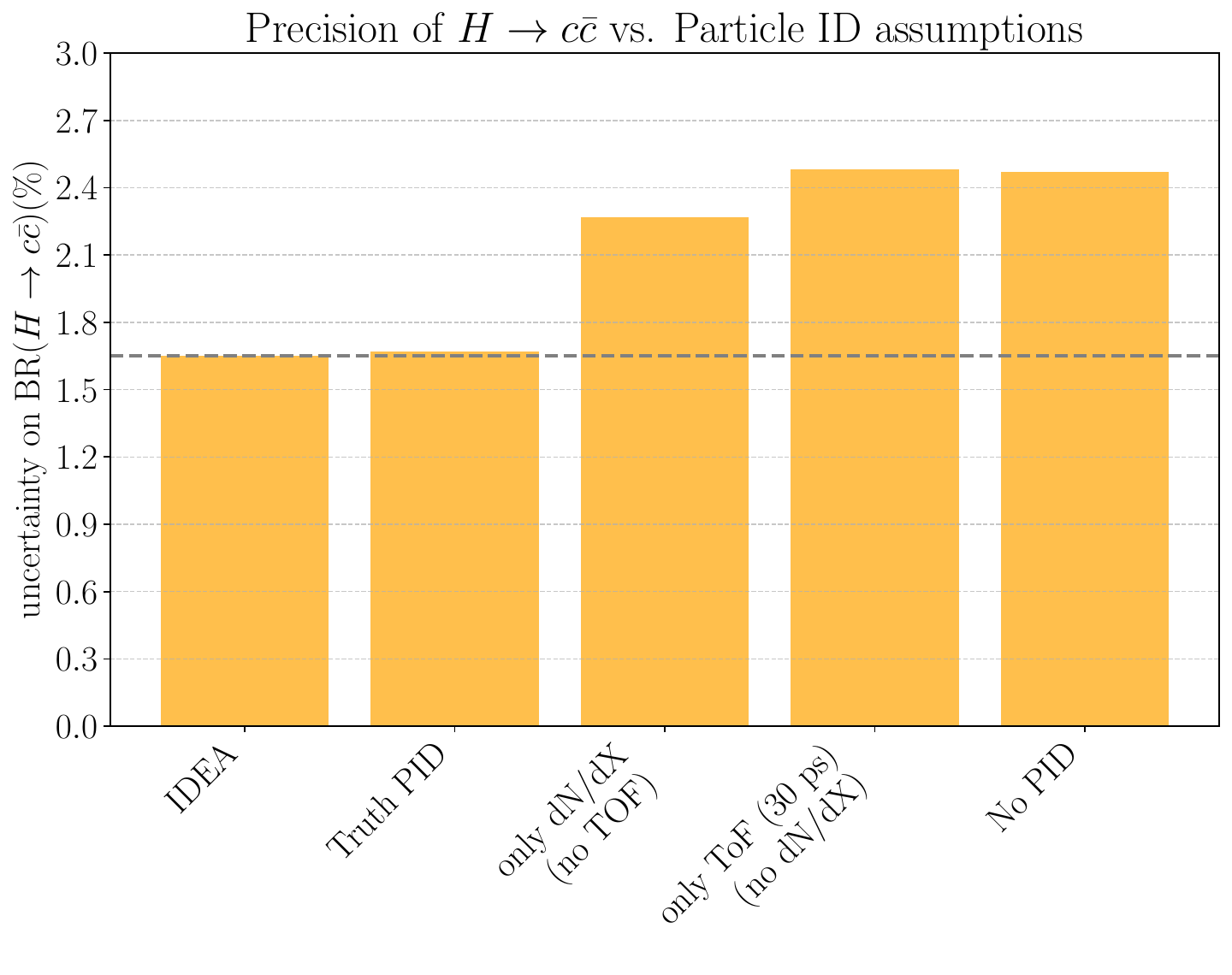}
\caption{Relative uncertainty in the $\PH \to \PQs \PAQs$ (left) and $\PH \to \PQc\PAQc$ (right) branching fractions, 
as expected for several assumptions on particle identification performance setups.}
\label{fig:Hss_versus_PID}
\end{figure}

The anticipated precision on the Higgs boson branching fractions to $\PQc \PAQc$ and $\PQs \PAQs$ has been explored for several particle identification capabilities 
and the result is shown in Fig.~\ref{fig:Hss_versus_PID}. 
Given that charged kaons from strange quark hadronisation feature a hard momentum spectrum, 
it is crucial to have an efficient $\PK/\PGp$ separation at large momenta, as provided by the d$N$/d$x$ cluster counting technique. 
It is remarkable that the expected precision on the measurement of the $\PH \to \PQs \PAQs$ branching fraction, 
obtained by combining d$N$/d$x$ with the time-of-flight technique, 
is almost identical to the asymptotic performance obtained with perfect PID capabilities. 
Conversely, if no PID is assumed, or only assuming ToF, the $\PH \to \PQs \PAQs$ branching fraction measurement cannot be performed, 
given an expected uncertainty in excess of 300\%. 
The PID also plays a non-negligible role in the $\PH \to \PQc\PAQc$ branching fraction measurement, 
by enabling the identification of secondary charged kaons produced in \PD meson decays, 
as shown in the right panel of Fig.~\ref{fig:Hss_versus_PID}. 
A comprehensive discussion on the impact of PID on flavour tagging performance can be found in Ref.\cite{sciandra_2024_gt813-z8602}.

\subsubsection{Separation of \texorpdfstring{\PKpm}{K+-} from \texorpdfstring{\PGppm}{pi+-} and measurement of \texorpdfstring{$\PQb \to \PQs \PGn \PAGn$}{b to s neutrino antineutrino} }

\begin{figure}[ht]
\centering
\includegraphics[width=0.495\textwidth]{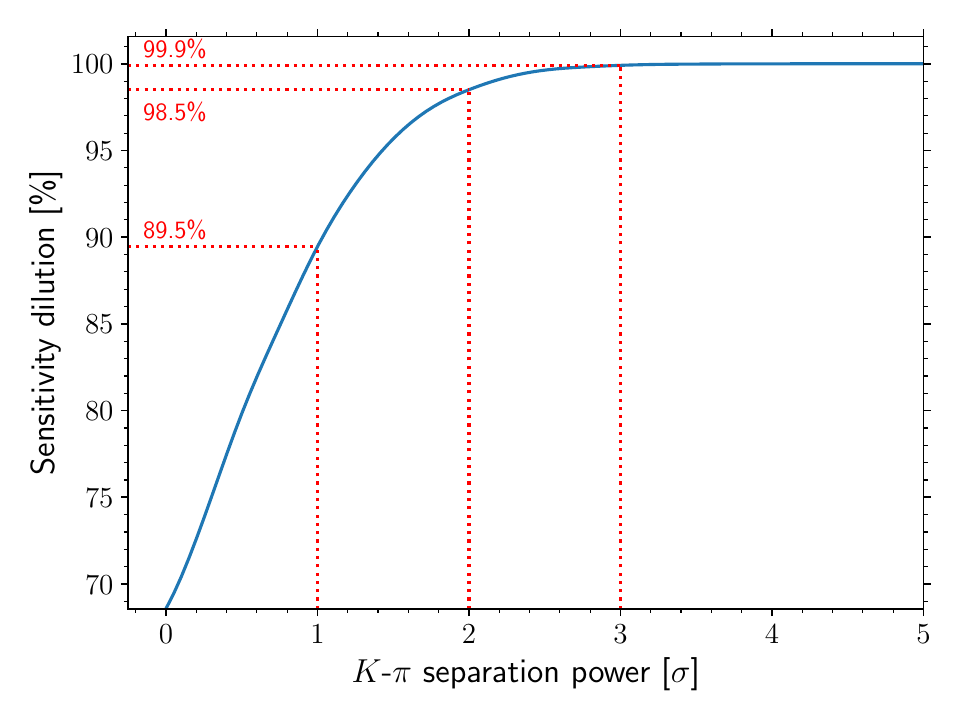} 
\includegraphics[width=0.495\textwidth]{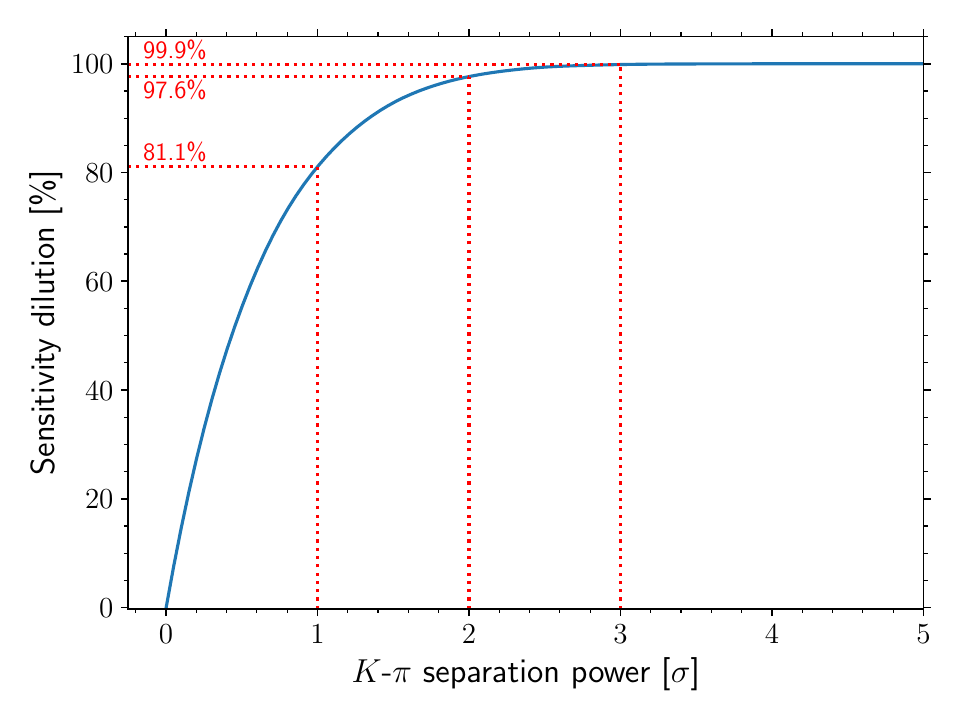}
\caption{Sensitivity to the $\PB \to \PK^{*} \PGn \PAGn$ (left) and $\PBs \to \PGf \PGn \PAGn$ (right) branching fractions, 
as a function of the $\PGp/K$ separation power, with respect to the sensitivity assuming perfect PID.}
\label{fig:PID_for_bsnunu}
\end{figure}

The trillions of $\PQb \PAQb$ events that will be collected, in a clean environment, 
during the Tera-\PZ run at FCC-ee makes it the ideal facility to study rare decays of \PQb~hadrons. 
Among those, the modes corresponding to the $\PQb \to \PQs \PGn \PAGn$ transition are of considerable interest within the flavour physics 
community\footnote{The first evidence for this transition has been recently obtained by the Belle~II Collaboration in the 
$\PBp \to \PKp \PGn \PAGn$ decay channel~\cite{BelleII-EPS2023}. 
In the future, with an integrated luminosity of 50\,ab$^{-1}$, 
Belle~II is expected to measure the corresponding branching fraction with ${\cal{O}}(10\%)$ experimental precision~\cite{Belle-II:2018jsg}.}. 
A first sensitivity study for FCC-ee, using the decays $\PB \to \PK^{*} \PGn \PAGn$ and $\PBs \to \PGf \PGn \PAGn$, is detailed in Ref.~\cite{Amhis:2023mpj}. 
According to the Standard Model, both decays should have branching fractions around $10^{-5}$. 
The current experimental upper limit on the branching fraction of $\PB \to \PK^{*} \PGn \PAGn$ is larger than this prediction by a factor of two, 
while for $\PBs \to \PGf \PGn \PAGn$, the upper limit is two orders of magnitude larger. 
The study looks for a $\PK^{*}$ or a \PGf resonance in a hemisphere with a large amount of missing energy and runs a multi-variate analysis, 
following the strategy described in Ref.~\cite{Zuo:2023dzn}. 
Assuming a perfect PID, the $\PB \to \PK^{*} \PGn \PAGn$ ($\PBs \to \PGf \PGn \PAGn$) signal can be selected with an efficiency of 3.7\% (7.4\%), 
for a signal-to-background ratio of 0.17 (0.13). 
With $6 \times 10^{12}$ \PZ bosons produced, 
this corresponds to a sensitivity of 0.5\% for $\PB \to \PK^{*} \PGn \PAGn$ and of 1.2\% for $\PBs \to \PGf \PGn \PAGn$. 
Here, sensitivity is defined as the relative precision expected on the branching fraction, computed as $\sqrt{S+B} / S$, 
where $S$ and $B$ are the expected numbers of signal and background events, respectively. 
The dependence of this sensitivity on the performance of the $\PGp/\PK$ separation that the detector would achieve is shown in Fig.~\ref{fig:PID_for_bsnunu}. 
With a separation power of 2\,$\sigma$, the loss in sensitivity, compared to what is expected with perfect PID, is marginal, 
but it degrades quickly as the performance worsens. 
The momentum of the kaons involved in these decays peaks at about 3\,GeV, but extends up to 15--20\,GeV, 
hence the PID performance offered by time-of-flight measurements alone is unlikely to be sufficient for this analysis.

\subsubsection{Separation of \texorpdfstring{\PKpm}{K+-} from \texorpdfstring{\PGppm}{pi+-} 
and measurement of \texorpdfstring{$\PBs \to \PDs \PK$}{Bs to Ds K} }  
\label{sec:PhysPerf_Bs2DsK}

The measurement of time-dependent CP asymmetries in the decay $\PBs \to \PDs \PK$ is a well-known method to extract the $\gamma$ angle of the CKM matrix, 
which is currently only known with a precision of 4~degrees. 
The prospects for this measurement at FCC-ee have been first studied in Ref.~\cite{Aleksan:2021gii}, using the $\PDs \to \PGf (\PK\PK) \PGp$ decay channel. 
The presence of the \PGf resonance and the absence of neutral particles among the final decay products makes this signal easy to reconstruct. 
A precision on the $\gamma$ angle of a few tenths of a degree was shown to be within reach at FCC-ee, using only this decay 
mode\footnote{A similar sensitivity is expected from the Phase-2 LHCb upgrade, with an integrated luminosity of 300\,fb$^{-1}$~\cite{Charles:2020dfl}. 
The FCC-ee sensitivity can be improved by including other \PDs decay modes, in particular those that include neutral particles (see Section~\ref{sec:PhysPerf:pi0_reco_for_flavour}).}.

\begin{figure}[ht]
\centering
\includegraphics[width=0.49\textwidth]{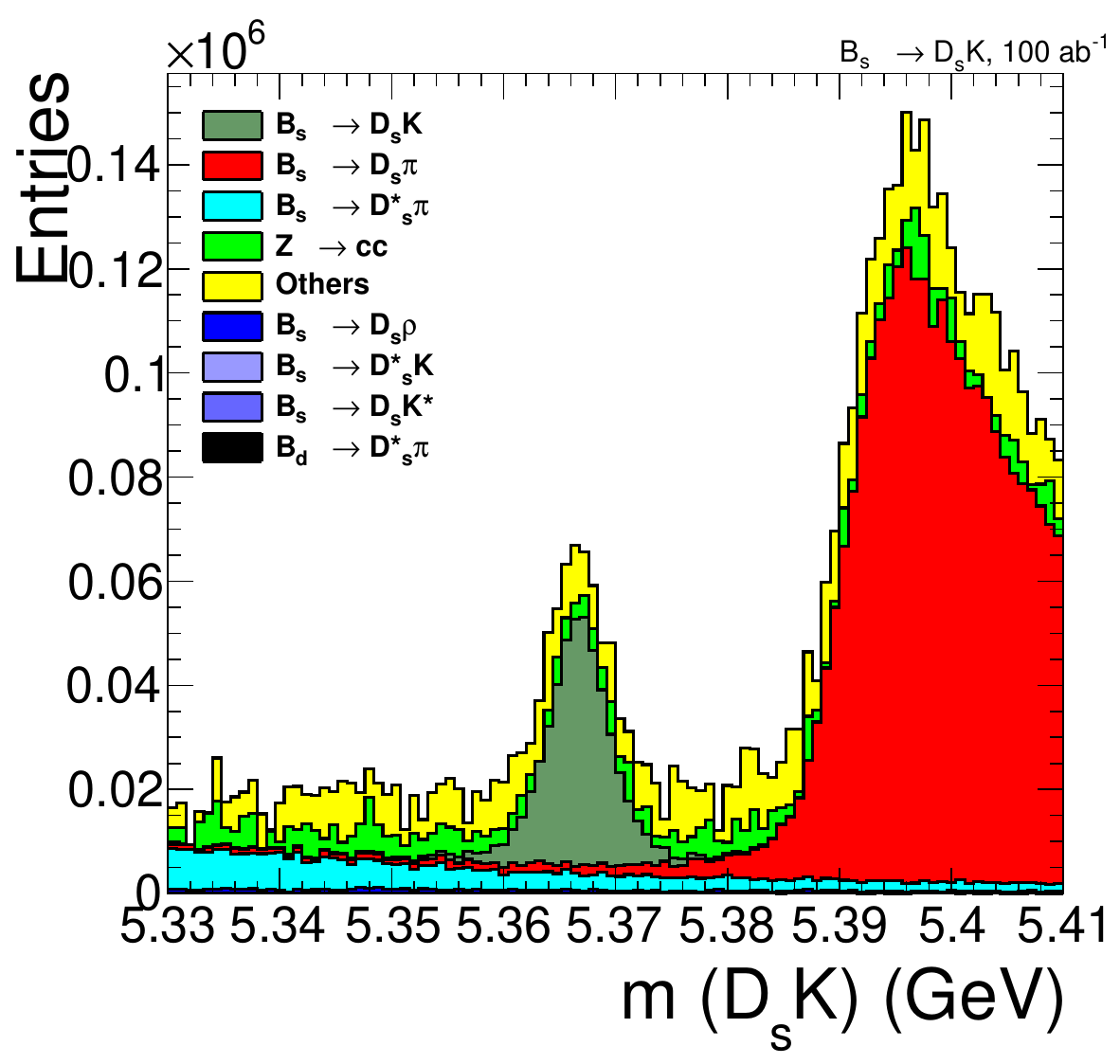}
\includegraphics[width=0.49\textwidth]{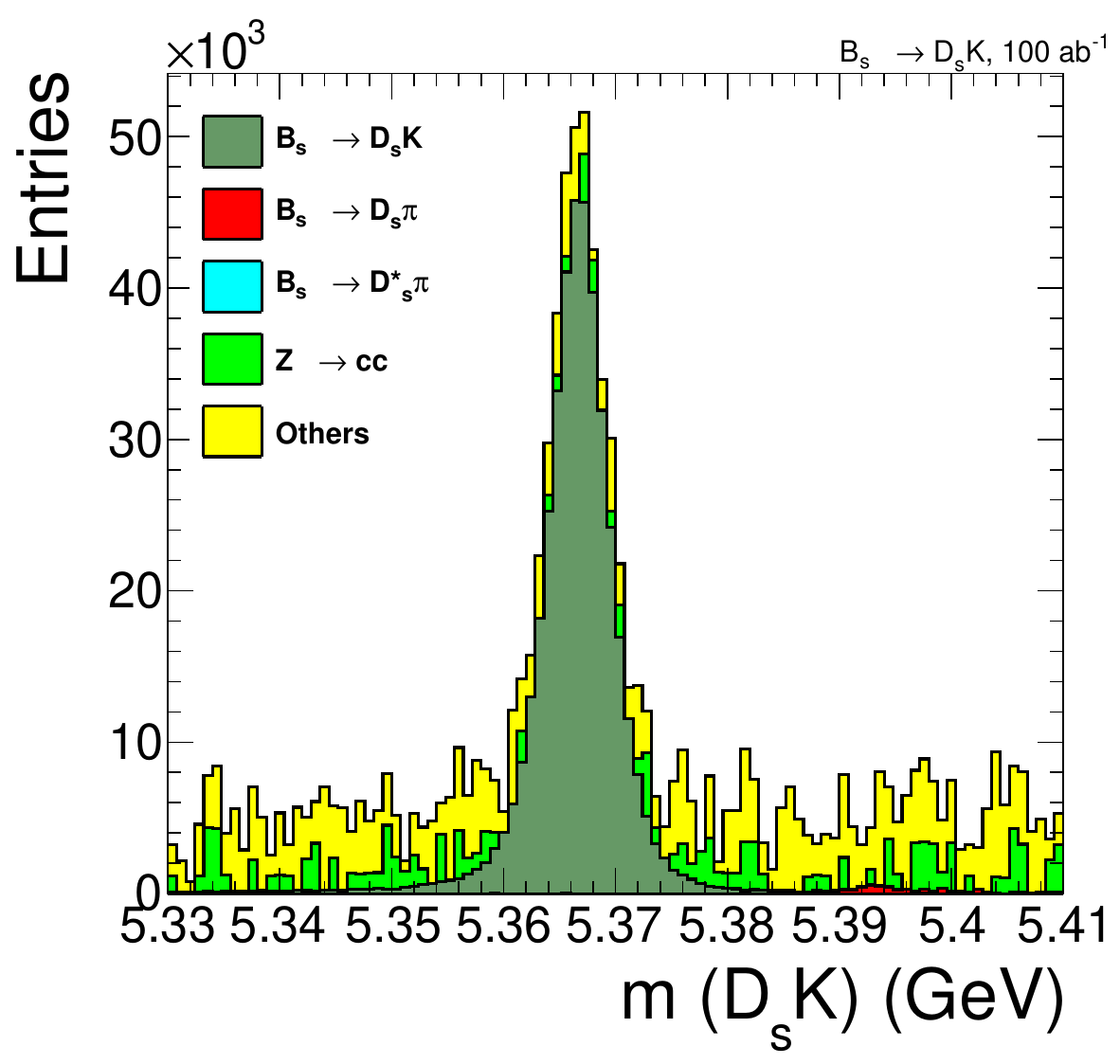}
\caption{Mass distributions of the \PBs candidates in the $\PBs \to \PDs \PK \to \PGf (\PK\PK) \PGp \PK$ decay channel, 
prior to any PID cut (left) and after requiring that the `bachelor' 
(the particle accompanying the reconstructed \PDs) must be identified as a kaon (right). 
The momentum of the `bachelor' is required to be larger than 1.5\,GeV.}
\label{fig:Bs2DsK}
\end{figure}

The study reported in Ref.~\cite{Aleksan:2021gii}, which considered a few exclusive background processes and used a simple parametrisation of the detector response, 
has been redone for this report with \textsc{Delphes} to simulate the response of the IDEA detector and accounting for the inclusive $\PZ \to \PQb \PAQb$ and $\PZ \to \PQc\PAQc$ backgrounds. 
The analysis starts by exploiting the mass resolution to reconstruct the \PGf candidates and, afterwards, the \PDs candidates, without any PID requirement. 
In a second step, \PBs candidates are built by fitting the \PDs and another track (the `bachelor') to a common vertex~\cite{BedeschiCode}. 
The invariant mass of the \PBs candidates satisfying minimal kinematic and vertex $\chi^2$ cuts is shown in the left panel of Fig.~\ref{fig:Bs2DsK}. 
The signal, shown in dark green, is very well separated from the ten times larger $\PBs \to \PDs \PGp$ background, shown in red. 
However, running the same selection over the inclusive $\PZ \to \PQb\PAQb$ sample shows that there are other sources of backgrounds that contaminate the signal mass peak, 
in addition to the exclusive backgrounds that were considered in Ref.~\cite{Aleksan:2021gii} and that are also shown in Fig.~\ref{fig:Bs2DsK}. 
The right panel of this figure shows that this background is largely suppressed by requiring that the `bachelor' particle must be identified as a kaon, 
which is done by using a PID cut that is 95\% efficient on the signal. 
The PID cut uses the ratio of the likelihoods for the `bachelor' to be a kaon or a pion and combines the d$N$/d$x$ measurement of the track, 
determined with a resolution of about 2\%, and the measurement of its TOF at 2\,m from the interaction point, with a 30\,ps resolution. 
Even for this mode with charged particles only, PID capabilities appear to be mandatory to extract the signal with a decent signal-to-background ratio. 
Since the momentum distribution of the kaons from $\PBs \to \PDs \PK$ decays is hard, time-of-flight measurements alone would not lead to an acceptable background suppression. 
For example, in a mass window that contains 95\% of the reconstructed \PBs candidates (the signal), 
the background contamination decreases from 48\%, without any PID, 
to 33\% if only the TOF measurement is used, 
and to 19\% if both the d$N$/d$x$ and TOF measurements are used (assuming the PID capabilities of IDEA). 
This last value would degrade to 24\% with a twice worse d$N$/d$x$ resolution.

\subsubsection{Work ahead}

\begin{itemize}

\item The measurement of time-dependent CP asymmetries in the \PBs system requires the flavour of the meson (\PBs or \PABs) to be tagged at the time of production. 
Various algorithms are typically combined to achieve this goal. 
A powerful method uses the charge of the kaons in the event that are not associated with the signal decay, which typically have momenta well below 10\,GeV.
The performance of flavour tagging is studied in order to provide requirements on PID at low momentum. 
For example, the determination of the angle $\gamma$ of the CKM matrix, from the $\PBs \to \PDs \PK$ process, 
or of the phase $\beta_s$, from the $\PBs \to \PJGy \, \PGf$ decay, could serve as benchmark measurements. 
The identification of very soft fragmentation kaons (produced when a \PQb (anti-)quark hadronises into a \PBs meson), 
for which the PID performance brought by TOF measurements is relevant, 
should also play a key role in the measurement of the branching fraction of the $\PBs \to \PGn \PAGn$ decay.

\item The determination of $V_{\PQc\PQs}$ from \PW decays may benchmark the performance of strange-quark tagging 
for working points with a much higher purity than that needed for the determination of the coupling of the Higgs boson to the strange quark. 
Alternatively, $V_{\PQc\PQs}$ may be measured in events where exclusive decays of charmed hadrons into kaons are selected, calling for a good $\PGp/\PK$ separation. 
More generally, electroweak measurements involving strange quarks, such as $R_{\PQs}$ or $A_\text{FB}^{\PQs}$, are studied for the requirements on the PID performance.

\item A very precise test of lepton flavour universality in the tau sector is provided by the simultaneous measurements, at the $10^{-5}$ level, 
of the tau lifetime and of the branching fraction of its decay to electrons (Section~\ref{sec:PhysPerf_Taulifetime}). 
The latter requires a $\Pe / \PGp$ separation controlled at the $10^{-5}$ level, 
which might be difficult to achieve from calorimetry measurements alone. 
The requirements on d$E$/d$x$ or d$N$/d$x$ measurements should also be studied in the context of $\Pe / \PGp$ separation.

\end{itemize}

\subsubsection{Preliminary conclusions}

\begin{itemize}

\item It is not only for flavour physics that charged-hadron PID is needed. 
In particular, the potential for constraining the coupling of the Higgs boson to the strange quark provides a strong motivation for having PID, up to high momenta. 
Similarly, powerful strange tagging offers new prospects for electroweak and top measurements in the strange sector, 
such as measurements of the $\PZ \to \PQs \PAQs$ branching fraction~\cite{Blekman:2024wyf} and forward-backward asymmetry,
or of the $V_{\PQt\PQs}$ CKM element via rare $\PQt \to \PW \PQs$ decays~\cite{giappichini_2024_te64m-zbm05}. 

\item The momentum range to be covered, extending from ${\cal{O}}$(150)\,MeV to 40\,GeV, is challenging. 
Current studies show that the $\PGp/\PK$ separation offered by the cluster counting approach, in the IDEA drift chamber, is probably adequate. 
A degradation of the resolution of the energy loss measurement by a factor of 2, with respect to what is expected from the baseline IDEA d$N$/d$x$ performance, 
would have a non-negligible impact on the determination of the Higgs coupling to strange quarks. 
Such a degradation would correspond to the expected d$E$/d$x$ performance.

\item In the absence of specific energy loss measurements in the tracker, 
there is a conceptual design for a compact RICH detector that looks promising and could comfortably cover the required momentum range. 
Detailed simulation studies are needed to evaluate the impact of such a detector on particle-flow reconstruction performance.

\end{itemize}

\subsection{Requirements for electromagnetic calorimetry}
\label{sec:PhysPerf_ECAL}

This section summarises the needs for the measurement of electromagnetic (EM) showers, which may, but does not have to, be performed in a dedicated electromagnetic calorimeter (ECAL).
In the baseline IDEA concept, the EM showers are measured using a finely segmented crystal ECAL with dual readout capabilities~\cite{Lucchini:2020bac}.  
In a variant of this concept, considered in Ref.~\cite{fcc-phys-cdr}, the electromagnetic and hadron showers are both measured in a dual readout (DR) calorimeter that uses scintillation and Cerenkov fibres. 
The CLD concept uses a high-granularity ECAL built as a sandwich of tungsten plates and silicon sensors. 
A high granularity noble liquid ECAL is also under consideration in ALLEGRO. 
These concepts have complementary strengths, ranging from an energy resolution of $\sigma_E / E \simeq 3\% / \sqrt{E}$ for the crystal-based ECAL 
to an extreme transverse resolution of $\sim 2 \times 2$\,mm$^2$ for the fibre DR calorimeter, if each fibre is equipped with a SiPM readout device, 
or to a high longitudinal segmentation for the Si/W or noble liquid solution. 
More details are given in Table~\ref{tab:ECAL_options}. 
The next paragraphs illustrate the requirements on electromagnetic calorimetry from representative analyses. 
Benchmark processes for which the performance is solely driven by the energy resolution are described first, 
followed by one case for which the transverse granularity is crucial, and by examples that place stringent requirements on both the energy resolution and the granularity.

\begin{table}[ht]
\centering
\caption{Expected energy resolution of the different electromagnetic calorimeter options considered in the studies of Ref.~\cite{Aleksa:2021ztd}.}
\label{tab:ECAL_options}
\begin{tabular}{l c c  c }
\toprule
        Technology & \multicolumn{2}{c}{EM energy resolution}\\
                   & stochastic term  & constant term       \\   
\midrule
        Highly granular Si/W based & 15--17\% & 1\% \\
        Dual readout fibre (ECAL$+$HCAL)  & 11\% & $<1$\%  \\
        Hybrid crystal (dual readout) &  3\% &  $<1$\%  \\
        Highly granular noble liquid based ECAL  & 8--10\% & $<1$\% \\
\bottomrule
\end{tabular}
\end{table}

\subsubsection{Energy resolution and monophoton final states}

For final states that consist of a single photon and nothing else in the detector, the main handle to suppress the backgrounds is the electromagnetic energy resolution.
Two physics cases have been considered for such final states: 
the determination of the $\PZ \PGne \PAGne$ coupling and the search for long-lived axion-like particles.

\paragraph{Measurement of the \texorpdfstring{\PZ}{Z} coupling to the \texorpdfstring{\PGne}{nu-e}}

\begin{figure}[ht]
\centering
\includegraphics[width=0.495\textwidth]{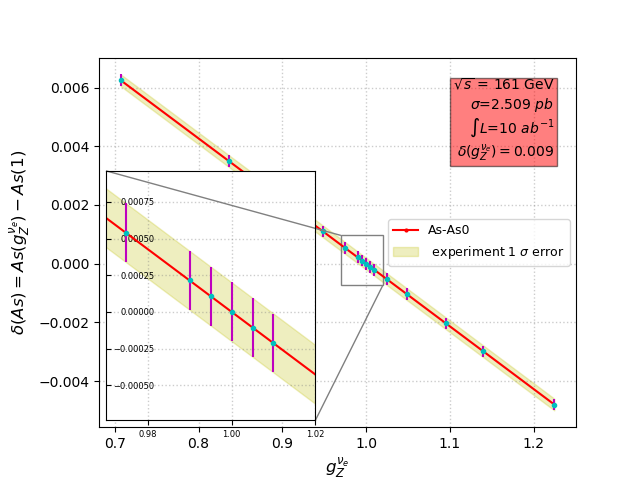} 
\includegraphics[width=0.495\textwidth]{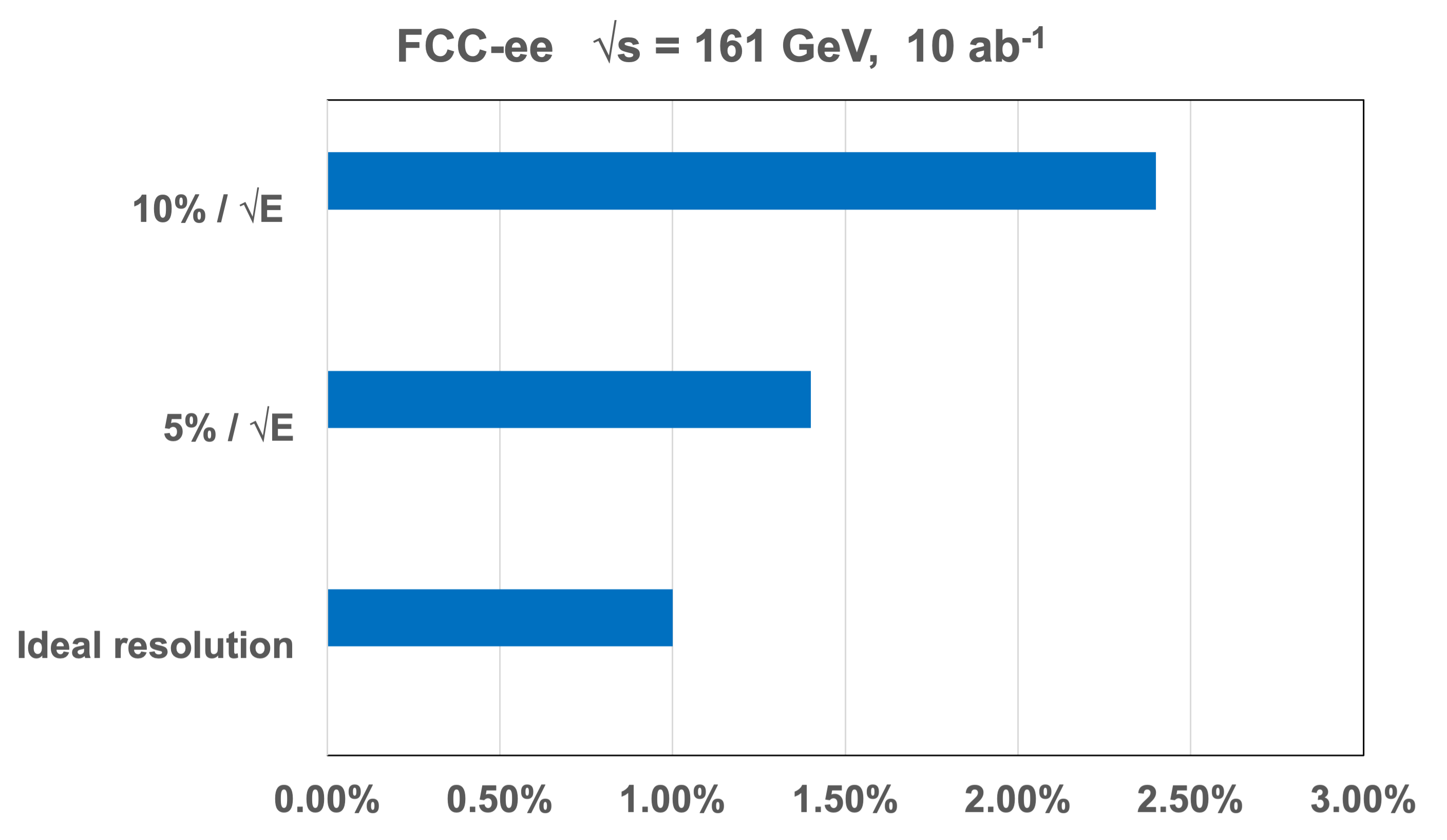}
\caption{Left: Asymmetry $A_s$ of the photon energy spectrum, at the $\PW\PW$ threshold and without detector effects,
after subtracting the SM asymmetry predicted for $g^{\PGne}_{\PZ} = 1$, as a function of the coupling. 
The yellow band represents the statistical uncertainty expected with an integrated luminosity of $10$\,ab$^{-1}$. 
Illustrative measurement points along the prediction, with error bars that span the uncertainty band, are also shown. 
Right: Statistical uncertainty on the coupling $g^{\PGne}_{\PZ}$ expected at the $\PW\PW$ threshold, 
for two example values of the stochastic term of the electromagnetic energy resolution (with a constant term of 0.2\% added in quadrature). 
The uncertainty that would be obtained with a perfect detector is also shown.}
\label{fig:nunugamma}
\end{figure}

As described in Ref.~\cite{Aleksan_2019}, the $\epem \to \PGne \PAGne \PGg$ process at FCC-ee can be used to improve the precision measurement of the poorly known coupling of the \PZ to the \PGne, 
currently known as $g^{\PGne}_{\PZ} = 1.06\pm 0.18$~\cite{ParticleDataGroup:2024cfk}.
This is possible because of the presence of the $t$-channel \PW exchange in the $\epem \to \PGne \PAGne \PGg$ process, interfering with $\epem \to \PZ(\PGne \PAGne) \PGg$. 
This interference slightly deforms the distribution of the missing mass $M_{\PGn \PAGn}$ (or, equivalently, of the photon energy) in the vicinity of the resonant peak, $M_{\PGn \PAGn} = M_{\PZ}$, 
increasing (decreasing) the cross section above (below) the mass peak.
The measurement is based on the asymmetry $A_s$ of the photon energy spectrum in an interval of about $\pm 1$\,GeV around the peak 
(where the photon energy is approximately equal to $\sqrt{s}/2 \times (1 - M^2_{\PZ} / s)$, i.e.\ 54\,GeV at $\sqrt{s} = 161$\,GeV)
using event samples generated with the \textsc{kkmc}~\cite{Arbuzov:2020coe,Arbuzov_2021} matrix element MC generator.   
At the $\PW\PW$ threshold, with an integrated luminosity of 10\,ab$^{-1}$ and without considering any detector effects, 
this asymmetry would be measured with a statistical uncertainty of $2 \times 10^{-4}$, which translates into a 1\%  statistical uncertainty on $g^{\PGne}_{\PZ}$, 
as shown on the left panel of Fig.~\ref{fig:nunugamma}.

To evaluate the impact of a more realistic detector, 
the study was redone including the resolution from a homogeneous calorimeter with an energy resolution of $\sigma(E_{\PGg})\,/\,E_{\PGg} = 0.05/\sqrt{E_{\PGg}} \,\oplus\, 0.002$, 
resulting in a 50\% degradation of the sensitivity to 1.4\%, which remains an excellent performance. 
Should the stochastic term be two times larger (i.e., $0.1/\sqrt{E}$, a value typical for a sampling calorimeter) the sensitivity would degrade to 2.4\%. 
The right panel of Fig.~\ref{fig:nunugamma} summarises the expected uncertainties. 
This study brings a clear constraint on the need for a very good calorimeter energy resolution and knowledge of photon energy calibration, 
to keep the uncertainty due to the calorimeter resolution at the per-cent level, as expected with a perfect detector.

\paragraph{Search for long-lived ALPs that decay outside the detector}

\begin{figure}[ht]
\centering
\includegraphics[width=0.55\textwidth]{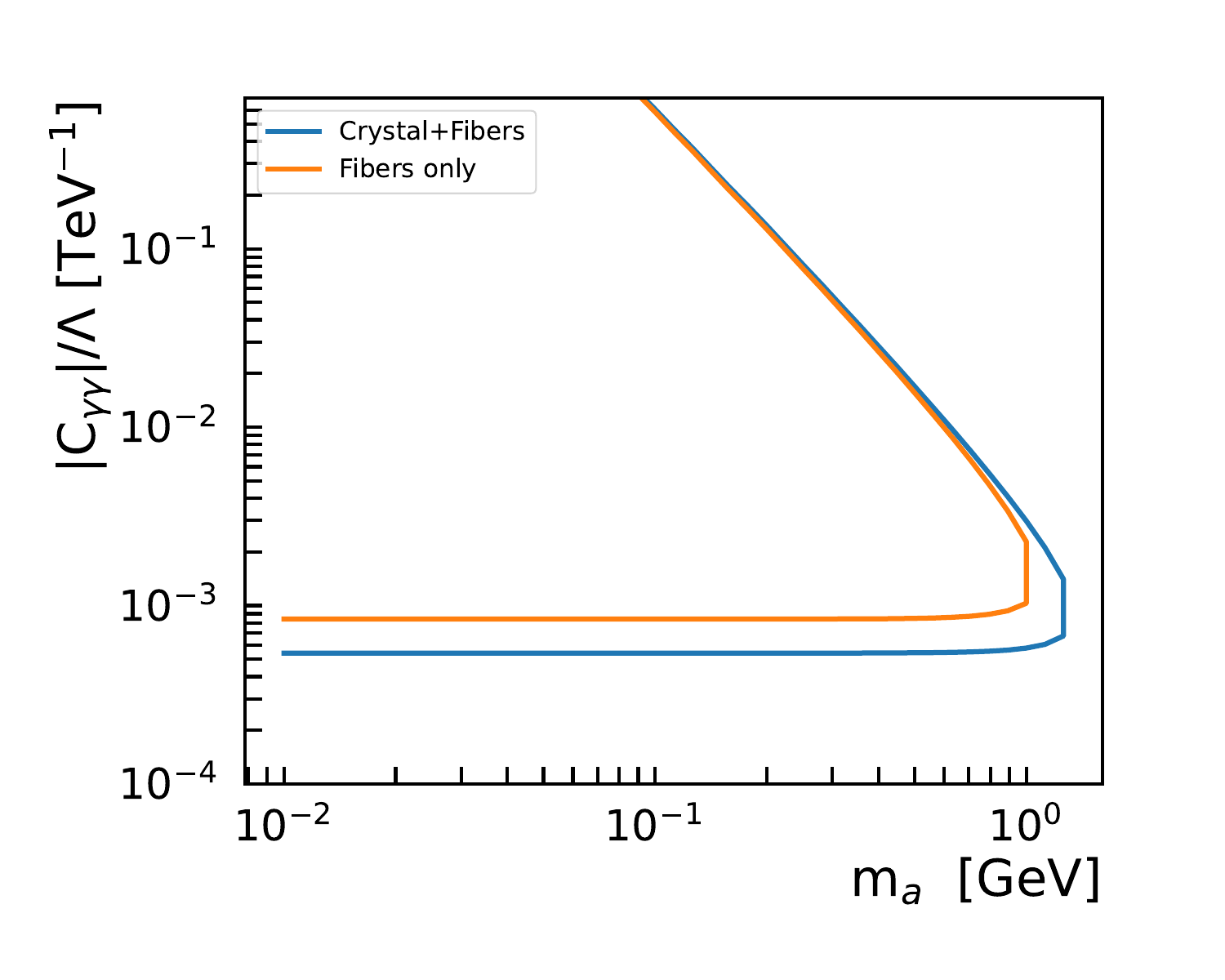}
\caption{Expected sensitivity, at the 2\,$\sigma$ level, of an FCC-ee experiment with a relative EM energy resolution 
of $3\% / \sqrt{E}$ (blue curve) or $14\% / \sqrt{E}$ (orange curve), in the parameter space spanned by the mass of the ALP and its coupling to two photons, 
from a search in the monophoton final state~\cite{note_ALP_Polesello}.}
\label{fig:PhysPerf_ALPs_monophotons}
\end{figure}

Very long-lived axion-like particles (ALPs) could be copiously produced at Tera-\PZ in \PZ decays 
($\epem \to \PZ / \PGg^{*} \to \axion + \PGg$). 
When the ALP $\axion$ decays outside the detector, the final state consists in a (monochromatic) single photon, with nothing else in the detector. 
The relevant mass range for such long-lived ALPs being below ${\cal{O}}$(1)\,GeV, the photon energy is about 45\,GeV.  
The sensitivity of FCC-ee to such particles has been studied in Ref.~\cite{note_ALP_Polesello}. 
Simple kinematic cuts allow the reducible background 
(dominated by the associated production of a photon with two fermions that escape detection) 
to be fully eliminated. 
Since the irreducible background due to $\epem \to \PGn \PAGn \PGg$ production 
rises very steeply when the photon energy decreases, 
the ECAL energy resolution is a key driver of the experimental sensitivity. 
Figure~\ref{fig:PhysPerf_ALPs_monophotons} shows the 2\,$\sigma$ sensitivities expected for a crystal-based ECAL 
and for a dual-readout calorimeter, the former allowing  significantly lower values of the ALP coupling to be probed. 
As was shown in Fig.~\ref{fig:axion}, this analysis would uniquely cover the range between $\lesssim$\,0.1 and $\sim$\,1\,GeV, 
which is beyond the reach of beam dump experiments.

\subsubsection{Energy resolution and decays of heavy flavoured hadrons into photons or \texorpdfstring{\PGpz}s}
\label{sec:PhysPerf:pi0_reco_for_flavour}

The FCC-ee flavour programme at the \PZ pole significantly expands beyond that of Belle~II and LHCb, as described in Ref.~\cite{Monteil:2021ith}. 
In particular, for all decay modes common to the \PBs and the \PBd mesons, and that involve neutral particles, 
the \PBd decays constitute an irreducible background to the \PBs signal. 
In that case, the mass resolution, driven by the electromagnetic calorimeter resolution, is the only handle to separate them.
Two examples have been looked into: 
the radiative decay of $\PB_{(\PQs)}$ into $\PK^* \PGg$ and the $\PBs \to \PDs \PK$ decay where the \PDs decays into $\PGf \PGr$.

\begin{figure}[ht]
\centering
\includegraphics[width=0.45\textwidth]{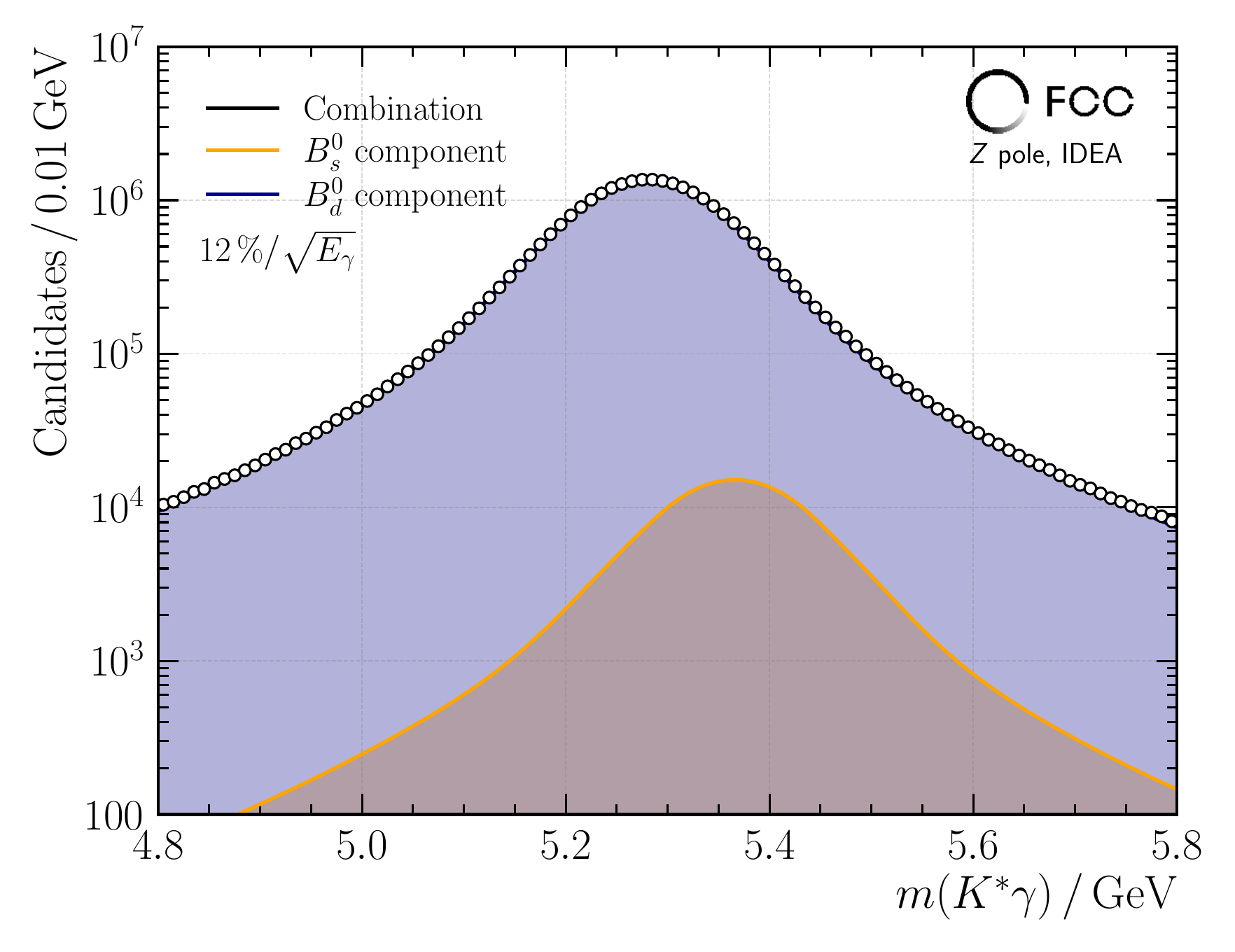}
\includegraphics[width=0.45\textwidth]{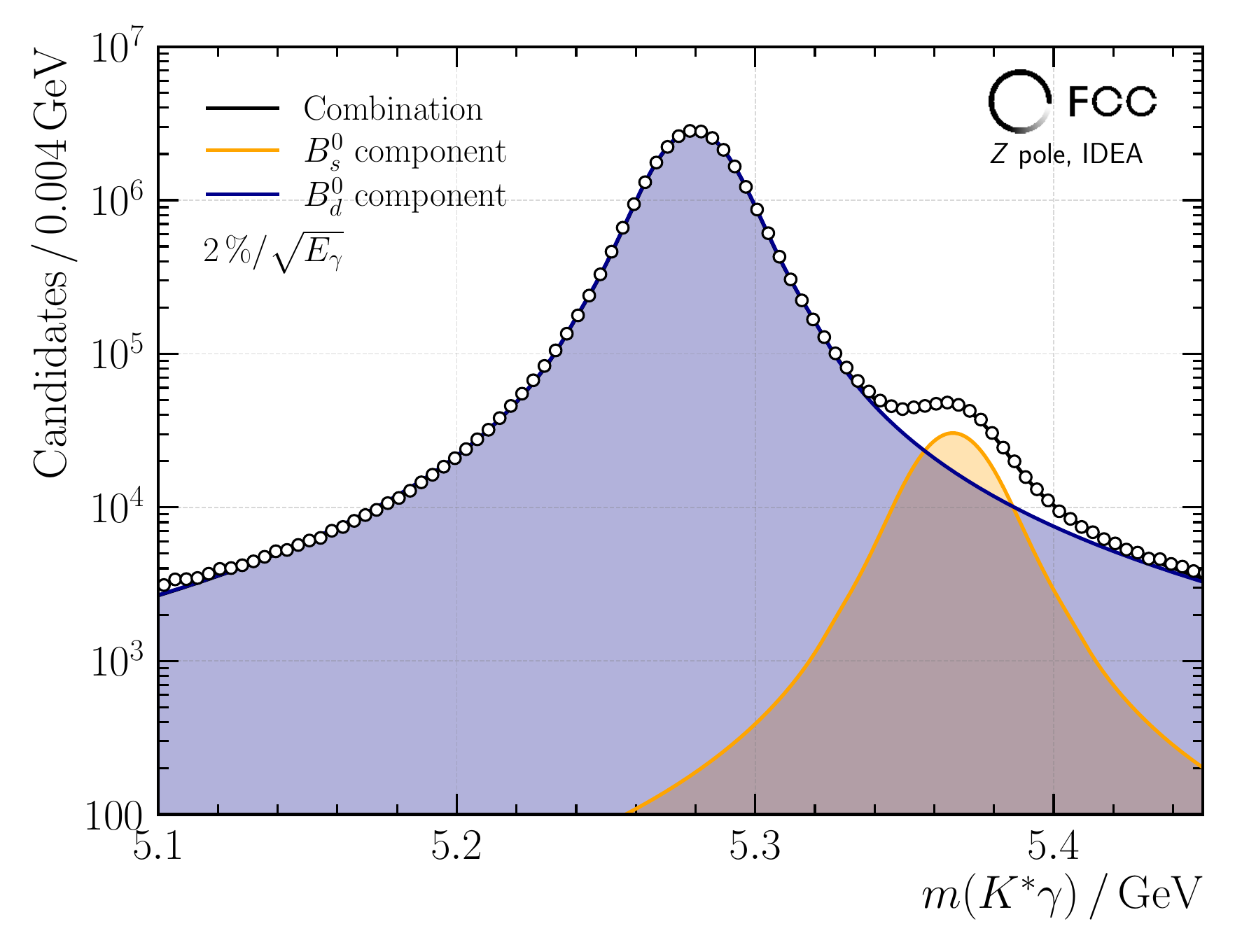}
\includegraphics[width=0.45\textwidth]{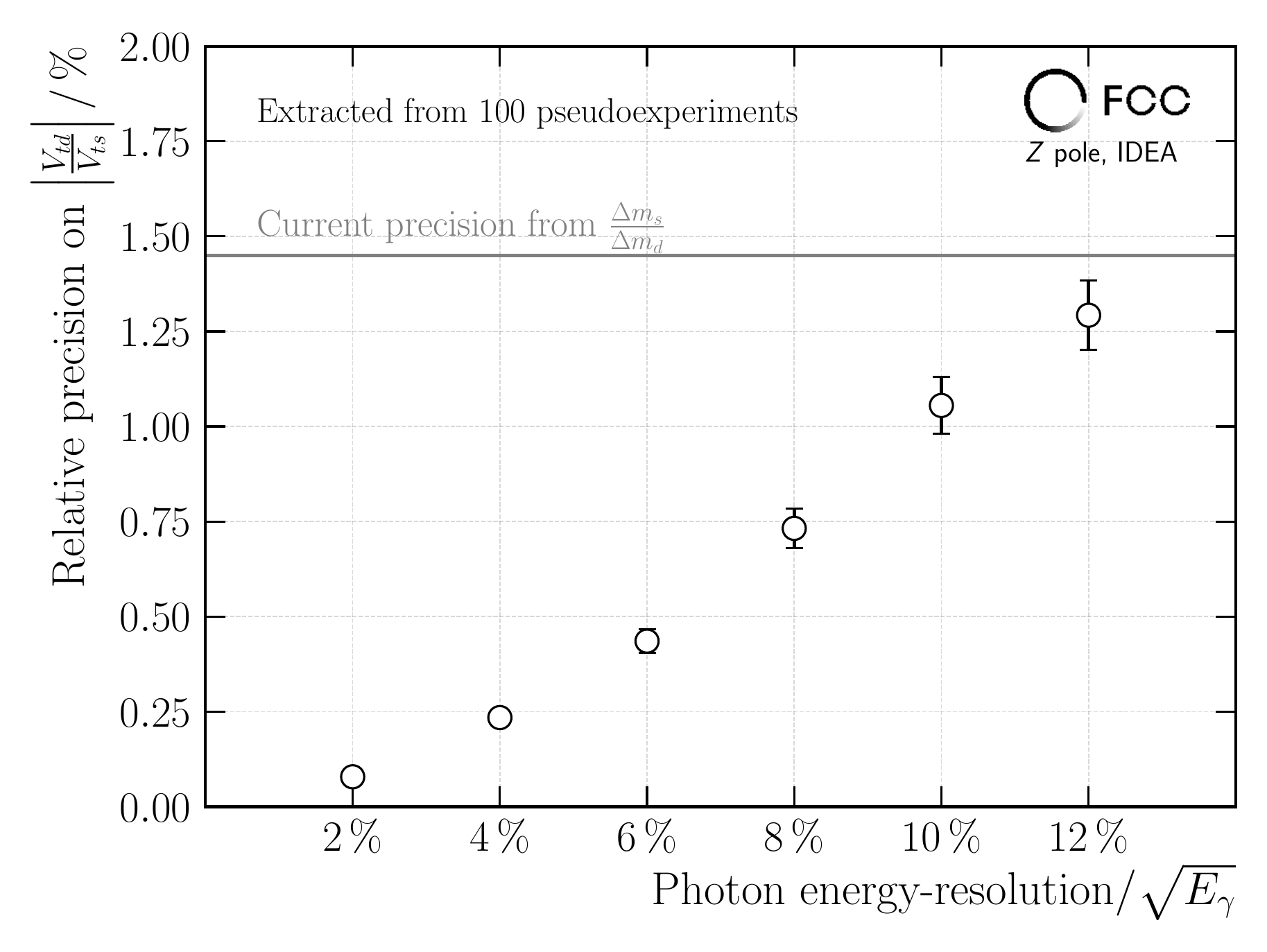}
\caption{Top: Mass distribution of \PB mesons decaying into $\PK^* \PGg$ with an electromagnetic energy resolution 
of $12\% / \sqrt{E}$ (left) or $2\% / \sqrt{E}$ (right). 
Bottom: Relative uncertainty on $|V_{\PQt\PQs} / V_{\PQt\PQd}|$ expected from the measurement of the ratio of the two branching fractions, 
as a function of the stochastic term of the EM energy resolution. The current uncertainty is shown by the horizontal line.}
\label{fig:KstarGamma}
\end{figure}

\paragraph{The radiative decay of the \texorpdfstring{$\PB_{(\PQs)}$}{Bs} meson to \texorpdfstring{$\PK^* \PGg$}{K* gamma}}

Radiative decays, such as $\PBd \to \PK^* \PGg$ ($\PQb \to \PQs \PGg$) and $\PBs \to  \PK^* \PGg$ ($\PQb \to \PQd \PGg$), 
are a sensitive probe for physics beyond the SM and can be used to place complementary constraints on the unitarity triangle.
The yet unobserved $\PBs \to \PK^* \PGg$ decay is an example of $\PQb \to \PQd$ transitions that 
an experiment at a high-luminosity \PZ factory can probably uniquely study, 
provided that it can benefit from a precise photon-energy measurement complementing an excellent reconstruction of displaced vertices.
A first study of the FCC potential to observe this decay has been reported in Ref.~\cite{note_KstarGamma}. 
It assumes perfect charged hadron particle and photon identifications and ignores background contributions.
The reconstructed charged kaon and pion corresponding to the $\PK^* \to \PK \PGp$ decay 
are combined with the generated photon (the energy of which is smeared to account for the detector resolution).
The \PBs signal yield is estimated from the CKM matrix elements and the corresponding hadronisation fractions $f_{\PQd}$ and $f_{\PQs}$, 
leading to the ratio $N_{\PBd} / N_{\PBs} \sim 92$.
The distribution of the resulting ${\PK}^\ast \PGg$ invariant mass is shown in Fig.~\ref{fig:KstarGamma} 
for two values of the stochastic term of the electromagnetic energy resolution.
A sufficiently-high photon-energy resolution is mandatory for the peak of the \PBs signal to become visible 
and distinguishable from the overwhelming $\rm B^0_d$ counterpart spectrum. 
With a stochastic term of 2\%, the ratio of the two  branching fractions could be extracted with a relative statistical uncertainty better than 0.1\%. Provided that the ratio of the form factors of the two decays is well known, 
this opens the way to a very precise direct measurement of the $|V_{\PQt\PQd} / V_{\PQt\PQs}|$ ratio, as shown in the bottom plot of Fig.~\ref{fig:KstarGamma}.
A measurement may still be possible with a resolution of $12 \% / \sqrt{E}$, 
but that would require an excellent control of the shapes of the mass distributions, 
in particular of the tails (assumed here to be perfectly known), 
and the precision would anyway degrade by at least an order of magnitude.

\paragraph{The \texorpdfstring{$\PBs \to \PDs \PK$}{Bs to Ds K} decay}

\begin{figure}[ht]
\centering
\includegraphics[width=0.495\textwidth]{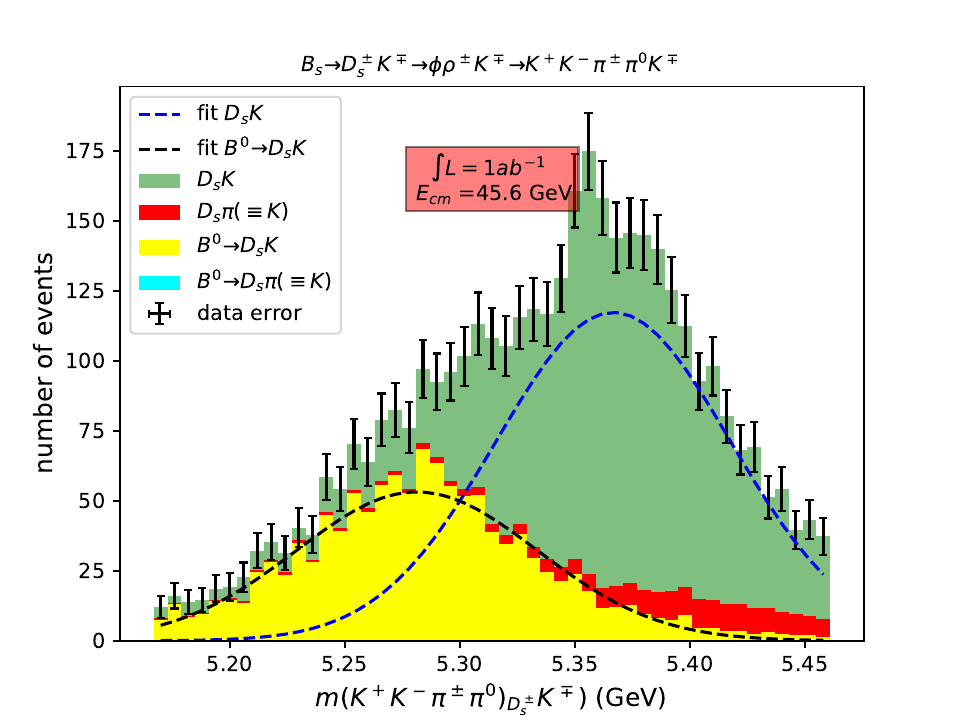} 
\includegraphics[width=0.495\textwidth]{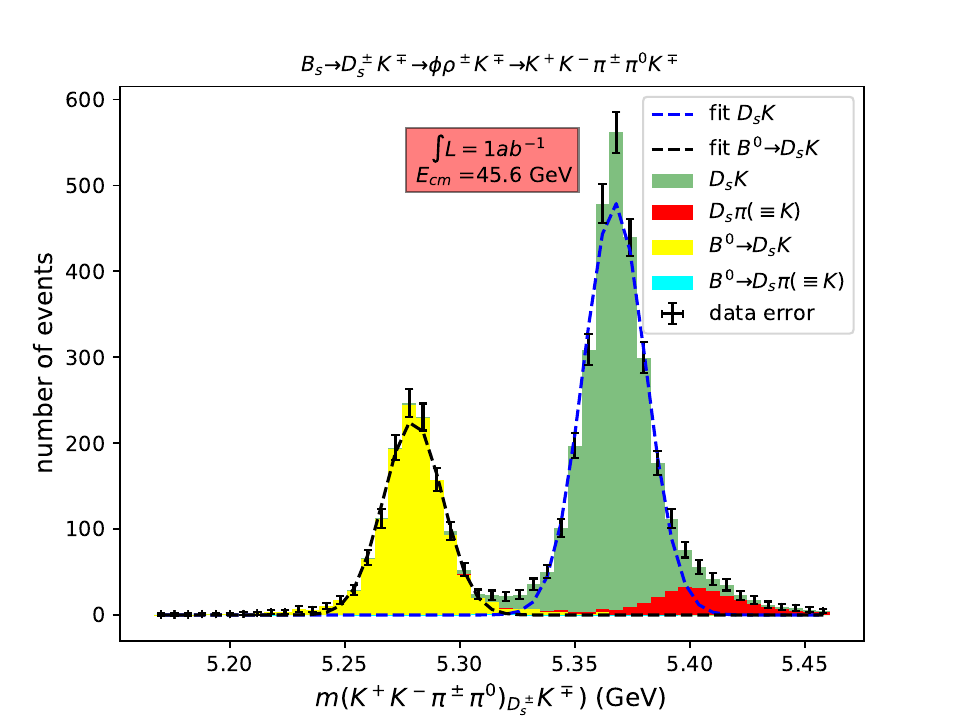} 
\caption{Distribution of the mass of the reconstructed \PBs system (in GeV) for the process $\PBs \to \PDs \PK \to \PGf\PGr \PK \to \PK\PK \PGp\PGpz \PK$, 
on top of the main backgrounds, for a calorimeter with an energy resolution of $15\% / \sqrt{E}$ (left) or $3\% / \sqrt{E}$ (right). 
PID is included. 
The resolution on the photon angles has a negligible effect on the resolution of the mass of the \PBs candidates.}
\label{fig:BsDsK}
\end{figure}

The copious \PBs decay modes to neutral particles are inaccessible at Belle~II and challenging at LHCb, and can be uniquely studied at FCC-ee. 
A few benchmark studies have been performed to extract the necessary detector requirements for PID (Section~\ref{sec:PhysPerf_PID}) and calorimetry. 
One of them is the study of the $\PBs \to \PDs \PK$ decay~\cite{AleksanFCCPhys4}, 
where the \PDs decays via $\PDps \to \PGf\PGr \to \PKp\PKm \PGpp\PGpz$, 
which could increase the size of the \PBs sample by a factor of 3 with respect to that obtained with the single mode studied in Section~\ref{sec:PhysPerf_PID}. 
Figure~\ref{fig:BsDsK} shows the impact on the mass resolution of the \PBs system when a crystal-like calorimeter is employed, 
improving the resolution of the calorimeter 
from $\sigma_E\,/\,E = 0.15/\sqrt{E} \,\oplus\, 0.005$, corresponding to a resolution of 51\,MeV, 
to $\sigma_E\,/\,E = 0.03/\sqrt{E} \,\oplus\, 0.005$, corresponding to 14\,MeV.

\subsubsection{Requirements on the geometrical acceptance for 
\texorpdfstring{$\epem \to \PGg\PGg$}{e+e- to gamma gamma} and \texorpdfstring{$\ell^+\ell^-$}{ell+ ell-} events at \texorpdfstring{\PZ}{Z}-pole energies}
\label{sec:PhysPerf_gammagamma}

The determination of the \PZ lineshape parameters requires the measurement of the hadron and lepton cross sections at and around the \PZ pole. 
With $1.5 \times 10^{12}$ \PZ bosons produced at each FCC-ee interaction point, statistical precisions of ${\cal{O}}$($10^{-6}$) are within reach. 
Critical limiting uncertainties come from the counting of lepton pairs ($\epem \to \ell^+\ell^-$, with $\ell = \PGm$, \PGt) 
and from the integrated luminosity measurement. 
At \mbox{FCC-ee}, the wide-angle diphoton process $\epem \to \PGg\PGg$ becomes statistically relevant and offers prospects for a safer determination 
(in terms of systematic uncertainties) of the integrated luminosity than the traditional low-angle Bhabha scattering, $\epem \to \epem$.

\begin{figure}[ht]
\centering
\includegraphics[width=0.50\textwidth]{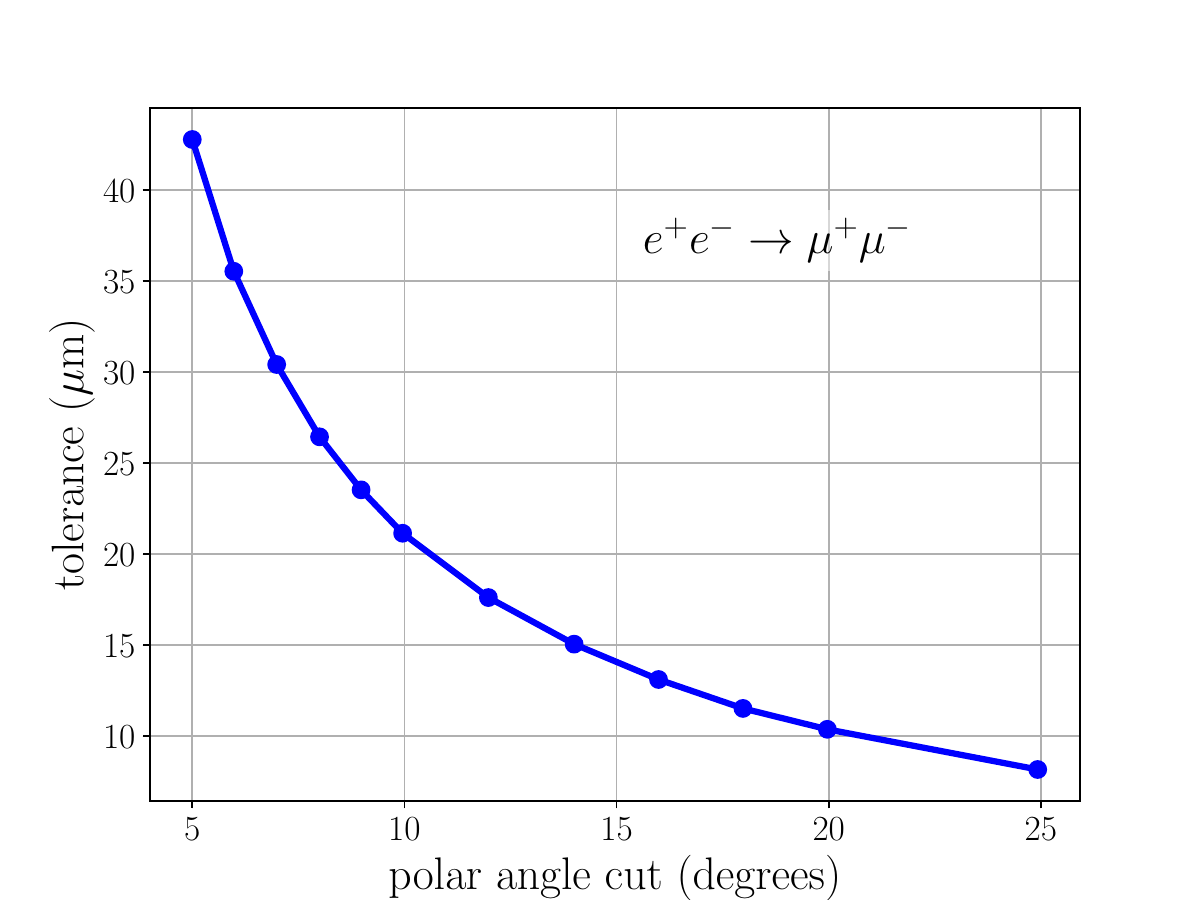}
\includegraphics[width=0.48\textwidth]{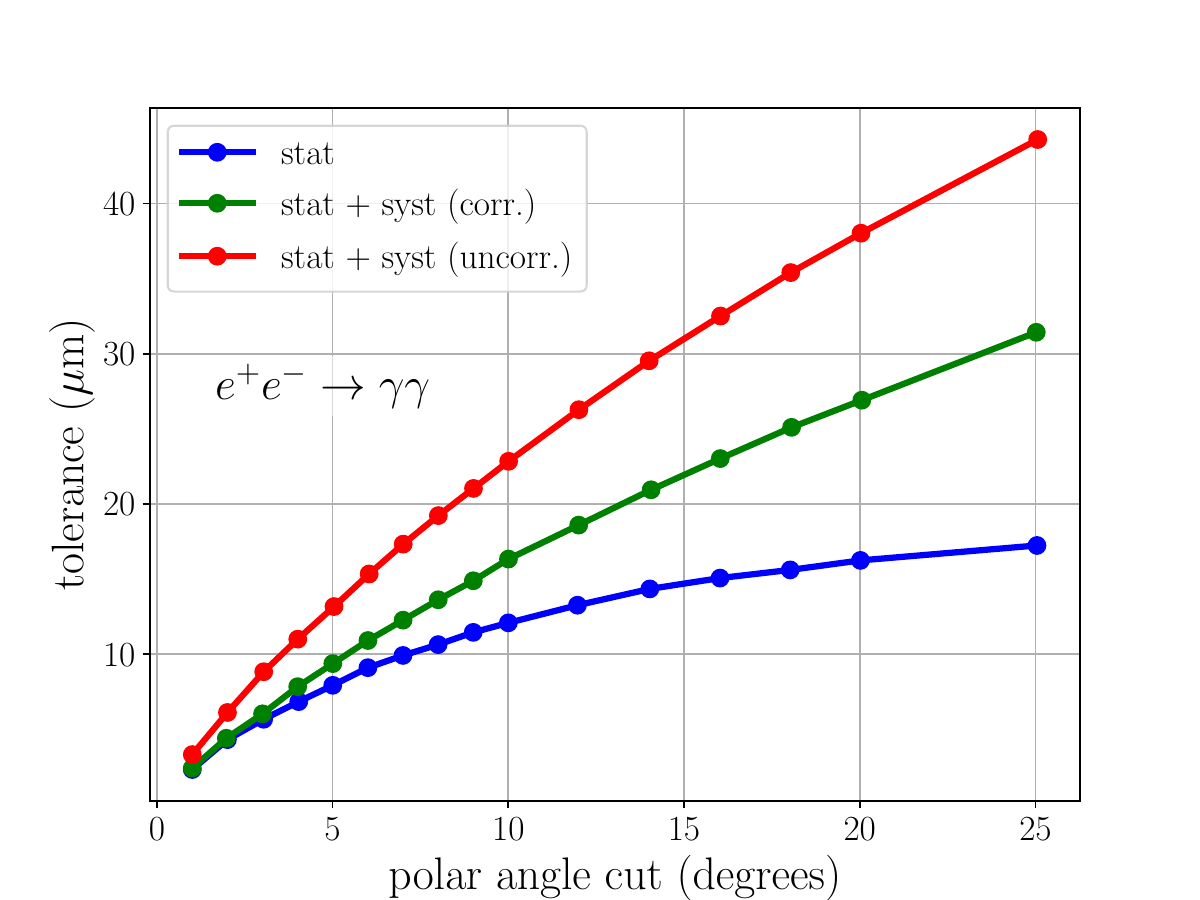}
\caption{Tolerance on the accuracy of the polar angle cut as a function of the polar angle cut in degrees, 
expressed as a radial accuracy for a detector situated at 2.5\,m from the IP in the centre-of-mass of the collision. 
Left: For dilepton events, to ensure a systematic precision of $2.2 \times 10^{-6}$. 
Right: For diphoton events, to ensure a systematic precision of $1.5 \times 10^{-5}$ (lower blue curve). 
Also indicated is the tolerance corresponding to having a systematic uncertainty equal to the statistical one, 
assuming either fully correlated (middle green curve) or uncorrelated (upper red curve) displacements in the two endcaps.}
\label{fig:tolerance}
\end{figure}

The definition of the geometrical acceptance (defined, e.g., by a given polar angle lower cut in the centre-of-mass frame of the collision) 
for both $\epem \to \PGg\PGg$ and $\epem \to \ell^+\ell^-$ is an obvious source of a potentially large systematic bias, which was dominant at LEP. 
Indeed, any global or local detector misalignment can generate a systematic bias on the acceptance. 
The tolerance on this bias can be calculated, 
e.g., either to match the resulting systematic uncertainty on the cross section to its statistical precision or to obtain a specific precision, 
as done in Ref.~\cite{Blondel202308} and displayed in Fig.~\ref{fig:tolerance} for a polar angle cut between 1 and 25~degrees.

The tolerance, expressed as the accuracy of the definition of the polar angle cut for a detector situated at 2.5\,m from the IP, is similar for dileptons and diphotons, 
and is in the 10--20\,$\mu$m range for a cut between 10 and 20~degrees, corresponding to a required polar angle accuracy of 4 to 8\,$\mu$rad.

Inspired by the work done for integrated luminosity measurements with low-angle Bhabha scattering at LEP, 
such a tight requirement calls for the design and the construction of the calorimeter (for diphotons) 
and the tracker (for dileptons) endcaps with a mechanical precision ensuring the specified tolerance of 10 to 20\,$\mu$m, 
complemented by dedicated test-beam measurements
and a survey/monitoring of the distance between the two endcaps with a precision of 50 to 100\,$\mu$m. 
While the task is  challenging, it should be kept in mind that for the luminosity monitor (LumiCal), situated at lower angles, 
the transverse (longitudinal) tolerance is of 1\,$\mu$m (50\,$\mu$m) at 1\,m of the luminous region, 
for a precision of $10^{-4}$ on the integrated luminosity measurement~\cite{Dam:2021sdj}. 

\begin{figure}[ht]
\centering
\includegraphics[width=0.6\textwidth]{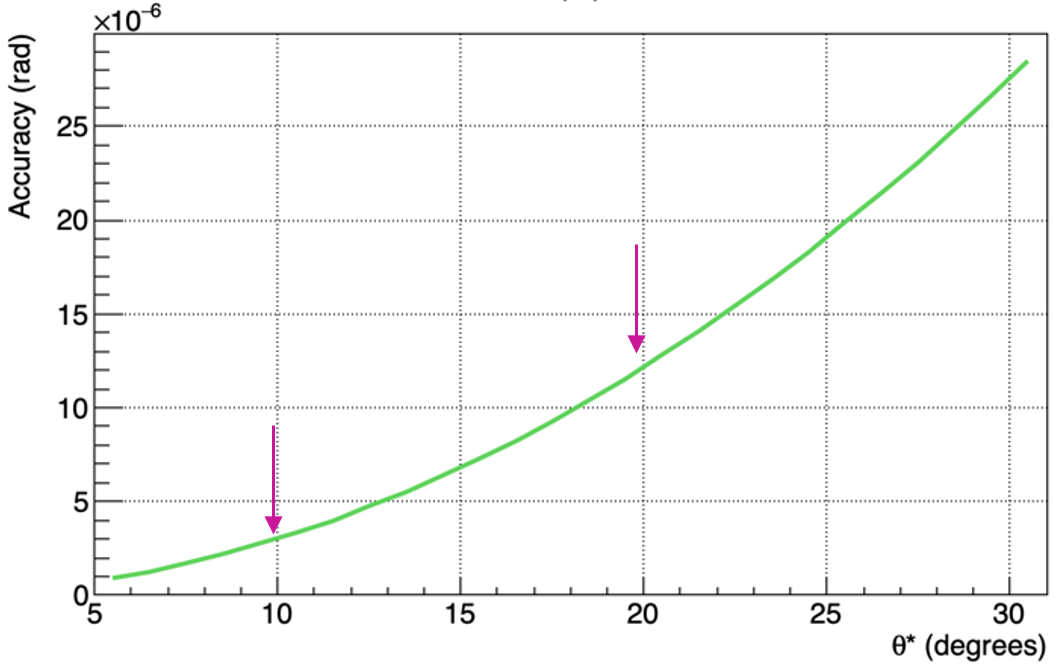} 
\caption{Analytically-predicted precision of the diphoton acceptance determination (in $\mu$rad) 
as a function of a polar angle cut applied in the event-by-event centre-of-mass frame, 
assuming a 100\,$\mu$m tolerance in the azimuthal direction, 
and exploiting the energy-momentum conservation 
and the constraints that the beam crossing angle and the longitudinal boost are the same (on average) for all events 
and constant in time.
This result is obtained with a readout cell of $1^\circ \times 1^\circ$ and position resolution of $\sigma_{x,y}=0.5$\,mm in the endcap calorimeter 
(commensurate with the various proposed options).}
\label{fig:PhotonAcc}
\end{figure}

With respect to the LEP case, FCC-ee offers an additional totally new and possibly synergistic approach to a precise definition of the acceptance. 
Indeed, the large and constant beam crossing angle of 30\,mrad provides an absolute angular scale to dilepton and diphoton events at all angles. 
Total energy and momentum conservation allows for the event-by-event determination of this crossing angle 
(together with the constant longitudinal boost resulting from the specific positioning of the accelerating RF cavities in the FCC tunnel) 
with excellent accuracy from the final state particle kinematic observables.
A global fit of all events, constraining the crossing angle and the longitudinal boost to be constant, 
would then provide an in-situ local `calibration' of final state particle angles and, in particular, of the acceptance cut. 
The required statistical precision was demonstrated analytically~\cite{JanotFCCWeek} to be within reach with the FCC-ee dilepton and diphoton samples, but the actual fit procedure remains to be 
consolidated\,\footnote{It was already shown in Ref.~\cite{Blondel:2019jmp} that this finite crossing angle and longitudinal boost 
could be used to find and monitor the directions of the $x$, $y$, and $z$ axes within a few $\mu$rad precision every hour at the \PZ pole with dimuon events.}. 
The sensitivity of this approach is proportional to the magnitude of the crossing angle, 
and could not be used with the LEP head-on collisions.
At FCC-ee, it was already shown~\cite{JanotFCCWeek} that a (mechanical or otherwise) tolerance of 100\,$\mu$m in the $r$-$\phi$ direction 
would allow an acceptance definition of 3 to 12\,$\mu$rad, 
irrespective of the (global or local) endcap radial or longitudinal displacements, 
consistent with the tolerances mentioned above, as displayed in Fig.~\ref{fig:PhotonAcc}. 
More recent developments seem to indicate that the required $\phi$ tolerance can also be reached in situ. 

Further work is needed to verify the feasibility of the second method beyond analytical computations, 
as well as to evaluate the effect of the beam crossing angle on the acceptance determination with the classical method. 
The large forward cross section of $\epem \to \epem$ events will open the possibility of mutual in-situ alignment of the tracker and the calorimeter, 
offering a cross-check of the acceptance cut. 
It will be important to check whether the second approach would also be suitable for the integrated luminosity measurement 
with low-angle Bhabha scattering in the luminosity monitors. 
The two approaches are expected to help each other, 
either by reducing the mechanical constraints of the former or by improving the convergence of the latter, and most likely both. 
Their combination is expected to provide an excellent understanding of the acceptance, 
a cross-check of their accuracy, and an explicit requirement on the position and angular resolutions of the tracker and of the calorimeter. 

\subsubsection{Sensitivity to charged lepton flavour violation: \texorpdfstring{$\PZ\to \PGm e$}{} and \texorpdfstring{$\PGt\to \PGm\PGg$}{}}

With its $6 \times 10^{12}$ \PZ bosons, FCC-ee enables precise tests of charged lepton flavour violation (cLFV) processes. 
The $\PZ \to \PGm \Pe$  process has an extremely clean signature --- a beam-energy electron recoiling back-to-back against a beam-energy muon --- 
with fully reducible background from the $2 \times 10^{11}$ $\PZ \to \PGt^+\PGt^-$ events. 
It is estimated~\cite{Dam:2021ibi} that FCC-ee will improve by 2 to 3 orders of magnitude 
the current limits of about $10^{-7}$ from the LHC~\cite{Atlas:2022uhq}. 
The most dangerous background comes from the `extreme bremsstrahlung' emission by a muon in a $\PZ \to \PGm\PGm$ event 
that creates a large deposit in the electromagnetic calorimeter, faking an electron. 
This effect could be limited by improving the ECAL energy resolution and having the possibility of longitudinal segmentation to veto showers starting later in the calorimeter. 

\begin{figure}[ht]
\centering
\includegraphics[width=0.4\textwidth]{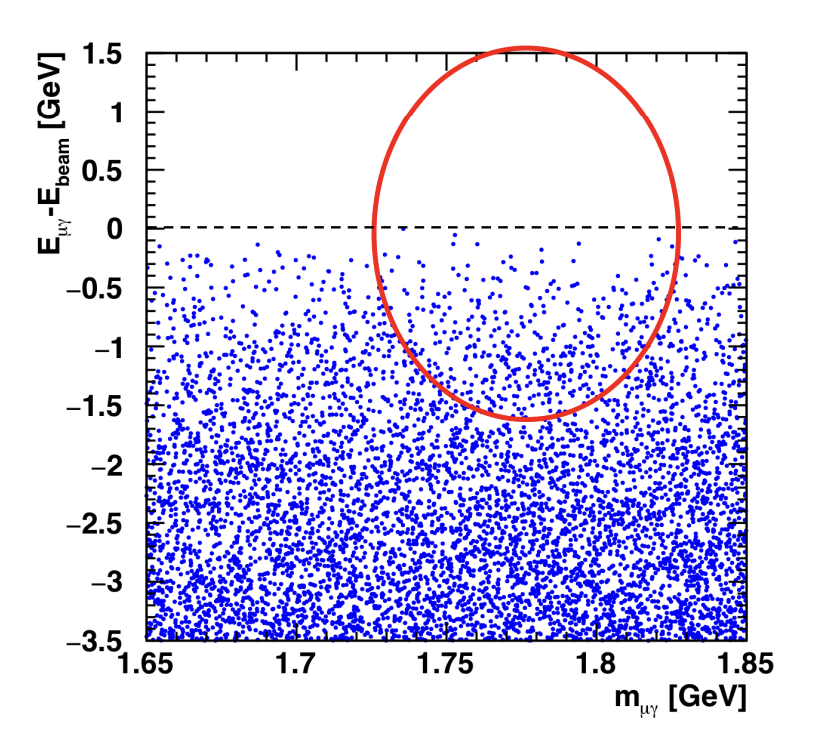}
\includegraphics[width=0.58\textwidth]{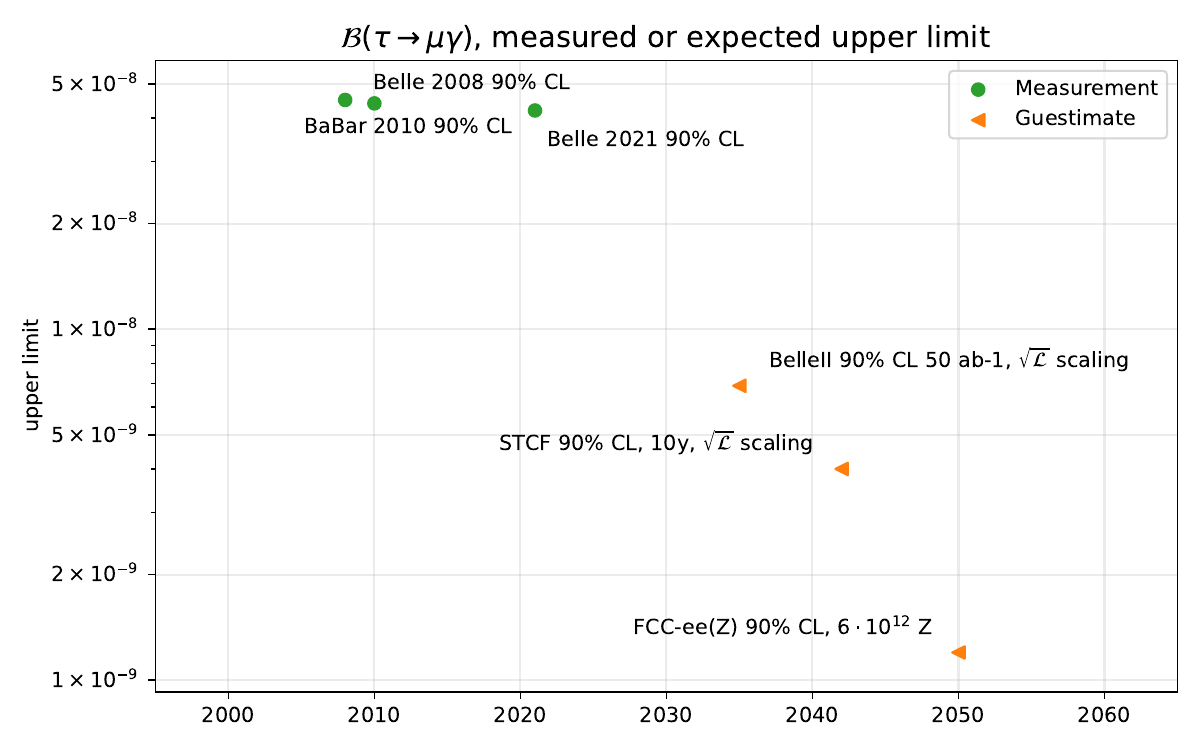}
\caption{Left: Reconstructed energy vs.\ mass for all $\PGm\PGg$ combinations in simulated $\PZ \to \PGt\PGt\PGg$ background events with one $\PGt \to \PGm \PAGn \PGn$ decay. 
The signal region corresponding to 2\,$\sigma$ resolution is indicated by the red ellipse. 
Right: Present and expected upper limits for ${\cal B}(\PGt\to\PGm\PGg)$~\cite{Lusiani:2024tau}. 
The dates of the future measurements are mainly chosen for plotting purposes. 
The expected limits for Belle~II and the Super Charm-Tau factories are conservative estimates, assuming that those searches will be mostly background-constrained.}
\label{fig:taumugamma}
\end{figure}

The sensitivity of FCC-ee to the LFV $\PGt \to \PGm \PGg$ decay was established in Refs.~\cite{Dam:2021ibi, Dam:2018rfz}
and the major detector effects on the measurement were evaluated. 
The analysis strategy requires the identification of a clear SM \PGt decay on one side (`tag') 
so that the search for the LFV decay is performed in the other hemisphere. 
The discriminant variables are the total energy and the invariant mass of the final state. 
As shown in Fig.~\ref{fig:taumugamma}~(left), the signal region is background dominated and 
the background density rises linearly away from the $E_{\PGm\PGg} - E_\text{beam}=0$ line. 
This implies that the sensitivity (upper limit) scales linearly with the $E_{\PGm\PGg}$ resolution, which is dominated by the ECAL energy resolution. 
With the full event sample of $4 \times 10^{11}$ \PGt decays, 
an expected 90\%~CL upper limit of $1.2\times 10^{-9}$ can be reached~\cite{Lusiani:2024tau}, 
as shown in Fig.~\ref{fig:taumugamma}~(right).

This benchmark is particularly interesting because it also shows a requirement for the photon position resolution, which contributes to the invariant mass calculation.  
In this example, where the calorimeter has a resolution of about $16.5\%/\sqrt{E}$ (as in CLD) and a position resolution of 2\,mm, 
the two contributions in the invariant mass calculation are similar. 
Therefore, for this measurement, while the sensitivity is found to scale linearly (or slightly more strongly) with the photon energy resolution, 
a fine-grained crystal calorimeter with a $3\%/\sqrt{E}$ energy resolution would be particularly valuable. 
This is especially true since it would also provide a significantly better spatial resolution, of around 1\,mm, 
leading to a fivefold improvement in sensitivity compared to the previous example.

\subsubsection{Bremsstrahlung recovery}

\begin{figure}[ht]
\centering
\includegraphics[width=0.45\textwidth]{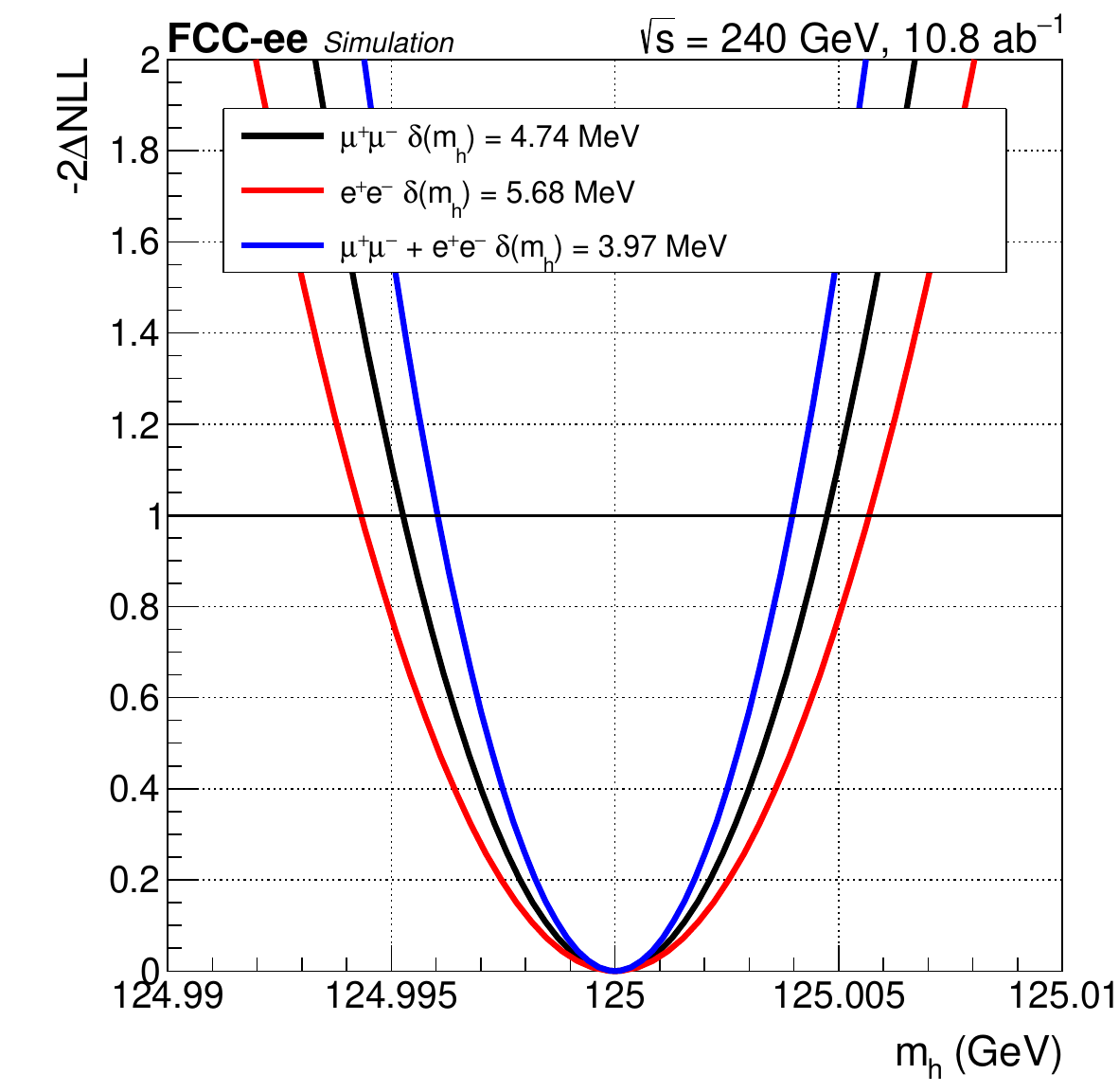} 
\caption{Likelihood scan of the Higgs boson mass with the recoil method, combining the muon and electron channels. 
The uncertainties correspond to the quadratic sum of the statistical and systematic terms.}
\label{fig:mH_recoil-ee}
\end{figure}

Even with an excellent EM calorimeter resolution, 
for 50\,GeV central electrons the resolution of a calorimetric measurement is worse than the resolution of the tracker determination of the momentum. 
Furthermore, the latter is not as good as that of muons since electrons emit bremsstrahlung radiation: 
even with the light tracker of IDEA, corresponding to $\sim$\,10\% of a radiation length in the central region, 
the effective momentum resolution of electron tracks is at least a factor of two worse than that of muons~\cite{note_bremRecovery}. 
Several measurements call for a better electron momentum resolution. 
For example, the determination of the Higgs boson mass from $\PZ(\PGm\PGm)\PH$ events (Section~\ref{sec:PhysPerf_HiggsRecoil}) 
can gain from exploiting the $\PZ \to \epem$ channel. 
A resolution improvement can be brought by the recovery of the bremsstrahlung photons in the calorimeter, 
if their energy can be added to the momentum of the track. 
An effective recovery requires that the showers of the electron and of the radiated photon(s) be disentangled by the reconstruction algorithm. 
For 50\,GeV central electrons, in a 2\,T magnetic field, photons radiated at small radii 
(where, in the IDEA tracker, most of the bremsstrahlung emissions happen) 
are separated by typically 2--3\,cm from the electron at the calorimeter entrance.  
The ability to separate showers that are so close to each other calls for a transverse granularity of $1 \times 1$\,cm$^2$ or better 
and may place some demands on the longitudinal segmentation, which helps to disentangle showers that partially overlap. 
Detailed \textsc{Geant} simulations and reconstruction algorithms are needed for a precise assessment of these constraints. 
Meanwhile, it is estimated that, with the recovery of bremsstrahlung photons, 
the effective resolution of electrons would be 25\% (45\%) worse than the muon resolution 
for a calorimeter EM energy resolution of $3\% / \sqrt{E}$ ($15 \% / \sqrt{E}$). 
More details can be found in Ref.~\cite{note_bremRecovery}. 
The \textsc{Delphes} simulations used for the studies reported in this document rely on this estimation. 
With this bremsstrahlung recovery, the $\PZ \to \epem$ channel significantly contributes to the determination of the Higgs boson mass from $\PZ\PH$ events, 
as shown in Fig.~\ref{fig:mH_recoil-ee}, improving the uncertainty by 22\% compared to what is obtained from the muon channel alone~\cite{HiggsNoteRecoil}.

\subsubsection{Prompt decays of ALPs \texorpdfstring{$\axion \to \PGg \PGg$}{a to gamma gamma}}

A search for ALPs produced in $\PZ \to \axion \PGg$ decays, complementary to the one previously mentioned, 
consists in looking for ALPs that decay promptly into two photons, leading to a three-photons final state. 
The analysis presented in Ref.~\cite{note_ALP_Polesello} compares the sensitivities that are expected with the dual-readout fibre calorimeter of IDEA 
and with the variant where crystals are placed in front of it.

The experimental sensitivity is driven by the ability to precisely reconstruct the invariant mass of the two photons coming from the ALP decay. 
For low masses, up to a few hundreds of MeV, the resolution on this diphoton mass is dominated by the resolution on the photon angles, 
hence by the precision with which the impact point of the photons at the calorimeter entrance is determined. 
This contribution decreases with increasing ALP mass and, above 10\,GeV, the resolution depends only on the energy resolution of the calorimeter. 
In addition, at low masses, below $\sim$\,5\,GeV, the two photons from the ALP decay may not be resolved in the detector but be `merged' into a single reconstructed object.
Detailed \textsc{Geant} simulations and reconstruction algorithms, still under development, are needed for a precise understanding of these effects. 
A simplified approach is adopted for the current sensitivity study, 
whereby the angular distance $\Delta \alpha$ between the two photons is used to determine whether the photons are resolved.
Taking into account the transverse granularity and the Moli{\`e}re radius of the detectors, 
it is assumed that photons with $\Delta \alpha > 20$\,mrad can be reconstructed in the crystal calorimeter, 
which corresponds to a separation of 4\,cm (four calorimeter cells) of the two photons at the calorimeter entrance. 
For the dual-readout calorimeter, photons are required to be separated by $\Delta \alpha > 10$\,mrad to be resolved, 
which corresponds to the distance between 10 fibres of the detector.

\begin{figure}[ht]
\centering
\includegraphics[width=0.5\textwidth]{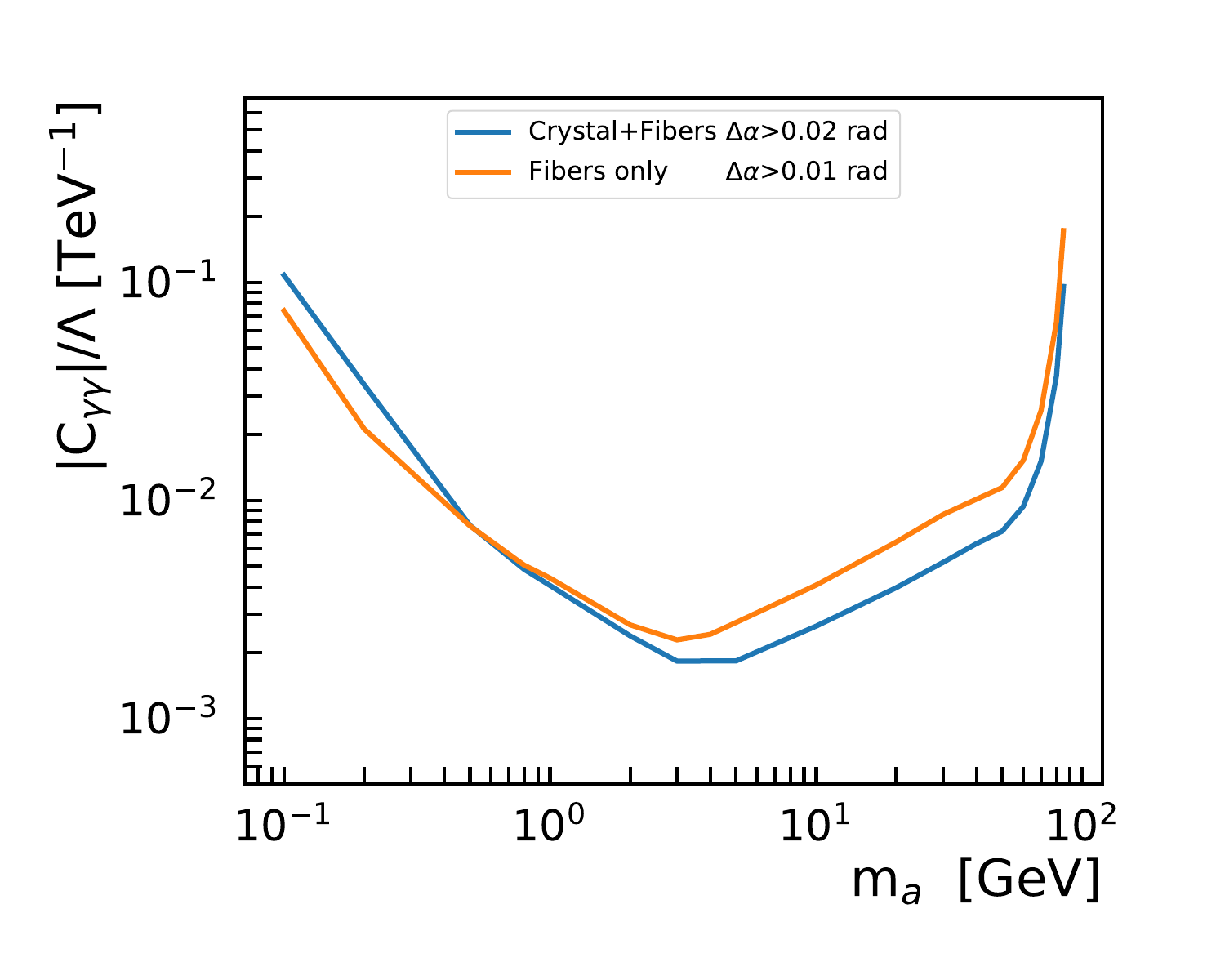} 
\caption{Expected sensitivity at the 2\,$\sigma$ level of a FCC-ee experiment, 
in the parameter space spanned by the mass of the ALP and its coupling to two photons, 
from a search for ALPs that decay promptly into two photons. 
In the legends, $\Delta \alpha$ denotes (in rad units) the angular separation down to which the calorimeter is assumed to be able to resolve two photons.}
\label{fig:ALPs_prompt}
\end{figure}

Taking into account the irreducible $\epem \to \PGg\PGg\PGg$ background, 
the experimental reach expected for the two calorimeter options is shown in Fig.~\ref{fig:ALPs_prompt}. 
The crystal calorimeter has a better sensitivity for high masses, where the sensitivity is driven by the energy resolution.
The impact point resolution is very similar for the two calorimeters, 
but the better granularity of the fibre calorimeter is expected to provide a better separation for collimated photons that emerge from the decay of ALPs at very low masses, 
resulting in a gain in sensitivity compared to that provided by the crystal detector.

As shown by Fig.~\ref{fig:ALPs_prompt}, the parameter space covered by this analysis, that focuses on prompt decays of ALPs, 
nicely complements the one explored by a search or quasi-stable ALPs in the monophoton final state (Fig.~\ref{fig:PhysPerf_ALPs_monophotons}). 
The sensitivity to ALPs with intermediate lifetimes, which are long-lived but decay inside the detector, is more difficult to assess. 
For ALPs that decay before reaching the calorimeter, it depends on the ability to reconstruct the displaced decay vertex of the ALP into two photons. 
For ALPs that decay inside the calorimeter, the sensitivity depends on the ability to identify late-starting showers. 
The requirements that these analyses will place on the longitudinal segmentation of the calorimeter, on the measurement of the timing of calorimeter signals, 
and on a potential preshower, remain to be studied.

\subsubsection{Requirements from \texorpdfstring{$\PGpz \to \PGg\PGg$}{pi0 to gamma gamma} reconstruction in \texorpdfstring{\PGt}{tau} decays}
\label{sec:PhysPerf_Pi0sFromTaus}

\begin{figure}[ht]
\centering
\includegraphics[width=\textwidth]{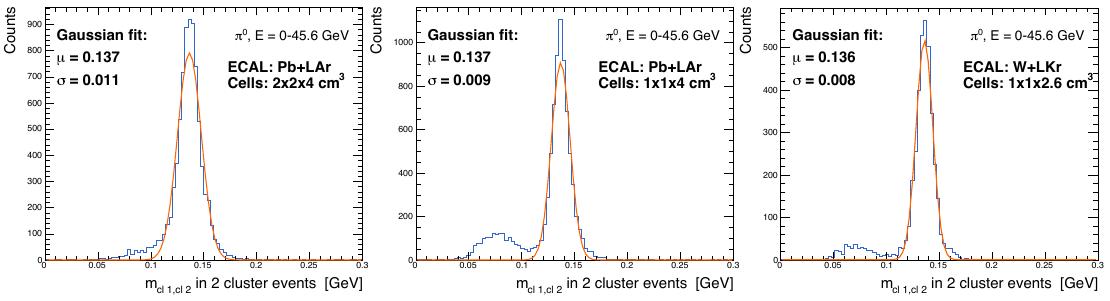} 
\caption{Invariant mass distributions of two-cluster \PGpz mesons for different noble liquid calorimeter configurations.} 
\label{fig:Pi0RecoComp}
\centering
\includegraphics[width=0.5\textwidth]{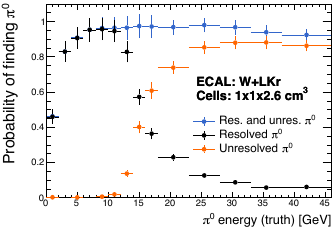} 
\caption{Probability of \PGpz identification as a function of its energy for a W$+$LKr calorimeter with a granularity of $1\times 1\times 2.6$\,cm$^3$.}
\label{fig:Pi0Reco}
\end{figure}

The reconstruction of $\PGpz \to \PGg\PGg$ decays is a crucial component for several measurements of the FCC-ee physics programme, 
such as the tau branching ratios or the tau polarisation. 
The kinematic of a $\PGt \to \PGr\PGn$ decay produced at the \PZ pole provides initial detector requirements~\cite{KWandallThesis}. 
For instance, the momentum of the \PGpz produced in the \PGr decay is quite soft and its decay photons are even softer.
The capability of identifying these low-energy photons requires a low noise and high energy resolution calorimeter. 
The energy resolution is important also because it affects the \PGpz identification itself, which relies on a cut on the diphoton invariant mass.
The granularity is another important aspect that plays a role in the separation of the two photons. 
The angular separation between the two photons, $\approx 2 m_{\PGpz} / E$, is typically 10\,mrad, 
which leads to a separation of 2\,cm at the calorimeter entrance; 
however, for the highest possible energy ($E_{\PGpz} = 45.6$\,GeV) the opening angle is reduced to 6\,mrad, 
corresponding to 1.2\,cm separation at the calorimeter 
entrance\footnote{A large tracking volume, optimised for track momentum resolution, 
is also beneficial for the identification of collimated objects such as \PGpz decays.}. 
The granularity is also needed for the correct separation of the photons from charged pions in the decays. 
A study has been done comparing the performance of different noble liquid calorimeters (LAr and LKr) in \PGpz events. 
Cases have been considered where the photons from the \PGpz decays can be reconstructed separately (resolved) 
or are reconstructed as a single EM object (unresolved).
As seen in Fig.~\ref{fig:Pi0RecoComp},
the \PGpz reconstruction efficiency and the \PGpz mass resolution are better in the LAr option with the finer granularity 
and improve further in the LKr option, given its smaller Moli\`ere radius.

Figure~\ref{fig:Pi0Reco} shows the \PGpz reconstruction efficiency, as a function of its energy, all the way to 45\,GeV. 
Above 40\,GeV, the transverse shape of an unresolved \PGpz electromagnetic shower resembles more and more that of a single photon 
and the identification efficiency progressively vanishes.

\subsubsection{Work ahead}
\begin{itemize}

\item The measurement of tau polarisation in \PZ decays, from which the tau and electron asymmetries, $A_{\PGt}$ and $A_{\Pe}$, are extracted, 
is a cornerstone of the Tera-\PZ physics programme. 
Experience from LEP shows that the precision of this measurement critically depends, in particular, on the design of the electromagnetic calorimeter. 
While the sizes of the event samples used by the four LEP experiments were similar, 
the uncertainties differed by nearly a factor of two from one experiment to the other. 
As noted above, a high calorimeter transverse granularity is indeed crucial to identify and precisely reconstruct the \PGpz mesons produced in \PGt decays. 
The precision that different detector concepts can achieve on this measurement needs to be studied with a full simulation description, 
which also includes, besides the spatial and energy detector resolution, the effect of fake photons. 
Work in this direction has started, as is shown in Section~\ref{sec:PhysPerf_FullSim}.

\item The reconstruction of the events on fully simulated detector concepts will be crucial to extract further requirements on the detector performance. 
For instance, since photon energy contributes 25\% of the total energy of a jet, 
the ability to reconstruct low-energy photons is very important in order to improve the jet energy resolution, 
which is crucial for Higgs boson hadronic final states, such as $\PH \to \Pg\Pg$. 
Moreover, in a study considering multi-jet topologies~\cite{Lucchini:2020bac} it has been shown 
that the absence of photon and \PGpz identification prior to the jet particle-flow reconstruction 
would worsen the resolution of the reconstructed hadronic objects in about one third of the cases. 
Preliminary studies will need to be updated once the reconstruction of full simulation detector concepts become available in the coming months. 

\item The measurement of long-lived particles decaying to photons relies on a precise standalone ECAL measurement of the direction of the shower. 
The impact of longitudinal segmentation on the pointing capabilities and on displaced vertex reconstruction performance needs to be studied in full simulation.

\end{itemize}

\subsubsection{Preliminary conclusions}

\begin{itemize}

\item An electromagnetic calorimeter with excellent energy resolution (a few \%) is required to optimise the expected precision on the \PZ to \PGne coupling 
via the $\epem \to \PGne \PAGne \PGg$ process, 
lepton flavour violating decays such as $\PZ  \to \PGm \Pe$ and $\PGt \to \Pe \PGg$, 
decays of heavy flavoured hadrons to photons or \PGpz, 
and ALPs searches. 

\item Excellent transverse granularity and resolution is needed for optimal Bremsstrahlung recovery and \PGpz identification. 
Longitudinal segmentation and pointing capabilities are required for displaced photon vertex identification, such as long-lived ALP decays. 

\item A knowledge of the acceptance at the level required to match the statistical precision on the $\epem \to \PGg \PGg$ cross section 
for the absolute luminosity determination requires that the calorimeter be constructed with a mechanical precision of few tens of microns. 

\item Reconstructing low energy photons from decays of \PGpz mesons, either produced in \PB or \PD decays or in $\PH \to \Pg\Pg$,
requires as little material in front the ECAL as possible, as well as a noise term as small as a few tens of MeV. 

\end{itemize}

\subsection{Requirements for the hadron calorimeter}
\label{sec:PhysPerf_HadCal}

To fully exploit the statistical power of FCC-ee, precise measurements of processes in their hadronic final states are crucial because of their typically large branching ratios. 
A clear advantage of $\epem$ collisions over hadron collisions, such as at the HL-LHC, 
is the almost negligible pile-up and the absence of underlying event and initial-state QCD radiation, 
as well as the precise knowledge of the total energy-momentum of the final state,
resulting in cleaner data and improved jet energy resolutions.
For FCC-ee, the possibility of having a full \textsc{Geant} simulation of the events in the new software has become available only recently, 
so that a state-of-the-art event reconstruction based on the particle flow approach is not yet validated for complete physics studies.  
Since the reconstruction algorithm significantly impacts the final resolution on hadronic objects, 
it is challenging to determine specific requirements for the hadron calorimeter performance from existing studies, performed with fast simulation \textsc{Delphes} samples. 
Nevertheless, valuable information can be derived regarding the overall impact that the performance of the calorimeter has on the accuracy of the measurement, 
profiting from the particle-flow inspired reconstruction available in \textsc{Delphes}.

\subsubsection{Reconstruction of Higgs boson hadronic final states}

The identification of $\PH \to jj$ signal events relies on the resolution of the dijet system mass, which is influenced by the performance of the calorimeter. 
Additionally, the precise separation of different flavours depends on the effectiveness of the flavour tagging algorithm, 
which also imposes requirements on the vertex detector and on particle identification, as discussed in Sections~\ref{sec:PhysPerf_Hcc} and~\ref{sec:PhysPerf_strangeTagging}. 
Assuming an ideal particle-flow algorithm, with perfect charged and neutral particle identification, 
the visible energy (and mass) resolution is dominated by the neutral hadron resolution, 
which is in turn driven by the HCAL stochastic term. 
A study has been performed to assess the impact of a degraded hadron calorimeter performance~\cite{sciandra_2024_gt813-z8602}, 
by worsening the HCAL stochastic term (50\% and 100\%) and then compare the resulting $\PH \to jj$ sensitivity to that of the IDEA detector baseline 
(which assumes dual-readout calorimetry with a 30\% stochastic term).

\begin{figure}[ht]
\centering
\includegraphics[width=0.7\textwidth]{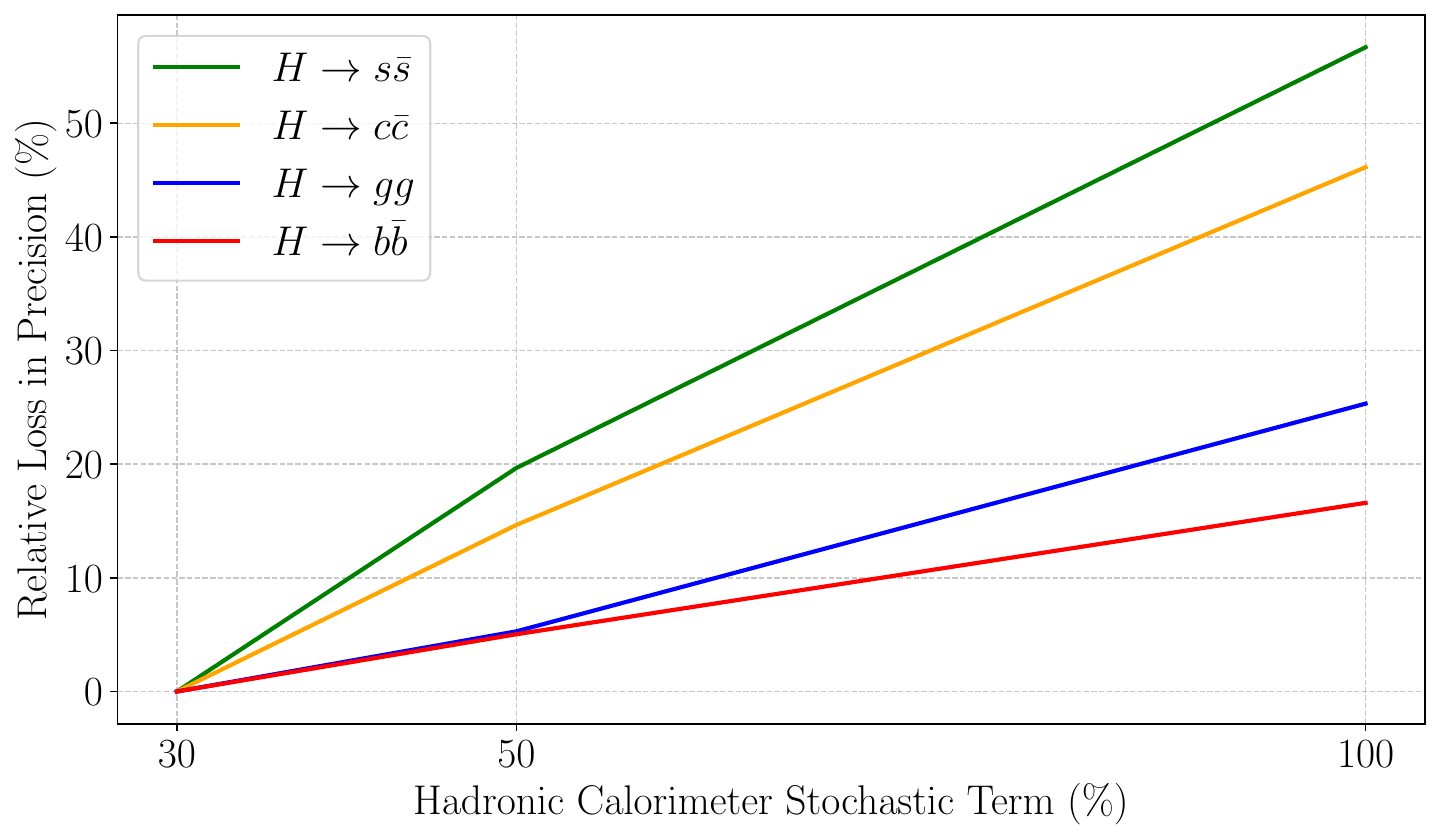} 
\caption{Expected precision degradation of the branching ratio measurements for the $\PH \to \PQb\PAQb$, $\PQc\PAQc$, $\PQs\PAQs$, and $\Pg\Pg$ decays 
as a function of the HCAL stochastic term. 
To guide the interpretation, the value 30\% would correspond to a dual-readout calorimeter as in the baseline IDEA simulation, 
50\% corresponds to an ATLAS-type calorimeter, and 100\% to a CMS-type calorimeter.}
\label{fig:HiggsBRPrec_vs_HadRes}
\end{figure}

The results are summarised in Fig.~\ref{fig:HiggsBRPrec_vs_HadRes} for several Higgs boson decay channels. 
When the neutral hadron energy stochastic term is degraded to 50\% (100\%), the expected precision of the measurement of the $\PH \to \PQs\PAQs$ branching ratio degrades by about 20\% (55\%).
The larger is the expected signal-to-background ratio in a given channel (the largest being for $\PH \to \PQb\PAQb$), 
the smaller is the effect of the degradation of the neutral hadron energy resolution on the expected precision. 
The visible mass resolution is, therefore, particularly important for accurately measuring the $\PH \to \PQc\PAQc$ and $\PH \to \PQs\PAQs$ branching fractions.

\subsubsection{Measurement of the Higgs boson invisible width}

As discussed in Section~\ref{sec:PhysPerf_HiggsRecoil}, FCC-ee offers a unique opportunity to use the `recoil method' for precise, model-independent measurements of Higgs boson properties. 
This method becomes particularly valuable when studying the Higgs boson decay into an invisible final state. 
The Standard Model process $\PH \to \PGn\PAGn\PGn\PAGn$ has a very small branching ratio, $1.06\times 10^{-3}$~\cite{deFlorian:2227475}, 
so that it is beyond the reach of the LHC programme and might only be observable at the FCC-hh. 
However, with possible extensions of the Standard Model, such as the inclusion of a Higgs portal, 
the Higgs boson width to invisible final states could be increased because of decays to non-SM particles, 
which could be potential dark matter candidates~\cite{Patt:2006fw, sym13122406}, making the study of this channel highly relevant.

In a recent study~\cite{HiggsInvisibleNote}, the \textsc{Delphes} simulation of the IDEA detector was used, 
along with the state-of-the-art flavour tagging algorithm \textsc{ParticleNetIDEA}, described above. 
To maximise the size of the event sample, the analysis considers not only \PZ decays to $\Pe\Pe$ and $\PGm\PGm$ but also to $\PQb\PAQb$, $\PQc\PAQc$, and $\PQq\PAQq$. 
The visible mass ($M_\text{vis}$) serves as the discriminant variable in the final fit (with no systematic uncertainties considered at this stage). 
The results indicate that, with an integrated luminosity of 10.8\,ab$^{-1}$ at $\sqrt{s} = 240$\,GeV and of 3\,ab$^{-1}$ at $\sqrt{s} = 365$\,GeV, a measurement of the SM branching ratio with a precision of 25\% could be achieved. 
Furthermore, when accounting for the possibility of exotic decays, the study results in the exclusion, at the 95\% CL, of branching fractions lower than 0.05\%, or a 5\,$\sigma$ observation if greater than 0.13\%. 
The $\PZ \to \PQq\PAQq$ decay channel was found to be the one with the best sensitivity, given its higher statistical power, and the run at $\sqrt{s} = 240$\,GeV has much better sensitivity than the highest energy run.
A preliminary study performed with CLD full simulation can also be found in Ref.\cite{sciandra_2024_gt813-z8602}.

\begin{figure}[ht]
\centering
\includegraphics[width=0.7\textwidth]{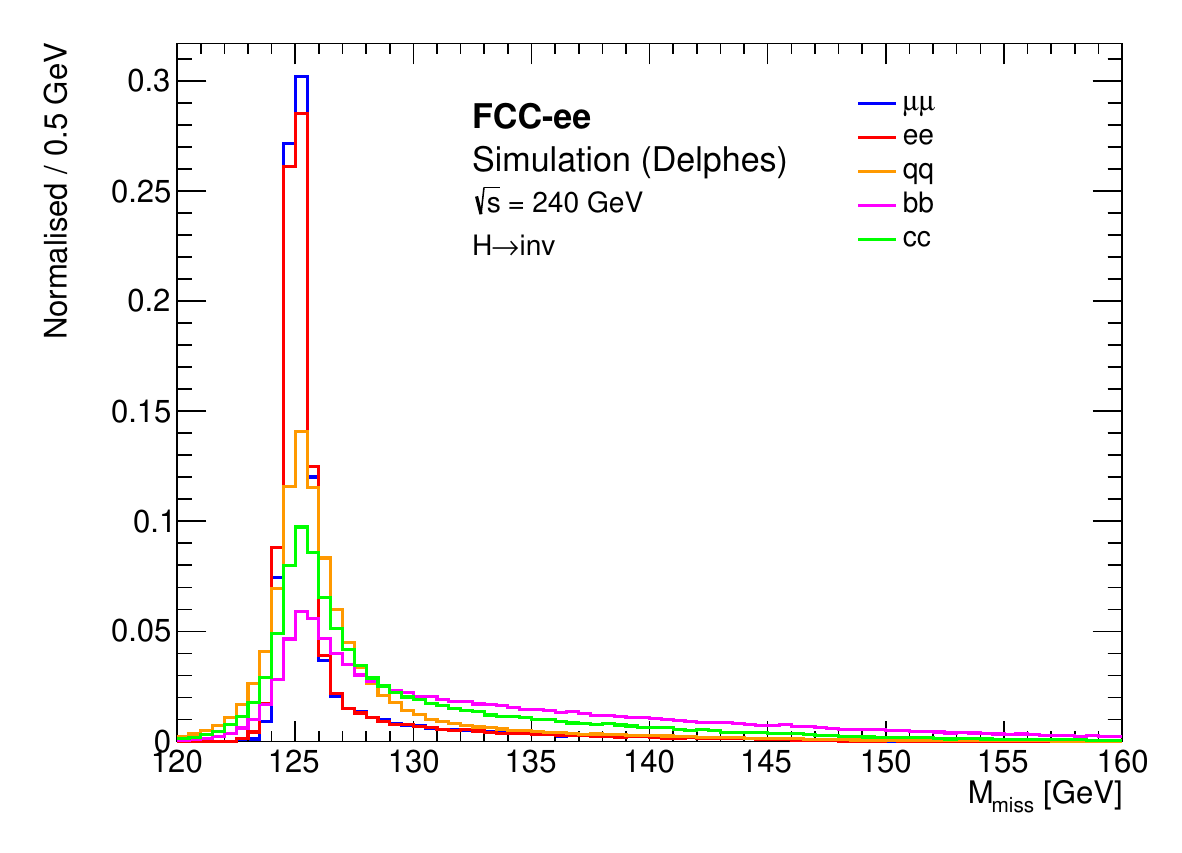} 
\caption{The $M_\text{miss}$ distribution for several \PZ decay channels, corresponding to the Higgs boson signal, in $\PH\PZ$ processes. 
The distributions are normalised to the same number of events and are shown after the $P_\text{miss}>10$\,GeV cut and the $M_\text{vis}$ selection around the \PZ mass.}
\label{fig:Hinv_Mmiss}
\end{figure}

\begin{figure}[t]
\centering
\includegraphics[width=0.8\textwidth]{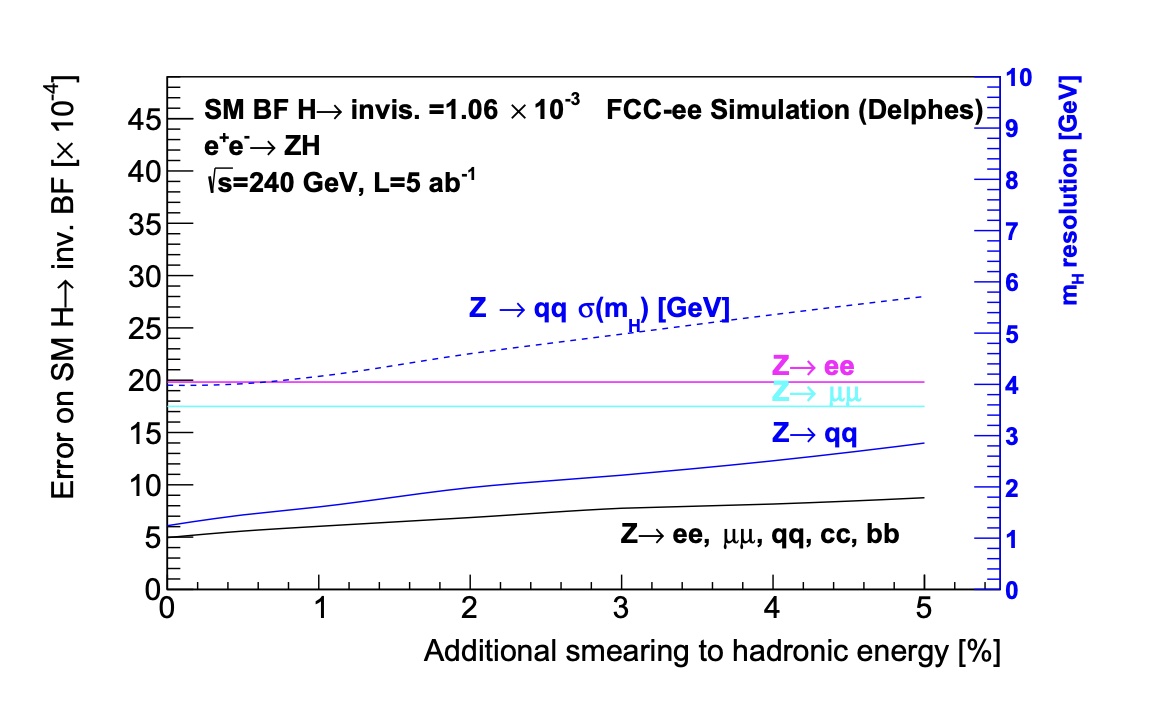} 
\caption{The effect of an additional smearing to the hadronic energy of the events on the expected uncertainty in a measurement of the branching fraction of the Higgs boson decay to invisible particles 
is shown for three individual channels and for all channels combined. 
The root mean squared of the reconstructed Higgs boson mass is also shown, as a dashed blue line (corresponding to the vertical axis on the right side).}
\label{fig:Hinv_resStudy}
\end{figure}

Figure~\ref{fig:Hinv_Mmiss} displays the normalised distribution of the missing mass, representing the Higgs boson, 
for the various \PZ decay modes, allowing the comparison of resolutions between the lepton and hadron channels. 
An investigation was also conducted into the effects of a poorer detector energy resolution on the measurement, 
applying a simple additional Gaussian smearing to the four-vector of the hadronic final state. 
As shown in Fig.~\ref{fig:Hinv_resStudy}, an additional smearing of 5\% led to a 130\% (80\%) increase in the uncertainty on the $\PQq\PAQq$ channel (combined result).
The blue dashed curve in Fig.~\ref{fig:Hinv_resStudy} represents the resolution of the reconstructed Higgs boson mass for the case where the \PZ boson decays into hadronic modes, as a function of the additional smearing. 
It is important to note that this quantity includes other effects from beam spread and ISR. 
In the future, it may be beneficial to use a better-defined quantity, such as the \PZ resolution itself, to further evaluate the dependence on the calorimeter performance. 

\subsubsection{Search for heavy neutral leptons}

It has been shown in Refs.~\cite{Ellis:2691414, Verhaaren_2022} that an $\epem$ machine like FCC-ee, with the potential to collect a vast event sample at the \PZ peak
holds significant discovery potential for feebly interacting particles, such as heavy neutral leptons (HNL). 
Such a large dataset, covering a phase space at low coupling strengths, is unmatched by other proposed future colliders and could potentially reach the lower bounds set by theory.
The study reported in Ref.~\cite{note_HNL_Polesello_Valle_mujj} focuses on the production of HNLs with masses between 20 and 80\,GeV at $\sqrt{s} = 91$\,GeV, 
with an integrated luminosity of $2.05\times 10^8$\,pb$^{-1}$, corresponding to $6\times 10^{12}$ $\epem \to \PZ$ events. 
In particular, the process $\epem \to N_{\PGm} \PAGnGm \to \PGm \PQq \PAQq^\prime \PAGnGm$ has been investigated. 
This process has a substantial branching ratio of about 50\% over the entire mass range of interest 
and complements other studies in the leptonic channel with long lifetimes~\cite{Verhaaren_2022,Abdullahi:2022jlv, BayNielsen:2017yws}.

\begin{figure}[ht]
\centering
\includegraphics[width=0.8\textwidth]{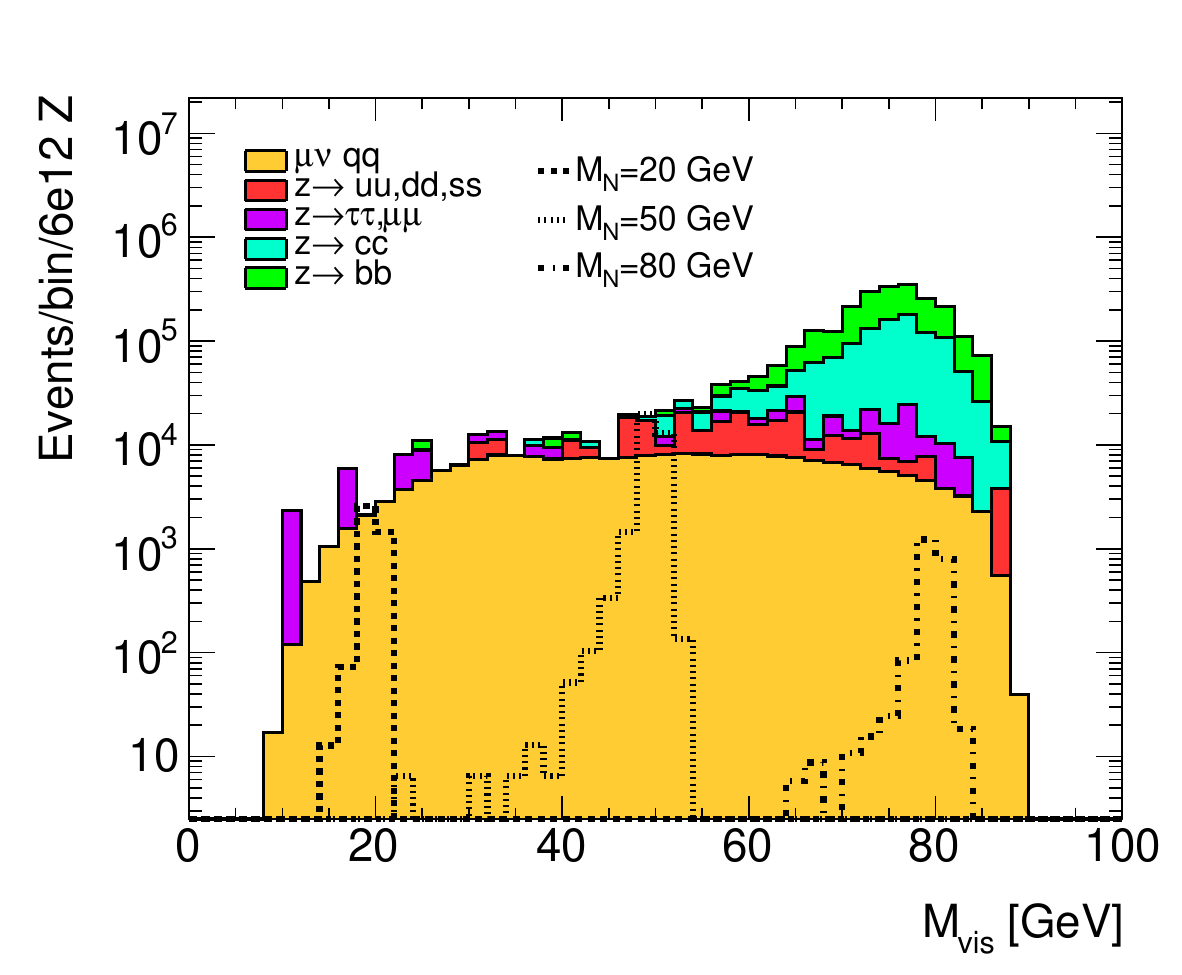} 
\caption{The $M_\text{vis}$ distribution for background and HNL signal events after a preliminary selection.}
\label{fig:HNLmujj_recoMass}
\end{figure}

In a search for HNLs that decay promptly in the detector, the analysis applies selections on muons, jets, and missing energy to minimise background contaminations. The discriminant variable used in this analysis is the visible mass, which corresponds to the HNL mass, as shown in Fig.~\ref{fig:HNLmujj_recoMass}.

\begin{figure}[ht]
\centering
\includegraphics[width=0.7\textwidth]{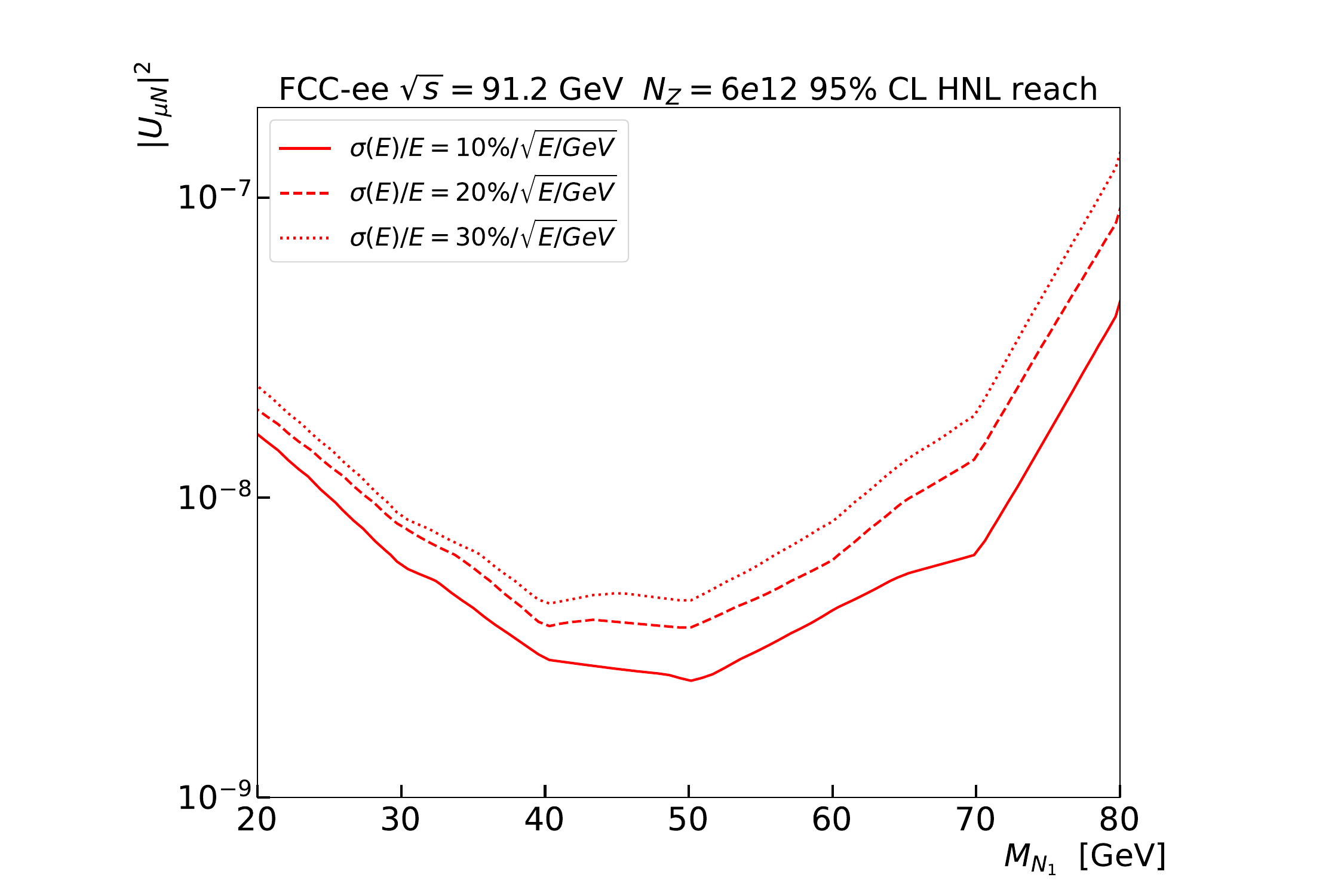} 
\caption{The 2\,$\sigma$ significance of an HNL signal for several visible mass width resolutions, from a search for HNLs that decay promptly into a muon and two jets.}
\label{fig:HNLmujj_ResScan}
\end{figure}

The final selection involves a sliding cut on the search HNL mass, taking into account the observed resolution ($\sigma = 20\% \sqrt{m_{\text{HNL}}}$). 
A study to evaluate the impact of varying the mass resolution has been performed and the result is shown in Fig.~\ref{fig:HNLmujj_ResScan}, 
indicating a minimal effect at low masses, where the background from $\PZ \to \PQq\PAQq$ events is negligible, 
but becoming more relevant at higher masses, where the background is higher.

\subsubsection{Work ahead}

\begin{itemize}

\item The total Higgs boson width measurement at $\sqrt{s} = 240$\,GeV is statistically limited by the precision on the $\PH \to \PZ\PZ^{*}$ partial width, 
which crucially depends on identifying all the hadronic \PZ decay modes. 
The separation of the \PW and \PZ mass peaks is driven by hadronic resolution and is required in order to suppress the overwhelming $\PH \to \PW\PW^{*}$ background. 
The current detector proposals satisfy this requirement in fast simulation studies, 
but this result has to be confirmed by constrained kinematic fits in full-simulation studies. 

\item These preliminary studies have been performed using a fast simulation of the detector with close to ideal particle-flow performance and can, therefore, 
only provide overall guidelines for the needed hadron calorimeter resolution.

Full-simulation performance studies with particle-flow reconstruction using integrated tracking, ECAL, and HCAL information, 
and kinematic fits constrained by the total energy and momentum conservation, 
are of paramount importance to consolidate the requirements on hadronic calorimeter technology, 
transverse and longitudinal granularity, and resolution. 

\end{itemize}

\subsubsection{Preliminary conclusions}

\begin{itemize}

\item The precision of the Higgs boson couplings to quarks and gluons, in particular the strange and charm Yukawa couplings, 
as well as the Higgs to invisible one, drives the hadron calorimeter requirements. 
Heavy neutral lepton hadronic decay modes also benefit from excellent visible energy resolution.

\item In particular, assuming an ideal particle-flow reconstruction setup with highly-efficient tracking and neutral hadron identification capabilities, 
the neutral hadron resolution drives the sensitivity to the strange Yukawa coupling. 
An excellent visible mass resolution, obtained through particle-flow reconstruction, is of paramount importance for the observation of this mode.  
\end{itemize}

\subsection{Requirements for the muon detector}
\label{sec:PhysPerf_muons}

\begin{figure}[ht]
\centering
\includegraphics[width=0.7\textwidth]{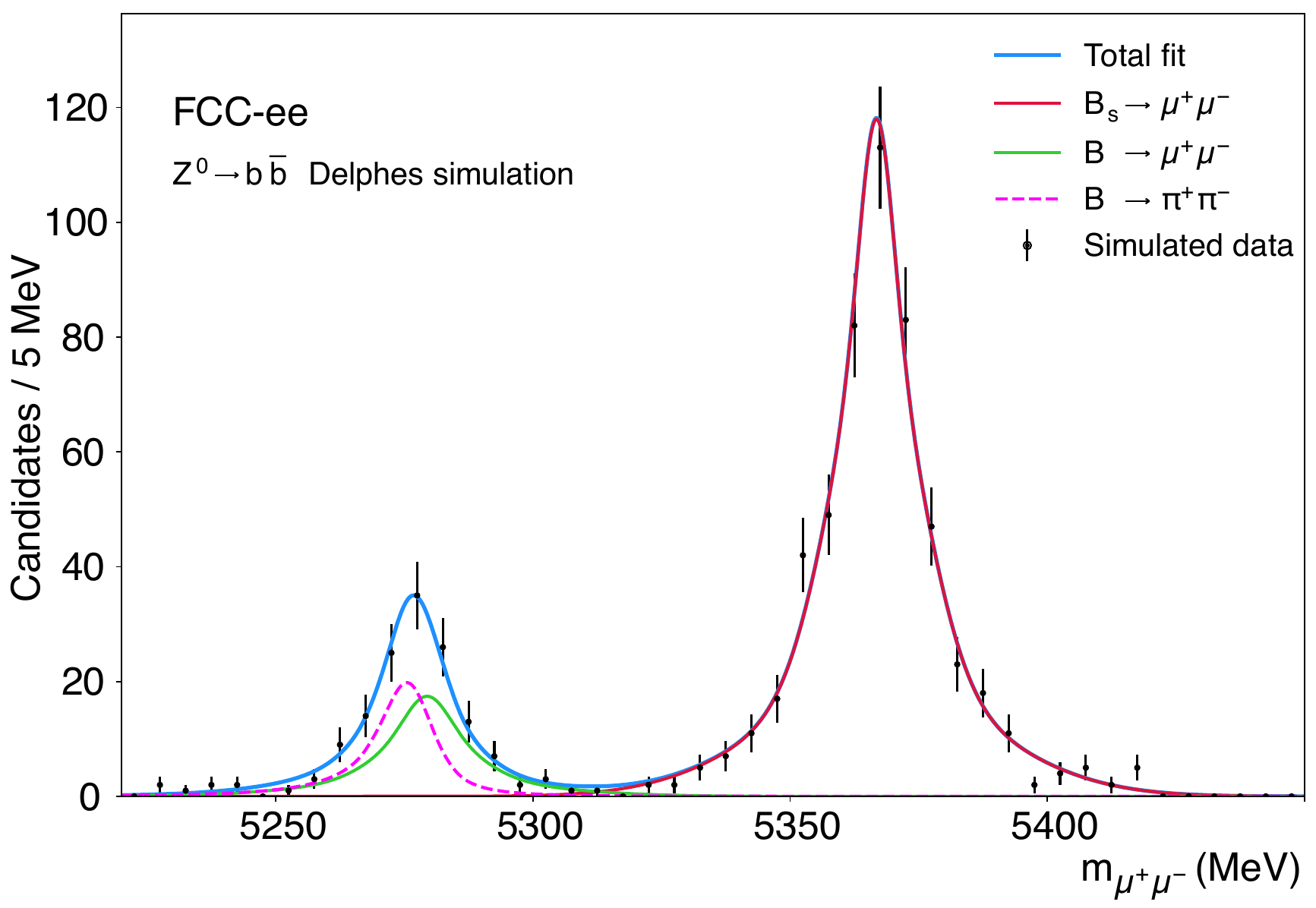} 
\caption{Mass distribution of $\PB \to \PGmp\PGmm$ and of $\PBs \to \PGmp\PGmm$ as expected from the IDEA detector, 
with an event sample of $5 \times 10^{12}$ \PZ decays. 
The background contribution from misidentified pions in $\PB \to \PGpp\PGpm$ decays is also shown, as the dashed curve. 
From Ref.~\cite{Monteil:2021ith}.}
\label{fig:muons_b2mumu}
\end{figure}

At the FCC-ee energies, the muon momentum reconstruction is entirely driven by the tracking system 
and the main role of the muon detector is to identify muons with a very high efficiency and purity, 
as well as to serve as `tail-catcher' for the hadron showers that may not be fully contained in the calorimeter. 
An important figure of merit is the probability that a pion be misidentified as a muon.
An example benchmark measurement to set a requirement on the control of the pion contamination is that of the ultra-rare $\PB \to \PGmp\PGmm$ decay. 
As illustrated in Fig.~\ref{fig:muons_b2mumu}, the excellent mass resolution of the IDEA detector offers a very good separation of the \PB and \PBs peaks. 
A significant background coming from $\PB \to \PGpp \PGpm$ decays is, however, present under the \PB peak. 
Assuming a double $\PGp \to \PGm$ mis-identification probability of $2 \times 10^{-5}$, 
this background contribution (dashed histogram in Fig.~\ref{fig:muons_b2mumu}) would be as large as the signal~\cite{Monteil:2021ith}. 

While the measurement of the track momentum in the muon detector alone does not improve (in the momentum range of interest) 
the excellent prompt muon momentum resolution already provided by the tracker, 
a good standalone resolution is useful to reduce the pion contamination and to identify pions that decay before the muon detector. 
In addition, the standalone performance of the muon detector is important in searches for long-lived particles that decay outside the tracker volume. 
The requirements on the standalone momentum resolution need to be quantified and 
they are likely to have a significant impact on the technologies that will be chosen for the muon system of an FCC-ee detector~\cite{Braibant:2021wts}.

The FCC-ee could uniquely probe yet another suppressed $\PQb \to \PQs$ transition, via the $\PBs \to \PGn \PAGn$ decay mode~\cite{Alonso-Alvarez:2023mgc}. 
The main backgrounds for this search are the $\PQb \to \PGt \PGn \mathrm{X}$ decays, 
where the \PGt decays semi-leptonically and the decay lepton escapes detection. 
This configuration can occur when the muon is forward or too soft to reach the muon detector. 
In this case, a calorimeter equipped to perform muon identification might help, together with a low detector magnetic field. 

\subsection{Precise timing measurements}

The motivations for timing measurements at FCC-ee are outlined in this section. 
Quantified requirements are still to be assessed for most of the use cases.

\subsubsection{Time-of-flight measurements}
\label{sec:PhysPerf_TOF}

As mentioned in Section~\ref{sec:PhysPerf_PID}, 
charged hadron particle identification relying on the specific energy loss must be complemented by other measurements in order to fill the gap around 1\,GeV, 
where the Bethe--Bloch energy loss function is close to its minimum. 
Time-of-flight measurements, for example in a layer of silicon sensors at 2\,m from the interaction point, with a non-challenging resolution $\sigma_t \simeq 100$\,ps, 
are adequate for this purpose (with a magnetic field of 2\,T or less). 
With a resolution of 30\,ps, such (standalone) TOF measurements would allow charged kaons to be separated from charged pions up to ${\cal{O}}$(3)\,GeV in momentum, 
while an excellent resolution of 10\,ps would be 
needed\,\footnote{Such a resolution is smaller than the time it takes, at the \PZ peak, for the two bunches to cross each other (about 36\,ps). 
Hence, to be able to exploit a 10\,ps resolution, the `event time', $t_0$, would need to be reconstructed. 
Simple algorithms should allow this $t_0$ to be reconstructed, for most SM processes, 
with a resolution of $\sigma_t / \sqrt{N}$ for events with $N$ primary tracks that reach the timing layer.} 
to extend this momentum range up to 5\,GeV.  

Such TOF information could also be obtained in a silicon-tungsten CLD-like ECAL, 
where measurements could be made in several layers, each with a resolution of, e.g., $\sim$\,100\,ps. 
Dedicated timing layers made of inorganic scintillator offer another solution. 
For example, the calorimeter design proposed in Ref.~\cite{Lucchini:2020bac} includes two such thin layers, upstream of the crystal-based ECAL mentioned earlier, which could provide a 20\,ps timing resolution.  

\begin{figure}[t]
\centering
\includegraphics[width=0.46\textwidth]{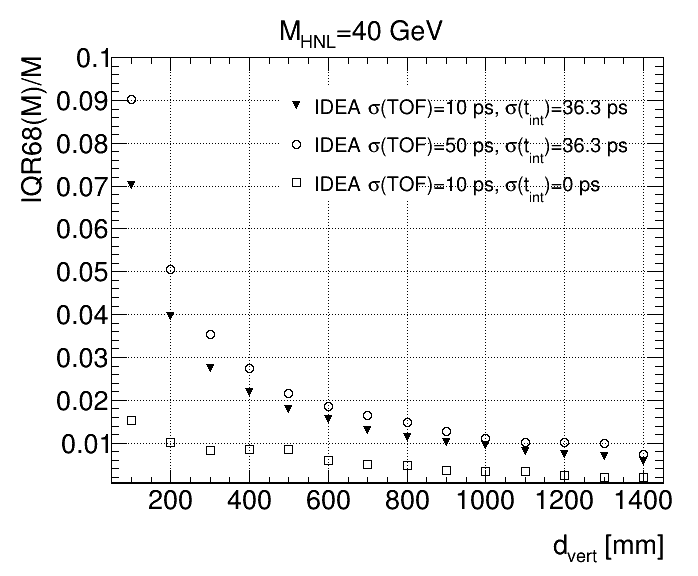} 
\includegraphics[width=0.5\textwidth]{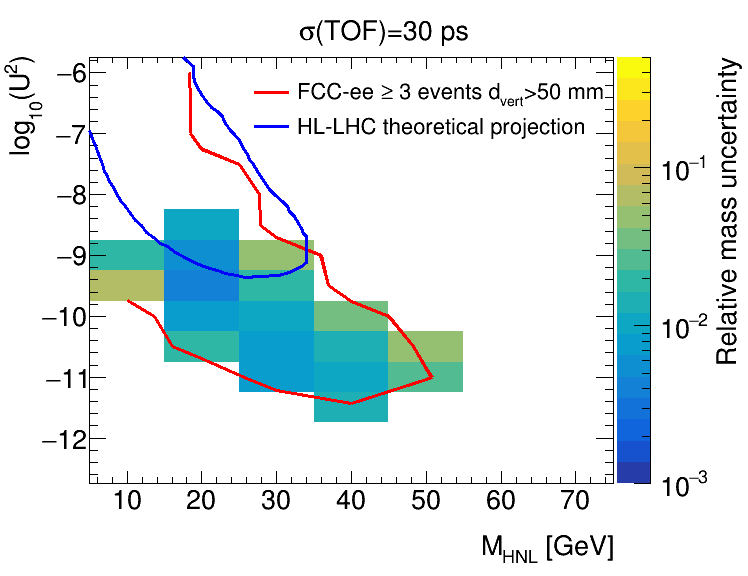}
\caption{Left: Relative mass resolution of the HNL as a function of its flight distance, for an HNL mass of 40\,GeV, 
using a reconstruction technique exploiting timing measurements; 
two hypotheses on the resolution of the timing measurements are considered ($\sigma_t = 10$ and 50\,ps). 
Right: Relative mass resolution expected at FCC-ee, as a function of the HNL mass and of its (squared) mixing with the $\PGnGm$, 
in the parameter space where FCC-ee could make a discovery, for a timing resolution of 30\,ps.}
\label{fig:HNL_timing}
\end{figure}

Another well-known use-case of TOF measurements is the determination of the mass and lifetime of new massive particles. 
An example is provided by feebly coupled heavy neutral leptons that could be copiously produced in $\PZ \to \PGn \mathrm{N}$ decays at Tera-\PZ. 
In such decays, the energy of the new lepton $\mathrm{N}$ depends only on its mass. 
In subsequent decays, such as $\mathrm{N} \to \PZ (\ell \ell) \PGn$ or $\mathrm{N} \to \PW(\PQq\PAQq^\prime) \ell$, 
the flight distance of $\mathrm{N}$ is obtained from the decay vertex of the \PZ boson 
and the TOF of the charged particles provides the time at which $\mathrm{N}$ decayed, 
from which both the lifetime and the velocity (hence the mass) of $\mathrm{N}$ can be extracted~\cite{Blondel:2022qqo}.
The need for a large tracking volume, as well as a timing resolution in the ballpark of a few tens of picoseconds, 
has so far placed the practical realisation of this idea out of realistic experimental reach.
A first study of the expected performance of this measurement technique has been presented in Ref.~\cite{Aleksan:2024hyq}, in the context of FCC-ee. 
It uses as a benchmark the process $\epem \to \mathrm{N}_{\PGm} \PAGnGm \to \PGm \PQq\PAQq^\prime \PAGnGm$, 
already considered in Section~\ref{sec:PhysPerf_HadCal}.
A simple algorithm has been developed that allows the reconstruction of the time when $\mathrm{N}$ decays, 
despite the fact that the masses of the charged particles produced in the hadronic \PW decay are, a priori, unknown. 
Since there is no way to determine the `event time' ($t_0$) of the collision for such interactions, with no primary tracks, 
it must be assumed that $\mathrm{N}$ was produced at the nominal beam crossing time. 
This leads to a smearing, of about 36\,ps, of the reconstructed time of flight of $\mathrm{N}$, from which its velocity is derived. 
Figure~\ref{fig:HNL_timing}~(left) shows the relative resolution with which the mass of the HNL can be reconstructed with this method, 
as a function of its flight distance, for an HNL mass of 40\,GeV and several assumptions on the resolution of the timing measurements, $\sigma_t$.
The resolution is defined as the width of the interval between the 16\% and the 84\% quantiles of the reconstructed mass distribution. 
The aforementioned smearing prevents reaching the ideal resolution that would be obtained with $\sigma_t = 10$\,ps, if the event $t_0$ was known, 
represented by the lowest curve in Fig.~\ref{fig:HNL_timing}~(left). 
From the two upper curves it can be seen that, provided $\sigma_t$ is smaller than 50\,ps, the degradation due to the detector resolution is limited to 20\%.
For an HNL of mass 40\,GeV that decays at 1.5\,m from the interaction point, this technique provides a mass resolution below 1\%, 
which is significantly better than what can be obtained by measuring the visible mass in the detector. 
The right panel of Fig.~\ref{fig:HNL_timing} shows that, in the parameter space that FCC-ee can probe and that extends beyond the sensitivity of the HL-LHC, 
timing measurements with $\sigma_t = 30$\,ps would allow the HNL mass to be reconstructed at the per-cent level.

It should be noted that, should the magnetic field be increased to 3\,T (as could be considered for the runs at $\sqrt{s} = 240$\,GeV and beyond), 
particles with a momentum of 1\,GeV may not reach the outer radius of the
tracker\footnote{Prompt particles with a curvature radius $\rho$ are confined to radial distances below 2\,$\rho$ and, hence, 
need to have a transverse momentum 
$\pt \, (\text{GeV}) > 0.3 \times R  \, (\mathrm{m}) \times B \, (\mathrm{T}) / 2$ 
to reach a barrel layer situated at a radius $R$. 
Particles with a momentum of 1\,GeV and a polar angle of 45$^{\circ}$ only reach $R \sim 1.6$\,m in a 3\,T magnetic field.} 
of about 2\,m. 
Hence, to provide PID capabilities in the momentum range around 1\,GeV, where the measurement of ionisation energy per unit length does not provide any separation, 
the TOF measurements should be made either at a smaller radius  
or when the particle eventually reaches the endcaps. 
In the latter case, in particular at normal incidence, 
it is essential to investigate the effects of multiple scattering and energy loss resulting from interactions with the beam pipe and the tracker material. 
While this may pose less of an issue for a gaseous tracker, it remains important to study this aspect in detail.
In the former case and for the IDEA detector, 
the outermost layer of the vertex detector, at a radius of 35\,cm from the beams, could be used for that purpose, at the expense of an increased material budget. 
With a time resolution of 30\,ps, such standalone TOF measurements would ensure a $\PK / \PGp$ separation larger than 3\,$\sigma$ in the whole momentum range 
where PID based on d$N$/d$x$ is ineffective. 
The impact that the larger material would have on the measurement of the track parameters remains to be studied. 
In a detector design with a full silicon tracker, the TOF measurements could be made in one of the layers of the outer tracker.

\subsubsection{Time measurements very close to the IP}

At FCC-ee, the distribution of the `event time' ($t_0$) of the collisions follows a Gaussian distribution to a very good approximation, 
with a width ranging between 36\,ps during the Tera-\PZ run and 6.5\,ps during the $\ttbar$ threshold run.
The arrival time of a particle in the innermost layer of the vertex detector depends only minimally on its nature (\PK, \PGp, \PGm, \Pe, etc.), 
such that a measurement of this time provides (for primary particles) a measurement of the particle production time and of the event time $t_0$. 
For a given resolution of each measurement, $\sigma_t$, the event $t_0$ can then be determined with a resolution of $\sigma_t / \sqrt{N}$, 
with $N$ being the number of charged primary particles crossing this layer. 
Measuring the event $t_0$ offers several advantages, as listed below.

\begin{itemize}

\item It provides a robust reference for some of the TOF measurements outlined above. 

\item The spread of the $t_0$ distribution being proportional to the beam energy spread, its measurement provides an independent determination of the latter.

\item Because of the crossing angle, the $t_0$ of a collision and the longitudinal position of the colliding particles within their respective bunch are correlated. 
With respect to a time origin, defined as the time when the centres of the two bunches coincide with the IP, collisions that occur early (late) always involve particles that are in the head (tail) of both bunches. 
Collisions that happen at a time close to the origin involve particles that are in the middle of the bunches. 
Splitting the events into early, central, and late collisions can provide relevant accelerator-related information. 
For example, the mass distribution of dimuon events at Tera-\PZ would provide a check that there is no unexpected difference in the centre-of-mass energy between head and tail collisions. 
Alternatively, the longitudinal boost distribution of dimuon events at $\sqrt{s} = 125$\,GeV would exhibit a correlation~\cite{JanotEPOLtalk} 
between the centre-of-mass energy and the longitudinal position that could improve the effective monochromatisation (Section~\ref{sec:epol}). 
Moreover, further checks of the beam-beam effects could be made, 
by exploiting the fact that these effects depend strongly on the longitudinal position of the colliding particles~\cite{myTalkatKrakow}.

\end{itemize}

Moreover, the measurement of the production time of the particles complements the spatial reconstruction of primary vertices in view of a direct pile-up measurement 
(expected to be at the per-mil level at Tera-\PZ). 
However, achieving precise timing measurements in the innermost layer of the VXD, without heavily compromising the material budget, will probably be a challenge.

\subsubsection{Time measurements in the calorimeters}

Besides the TOF measurements mentioned earlier, time measurements in the calorimeters provide handles to exploit the shower development in space and time. 
The hadronic energy reconstruction, in particular in a non-compensating calorimeter, as in the CLD design, can benefit greatly from such measurements, 
which allow the delayed signal from neutrons to be identified. 
The possible benefit of timing for pattern recognition in high-granularity imaging calorimeters remains to be studied in detail. 

Finally, in the DR calorimeter of IDEA, a Fourier transform of the full signal from the SiPMs reading out the fibres would provide high precision (better than 100\,ps) timing for the individual fibres and, 
in turn, longitudinal segmentation; 
the potential benefit in a particle-flow reconstruction algorithm adapted to the hybrid segmented crystal and fibre dual-readout calorimeter of Ref.~\cite{Lucchini:2020bac} remains to be studied. 
It could also benefit the identification of low energy neutral hadrons, potentially providing neutron vs.\ \PKL separation, 
assuming that the initial time $t_0$ is known. 
However, a concrete implementation has to be studied in detail.

\subsection{Selected studies with full simulation}
\label{sec:PhysPerf_FullSim}

Most of the requirements discussed in the previous sections have been obtained using fast parametrised simulations. 
While a full simulation and event reconstruction with every detector concept is beyond the scope of this feasibility study, 
a few selected preliminary reconstruction performance and physics results obtained with full simulation are presented in this section. 
With the exception of ML-based tracking, all of them use the reconstruction of the CLD detector concept. 
A high-level machine-learning-based particle-flow reconstruction algorithm is first discussed. 
An initial implementation of a full simulation flavour tagging algorithm in CLD is described next and the achieved performance is compared with the fast simulation results. 
Finally, the Higgs mass measurement using the recoil mass method and the \PGt polarisation measurement at the \PZ pole are addressed.

\subsubsection{Machine-learning event reconstruction}

The Higgs boson decays preferentially into hadrons. 
Therefore, the optimal identification and reconstruction of hadronic jets is crucial for measuring Higgs boson properties. 
In particular, resolving the mass peaks from $\PW, \PZ \to jj$ hadronic decays at the 3\,$\sigma$ level 
requires reconstructing the dijet masses with a resolution $\sigma_m$ at the $\Gamma / m \approx 2.5\%$ level, 
where $\Gamma$ is the natural \PW or \PZ width. 
Moreover, the measurement of rare hadronic Higgs decay rates, such as $\PH \to \PQc\PAQc$ and $\PH\to \PQs\PAQs$, 
requires excellent dijet mass resolution, since the relative uncertainty on these modes scales with $\sqrt{\sigma_m}$, 
as discussed in Section~\ref{sec:PhysPerf_HadCal}.

The visible energy of hadronic jets is roughly distributed as follows: 
$f_{\pm} = 65\%$ from charged hadrons (\PGppm, \PKpm, \ldots), 
$f_{\PGg} = 25\%$ from photons originating from \PGpz decays, 
and $f_\mathrm{n} = 10\%$ from neutral hadrons (\Pn, \PKL, \ldots). 
The particle-flow approach relies on the tracking system to measure the momenta of charged particles, benefiting from its superior resolution, compared to the calorimeters, 
while the energies of photons and neutral hadrons are measured with the calorimeters. 
In an ideal PF algorithm, the visible mass resolution is dominated by the HCAL resolution, despite the neutral hadron energy content being subdominant. 
Achieving the best mass resolution requires efficiently identifying neutral hadrons and disentangling them from mismeasured clusters originating from charged hadron showers.

An optimal PF algorithm requires nearly 100\% tracking efficiency and purity over the full momentum range for an accurate measurement of the charged energy component. 
Fine transverse and longitudinal calorimeter segmentation is necessary for optimal geometrical matching of extrapolated track trajectories to electromagnetic and hadron showers. 
Additionally, low-noise high-resolution calorimetry and efficient clustering algorithms are essential for identifying and reconstructing individual photon and neutral hadron showers. 
Excellent electromagnetic and hadron calorimeter resolutions are crucial for comparing tracking and calorimeter energy measurements after geometrical matching. 
A low material budget in the tracker and before the calorimeter (e.g., pre-shower or solenoidal magnet) is desirable to minimise nuclear interactions and conversions 
that could compromise track, photon, and neutral hadron reconstruction efficiency. 
Moreover, the PF reconstruction performance depends critically on the transverse and longitudinal granularity for optimal track-cluster matching and neutral hadron identification. 
Finally, maximising the hermeticity and uniformity of the detector design is crucial for PF reconstruction.

Electron and muon identification and momentum determination, whether isolated or within jets, also benefit from a holistic treatment of all detector components~\cite{ALEPH:1994ayc,CMS:2017yfk}. 
For example, as shown in Section~\ref{sec:PhysPerf_ECAL}, photon bremsstrahlung emissions within the tracker requires combining tracking and ECAL information. 
Similarly, muons are identified by combining data from tracking, calorimetry, and muon detectors. A
comprehensive PF reconstruction should encompass a wide range of tasks, 
aiming at providing an exhaustive list of identified final-state particles produced in the collisions and interacting with the detector, 
along with accurate measurements of their momenta and origin vertices.

\begin{figure}[t]
\centering
\includegraphics[width=0.495\textwidth]{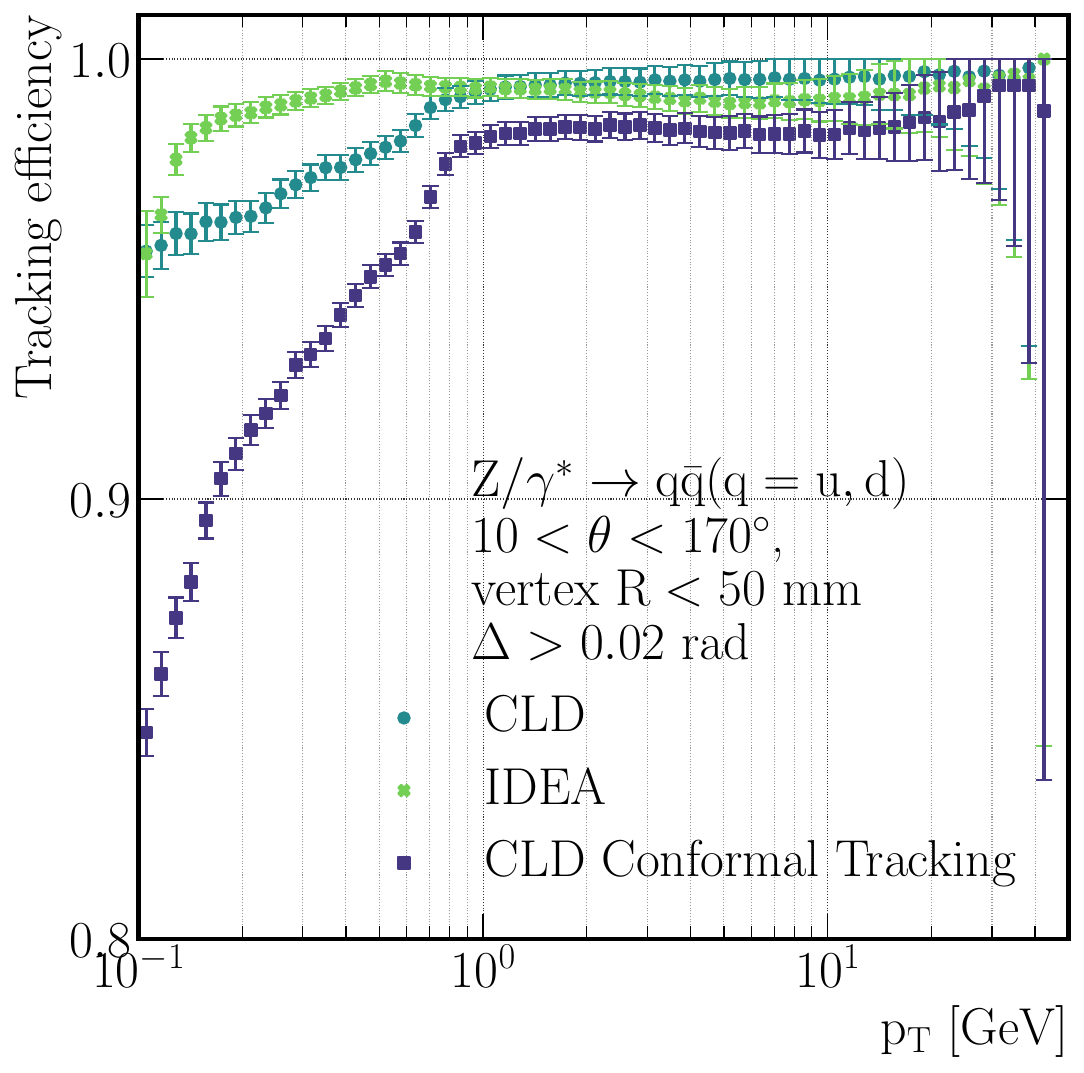} 
\includegraphics[width=0.495\textwidth]{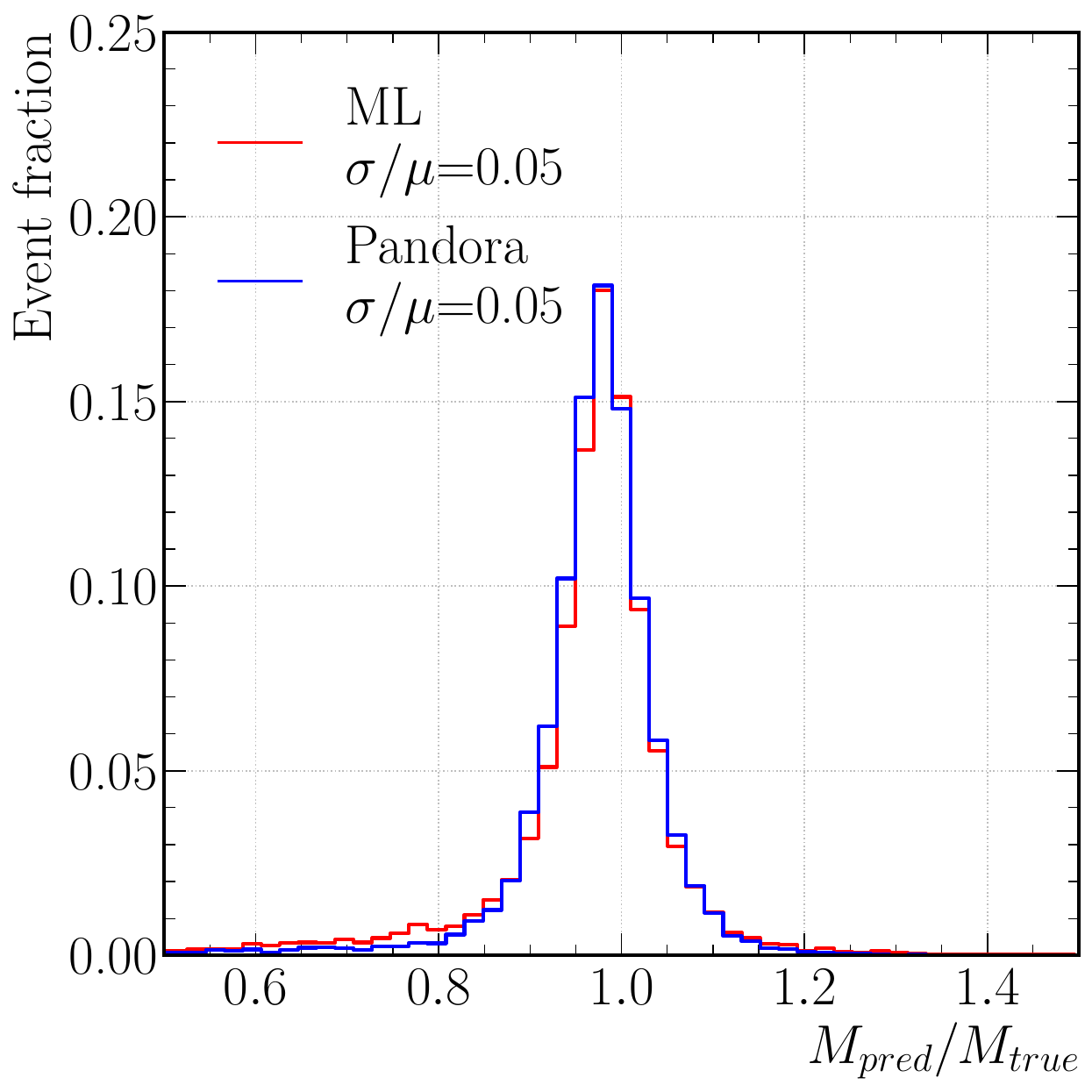}
\caption{Left: Tracking efficiency in CLD and IDEA using the classical (Conformal Tracking) and ML-based (\textsc{ggtf}) approaches. 
Right: Visible mass resolution in $\PH \to \PQs\PAQs$ events using the classic \textsc{Pandora} reconstruction and the ML-based particle flow approach.}
\label{fig:fullsim_mlreco}
\end{figure}

A novel approach to full event reconstruction based on graph neural network (GNN) methods is described in the following. 
The goal is to improve every aspect of the reconstruction process, from efficiency to resolution, 
and to enable rapid iterations over various detector concepts and geometries, so as to estimate performance and, hence, optimise detector technologies and design options. 
The effectiveness of this approach is shown in Fig.~\ref{fig:fullsim_mlreco}. 
The left panel displays the tracking efficiency achieved with machine learning (ML) methods using the \textsc{ggtf} algorithm~\cite{MLTRACKING_NOTE} 
for the CLD and IDEA detectors in $\PZ \to jj$ events at the \PZ pole. 
The ML approach significantly improves the CLD performance with respect to the standard tracking 
(a similar comparison for the IDEA case cannot be made because the baseline reference is missing). 
It also shows that IDEA, with its superior number of track measurements, can reconstruct much lower transverse momentum tracks, 
which are crucial for optimal PF reconstruction since a significant portion of the visible jet energy is carried by soft charged particles. 
The right panel illustrates the visible mass resolution in $\PZ\PH \to \PGn \PAGn \PQs \PAQs$ events at $\sqrt{s} = 240$\,GeV 
obtained with ML-based PF (MLPF) compared to the \textsc{Pandora} particle-flow reconstruction~\cite{pandorapfa,pandorasdk}, showing similar performance. 
This result was obtained using GNN-based PF reconstruction, described in more detail in Ref.~\cite{MLPF_NOTE}, 
taking as input reconstructed classical tracks and low-level calorimeter hit reconstruction. 
Although this is work in progress, it indicates that the established PF algorithm performance can already be reproduced with this approach. 
Further improvements are foreseeable by optimising the clustering step, 
enhancing the neutral hadron identification efficiency, and making use of further optimised tracking algorithms.

\subsubsection{Jet flavour tagging}

The first investigation of jet-flavour tagging performance with a full simulation of the CLD detector concept at FCC-ee is briefly presented here, 
and discussed in more details in Ref.~\cite{TAGGINGFULLSIM_NOTE}. 
As discussed in Sections~\ref{sec:PhysPerf_Hcc} and~\ref{sec:PhysPerf_strangeTagging}, 
jet-flavour tagging is essential for precision measurements of the Higgs couplings to quarks and gluons. 
A modified \textsc{Delphes} configuration simulates the CLD detector geometry and resolutions in fast simulation, referred to as the \emph{CLD fast simulation}. 
The full simulation uses the CLD conformal tracking (the baseline tracking reconstruction in CLD) to reconstruct charged particles. 
The \textsc{Pandora} particle flow algorithm~\cite{pandorapfa,pandorasdk} is used for global event reconstruction, to build a particle-level view of the event. 

Jets are clustered with the \kt Durham algorithm~\cite{Catani:1991hj} using $N = 2$ exclusive clustering on 
$\epem \to \PZ\PH \to \PGn \PAGn j j$ events. 
The input to jet clustering are the reconstructed tracks and the neutral particles reconstructed by \textsc{Pandora}. 
The jet flavour tagger~\cite{Bedeschi:2022rnj} is then applied to assign each jet a probability of originating from a given flavour among seven possible categories: 
\Pg, \PQu, \PQd, \PQs, \PQc, \PQb, and \PGt. 
The tagger uses the information of all the input particles within a jet, including kinematic, displacement, and particle identification properties. 
Adding direct vertex information (positions and invariant masses of secondary vertices) to the network does not improve performance, 
indicating that the model effectively learns vertexing from the track displacement, the charged particle content, and the kinematic variables. 

\begin{figure}[ht]
\centering
\includegraphics[width=0.48\textwidth]{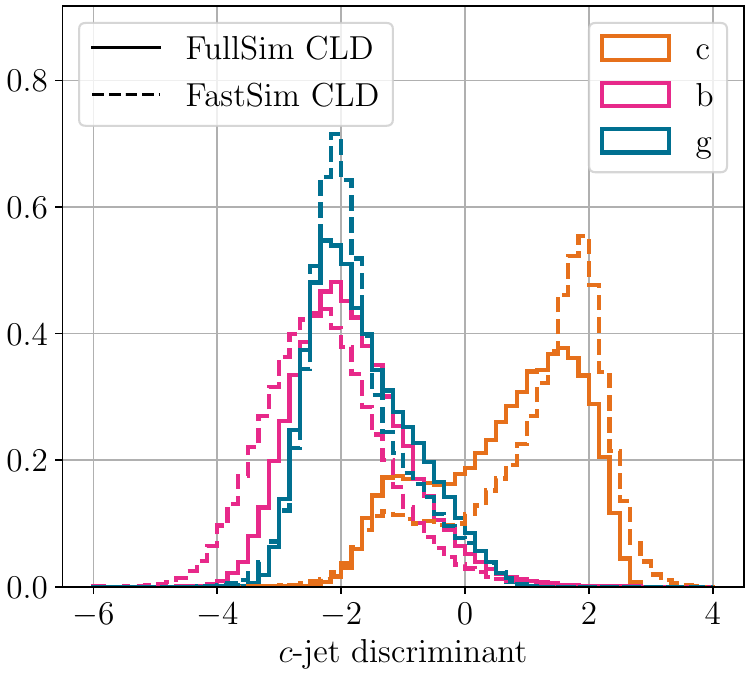} 
\includegraphics[width=0.48\textwidth]{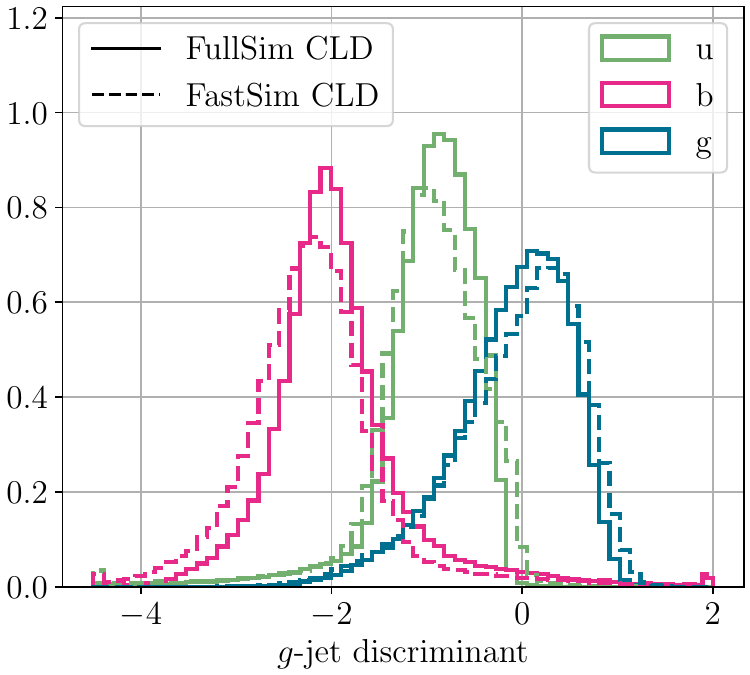} 
\caption{Flavour tagging output in the full (solid) and fast (dashed) CLD simulation for charm (left) and gluon (right) multi-variate discriminants for various jet species.}
\label{fig:fullsim_tagging}
\end{figure}

The implementation of the algorithm in \textsc{Delphes} and its performance are described in detail in Ref.~\cite{Bedeschi:2022rnj}. 
Both fast and full simulation datasets are trained using identical input features. 
For illustration purposes, Fig.~\ref{fig:fullsim_tagging} shows the charm and gluon score, 
defined as the monotonic transformation of the tagging probability $\log(P\,/\,(1-P))$. 
The fast- and full-simulation output scores agree reasonably well. 
Additional jet tagging scores and performance figures can be found in Ref.~\cite{TAGGINGFULLSIM_NOTE}.

It is observed that fast simulation features better performance across all tagging categories. 
For instance, in \PQc-tagging, the efficiency of correctly identifying \PQc-jets vs.\ misidentifying light-quark jets decreases from 80\% in fast simulation to 70\% in full simulation, 
for a 1\% light quark misidentification rate. 
This loss in performance is under investigation but can be largely explained by track reconstruction inefficiencies in full simulation and a sub-optimal particle-flow algorithm parameter tuning.

This study shows that initial jet-flavour tagging performance in full simulation in CLD is sub-optimal. 
Future work should focus on improving CLD reconstruction to mitigate the identified limitations, such as low tracking efficiency at low momentum and large neutral hadron mis-identification probability. 
In addition, exploring the use of energy loss per unit length (d$E$/d$x$) information from the silicon tracker may provide additional particle identification capabilities, 
which can be useful for strange jet identification and can also provide additional exploitable information for charm, bottom, and tau tagging. 
Making progress in these areas is crucial to achieve the FCC-ee precision Higgs physics goals.

\subsubsection{Higgs boson mass determination}

Measuring the Higgs boson mass precisely at FCC-ee is important because it constitutes a limiting parametric uncertainty in the calculations of the branching ratios of the Higgs boson decay channels. 
Achieving a precision of 4\,MeV makes this uncertainty negligible and leads to more accurate predictions of the Higgs boson decay properties. 
Also, knowing $m_{\PH}$ with a precision comparable to the Higgs boson natural width 
is the minimum requirement for potentially measuring the electron Yukawa coupling at $\sqrt{s} = 125$\,GeV, 
the Higgs boson pole (Section~\ref{sec:eYukawa}).

\begin{figure}[ht]
\centering
\includegraphics[width=0.48\textwidth]{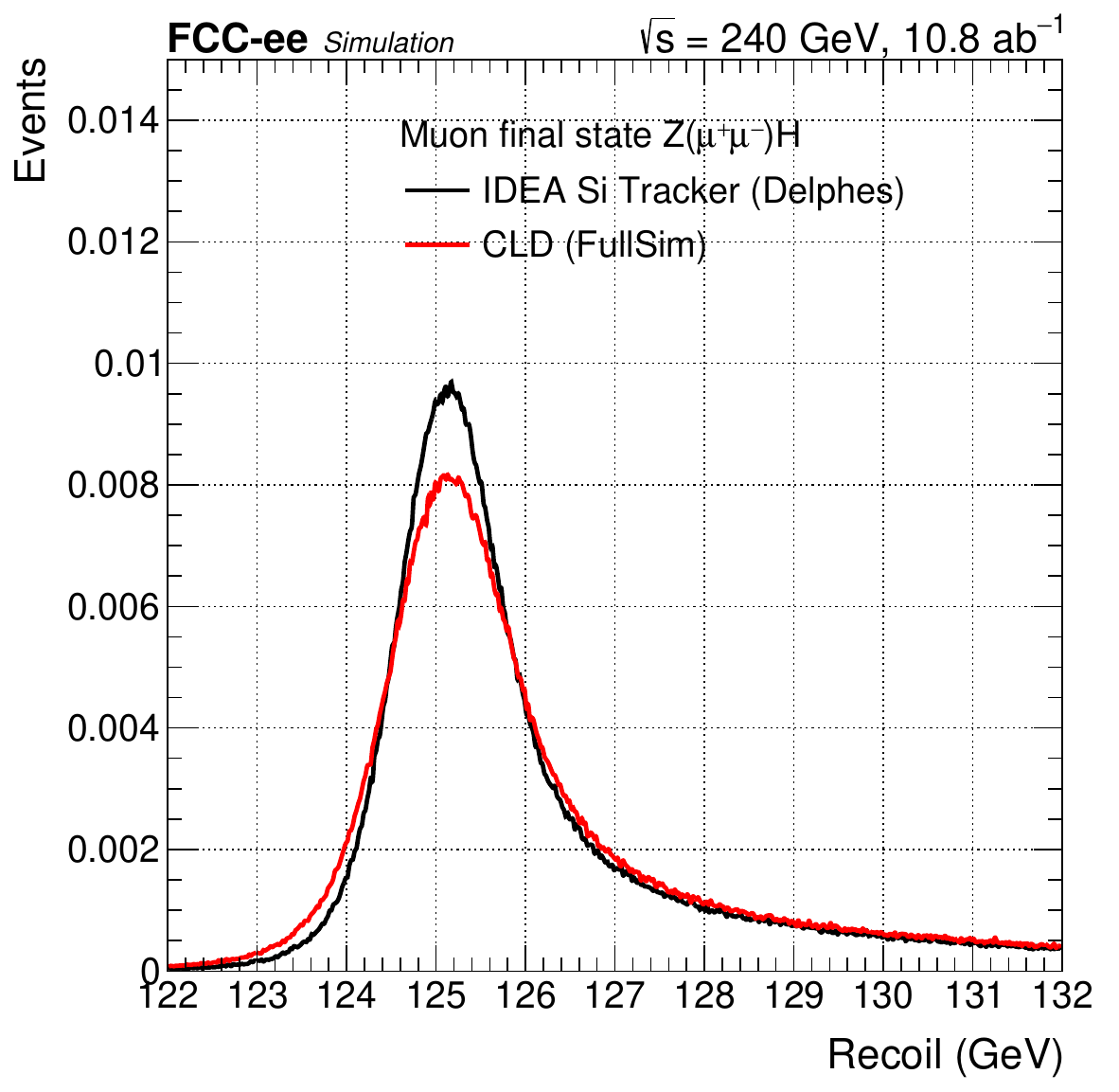}    
\includegraphics[width=0.48\textwidth]{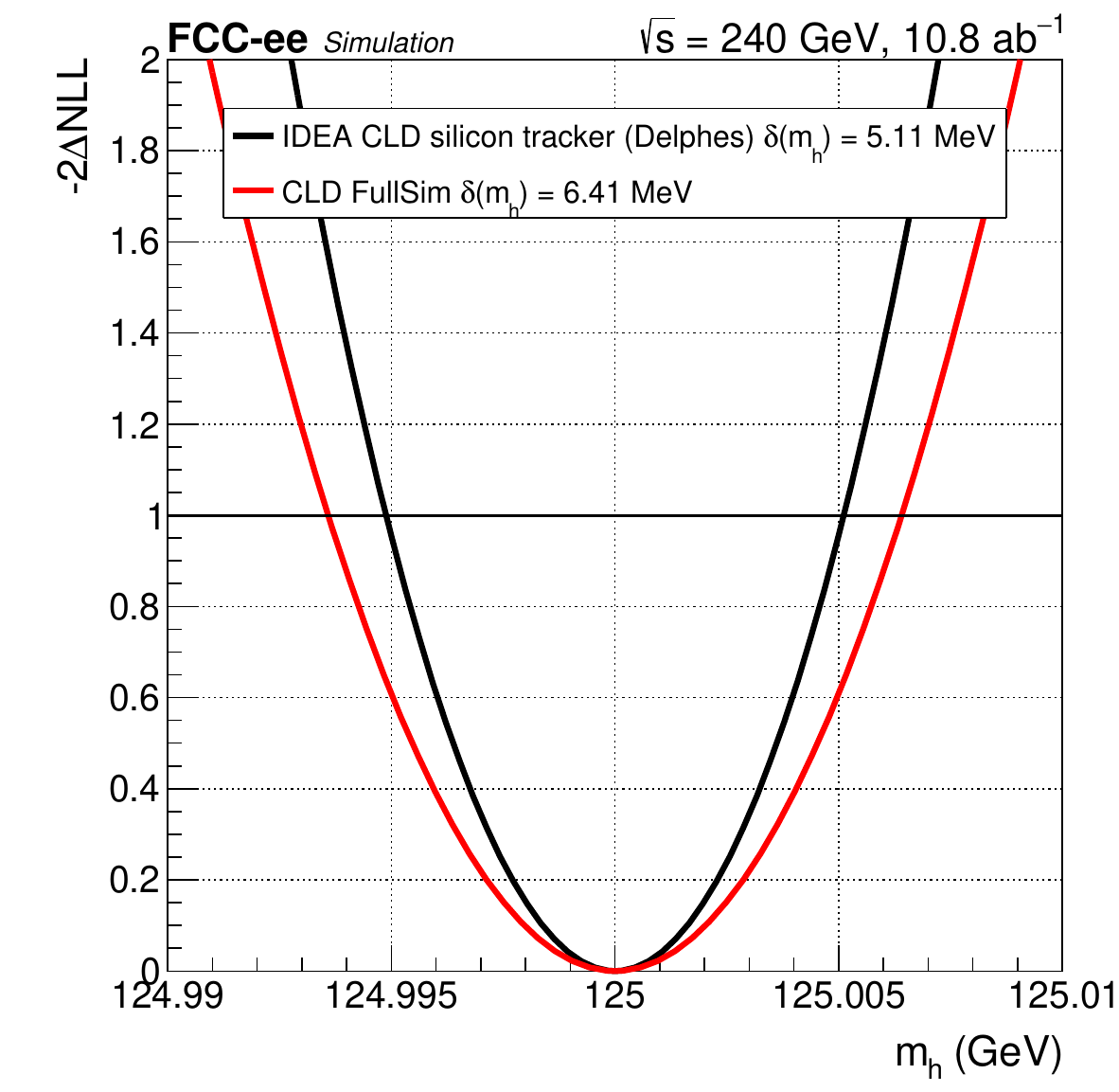}
\caption{Higgs boson mass determination using the recoil mass method in the dimuon channel in the full (red curves) and fast (red curves) simulation. 
Left: The $m_\text{recoil}$ distribution. 
Right: Likelihood scan of the Higgs boson mass and corresponding precision.}
\label{fig:fullsim_mh}
\end{figure}

The Higgs boson mass measurement prospects have been studied in full simulation using the CLD detector~\cite{HiggsNoteRecoil}. 
Track reconstruction and muon identification are the main event reconstruction features required for this measurement, 
which allows a straightforward comparison to fast simulation. 
The Higgs boson mass recoil method and the resulting sensitivity obtained with fast simulation in the dimuon and dielectron channels 
are extensively discussed in Sections~\ref{sec:PhysPerf_HiggsRecoil} and~\ref{sec:PhysPerf_ECAL}, 
assuming the IDEA detector performance with a gaseous drift chamber or, alternatively, a CLD-like silicon tracker. 
The driving experimental factor for the sensitivity is the track momentum resolution, ultimately limited by the multiple scattering induced by the beam pipe and tracker material. 
A gaseous tracking device is, therefore, preferred for this measurement. 
In this study, the recoil mass method has been applied to the CLD silicon tracking full simulation and compared to the CLD-like fast simulation results, 
as shown in Fig.~\ref{fig:fullsim_mh}. 
The recoil mass distribution obtained in $\epem \to \PZ\PH \to \PGmp\PGmm \PH$ events is shown on the left panel; 
it agrees at the 15\% level with the fast simulation distribution. 
This small discrepancy is most likely explained by a small difference in the tracker material description and is currently under investigation. 
The right panel shows the corresponding precision on the Higgs boson mass determination, 
obtained from a likelihood scan using different mass hypotheses, yielding a precision of 6.4\,MeV for full simulation, 
compared to 5.1\,MeV in fast simulation.

This study shows that the full simulation of the CLD detector is available to perform a complete physics benchmark study. 
Further work is needed to understand the discrepancy in the observed precision between fast and full simulation, 
which may also point to shortcomings in the full event simulation and reconstruction. 
Extending the study to the dielectron final state, which requires the implementation of photon bremsstrahlung recovery to fully reconstruct the final-state electron momentum, 
will be an important step forward in the understanding of the CLD detector performance in this important measurement.

\subsubsection{Tau polarisation}

The \PZ-pole running will provide an unprecedented sample of $2 \times 10^{11}$ clean $\PGtp\PGtm$ events
and a unique opportunity for the determination of the tau polarisation,
which is in turn very sensitive to the electroweak couplings of the \PZ boson ($\sin^2 \theta_{\PW}$). 
The tau polarisation is defined as 
\mbox{$\mathcal{P}_{\PGt} = (\sigma_+ - \sigma_-)/(\sigma_+ + \sigma_-)$}, 
where $\sigma_+$ and $\sigma_-$ are the production cross sections of left-handed and right-handed taus, respectively. 
It can be expressed as a function of the electron and tau asymmetry parameters, $\mathcal{A}_{\Pe}$ and $\mathcal{A}_{\PGt}$, 
and it depends on the angle $\theta$ between the tau momentum and the electron beam in the centre-of-mass frame of the collision:
\begin{equation}
\mathcal{P}_{\PGt}(\cos\theta) = 
- \frac{\mathcal{A}_{\PGt} (1+\cos^2 \theta) + 2 \mathcal{A}_{\Pe} \cos\theta} 
{1+ \cos^2\theta + 2 \mathcal{A}_{\Pe} \mathcal{A}_{\PGt} \cos\theta} \, .
\end{equation}

The measurement of $\mathcal{P}_{\PGt}(\cos\theta)$ allows the determination of $\mathcal{A}_{\PGt}$ and $\mathcal{A}_{\Pe}$, 
as a test of the universality of the \PZ boson couplings to electrons and taus. 
Integrating over $\cos\theta$, one obtains $\mathcal{P}_{\PGt}^\text{total} = -\mathcal{A}_{\PGt}$. 
At LEP, the uncertainty in $\mathcal{A}_{\Pe}$ was statistically dominated~\cite{ALEPH:2005ab}, while for $\mathcal{A}_{\PGt}$ systematic and statistical uncertainties were at the same level. 
The large FCC-ee data samples and advanced detector capabilities are expected to significantly reduce these uncertainties. 
Assuming a factor 10 improvement in the systematic uncertainty with respect to LEP, the systematic uncertainty in $\mathcal{A}_{\PGt}$ could be as low as 0.02\%, 
with $\mathcal{A}_{\Pe}$ having an even smaller uncertainty.

To confirm our expectations that this precision can be achieved, 
a full simulation study of the CLD detector is underway, focusing on tau reconstruction and identification. 
Tau identification relies on reconstructing charged hadrons and photons from $\PGpz \to \PGg\PGg$ decays. 
About 65\% of tau decays are hadronic, decaying to one or three charged hadrons accompanied by photons from \PGpz decays. 
An algorithm based on the \textsc{Pandora} particle-flow candidates has been developed to reconstruct the main decay modes: 
$\PGt \to \PGp\PGn$, $\PGt \to \PGr\PGn \to \PGp\PGpz \PGn$, and $\PGt \to \Pa_1\PGn$. 
In parallel, a GNN method is being developed to improve the identification performance. 

A first analysis limited to $\PZ \to \PGt\PGt$ events in which one of the taus decays leptonically in one hemisphere and the other decays hadronically in the other hemisphere 
has been performed~\cite{alcaraz_maestre_2024_qnyyh-skw71}. 

\begin{figure}[ht]
\centering
\includegraphics[width=0.48\textwidth]{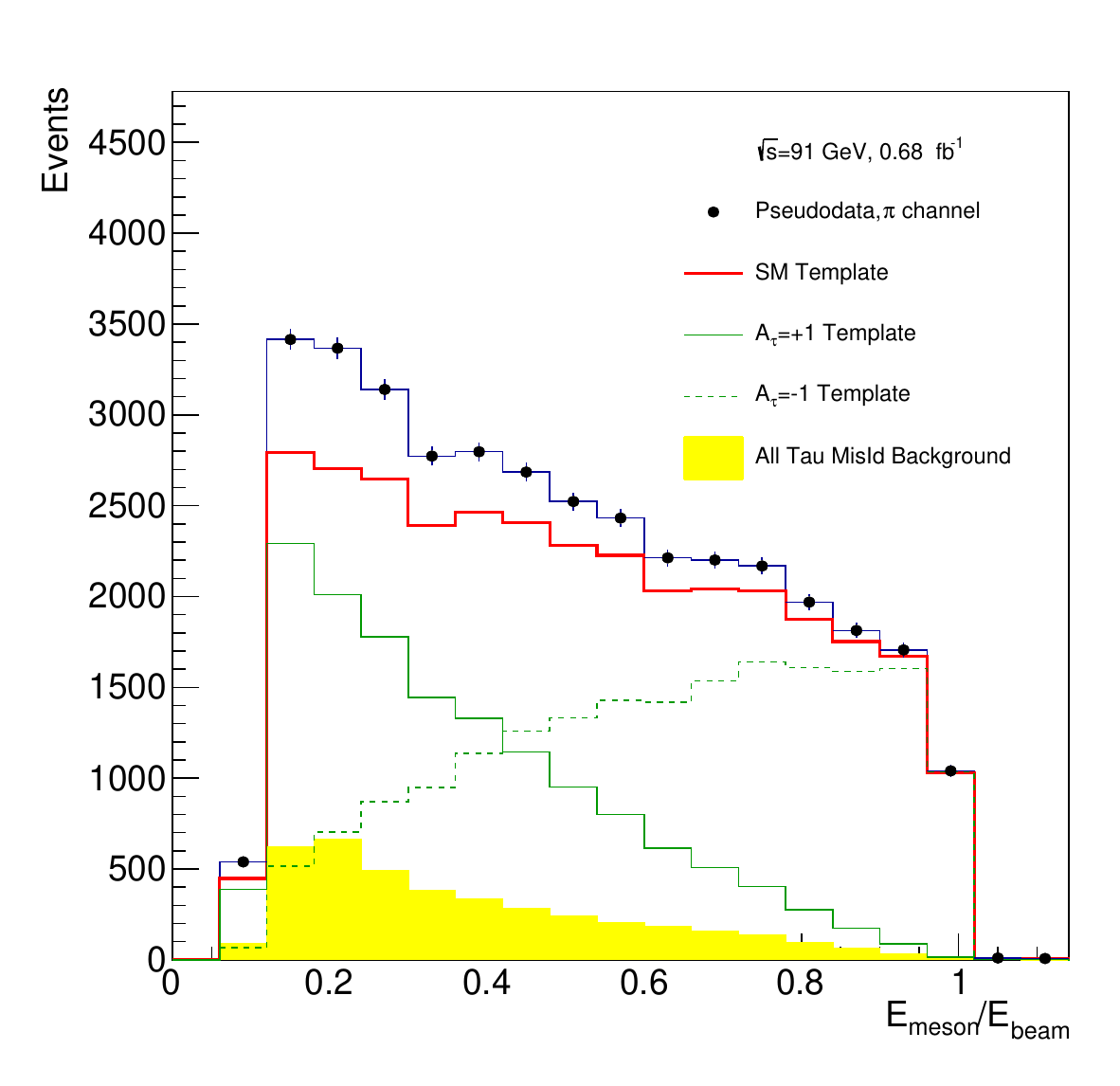} \\
\includegraphics[width=0.48\textwidth]{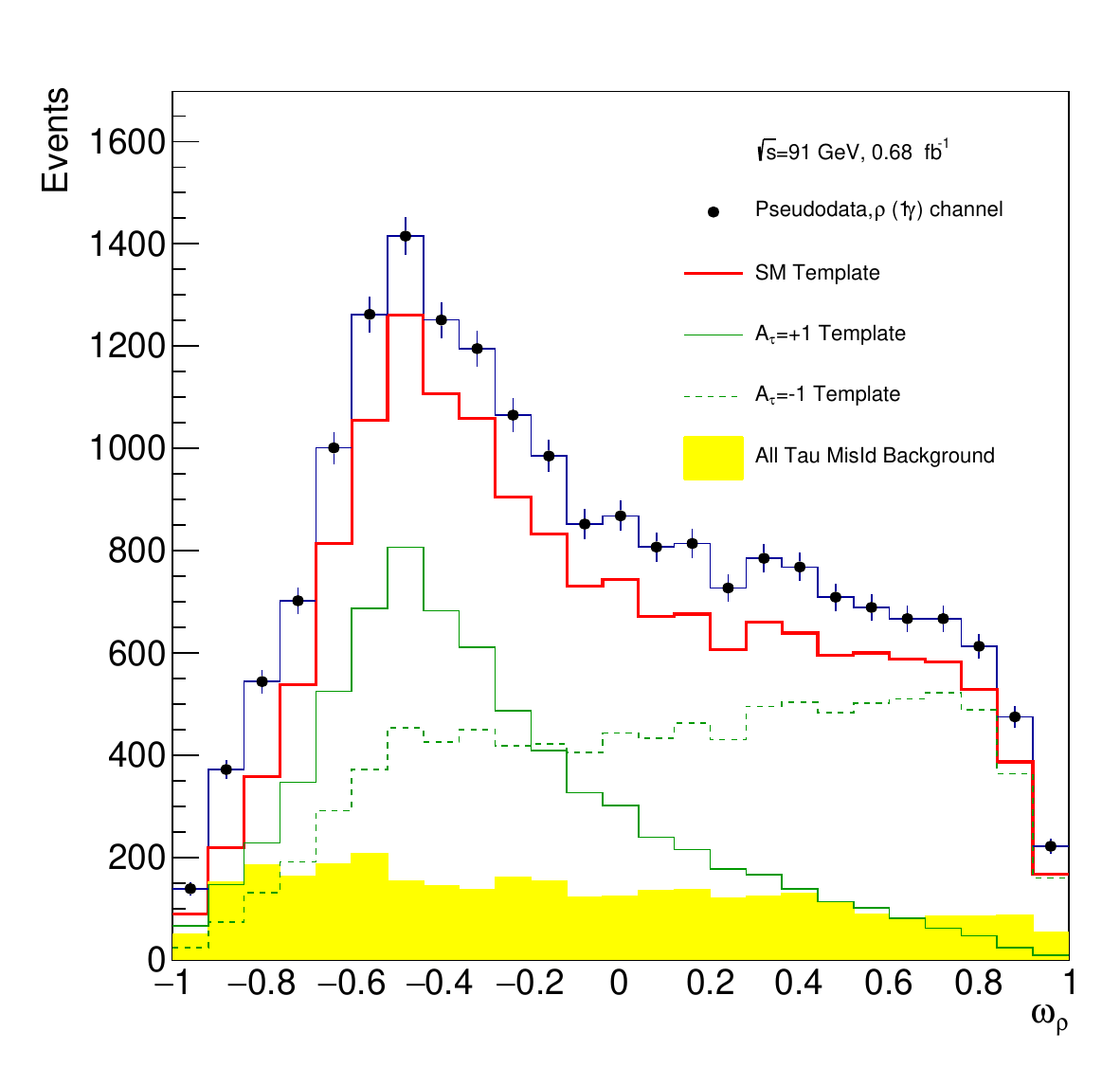}
\includegraphics[width=0.48\textwidth]{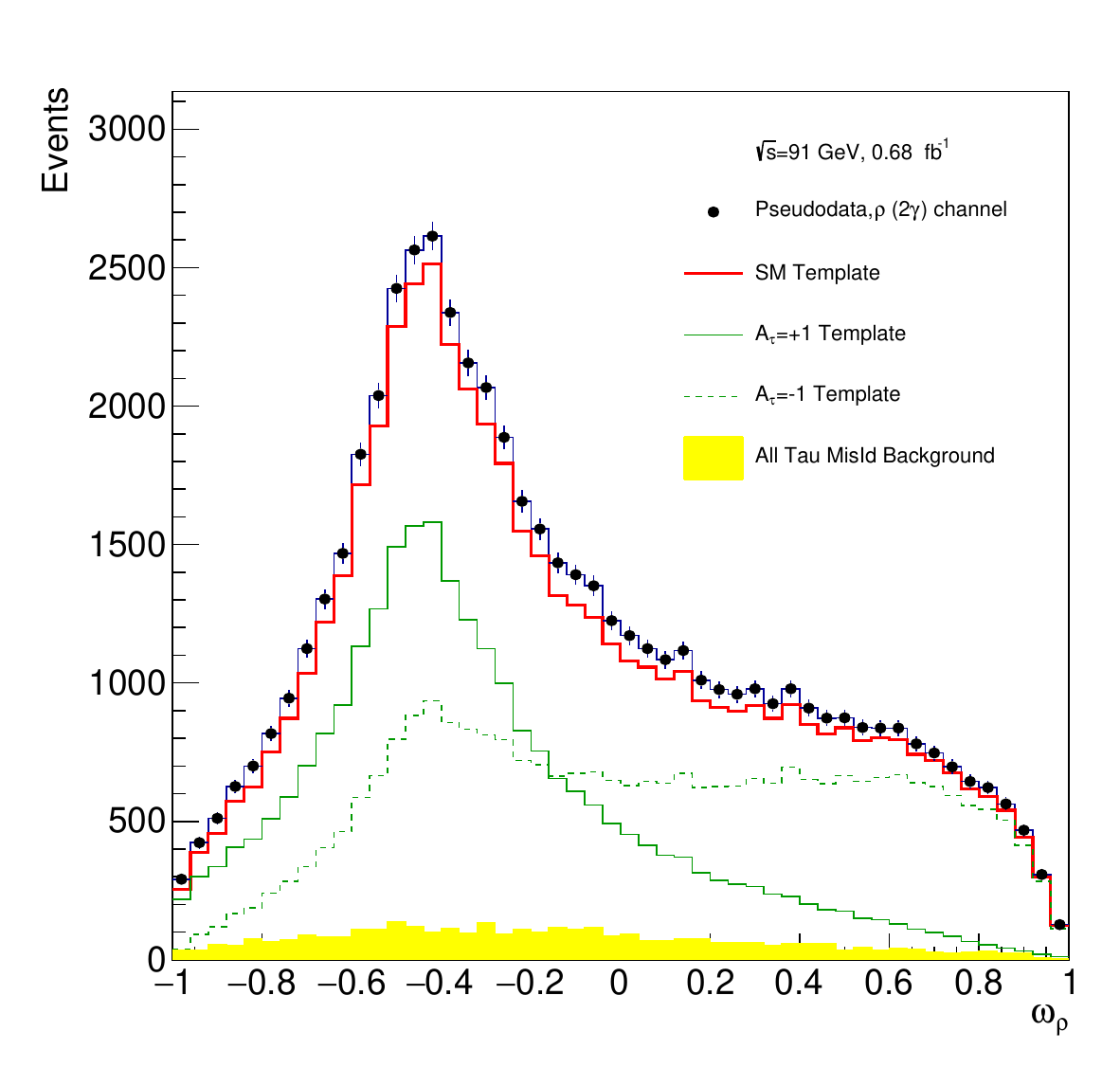}    
\caption{Optimal variables used to extract the \PGt polarisation in three decay modes: 
\PGp (top), \PGr with one identified photon (bottom left), and \PGr with two identified photons (bottom right).}
\label{fig:fullsim_taupol}
\end{figure}

Figure~\ref{fig:fullsim_taupol} shows the optimal variables for the three main configurations considered in the analysis. 
The first mode is designed to select $\PGt \to \PGp \PGn$ decays. 
In this case the optimal variable is directly the energy fraction  $x_{\PGp} = {E_{\PGp}} \,/\, {E_\text{beam}}$. 
The other two configurations correspond to the $\PGt \to \PGr \PGn \to \PGp \PGpz\PGn$ channel with the reconstruction of either one or both of the \PGpz decay photons. 
Here the optimal observable at LEP was defined to be
\begin{equation}
\omega_{\PGr} = \frac{W_+(\theta^*, \psi) - W_-(\theta^*, \psi)}{W_+(\theta^*, \psi) + W_-(\theta^*, \psi)} \, ,
\end{equation}
where $W_+$ and $W_-$ represent the angular distributions of the \PGr decay products for different helicity states, 
and $\theta^*$ and $\psi$ are angles describing the decay products in the \PGt rest frame~\cite{Nikolic:1996nj}. 
The reweighted $\mathcal{P}_{\PGt} = \pm 1$ signal templates are shown in comparison to the SM prediction. 
Additionally, the backgrounds coming from tau misidentification are also shown. 
No other backgrounds are considered at this stage. 
Further studies of the full simulation samples and optimisation of the reconstruction algorithm are in progress. 

An analysis aiming to extract $\mathcal{A}_{\PGt}$ was performed using a log-likelihood fit of the optimal variable in the \PGr channel with two resolved photons. 
For an integrated luminosity of $17\,\text{ab}^{-1}$, corresponding to data collected by one FCC-ee experiment in a single year, 
the statistical uncertainty achieved is $7 \times 10^{-5}$. 
Future work includes refining the tau reconstruction algorithms, expanding the analysis to cover more decay channels and backgrounds, 
and performing a comprehensive extraction of the polarisation parameters across multiple $\cos\theta$ bins. 
In addition, focus will be placed on deriving explicit calorimeter requirements to reduce 
the dominant systematic uncertainties reflecting the \PGpz and fake photon identification efficiencies 
as close as possible to (or below) the statistical uncertainties.

\subsubsection{Summary of detector requirements}

The FCC-ee detectors are expected to perform precise measurements across a wide range of physics processes, 
requiring optimised designs. 
This summary provides a simplified overview of the key requirements, but it is important to recognise that these specifications often involve trade-offs, 
which should and will be studied in full simulation. 
For instance, incorporating a RICH or time-of-flight detector for particle identification may negatively impact the resolution of visible energy measurements. 
Identifying low-momentum muons effectively requires smaller detectors or a reduced magnetic field, 
which compromises the ability to accurately track high-momentum particles, impacting the overall physics programme. 
Finally, it should be noted that over-ambitious requirements could compromise the overall feasibility and the reliable operation of the detector systems. 
It is therefore of the utmost importance that the continued and creative effort to soften certain detector requirements with in-situ methods be consolidated and amplified.

\begin{table}[ht]
\renewcommand{\arraystretch}{2}
\centering
\caption{Summary of detector requirements.}
\label{tab:requirements}
\small
\begin{tabular}{lccc}
\toprule
& \textbf{Aggressive} & \textbf{Conservative} & \textbf{Comments} \\
\midrule
\textbf{Beampipe} & $X/X_0 < 0.5$\% & $X/X_0 < 1$\% & $\PB \to \PK^{\ast} \PGt \PGt$\\
\hline
\multirow{2}{*}{\textbf{Vertex}} & 
\makecell{$\sigma(d_0) = 3 \oplus 15\,/\,(p \, \sin^{3/2} \,\theta)$\,$\mu$m \\ $X/X_0 < 1$\%} & --  & \makecell{$\PB \to \PK^{\ast} \PGt \PGt$ \\ $R_{\PQc}$} \\
\cline{2-4}
& $\delta L = 5$\,ppm & -- & $\delta \tau_{\PGt} < 10$\,ppm\\
\hline
\multirow{2}{*}{\textbf{Tracking}} & 
\makecell{${\sigma_p}/{p} < 0.1$\% \\ for $\mathcal{O}$(50)\,GeV tracks} & 
\makecell{${\sigma_p}/{p} < 0.2$\% \\ for $\mathcal{O}$(50)\,GeV tracks} & 
\makecell{$\delta M_{\PH} = 4$\,\MeV \\ $\delta \Gamma_{\PZ} = 15$\,keV \\ $\PZ \to \PGt \PGm$} \\
\cline{2-4}
 & t.b.d.\ & $\sigma_\theta < 0.1$\,mrad &  $\delta \Gamma_{\PZ} (\text{BES}) < 10$\,keV \\
\hline
\multirow{3}{*}{\textbf{ECAL}} & 
${\sigma_E}/{E} = {3\%}/{\sqrt{E}}$ & ${\sigma_E}/{E} = {10\%}/{\sqrt{E}}$ & 
\makecell{$\PZ \to \PGne\PAGne$ coupling, \\ B physics, ALPs} \\
\cline{2-4} & 
\makecell{$\Delta x \times \Delta y =$ \\ $2 \times 2$\,mm$^2$} & 
\makecell{$\Delta x \times \Delta y =$ \\ $5 \times 5$\,mm$^2$} & 
\makecell{\PGt polarisation \\ boosted \PGpz decays \\ bremsstrahlung recovery }\\
\cline{2-4}
& \makecell{$\delta z = 100$\,\micron, \\ $\delta R_{\text{min}} = 10$\,\micron\ ($\theta = 20^\circ$)}
& \makecell{In-situ constraint with \\ dilepton/diphoton events} & 
\makecell{alignment tolerance for \\ $\delta \lumi = 10^{-5}$ with $\PGg\PGg$ events} \\ 
\hline
\multirow{2}{*}{\textbf{HCAL}} & 
${\sigma_E}/{E} = {30\%}/{\sqrt{E}}$ & ${\sigma_E}/{E} = {50\%}/{\sqrt{E}}$ & 
\makecell{$\PH \to \PQs\PAQs$, $\PQc\PAQc$, $\Pg\Pg$, invisible \\ HNLs} \\
\cline{2-4} & 
\makecell{$\Delta x \times \Delta y =$ \\ $2 \times 2$\,mm$^2$} & 
\makecell{$\Delta x \times \Delta y =$ \\ $20 \times 20$\,mm$^2$} & 
$\PH \to \PQs\PAQs$, $\PQc\PAQc$, $\Pg\Pg$\\
\hline
\textbf{Muons} & low momentum ($p < 1$\,GeV) ID & -- & $\PBs \to \PGn \PAGn$ \\
\hline
\textbf{Particle ID} & \makecell{3\,$\sigma$ $\PK/\PGp$ \\ $p < 40$\,GeV} & \makecell{3\,$\sigma$ $\PK/\PGp$ \\ $p < 30$\,GeV} & \makecell{$\PH \to \PQs\PAQs$ \\ $\PQb \to \PQs \PGn\PAGn$, \ldots}\\
\hline
\textbf{LumiCal} & \makecell{tolerance $\delta z = 100$\,\micron, $\delta R_{\text{min}} = 1$\,\micron \\ 
acceptance 50--100 mrad} & -- 
& \makecell{$\delta \lumi = 10^{-4}$ target \\ (Bhabha)} \\
\hline
\textbf{Acceptance} & 100\,mrad  & -- & \makecell{$\epem \to \PGg \PGg$ \\ $\epem \to \epem \PGtp \PGtm (\PQc\PAQc)$} \\
\bottomrule
\end{tabular}
\end{table}

An incomplete list of requirements is summarised in Table~\ref{tab:requirements}.
For physics channels involving charm, bottom, and tau production, the beam-pipe is required to have a minimal material budget, 
particularly critical for processes such as the $\PB \to \PK^\ast \PGt \PGt$ decay. 
The vertex detector must achieve a spatial resolution of 3\,$\mu$m or better and a material thickness below 1\% of a radiation length, 
to improve the precision of certain important measurements, 
such as the measurement of $R_{\PQc}$ at the \PZ pole or that of the branching fraction of the aforementioned $\PB \to \PK^\ast \PGt \PGt$ decay. 
The \PGt lifetime measurement requires the knowledge of the length scale of the vertex detector to be monitored with optical techniques with a precision of $\delta \tau_{\PGt} < 10$\,ppm. 
The momentum resolution in the tracking systems, $\sigma_p\,/\,p$, needs to be better than 0.2\% for ${\cal{O}}$(50)\,GeV tracks 
in order to achieve a precision of 4\,MeV on the Higgs boson mass. 
Achieving a precision of ${\cal{O}}$(10)\,keV on the \PZ width requires an exceptional track momentum resolution for such tracks, better than 0.1\%.
It also demands a precise knowledge of the BES, for which an angular tracking resolution $\sigma_\theta$ better than 0.1\,mrad is needed.
For flavour physics at the \PZ peak, where low momentum tracks are involved, 
a low mass gaseous tracker is advantageous since the momentum resolution is minimally affected by multiple scattering.
Having a highly-efficient and pure tracking from 100\,MeV to tens of GeV is of crucial importance to improve the dijet energy and mass resolution with particle-flow reconstruction.

Electromagnetic calorimetry with excellent energy resolution (few \%) is required to improve the precision 
on the $\PZ \to \PGne \PAGne$ coupling via the $\epem \to \PGne \PAGne \PGg$ process, 
on lepton flavour violating decays such as the $\PZ \to \PGm \Pe$ and $\PGt \to \PGm \PGg$ decays, 
on decays of heavy flavoured hadrons to photons or to \PGpz mesons, 
and on ALPs searches. 
The electromagnetic calorimeters must provide an energy resolution of $\sigma_E\,/\,E = 3\%\,/\,\sqrt{E}$
for aggressive scenarios and $10\%\,/\,\sqrt{E}$ for conservative ones, 
with spatial granularity between $2 \times 2$\,mm$^2$ and $5 \times 5$\,mm$^2$. 
High granularity is crucial for \PGt polarisation measurements, boosted \PGpz decays, and bremsstrahlung photon recovery. 
Additionally, alignment tolerances of $\delta z = 100$\,$\mu$m and $\delta R_\text{min} = 10$\,$\mu$m (measured at a 20$^\circ$ polar angle) 
are essential for a precise calibration of the acceptance or, equivalently, of the luminosity. 
The large dilepton and diphoton samples will be instrumental to ascertain in-situ the acceptance determination.

A good particle identification performance is required to achieve 3\,$\sigma$ $\PK/\PGp$ separation up to 30\,GeV 
and is essential for studying $\PH \to \PQs\PAQs$ decays, to study rare decays such as $\PQb \to \PQs \PGn \PAGn$, and, more generally, to maximise the FCC-ee flavour physics potential.

Hadron calorimeters with sufficient resolution and granularity are crucial for the success of the Higgs boson physics programme. 
In particular, the expected precision in the Higgs couplings to quarks and gluons is driven by the particle-flow global event reconstruction capabilities and the visible mass resolution. 
The hadron calorimeters must feature a stochastic term of less than 30\% in the aggressive scenario or less than 50\% in the more conservative scenarios. 
The transverse granularity should range from $2 \times 2$\,mm$^2$ to $20 \times 20$\,mm$^2$. 
The luminosity monitors must have a spatial tolerance of, at least, $\delta z = 100$\,$\mu$m and $\delta R_\text{min} = 1$\,$\mu$m, with an acceptance range of 50--100\,mrad. 
The detector must ensure optimal hermeticity. 
Requiring no gaps (and possibly a small overlap) between the LumiCal and the combined tracker plus calorimeter system 
sets the requirement for the tracker and calorimeter acceptance at 100\,mrad. 
Forward coverage is necessary for the detection and identification of photons produced in the $\epem \to \PGg \PGg$ process, 
which promises to provide an independent measurement of the luminosity. 
Processes involving the production of forward electrons, for example $\epem \to \epem \PGtp \PGtm (\PQc\PAQc)$ might also benefit from even more forward coverage.

Finally, the muon systems will be used mainly to identify muons, given that their momenta will be optimally measured by the tracking system at FCC-ee. 
They should provide excellent identification for pion rejection and standalone momentum measurement for long-lived particle searches. 

\subsection{Outlook}

Both software and analysis tools have matured during the Feasibility Study, 
and the common software framework (Chapter~\ref{sec:software}) was instrumental 
in developing a number of key physics case studies in view of extracting some of the necessary detector requirements. 
So far, most of the studies have been
performed with a fast simulation of the response of a few detector benchmarks, 
similar to those presented in the Conceptual Design Report. 
Variations around these baseline concepts were applied 
to quantify the generic sensitivity dependence of key observables on the detector performance. 
In the next phase of the study, 
more explicit requirements will be obtained with a (full) simulation, reconstruction, and analysis strategy, 
optimised to the specific capabilities of the considered hardware, 
when the current R\&D efforts start converging in concrete and definite detector proposals. 

Matching experimental systematic uncertainties to the FCC-ee statistical precision 
will remain a critical objective of the entire physics programme. 
This perspective constitutes a superb opportunity for detector designers. 
It also continuously generates new and creative ideas from the analysts to control uncertainties from the data themselves: 
past experience has shown that a careful analysis of systematic effects and corrections almost always
boils down to a statistical problem. 
This strategy will be generalised and will lead to a better precision for most observables, 
owing to the very large event samples expected at FCC-ee.   

With $6 \times 10^{12}$ \PZ and $3 \times 10^8$ \PW pairs, 
the leading challenge for the detector and analysis designs will continue to be the precision 
target with which the electroweak observables (Table~\ref{tab:EWPO}) can be measured.
Work on the hadronic, dilepton, and diphoton cross-section measurements has started, 
but much remains to be done, 
for example, to consolidate the in-situ acceptance determination method for lepton pairs 
and for the luminosity measurement with photon pairs and low-angle Bhabha scattering. 
The field of measurements at FCC-ee is huge, even when limited to the benchmark processes listed in
Ref.~\cite{SnowmassBenchmarkProcesses}. 
An even stronger connection between theorists and experimentalists will help streamlining and refining 
the list of observables and the parameter space to be explored. 
Clearly, the physics requirements will continue to be the driving factor in all efforts to design 
the most technologically-advanced (but also the most realistic) FCC-ee detector concepts, 
to ensure the full coverage of this challenging and promising physics programme.

\cleardoublepage
\section{Machine-detector interface}
\label{sec:mdi}
The FCC-ee interaction region (IR) is designed to reach the highest luminosities 
at all centre-of-mass energies, from the \PZ pole to the $\ttbar$ threshold. 
This design is based on the crab-waist collision scheme, 
with nano-beams at the interaction point (IP), 
large horizontal crossing angle, and crab-waist sextupoles~\cite{Raimondi:2007vi}. 
The Machine Detector Interface (MDI) of FCC-ee has a 
compact and complex design~\cite{FCC-PED-FSR-Vol2,Boscolo:2024bxm} 
that fulfils constraints imposed by 
the machine~\cite{FCC-PED-FSR-Vol2} and detector requirements (Chapter~\ref{sec:requirements}). 

The main beam parameters are recalled in Table~\ref{tab:IR_parameters},
for the four main operational centre-of-mass energies.
In FCC-ee, the two beams circulate in different vacuum chambers, which merge at 1.3\,m from the IP. 
The distance between the face of the  superconducting final focus quadrupole (FFQ) 
and the IP (${\ell^{*}}$) is 2.2\,m, 
well inside the detector volume (Section~\ref{sec:concepts}). 
In addition to the optics constraints on the IR layout, 
physics considerations strongly advocate for hermetic detectors 
and for constraining
the accelerator components within a cone of 100\,mrad from the IP, 
along the $z$ axis\footnote{The coordinate system of the detector 
has the origin centred at the nominal collision point. 
The $z$ axis is defined as the bisector of the axes of the incoming and outgoing beams, 
ideally in the direction of the axis of the experiment solenoid. 
The $x$ axis is in the plane subtended by the two beams and pointing away from the centre of FCC. 
In this way, the positron beam is travelling towards positive values of $z$ (mainly) and $x$ (subordinately). 
Perpendicular to the $(x,z)$-plane, the $y$ axis points upwards. 
The polar angle, $\theta$, is then measured with respect to the $z$ axis 
and the azimuthal angle, $\phi$, with respect to the $x$ axis in the $(x,y)$ plane. 
The radial coordinate, $r=\sqrt{x^2+y^2}$, is the distance from the $z$ axis.}.
These requirements demand a compact MDI design with tight space constraints.

\begin{table}[ht]
\centering
\caption{Key collider parameters for the FCC-ee IRs with 4 IPs.
The bunch length, $\sigma_{z}$, is different for non-colliding bunches (determined by `synchrotron radiation', SR) 
and colliding bunches (determined by `beamstrahlung', BS); 
$\sigma_{x}^{*}$ and $\sigma_{y}^{*}$ denote the bunch sizes at the IP in the (horizontal and vertical, respectively) 
transverse directions, while $\sigma_{\delta}$ is the relative beam energy spread.}
\label{tab:IR_parameters}
\setlength{\tabcolsep}{2.5pt}
\renewcommand{\arraystretch}{0.8}
\begin{tabular}{l c c c c}
\toprule
     &  \PZ   & $\PWp\PWm$ & $\PZ\PH$ & $\ttbar$ \\ \midrule
Beam energy (GeV)              & 45.6  & 80 & 120 & 182.5  \\
Luminosity / IP ($10^{34}$\,cm$^{-2}$s$^{-1}$) & 145 & 20 & 7.5 & 1.41 \\
Beam current (mA)              & 1\,294  &  135  & 26.8  & 5.1  \\
Bunch number / beam            & 11\,200  &  1\,852 & 300   & 64      \\
Bunch spacing (ns)             & 27 & 163 & 1\,008 & 4\,725 \\
$\sigma_{x}^{*}$ ($\mu$m)      & 9.5  & 21.8 & 12.6 & 36.9  \\
$\sigma_{y}^{*}$ (nm)          & 40.1   &  44.7 & 31.6 &  43.6  \\
$\sigma_{z}$ (mm) SR / BS      & 4.7 / 14.6 & 3.46 / 5.28 & 3.26 / 5.59 & 1.91 / 2.33 \\
$\sigma_{\delta}$ (\%) SR / BS & ~0.039 / 0.121~ & ~0.069 / 0.105~ &  ~0.102 / 0.176~ &  ~0.151 / 0.184~ \\
\bottomrule
\end{tabular}
\end{table}

The crab-waist scheme requires a large horizontal crossing angle of 30\,mrad. 
The incoming beams point straight to the IP, 
while the outgoing beam trajectories are strongly bent from the IP, 
so that the beams can successfully merge back close to the opposite ring~\cite{Oide:2016mkm}. 
This scheme ensures that most of the synchrotron radiation (SR) generated at the IR magnetic elements 
does not strike the IR central beam pipe.
The intense radiation emitted during the collision in the electromagnetic field of the opposite beam, 
known as beamstrahlung (BS), is mostly collinear with the outgoing beams, similarly to the SR~\cite{Boscolo:2023grv}. 
Both the BS and SR photons are stopped at about 500\,m downstream of the IP, 
in dedicated dumps~\cite{frasca:ipac2024-tupc66}.
To minimise the level of SR photons that reach the detectors, 
the closest bending magnet is located at more than 100\,m from the IP 
and those located up to 500\,m have a SR critical energy below 100\,keV.

To counteract the beam rotation and deflection 
that would be caused by the simultaneous effects of the detector magnetic field of 2\,T and the crossing angle, 
a compensating solenoid delivering a magnetic field of $-5$\,T is placed at 1.23\,m from the IP, 
cancelling out the longitudinal magnetic field integral along the $z$ axis 
from the last focusing quadrupole to the IP.
Consequently, the front face of the luminosity calorimeter (LumiCal), 
placed in front of the compensating solenoid, is only 1\,m from the IP. 
The LumiCal is centred around the outgoing beam direction 
and measures the integrated luminosity from the rate of low-angle Bhabha events, $\epem\to\epem$.
The first layer of the vertex detector must be placed as close as possible to the IP 
to optimise the precision of the primary and secondary vertex position determination, 
with direct impact on the efficiency and purity of flavour tagging algorithms. 
The smallest affordable distance is set by the central beam pipe radius of 1\,cm. 
The length of the vertex detector (1.86\,m) is chosen to cover the angular region $|\cos\theta| < 0.99$. 
A lightweight mechanical structure is designed for its support.
To avoid any material in front of the LumiCal, 
all other detector elements must be placed above 110\,mrad with respect to the $z$ axis.

\begin{figure}[ht]
\centering
\includegraphics[width=0.75\linewidth]{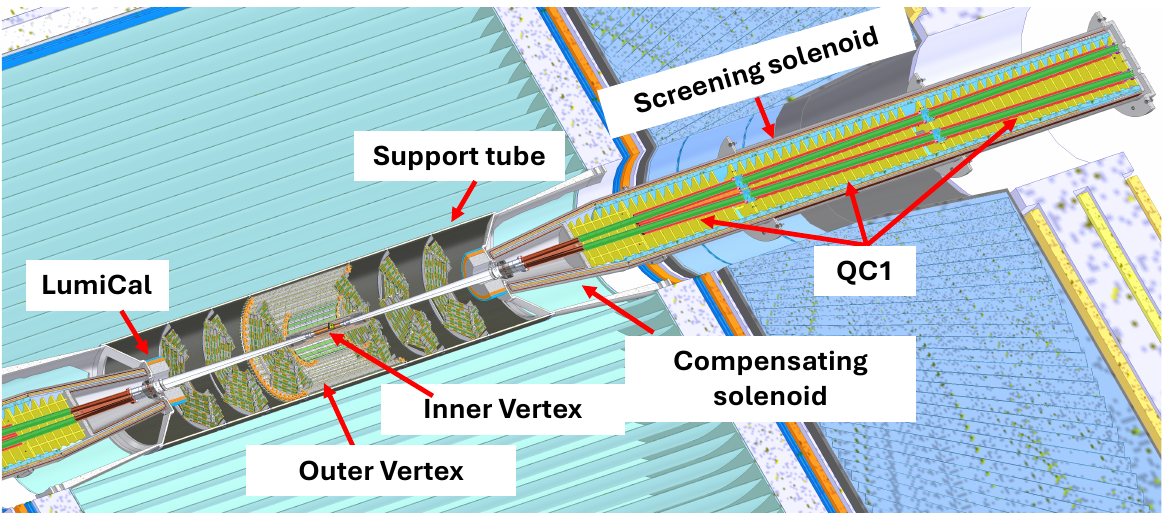}
\caption{Layout of the interaction region. 
The support tube allows the integration of the luminosity calorimeter (LumiCal) and the vertex detector. 
The three segments of the final focus quadrupoles (QC1) are shown with the screening and compensating solenoids.}  
\label{fig:IR}
\end{figure}

An overall design of the interaction region is shown in Fig.~\ref{fig:IR}.
The next sections describe the main features of the MDI. 
A more detailed discussion can be found in Ref.~\cite{boscolo_2023_p4vnt-2va28}.

\subsection{Interaction region layout}

The mechanical layout of the IR comprises an 18\,cm long central beam pipe with a 10\,mm internal radius, 
a pair of ellipto-conical beam pipes about 1064\,mm long on either side, 
a silicon vertex detector covering the radial range from 13.7 to 315\,mm, 
and the LumiCal at 1074\,mm from the IP, 190\,mm thick, covering the angular range between 50 and 134\,mrad.
All these elements are held in place by a lightweight carbon fibre and honeycomb rigid structure support tube,
which allows the overall integration before the insertion in the experiment~\cite{Boscolo:2023hwo}.
The vacuum chambers are made of AlBeMet162, an alloy of Aluminium (38\%) and Beryllium (62\%),  
chosen for its high elastic modulus and low density. 

\begin{figure}[t!]
\centering
\includegraphics[width=0.45\linewidth]{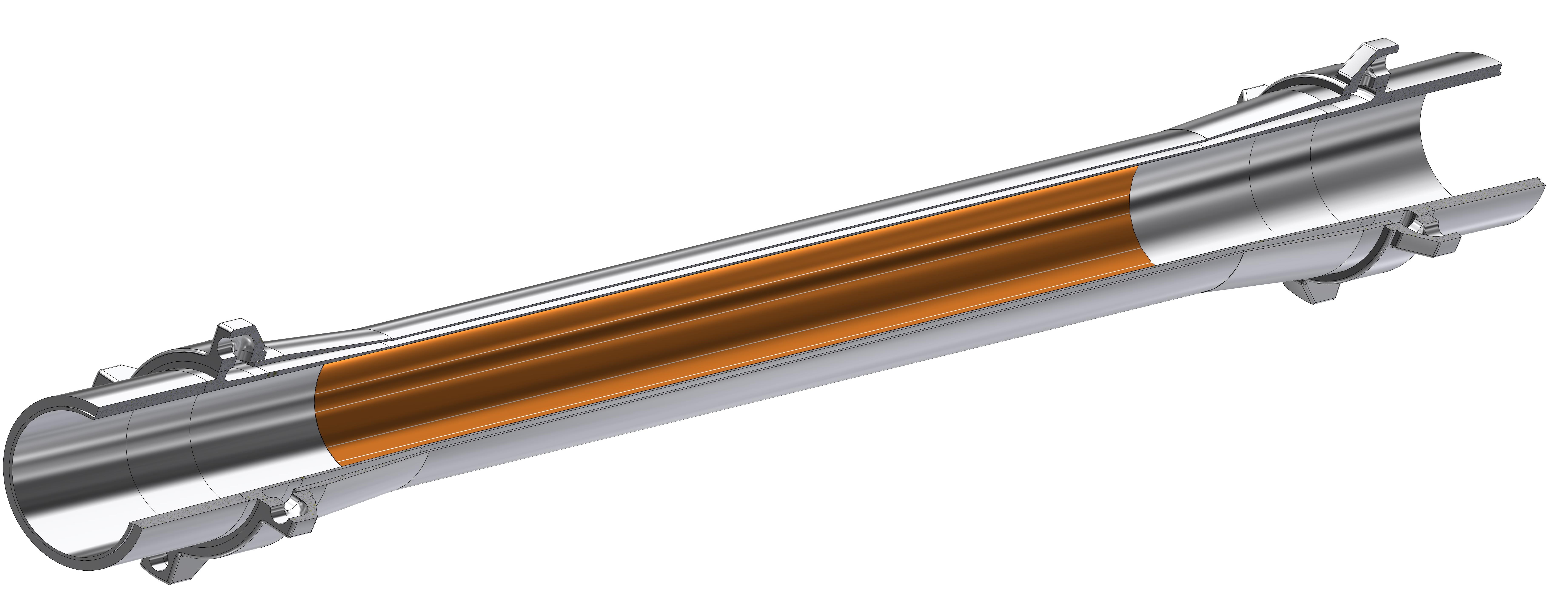}
\includegraphics[width=0.39\linewidth]{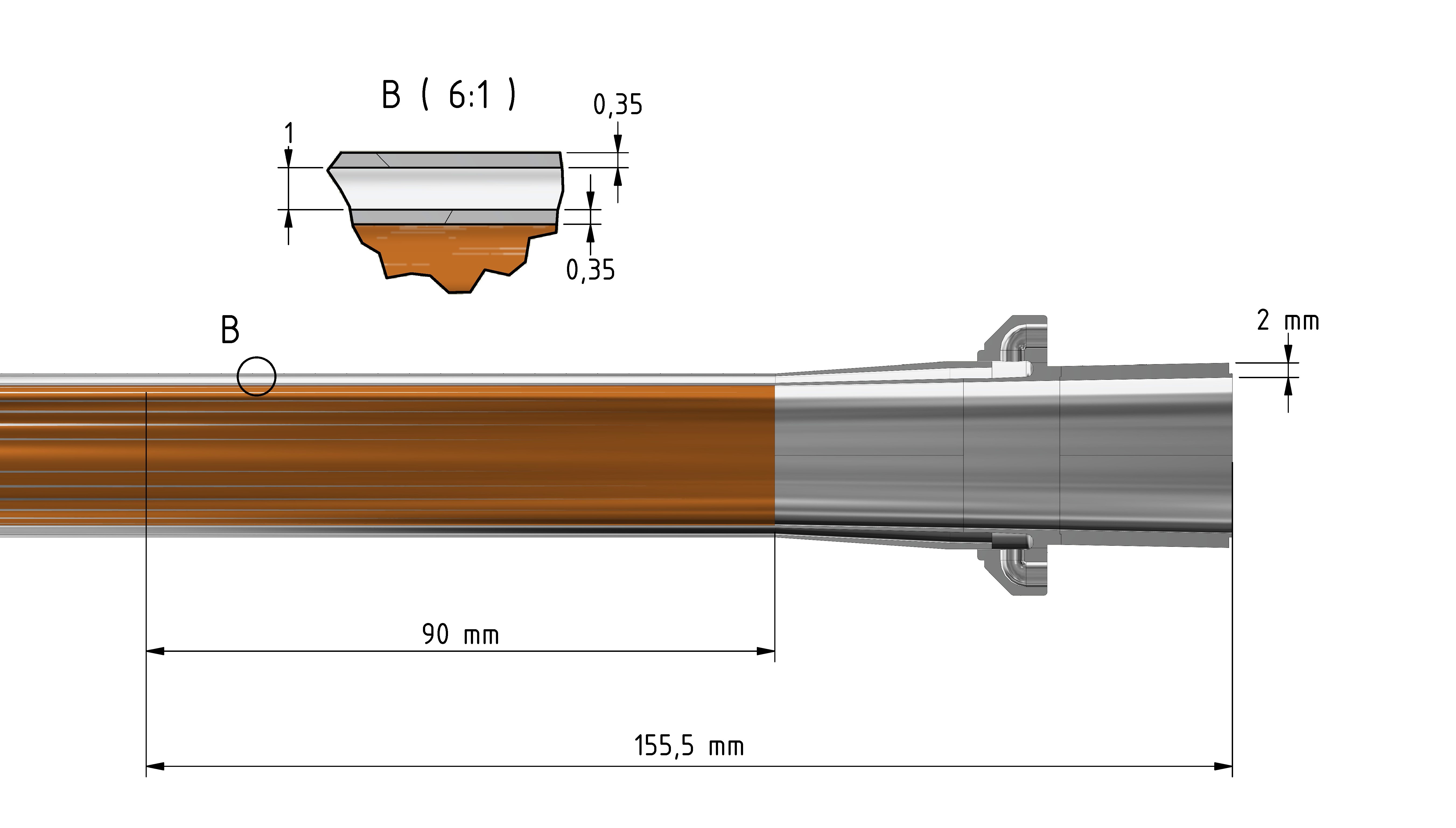}
\caption{Left: The central chamber in AlBeMet162 with its cooling inlets and outlets, 
and its internal gold coating layer. 
Right: Chamber cross section and zoom on the cooling channel for the paraffin flow.}  
\label{fig4:central}
\centering
\includegraphics[width=0.6\textwidth]{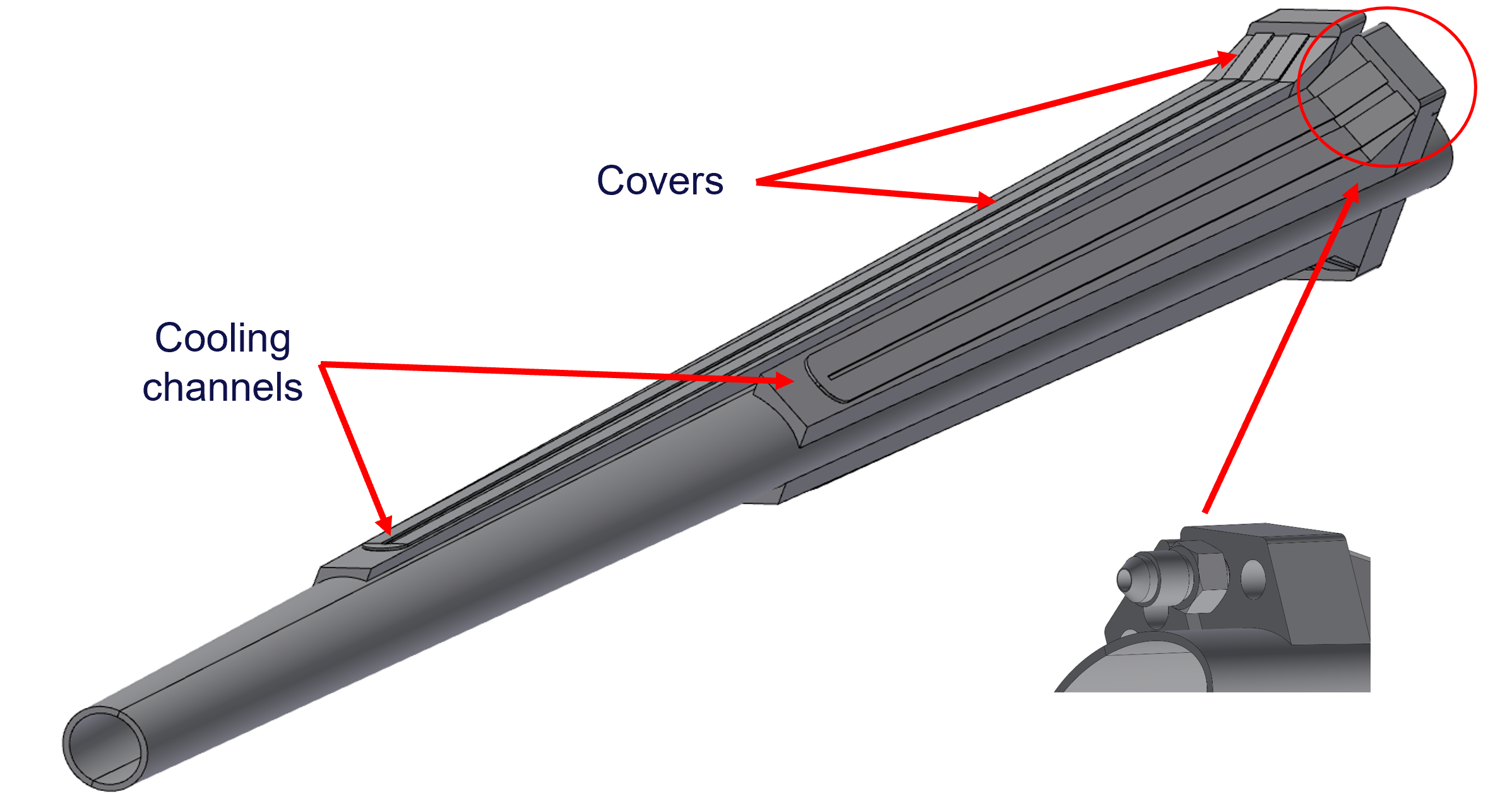}
\caption{Ellipto-conical vacuum chamber.}
\label{fig5:ellipto_chamber}
\centering
\includegraphics[width=0.45\linewidth]{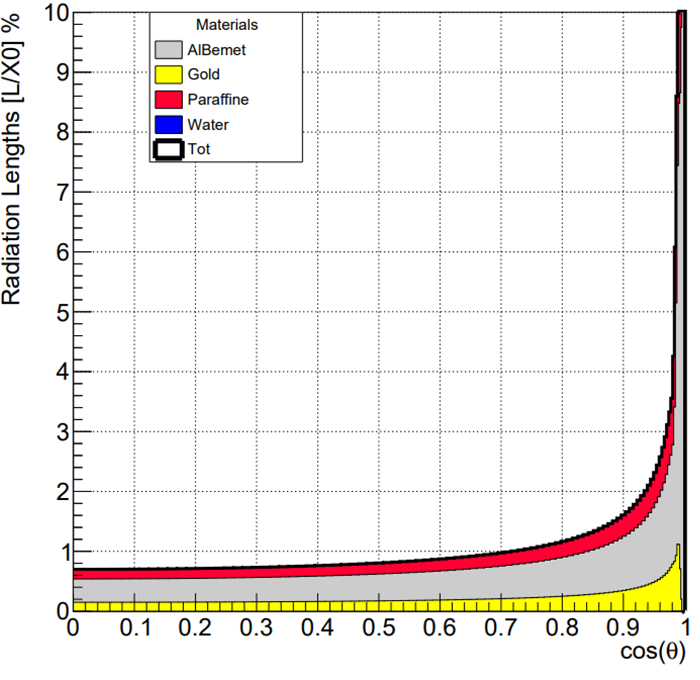}
\includegraphics[width=0.5\linewidth]{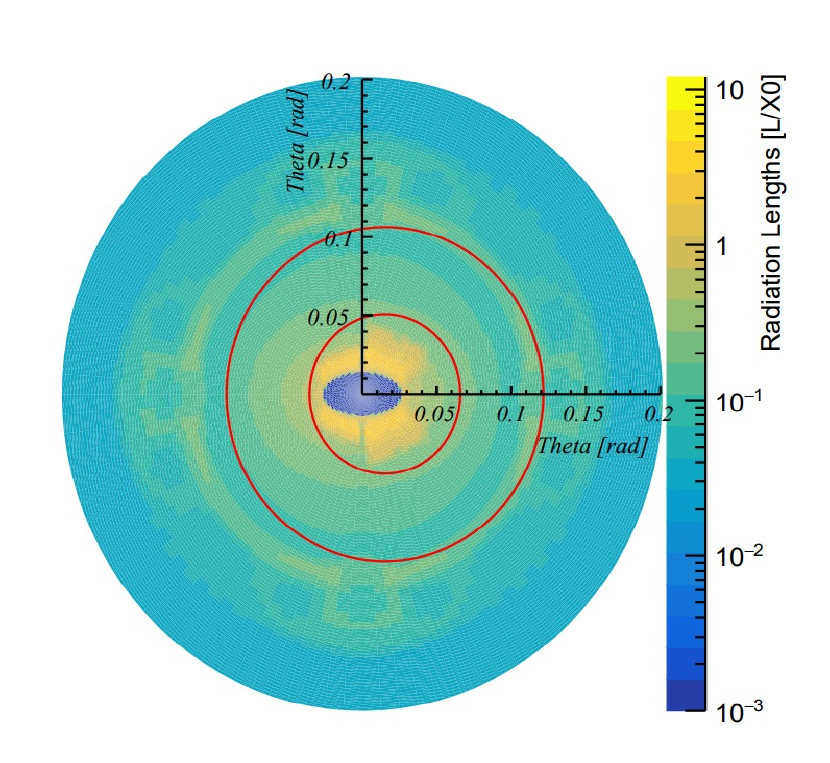}
\caption{Left: Material budget of the beam pipe, in per cent of a radiation length,
as a function of the $\cos\theta$ polar angular.
Right: Distribution of the material budget of the beam pipe in the transverse plane,
in the $0 < \theta < 0.2$\,rad angular range, at the entrance of the LumiCal. 
The red lines represent the LumiCal angular coverage.}
\label{fig:material_budget_Gold_5mum}
\end{figure}

The central beam pipe has a double layer structure, 
cooled by liquid paraffin flowing between the two layers, 
as displayed in Fig.~\ref{fig4:central}. 
A cross-sectional view of the central beam pipe and of the cooling manifolds, 
also made of AlBeMet162, is shown in Fig.~\ref{fig4:central} as well.
The double-layer structure is made of two concentric cylinders, 
each one 0.35\,mm thick and assembled with 1\,mm gap for the paraffin flow, 
thus bringing the effective diameter of the central beam pipe and its cooling layer to 23.4\,mm.
An internal 5\,$\mu$m coating layer of gold ensures a good electrical conductivity 
to minimise the beam heat load to nearly 60\,W~\cite{Novokhatski:2023wyf}
and to shield the vertex detector from residual high energy SR photons.

The ellipto-conical vacuum chamber, shown in Fig.~\ref{fig5:ellipto_chamber},
extends between 90 and 1154.5\,mm from the IP. 
Following a short transition from the central chamber, its thickness remains at a constant value of 2\,mm. 
Water flows in the cooling channels to extract an expected heat load of about 130\,W; 
an asymmetric design is needed to match the angular acceptance of the luminosity calorimeter.

A thermo-structural analysis has been performed to calculate the temperature distribution, 
stress, strain and displacement of the beam pipes.
The maximum temperature of the central chamber reaches 29\,$^{\circ}$C, 
cooled with paraffin entering at 18\,$^{\circ}$C , 
while it amounts to 50\,$^{\circ}$C for the conical chamber cooled with water entering at 16\,$^{\circ}$C. 
The maximum stress has been calculated considering the constraint configurations of a cantilevered support, 
resulting in a maximum displacement of 0.5\,mm 
and a maximum stress ten times lower than the AlBeMet162 yield strength (193\,MPa).
The resulting material budget distribution for the vacuum chamber 
is shown in Fig.~\ref{fig:material_budget_Gold_5mum} as a function of the polar angle.
The material encountered by a particle produced at normal incidence corresponds to 0.68\% of a radiation length ($X_0$). 

Several vertex detector designs are described in Section~\ref{sec:concepts}. 
The integration study performed here is based on the engineered version currently adopted by the IDEA concept (Section~\ref{sec:IDEA_vertex}). 
This version features two main subsystems, of active elements based on 50\,$\mu$m thick `Monolithic Active Pixel Sensors': 
an inner vertex detector, composed of three barrel layers at a radial distance between 13.7 and 35.6\,mm, 
covering an angular acceptance of about $|\cos\theta| < 0.99$; 
and an outer vertex detector, composed of two barrel layers at 13 and 31.5\,cm, and three disks on either side of the IP.

\begin{figure}[ht]
\centering
\includegraphics[width=0.8\linewidth]{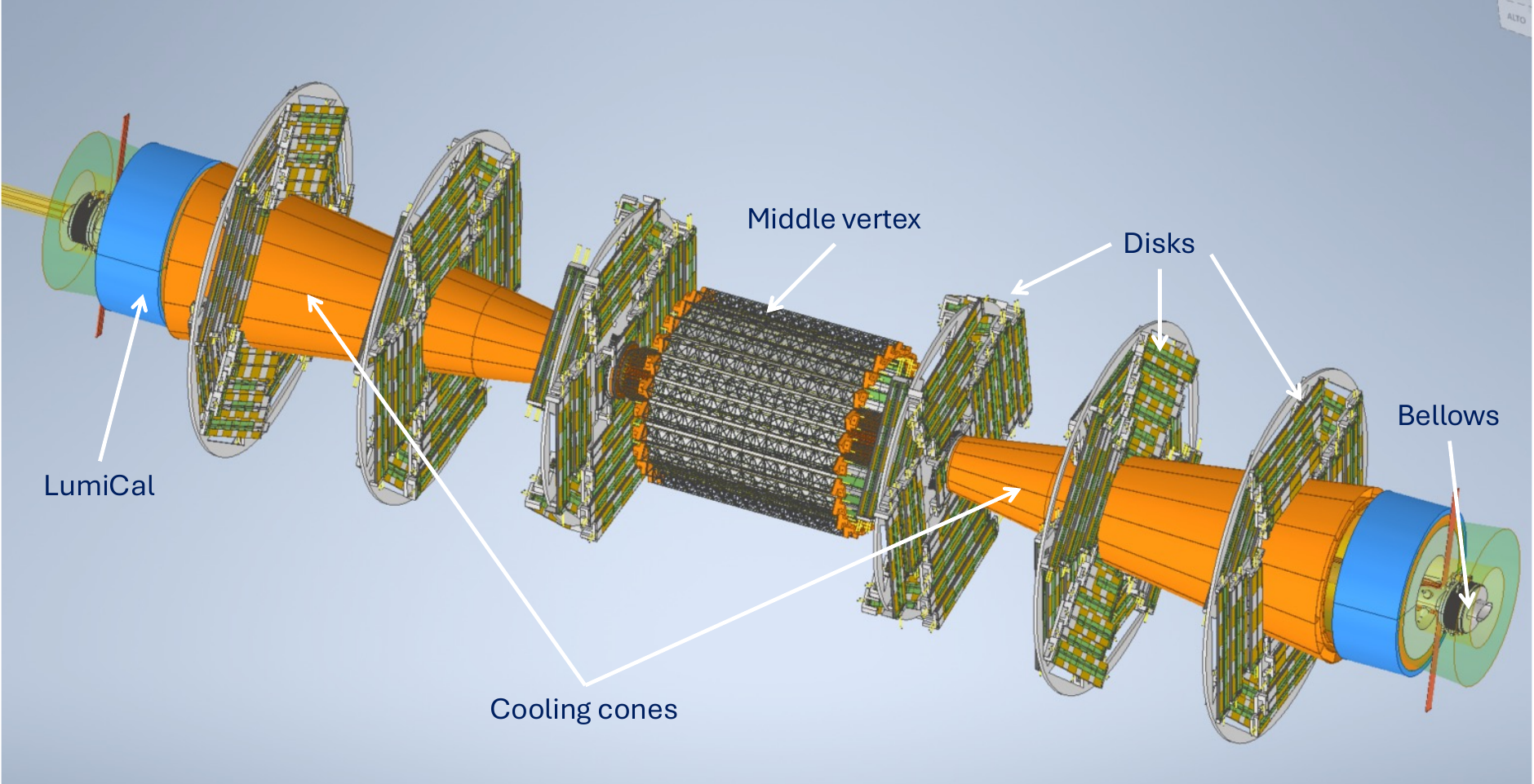}
\caption{Layout of the vertex detector cooling cones assembly, together with the main elements to be integrated around it.}
\label{fig:cooling_cones}
\end{figure}

\begin{figure}[ht]
\centering
\resizebox{\textwidth}{!}{\includegraphics{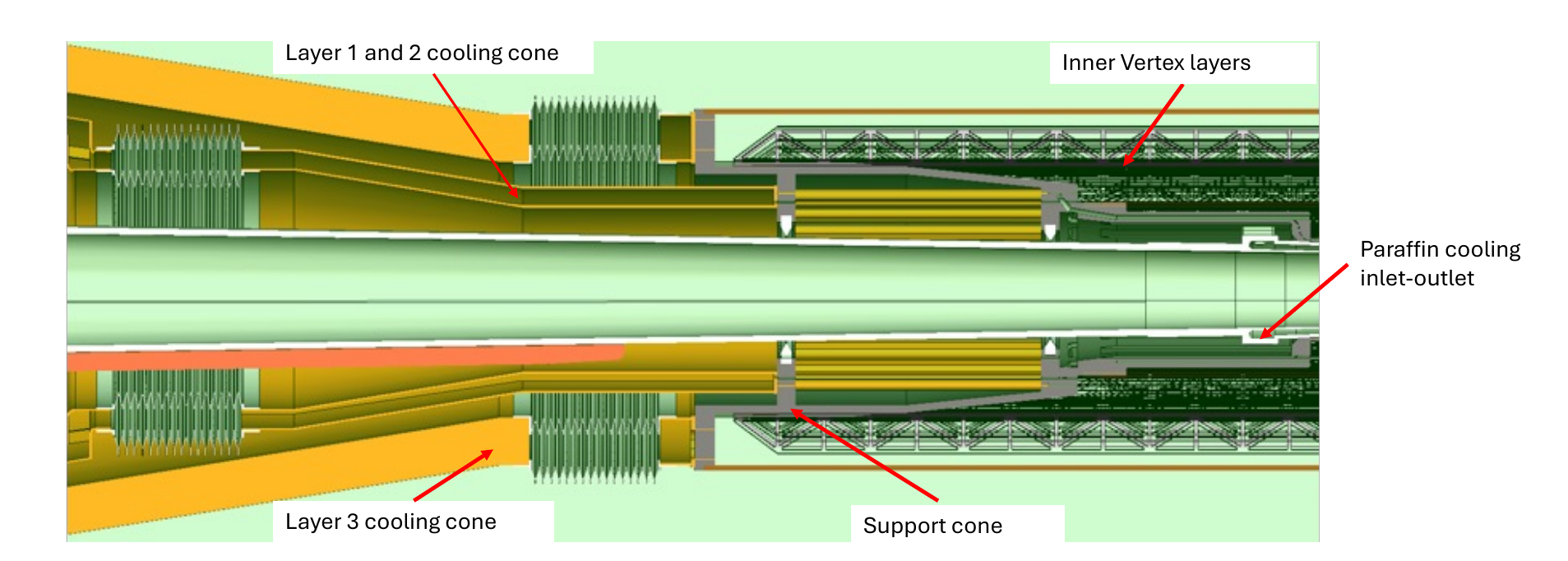}}
\caption{Longitudinal section of the beam pipe and inner vertex layers. 
The dark gray object is the conical support of the vertex detector, which is supported by the conical beam pipe. 
The inlet/outlet paraffin of the central chamber cooling manifolds are visible at the edge of the support cone.
The orange structures represent the cooling cones.}
\label{fig:zoom_vtx_beampipe}
\end{figure}

The inner vertex detector is cooled by flowing gas (air or helium) through a system of carbon fibre cones, 
shown in Fig.~\ref{fig:cooling_cones}, 
that forces gas convection inside the detector volume.
The same carbon fibre structure supports the power and readout circuits.
The inner vertex detector is mounted on top of the conical vacuum chamber by means of two thin peek-based rings, 
anchored on either side at about 170\,mm from the IP and soldered on a conical carbon fibre structure supporting three layers of silicon vertex ladders, 
also integrating the central beam pipe cooling circuits (Fig.~\ref{fig:zoom_vtx_beampipe}). 
The outer vertex detector and disks are mounted on the inner surface of the support cylinder, by means of insert rings, and cooled by water pipes.

The luminosity monitor is described in Section~\ref{sec:LumiCal}.
It consists of two cylindrical devices, placed at about 1\,m away from the IP, on either side of the IP.
Each device spans a sensitive radial coverage between 54 and 115\,mm from the beamline plus a service region from 115 to 145\,mm, 
which houses the electronics readout, cables, and cooling. 
In order to achieve the desired accuracy of $10^{-4}$ in the measurement of the luminosity, 
the calorimeter has a stringent requirement on the knowledge of its boundaries.
In particular, the relative positioning of the two sides along the $z$ axis needs to be known with a precision of $\pm 100$\,$\mu$m and, 
once assembled, the calorimeter must be a rigid hollow cylinder. 
This is guaranteed by dimensioning the bellows of the beam pipe such that their external dimensions are smaller than the internal calorimeter bore
and such that it could possibly slide inside.
The beam pipe has been carefully designed to minimise the material effects on particles reaching the luminosity calorimeter. 
In fact, the material budget ranges from approximately 0.07\,$X_0$ to 0.5\,$X_0$ (Fig.~\ref{fig:material_budget_Gold_5mum}). 
The maximum values, driven by the AlBeMet cooling manifolds, are at a safe distance from the 50\,mrad LumiCal acceptance cone. 
The design of the cooling manifolds has been optimised to limit the effects of showers originating from electrons scattering in the beam pipe. 
Ongoing studies show that the impact on the luminosity measurement is limited.

The IR magnet system~\cite{FCC-PED-FSR-Vol2} includes the final focus quadrupoles, the compensating solenoid, the screening solenoid, 
and magnetic correctors for IR tuning. 
The limited space available constitutes a challenge to the design of the IR magnet system,
currently envisaged to fit in an envelope of 100\,mrad around the $z$ axis.
The system design includes a cryostat partially located inside the detector, 
surrounding the compensating solenoid and the first final focus quadrupole (QC1),
which is itself embedded in the screening solenoid (for the part inside the detector).
It also includes the second final focus quadrupole (QC2), starting 0.30\,m upstream of QC1.

\subsection{Integration and alignment}

\begin{figure}[h!]
\centering
\includegraphics[width=0.85\linewidth]{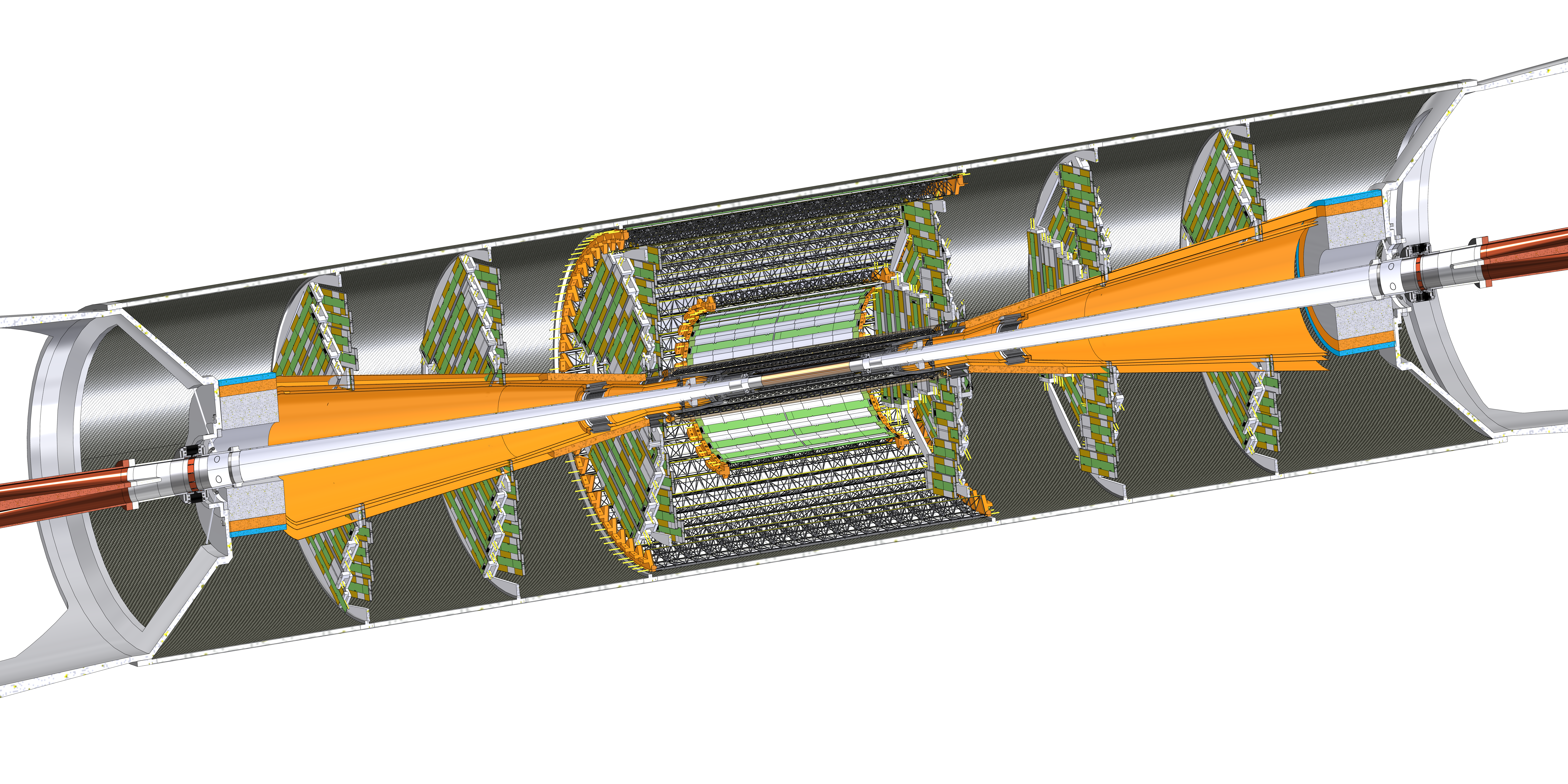}
\caption{Support tube showing the beam pipes and bellows, the vertex detector with air-cooling cones, the luminosity detector, 
and, in brown, the vacuum chambers internal to the cryostat (not shown).}
\label{fig:support_cyl}
\end{figure}

The accelerator and detector components that are placed within $\pm 1.5$\,m from the IP are mounted in a single rigid structure, 
which provides a cantilevered support for the pipe and avoids loads on the thin-walled central chamber during assembly. 
This rigid structure also supports the LumiCal and the vertex detector (Fig.~\ref{fig:support_cyl}).
The support tube is an empty cylindrical structure made of a 4\,mm honeycomb structure interleaved within two 1\,mm thick carbon fibre walls. 
It is longitudinally split in two halves and is complemented by two aluminium flanges and two endcaps that support the LumiCal and the beam pipe. 
Six aluminium ribs are fixed inside the tube in order to support the outer vertex layers. 

A structural analysis, performed to calculate the stress and displacement of each part of the support tube
taking into account the estimated weights of detector elements, including services material,
shows that the structural resistance is safely respected.

The insertion of the support tube in the detector is foreseen either with a few sleds or with longitudinal rails fixed to its external surface. 
A possible option could then be to slide these sleds (rails) inside hollow carbon fibre rails, 
permanently fixed on the inside wall of the tracker to guarantee the structural rigidity while inserting the support tube. 
The corresponding calculations and simulations are beyond the scope of this feasibility study.

The alignment and monitoring device~\cite{Watrelot:2894663} is composed of three main subsystems:

\begin{itemize} 

\item The deformation monitoring system is capable of monitoring the shape of the screening solenoid support, 
which is then used as a reference~\cite{Watrelot_2023}. 
It exploits 
in-line multiplexed and distributed frequency scanning interferometry (IMD-FSI)~\cite{10.1117/12.2529157} 
to monitor sections of optical fibres firmly installed on the inner surface of the screening solenoid support.
This system occupies no more than 5\,cm$^{3}$ within the assemblage and requires an interface with the endcap of the first final focus quadrupole.

\item At the end of each fibre used in the deformation monitoring system, 
a mirror is installed to redirect the laser beam towards the centre of the assemblage, 
targeting the final focusing quadrupoles, the beam position monitors (BPMs), the LumiCal, 
and any other component that requires position monitoring. 
This subsystem is very similar to the FSI heads installed on the low-$\beta$ quadrupoles in the HL-LHC MDI~\cite{MainaudDurand:2667534}. 
The corresponding distance measurements monitor the position of the inner components relative to the cryostat. 
 
\item Finally, to ensure the alignment of both sides of the MDI with respect to each other, 
a long-range alignment system is foreseen, also based on FSI but with a different optical setup to enable longer-distance measurements. 

\end{itemize} 

A dense network of such measurements around the detector provides a continuous link between the endcaps of the QC1 on each side of the MDI, 
which serve as the reference surface for the cylinder monitored by the deformation system. 
Reference points are placed on the supporting feet of the detector, 
easy to access for a technician or a robot, 
and used to link to the cavern survey network and the rest of the accelerator.

A network of Frequency Scanning Interferometry is proposed for the alignment of the inner tracker. 
This network will be implemented similarly to what has been done for the ATLAS SemiConductor Tracker~\cite{gibson2006multi}, 
but with the latest version of the technology, more compact and precise. 
This system also allows the measurement of the positions of the LumiCal and of the beam pipe relative to the FFQs.

\subsection{Maintenance and detector opening}

Periodic (or exceptional) detector maintenance requires, in general, access to the internal subsystems, known as `opening the detector'. 
The design of the MDI region must carefully anticipate the intertwined requirements from the civil engineering, the machine, and the detector.  
Three opening scenarios, with their own advantages and drawbacks, have been examined.

\begin{figure}[ht]
\centering
\includegraphics[width=0.49\textwidth]{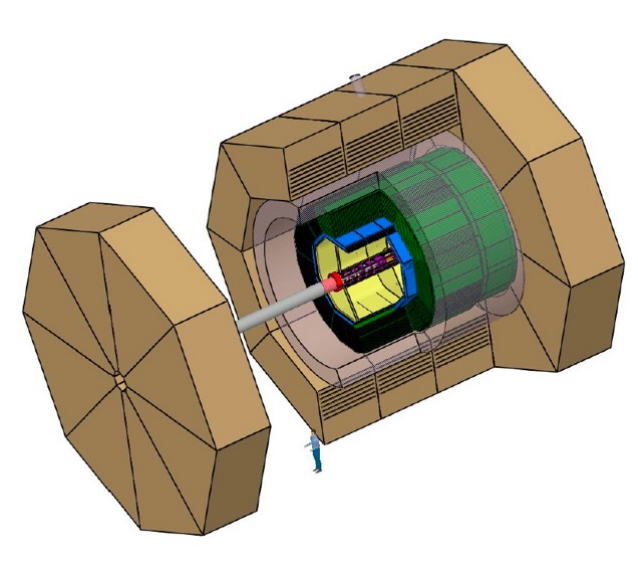}
\includegraphics[width=0.49\textwidth]{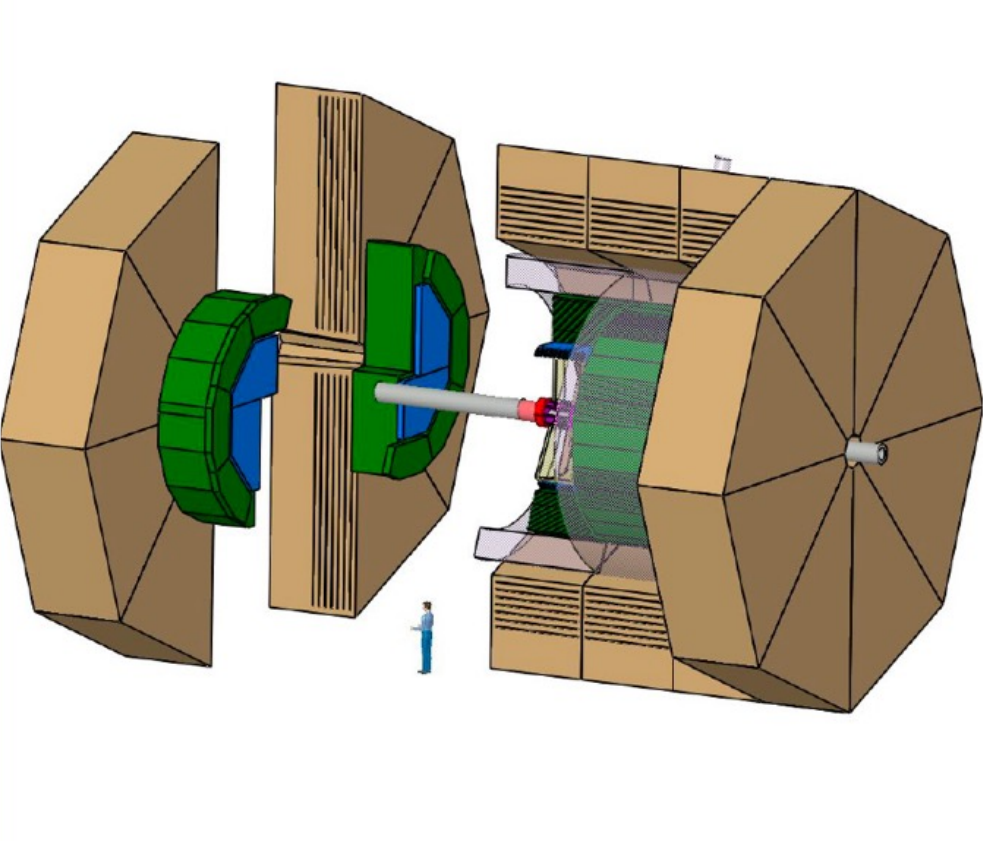}
\caption{Longitudinal (left) and short longitudinal plus transversal endcap (right) detector opening}
\label{fig:Detector_openings}
\end{figure}

\begin{enumerate}

\item In a first scenario, the two endcap calorimeters are moved along the $z$ axis to disengage them from the barrel, 
as displayed in the left panel of Fig.~\ref{fig:Detector_openings}, 
by a couple metres if only access to the central tracker is needed 
and by up to 7\,m if the tube supporting the LumiCal, the vertex detector, and the central vacuum chamber requires extraction (and re-insertion). 
The mechanical stability of the final focus quadrupoles requires rigid supports, as close as possible to the detector boundaries. 
These supports, if fixed, would make the longitudinal opening of the detector endcaps  impossible. 
Removable supports would then need to be designed to allow safe removal of the FFQs, 
and a quick re-installation and alignment following the access to the inner detectors. 
In this scenario, in addition to the re-alignment issue, the beam-pipe vacuum is broken. 
As the removal of the FFQs from inside the detector would first require the removal of other machine elements just behind them, 
this scenario can be envisaged only for medium or long machine shutdowns (several weeks or months), 
as done with the BELLE~II detector at SuperKEKB.

\item A second option involves vertically splitting the detector endcaps, allowing opening without removing the FFQs. 
In this scenario, the two split endcaps are first moved longitudinally for about 2\,m and
then moved from the beamline in the transverse direction, as shown in the right panel of Fig.~\ref{fig:Detector_openings}. 
The FFQs could stay cold, with the beam pipe under vacuum. 
This scenario could be very effective in the case of a quick access. 
The only constraint from the machine side is to keep all the services needed by the FFQs 
(supports, vacuum, cryogenics, powering) 
in the shadow of the FFQs, i.e., inside the detector forward acceptance cone. 
The main drawback is the serious impact on the detector acceptance, 
as splitting the endcaps creates a dead zone in the vertical plane 
and may imply a complete recalibration of the calorimeter angular acceptance determination.
    
\item A third, currently preferred, possibility consists of moving the complete detector along the $x$ axis, away from the beamline. 
As in the first scenario, this requires a disconnection of the internal FFQs from the rest of the machine, 
but allows them to be kept inside the detector. 
The FFQs can then be easily removed without touching any other machine element, 
before proceeding to a longitudinal opening of the endcaps without any obstacle. 
The detector integrity is preserved and, although the beam vacuum is broken and a re-alignment of the FFQs is needed, 
relatively short access periods (few weeks) can be envisioned. 
As the FCC-ee detectors have a diameter of typically 12\,m, 
this scenario requires enough transverse free space on one side of the detector, 
which is currently not available in the two small experiment caverns (Section~\ref{sec:caverns}). 
Another location for the balconies (in green on Fig.~\ref{small_cavern}, Chapter~\ref{sec:caverns}) may have to be found in these caverns, 
e.g., by placing all of them on the other side of the detector, 
and a small additional alcove (typically 25\,m long and a few metres deep) might also be needed, 
depending on the exact transverse size of the detector. 
Designing the small caverns so that their axis can be transversely displaced from the beamline by a few metres, 
if compatible with the size of the specialised FCC-hh detectors (Section~\ref{sec:caverns}), 
would alleviate the need for such an alcove.

\end{enumerate}

Technical mitigation solutions for the drawbacks of these three scenarios will be investigated in detail during the next phase of the study.

\subsection{Beam-induced backgrounds in the detectors}
\label{sec:MDI_backgrounds}

The beam-induced backgrounds in the detectors arise from two categories of sources: 
those generated by processes involving only a single beam, 
and those generated by the interactions between the two beams at each IP, usually called luminosity backgrounds.

Although the collimation system and absorbers remove the bulk of the particles produced by single beams 
(electrons/positrons and photons) 
before they can hit the detectors, 
some background may remain because the particles can scatter off them. 
Elastic and inelastic interactions of the beam with the molecules of the residual gas produce scattered particles, 
deviating from the original orbits, as well as photons from inverse-Compton scattering. 
Showers may also be produced for a very small fraction of events. 

Intra-beam particle interactions, also known as Touschek scattering, are negligible at FCC-ee, 
given that they are suppressed by a large energy factor in comparison to lower energy colliders, 
such as SuperKEKB and DA$\Phi$NE. 
Synchrotron radiation~\cite{Andre:2024bfm} has been simulated with \textsc{Bdsim}~\cite{Nevay} and the \textsc{Geant4}~\cite{Allison} toolkit. 
The bulk of SR is emitted almost collinearly with the beam, 
and the IR optics keeps it away from the detector sensitive layers. 
Magnet misalignments, imperfections, and beam tails, however, may cause SR to reach the detector. 
Studies are ongoing to optimise the design and the position of dedicated masks 
(a few metres upstream of the detector) 
in order to minimise the effects of SR in the detector.

A variety of processes such as space-charge effects, beam instabilities, interaction with residual gas, 
intra-beam scattering, magnet misalignments, and magnetic field errors can lead to the population of a beam halo, 
potentially leading to beam particle losses.
These particles are mostly intercepted by the collimators located in the dedicated section of the collider~\cite{FCC-PED-FSR-Vol2}, 
following the approach used for LHC~\cite{bruce2014simulations}.  
The relevant studies have been performed with the \textsc{Xsuite-Bdsim} simulation tool~\cite{Abramov:2024uas,Broggi:2024ocw}.
The vast majority of the beam halo losses in the MDI are intercepted by the SR collimators,
so that no losses are observed beyond the last SR collimators before the IPs, in all cases.
The background contribution from beam halo particles that leak from the beam halo collimation system is,
therefore, not expected to be an issue. 
Tertiary collimators have been included upstream of the SR collimators, 
strongly further reducing the effects of residual losses.

Beam-gas interactions depend on the residual gas pressure profile in the rings and on the outgassing materials inside the vacuum chambers. 
Simulations of the beam-gas interactions have been performed with \textsc{Molflow+}~\cite{Kersevan:IPAC2019-TUPMP037}, 
with a residual gas pressure profile resulting from 1\,h beam conditioning at a full nominal current of 1.27\,A at the \PZ pole. 
Within $\pm 500$\,m from the IP, 
the gas composition is 85\% hydrogen, 10\% carbon monoxide, and 5\% carbon dioxide, 
at an average pressure of $10^{-11}$\,mbar and spikes of $10^{-8}$\,mbar in the vicinity of the SR absorbers. 
This vacuum level, conservatively estimated at the beginning of the machine commissioning, is expected to progressively improve over time. 
Preliminary results obtained with \textsc{Fluka} indicate that the dose and fluence due to beam-gas interactions within $\pm 500$\,m from the IP
in the \PZ pole run are not the leading contributors: 
for a pressure profile resulting from only one hour of beam conditioning, 
the estimated dose and fluence in the innermost vertex detector layers are as low as 3\,kGy/year 
and $\sim 7 \times 10^{11}$\,cm$^{-2}$/year, further decreasing with increasing conditioning.

\begin{figure}[ht]
\centering
\includegraphics[width=0.4\linewidth]{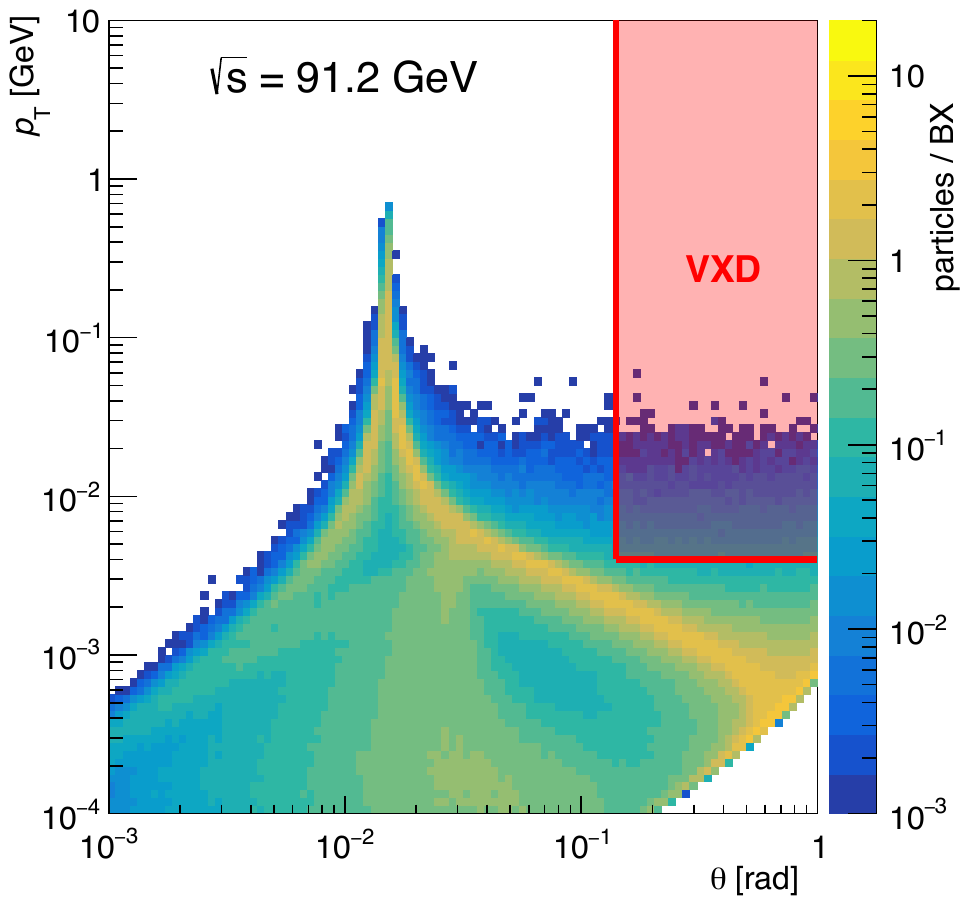}
\includegraphics[width=0.4\linewidth]{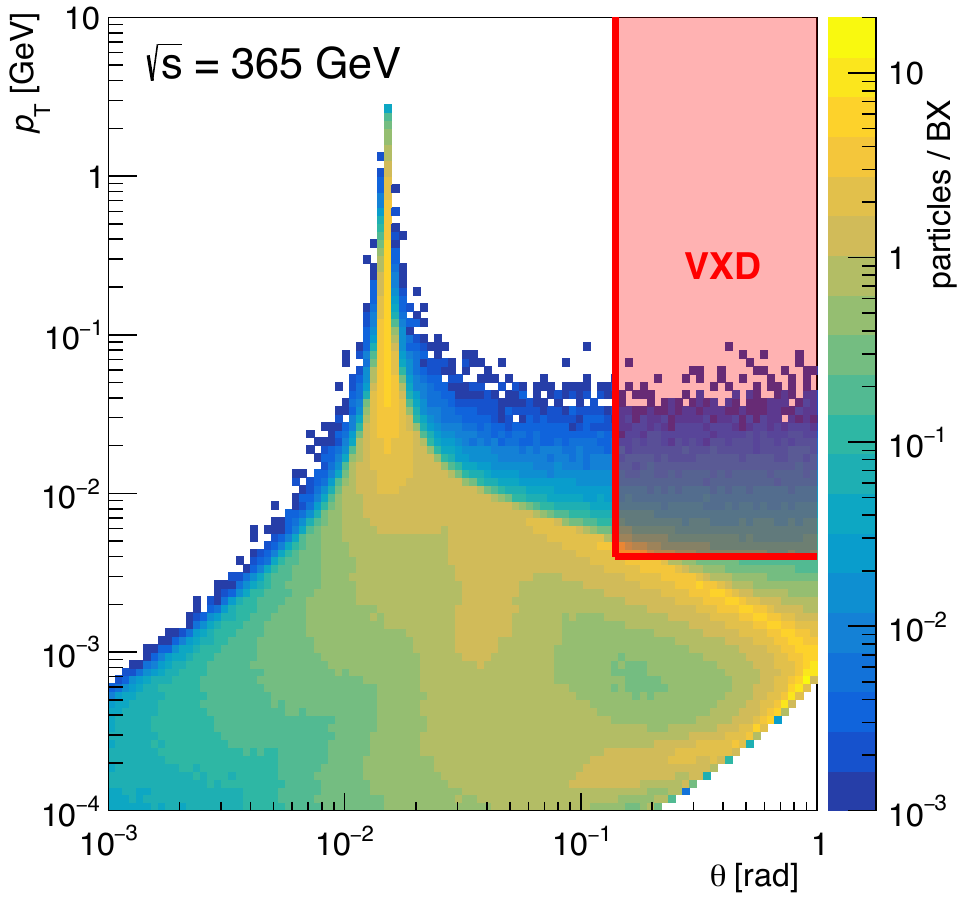}
\caption{Number of IPC particles produced for each FCC-ee bunch crossing, 
as a function of their transverse momentum and polar angle, 
for the \PZ (left) and $\ttbar$ (right) working points. 
The area limited by the red lines represents the acceptance of the vertex detector.}
\label{fig:ipc}
\end{figure}

The most relevant luminosity background arises from incoherent pairs creation (IPC)~\cite{Ciarma:2023iat}, 
a class of interactions between two photons emitted by a single electron and a single positron at the IP, $\epem \to (\epem) \epem$. 
The simulation of these backgrounds has been performed with \textsc{GuineaPig}~\cite{Schulte:1998au} 
and interfaced to \textsc{Key4HEP} (Section~\ref{sec:software}). 
The resulting $\epem$ pairs are predominantly produced in the forward direction, with small transverse momenta. 
Although the production cross section is large, only a small fraction of these particles are in the detector acceptance, 
as shown in Fig.~\ref{fig:ipc} for the \PZ-pole and $\ttbar$ runs;
a few others hit the region where the two beam pipes bifurcate (crotch) and scatter off the detector.

The largest effect occurs in the \PZ-pole run, given its instantaneous luminosity, the highest of the FCC-ee programme.
In the innermost layer of the IDEA vertex detector,
the hit rate reaches about 70\,MHz/cm$^2$, for an average of 5 pixels per cluster, 
which imposes a rather stringent requirement on the readout electronics of the order of 200\,MHz/cm$^2$
with, conservatively, a safety factor of~3.
In the IDEA drift chamber, 
where the pairs created by several successive bunch crossings need to be integrated during the drift time (400\,ns), 
the occupancy is about 7\%. 
Incoherent pairs may also reach larger radii and their effects have been studied for the ALLEGRO liquid Argon electromagnetic calorimeter. 
The first layers of the calorimeter are the most affected, 
but the effect can be largely mitigated by setting a minimum readout energy threshold of 20\% of the energy released by a minimum ionising particle, 
resulting in an occupancy of 0.03\% in the barrel and 0.2\% in the endcaps.

Radiative Bhabha (RB) scattering, $\epem \to \epem \PGg$, 
is another important luminosity background that may severely affect the superconducting FFQs. 
Radiative Bhabha events were generated with \textsc{BBBrem}~\cite{KLEISS1994372} and \textsc{GuineaPig}, 
and processed by \textsc{Fluka}~\cite{FLUKA, FLUKA:2015, FLUKA2021}. 
Their study shows that a substantial annual dose of power is deposited in QC1, 
which necessitates a tungsten shielding of approximately 2\,mm around the beam pipe to protect the superconducting magnets.
This shielding decreases the annual dose power deposition peak value to about 3\,MGy (a reduction by an order of magnitude)
and the deposited power density to about 1.5\,mW/cm$^3$.
These values are compatible with the design dose limit of 30\,MGy for the full magnet lifetime and the quench limit of 10--20\,mW/cm$^3$ adopted at the LHC.
Other radiation sources contributing to the energy deposition in QC1, 
like beam-gas scattering, incoherent pair creation or synchrotron radiation, 
still need to be carefully evaluated at the FFQs.
They are expected, however, to be much more benign than radiative Bhabha events.
The integration of this shielding with the design of QC1 will be performed in the next phase of the study.

\begin{figure}[ht]
\centering
\includegraphics[width=0.99\linewidth]{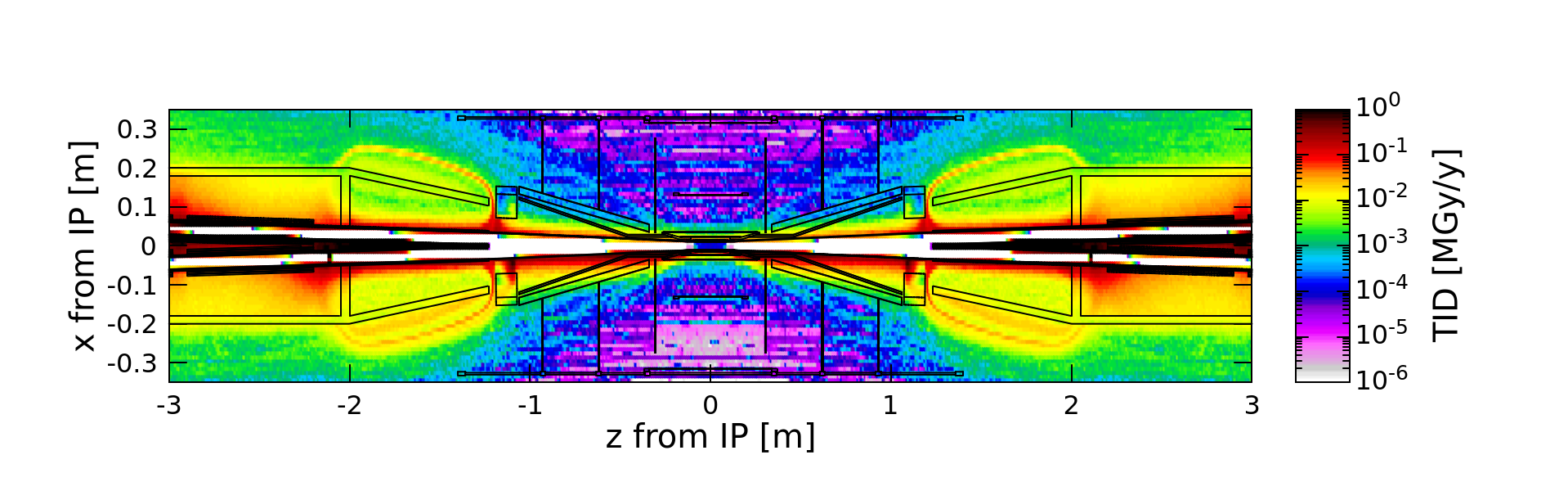}
\includegraphics[width=0.99\linewidth]{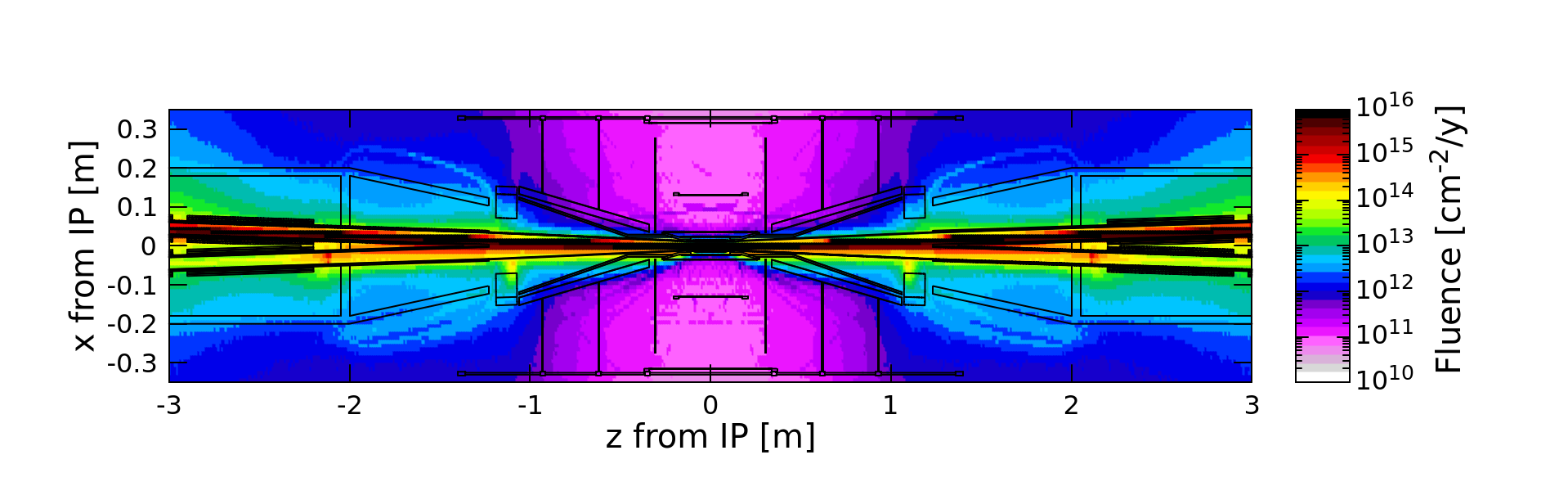}
\caption{Total ionisation dose (top) and 1~MeV\,n$_\text{eq}$ fluence (bottom) in the interaction region of the IDEA detector.}
\label{fig:TID_IR}
\end{figure}

The total ionisation dose (TID) and the 1\,MeV neutron-equivalent (n$_\text{eq}$) fluence 
from the main radiation sources in the \PZ-pole operational mode (i.e., RB and IPC) 
have been estimated in the IDEA interaction region with \textsc{Fluka}, 
and are displayed in Fig.~\ref{fig:TID_IR}.
The peak annual dose and fluence in the inner vertex detector are at the level of a few tens of kGy and a few 10$^{13}$\,cm$^{-2}$, 
respectively, for the innermost layers. 
These numbers are compatible with most of the technologies currently under consideration for the monolithic active pixel sensors. 
At higher centre-of-mass energies, the TID and fluence are expected to be smaller, given the reduced instantaneous luminosity. 

\subsection{Implementation tests and prototyping}

Design and simulation studies must be complemented by actual prototyping and implementation tests, 
to develop dedicated setups and evaluate technical solutions to the integration challenges. 
The following activities are proposed or planned; some are already ongoing.

\begin{itemize}
    
\item The realisation of a full scale mock-up of the beam pipes of the IR is being carried out at INFN (LNF and Pisa). 
This study includes the integration of the vertex and LumiCal detectors in the support tube, 
with particular emphasis on the paraffin and water cooling of the pipes and on the air cooling of the inner vertex detector. 
This activity is fully integrated with the R\&D carried out in the ECFA DRD8 Working Group~\cite{DRD8}.

\item The experimental validation of the alignment system of the FFQ is ongoing with the FSI system at CERN, 
using a 1:2 mock-up of the beam pipes and cryostat.

\item The design of the screening anti-solenoid and
and the production of QC1 corrector prototypes 
are proposed at the Brookhaven National Laboratory (BNL), 
leveraging their `direct winding technology'~\cite{Parker:2012pla}.

\item The fabrication of a High Temperature Superconducting (HTS) magnet is proposed by LAPP Annecy for the final quadrupole QC1, 
together with its integration with the water-cooled beam-pipe.

\end{itemize}

In summary, a lot of progress has been made in the past years. 
The layout has been significantly improved by reducing the material budget in front of the LumiCal 
and also through a careful design of the cooling manifolds, re-engineered in AlBeMet162 material. 
The feasibility of the inner vertex detector cooling has been validated 
and the routings of its services have been designed and integrated with those of the beam pipe.
More studies are on the way to optimise the design and confirm its validity experimentally. 
A careful analysis of the detector maintenance and FFQ integration has been performed, 
leading to the identification of the main optimisation issues to be tackled during the next phase of the study. 
An alignment strategy has been outlined, based on the technology planned for HL-LHC. 
Finally, the study of the beam-induced backgrounds and the evaluation of the main radiation doses and fluences in the IR region 
led to the development of mitigation measures, like the design of the collimation system and SR shielding.

\subsection{Outlook}

Studies carried out during the Conceptual Design Study and the Feasibility Study have converged to a baseline MDI design with satisfactory overall performance. 
Further investigations to consolidate this design and to improve its performance will be performed. 
Refinements will also be needed when actual detector proposals are finalised.   

Several options are currently considered for the QC1 cryostat design. 
Most importantly, a choice will be made for the operating temperature of the cooling scheme, 
among three options: 
1.9-2.1\,K in pressurised He~II; 
4.5\,K for supercritical He; 
or 10--20\,K for He gas forced flow. 
Another important design decision concerns the dimensions of the cryostat, 
which currently covers a cone of 100\,mrad around the $z$ axis. 
While the forward acceptance of the detector calorimetry would benefit from a smaller cryostat,
the operability of the collider, and the integrated luminosity, would advocate for more relaxed dimensions, 
calling for a complete optimisation study.  

In the baseline MDI design, 
the detector solenoid field of 2\,T is `locally' compensated with $-5$\,T solenoids placed within $\pm\ell^*$ around the IP. 
The performance of an alternative `non-local' scheme, 
with $-2$\,T solenoids positioned outside the detector at about $\pm 20$\,m from the IP~\cite{Ciarma:2024mri}, 
will be studied and compared to that of the baseline. 
First observations show that the non-local scheme might achieve a better residual vertical emittance, 
opening the possibility to increase the strength of the detector solenoid field to 3\,T. 
An adverse consequence of the non-local scheme seems to be a higher spin depolarisation, 
so that further investigations are needed to preserve the possibility of beam-energy calibration with resonant depolarisation (Chapter~\ref{sec:epol}). 

The main beam-induced background sources have been studied during the Feasibility Study, 
with the exception of the injection backgrounds and thermal photons, which will have to be included. 
The studies of the impact of all backgrounds in the data taking and in the physics performance will be finalised, 
and proposed mitigation solutions will be consolidated. 
To this aim, the \textsc{Fluka} software interface between the simulation of the machine backgrounds and the simulation of the detector will be refined. 
A deeper understanding will be sought of the synchrotron radiation background mitigation, 
in particular the positioning of the SR masks or the tolerances of the machine elements that might impact the detectors, 
such as the alignment of the FFQs, the collimators, and the vacuum system.  
More generally, the effectiveness of the shielding elements to protect the detector and the superconducting FFQ magnets will be thoroughly evaluated.

As the design of the sub-detectors (e.g., vertex detectors, luminometers, superconducting pipes, supporting structures) 
and of the accelerator components (e.g., beam position monitors, vacuum pumps, remote vacuum connections, cryostat supporting structures) 
get finalised, their integration, as well as the impact on the maintenance and assembly of the interaction region, will be established in more detail. 
Finally, methods for the alignment, the opening, and the maintenance of the detector and accelerator elements 
will be developed, and possible consequences on the cavern dimensions will be evaluated.

\cleardoublepage
\section{Detector concepts and systems}
\label{sec:concepts}
The development of detectors for FCC-ee represents a challenge that drives technological developments beyond the present state of the art. 
The main challenge arises from the richness of the physics programme and the large number of events, especially at the \PZ pole run. 
Matching the experimental accuracy to the statistical precision and the detector configuration to the variety of channels and discovery cases, 
lead to performance requirements (Chapter~\ref{sec:requirements}) that exceed those studied for TeV-class linear colliders over the past decades.
To preserve the low emittance of the FCC-ee beams during the \PZ-pole run, 
the strength of the detector magnetic field crossed by the beams under an angle of 15\,mrad is limited to 2\,T (Chapter~\ref{sec:mdi}). 
This constraint calls for larger tracking volumes that can, in part, be compensated by shallower calorimeter systems,
given the FCC-ee emphasis on the lower energies.
In addition, with continuous bunch crossings at rates of up to 50\,MHz, the front-end electronics needs to be permanently powered 
and, given the much higher bandwidth of data to be recorded, 
the instantaneous power (and cooling) demand is also higher than in cases where power pulsing can be applied.
Meeting these demands while preserving the material budget of the tracking detectors, 
the compactness of the calorimeters, and therefore their performance, requires further developments. 
In order to fully exploit the excellent precision potential, in particular at the \PZ resonance, 
events must be recorded at rates of about 100--200\,kHz, 
comparable to first-level trigger rates in the upgraded HL-LHC detectors. 

These challenges motivate the development of new detector technologies, 
followed up in the framework of newly created CERN-anchored detector R\&D Collaborations (DRDs). 
This framework aims at maximising synergies with parallel or consecutive developments (`stepping stones'), 
such as the development of an ultra-lightweight Si-based tracking system for the ALICE experiment. 
Such innovative R\&D efforts towards FCC-ee will be intensified in the coming years, 
as the major detector upgrades of the ATLAS and CMS multi-purpose experiments come to an end. 

This chapter presents an overview of the currently proposed technologies for detector systems, and the status and plans for R\&D. 
Detector concepts that, in this context, 
denote consistent and preliminary arrangements of detector systems in a full experiment, 
informed by engineering considerations and realised in a simulation software suite, are also introduced. 
These detector concepts may or may not become proposals for future FCC-ee detectors, 
as the outcome of the ongoing and future innovative R\&D efforts might well motivate new concepts 
and change the overall landscape very significantly. 
Detector concepts are, however, indispensable to guide R\&D efforts, 
to reveal typical system integration constraints, 
to optimise the design and technology choices of detector systems, 
and to study their impact on the overall physics performance in the interplay of all components, 
e.g., for flavour tagging or particle-flow reconstruction. 

Some performance studies require a detailed implementation of the detector characteristics in full \textsc{Geant4} simulations. 
The related software developments have, therefore, been a strong focus of the detector concept work in the feasibility study, 
and will have to intensify during the next phase of the study. 
In order to optimise the use of still sparse human and financial resources, and to maximise interchangeability, 
the software framework \textsc{Key4hep} (Chapter~\ref{sec:software}) is used throughout the FCC detector studies. 
For example, 
the studies for the integration of the vertex detector and beam pipe into the interaction region (Chapter~\ref{sec:mdi})
are to be considered a generic proof-of-principle for the machine-detector interface of all four experiments.  
In general, the combination of technologies for tracking, calorimetry, and particle identification is still very much open.
For example, the ALLEGRO concept considers both gaseous and silicon-based tracking. 
The \textsc{Key4hep} framework~\cite{SPack-CHEP2024-K4H} allows for different combinations in a `plug-and-play' approach, 
without duplicating the sub-detector developments and simulation implementations for each concept. 

The presently considered detector concepts (CLD/ILD, IDEA, and ALLEGRO) are briefly described, 
before a discussion of the sub-system developments, within or beyond these concepts. 
The integrated luminosity determination is discussed at the end of the chapter. 

\subsection{Detector concepts}
\label{sec:DetCon}

A fundamental difference among the current concepts derives from their different approaches to calorimetry, 
as this choice profoundly impacts the overall detector architecture. 
All concepts aim for a calorimeter segmentation that allows for particle separation within jets, 
and consequently for particle-flow reconstruction, 
but with different balances between spatial and energy resolutions. 
Hadron-shower tracking demands the full calorimeter to be inside the coil, 
whilst such an arrangement is at variance with the larger depth required by fibre-based dual read-out techniques, 
or with the cryogenic infrastructure requirements of liquid noble gases.

The coordinate system used for the description of detector concepts is as follows.
The $z$ axis is defined as the bisector of the axes of the incoming and outgoing beams. 
The $x$ axis is in the plane subtended by the two beams and pointing away from the centre of the FCC-ee ring. 
In this way, the positron beam travels towards positive values of $z$ (mainly) and $x$ (subordinately). 
Perpendicular to the $(x,z)$ plane, the $y$ axis points upwards. 
The polar angle, $\theta$, is measured with respect to the $z$ axis 
and the azimuthal angle, $\phi$, with respect to the $x$ axis in the $(x,y)$ plane. 
The radial coordinate, $r = \sqrt{x^2+y^2}$, is the distance from the $z$ axis.

\subsection{The CLD and ILD detector concepts}
\label{sec:CLDDetCon}

\begin{figure}[h]
\centering
\includegraphics[width=0.51\textwidth]{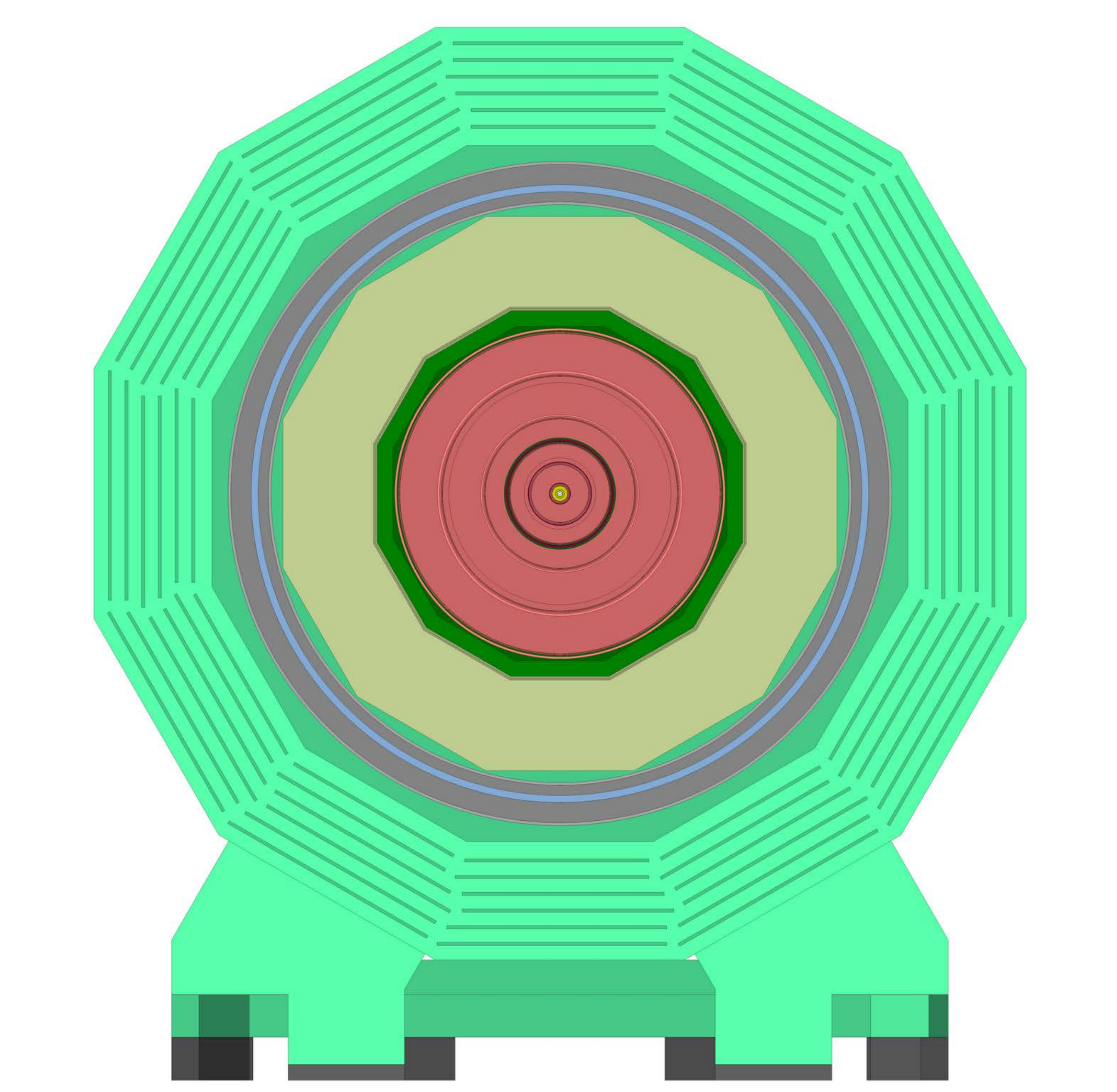} \hfill
\includegraphics[width=0.48\textwidth]{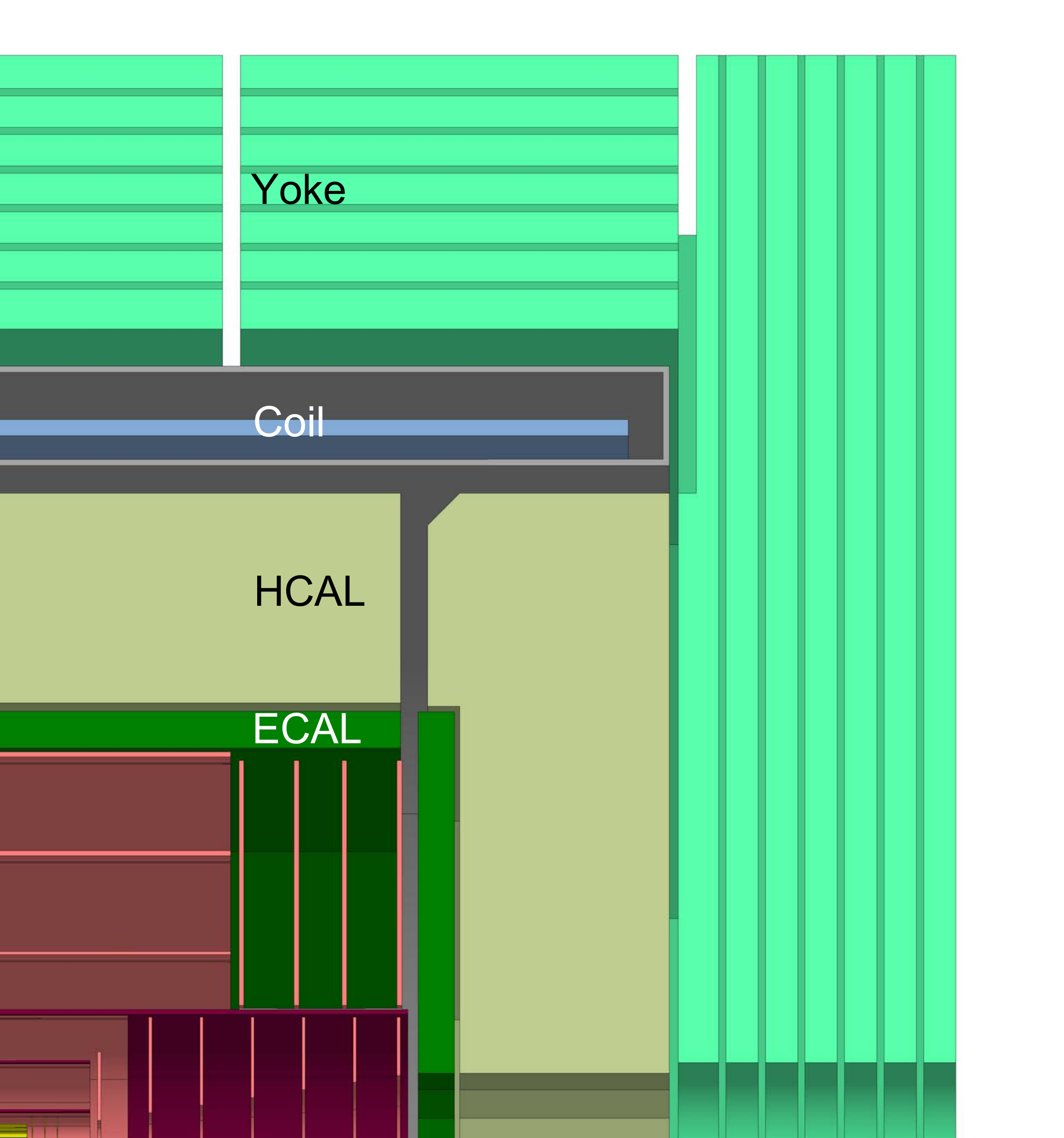}
\caption{The CLD concept detector: 
end-view cut through (left) and longitudinal cross section of the top right quadrant (right). 
The detector has an overall diameter of 12.0\,m and an overall length of 10.6\,m.}
\label{fig:CLD:detector}
\end{figure}

The CLIC-like detector CLD~\cite{Bacchetta:2019fmz}, illustrated in Fig.~\ref{fig:CLD:detector}, 
is an adaption to the experimental conditions at FCC-ee of the CLIC detector model~\cite{Linssen:2012hp,CLICdp:2018vnx}, 
itself developed from the ILD~\cite{Behnke:2013lya} and SiD~\cite{Behnke:2013lya} concepts, 
originally designed for linear colliders.
Both ILD and SiD, and consequently CLD, have a design driven by very-high-granularity calorimeters. 
They feature electromagnetic and hadron calorimeters (ECAL and HCAL), 
both located inside a solenoidal coil, 
to ensure continuous topological reconstruction of shower evolution.  
To achieve a fine 3D-imaging segmentation, 
their front-end read-out electronics is embedded in the active volumes of the sandwich structures. 
The CLD ECAL has tungsten as the absorber and is read out with silicon sensors, 
while the steel HCAL has scintillator tiles directly coupled to silicon photo-multipliers (SiPMs) as the active medium. 
These technologies were originally developed by the CALICE Collaboration~\cite{Sefkow:2015hna} 
and are presently being applied at large scale in the CMS high-granularity calorimeter (HGCAL) upgrade~\cite{CMS:2017jpq}. 

With respect to the original CLIC detector, 
CLD has an interaction region adapted to the larger crossing angle and 
the more invasive machine-detector interface (Chapter~\ref{sec:mdi}), 
a bigger tracking volume to compensate for the magnetic field limitation of 2\,T, 
and the calorimeter depth adapted to the FCC-ee energy range. 
The tracking system of CLD features a silicon pixel vertex detector (VXD) and a silicon tracker.

A full simulation suite, including full-event reconstruction, 
already exists and provides realistic estimates of full-detector performance figures of merit, 
such as jet resolutions, flavour-tagging efficiencies and purities, background levels, etc. 
Recently, to provide additional handles on particle identification, 
a novel ring-imaging Cherenkov system, ARC~\cite{Forty_ARC, Tat_ARC, forty_2024_6g0gs-7kw30}, 
has been suggested as an extra barrel layer between the tracker and the ECAL;
the tracking system has been correspondingly re-optimised, on the basis of full simulations.

The ILD detector concept is also being studied for a tentative adaptation to FCC-ee. 
It is similarly based on highly granular calorimeter technologies with embedded electronics, 
including silicon pads, monolithic active pixel sensors (MAPS), scintillator strips for the ECAL, 
and scintillator tiles and various gaseous technologies (RPCs, MPGDs) for the HCAL.
Like in the CLD concept, both sections are placed inside the magnet coil. 
The major distinction with CLD is that ILD focuses on a radically different tracking choice, 
a gaseous time projection chamber (TPC), 
favoured for its transparency and particle identification capabilities,
complemented by a surrounding silicon layer to maximise momentum and timing precision. 

The long signal collection times of a TPC, together with the beam-induced background, 
which scales with the high bunch crossing rate, 
raises the question of how reliably a TPC can be operated at FCC-ee, in particular during the Tera-\PZ run. 
This issue is currently being investigated with full-simulation studies using a plug-and-play combination of CLD and ILD, 
which merges the interaction region and inner tracking system of CLD with the ILD TPC and calorimeters.
First results~\cite{DJeansECFA-Paris} indicate that there are distortions due to space-charge effects in the centimetre range. 
Work is ongoing to mitigate these effects 
and possibly demonstrate the feasibility of reaching the performance goals in the \PZ-pole run conditions.

\subsection{The IDEA detector concept} 
\label{sec:IDEADetCon}

\begin{figure}[ht]
\centering
\includegraphics[width=0.75\textwidth]{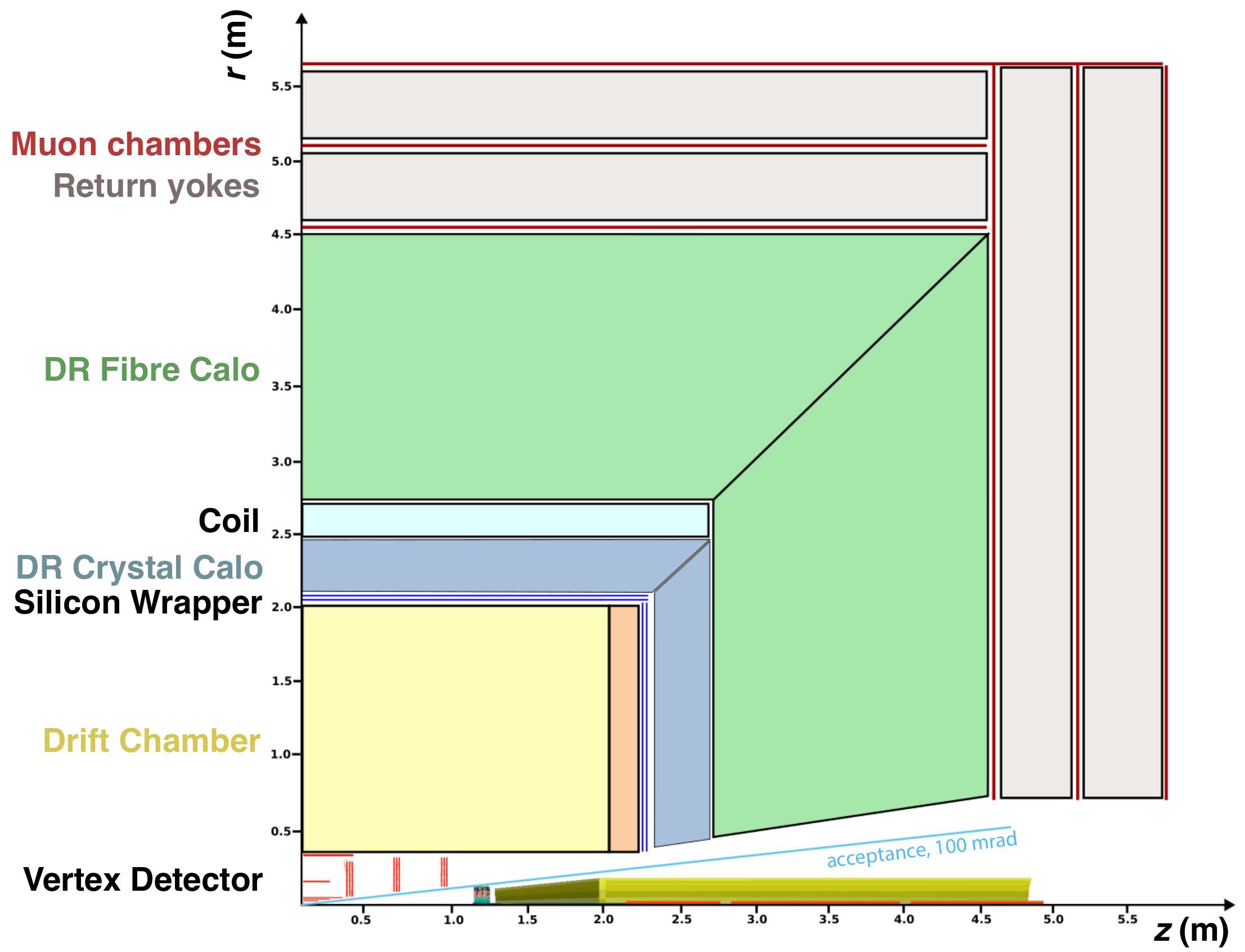}
\caption{Longitudinal cross section of the top right quadrant of the IDEA concept detector.
The LumiCal, the compensating solenoid, and the final-focus quadrupole are displayed along the $z$ axis, 
below the 100\,mrad acceptance line.}
\label{fig:IDEA:detector}
\end{figure}

The Innovative Detector for $\epem$ Accelerator (IDEA)~\cite{Antonello:2020tzq,Gaudio:2022jve,IDEAStudyGroup:2025gbt},
illustrated in Fig.~\ref{fig:IDEA:detector}, 
is a detector concept specifically designed for FCC-ee, introduced during the Conceptual Design Study~\cite{fcc-ee-cdr}. 
It is based on a silicon pixel vertex detector, a gaseous central tracker, 
and a crystal-based electromagnetic calorimeter placed inside a superconducting solenoidal magnet, 
surrounded by a dual-readout fibre calorimeter and completed by a muon detection system 
placed in the iron yoke that closes the magnetic field. 

The tracker is composed of a silicon pixel vertex detector,
a large-volume extremely-light short-drift wire chamber for central tracking and particle identification, 
and a surrounding wrapper made from silicon micro-strip sensors for improved momentum resolution. 
The wire chamber provides up to 112 space-point measurements along a charged particle trajectory 
with particle-identification capabilities provided by the cluster-counting technique. 
Energy resolution for electrons and photons is provided by a finely-segmented crystal electromagnetic calorimeter. 
The particle identification capabilities of the drift chamber are complemented by a time-of-flight measurement, 
provided either by LGAD technologies in the Si wrapper or by a first layer of LYSO crystals in the ECAL. 
Outside a thin (about 30\,cm thick) low-mass (corresponding to less than 0.8\% of a radiation length) superconducting solenoid, 
the calorimeter system is completed by a dual readout fibre calorimeter, 
providing enhanced energy precision for isolated hadrons and hadron showers, 
and a return path for the magnetic field. 
The muon detection system is based on the \mbox{$\mu$-Rwell} detectors, 
a recent micro-pattern gas detector that provides a space resolution of a few hundred microns, 
giving the possibility of reconstructing secondary vertices at a large distance from the interaction point. 

\subsection{The ALLEGRO detector concept}
\label{sec:ALLEGRODetCon}

\begin{figure}[ht]
\centering
\includegraphics[width=0.6\textwidth]{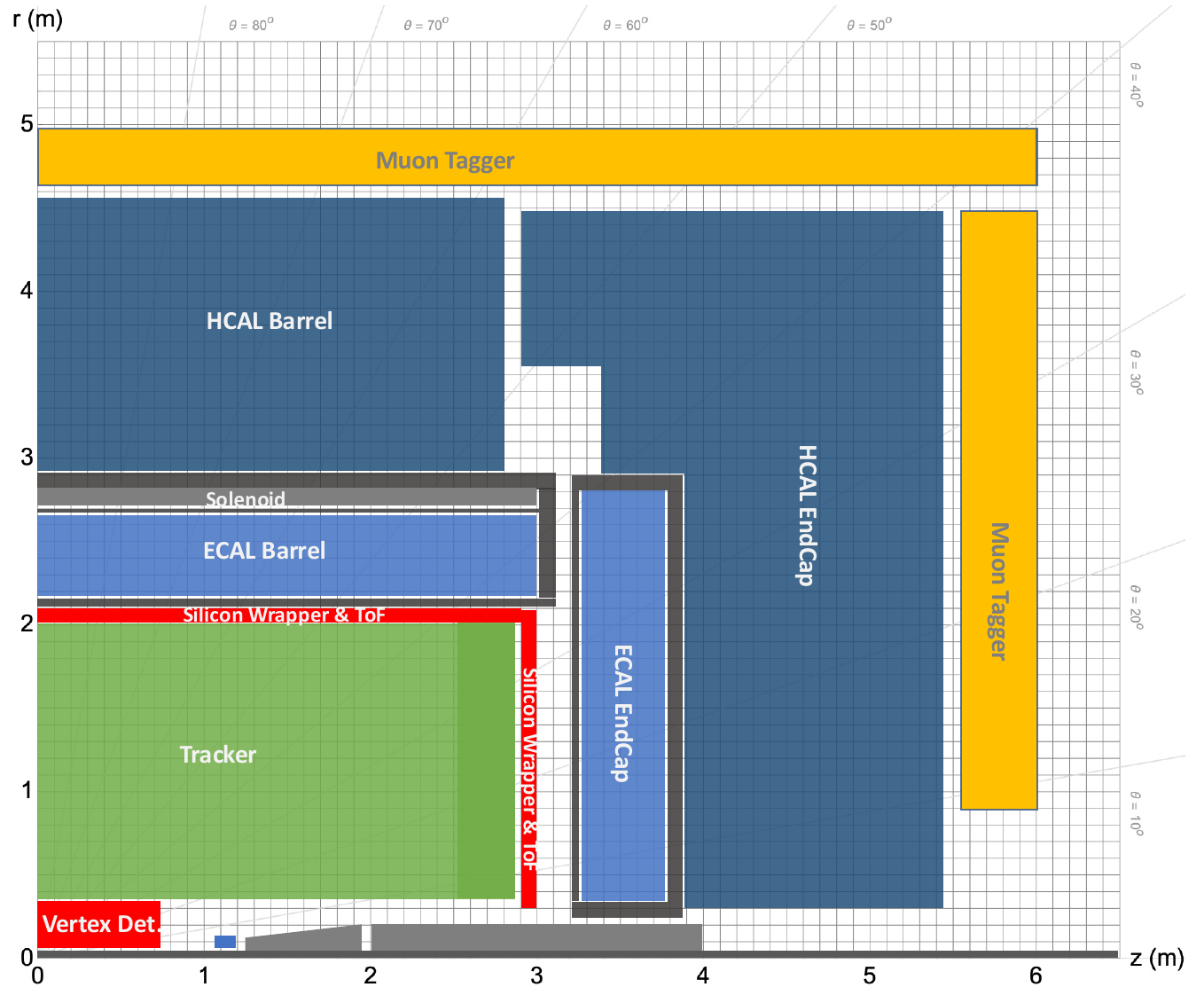}
\caption{Longitudinal cross section of the top right quadrant of the ALLEGRO concept detector.
The LumiCal, the compensating solenoid, and the final-focus quadrupole are displayed along the $z$ axis.}
\label{fig:allegro_concept}
\end{figure}

The ALLEGRO (A Lepton-Lepton collider Experiment with Granular Read-Out) detector concept 
is relatively recent and under rapid development.
Its design, illustrated in Fig.~\ref{fig:allegro_concept}, 
is articulated around a high-granularity noble liquid electromagnetic calorimeter in a cryostat made from novel lightweight materials, 
thus taking advantage of the excellent stability of this technology, 
and it capitalises on recent progress in electronics integration and advanced mechanical structures. 
The inherent linearity, stability, and uniformity of noble liquid calorimeters 
allow exquisite control of systematic uncertainties, a unique asset for high-precision measurements. 

The tracking system consists of a silicon vertex detector and a main tracker that remains to be identified. 
The silicon vertex detector is expected to use  MAPS or DMAPS technology, 
with the possible inclusion of an LGAD layer for precise timing measurements. 
Both a drift chamber and a straw layer are considered for the gaseous tracker option.
If a gaseous tracker is chosen, a silicon wrapper is foreseen at the outer periphery of the tracking volume,
to provide precise track measurements at the entrance of the calorimeter
and possibly a precise measurement of the time of flight by using LGAD technology.
A fully silicon-based tracking system, like in CLD, is also being considered 
and is used in full-simulation particle-flow studies. 

A high-granularity-sampling noble liquid ECAL surrounds the tracker. 
The options considered consist of lead (or tungsten) and liquid argon, or tungsten and liquid krypton.  
The use of innovative, low-power, cold front-end electronics, placed inside the cryostat, could provide a noiseless readout, 
but the option of placing low-noise electronics outside the cryostat is also being studied. 
The thin, lightweight coil is located around the ECAL, within the same cryostat. 
The high-granularity HCAL
is made of steel absorber plates interleaved with scintillator tiles, 
corresponding to at least 8 interaction lengths.
Two design options are being considered, 
one with SiPMs directly coupled to the scintillating tiles (as for CLD), 
and the other with wavelength-shifting (WLS) fibres used to guide the light to SiPMs located outside the detector (as for the ATLAS HCAL).
The HCAL also acts as a return yoke for the magnetic field.

The geometry described above, with a barrel and two end-caps, is considered as a `baseline'.
An alternative layout is also being considered, 
with a somewhat longer barrel and the end-caps replaced by smaller radius end-plugs fitting in the barrel opening. 
This solution would provide several engineering advantages, 
including the routing of the cables in and out of the barrel cryostat
and the reduction of the gap-widening issue in the end-cap ECAL modules. 

For the muon system, several options are under consideration,
including drift tubes or chambers, resistive plate chambers (RPCs), and micromegas (MM). 
Here, the development of eco-friendly but performant gas mixtures is a compelling line of R\&D.

\subsection{Vertex detectors}
\label{sec:IDEA_vertex}

Vertex detector design is undergoing a strong technological development that can be summarised by the terms \emph{lighter}, \emph{closer}, and \emph{more precise}.  
The development is driven by upgrade activities in Belle~II and in the LHC experiments. 
The second and third generation ALICE vertex detectors, ITS2 and ITS3~\cite{Mager:2747243}, are of particular relevance for FCC-ee,
with many common conditions and requirements, such as moderate radiation environments and the need for high resolution and low multiple scattering. 
The technology of choice, CMOS MAPS, can be thinned down to a thickness of 50\,$\mu$m or less, 
so that the sensors can ultimately be bent into cylindrical layers around the beam pipe with very little support material. 

A comprehensive description of the silicon vertex detector, considered for the IDEA and ALLEGRO detector concepts, is presented in this section. 
This detector element is currently the most advanced from an engineering point of view, 
being already integrated in the complex Machine Detector Interface layout (Chapter~\ref{sec:mdi}). 
A much briefer description of the vertex detector considered for CLD follows.

The vertex detector considered for IDEA and ALLEGRO features two main subsystems,  
whose active elements are based on 50\,$\mu$m thick MAPS:
\begin{itemize}
\item an inner detector, 
located close to the beam pipe, at radii between 13.7 and 35.6\,mm, 
covering an angular acceptance of about $|\cos\theta| < 0.99$;
\item an outer detector, located at radii between 130 and 315\,mm, 
composed of a middle barrel, an outer barrel, and three disks on each end-cap.
\end{itemize}

\begin{figure}[ht]
\centering
\includegraphics[width=0.995\linewidth]{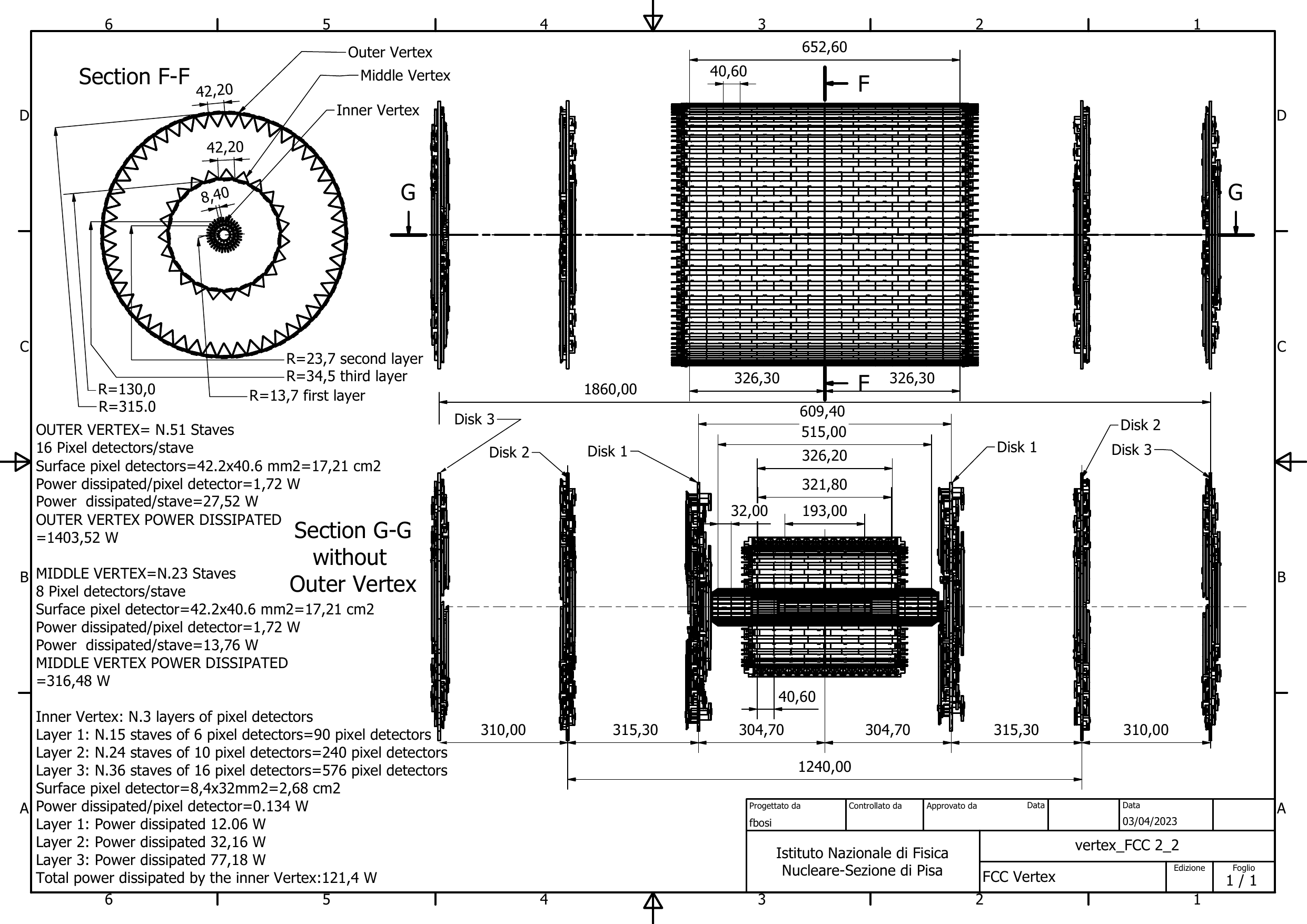}
\caption{Engineered layout and main characteristics of the baseline vertex detector for the IDEA and ALLEGRO detector concepts. 
Top-left: Cross sectional view showing the barrel layers. 
Top-right: Longitudinal view showing the outer barrel and disks.
Bottom-right: Longitudinal view with the inner and middle barrels, as well as the disks. 
All dimensions are in mm. 
The bottom-left panel lists the main elements for the outer, middle, and inner barrel detectors,
together with the dissipated power.}
\label{fig:Vertex_all}
\end{figure}

Two layouts are being explored for the inner vertex detector.
The baseline layout uses a traditional approach, 
with modules mounted on a carbon fibre support structure.
An alternative, less advanced, ultra-light layout is based on curved detectors, 
using a concept similar to the ALICE ITS3~\cite{Mager:2747243}.
The baseline layout of the vertex detector is displayed in Fig.~\ref{fig:Vertex_all}.

\begin{figure}[ht]
\centering
\includegraphics[width=0.99\linewidth]{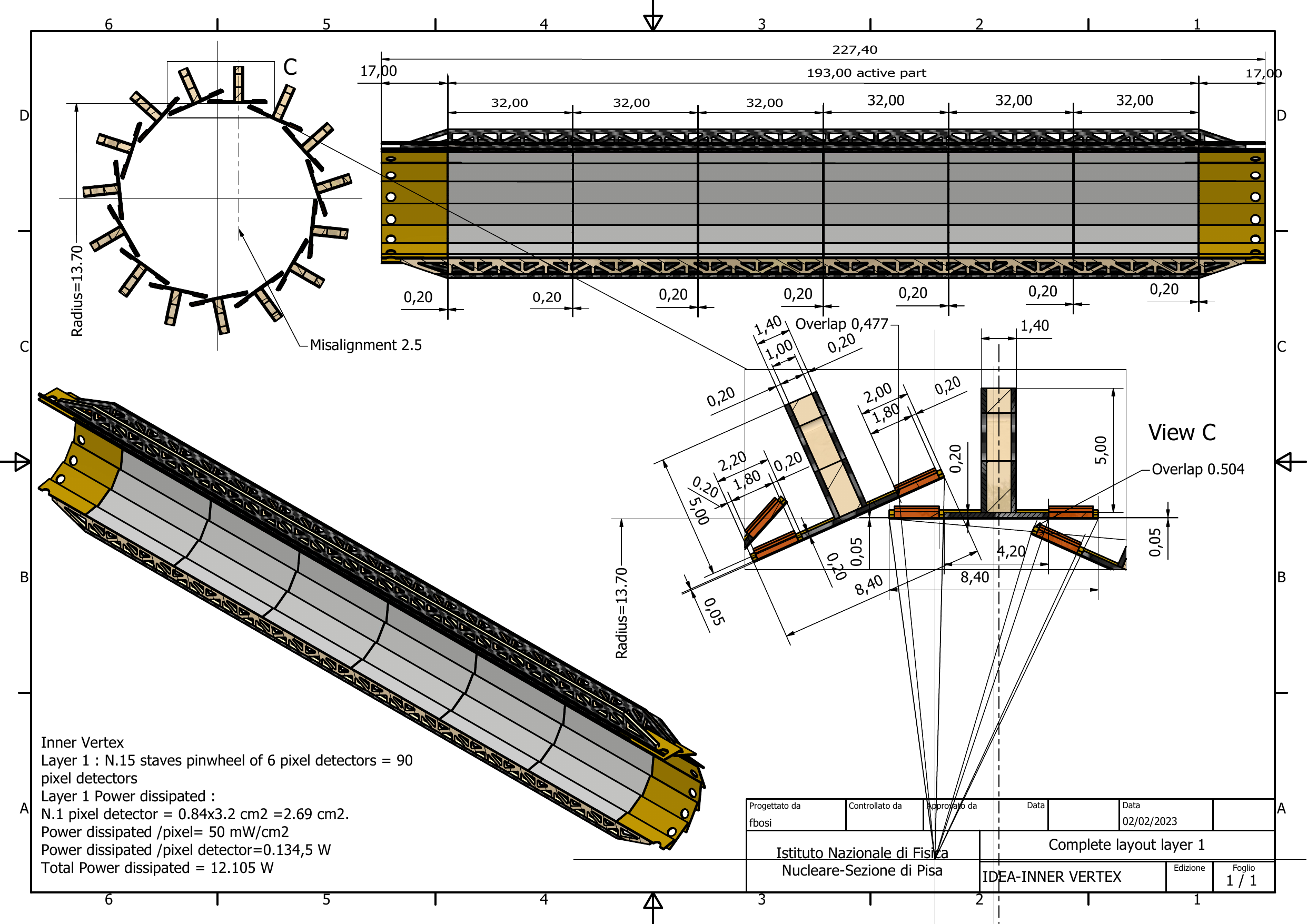}
\caption{First layer of the baseline inner vertex detector for the IDEA and ALLEGRO detector concepts.
Transverse (top-left) and longitudinal (top-right) cross sections of the assembly;
3D view (bottom-left); 
and detailed view showing the overlaps and the different structures (bottom-right). 
All dimensions are in mm.
The 50\,$\mu$m thick MAPS sensors face the centre of the structure. 
The brown structures are the buses and the golden parts are the electronic hybrid circuits for readout.}
\label{fig:layer1}
\end{figure}

In the baseline layout, the inner vertex detector is composed of three concentric barrel layers 
mounted on a carbon fibre structure. 
The elementary unit is a module of $32 \times 8.4$\,mm$^2$ $z \times r\phi$ dimensions. 
Each module has two chips abutted together in $z$, inspired from the ARCADIA R\&D~\cite{ARCADIA}. 
The chip active area features $640\,(z) \times 256\,(r\phi)$ pixels 
of $25 \times 25$\,$\mu$m$^2$ area. 
A 2\,mm inactive space is envisaged in $r\phi$, corresponding to the periphery of the chip. 
A conservative power consumption of 50\,mW/cm$^2$ is assumed, 
based on the $\sim$\,30\,mW/cm$^2$ measured for the current ARCADIA prototype, read out at 100\,MHz/cm$^2$, 
and considering the higher expected FCC-ee data rate.
In order to allow for a 2\,mm radial clearance in view of its insertion, 
the first layer is located at a radius of 13.7\,mm. 
The length is constrained by the central beam pipe cooling manifolds. 
The first layer comprises 15 staves with 6 modules each, along $z$. 
The staves overlap in $\phi$, as shown in Fig.~\ref{fig:layer1}, to allow internal alignment.

A lightweight support on each ladder provides rigidity and facilitates the mounting of the MAPS. 
The structure is made of thin carbon fibre walls interleaved with Rohacell~\cite{Rohacell}, 
which holds the sensors (facing the beam pipe) 
and two 1.8\,mm wide buses (one for data, the other for power) at the edges of the structure.
The thickness of the bus comprises 200\,$\mu$m Kapton and 50\,$\mu$m aluminium, for a total of 0.09\%\,$X_0$. 
The ladders are arranged in a pin-wheel geometry.

\begin{figure}[ht]
\centering
\includegraphics[width=0.46\textwidth]{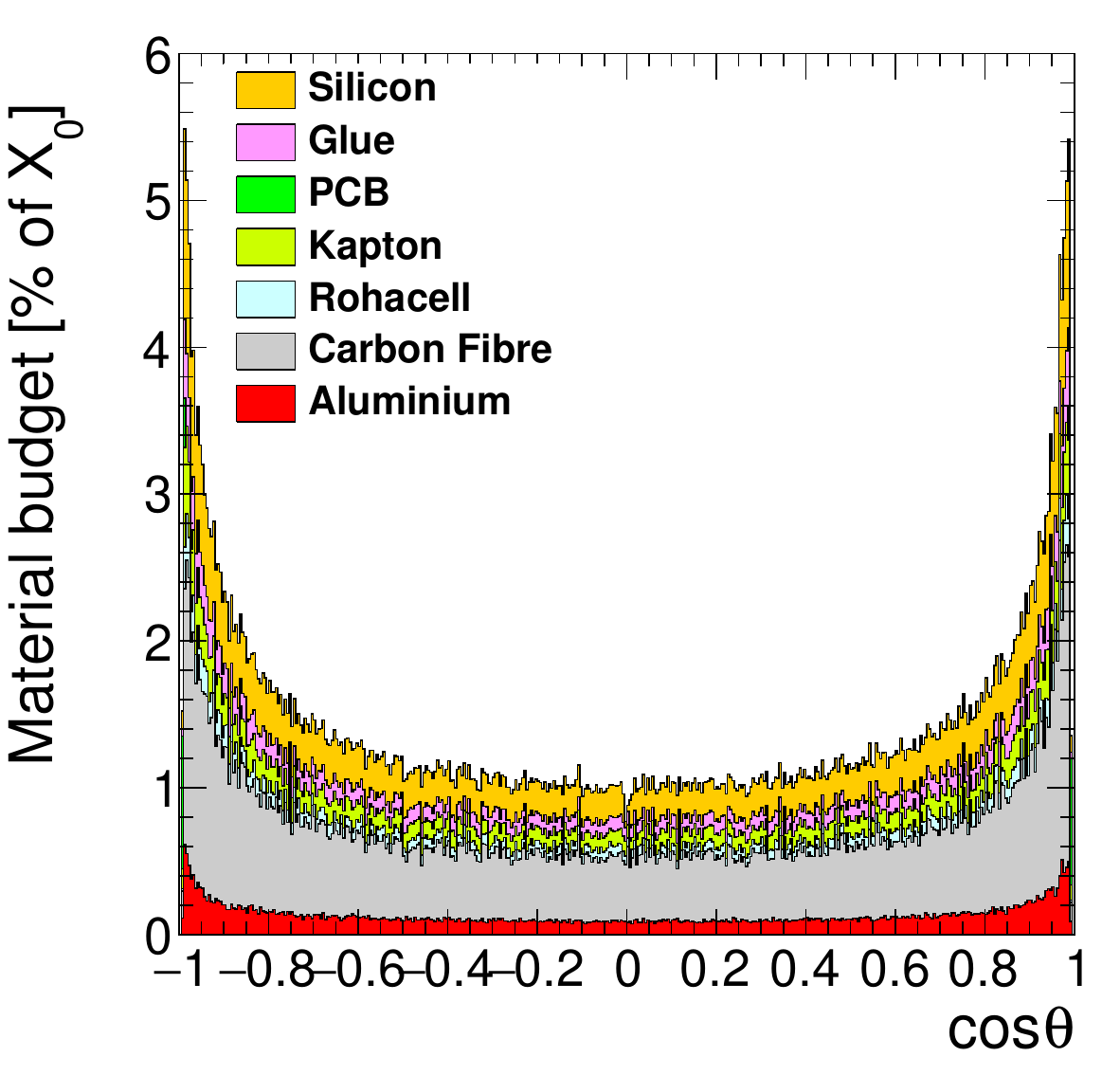}
\includegraphics[width=0.46\textwidth]{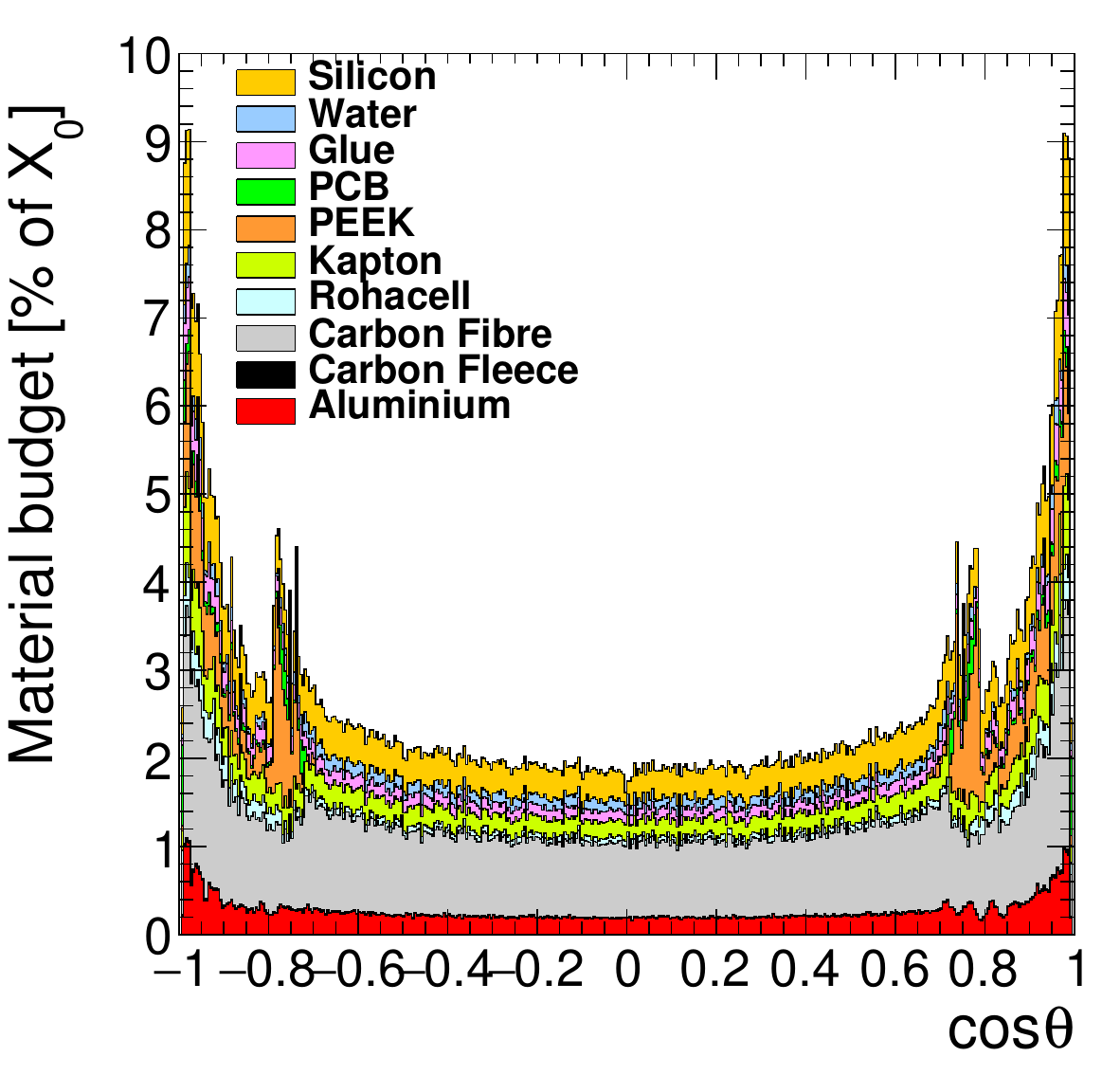}
\caption{Material budget (expressed as a percentage of a radiation length) 
for the silicon vertex detector considered for ALLEGRO and IDEA for the inner vertex detector (left) 
and for the complete vertex detector (right), 
in the baseline design, as a function of the cosine of the polar angle.}
\label{fig:baseline_vtx_material_budget}
\end{figure}

The second layer, with a structure similar to the first, 
is made of 24 ladders of 10 modules each, at a radius of 23.7\,mm, 
arranged in a pin-wheel geometry opposite in orientation to that of the first layer, 
to mitigate possible charge-dependent effects on charged-particle track reconstruction.
The third layer has 36 ladders, each composed of 16 modules. 
The ladders are arranged in a $\phi$-symmetric lampshade fashion: 
half of the ladders are located at a radius of 34\,mm, the other half at 35.6\,mm. 
Each layer contributes $0.25\%$\,$X_0$ at normal incidence, 
about one fifth of it coming from the overlap between staves of the same layer.
The vertex detector layers are mounted on conical carbon fibre support structures 
using two rings of peek material for thermal isolation during bakeout.
The conical support structures also guide cooling gas inside the detector volume 
and support the power and readout circuits.
The middle barrel and the disks, together with the support cones, 
are shown in Fig.~\ref{fig:cooling_cones} (Chapter~\ref{sec:mdi}). 
The amount of detector material crossed by a particle originating from the main IP 
(`material budget') is shown in Fig.~\ref{fig:baseline_vtx_material_budget}, 
for the inner vertex layers and for the complete vertex detector.

The main geometric dimensions of the vertex detector, 
together with the power dissipated by each layer, 
are reported in Table~\ref{tab:vertex_details}. 
The inner vertex detector is planned to be cooled with gas passing through channels embedded in the conical support structure. 
Both atmospheric air and helium are considered as coolant gases. 
The gas is contained in a cylindrical, 200\,$\mu$m thick, carbon fibre envelope surrounding the detector.
The cooling performance has been analysed using computational fluid dynamics (CFD) simulation models, 
and the result is that the largest temperature difference between the modules along a stave 
is less than 15\,$^\circ$C.
Although this is considered an acceptable level,
further optimisations are ongoing to reduce this temperature difference.
A mechanical vibration analysis based on Finite Element Analysis (FEA) in ANSYS
showed a maximum displacement amplitude in the radial direction of about 1.5\,$\mu$m 
for a nominal air flow of 0.7\,g/s,
resulting in a negligible effect on the impact-parameter resolution.

\begin{table}[t]
\caption{Main parameters of the baseline vertex detector considered for the IDEA and ALLEGRO detector concepts.}
\centering
\footnotesize
\begin{tabular}{l c c c c c c}
\toprule
Subsystem & Layer ID & Radius (mm) & $|z|$ (mm) & No.\ staves & No.\ modules / stave & Power (W) \\
\midrule
Inner barrel & 1 & 13.7 &  $<96.5$ & 15 & 6 & 12\\
Inner barrel & 2 & 23.7 &  $<160.9$ & 24 & 10 & 32\\
Inner barrel & 3 & 34 and 35.6 & $<257.5$ & 36 & 16 & 77 \\
\midrule
Middle barrel & 1 & 130 & $<163.1$ & 23 & 8 & 316\\
Outer barrel & 2 & 315 &  $<326.3$ &  51 & 16  & 1400\\
\midrule
Disks & 1 & $34.5 < r < 275$ & 304.7 & 56 & 2--6 & 135\\
Disks & 2 & $70 < r < 315$ & 620 &   48 & 3--7 & 420\\
Disks & 3 & $105 < r < 315$ & 930 &  40 & 4--7 & 370\\
\bottomrule
\end{tabular}
\label{tab:vertex_details}
\end{table}

\begin{figure}[ht]
\centering
\includegraphics[width=0.95\linewidth]{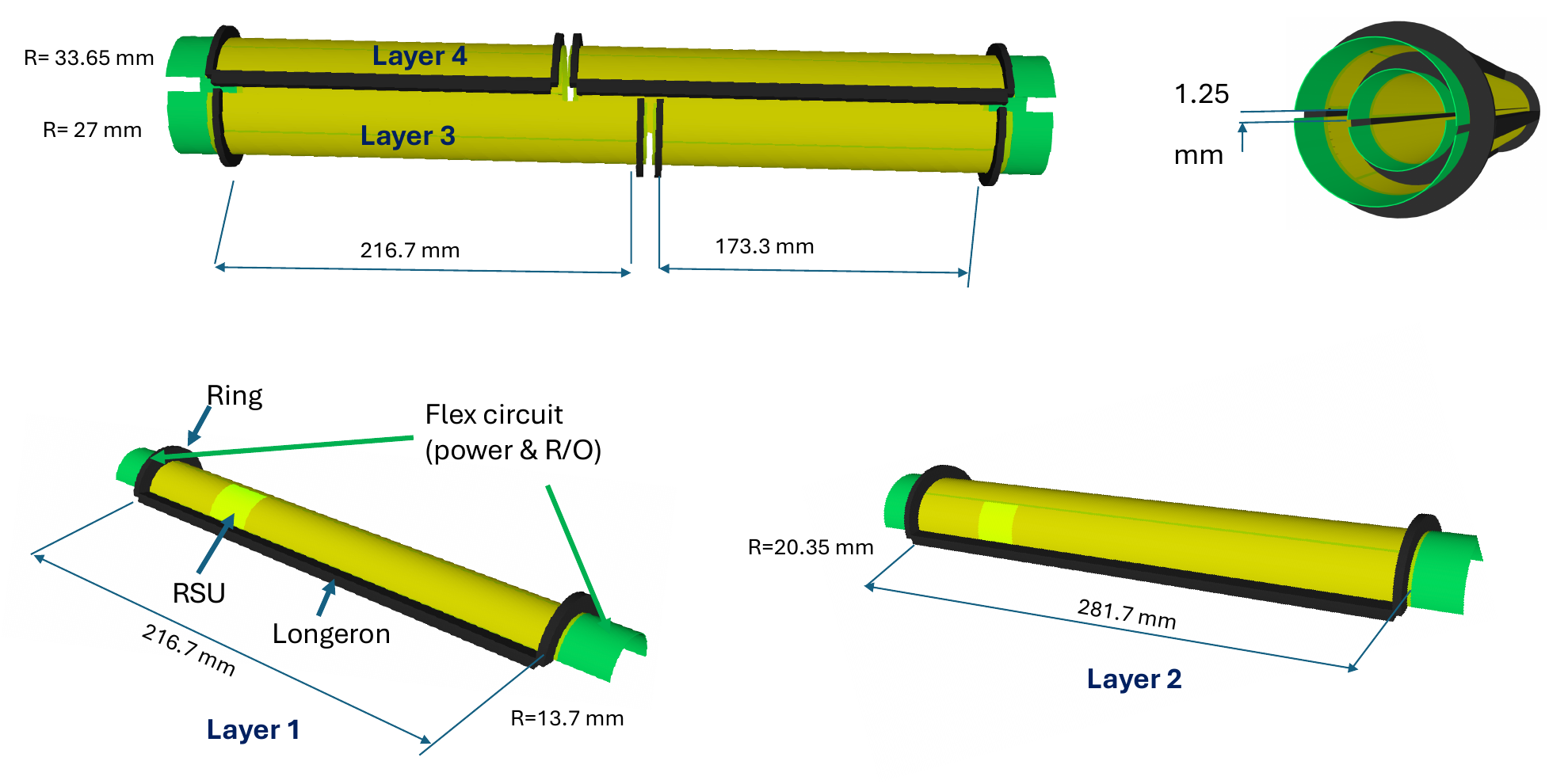}
\caption{Ultra-light inner vertex detector layout for the IDEA and ALLEGRO detector concepts. 
The four half-layers are shown separately in the top-left and in the bottom. 
The arrangement of the first two layers is shown in the top-right view.}
\label{fig:curved_vertex}
\end{figure}

\begin{figure}[ht]
\centering
\includegraphics[width=0.46\textwidth]{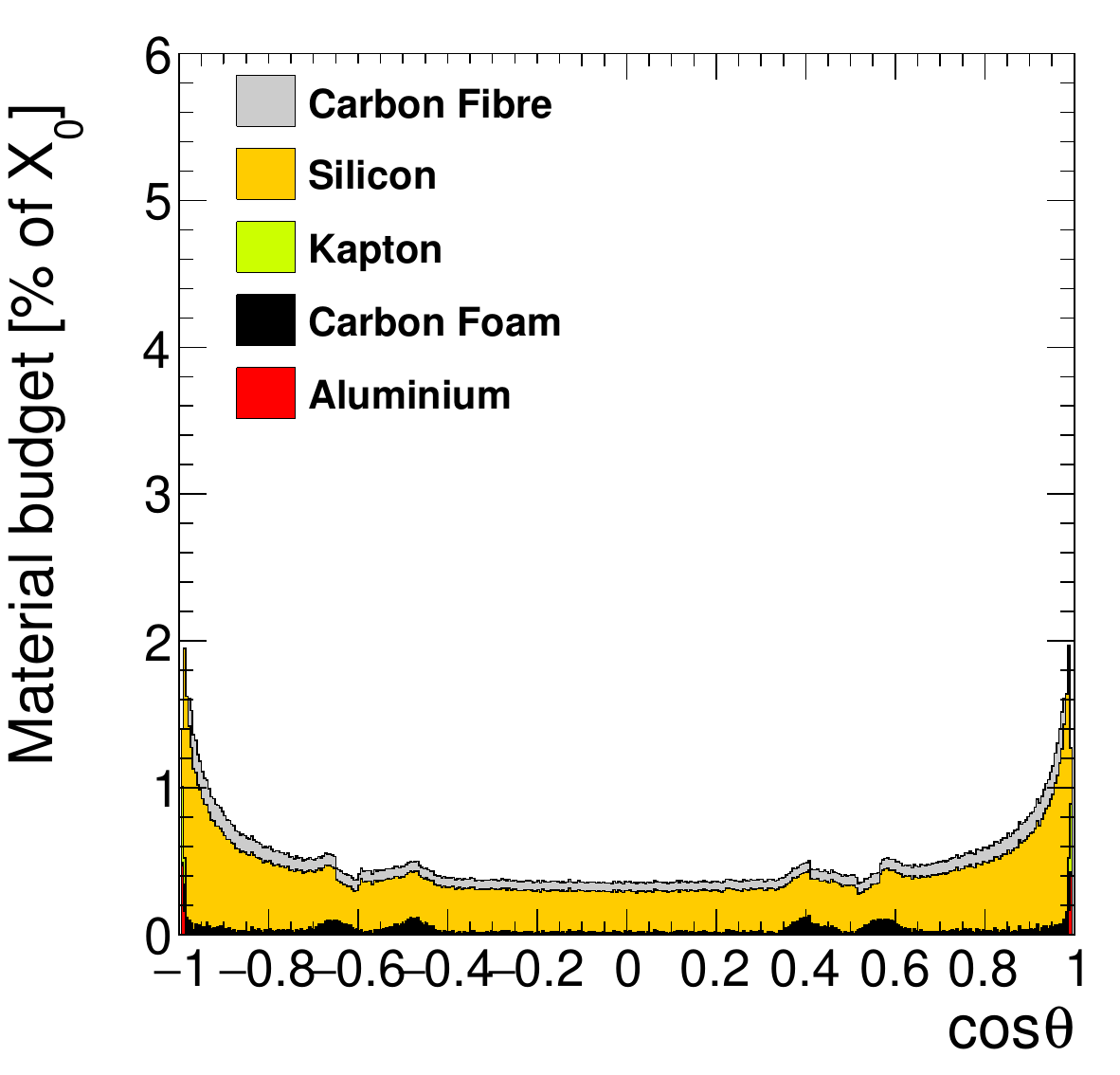}
\includegraphics[width=0.46\textwidth]{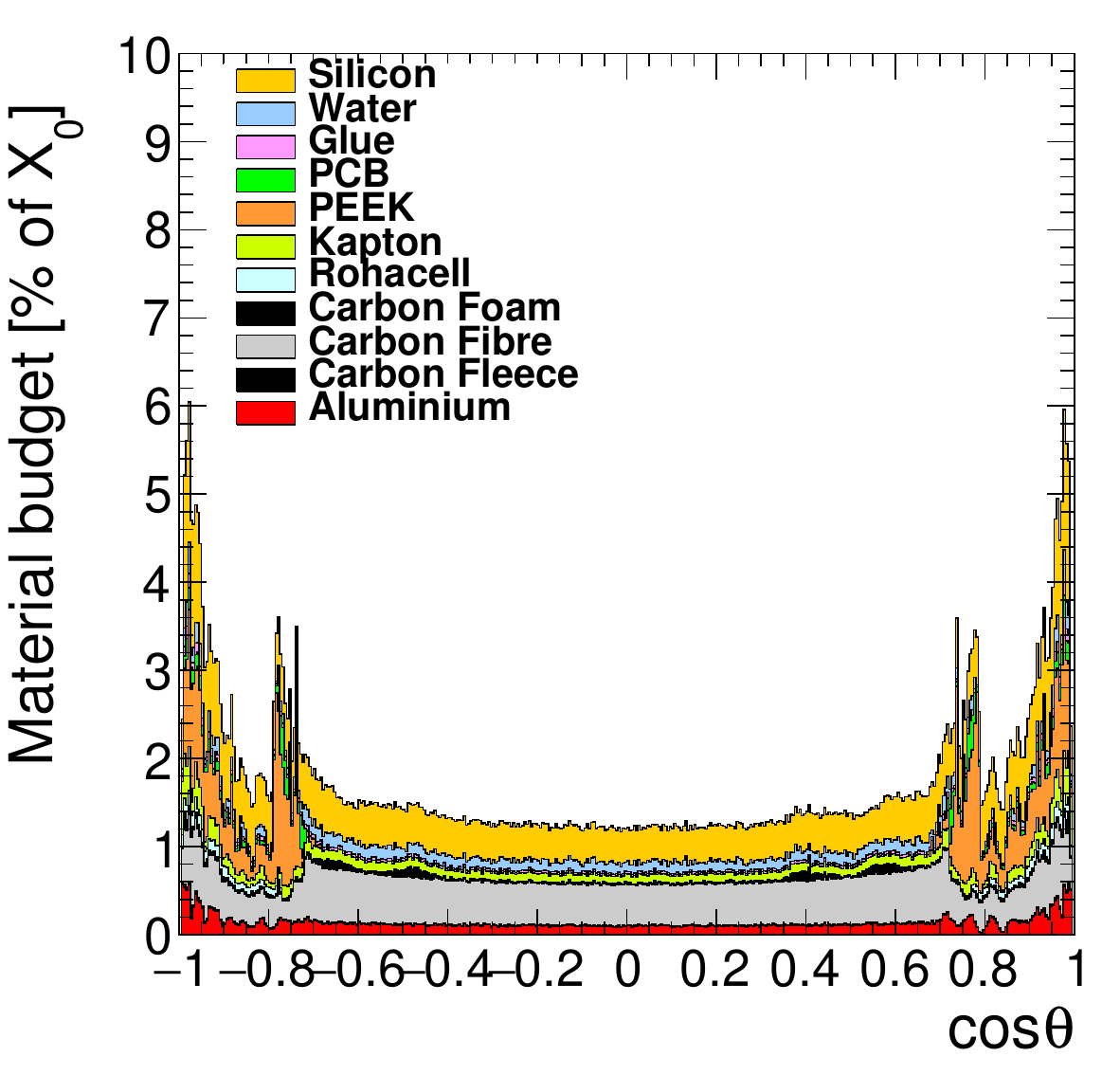}
\caption{Material budget (expressed as a percentage of a radiation length) 
for the silicon vertex detectors considered for ALLEGRO and IDEA, 
for the inner vertex detector (left) 
and for the complete vertex detector in the ultra-light design (right), 
as a function of the cosine of the polar angle.}
\label{fig:UL_vtx_material_budget}
\end{figure}

A second, ultra-light layout for the inner vertex detector is being explored, albeit with a less advanced system engineering. 
The layout is based on a concept similar to the ALICE ITS3 curved sensor technology, 
which facilitates a self-supporting structure with essentially no material besides that of the Si sensors. 
The ITS3 uses a stitching technique to form wafer-scale sensors from multiple repeated sensor units (RSUs). 
Each layer comprises two half-cylindrical sensors featuring ten RSUs in $z$ 
and three, four, or five in $\phi$ for the first, second, and third layers, respectively.
Unlike ITS3, a vertex detector for FCC-ee must have the largest possible polar-angle coverage. 
The inner vertex detector has four layers, 
of which the first three are arranged to have approximately the same angular acceptance 
while the fourth has the same length as the third. 
The first layer, at 13.7\,mm radius, uses two $\phi$ rows of 10 RSUs in $z$ and is supported by only two carbon foam longerons and rings. 
The spacing between the two half-shells is 1.25\,mm, thus leaving a gap in the $\phi$ acceptance. 
This gap is recovered by a rotation in $\phi$ of 1.25\,mm of the second layer, although with a worse impact parameter resolution. 
The second layer is composed of three $\phi$ rows of 13~RSUs. 
The first two layers are read out and powered from both ends. 
For the third and fourth layers, the use of two sensors in $z$ per half-layer is foreseen to circumvent the limitation of the \mbox{12-inch} wafer diameter. 
The third layer uses 8 layers on the $-z$ side and 10 on the $+z$ side, 
while the fourth layer has a reversed distribution (10 on the $-z$ side and 8 on the $+z$ side);
in this way, the acceptance gap in $z$ of one layer, of about 10\,mm, is covered by the other. 
For these layers, the readout occurs only at the far ends in $z$. 
This layout design is illustrated in Fig.~\ref{fig:curved_vertex}. 
Integration studies are ongoing to validate its technical feasibility.
The material budget is very much reduced with respect to the more classic design, 
as shown in Fig.~\ref{fig:UL_vtx_material_budget}. 


\begin{figure}[ht]
\centering
\includegraphics[width=0.49\textwidth]{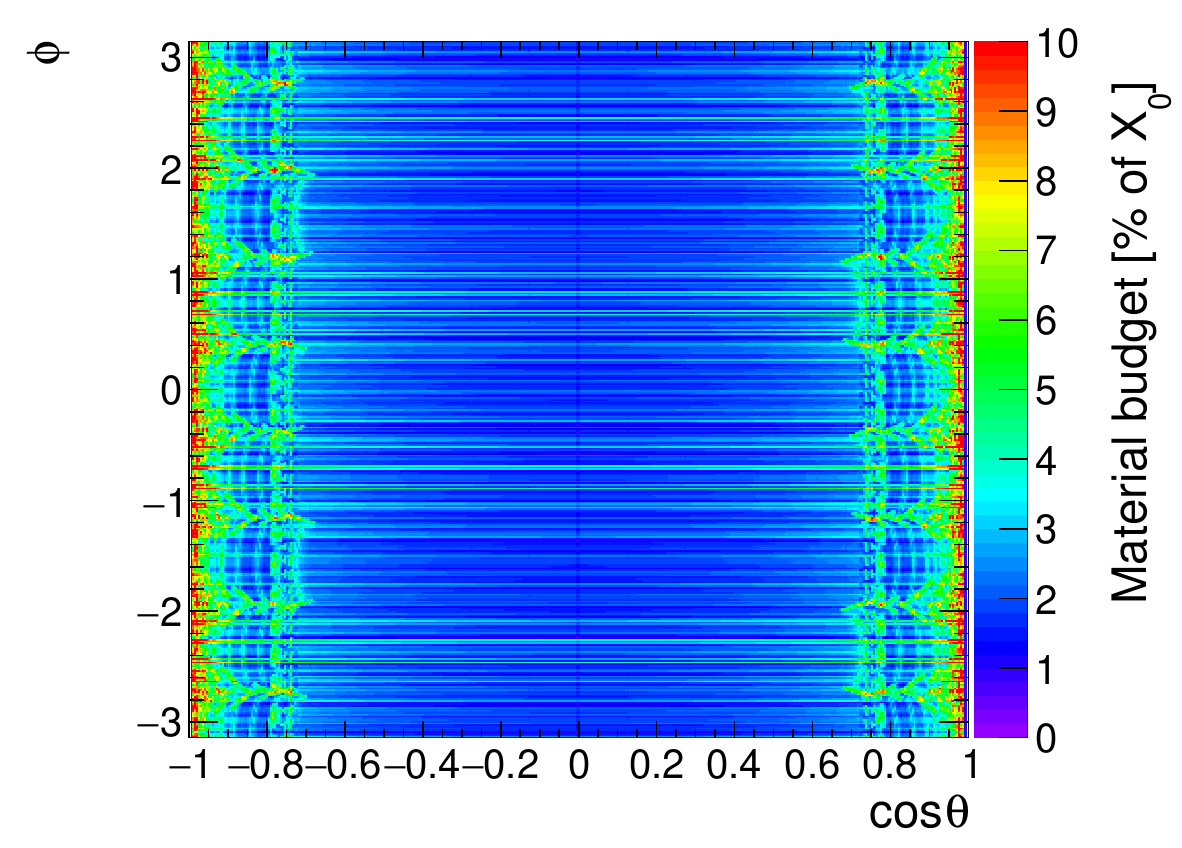}
\includegraphics[width=0.49\textwidth]{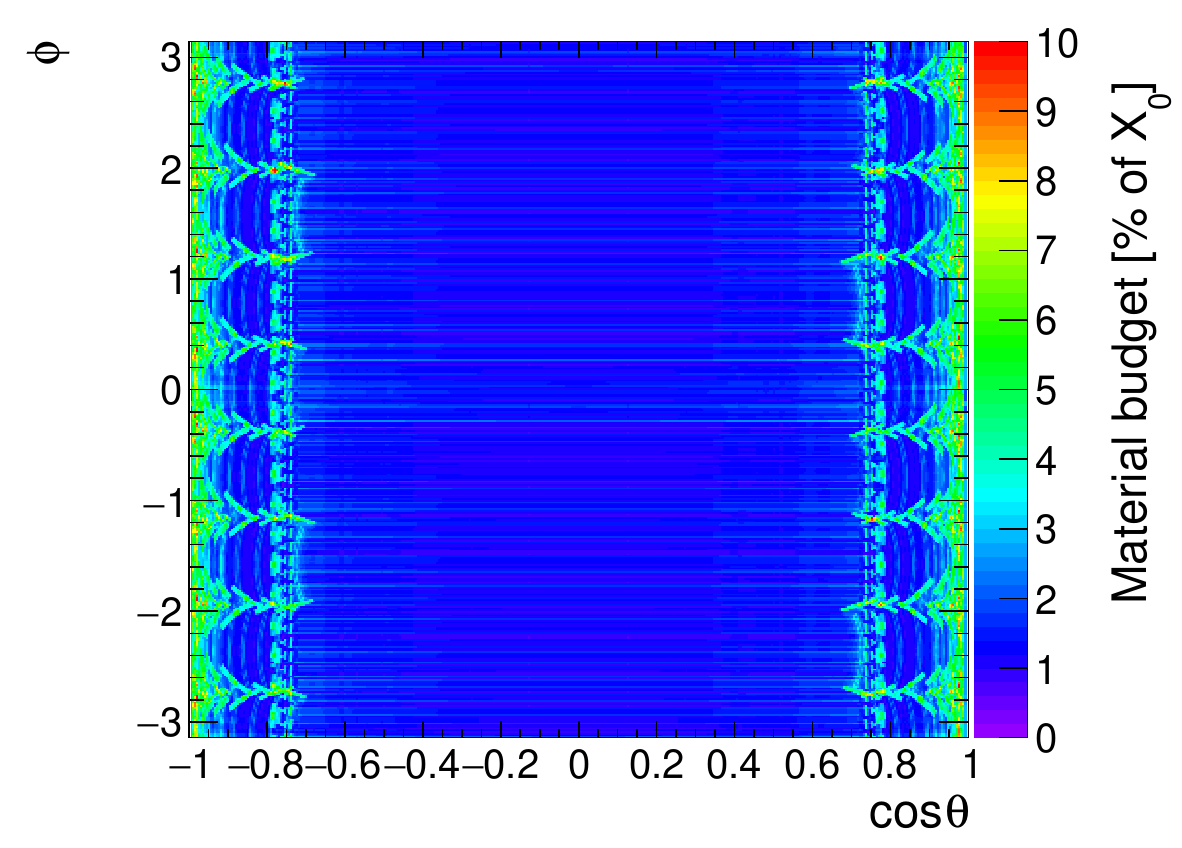}
\caption{Material budget (expressed as a percentage of a radiation length) 
for the silicon vertex detectors considered for ALLEGRO and IDEA in the baseline (left) and the ultra-light designs (right), 
for the complete vertex detector, as a function of the azimuthal angle vs.\ the cosine of the polar angle.}
\label{fig:vtx_material_budget_2D}
\end{figure}

A single layer contributes only about 0.07\%\,$X_0$, a reduction by about a factor three compared to the traditional design.
Furthermore, the material is more uniformly distributed in $\phi$, 
given the absence of overlapping structures in the same layer, 
as can be seen by comparing the left and right panels of Fig.~\ref{fig:vtx_material_budget_2D}.

For both alternatives, the outer vertex detector is composed of two barrel layers and three disks on each side. 
The elementary unit is a module of area $40.6\,(z) \times 42.2\,(r\phi)$\,mm$^2$. 
Each module has four hybrid pixel sensor chips, inspired by the ATLASPIX3 R\&D~\cite{ATLASPIX3}. 
The 50\,$\mu$m thick sensor has $372 \times 132$ pixels of $50 \times 150$\,$\mu$m$^2$ area. 
The power consumption is assumed to be 100\,mW/cm$^2$, 
half of the level observed at the current stage of development, 
but still too high (by at least a factor of two) to be handled by air cooling.
The modules are placed on a lightweight triangular truss structure. 
The staves are mechanical structures that hold the modules.
A stave is composed of a carbon-fibre multilayered structure, 
consisting of 120\,$\mu$m thick carbon fibre KDU13 and two carbon fleeces of 65\,$\mu$m in total, 
to support two 90\,$\mu$m thick polymide tubes of 2.2\,mm diameter, where demineralised cooling water circulates. 
An electronic bus, providing power distribution and readout and control signals, 
runs along the entire stave length and is placed on top of the modules. 
It is terminated, at the end of both sides, by a hybrid circuit.

The middle barrel layer is positioned at a radius of 130\,mm and is composed of 22 ladders, each with 8 modules.
The outer barrel layer is placed at a radius of 315\,mm and is composed of 51 ladders, of 16 modules each.
The outer and middle barrels are supported by a flange attached to an external support tube. 
The flange also supports the two innermost disks.
Three disks per side, located at $|z|$ = 304.7, 620, and 930\,mm, complete the outer vertex tracker. 
The inner disk is located inside the barrel.
Each disk is composed of eight petals, 
four facing the IP and four facing the opposite direction,
made of modules of the same type as those of the barrels.
The support structure of each disk is made of a sandwich of thin carbon fibre walls (each 0.3\,mm thick) 
interleaved with 5.4\,mm thick Rohacell~\cite{Rohacell}. 
The material budget of the outer vertex detector is about 1\% of a radiation length at normal incidence, 
increasing to about 3\% at the edges of the middle and outer barrel, because of the supports and hybrid circuits.

The vertex detector of the CLD concept~\cite{Bacchetta:2019fmz}, composed of a cylindrical barrel and forward disks, 
is located at radii below 112\,mm. 
The layout is based on double layers of ultralight MAPS fixed on a common support structure that includes cooling circuits. 
The 125\,mm long barrel comprises three double layers (0.63\%\,$X_0$ each) at radii 12.5--13.5, 37--38, and 57--58\,mm. 
Three disks on each side (0.70\%\,$X_0$ each), at distances from the IP of 160, 230, and 300\,mm, complete the detector. 
For simulation studies, a point resolution of $3 \times 3$\,$\mu$m$^2$ is assumed.

Full simulation studies~\cite{Sadowski:2024} with the smaller radius beam pipe 
introduced since the publication of Ref.~\cite{Bacchetta:2019fmz} 
(the inner radius decreased from 15 to 10\,mm)
and a corresponding reduction of the radius of the inner barrel layer 
confirm earlier conclusions from fast simulations~\cite{Bacchetta:2019fmz} 
of a better impact parameter resolution and a significantly improved flavour-tagging performance. 
Even with a simultaneous increase of the beam-pipe material budget from 0.45\% to 0.61\%\,$X_0$, 
the studies show, for example, an impact parameter resolution improved by 20\% for 10\,GeV tracks.

\subsection{Main tracking systems}
\label{sec:Detectors_MainTracking}

\subsubsection{Silicon tracker}

\begin{figure}[ht]
\centering
\includegraphics[width=0.60\textwidth]{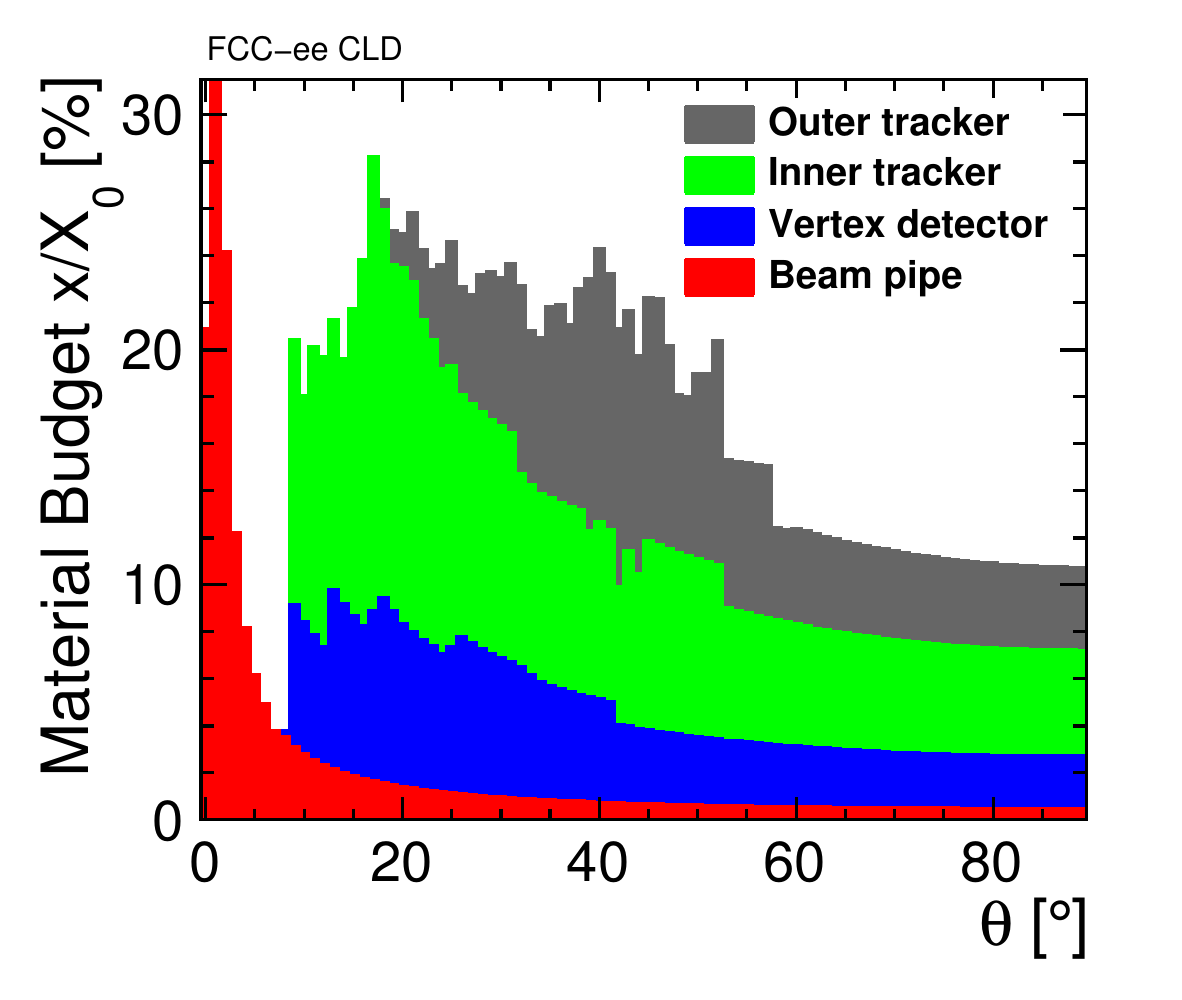}
\caption{Stacked material budget (expressed as a percentage of a radiation length) 
of the different parts of the CLD tracking system (plus the beam pipe), 
as a function of the polar angle and averaged over the azimuthal angle, 
including contributions from sensitive layers, cables, supports, and cooling.}
\label{fig:CLD:TrackerMaterialBudget}
\end{figure}

The CLD concept features an all-silicon tracker,
complementing the vertex detector (briefly described at the end of the previous section) 
with a `main tracker', itself divided into inner and outer sections 
by a lightweight (1.25\%\,$X_0$) carbon-fibre support tube, at a radius of 690\,mm.
The inner tracker comprises three barrel layers and seven forward disks on each end-cap,
while the outer tracker has three more barrel layers and four disks on each side. 
The material budget of the modules 
(200\,$\mu$m of silicon sensors, or 0.21\%\,$X_0$, including electronics),
plus cooling and connectivity, 
is estimated to range between 1.1\% and 1.3\%\,$X_0$. 
In addition, carbon-fibre support structures, which differ from layer to layer, 
contribute between 0.13\% and 0.37\%\,$X_0$.
For simulation studies, 
point resolutions of $5 \times 5$\,$\mu$m$^2$ and $7 \times 90$\,$\mu$m$^2$
are assumed for the innermost disk of the inner tracker and for all other layers, respectively.
The material budget of the complete CLD tracking system,
including the vertex detector and the main tracker (plus the beam pipe),
is shown in Fig.~\ref{fig:CLD:TrackerMaterialBudget}.

\begin{figure}[ht]
\centering
\includegraphics[width=0.49\textwidth]{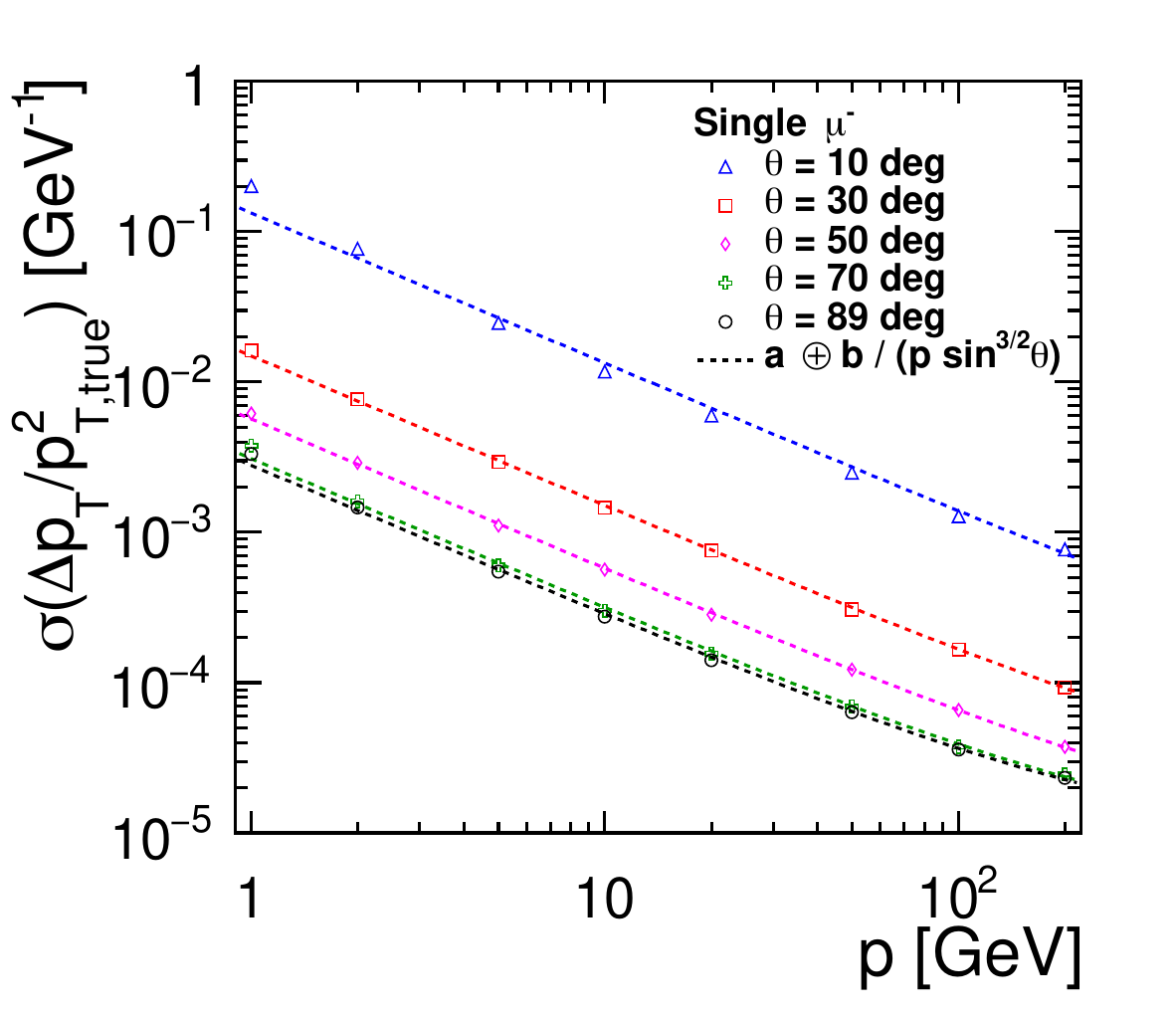}
\includegraphics[width=0.49\textwidth]{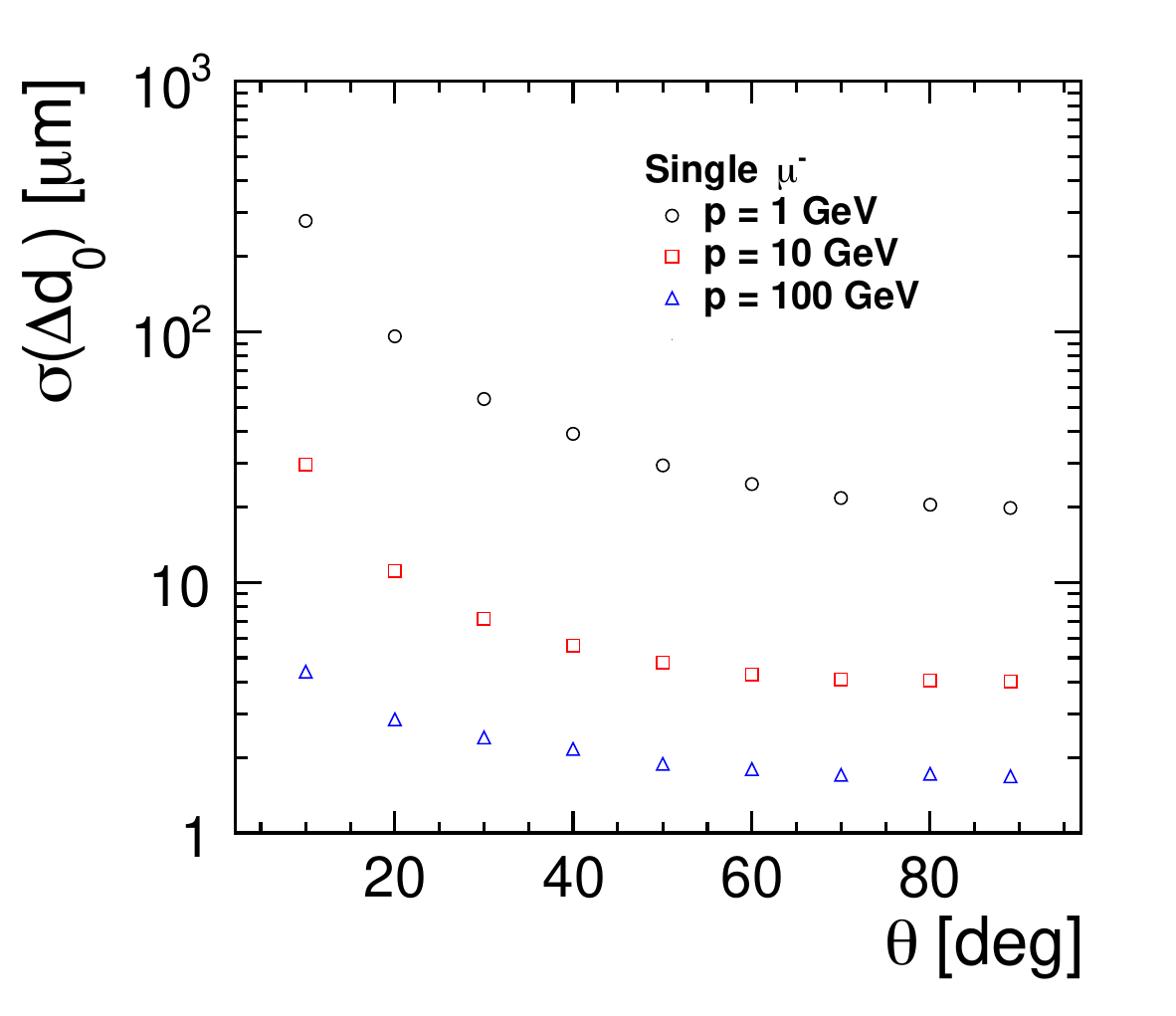}
\includegraphics[width=0.49\textwidth]{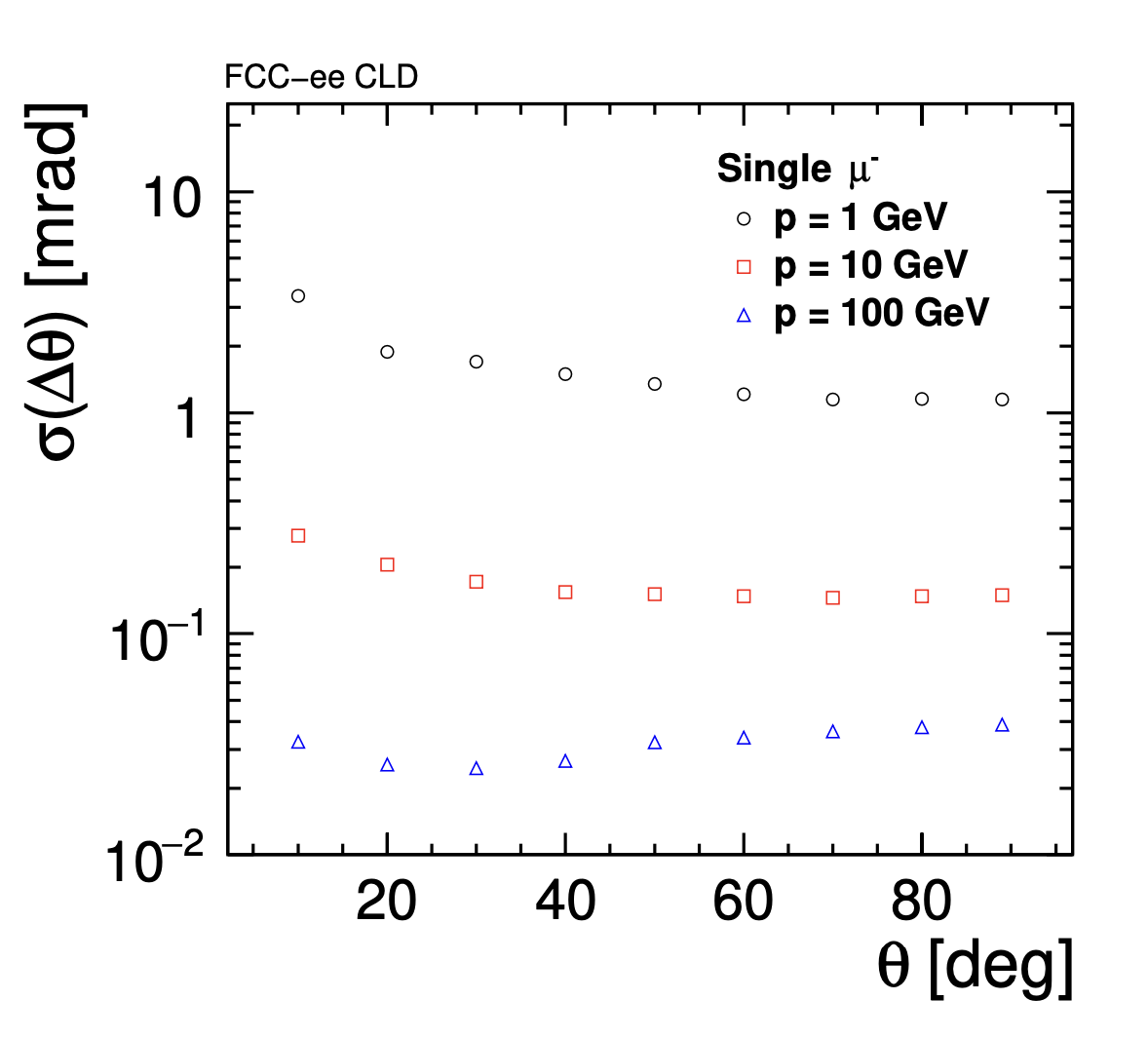}
\includegraphics[width=0.49\textwidth]{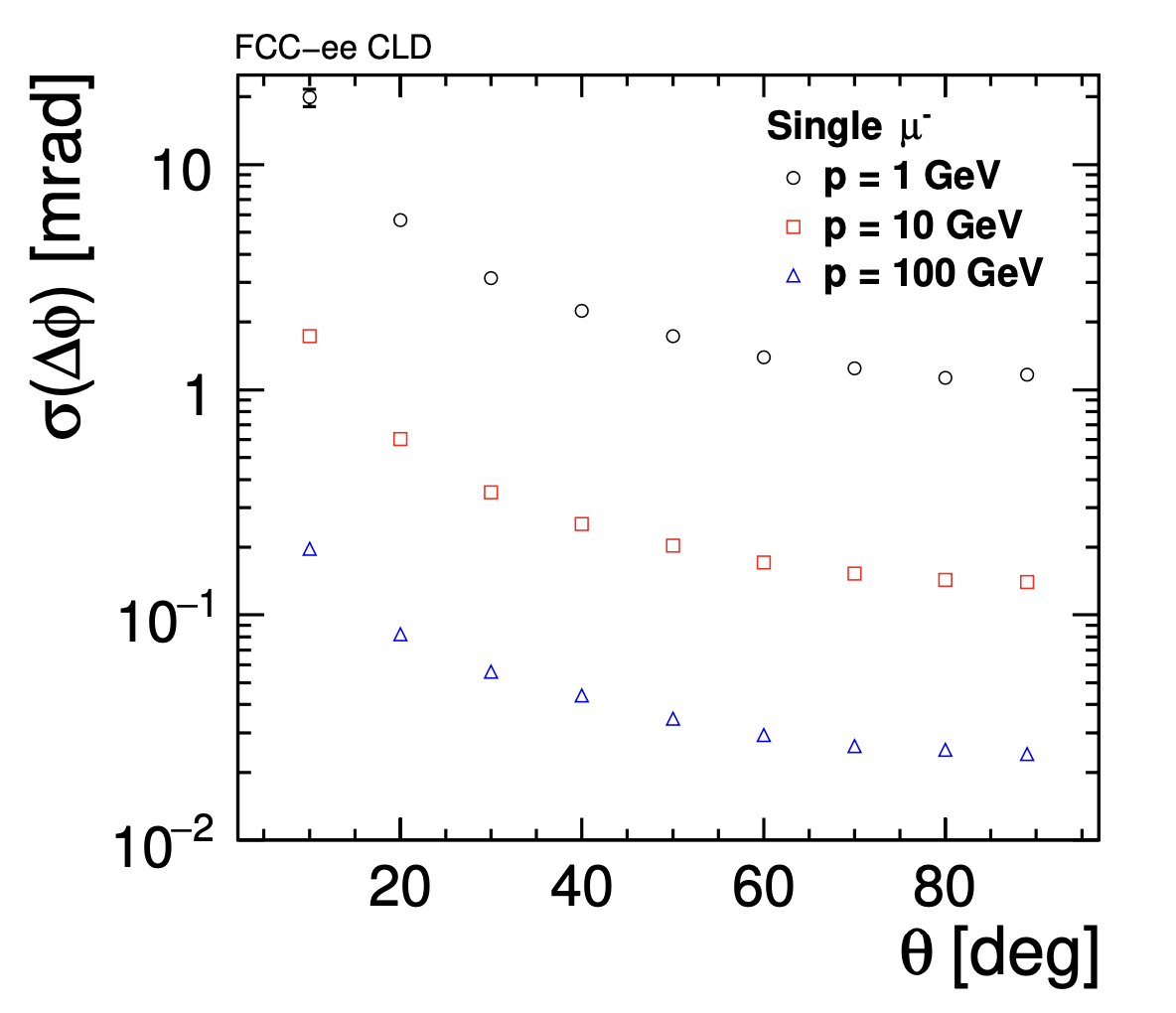}
\caption{Top-left: Transverse momentum resolution for single muons as a function of momentum at five fixed polar angle values.
Polar-angle dependence of the impact parameter resolution in the transverse plane (top-right), 
of the polar angle resolution (bottom-left), 
and of the azimuthal angle resolutions (bottom-right), 
for muon momenta of 1, 10, and 100\,GeV.
From Ref.~\cite{Bacchetta:2019fmz}.}
\label{fig:CLD:resolutions}
\end{figure}

Full-simulation studies to assess the performance of the CLD tracker are reported in Ref.~\cite{Bacchetta:2019fmz}. 
Figure~\ref{fig:CLD:resolutions} shows some essential performance figures for single muons: 
the momentum and impact parameter resolutions, 
as well as the polar and azimuthal angular resolutions. 
For $\pt > 100$\,GeV, the momentum resolution curves start to flatten out, 
reaching an asymptotic value of $\delta (1/\pt) \simeq 1 \times 10^{-5}$\,GeV$^{-1}$. 
For lower momenta, the momentum resolution is dominated by multiple scattering.
More studies are needed to further reduce the material budget, 
in particular for the support structures and services.

A smaller tracker outer radius would, in particular, allow for the insertion of a compact RICH detector, 
such as the ARC detector (Section~\ref{sec:ARC}), 
which could provide improved particle identification performance at the cost of a worse momentum resolution. 
A recent full-simulation study~\cite{Sadowski:2024} confirms fast-simulation indications 
that a reduction of the tracker outer radius from 2.31 to 1.80\,m leads to a degradation of the momentum resolution by about 15\%, 
showing that the momentum resolution scales roughly like $1/R$, where $R$ is the outer tracker radius. 
This observation is compatible with expectations in cases where multiple scattering is the dominant effect
(otherwise, a $1/R^2$ dependence would be expected). 
The study also confirms that an increase of the magnetic field strength from 2 to 3\,T 
would lead to a $\sim$\,30\% improvement of the \pt resolution. 

\subsubsection{Drift chamber}

The IDEA concept features the InTrEPId (Inner Tracking Equipped with Particle Identification) drift chamber, 
designed to provide tracking with high-precision momentum measurement and particle identification by cluster counting. 
A high transparency is obtained by a novel approach for the wiring and assembly procedures~\cite{chiarello}. 
The total amount of material is about 1.6\%\,$X_0$ at normal incidence (dominated by the outer container material), 
increasing to about 5.0\%\,$X_0$ in the forward direction 
(the end plates instrumented with front-end electronics contributing 75\% of that value). 
The drift chamber is a single-volume, high-granularity, all-stereo, short-drift, low-mass cylindrical wire chamber, 
co-axial with the 2\,T magnetic field. 
It extends from an inner radius of 0.35\,m to an outer radius of 2\,m, has a length of 4\,m along the $z$ axis, 
and consists of 112 co-axial layers, at alternating-sign stereo angles, arranged in 24 identical azimuthal sectors. 
The angular coverage extends down to $\sim$\,13$^\circ$ (227\,mrad). 
The size of the approximately square cells ranges between 12 and 14.5\,mm, for a total of 56\,448 drift cells.
The challenges potentially arising from a large number of wires (about 350\,000 in total) 
are addressed by the design of the wiring procedures, 
successfully exploited during the recent construction of the MEG2 drift chamber~\cite{chiappini2}. 

A very light gas mixture, 90\%~He and 10\%~iC$_4$H$_{10}$, corresponding to a maximum drift time of $\sim$\,400\,ns, 
is expected to be used in operation. 
The number of ionisation clusters generated by a minimum ionising particle is about 12.5\,cm$^{-1}$, 
allowing cluster counting/timing techniques to be exploited,
to improve both the spatial resolution (better than 100\,$\mu$m)
and the particle identification resolution ($\sim$\,3\%).
A spatial resolution around 120\,$\mu$m has been achieved in the 7\,mm cell size MEG2 drift chamber 
with the same gas mixture and very similar electrostatic configuration~\cite{afanaciev}. 
An even better spatial resolution is expected in InTrEPId, 
with the application of cluster timing techniques on the one hand and with the longer drift on the other, 
which mitigates, on average, the effects of the short-drift stochastic behaviour.

\begin{figure}[ht]
\centering
\includegraphics[width=0.5\linewidth]{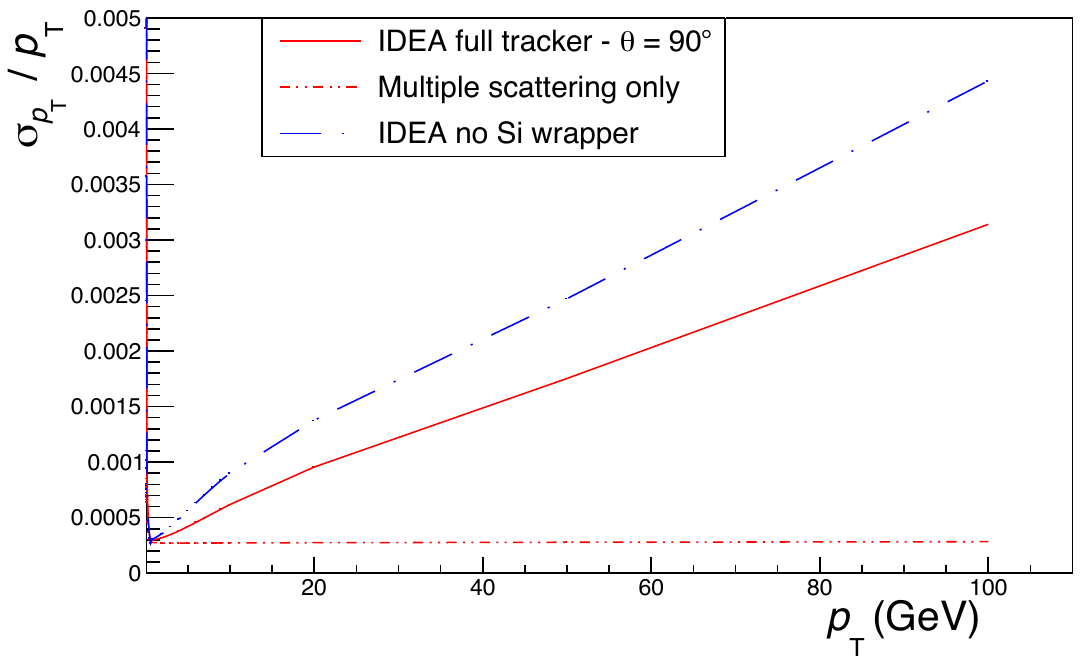}
\caption{Transverse momentum resolution as a function of \pt, for tracks at a polar angle of 90$^\circ$,
in the IDEA tracking system (full red curve). 
The contributions from multiple scattering (dot-dashed red curve) and the impact of the silicon wrapper (dash-dotted blue curve) are also illustrated.}
\label{fig:dch1}
\end{figure}

Fast simulation studies have been performed with \textsc{Delphes} to evaluate the tracking performance, 
with the vertex detector, the drift chamber, and a surrounding double layer of silicon microstrip detectors. 
This `Si wrapper' provides an additional accurate space point with an assumed resolution of $7 (r\phi) \times 90 (z)$\,$\mu$m$^2$, 
and a precise definition of the tracker acceptance. 
Details of ionisation clustering to exploit the cluster counting/timing technique were not simulated, 
conservatively limiting the spatial resolution of the drift chamber to 100\,$\mu$m. 
The resulting momentum resolution is displayed in Fig.~\ref{fig:dch1}. 
For transverse momenta in excess of $\sim$\,20\,GeV, 
the resolution is well described by $\delta \pt / \pt^2 = 3 \times 10^{-5}$\,GeV$^{-1}$.
Angular resolutions better than 0.1\,mrad are obtained in both the azimuthal and polar angles. 

\begin{figure}[ht]
\centering
\includegraphics[width=0.55\linewidth]{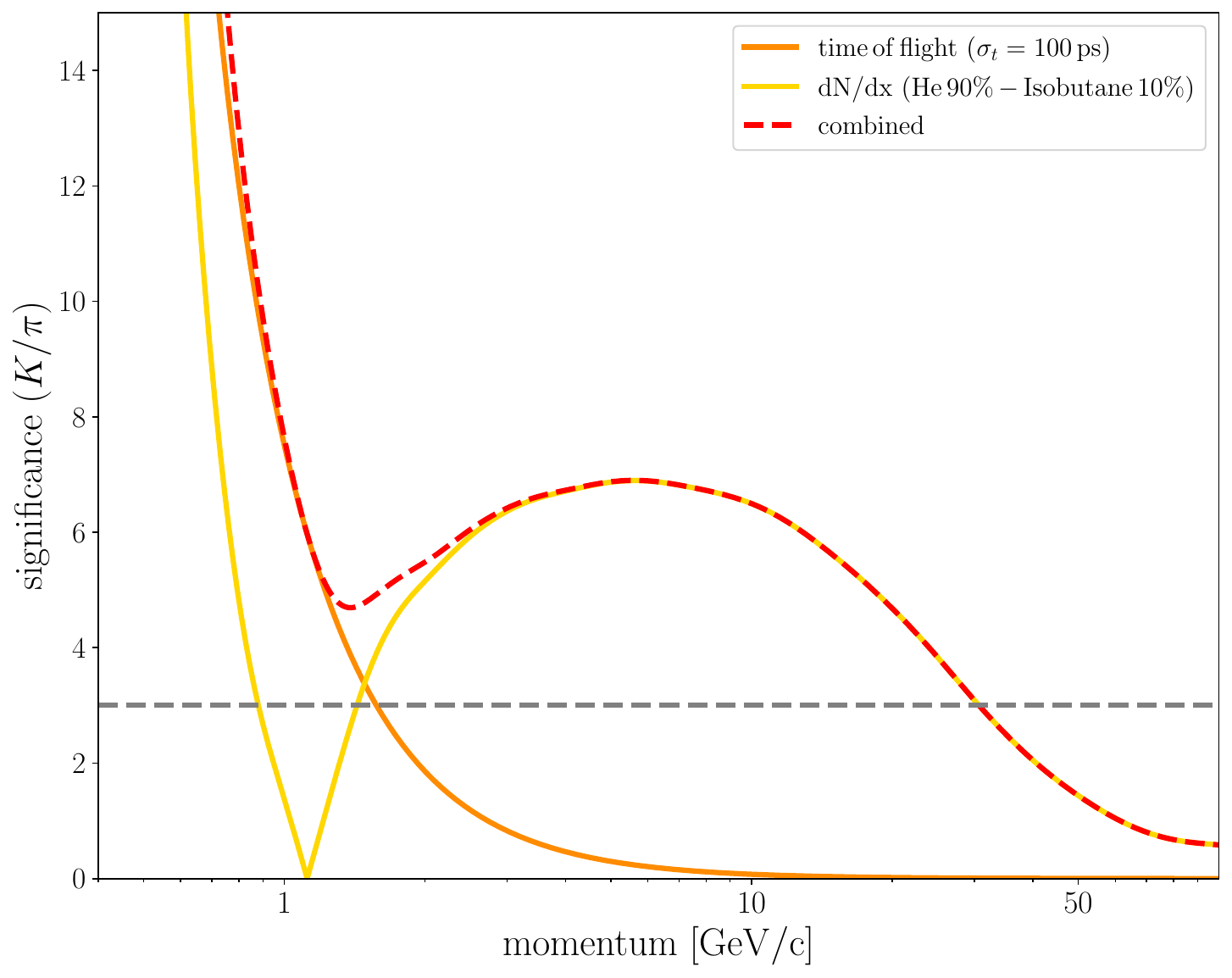}
\caption{Significance (in standard deviations) of the $\PGp/\PK$ separation in the IDEA drift chamber, 
from a fast \textsc{Delphes} simulation study, 
with the cluster counting performance parametrised from \textsc{Garfield++} results (yellow curve).
A better than 3\,$\sigma$ separation is obtained up to momenta $\sim$\,30\,GeV. 
The $p < 1.6$\,GeV region is covered by a time-of-flight system, 
extending over a distance of 2\,m and with a timing resolution assumed to be 100\,ps (orange curve). 
The combination of the drift chamber and the time-of-flight system is represented by the dashed red curve.}
\label{fig:dch4}
\end{figure}

Cluster counting and timing provide a $\PGp/\PK$ separation better than three standard deviations up to momenta about 30\,GeV, 
except in a narrow gap between 0.9 and 1.6\,GeV, where the Bethe--Bloch energy-loss curves for the two particle types cross. 
These values were obtained with a fast simulation study that parametrised \textsc{Garfield++}~\cite{garfield} results within \textsc{Delphes}. 
As illustrated in Fig.~\ref{fig:dch4}, the gap can be adequately covered with a time-of-flight measurement over a distance of 2\,m, 
with a non-challenging resolution of $\mathcal{O}(100)$\,ps.

\subsubsection{Straw tracker}

A straw tracker with thin Mylar walls may also be a promising option to meet the stringent FCC-ee detector requirements,
when combined with a pixel detector near the beam pipe and a silicon wrapper as an outer layer. 
A straw tracker can provide a spatial resolution of 100--120\,$\mu$m per straw 
and a low material budget of 1.2\%\,$X_0$ at $\theta = 90^{\circ}$ for 100 layers of straws 
made with 12\,$\mu$m-thick Mylar walls. 
With ${\cal{O}}(100)$ hits per track, 
such a straw tracker is well suited for both pattern recognition and searches for long-lived particles. 
In addition, it can provide excellent particle identification capabilities, 
such as $\PGp/\PK$ and $\PK/\Pp$ discrimination, 
across a wide momentum range and could also be used for triggering purposes, 
making it valuable for both collider data taking and cosmic ray studies.

Each straw functions as an independent channel, providing significant flexibility for design optimisation. 
The operational robustness also benefits from the fact that each straw can be easily disconnected if its wire breaks. 
The electric charges generated within a straw remain confined to that specific unit. 
The electric field within a straw is radially symmetric, 
making the resolution independent of the particle incident angle and leading to a good single-hit resolution. 
Straws with different radii can be used in various detector regions to optimise hit occupancy, 
material budget, and channel counting. 
The gas mixture can be optimised to enhance particle identification capabilities 
and different gas mixtures can be simultaneously used for straws in different layers.

\begin{figure}[t]
\centering
\includegraphics[width=0.54\linewidth]{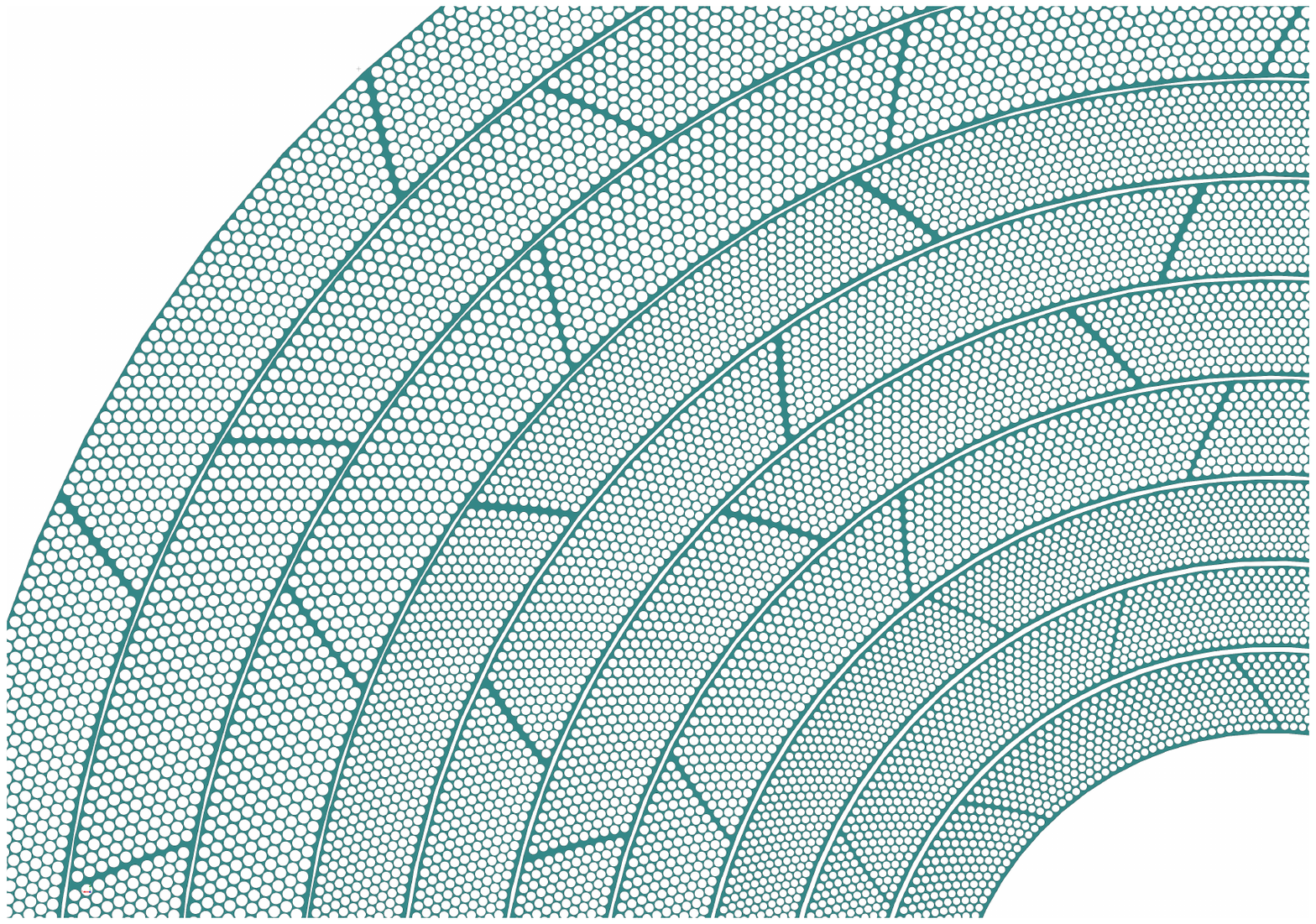}
\includegraphics[width=0.4\linewidth]{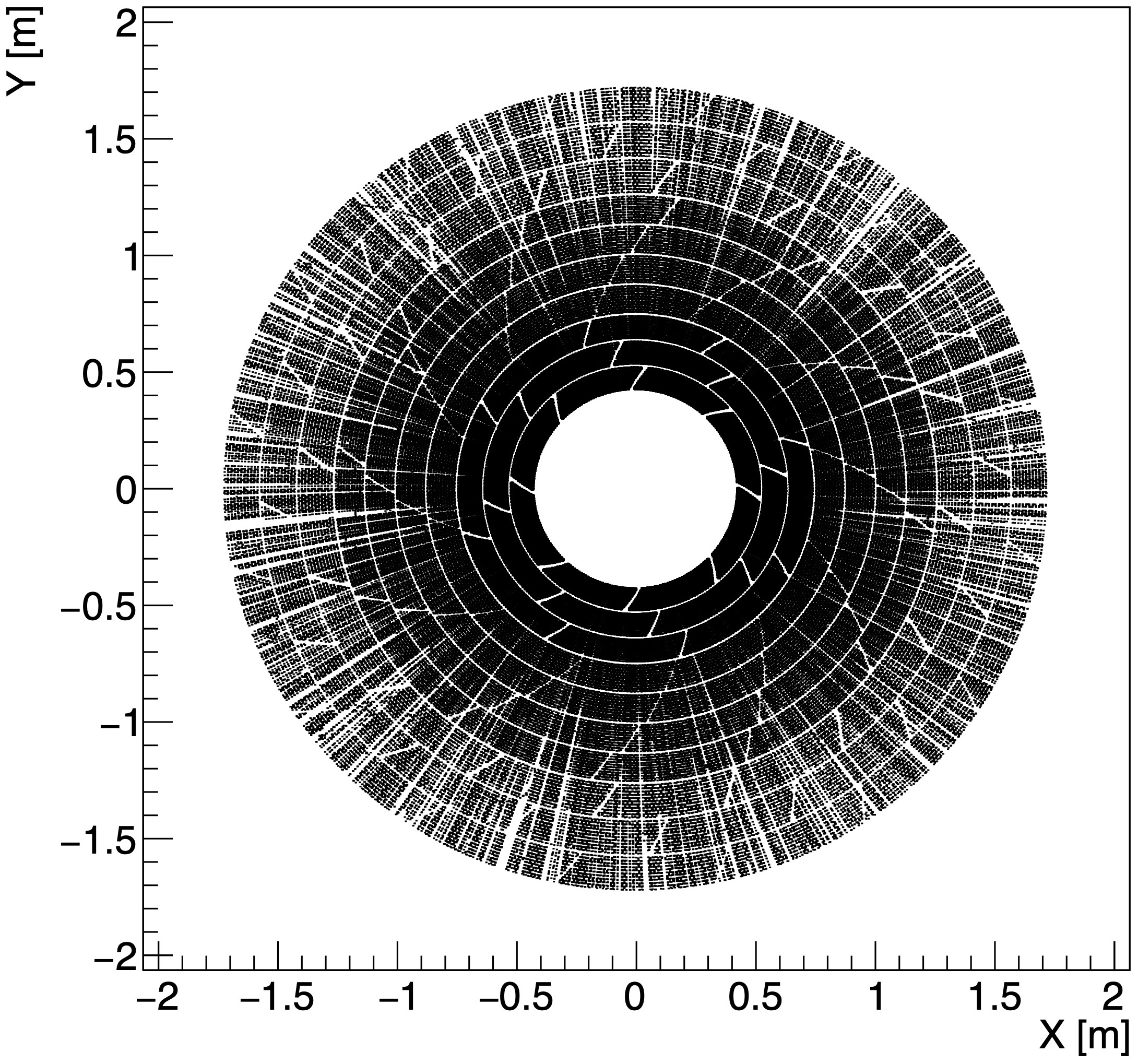}
\caption{(Left) An example layout of the straw tracker with ten superlayers, 
each superlayer containing ten sublayers.
(Right) Simulated hits from 5000 muons, revealing the layout of the straw tracker in the \textsc{Geant4} simulation.}
\label{fig:straw_tracker}
\end{figure}

Figure~\ref{fig:straw_tracker} shows an example layout of a straw tracker implemented in a \textsc{Geant4} simulation, 
featuring ${\cal{O}}(60\,000)$ straws with a diameter of 1 to 1.5\,cm and a length of 5\,m,
arranged in ten concentric superlayers. 
Straws are arranged in both axial and stereo layers with a stereo angle of a few degrees 
to determine the hit position with an accuracy of a few mm along the $z$ direction. 
Initial simulations~\cite{StrawTrackerRnD} predict similar momentum resolutions as with the InTrEPId drift chamber. 

Significant R\&D is required to realise such a detector design. 
Close collaboration with straw manufacturers is essential to produce ${\cal{O}}(12)$\,$\mu$m-thick-walled aluminium-coated straws with high yields. 
Optimising the detector layout and developing a robust mechanical design for the end-plates and support structure of these 5\,m-long straws are critical tasks. 
Further investigations into gas mixtures, \textsc{Garfield} simulations, and front-end electronics are important for d$E$/d$x$ and d$N$/d$x$ measurements. 
Fast algorithms need to be developed and implemented in the front-end ASIC or FPGA in order to effectively identify individual clusters.
In the coming years, prototype chambers will be built and tested with cosmic-ray muons and with beams, in view of validating the simulated performance.

\subsubsection{Time projection chamber}

Time projection chambers (TPCs) provide continuous 3D tracking over a large volume with minimal material interference, 
while also enabling particle identification via energy deposition in the gas. 
The gas choice is critical to maximise ionisation signal yield and minimise transverse diffusion, 
which strongly depends on the magnetic field strength and on the drift velocity. 
Minimising transverse diffusion calls for a `hot gas', commonly achieved so far with a small admixture of CF$_4$. 
Given that CF$_4$ has a strong greenhouse effect, a replacement gas will have to be identified. 
For particle identification via cluster counting, d$N$/d$x$, small, sub-millimetre pad sizes are needed, in which case digital readout is sufficient.

Mechanical alignment must be precise, within a few tens of microns, to avoid systematic errors. 
Electric and magnetic fields must be parallel, to prevent $E \times B$ distortions, calling for a highly uniform magnetic field. 
Minimising space-charge build-up is essential, as transverse electric field components distort electron drift paths. 
Hence, beam backgrounds 
(such as low-energy X-rays and muons from beam-halo interactions, both generating low-\pt particles that curl and deposit significant ionisation) 
must be controlled, or corrective strategies must be implemented. 
First estimates of the effect of beamstrahlung backgrounds were recently presented~\cite{DJeansECFA-Paris}, 
with the conclusion that distortions expected with the FCC-ee Tera-\PZ luminosities are at the same level as those observed in ALICE.
Ion feedback, where ions from the amplification process drift towards the cathode (over $\sim$\,0.5\,s), must be carefully managed. 
Operating at low gain with effective passive backflow mitigation, 
such as double misaligned meshes or graphene filters, would be necessary. 
If space charge effects are unavoidable, correction techniques (built on the experience of ALICE) should be employed.
The operability of a TPC at FCC-ee luminosities is still under discussion in the community and is being actively investigated. 

\subsection{Particle identification}

The identification of charged hadrons (\PGp, \PK, \Pp) 
significantly enhances the physics potential of experiments at FCC-ee, as discussed in Section~\ref{sec:PhysPerf_PID}.  
In particular, it improves the flavour-tagging performance in the selection of 
Higgs boson decays to $\PQc\PAQc$ or $\PQs\PAQs$, 
and is crucial for the extensive flavour physics programme with the \PZ-pole data. 
From simulation studies, the momentum region to be covered is up to a few tens of GeV.
For experiments with gaseous trackers, 
the specific energy loss (d$E$/d$x$) provides separation power up to moderate momenta, 
in conjunction with time-of-flight to cover the overlap region around 1\,GeV, where the d$E$/d$x$ separation is poor. 
As already mentioned for drift chambers or straw trackers, the performance can be enhanced with cluster counting, d$N$/d$x$. 
For experiments with silicon-based tracking, however, neither d$E$/d$x$ nor d$N$/d$x$ would provide the required level of performance. 
There, ring-imaging Cherenkov (RICH) detectors could deliver particle identification over a very wide momentum range. 
The following two sections describe the use of precision timing and RICH detectors 
to enable the required particle identification performance. 

\subsubsection{Precision timing detector}

The identification of charged hadrons based on their specific energy loss 
must be complemented by other measurements to fill the gap around 1\,GeV, 
where the Bethe--Bloch energy-loss curves for the different particle types cross over. 
Time-of-flight measurements, for example in a layer of silicon sensors at 2\,m from the interaction point, 
with a non-challenging resolution of $\sim$\,100\,ps, are adequate for this purpose. 
With a resolution of 30\,ps, 
time-of-flight measurements alone would allow kaons to be separated from pions up to ${\cal{O}}(3)$\,GeV in momentum, 
while an excellent resolution of 10\,ps would be needed to extend this momentum range up to 5\,GeV.
Such TOF information could also be obtained in a silicon-tungsten ECAL, 
where measurements, each of a resolution of, e.g., 100\,ps, could be made in several layers.  
Dedicated timing layers made of inorganic scintillator offer another solution. 
An example is the segmented crystal ECAL design of MAXICC (described in Section~\ref{sec:ECALs}), 
which includes two such thin layers, potentially providing a 20\,ps timing resolution. 

Because of the finite length of the colliding bunches, the `event time', $t_0$, has a natural spread of about 36\,ps at the \PZ pole. 
To exploit timing resolutions better than that value, $t_0$ needs to be determined on a event-by-event basis. 
In events with a high charged-particle multiplicity, 
it is possible to reconstruct $t_0$ from the timing measurement of the other tracks. 
For lower multiplicity events, e.g., $\PZ \to \PGt\PGt$, this method is less effective. 
As an alternative, an additional timing measurement close to the interaction point could be established, 
possibly in the form of an LGAD silicon sensor layer at low radius. 
This solution would allow a direct track-by-track time-of-flight measurement independent of $t_0$. 
Studies are needed in order to quantify the implications of such a layer in terms of 
additional material for readout electronics and cooling.

\subsubsection{Compact RICH detector}
\label{sec:ARC}

Studies are underway to develop a compact RICH detector for inclusion in CLD, 
with the goal of fitting it in a 20\,cm radial envelope and contributing less than 10\% of a radiation length to the material budget, 
hence limiting its impact on the other subsystems of the experiment, such as tracking and calorimetry. 
It is enabled by the new generation of photodetectors, in particular silicon photomultipliers (SiPM), 
that hold the promise of providing high detection efficiency and spatial granularity in a low thickness. 
A focusing geometry has been developed, 
where the surface of the detector is tiled with a large number of similar RICH detector elements, 
in a cellular approach, a concept named ARC, for `Array of RICH Cells'~\cite{Forty_ARC, Tat_ARC, forty_2024_6g0gs-7kw30}.

\begin{figure}[t]
\centering
\includegraphics[width=0.5\linewidth]{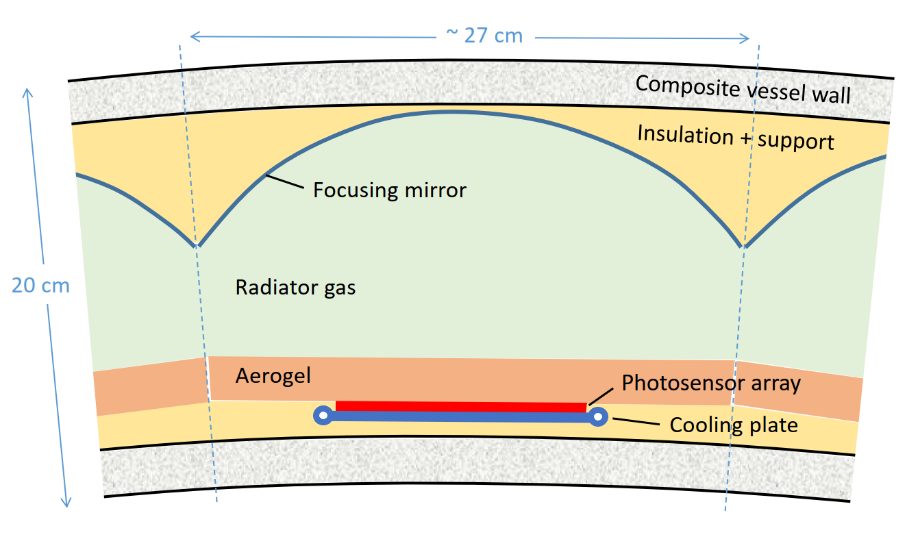} \\ 
\includegraphics[width=0.38\linewidth]{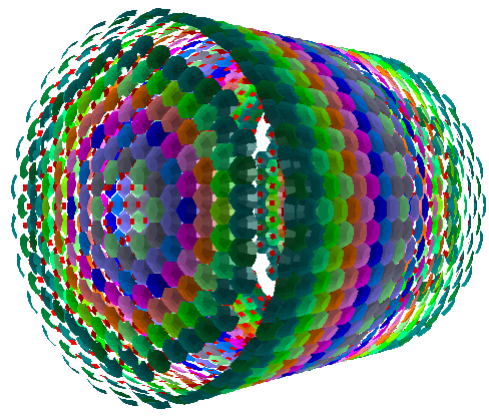} \hglue5mm
\includegraphics[width=0.45\linewidth]{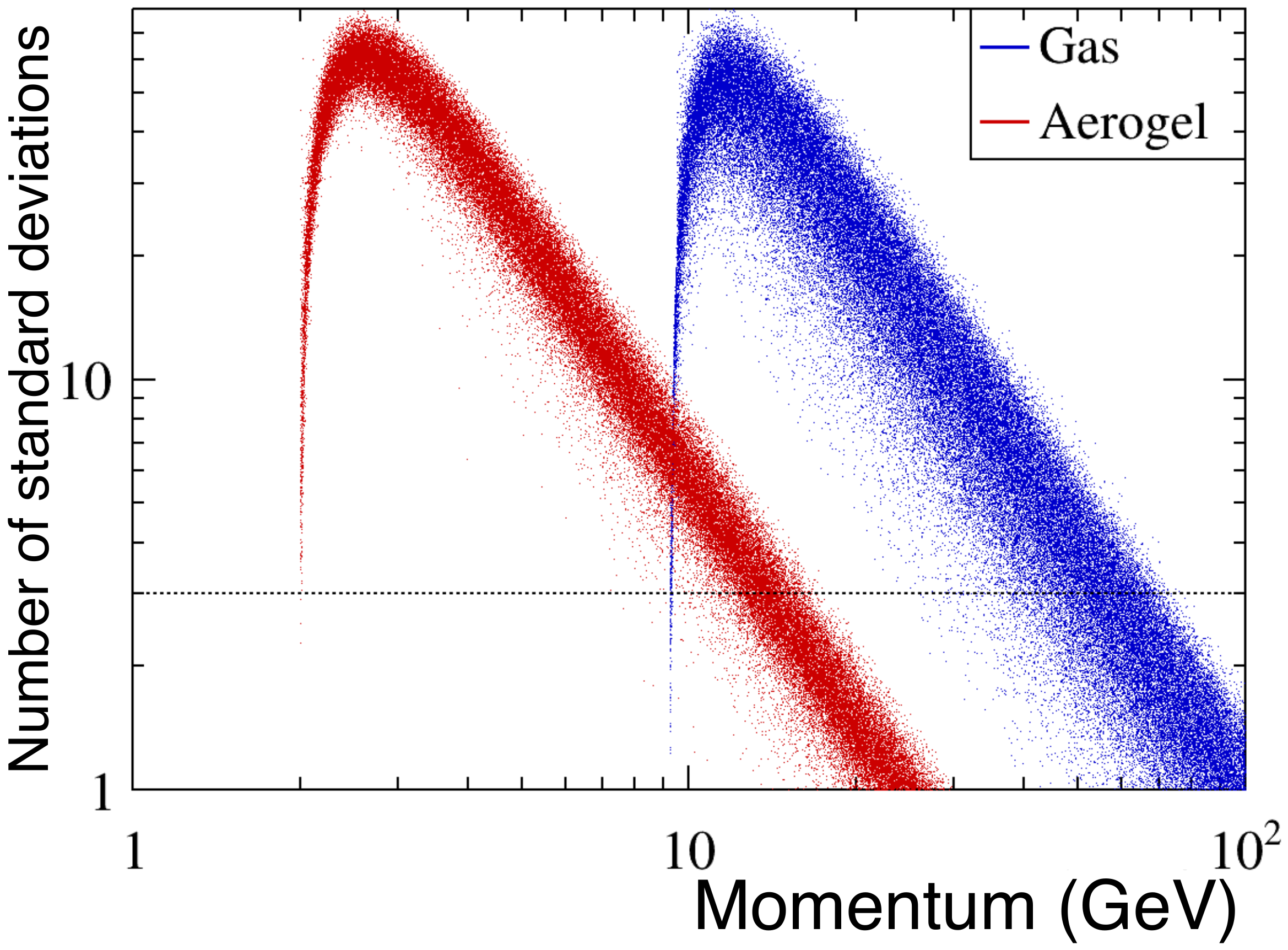}
\caption{Top: Schematic cross section through the ARC barrel detector, 
showing the various components of a single cell.
Bottom left: ARC barrel and end-cap as described in the detector geometry of the simulation,
with colours used to distinguish the cells. 
Bottom right: Momentum dependence of ARC's \PK/\PGp separation (in standard deviations), 
with two radiator options.}
\label{fig:ARC}
\end{figure}

The components of an ARC cell are shown in the top panel of Fig.~\ref{fig:ARC}. 
The lightweight vessel is made of carbon-fibre composite, 
with a wall thickness that depends on whether the contained gaseous radiator needs to be pressurised. 
The RICH uses dual radiators, silica aerogel and gas, 
with Cherenkov photons from both radiators focused via a spherical mirror onto a common SiPM detector plane. 
The baseline choice for gaseous radiator is C$_{4}$F$_{10}$ at atmospheric pressure, 
given its attractive optical properties, as used for example in the RICH1 detector of LHCb, 
aiming for a leak-free closed circulation system. 
With the high photon-detection efficiency achievable with SiPMs, 
this option can provide a sufficient number of detected photons, about 16 for a high-momentum particle, 
despite the limited radiator length of only around 15\,cm. 
On the timescale of FCC, however, such fluorocarbons may be banned, 
so that R\&D is required to find alternative gases, such as xenon with mild pressurisation, 
to ensure a sufficient yield of detected photons. 
The aerogel radiator extends the particle identification performance to low momenta, 
while also serving as an efficient thermal insulator, 
separating the gas radiator from the SiPM detector plane, 
which is likely to need cooling to limit the dark-count rate 
(although the option of suppressing noise with timing cuts will also be investigated). 
This insulator would allow the SiPMs to be operated at around $-40$\,$^\circ$C, 
e.g., with mixed-phase CO$_2$ circulation, 
while maintaining the radiator gas at room temperature to prevent condensation.

The performance of such a RICH concept has been evaluated with full simulation studies~\cite{Delgado_ARC}. 
The geometry of each cell, in terms of mirror and photodetector position and angle, 
depends on the position of the cell in the detector.  
Given symmetry considerations, there are about 40 unique cell geometries across the detector, 
which have been optimised in an automated procedure. 
The resulting detector description has been implemented in the \textsc{Key4Hep} software framework (Chapter~\ref{sec:software}), 
with \textsc{DD4hep} for the geometry and \textsc{Geant4} for the simulation, 
as illustrated in the bottom-left panel of Fig.~\ref{fig:ARC}. 
Taking the integration into CLD as an example, 
10\% of the original tracker volume has been reassigned to ARC, 
the tracker dimensions being adapted to fit in the remaining space. 
Studies are underway to develop pattern recognition and particle identification algorithms, 
as well as to check the impact on the experiment, e.g., on the performance of the tracking and particle-flow calorimetry. 
Meanwhile, the performance has been studied using standalone ray-tracing software, 
as shown in the bottom-right panel of Fig.~\ref{fig:ARC}.
The overall resolution per detected photon is around 2\,mrad, 
balanced between contributions from the chromatic dispersion, focusing errors, and pixel size, 
where mm-scale pixelisation of the SiPMs has been assumed. 
Excellent \PK/\PGp separation is achieved for momenta above 2\,GeV. 
For lower momenta, a TOF system would provide complementary information. 
The development of ARC is established as a task within the newly formed DRD4 collaboration~\cite{Easo_DRD4}, 
with the milestone of a full conceptual design within a year 
and a prototype of a single ARC cell to be delivered within three years. 
Such a prototype will provide an excellent test-bed for the study of the key developments required for this detector,
as well as the possibility of including RICH detectors as integral parts of the FCC-ee tracking detectors.

\subsection{Electromagnetic calorimeters}
\label{sec:ECALs}

\subsubsection{Silicon pads, MAPS, scintillator strips}

Electromagnetic calorimeter technologies with embedded front-end electronics 
and high 3-dimensional segmentation have been studied extensively~\cite{Sefkow:2015hna}. 
They include silicon pads, MAPS sensors and scintillator strips, 
and capitalise on the possibility of power-pulsing and overall modest bandwidth requirements. 
Adapting them to a circular collider such as FCC-ee, with continuous readout and high rates, 
poses challenges for data concentration, powering, and cooling, 
requiring a full re-optimisation to maximally preserve their compactness. 
Silicon diodes are currently implemented in the CMS end-cap calorimeter upgrade, 
but the FCC-ee goals in terms of compactness (Moli\`ere radius), 
energy resolution, and granularity are considerably more ambitious. 

Related R\&D has begun, in the framework of the DRD6 Collaboration. 
As a first step, quantitative requirements are being studied in simulations. 
Tools have been developed to evaluate hit occupancy, bandwidth and, 
with model assumptions on the embedded electronics, 
power consumption as a function of position in the detector~\cite{Hassouna:2024frw}.

\subsubsection{Noble liquid}
 
The proposed ALLEGRO noble-liquid electromagnetic calorimeter is a sampling detector 
using liquid argon as active medium and lead/steel absorbers as passive material. 
The absorbers and electrodes are straight, and inclined (in the barrel) in the ($x,y$) plane 
by 50$^\circ$ with respect to the radial direction. 
The detector has both transverse and longitudinal segmentation.
Alternative design options include liquid krypton as active medium and tungsten for passive material, 
as well as trapezoidal absorbers to maintain the sampling fraction constant as a function of depth.

The calorimeter is located inside a cryostat, 
made of 33.8\,mm-thick aluminium in the current simulations 
while novel, lightweight materials are being considered (Section~\ref{sec:Cryostat}).
The 2\,mm-thick absorber plates are composed of a sandwich of lead (1.8\,mm) with steel sheets glued on either side.
Between adjacent absorbers are two equal-size liquid argon gaps, 
separated from each other by a 1.2\,mm-thick electrode, realised as a multilayer PCB.
The use of multilayer PCBs as readout electrodes allows for great flexibility in the choice of granularity as a function of depth.
In the present geometry, the liquid argon gaps have thicknesses, 
projected along the direction orthogonal to the electrode,
of $2 \times 1.25$\,mm at the inner radius and $2 \times 2.43$\,mm at the outer radius.
The electrodes are segmented longitudinally into 11 layers.
The first layer, about half as thick (1.5\,cm) as the others (3 to 5\,cm), is used as a presampler. 
To achieve a $\phi$-uniform response of that first layer, 
the absorber does not contain lead, to form a `LAr-only' homogeneous presampler.
The sampling fraction is about 38\% in the presampler layer 
and increases from 13.5\% to 17.3\% throughout the other 10 layers.
A depth of 22\,$X_0$ is achieved in a thickness of 40\,cm for the Pb--LAr solution 
and about two thirds of that for denser solutions based on W and LKr.

The segmentation in $\phi$ is naturally provided by the absorbers.
The baseline is to read out two gaps together, forming a single cell spanning $\Delta\phi = 2\pi/768 = 8$\,mrad. 
The segmentation along the electrode and in $\theta$ is formed by cells on the readout electrode.  
It amounts to $\Delta\theta = 0.56^\circ$, except in one of the first layers, 
where a four-times finer granularity ($\Delta\theta = 0.14^\circ$) is implemented, 
to discriminate between single showers from prompt photon and collimated pairs of showers from neutral pion decays. 
The total number of readout channels for the barrel part of the calorimeter is around 2~million.
The expected sampling term of the resolution ranges from $\sim$\,7.5\%/$\sqrt{E}$ for a solution based on LAr, 
down to below 5\%/$\sqrt{E}$, for a denser solution based on LKr.

An adapted geometry is needed for the detector end-caps. 
The inclined absorber and electrode planes of the barrel calorimeter are replaced by `blades' arranged in a turbine-like structure. 
In order to avoid having too large variations of gap sizes and sampling fractions between the inner and outer radii, 
the detector is assembled as three nested wheels, with similar gap sizes at the inner radius of each wheel.

\begin{figure}[h]
\centering
\includegraphics[width=0.45\textwidth]{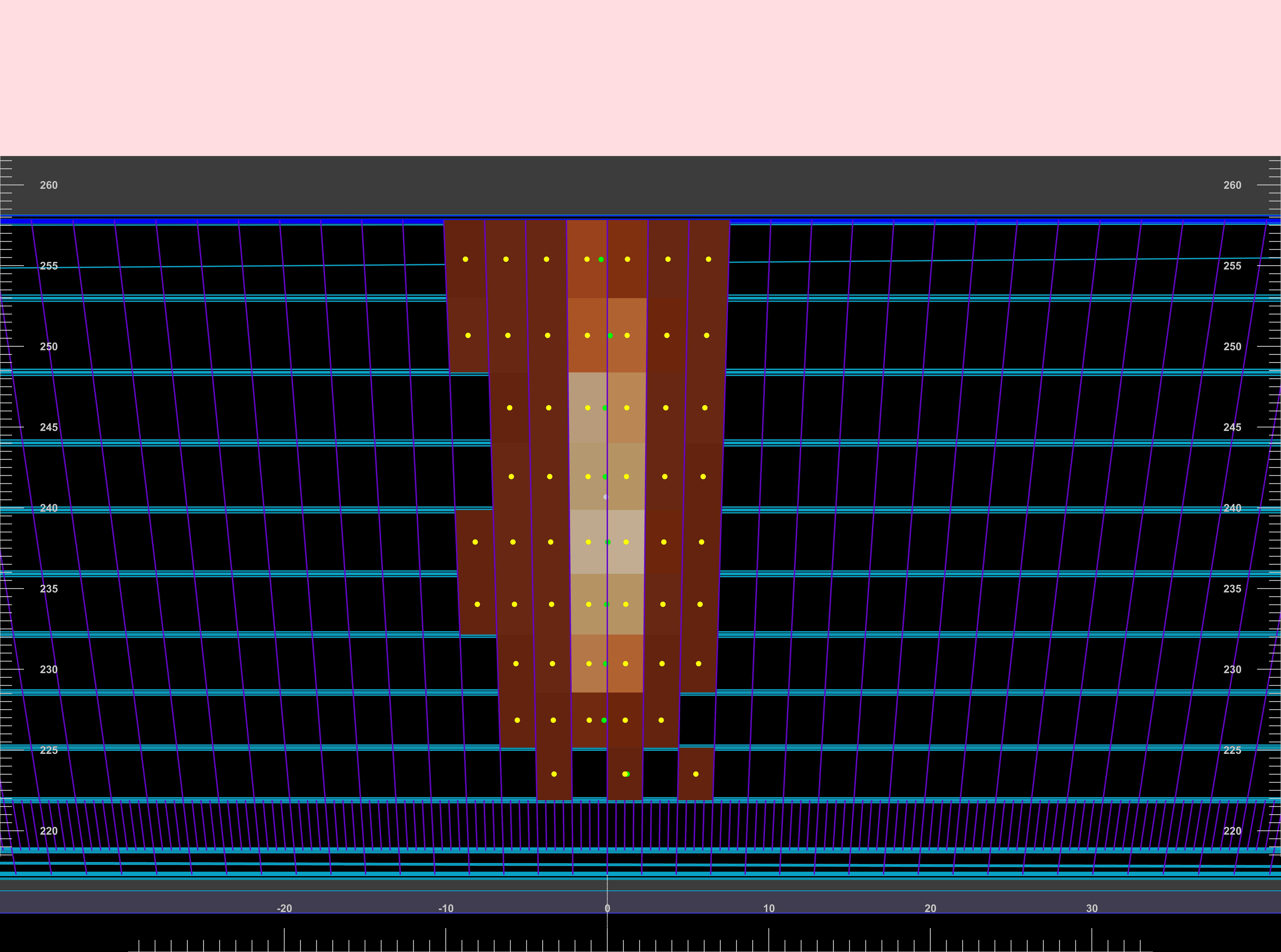}
\includegraphics[width=0.45\textwidth]{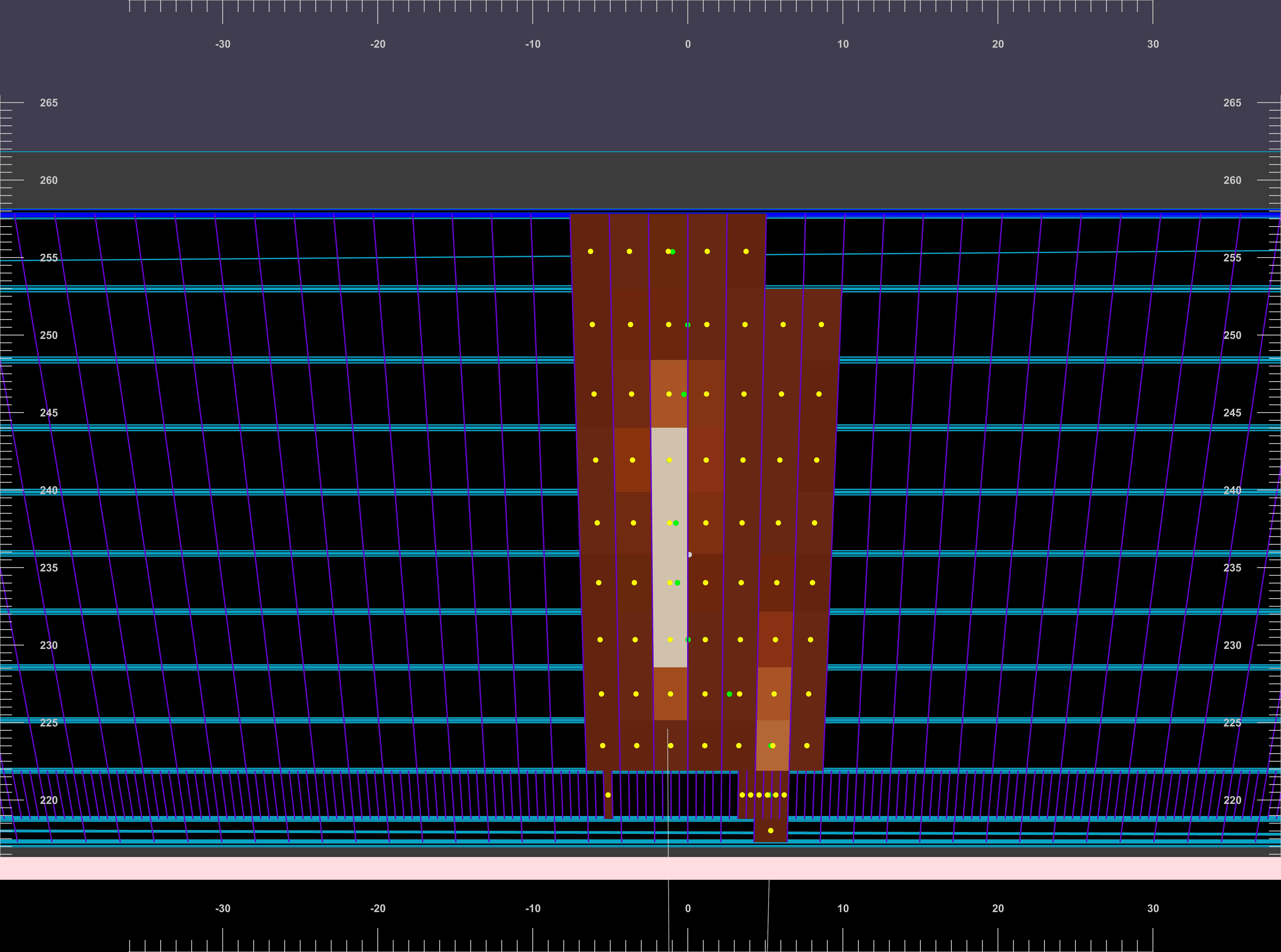}
\caption{Event displays in the ($r,z$) view for clusters reconstructed in the ECAL barrel of ALLEGRO, 
for a photon (left) and a $\PGpz \to \PGg\PGg$ (right), 
both coming from the main IP at normal incidence with an energy of 10\,GeV. 
The particles impinge on the detector from the bottom. 
Clustered cells are shown in brown, with lighter tones corresponding to higher cell energies.
The yellow dots are placed in the cell centres. 
Noise is not simulated.
In this simulation, the strip layer is the first one.}
\label{fig:ALLEGRO_ECAL_eventdisplays}
\end{figure}

The geometries of both the barrel and the end-cap sections are implemented 
in the \textsc{Geant4} full simulation of the \textsc{Key4Hep} software framework. 
Algorithms for digitisation and clustering (fixed-size sliding-window and topological clusters) are also implemented. 
Example displays of reconstructed clusters from single-photon and single-\PGpz events are shown in Fig.~\ref{fig:ALLEGRO_ECAL_eventdisplays}.

\begin{figure}[h]
\centering
\includegraphics[width=0.4\textwidth]{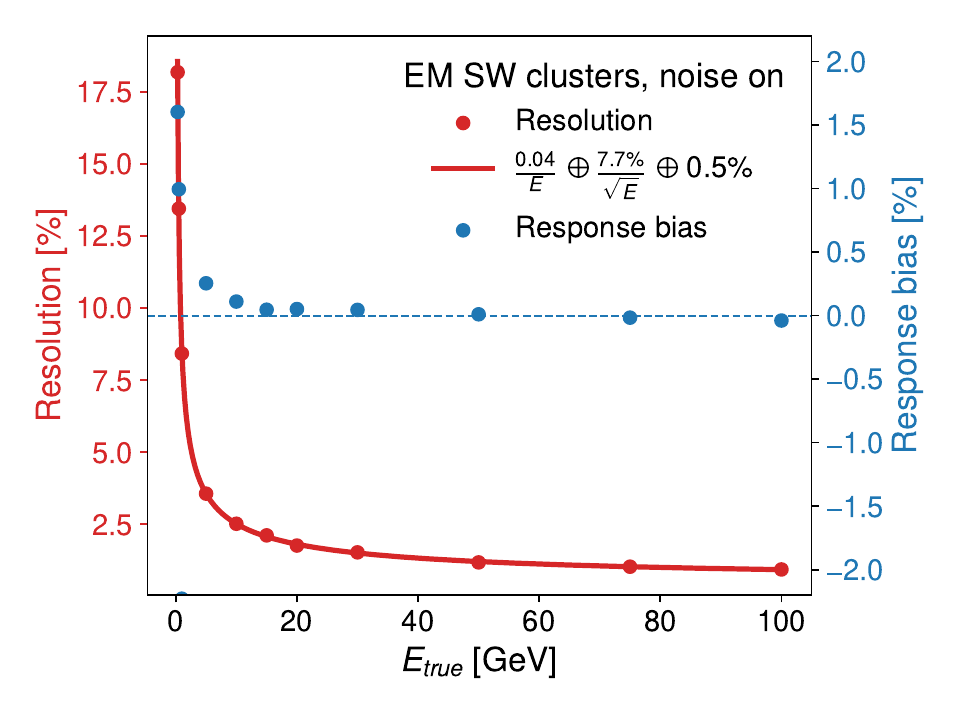}
\includegraphics[width=0.5\textwidth]{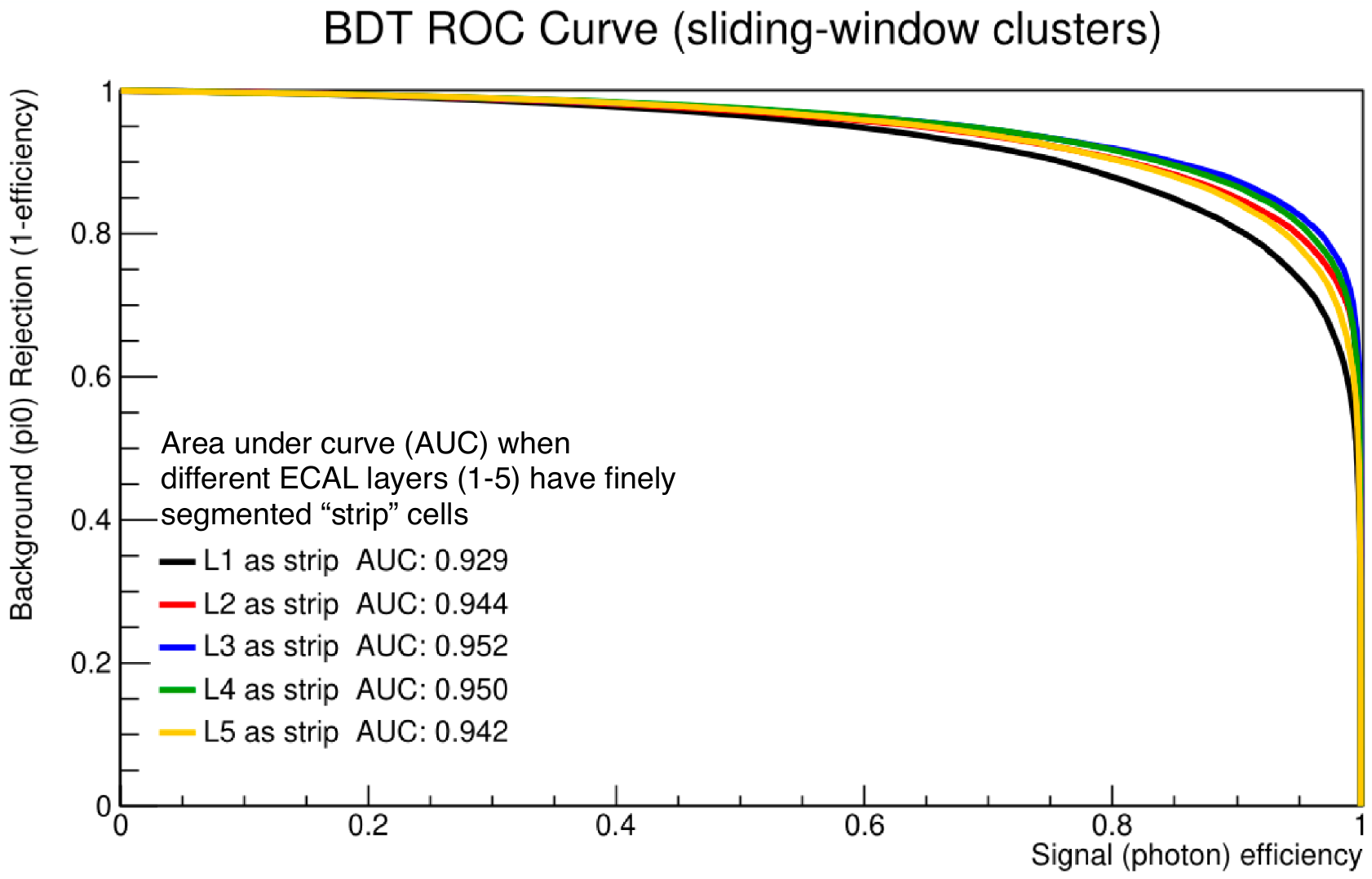}
\caption{Left: Simulated energy response and resolution of the LAr ECAL barrel calorimeter, including noise, 
for reconstructed (sliding window) photon clusters as a function of the true photon energy.
Right: ROC curves of a BDT trained to discriminate between photons and neutral pions with momenta between~1 and~100\,GeV, using input features calculated from the clustered cell energies. }
\label{fig:ALLEGRO_ECAL_performance}
\end{figure}

Examples of performance studies are illustrated in Fig.~\ref{fig:ALLEGRO_ECAL_performance}, 
where the energy resolution and response expected with the (LAr plus tungsten) baseline model are shown, 
together with the ROC curves for a simple BDT-based photon/\PGpz discrimination algorithm 
exploiting the shower shapes computed from the cell energies. 
The latter, which shows the area-under-curve (AUC) of the efficiency vs.\ rejection curve of the BDT algorithm 
for various alternative configurations corresponding to different positions of the strip layer, 
demonstrates how the simulation can help design choices.

Sustained R\&D on noble liquid calorimeters is carried out in the framework of the WP2 of DRD6. 
Ongoing activities focus on the optimisation of the detector material, geometry and segmentation, 
overall detector mechanical structure and integration, readout electronics, 
as well as on prototyping and testing the various detector elements. 
On a longer timescale, the plan is to build a prototype and test it with beam, around 2027--2028.

\subsubsection{Segmented crystals with dual-readout}
\label{sec:Segmented-crystals}

The Maximum Information Crystal Calorimeter (MAXICC) 
is a homogeneous electromagnetic calorimeter concept optimised for FCC-ee,
made of longitudinally segmented crystals with embedded dual-readout and precise timing capabilities. 
An overview of its design can be found in Ref.~\cite{Lucchini:2020bac}.
This calorimeter offers an electromagnetic energy resolution of $3\%/\sqrt{E}$, 
crucial for studies of heavy-flavour physics with low-energy final-state photons~\cite{RoyAleksan} 
and to improve the resolution of the mass recoiling against $\PZ \to \epem$ in $\PZ\PH$ events with a recovery of bremsstrahlung photons. 
Furthermore, it enables an efficient clustering of photon pairs from \PGpz decays, 
which can effectively reduce the splitting of the two photons across different jets in multi-jet events~\cite{Lucchini:2020bac}.

\begin{figure}[ht]
\centering
\begin{minipage}[t]{0.48\textwidth}
\includegraphics[width=\linewidth]{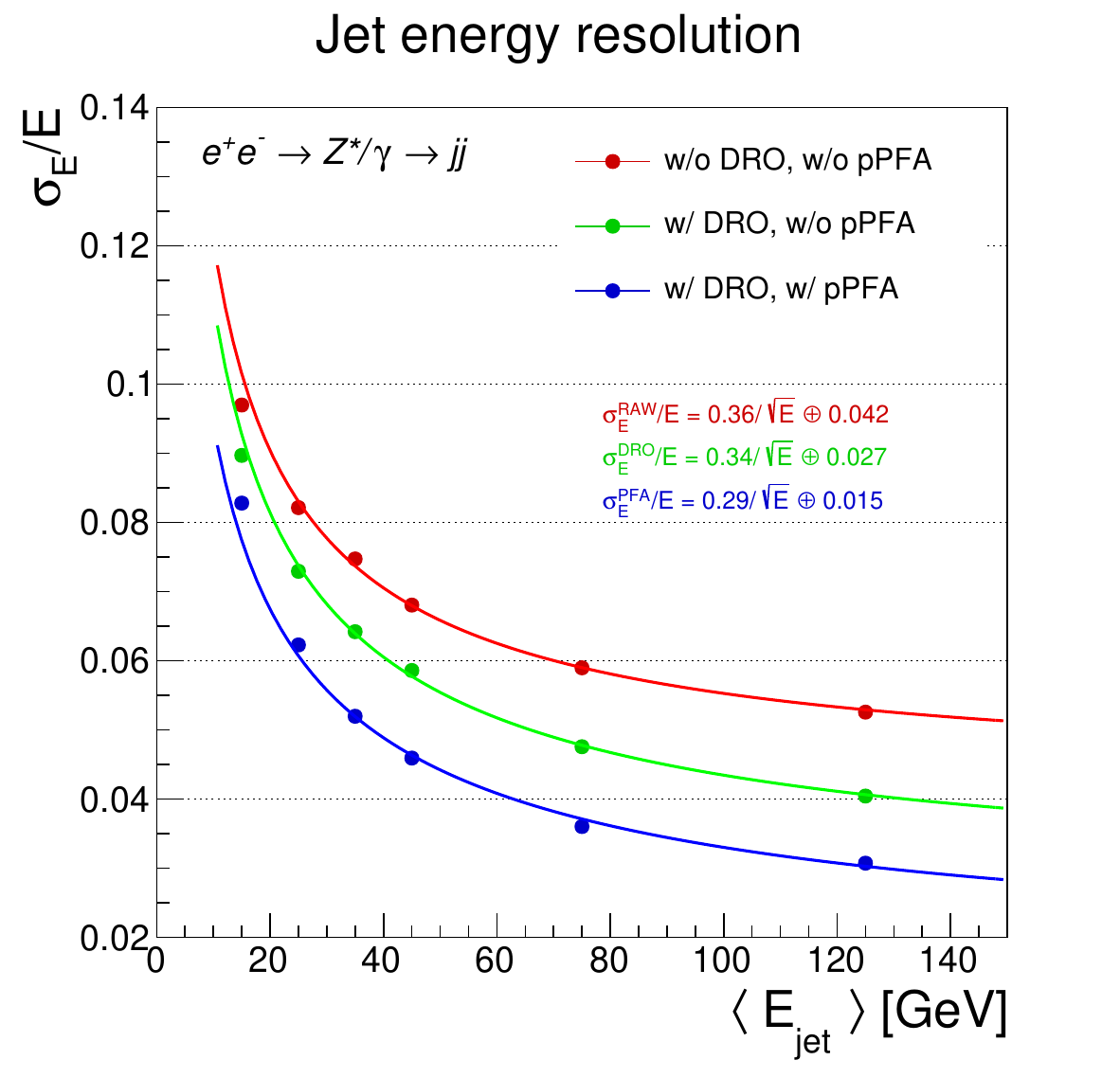}
\caption{Jet energy resolution vs.\ 
jet energy, with or without a dual-readout particle flow algorithm.
}
\label{fig:drpfa_jetres}
\end{minipage}    
\hfill
\begin{minipage}[t]{0.48\textwidth}
\includegraphics[width=\linewidth]{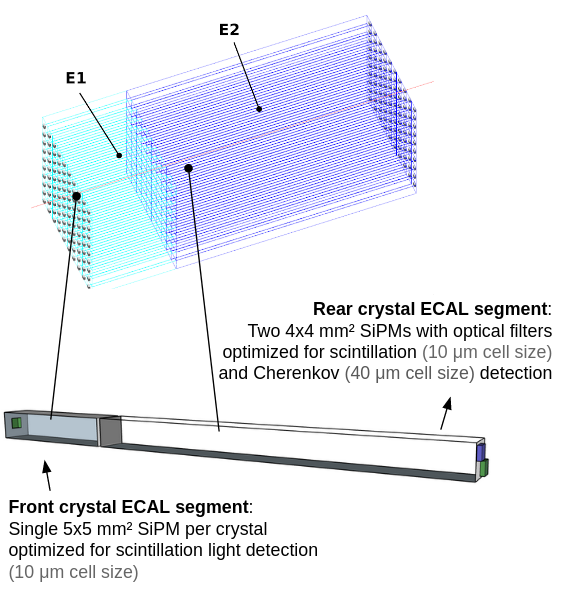}
\caption{Segmented crystal calorimeter prototype cell and module with SiPM dual-readout.}
\label{fig:maxicc}
\end{minipage}
\end{figure}

The MAXICC calorimeter design is optimised for integration with a dual-readout hadron calorimeter section 
to provide an energy resolution for charged and neutral hadrons of about $27\%/\sqrt{E} \oplus 2\%$,
by correcting each shower individually using its estimated electromagnetic fraction, 
simultaneously measured with different techniques, 
thereby reducing the impact of shower fluctuations on the resolution. 
The inclusion of longitudinal segmentation and the enhancement of transverse granularity with $\sim$\,1\,cm$^2$ cells 
(compared to the previous generation of crystal calorimeters) 
provide a powerful handle for particle identification 
and association of charged-particle tracks with calorimeter clusters in particle-flow jet reconstruction, 
with the potential to achieve an energy resolution of 4.5\% for 45\,GeV jets~\cite{Lucchini_2022}, 
as shown in Fig.~\ref{fig:drpfa_jetres}.
A complete simulation of the detector integrated with other IDEA sub-detectors,
within the \textsc{Key4hep} framework, 
is in progress, to allow more detailed physics studies~\cite{WonyongProceeding}.

As shown in Fig.~\ref{fig:maxicc}, 
the baseline calorimeter design features two longitudinal layers of high-density crystals (such as PWO or BGO), 
with 6\,$X_0$ in the front layer and 16\,$X_0$ in the rear layer, 
for a total of 22\,$X_0$ to ensure full containment of electromagnetic showers. 
The light from the crystals is read out with silicon photomultipliers (SiPMs) of active area 
in the range between $4 \times 4$ and $6 \times 6$\,mm$^2$. 
One SiPM is glued on the front side of the front layer and two are glued on the rear face of the rear layer. 
One of the rear SiPMs embeds an optical filter that cuts out the scintillation light based on wavelength discrimination 
and detects a pure Cherenkov signal for dual-readout corrections.  
An alternate approach also uses pulse shape analysis to differentiate scintillation and Cherenkov components.  
The front-end boards, electronics, and cooling services are located in the front and rear parts of the calorimeter 
to minimise the impact of passive material on energy resolution.

The MAXICC calorimeter can provide a time resolution better than 30\,ps 
for electromagnetic showers with energy of 20\,GeV or higher,
as obtained from beam tests with similar calorimeter prototypes~\cite{FERRI2020162159}.
In addition, two thin and highly segmented layers of LYSO:Ce crystals, for a total of 1\,$X_{0}$, 
could be added in front of the PWO crystals for time tagging of minimum ionising particles, 
using a technology similar to that used by the CMS MTD~\cite{CMS_MTD_TDR},
with a time resolution of 20\,ps.

The required technological R\&D, prototyping, and simulation are progressing steadily, 
through a coordinated effort of many international collaborators within the DRD6 WP3 (task~3.1.2). 
Single calorimetric cells have been tested with beam in 2024 and full containment calorimeter prototypes 
(one using PWO and the other using BGO)
are being built for beam tests in 2025. 
Each prototype will instrument $\sim$\,100--200 channels for readout,
covering a transverse size of at least five times the Moli\`ere radius and a depth of $22\,X_0$.  

\subsubsection{The GRAiNITA ECAL}

The GRAiNITA concept features a novel electromagnetic calorimetry design,
with a detection volume filled with millimetric grains of high-$Z$ and high-density inorganic scintillator crystals 
immersed in a bath of transparent high-density liquid. 
The multiple refractions of the light on the grains ensure the stochastic confinement of the light, 
as in the LiquidO detection technique~\cite{LiquidO:2019mxd}. 
The scintillation light can be collected towards the photodetectors by means of WLS fibres, 
regularly distributed in the detection volume as for a conventional Shashlik detector, 
allowing a potentially-large transverse granularity. 
This extremely-fine sampling of the electromagnetic shower promises an excellent energy resolution,
comparable to what is known from crystal calorimeters.

The main high-$Z$ and high-density inorganic scintillator crystal considered for GRAiNITA is ZnWO$_4$~\cite{bib:IEEE2022}, 
providing a light yield of $\sim$\,10\,000 photons per MeV. 
A considerable advantage is that ZnWO$_4$ can be successfully grown in the form of transparent granules of the desired size, 
with the method of spontaneous crystallisation from a flux melt~\cite{bib:IEEE2022}, significantly reducing the cost of the calorimeter. 
Such a high energy-resolution granular calorimeter would then meet the requirements,
from physics and otherwise, of an FCC-ee experiment. 
It would, moreover, be particularly useful to address a comprehensive flavour physics programme 
with rare decays involving photons or neutral pions in the final state. 
Given that the \PZ production rate at FCC-ee is of the order of 100\,kHz
and since these events should typically hit less than 1\% of the calorimeter cells, 
the 20\,$\mu$s signal decay time of the ZnWO$_4$ should not be a concern. 
Nevertheless, the possible use of other types of crystals will be investigated. 

\begin{figure}[ht]
\centering
\includegraphics[width=.45\textwidth]{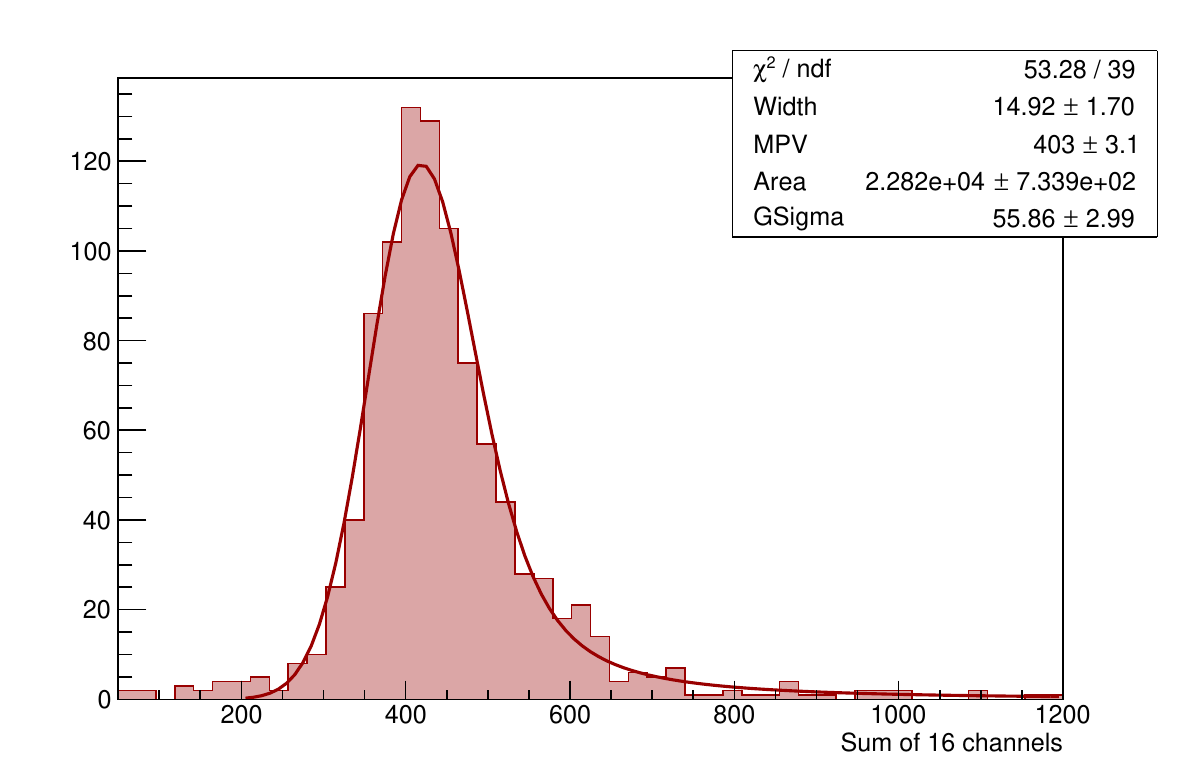}
\includegraphics[width=.45\textwidth]{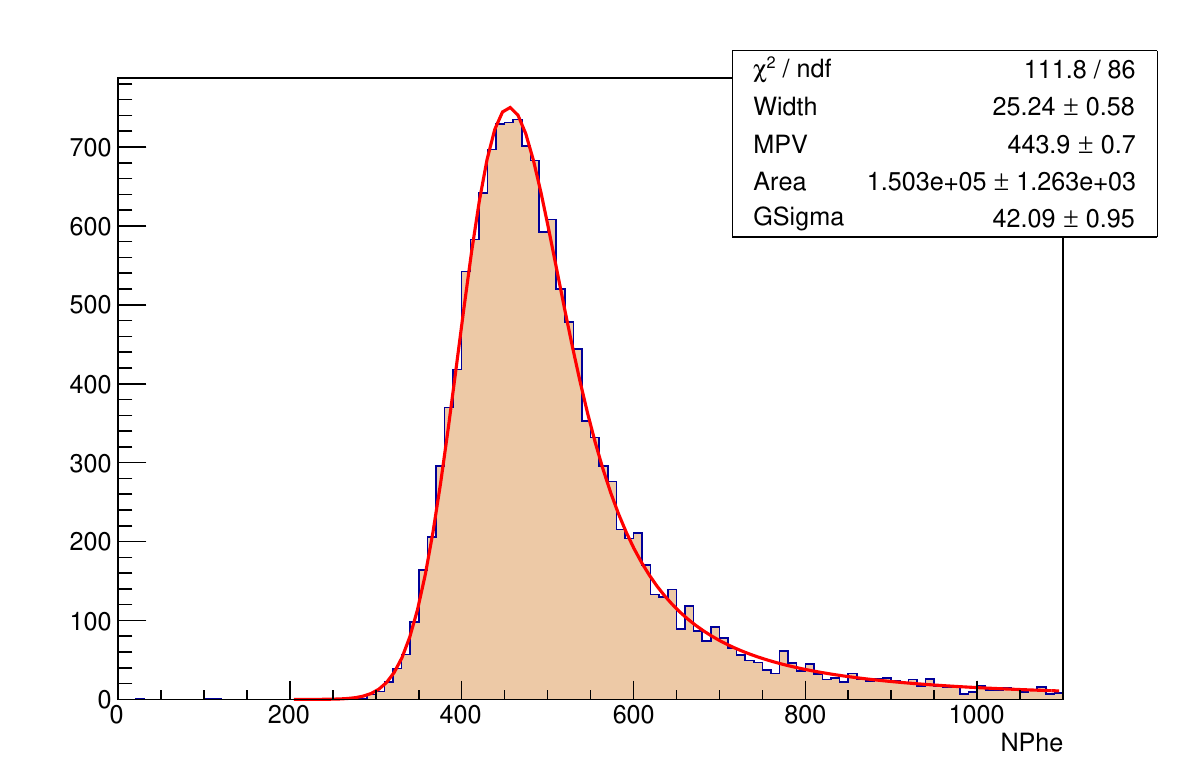}
\caption{Number of photo-electrons recorded in a small GRAiNITA prototype, 
where the ZnWO$_4$ grains are immersed in a heavy liquid to increase the density of the medium. 
Left: Cosmic-ray muons with ZnWO$_4$ grains immersed in ethylene-glycol (from Ref.~\cite{Barsuk_2024}). 
Right: Muon tracks from a beam test at CERN with ZnWO$_4$ grains 
immersed in LST~Fastloat (density = 2.8\,g/cm$^3$)~\cite{LST_Fastloat}.}
\label{fig:NPHE}
\end{figure}

Measurements with a small prototype ($2.8 \times 2.8 \times 5.5$\,cm$^3$) equipped with a green LED confirm that, as expected, 
the signal remains confined in the vicinity of its production point~\cite{Barsuk_2024}. 
From the analysis of cosmic-ray muons~\cite{Barsuk_2024} and of data from a muon beam test performed in summer 2024 at CERN with the same small prototype, 
it is clear that the collection of about 10\,000 photo-electrons per GeV is within reach with a GRAiNITA-like calorimeter. 
This result paves the way for a statistical fluctuation of $1\%/\sqrt{E}$ on the energy resolution from photo-electrons (Fig.~\ref{fig:NPHE}). 
The next steps, following this proof-of-principle, include the system aspects, including a mechanical configuration and an electronics read-out concept.

\subsection{Hadron calorimeters}

\subsubsection{Scintillator tiles}

\begin{figure}[h]
\centering
\includegraphics[width=0.5\linewidth]{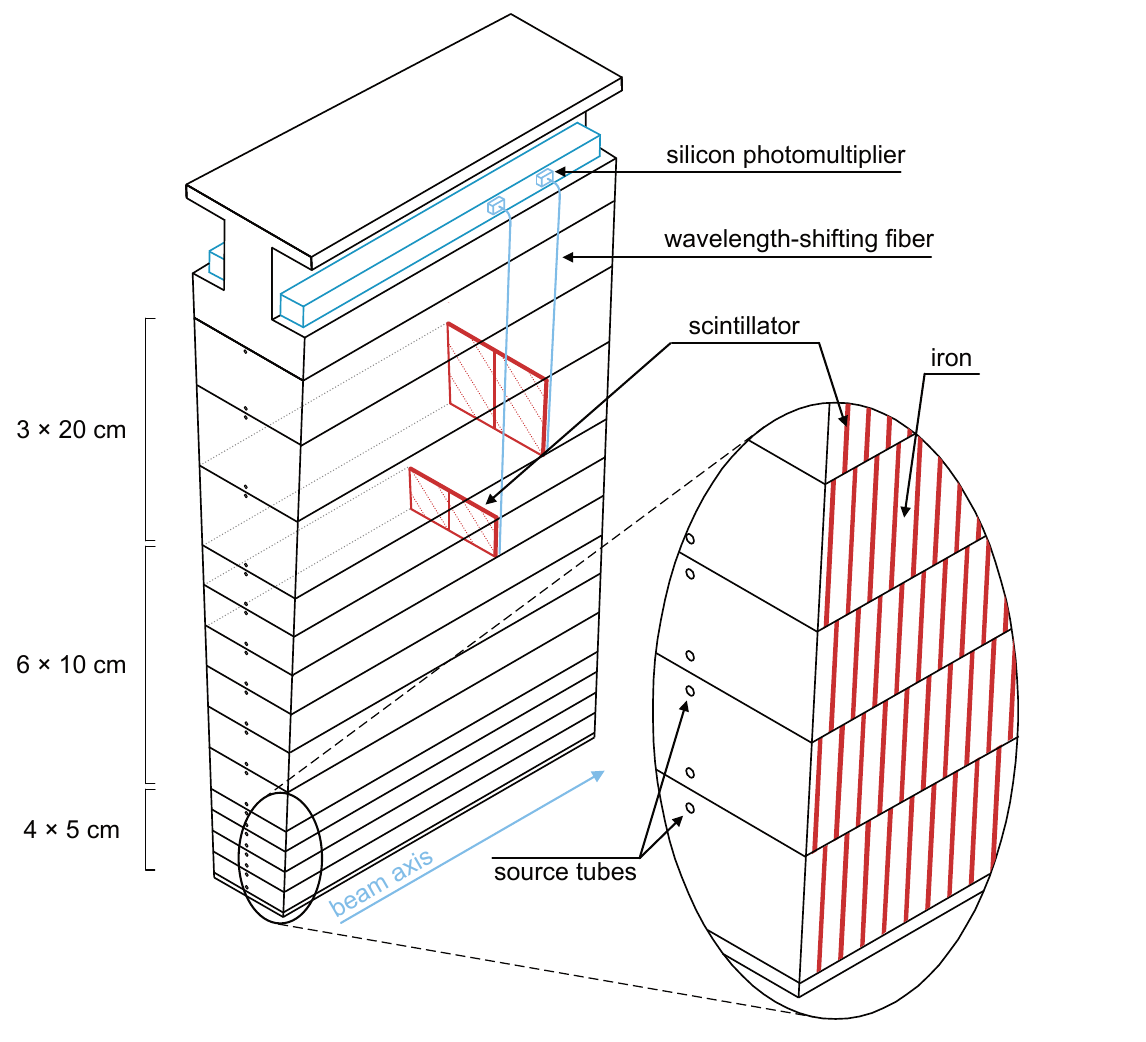}
\caption{HCAL tile module concept. 
Each module consists of 13 radial layers of scintillating tiles with different radial extension (5, 10, and 20\,cm), 
and each radial layer is divided into two tiles in azimuth (hatched red). 
Scintillation light is brought to SiPMs at the outer radius via WLS fibres. 
In the $z$ direction, tiles are read out in groups of 3 or~4.}
\label{fig:HCAL-Tile-module}
\end{figure}

One of the considered hadron calorimeter designs is a non-compensating sampling calorimeter based on the ATLAS TileCal~\cite{ATLAS:2024bxb}, 
with steel as absorber and plastic scintillating tiles as active material. 
The light from the scintillator is collected and transported to the SiPMs 
located outside the calorimeter via WLS fibres, as shown in Fig.~\ref{fig:HCAL-Tile-module}. 

The radial orientation of the tiles and the light collection via the WLS fibres along the tile edges allow a hermetic implementation in $\phi$.
The location of the photodetectors and their associated electronics in the modules' girder at the outside radius 
makes them serviceable during collider shutdown periods. 
It also embeds the calibration system with movable radioactive $^{137}$Cs photon sources that can pass through all calorimeter cells~\cite{Blanchot:2020lyh}.

The proposed design uses 5\,mm low-carbon steel absorber plates interleaved with 3\,mm plastic scintillating tiles. 
The barrel is segmented into 128 modules in $\phi$ and 13 radial layers 
(of different depths: four of 5\,cm, six of 10\,cm, and three of 20\,cm, as shown in Fig.~\ref{fig:HCAL-Tile-module}). 
Each layer is divided in $\phi$ in two longitudinal read-out compartments. 
In each compartment, the tiles are read out in groups of three or four along the $z$ axis, 
which leads to a cell granularity $\Delta\phi \times \Delta\theta$ of $24.5 \times 22$\,mrad$^2$ at normal incidence. 
Steel-scintillator technology is also being considered for the end-cap hadron calorimeter. 

\begin{figure}[ht]
\centering
\includegraphics[width=0.5\linewidth]{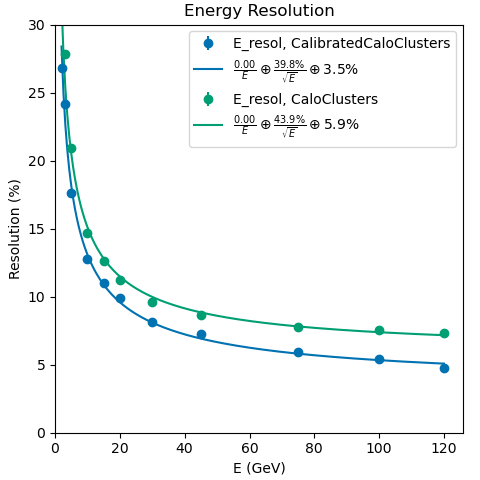}
\caption{HCAL tile energy resolution for hadron clusters.}
\label{fig:HCAL-Tile-resolution}
\end{figure}

The performance of the tile HCAL, evaluated with a sliding window clustering algorithm, is shown in Fig.~\ref{fig:HCAL-Tile-resolution}. 
The optimisation of the calorimeter details, including the choice of the absorber and scintillating materials, as well as their segmentation, is ongoing.  
These R\&D studies are coordinated by the DRD6 WP3 (task~3.3.2), including the characterisation of perspective scintillating materials, 
like PEN and PET~\cite{CondeMuino:2023uzm}, and the light collection from scintillating tiles via WLS fibres with SiPM photodetectors~\cite{Schliwinski:2687718}.

A complementary approach is also being explored, 
with the so-called SiPM-on-Tile technology and with front-end electronics embedded into the active layers of a steel sandwich structure. 
The SiPMs are integrated into the electronics PCBs, which also hold read-out ASICs, power distribution, and a LED-based calibration system. 
Reflector-wrapped scintillator tiles are glued onto the PCB, with a dimple leaving space for the SiPMs underneath, which read each tile individually.

While the ATLAS-style geometry offers easier access and space for service routing, 
including cooling for a calorimeter placed outside the coil, 
the SiPM-on-Tile approach is the natural choice for calorimeter systems entirely inside the solenoid, such as in the CLD concept, 
where radial compactness (especially in the barrel) is strictly mandated by cost considerations. 
The technology was originally developed by the CALICE Collaboration;
a scalable prototype with 22\,000 channels was built and has been successfully tested in beams since 2018~\cite{CALICE:2022uwn}.
The main challenge in the adaptation to FCC-ee lies in coping with the continuous readout 
(i.e., without power-pulsing) 
and with the much higher rate, bandwidth, power, and cooling requirements, 
without compromising granularity and compactness. 

The R\&D  has common challenges with other concepts based on embedded electronics 
and is pursued in the framework of the WP1 of the DRD6 Collaboration. 
The SiPM-on-Tile technology is currently being applied in a large scale in the CMS HGCAL, a detector with 280\,000 tiles.
Although the CMS HGCAL has to address radiation hardness and other challenges not present at FCC-ee, 
it has more relaxed performance and compactness requirements, 
and remains more than an order of magnitude smaller in terms of channel count than, e.g., the CLD HCAL.

\subsubsection{Gaseous detectors}

Gaseous detector technologies with two-dimensional transverse segmentation are also pursued as read-out options for hadron calorimeters. 
Proof-of-principle elements have been tested using GEM and micromegas detectors, 
and larger prototypes with glass RPCs have been built and successfully tested, 
with up to 500\,000 channels~\cite{Baulieu:2015pfa,Adams:2016por}. 
Granularities finer than with tiles are easily achievable, 
but to maintain the designs cost-effective and to limit read-out power consumption, 
trade-offs must be found between energy and space information. 
The channel amplitude is encoded in one or two bits only, hence the name digital or semi-digital HCAL.

The related R\&D is pursued both in the DRD6 (`calorimeters') and in the DRD1 (`gas detectors') projects. 
The challenges of data concentration, powering, and cooling are shared with other ECAL and HCAL technologies with embedded readout. 
The quest for improving the detection technologies themselves, in terms of rate capability and stability, 
is common with R\&D for muon detectors, 
and both face the imperative of replacing the currently used gas mixtures with high global warming factors 
by more eco-friendly solutions. 

\subsubsection{Dual readout}

A fibre-sampling Dual-Readout (DR) calorimeter~\cite{dualreadoutfibrecalo} can, in principle, 
be used as a single system for electromagnetic and hadron showers. 
This option is considered in the IDEA concept, 
with a calorimetric section made of a dual-readout, longitudinally unsegmented and fully projective, fibre-sampling calorimeter, 
providing both electromagnetic and hadron shower measurements.
Another option is to use this technology as a hadron calorimeter behind an electromagnetic section based on crystals, 
as discussed in Section~\ref{sec:Segmented-crystals}.

Alternate rows of scintillating and Cherenkov (clear) fibres are inserted in capillary tubes made of brass or iron. 
Each fibre is individually read out with a SiPM. 
In order to significantly reduce the number of readout channels, the analogue grouping of 4--8 SiPM signals is foreseen. 
The optimal granularity is different if the detector has to provide both electromagnetic and hadronic shower measurements or just the latter. 
The present choice (1\,mm fibres in 2\,mm capillary tubes) is designed to address both.

\begin{figure}[ht]
\centering
\includegraphics[width=0.5\textwidth]{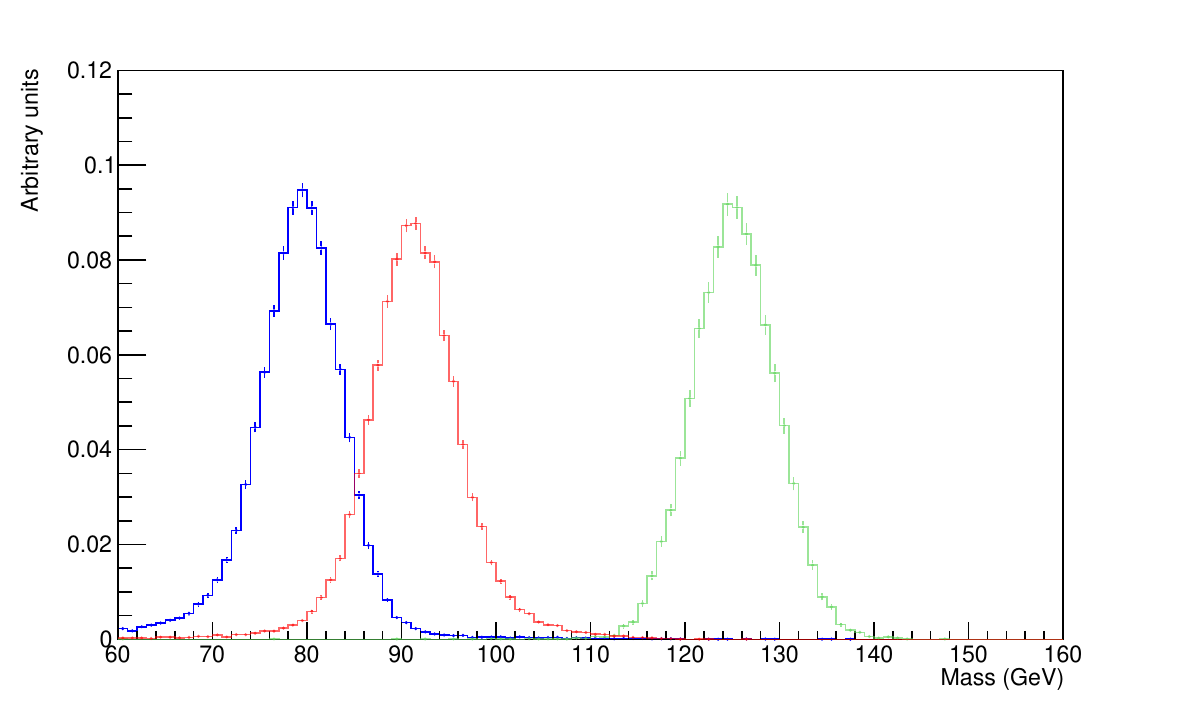}
\caption{From Ref.~\cite{Pezzotti:2021jxh}. Invariant mass distributions of \PW (blue), \PZ (red), and \PH (green) hadronic decays, from $\epem \to \PWp\PWm$ and $\PZ\PH$ events
fully simulated  and reconstructed at $\sqrt{s} = 240$\,GeV in a dual-readout fibre-sampling calorimeter.
To highlight the detector performance, semileptonic \PQb decays are excluded. 
}
\label{fig:DR-WZH2jj}
\end{figure}

The hadron energy resolution is estimated to be about $30\%/\sqrt{E}$, with a negligible constant term. 
The expected separation reachable for the \PW, \PZ and \PH hadronic decays is shown in Fig.~\ref{fig:DR-WZH2jj},
obtained with a stand-alone simulation and subsequent reconstruction of the DR calorimeter, 
and excluding semi-leptonic decays~\cite{Pezzotti:2021jxh}.
Similar studies show that this detector can provide excellent stand-alone electron-pion separation.
Moreover, the RD52 lead prototype obtained an efficiency of about 99\% in identifying electron showers, 
for a rejection ratio of charged-pion showers higher than 500~\cite{Akchurin:2014zna}.

Advances in solid-state light sensors, such as SiPMs, have opened the way for high transverse granularity,
which gives the detector the capability to determine shower angular positions at the mrad level, or better. 
In the present design, without a crystal-based electromagnetic calorimeter in front, 
the 1\,mm diameter fibres are placed in an iron absorber matrix, at a distance (apex to apex) of 1.5--2 mm. 
The lateral segmentation could be pushed down to the mm level, 
largely enhancing the resolving power for close-by showers, with a significant impact expected, for example, 
on the isolation and reconstruction of $\PGt \to \PGr \PGn$ final states. 
For the setup with crystals in front, the segmentation of the DR sections remains to be optimised.

The high photo-detection efficiency of SiPMs should lead to light yields of $\mathcal{O}(100)$ photo-electrons per GeV, 
for both the scintillation and Cherenkov signals, 
which guarantees a stand-alone EM resolution close to $10\%/\sqrt{E}$. 
Readout ASICs providing time information with about 100\,ps resolution 
may allow the reconstruction of the longitudinal shower position with a resolution of about 5\,cm.

On the other hand, the large number and density of channels require an innovative readout architecture for efficient information extraction. 
Both charge integration and waveform sampling ASICs are available on the market and candidates for early testing have been identified.
A first implementation of a scalable readout system is well-advanced.
Looking further ahead, digital SiPMs (dSiPMs) should allow a significant simplification of the readout architecture, 
but the technology is not yet sufficiently mature. 
A specific R\&D programme has been approved for the 2025--2026 period.

The mechanical assembly and integration of a system with $\mathcal{O}(10^8)$ sensitive elements 
require the development of a robust and engineered procedure. 
A scalable mechanical solution has been conceived for both non-projective and projective modules. 
A small ($\sim$\,$10 \times 10 \times 100$\,cm$^3$) EM prototype has been built,
based on the gluing of capillary tubes.
This procedure is being used for the construction of a hadron calorimeter prototype 
(of size $\sim$\,$60 \times 60 \times 250$\,cm$^3$) that should be completed by mid 2025. 
Alternative approaches, for both mechanical assembly and readout architecture, are being investigated within DRD6.

\begin{figure}[h]
\centering
\includegraphics[width=0.45\textwidth]{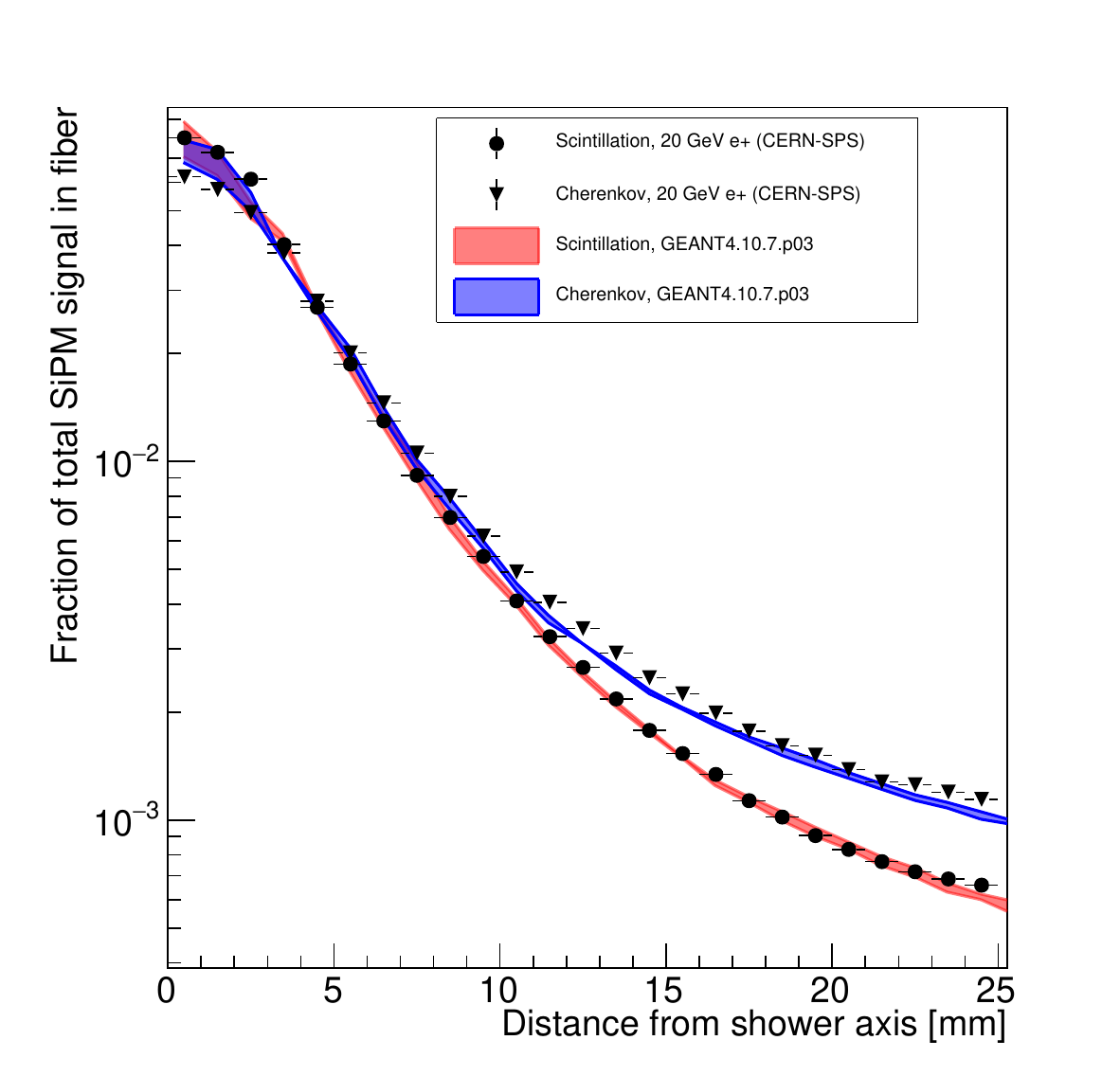}
\caption{Lateral shower profile of positrons with an energy of 20\,GeV. 
The black circles (triangles) refer to the scintillation (Cherenkov) options. 
The red and blue bands correspond to the simulation predictions for the scintillation and Cherenkov signal, respectively.}
\label{fig:emShowerProfile}
\end{figure}

The EM prototype was tested with beam at the CERN SPS and the results (Fig.~\ref{fig:emShowerProfile}) show that simulations model the instrumental effects well.
The performance in the reconstruction of the properties of both hadron and EM showers
is sufficient to retain the possibility of an integrated dual-readout solution based on ﬁbre-sampling only, 
for the calorimetric system of an FCC-ee experiment. 
The huge amount of information made available by the ﬁbre SiPM readout is well suited to take advantage of deep-learning algorithms. 

\subsection{Coil}

\begin{figure}[ht]
\centering
\includegraphics[width=0.47\linewidth]{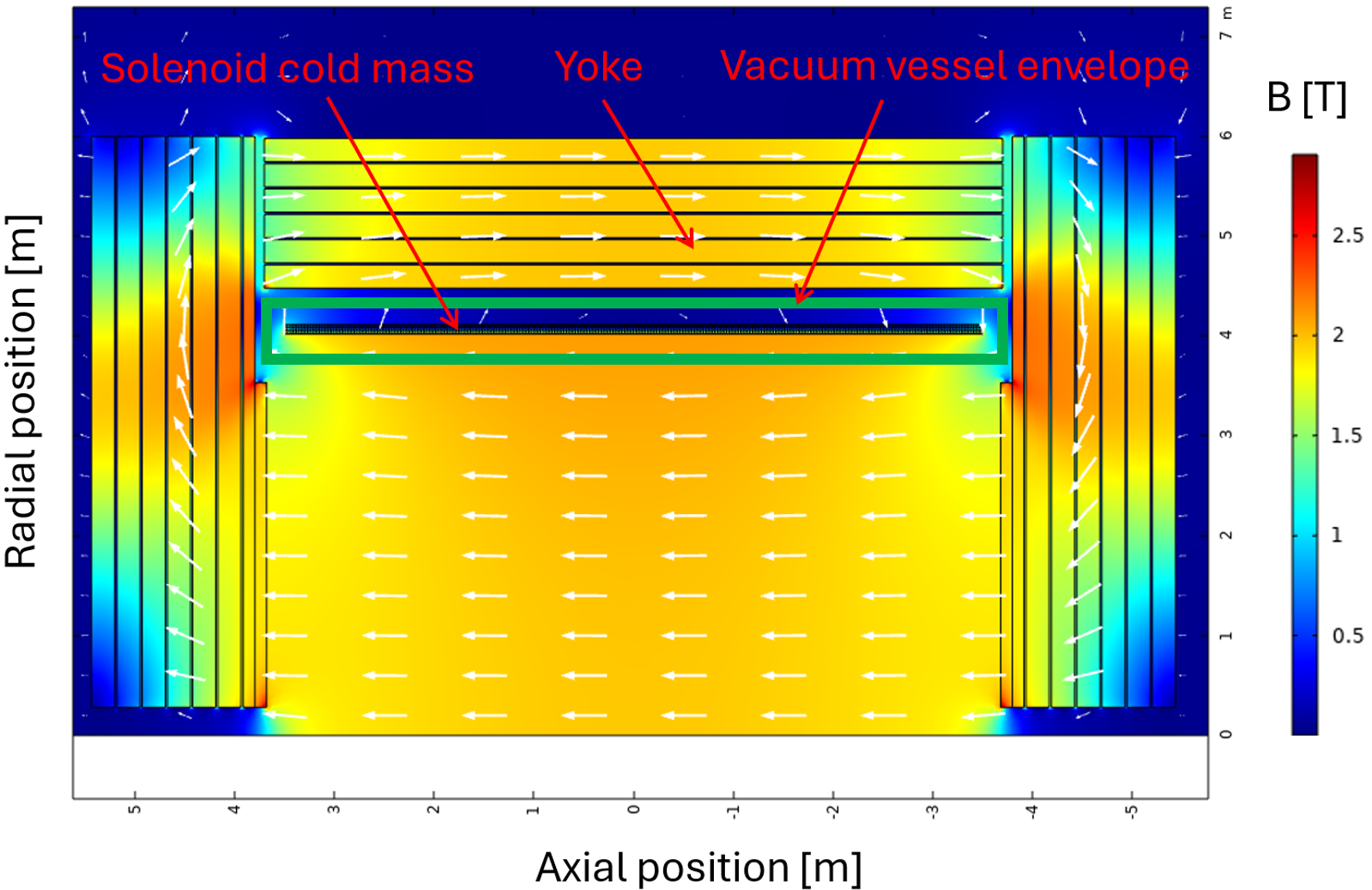}
\includegraphics[width=0.52\linewidth]{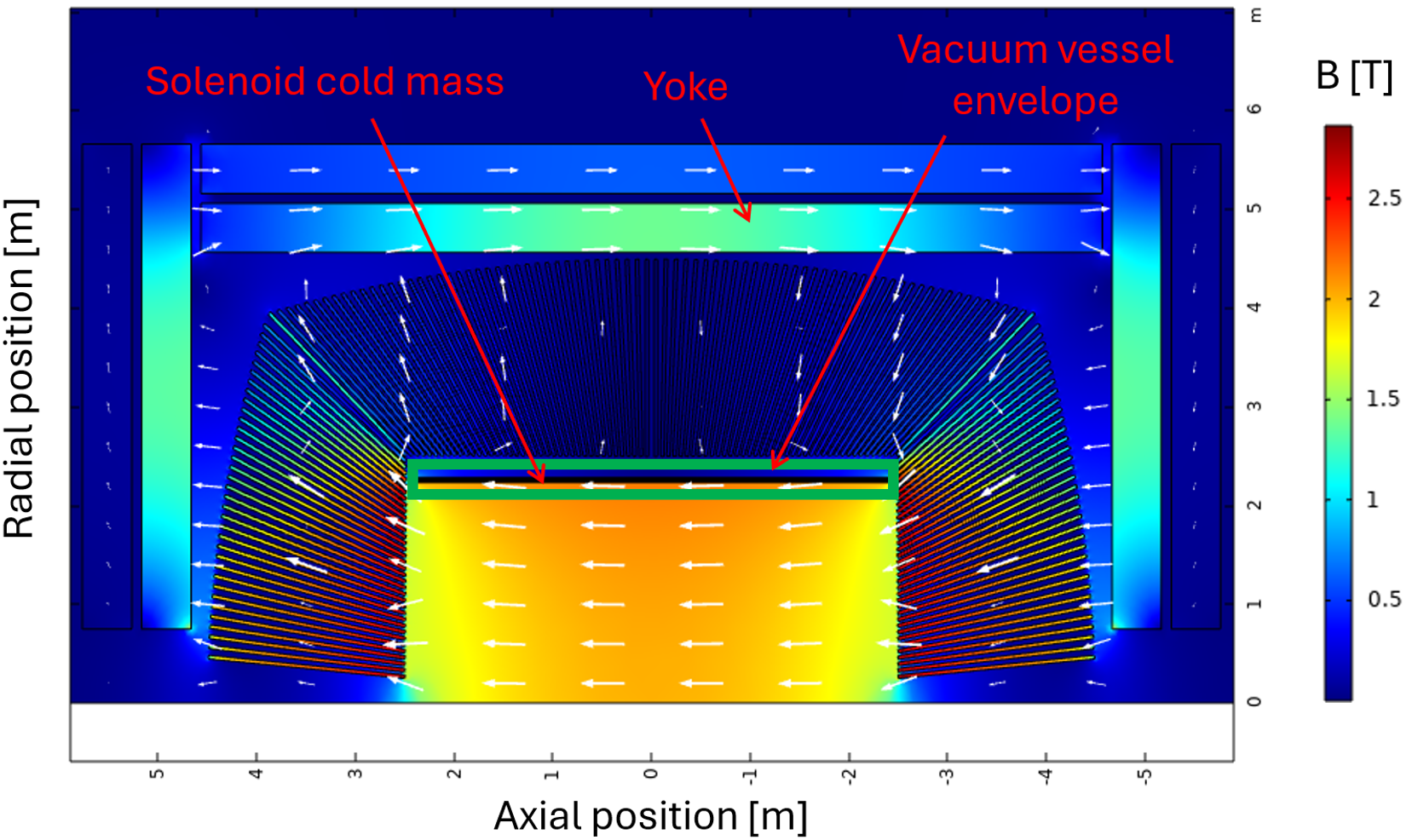}
\caption{Field maps of the CLD (left) and IDEA (right) detector magnets in the ($r,z$) view, featuring 2\,T in the free bore.}
\label{fig:CLD+IDEASolenoid}
\end{figure}

Conceptual studies of 2\,T superconducting solenoid variants have been concluded for the CLD, IDEA, and ALLEGRO FCC-ee detector concepts. 
From a conceptual perspective, the CLD solenoid is similar to that of CMS, 
with the calorimeters located inside the solenoid bore. 
A relatively high energy density is targeted, for the purpose of limiting the weight of the cold mass, with various benefits, 
in particular regarding the overall cost and easier transport from the manufacturer to CERN. 
For the IDEA and ALLEGRO concepts, the solenoid is located inside the hadron calorimeter and the free-bore diameter is substantially smaller. 
The advantage of this variant is that the cold-mass weight is reduced, 
albeit at the cost of requiring high particle transparency of the cold mass, thermal shields, and vacuum vessel. 
The design of thin, low-mass magnet systems is discussed in Ref.~\cite{fcc-ee-cdr}. 
Representative field maps are shown in Fig.~\ref{fig:CLD+IDEASolenoid} 
and typical design parameters are shown in Table \ref{tab:solenoidVariants}.

\begin{table}[ht] 
\caption{Overview of typical design parameters of the different solenoid variants.}
\label{tab:solenoidVariants}
\centering
\begin{tabular}{ c c c c}
\toprule
Variant & Stored magnetic & Cold mass & Energy density \\
        & energy (MJ) & weight (t) & (kJ/kg) \\
\midrule
CLD & 590 & 52 & 11.4 \\\
IDEA & 130 & 10.6 & 12.3 \\
Allegro & 250 & 22 & 11.4 \\
\bottomrule
\end{tabular}
\end{table}

For each variant, a relatively high energy density, of about 12\,kJ/kg, is targeted, 
a value comparable to what was previously shown in the CMS solenoid and the Bess--Polar solenoid~\cite{Yamamoto:2002dwn}. 
This energy density poses challenges for quench protection and cold-mass mechanics but, 
relying on technological solutions previously demonstrated for the ATLAS, CMS, and Bess--Polar solenoids, 
no setbacks were identified. 
The baseline is to build the solenoids with reinforced aluminium-stabilised low-temperature superconductors. 
These conductors have been successfully applied in many detector magnet projects worldwide over the past decades 
and they meet the technical requirements of the FCC detector magnets, 
regarding the field ranges, the free bores, the transparency to particles, and the mechanics and stability. 
The preferred technology is co-extrusion of Nb-Ti/Cu Rutherford cable in a high purity aluminium stabiliser. 
This technology offers the best characteristics for coil quench protection and coil stability, with an indirect cooling. 
Strong mechanical properties can be reached using either Ni-doped high-purity aluminium stabiliser, 
or aluminium alloy profiles electron-beam-welded to the coextruded aluminium-stabilised superconductor. 
It should be noted that this assumes the commercial availability of reinforced aluminium-stabilised Nb-Ti conductor. 
At the time of the conclusion of the conceptual cold-mass studies, 
it became clear that companies that had historically made this conductor type commercially available had discontinued production.

As an alternative to aluminium-stabilised Nb-Ti conductor technology, preliminary studies were made 
concerning the implications of aluminium-stabilised high-temperature-superconducting (HTS) conductors. 
The use of HTS brings interesting benefits, 
such as potentially allowing a higher operating temperature and, thus, reduced cryogenic operating costs.
The implications for capital costs, cold mass mechanics, quench detection and protection are, however, 
not yet fully understood and characterised. 
This technology requires further R\&D before it can be concluded that it is a reliable alternative to the Nb-Ti conductor.
Two manufacturing methods are considered for the aluminium-stabilised HTS: 
co-extrusion in high-purity aluminium profiles and soft soldering in copper-plated aluminium profiles. 
As the HTS are produced in thin tapes, several HTS conductor technologies are envisaged: 
stacks of tapes, Roebel cables, and conductor on round core (CORC\textsuperscript{\textregistered}) conductors.

An Experimental Magnets Committee was established by CERN in 2023 to address the R\&D needs of detector magnet projects 
and to resolve identified issues regarding the availability of technologies and facilities needed 
to manufacture superconductors for detector magnets. 
An ongoing programme is targeting to establish access to a co-extrusion line that would allow R\&D and prototyping, 
both with low- and high-temperature aluminium-stabilised superconductors. 
In addition to the CERN effort, specific R\&D for an HTS solenoid for IDEA/ALLEGRO is currently in progress at INFN-LASA. 

\subsection{Cryostat}
\label{sec:Cryostat}

\begin{figure}[ht]
\centering
\includegraphics[width=0.95\linewidth]{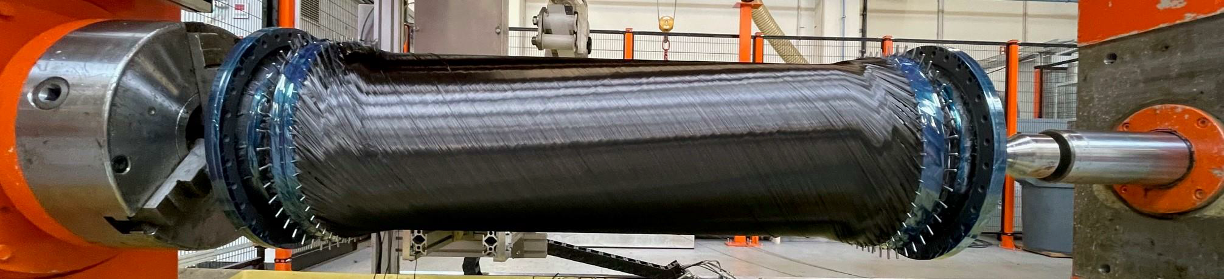}
\caption{Full liner-less tank demonstrator, 1\,m length, 0.3\,m diameter, and 5\,mm wall thickness. 
Manufacturing process: filament wet-winding, non-crossing, out-of-autoclave curing.}
\label{fig:CarbonCryostatPrototype}
\end{figure}

The use of carbon composite technologies for cryostat construction is being studied in an R\&D programme in the CERN EP Department. 
In collaboration with industry, a 1\,m carbon composite cryostat prototype has recently been developed, 
as shown in Fig.~\ref{fig:CarbonCryostatPrototype}. 
The manufacturing procedure, based on a wet filament winding process, 
ensures helium leak-tightness at cryogenic temperatures without the need for a metallic liner.

Recent developments to support large-scale production (larger than 5\,m in diameter and length), 
include the refinement of the winding process and the development of a toughened epoxy resin to enhance resistance to microcracking.
Additionally, a novel technique has been developed to wind each ply of material without fibre crossings, 
minimising porosity and further improving microcrack resistance. 
Promising helium leak tests at low temperatures indicate that strict cryogenic standards can be met for detector operations. 

The ability of cryostat vessels to resist structural instability (buckling) under compressive loads poses an additional challenge. 
The use of a full carbon honeycomb as a replacement for aluminium honeycomb is being explored. 
Initial prototypes have undergone extensive testing, including thermal expansion and dimensional stability assessments, 
underscoring the suitability of carbon honeycomb for large, buckling-sensitive cryostat walls. 

The combination of a leak-tight carbon wall with the full carbon honeycomb core 
represents a promising design direction for next-generation HEP cryostats. 
A preliminary study, using the ALLEGRO detector concept as an example, revealed 
a 64\% reduction in the material budget and a 30\% reduction in wall thickness for buckling-sensitive cryostat walls 
compared to an aluminium sandwich design, 
with a thermal expansion coefficient one order of magnitude lower while maintaining comparable mechanical properties.

\subsection{Muon system}

Several technologies are under consideration for use in detector muon systems at FCC-ee, 
including $\mu$-RWELL~\cite{MUON:micro-RWELL}, resistive plate chambers~\cite{SANTONICO1981377}, 
resistive Micromegas~\cite{ALEXOPOULOS2011110}, scintillators, drift tubes, and combinations thereof. 
Apart from muon identification, these systems can offer capabilities for tail catching of calorimeter showers, 
identification of long-lived particles by providing tracking with good position resolution, or independent triggers.

Three $\mu$-RWELL layers are currently implemented in the IDEA design, in both the barrel and end-cap regions, 
housed within the iron yoke that closes the detector magnetic field. 
The $\mu$-RWELL chambers provide 2D space point measurements with a few hundred micron precision. 
In order to benefit from industrial production capabilities of this technology, 
a modular design is adopted with basic $\mu$-RWELL `tiles' with an active area of $50 \times 50$\,cm$^2$
and a readout strip pitch of about 1\,mm, 
enabling a sufficient spatial resolution of about 400\,$\mu$m while limiting the number of readout channels to 1000 per tile.
The choices of detector tile size, strip pitch and width are a compromise between 
the largest $\mu$-RWELL detector that can be industrially mass-produced 
and the maximum input detector capacitance that can be tolerated in terms of signal over noise ratio by the front-end electronics.

Table~\ref{tab:barrel-muon} lists the dimensions, the number of basic $\mu$-RWELL tiles, 
and the readout channels of the three muon stations of the IDEA detector. 
In total, including the barrel and the end-cap muon stations, 
the system comprises about 5800 $\mu$-RWELL tiles with about 6 million readout channels. 

\begin{table}[ht]
\centering
\caption{Dimensions of the three IDEA barrel muon stations, 
together with the number of detector tiles and the corresponding number of readout channels.}
\label{tab:barrel-muon}
\begin{tabular}{@{}cccccccc@{}}
\toprule
Station &
  \begin{tabular}[c]{@{}c@{}}Radius\\ {(}m{)}\end{tabular} &
  \begin{tabular}[c]{@{}c@{}}Length\\ {(}m{)}\end{tabular} &
  \begin{tabular}[c]{@{}c@{}}Strip pitch\\ {(}mm{)}\end{tabular} &
  \begin{tabular}[c]{@{}c@{}}Strip length\\ {(}mm{)}\end{tabular} &
  \begin{tabular}[c]{@{}c@{}}Area\\ {(}m$^2${)}\end{tabular} &
  \begin{tabular}[c]{@{}c@{}}\# of\\ tiles\end{tabular} &
Channels \\ 
\midrule
1 & 4.52 & 9.0   & 1 & 500 & 260 & 1040 & 1\,040\,000 \\
2 & 4.88 & 9.0   & 1 & 500 & 280 & 1120 & 1\,120\,000 \\
3 & 5.24 & 10.52 & 1 & 500 & 350 & 1400 & 1\,400\,000 \\ 
\bottomrule
\end{tabular}
\end{table}

The geometries and material of both the barrel and the end-caps have been implemented in the \textsc{Geant4} full simulation of the muon system. 
The implementation of algorithms for digitisation and clustering will follow in the next step of the study. 
Since the $\mu$-RWELL technology has not yet been used to build a full detector system, 
a vigorous R\&D programme to study integration issues will be carried out in the coming years. 
The detailed layout of the detector, together with all its services, will have to be accurately developed. 
The optimisation of the basic $\mu$-RWELL tile, including gas gap, diamond like carbon layer resistivity, and gas amplification, 
has started~\cite{MUON:test-beam,MUON:resistivity} and is expected to be finalised to match the requirements of the IDEA muon system by 2027. 
The $\mu$-RWELL R\&D is performed in synergy with the WP1 of DRD1~\cite{MUON:DRD1}. 
Another important aspect of this R\&D programme is the design and development of dedicated front-end electronics based on a custom-made ASIC.

\subsection{Luminosity measurement}
\label{sec:lumi}

For the precise measurement of cross sections, the integrated luminosity must be determined with high precision. 
Ambitious goals have been formulated on the measurement precision: 
$10^{-4}$ or better on the \emph{absolute} luminosity 
and $5 \times 10^{-5}$ or better on the \emph{relative} luminosity between energy scan points. 
At FCC-ee, the wide-angle diphoton process, $\epem \to \PGg\PGg$, is statistically relevant 
and provides a promising complement to the traditional low-angle Bhabha scattering process, $\epem \to \epem$.
For both processes, the definition of the geometrical precision constitutes an important source of systematic uncertainty.

As discussed in Section~\ref{sec:PhysPerf_gammagamma}, 
the $\PGg\PGg$ final state places a requirement on the accuracy of the lower limit of the polar angle acceptance at the 8\,$\mu$rad level, 
corresponding to 20\,$\mu$m at 2.5\,m from the IP. 
For each of the proposed ECAL solutions, R\&D is needed on how to obtain such construction precision. 
Complementary methods to determine the acceptance in-situ with this precision or better 
are under development and alluded to in Section~\ref{sec:PhysPerf_gammagamma}.
Likewise, studies are needed on how to ensure that the efficiency of observing the $\PGg\PGg$ final state 
can be understood to the required precision over the wide detector acceptance.

For the measurement of low-angle Bhabha scattering, 
the luminosity calorimeters (LumiCals) are located close to the beam pipe, 
where they are neighbouring the complex Machine Detector Interface region (Chapter~\ref{sec:mdi}). 
Placed in front of the compensating solenoids, the LumiCals are at only $\sim$\,1.1\,m from the IP. 
In this region, space is severely limited and a very compact design is called for. 
The Bhabha cross section falls very steeply, as $1/\theta^3$, with $\theta$ being the scattering angle, 
resulting in very challenging tolerances on the definition of the geometrical precision at the $\mathcal{O}$(1\,$\mu$m) level.

\subsubsection{LumiCal design}
\label{sec:LumiCal}

\begin{figure}[ht]
\centering
\includegraphics[width=0.8\textwidth]{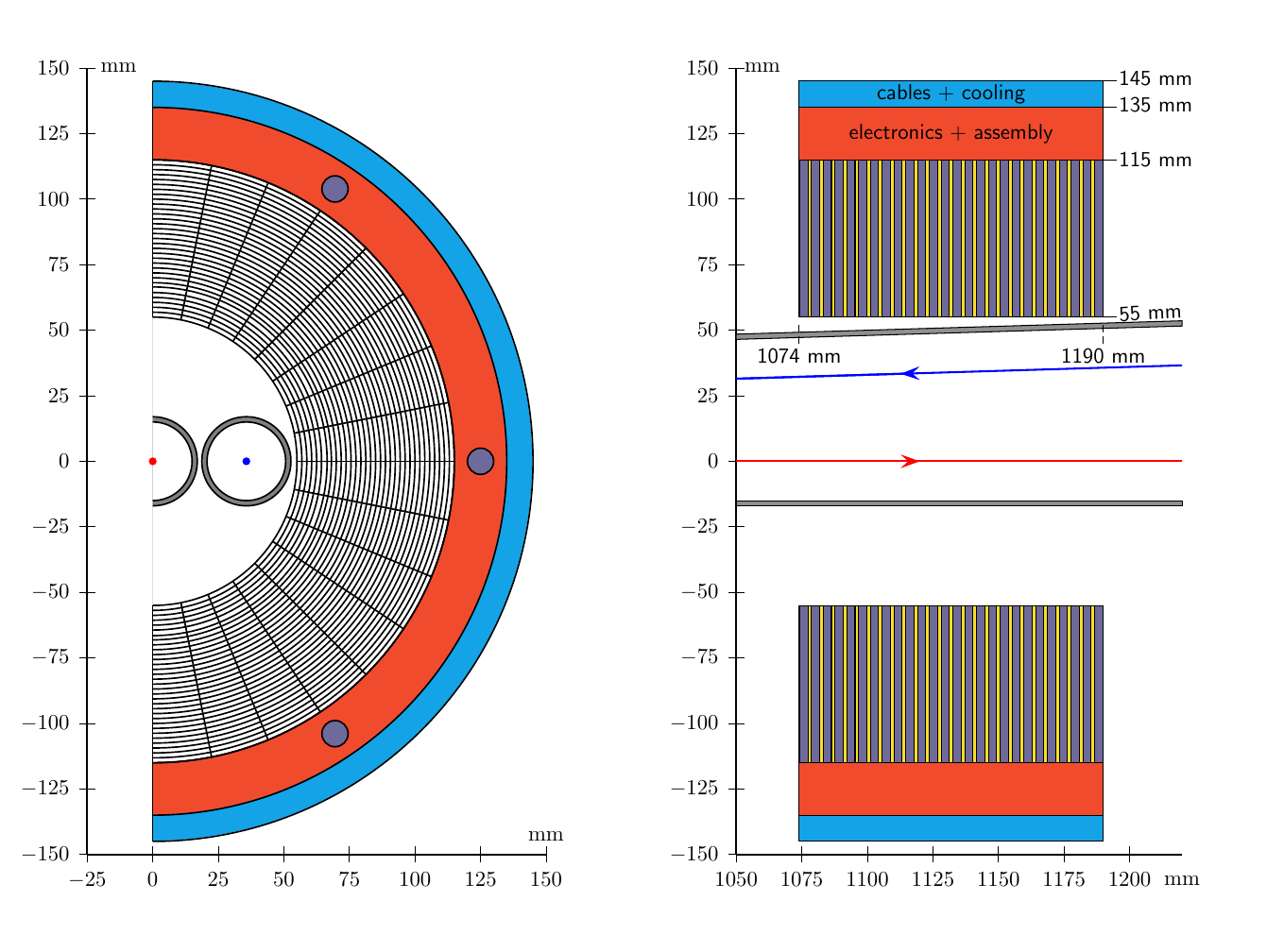}
\caption{Front view (left) and top view (right) of the luminosity calorimeter, 
centred around the outgoing beam direction (shown as a red arrow). 
A possible  segmentation of the Si pads is seen in the front view. 
Surrounding the sensitive region are regions for front-end electronics (red), and for cables and cooling (blue). 
The top view shows the interleaved silicon-tungsten layer structure.}
\label{fig:luminometer}
\end{figure}

Based on experience from LEP~\cite{OPAL:1999clt,Bederede:1995pc}
and from past linear collider studies \cite{Behnke:2013lya,CLIC:2016zwp,Abramowicz:2018vwb}, 
compact SiW sandwich calorimeters are proposed as luminosity monitors for the measurement of low-angle Bhabha scattering. 
The calorimeters are designed as cylindrical devices assembled from stacks of identical SiW layers. 
This simple geometry facilitates the control of construction and metrology tolerances to the necessary micron level, 
as emphasised by the OPAL experiment~\cite{OPAL:1999clt}.
The monitors are centred around the outgoing beam directions in the severely limited space available,
in front of the compensating solenoids, which extend to $|z| \simeq 1.2$\,m from the IP. 
The monitor design is illustrated in Fig.~\ref{fig:luminometer}, where the main physical dimensions are also given. 
The design includes 25 layers, each comprising a 3.5\,mm tungsten plate, equivalent to one radiation length, 
and a Si-sensor plane inserted in the 1\,mm gap. 
In the transverse plane, the Si sensors are finely partitioned into pads. 
The proposed segmentation features 32 divisions, both radially and azimuthally, 
for 1024 readout channels per layer, or 25\,600 channels in total for each calorimeter. 
A solution with a four times finer azimuthal segmentation is also under consideration.

A 30\,mm uninstrumented region at the outer circumference is reserved for services (front-end electronics, cables, and cooling), 
as well as for the physical structures (likely including precision dowels and bolts) needed for the assembly of the SiW sandwich. 
Overall, the proposed design is very compact, each calorimeter weighing only about 65\,kg. 
An example of the integration of the LumiCals with the MDI and the detector system 
is shown in Figs.~\ref{fig:cooling_cones} and~\ref{fig:support_cyl}, in Chapter~\ref{sec:mdi}.

The Si-sensor pads are connected to front-end electronics positioned at radii immediately outside the sensors. 
To minimise the occurrence of pile-up events, it is desirable to read out the sensors at the bunch-crossing rate. 
This constraint calls for the development of readout electronics with an $\mathcal{O}$(20\,ns) shaping time. 
From experience with similar electronics development~\cite{FCAL:FLAME}, 
a power budget of 5\,mW per readout channel has been estimated, for a total of 130\,W per calorimeter, to be removed by cooling. 
For the required geometric precision, the temperature needs to be controlled to within a tolerance of 1\,$^\circ$C.

The goal of $10^{-4}$ precision on the absolute luminosity measurements translates into a required precision of 
$\mathcal{O}$(1\,$\mu$m) on the radial dimensions of the calorimeters 
and of $\mathcal{O}$(100\,$\mu$m) on the distance between the two calorimeters.

First full simulation studies of the LumiCals have been performed using 45.6\,GeV electrons. 
They show that the proposed geometry covers the polar angle region between 53 and 98\,mrad for fully contained showers, 
corresponding to a cross section of 28\,nb at the \PZ-pole energy. 
With a sampling fraction of 1.1\%, the calorimeters have an energy resolution of 3.2\%, corresponding to $22\%/\sqrt{E}$. 
The intrinsic resolution on the radial coordinate of showers is found to be 75\,$\mu$m at a $z = 1100$\,mm reference plane, 
considerably better than the contribution of about 120\,$\mu$m 
from multiple scattering of the primary electrons in the 2\,mm of beam pipe material traversed at shallow angles.

The feasibility of the proposed design has been demonstrated in part, over the last two decades, 
by work of the FCAL R\&D Collaboration, which had been specialising in technologies for very-forward calorimeters at linear colliders. 
A five-layer prototype with a similar sandwich structure was built and successfully tested in a beam test~\cite{Abramowicz:2018vwb}, 
and a dedicated readout system based on a custom-designed readout chip was prepared~\cite{FCAL:FLAME,FCAL:FLAME2}. 

Much work is needed towards a consolidated LumiCal design that satisfies the many severe requirements. 
Important future steps include:

\begin{itemize}
\item 
Engineering-level study of the proposed detector assembly method, which involves precision dowels and through-going bolts. 
This study must take into account the required $\mathcal{O}$(1\,$\mu$m) geometrical precision on the radial coordinate.
\item 
Realistic estimate of the space needed for services at radii outside the sensitive region. 
This region shall be kept as transparent as possible by the use of lightweight materials, to minimise particle interactions.
\item 
Design of a procedure for maintaining and monitoring the geometric accuracy of the monitors via precise metrology and alignment.
\item 
Design of a cooling system allowing control of a constant and uniform temperature over the monitors. 
The required tolerances have to be developed.
\item 
Re-evaluation of the expected radiation dose based on the final collider parameters 
and assessment of whether existing sensor technologies are adequate or further sensor R\&D is required.
\item 
Design of compact low-power readout electronics that preferentially allows readout at the 50\,MHz bunch-crossing rate, 
including transmission of signals away from the detectors. 
The system developed by the FCAL Collaboration may be a good starting point.
\item
Further full simulation studies to optimise the design of the detector.
\end{itemize}

\subsection{Outlook}

During the Feasibility Study, 
the detector designer community confirmed and enhanced its engagement and commitment 
to support a few complete detector concepts 
and to develop innovative detectors capable of delivering the required physics performance. 
The FCC-ee detectors have to meet unprecedented performance requirements 
in order to match the expected statistical precision and discovery physics potential. 
These requirements are, in many respects, complementary to what was needed at HL-LHC: 
while radiation tolerance is not an issue, with few exceptions close to the beams, 
the demands on precision, efficiency, purity, transparency of trackers, 
and compactness of calorimeters are considerably more challenging. 
Even though the readout bandwidth requirements appear relaxed when compared to the situation at HL-LHC, 
the required rate capabilities far exceed those from past linear collider studies. 
These constraints significantly magnify the system design and integration challenges 
especially if material budget and dead space are to be kept at minimum.

The detector R\&D community has undergone a restructuring process, 
leading to the formation of DRD Collaborations to address these demands. 
While it is clear that these Collaborations will only be able to unfold their potential 
with an adequate ramp-up of resources, 
it is also important to emphasise that they must benefit from proper guidance 
through the detector conceptual activities in the FCC effort. 
In particular, it is crucial to continue to develop and maintain a powerful common software framework 
(Chapter~\ref{sec:software}), including full-simulation tools, 
in view of establishing and evolving the link between the detector concepts, on the one hand, 
and technological R\&D, prototype design and construction, and ultimately test-beam data analysis, on the other. 
It will be essential that this common framework be adopted by the DRD teams, 
to optimise the subsystem inclusion in the FCC-ee detector simulation studies. 
Indeed, realistic geometries and digitizers will be essential to assess physics performance, 
as well as to compare with prototypes and test-beam data.

The detector community is well connected worldwide. 
The US and European roadmap processes, for example, 
have been well informed of each other and led to the formulation of overarching themes that are aligned to a large extent. 
This process is reflected in the ongoing integration of international groups and projects into the new DRD structure. 
Many of the proposals submitted to the DRD Collaborations target future Higgs factories and, 
among those, many specifically target FCC-ee. 
Given the current FCC-ee timeline, the R\&D in the 2025--2027 period will give priority to conceptual and component studies, 
with system aspects in focus from the beginning, 
preparing the way to the possible development of prototypes and demonstrators 
during the subsequent three-year period (2028--2030), 
followed by the construction and test of realistic modules by 2035. 

The next phase of the FCC study will first capitalise on the developments made during the Feasibility Study 
regarding the full-simulation of the existing proto-detector concepts, 
to confront their simulated performance with the detector requirements from physics. 
This work will clarify if those physics-driven requirements are matched by the current proto-detector concepts 
and will point the way to alternative detector solutions.
Different detector configurations will then be studied to optimise both performance and cost. 
In parallel, the links to the DRD Collaborations will be tightened. 
Work towards the common goal of developing FCC-ee detector concepts will be encouraged, 
by giving the R\&D groups the necessary inputs and requirements from the FCC Feasibility Study, 
and by identifying their needs in terms of guidance and software tools. 

Several challenges, specifically needed for FCC-ee but maybe not included in the DRD programme, will also need to be addressed. 
For example:
\begin{itemize}
\item How to monitor the time-stability of the magnetic field map in the experiments at the $10^{-7}$ level 
(taking into account all magnetic elements, including compensating solenoids, etc.)?
\item How to achieve, possibly using a set of geometry systems, 
the accuracy on the knowledge of the alignment and surveying fiducials required for precision measurements 
(such as the luminosity measurement at low angles, the dilepton/diphoton cross section measurements at large angles, 
or the tau lepton lifetime measurement)?
\item Can cosmic-ray muons continue to be used for the global alignment of FCC detectors, 
despite the fact that they will operate in much deeper caverns than those of previous colliders?
\item Are there new detector technologies or configurations that would be particularly well-suited for specific FCC-ee measurements 
or that would be beneficial in view of improving the performance-to-cost ratio?
\end{itemize}

In the next few years, regular workshops will be organised with the DRD Collaborations, 
to evaluate possible design choices for detector subsystems, 
including their simulation, optimisation, demonstrators, and key engineering aspects, 
as well as to foster interactions and cross-fertilisation between interested institutes. 
Issues ranging from low-level readout architectures and engineering resources 
to larger-scale system integration, full-simulation-based optimisation, and high-level event reconstruction, 
will be addressed.
The goal will be to quantify the cost and performance (including flexibility aspects) of each specific design choice, 
always retaining a physics-driven perspective,
towards the timely delivery of conceptual design reports and full detector concepts proposals shortly after 2030. 
Reaching this ambitious goal will require an appropriate re-organisation 
and a significant injection of more (human and financial) resources in the project.

\cleardoublepage
\section{FCC cavern infrastructure}
\label{sec:caverns}
\subsection{Introduction}

Four experiment caverns are foreseen at FCC. 
During the FCC-ee operation, all these caverns will house general-purpose detectors, 
with various degrees of specialisation, aimed at exploiting the full physics potential of this collider.
During the \mbox{FCC-hh} operation, a layout similar to the present LHC is foreseen: 
two caverns will house general-purpose detectors that cover the entirety of the proton-proton physics programme, 
similar to ATLAS and CMS, and the other two will house detectors more specialised to dedicated physics topics, 
on the model of ALICE and LHCb. 
It would be highly inefficient, impractical, and costly to modify the caverns at the end of the FCC-ee phase.
A common cavern infrastructure that will serve equally well FCC-ee and FCC-hh must therefore be designed ahead of time.

Most aspects of the FCC experimental cavern infrastructure are defined by requirements for the FCC-hh detectors, 
given their much larger size, the extreme radiation environment and related shielding, as well as larger magnetic stray fields. 
A reference detector, conceived as a general-purpose detector able to fully exploit the physics potential of FCC-hh, 
was studied in detail in Ref.~\cite{Mangano:2022ukr}. 
A cavern size of L66\,m $\times$ W35\,m $\times$ H35\,m with a shaft diameter of 18\,m 
is deemed sufficient to install and house such a detector. 
The cross section of this cavern is very similar to that of the ATLAS experiment. 
Two caverns of this size and two smaller caverns for the specialised experiments have therefore been assumed 
to define the corresponding civil-engineering operations and costs.

To evaluate their size, the two smaller caverns were assumed to house 
a detector specialised in \PB-physics, on the one hand, and 
a detector specialised in heavy-ion physics, on the other. 
Preliminary studies~\cite{Chobanova:2020vmx,Prouvetalk} have shown that 
a detector with transverse and longitudinal sizes not significantly larger than the dimensions of LHCb 
(10\,m and 20\,m, respectively) 
would have the same acceptance and resolutions as LHCb, in spite of the much larger energies. 
A more compact design, 
inspired from that of the BTeV detector~\cite{Kulyavtsev:1999tcg},
would allow a two-arm spectrometer to be built in the same spatial envelope. 
Similarly, a heavy-ion experiment would need to be more granular and have better performance, 
but its overall size would not have to be different from that of ALICE
(the ALICE cavern dimensions are L53.5\,m $\times$ W15.5\,m $\times$ H22.7\,m).

A transverse cavern size similar to that of the CMS experiment would therefore be sufficient for both specialised detectors, 
leading to a L66\,m $\times$ W25\,m $\times$ H24.5\,m cavern and a shaft diameter of 15\,m for those two caverns. 
Such four caverns will definitely be more than adequate to house the FCC-ee detectors, 
which have typical diameters and lengths of 12--14\,m. 
The possibility of detector opening for maintenance in these caverns, especially the smaller ones, 
still need to be fully understood and ascertained.

In the following sections, 
the specifications and the layout of the FCC-hh reference detector are summarised, 
and more details on the cavern layout are given.

\subsection{The FCC-hh reference detector}

The FCC-hh parameters allow the direct exploration of particles with mass up to 
around 40--50\,TeV~\cite{Golling:2016gvc}, 
approximately an order of magnitude over the LHC sensitivity to heavy resonant states. 
During its lifetime, the FCC-hh is also expected to produce trillions of top quarks and tens of billions of Higgs bosons, 
allowing for rich SM precision measurement campaign~\cite{Contino:2016spe,fcc-phys-cdr} 
and rare-decay study programme~\cite{dEnterria:2023wjq}. 
Most importantly, FCC-hh is the only machine under study today that can provide 
a few per-cent level measurement of the Higgs boson self-coupling~\cite{Contino:2016spe,Mangano:2020sao}. 

An experimental apparatus that operates at FCC-hh must, hence, 
perform optimally on two main fronts (see Ref.~\cite{Mangano:2022ukr} for a thorough discussion). 
Firstly, physics at the EW and Higgs boson scale will produce objects in the detector with momenta in the \pt range 20--100\,GeV; 
the LHC detectors were built to produce an optimal response in such an energy range. 
Secondly, a new regime, at the energy frontier, 
will be characterised by the energy scale of decay products originating from highly boosted SM particles 
and heavy resonances (potentially with mass as high as 50\,TeV). 
An FCC-hh detector must therefore be capable of reconstructing leptons, jets, top quarks, 
and Higgs and \PW/\PZ bosons with momenta as large as $\pt = 20$\,TeV. 
In short, the detector must provide accurate measurements in the full energy range between tens of~GeV and tens of~TeV.

\begin{figure}[ht]
\centering
  \includegraphics[width=0.85\textwidth]{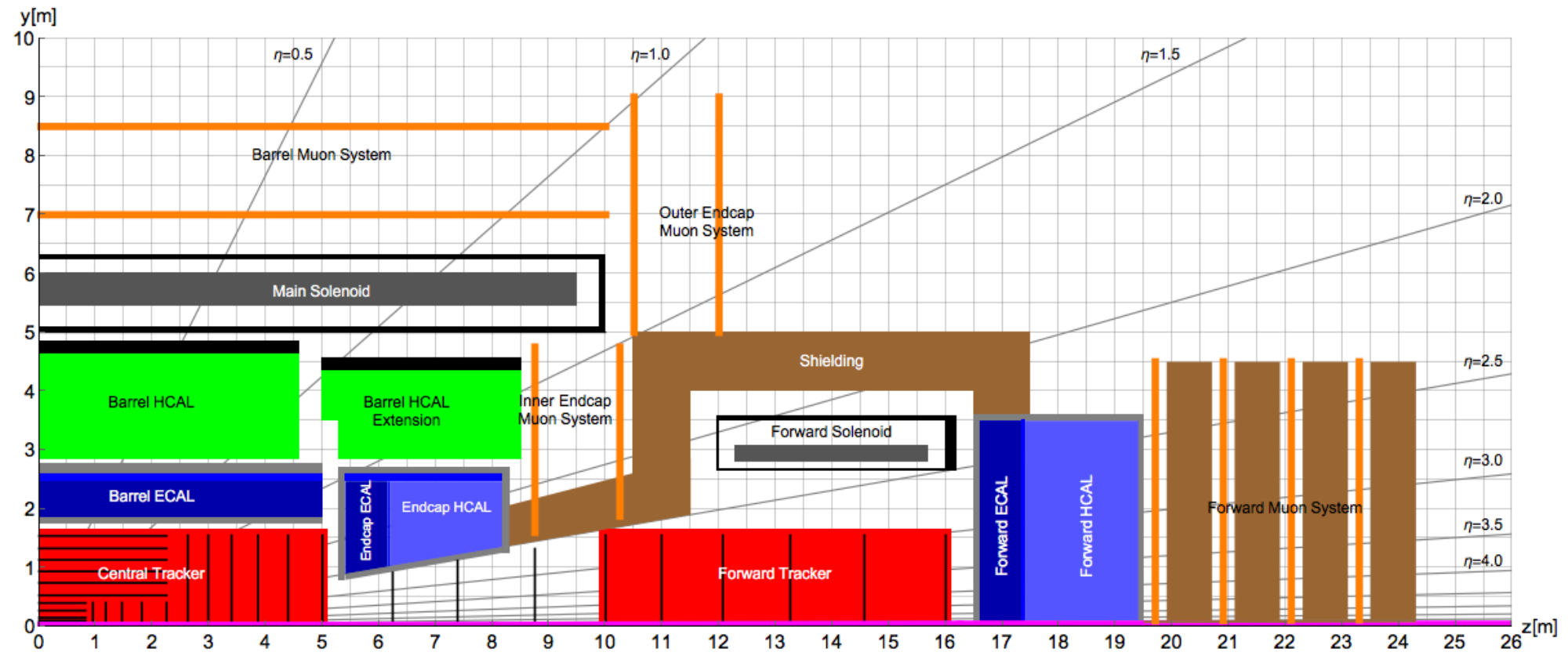}
  \includegraphics[width=0.85\textwidth]{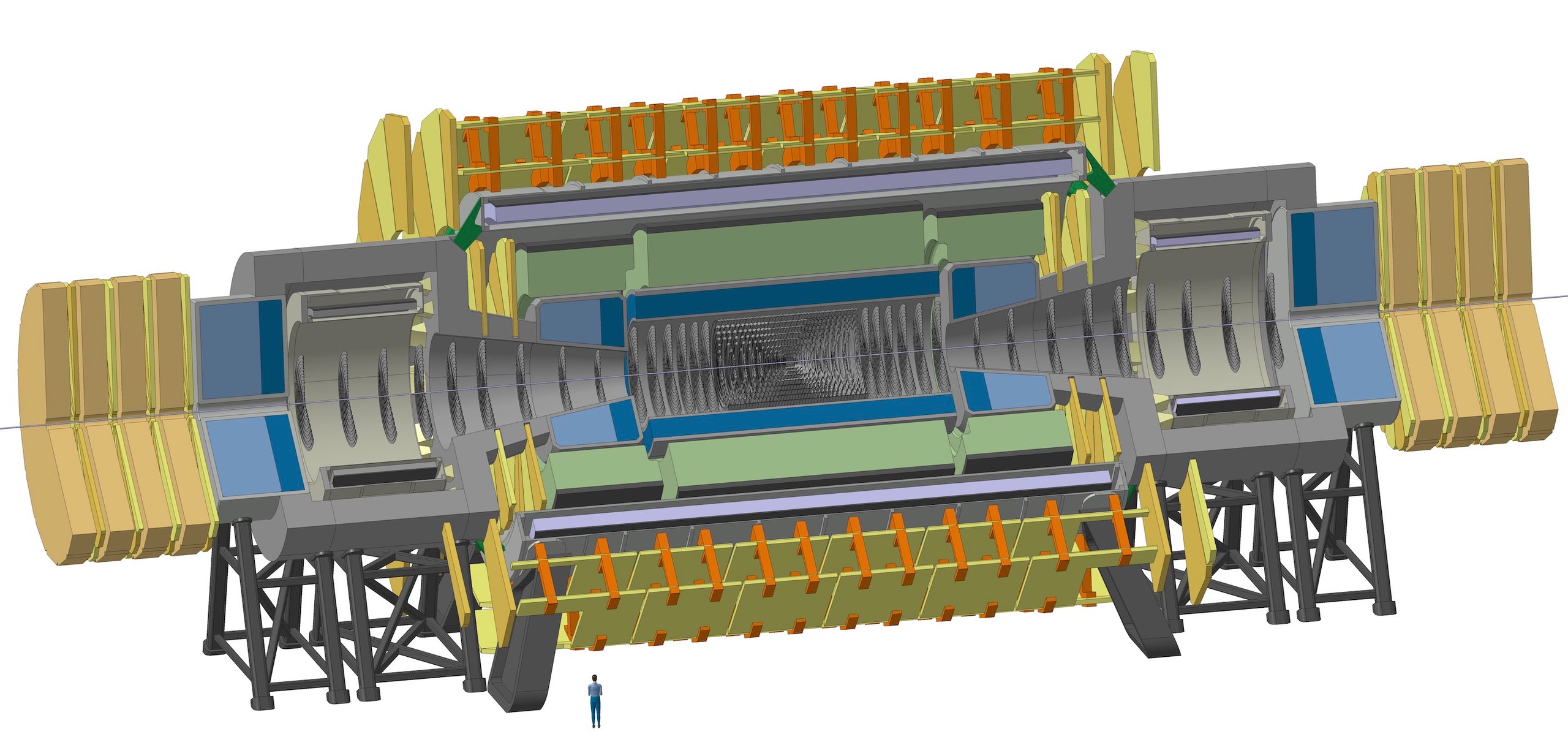}
  \caption{Schematic of the FCC-hh reference detector.}
  \label{FCChh-layout}
\end{figure}

Figure~\ref{FCChh-layout} shows the layout of the FCC-hh reference detector. 
This detector concept does not represent the final design, but rather a concrete example that suits the performance and physics requirements, 
and allows the identification of areas where dedicated further R\&D efforts are needed. 
The detector has a diameter of 20\,m and a length of 50\,m, comparable to the dimensions of the ATLAS detector but much heavier. 
The central detector, with a pseudorapidity coverage of $|\eta| < 2.5$, 
houses the tracking, electromagnetic calorimetry, and hadron calorimetry, inside a 4\,T solenoid with a free bore diameter of 10\,m. 
In order to reach the required performance over the $2.5 < |\eta| < 6$ pseudorapidity range, 
the forward parts of the detector are displaced along the beam axis by 10\,m from the interaction point. 
Two forward magnet coils with an inner bore of 5\,m provide the required bending power. 
These forward magnets are also solenoids with 4\,T field, 
providing a total solenoid volume of 32\,m length for high precision momentum spectroscopy up to $|\eta| \approx 4$ 
and tracking up to $|\eta| \approx 6$. 
As an option, replacing the forward solenoids with two forward dipoles is also being considered. 
The tracker cavity has a radius of 1.7\,m with the outermost layer at around 1.6\,m from the beamline in the central and forward regions, 
providing the full spectrometer arm up to $|\eta| = 3$. 
The electromagnetic calorimeter (ECAL) has a thickness of around 30 radiation lengths ($X_{0}$). 
The overall calorimeter thickness, including the hadron calorimeter (HCAL), 
exceeds 10.5 nuclear interaction lengths ($\lambda$), 
ensuring 98\,\% containment of high energy hadron showers and limiting punch-through to the muon system. 
The ECAL is based on liquid argon (LAr) given its intrinsic radiation hardness. 
The barrel HCAL is a scintillating tile calorimeter with steel and lead absorbers 
that uses wavelength shifting (WLS) fibres and silicon photomultipliers (SiPMs) for the readout. 
It is divided into a `central barrel' and two `extended barrels'. 
The HCALs for the endcap and forward regions are also based on LAr. 
The requirement of calorimetry acceptance up to $|\eta| \approx 6$ translates into 
an inner active radius of only 8\,cm at a $z$-distance of 16.6\,m from the IP. 
The transverse and longitudinal segmentation of both the electromagnetic and hadronic calorimeters 
is $\sim$\,4 times finer than the present ATLAS calorimeters.

The muon system is placed outside the magnet coils. 
The barrel muon system provides muon identification, stand-alone muon spectroscopy by measurement of the angle of the muon track, 
as well as precision muon spectroscopy by combining the tracker measurement with the measured space points in the muon chambers. 
The forward muon system can only provide spectroscopy if forward dipoles are used. 

\begin{figure}[ht]
\centering
\includegraphics[width=0.6\textwidth]{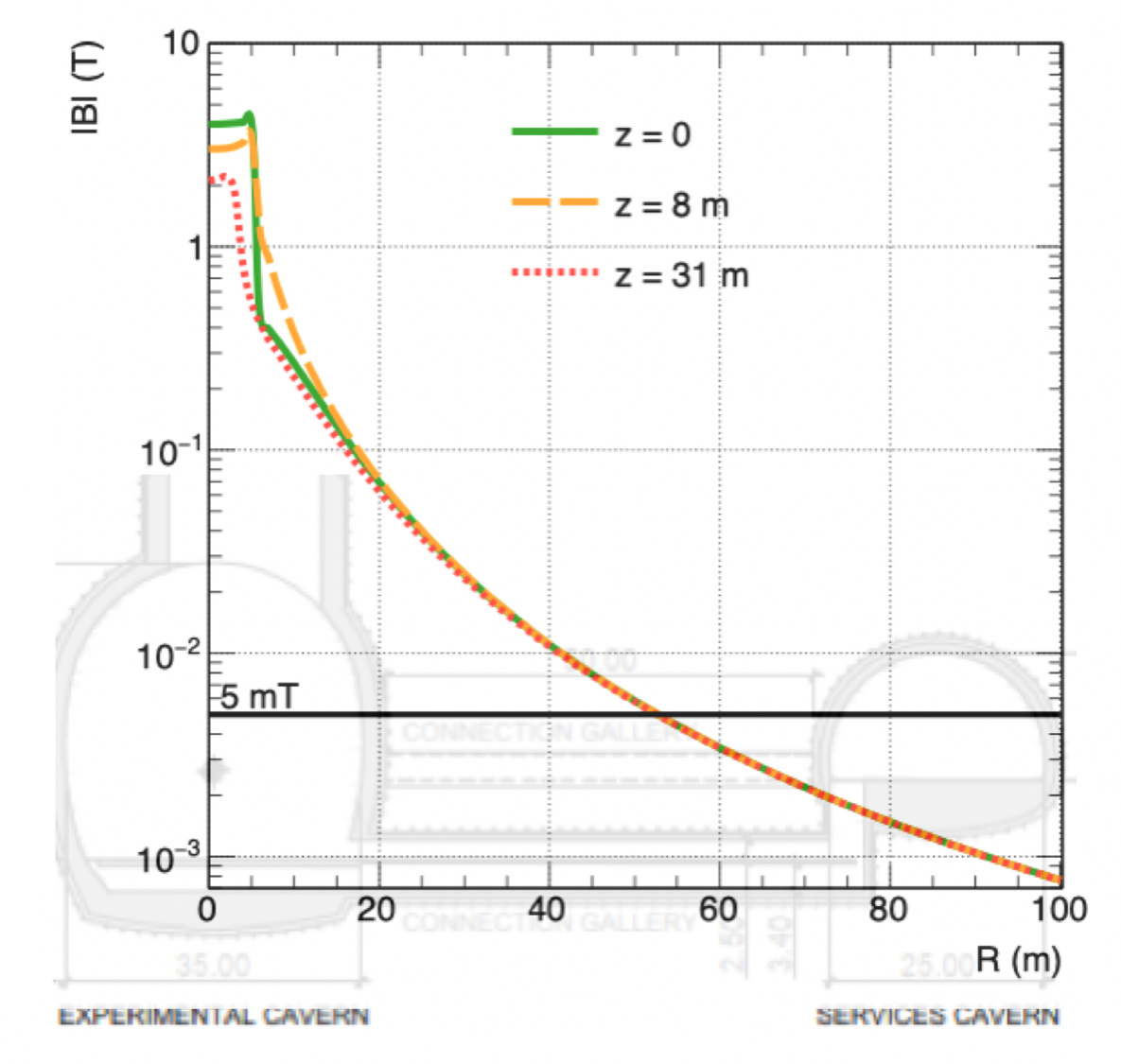}
\caption{Radial magnetic field for the FCC-hh reference detector, without magnet yoke. 
The stray field is around 100\,mT at the cavern wall, 
while in the service cavern (50\,m from the experimental cavern wall), it amounts to less than 3\,mT.}
\label{bfield}
\end{figure}

The reference detector does not assume a magnet yoke, i.e., there is no shielding of the magnetic field.  
Figure~\ref{bfield} shows the magnetic stray field as a function of radial distance from the beamline. 
The stray field still amounts to 100\,mT at the cavern wall ($R=17.5$\,m), 
which has to be considered for the implementation of access structures and services. 
The 5\,mT level is achieved at a distance of 55\,m from the beamline, still outside of the service cavern. 
Clearly, it would be more convenient to have a shield for the magnetic field, 
but to arrive at a level of 5\,mT at the cavern wall, which was the specification for CMS, 
an iron yoke of 4\,m thickness would be required, weighing a total of 52\,kt. 

\begin{figure}[t]
\centering
\includegraphics[width=0.9\textwidth]{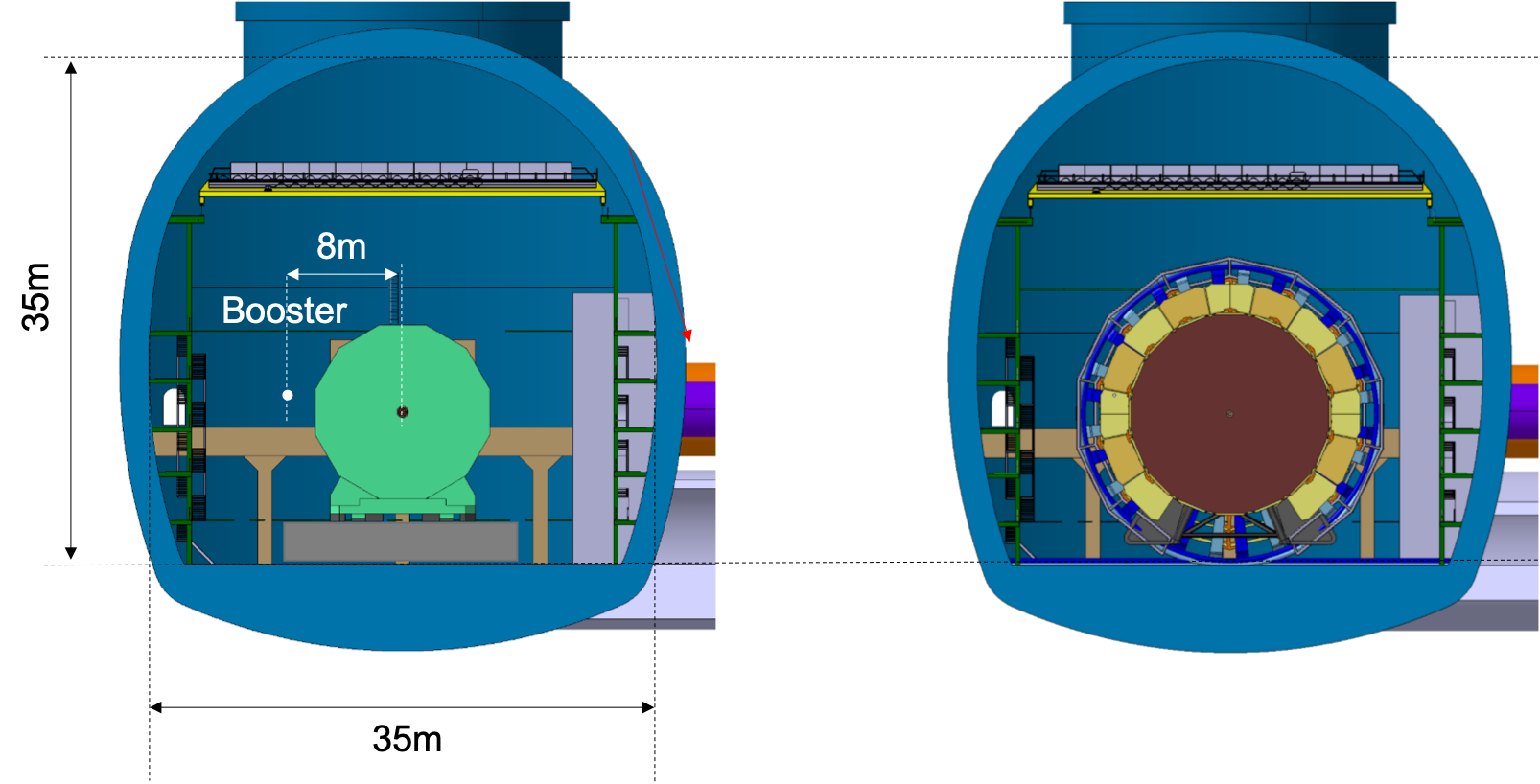}
\caption{An FCC-ee detector (left) and the FCC-hh reference detector (right) 
in a large cavern of 35\,m $\times$ 35\,m cross section.}
\label{large_cavern}
\end{figure}

\begin{figure}[t]
\centering
\includegraphics[width=0.9\textwidth]{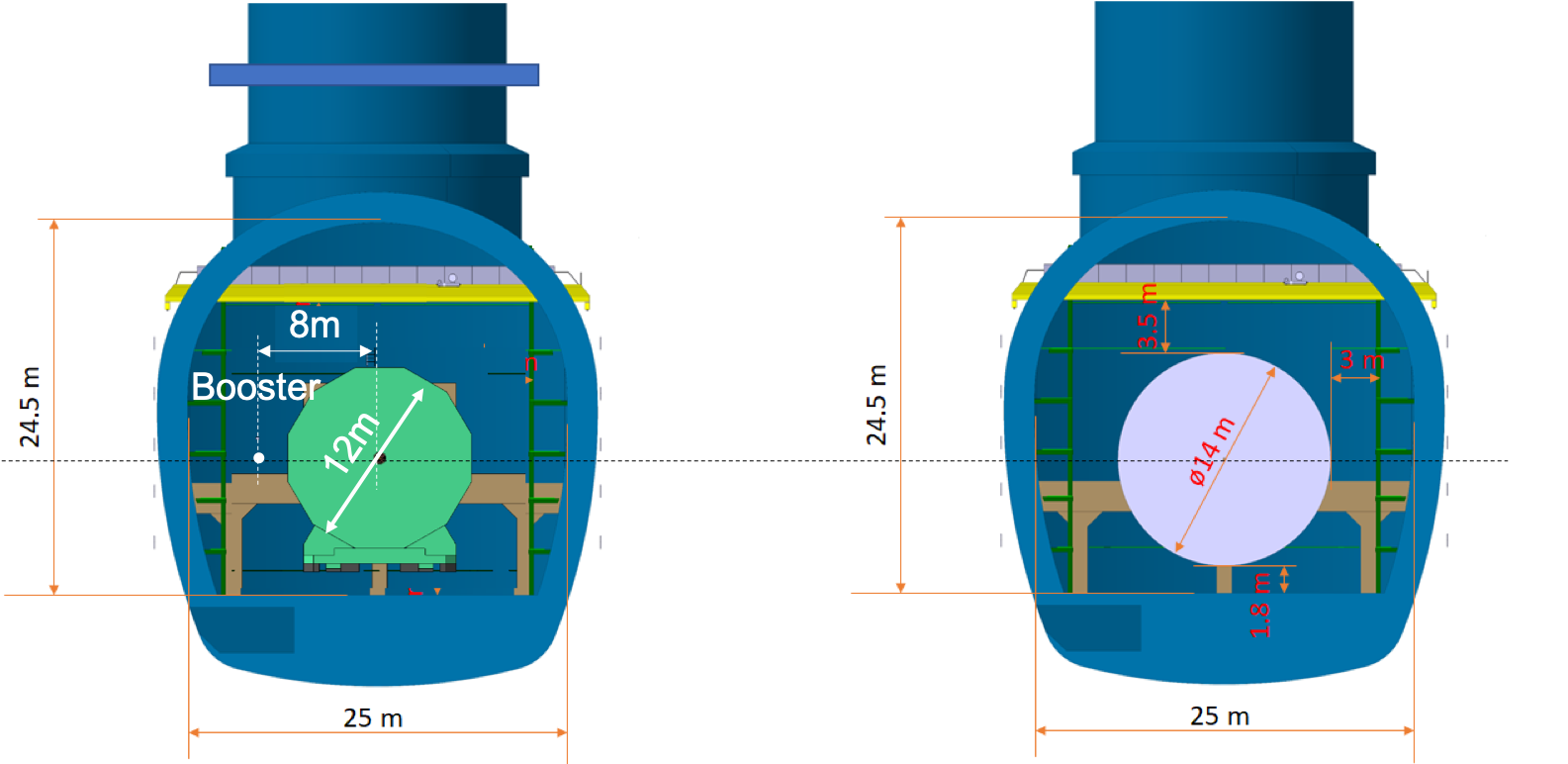}
\caption{An FCC-ee detector (left) and the FCC-hh reference detector (right) 
in a small cavern of 24.5\,m $\times$ 25\,m cross section.}
\label{small_cavern}
\end{figure}

As can be seen in Fig.~\ref{FCChh-layout}, 
embedding the muon system inside a yoke of 4\,m radial thickness would lead to a total detector radius of around 11\,m, 
similar to the ATLAS detector.  
Such a yoke is probably excessive in terms of cost and construction. 
With the radial extent of around 10\,m for the \mbox{FCC-hh} reference detector, a cavern cross section of 35\,m $\times$ 35\,m is appropriate. 
If needed, such a cavern could even house the magnet yoke.  
For illustration, 
Fig.~\ref{large_cavern} shows an FCC-ee detector and the \mbox{FCC-hh} reference detector in this large cavern, while
Fig.~\ref{small_cavern} shows an FCC-ee detector and a specialised \mbox{FCC-hh} detector of 14 m-diameter, 
similar to CMS, in the smaller caverns.


\begin{figure}[ht]
\centering
\includegraphics[width=0.95\textwidth]{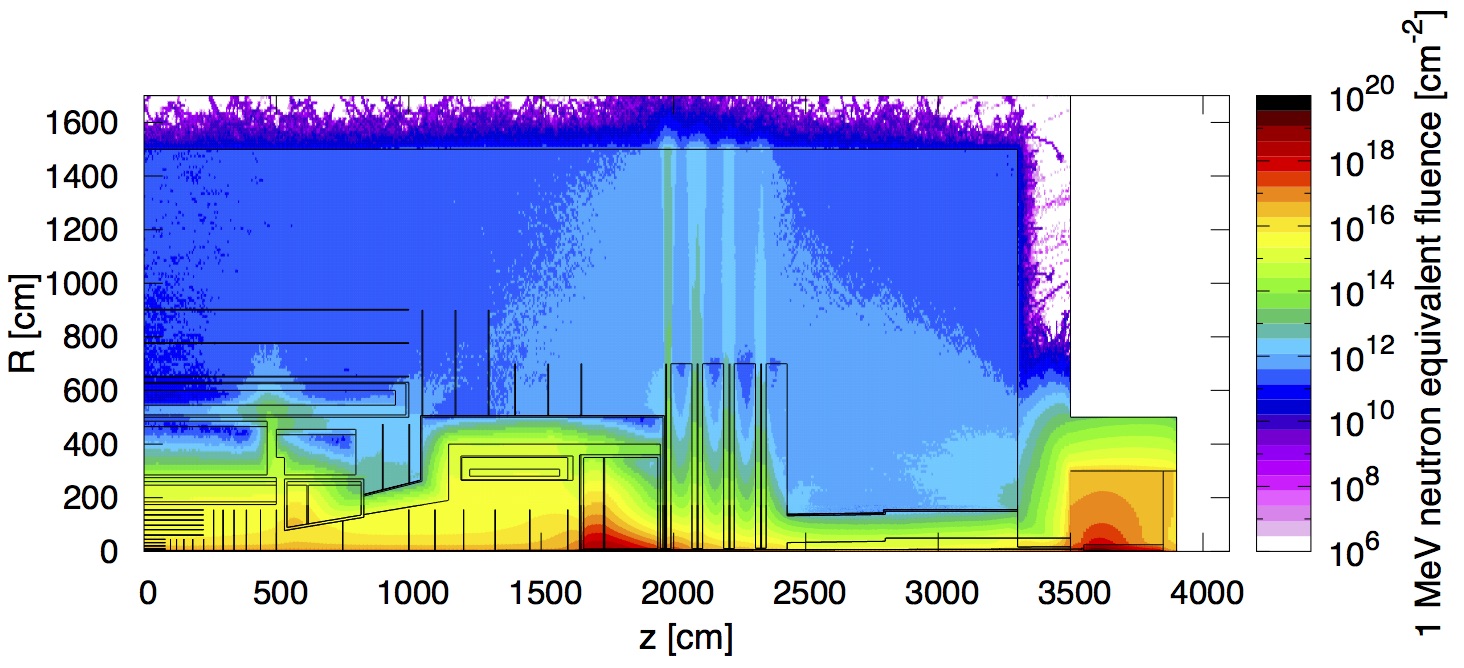}
\caption{Radiation load for the FCC-hh detector and the final focusing magnets in the machine tunnel.}
\label{radiation}
\end{figure}

In FCC-hh, the longitudinal extent of the detector is limited by the machine elements and the related shielding. 
The last magnetic elements in the final focusing magnets of FCC-hh are at $\ell^* = 40$\,m from the interaction point. 
In order to shield these magnets from the collision debris originating from the interaction point, 
the so-called `TAS absorbers' are placed at a distance of around 35\,m. 
The absorption of these high energy hadrons produces a significant amount of neutrons that `diffuse' out of this absorber, 
as can be appreciated in Fig.~\ref{radiation}. 
The concrete cavern wall at $z = 33$\,m stops these neutrons from entering the cavern, 
defining the maximum possible cavern length to be 66\,m.
Altogether, these basic considerations converge to a cavern size of L66\,m $\times$ W35\,m $\times$ H35\,m for the FCC-hh reference detector.

\subsection{FCC cavern infrastructure}

\begin{figure}[h]
\centering
\includegraphics[width=0.9\textwidth]{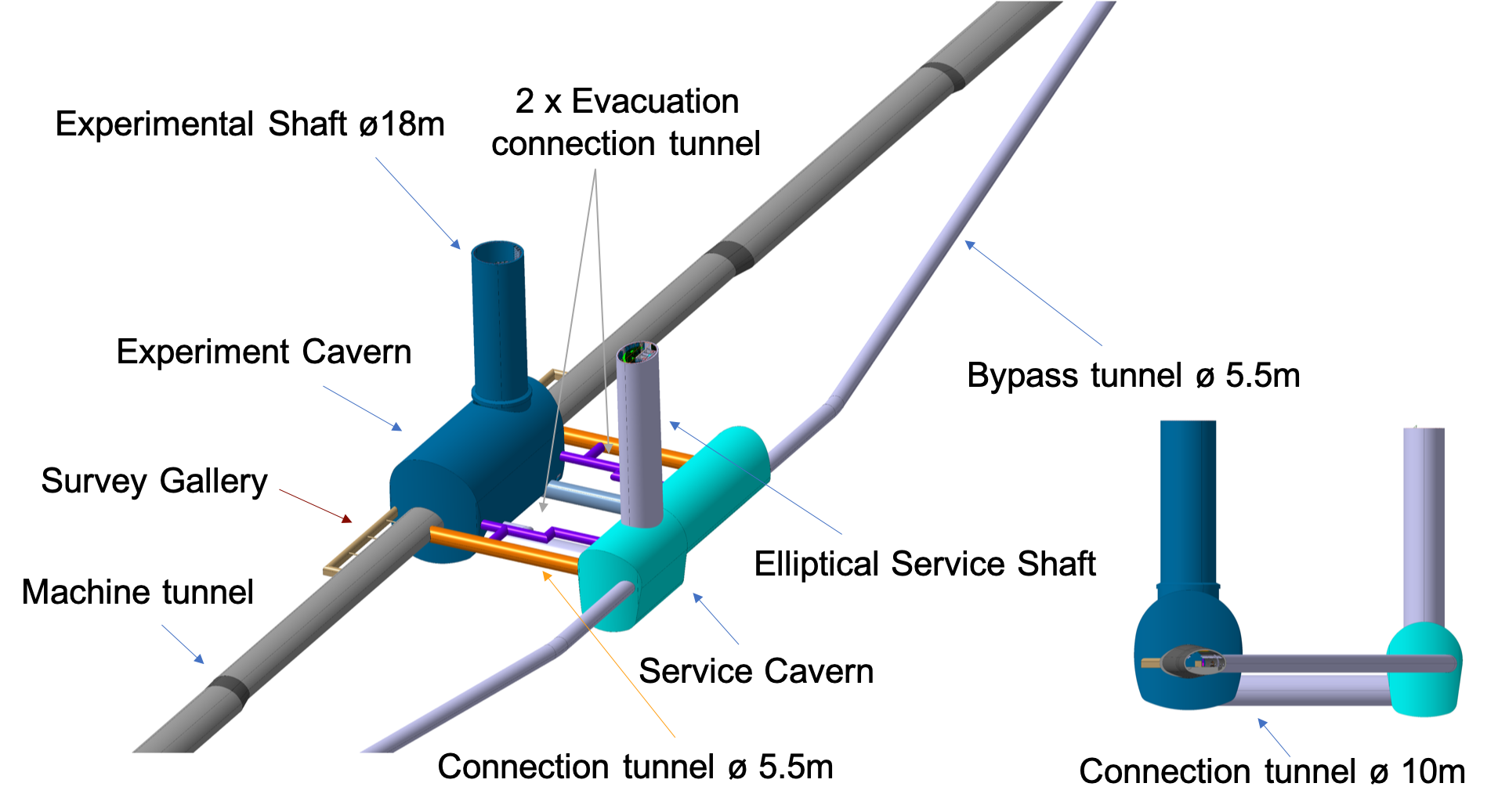}
\caption{FCC-hh cavern infrastructure.}
\label{infrastructure}
\end{figure}

The FCC-hh cavern infrastructure is shown in Fig.~\ref{infrastructure}. 
Since the service cavern is at a distance of around 50\,m from the detector cavern and between 60 and 70\,m away from the beamline, 
the stray field therein is below 5\,mT (Fig.~\ref{bfield}) and standard equipment can be placed at this location.
Besides, the radiation shielding provided by this distance suffices to allow permanent access to the service cavern. 
A main shaft of 18\,m diameter gives access from the surface to the experiment cavern. 
It houses only some ventilation pipes but no other services or elevators, and is only foreseen for the transport of equipment. 
The personnel access and the detector services are housed in the service shaft, which connects to the service cavern. 
Three connection tunnels, represented in Fig.~\ref{infrastructure}, connect the service and experiment caverns: 
a central one, of 10\,m diameter, for services, and two others, of 5.5\,m diameter, placed on each of the two ends of the detector. 
They are complemented by two evacuation tunnels, also represented in the figure.

The principal dimensions of the FCC cavern infrastructure are summarised in Table~\ref{t1}.

\begin{table}[ht]
\caption{Requirements and key numbers for the FCC-ee and FCC-hh detectors.}
\label{t1}
\centering
  \begin{tabular}{l c}
  \hline
    Item &  \\
    \hline
     Size of large caverns & L66\,m W35\,m H35\,m  \\
     Shaft diameter of large caverns & 18\,m \\
     Size of small caverns & L66\,m W25\,m H24.5\,m  \\
     Shaft diameter of small caverns & 14\,m \\
     \hline
     FCC-hh reference detector diameter without yoke & 20\,m \\
     FCC-hh reference detector diameter with yoke & 22\,m \\
     Stray field at cavern wall ($R = 17.5$\,m) & 100\,mT \\
     Distance for stray field $<5$\,mT & 55\,m \\
     \hline
     FCC-ee detector diameter & 12--14\,m \\
    \hline
  \end{tabular}
\end{table}

\cleardoublepage
\section{Software and computing}
\label{sec:software}
Ever since the launch of the FCC conceptual design study in 2014, 
one of the primary objectives has been to design and develop a unified, 
synergistic, and adaptable software `ecosystem', 
to serve as a foundation for all future experiments at FCC-ee and FCC-hh. 
With its data processing framework and all the necessary tools 
inspired by the advanced software of running experiments and ongoing R\&D initiatives, 
the resulting ecosystem is aiming at addressing all future experiment's use cases, 
based on modern software technology. 
While this transformative approach required significant time to gain general acceptance, 
and even if the level of completeness is not anywhere close to that of running experiments, 
this new ecosystem is now regularly used by the FCC particle physicists in their daily work. 
The performance of some tools, such as flavour tagging algorithms, 
already far surpasses that of the algorithms used for past linear collider studies~\cite{ilcsoft}. 
Needless to say, further work is needed to fully enable all studies targeted by the FCC-ee scientific programme. 
For example, the following tasks (among many other necessary developments) 
were identified as priorities for the FCC Feasibility Study: 
\begin{itemize}

\item Proceed with the full and parametrised simulations (and their interplay) 
of the sub-detectors currently considered (Section~\ref{sec:concepts}), 
including digitisation and sub-detector interchange with a `plug-and-play' framework, 
towards enabling the study of the performance (Section~\ref{sec:requirements}) of a large variety of detector concepts.  

\item Develop and implement the various reconstruction and analysis tools for use by all collaborators, 
reaping the benefits from LHC experience and, whenever relevant, from past linear collider studies.

\item Complete the simulation of the interaction region,
also known as machine detector interface (MDI, Section~\ref{sec:mdi}),
in view of the final evaluation and mitigation of beam-related backgrounds in the detectors. 
This task includes streamlining the access to the Monte Carlo codes that simulate the relevant backgrounds.

\item Provide the technology needed to improve the accuracy of the simulation of beam-related quantities 
(such as the beam-energy spread, crossing-angle spread, bunch length and transverse size, 
final state particle deflection by the colliding bunches, etc.) 
as well as initial state and final state radiation (and their interference) 
to match the statistical precision expected at FCC-ee, in particular at the \PZ pole. 
Ensure that these technologies are usable in all Monte Carlo generator codes relevant for FCC-ee, 
in close interaction with the code authors.

\item Continue to drive the development of the common software framework (\textsc{Key4hep}) 
and common data format (\textsc{edm4hep}), 
ensuring that they meet the requirements of FCC, 
both in terms of functionality and of infrastructure building, testing and deployment.

\item Evaluate the need for computing resources and proceed with regular simulated data production. 
The computing needs for FCC-ee are dominated by the \PZ-pole run, both online and offline.
The corresponding demands are significant, even when compared to HL-LHC. 
By the time FCC-ee operations start, 
it will be crucial to leverage the advances and insights gained from HL-LHC, 
also in terms of resource sustainability. 
Even during the next phase of the study, 
the needs for fully simulated and reconstructed event samples, and for their analysis, 
are potentially challenging and require detailed identification, evaluation, and purchase of the necessary computing resources. 
Coordination with resource providers, which include CERN, WLCG grid sites involved in FCC research, and HPC centres, 
is required to identify ways to guarantee access to these resources.

\item Last but not least, 
provide the users with detailed documentation and regular pedagogical tutorials, 
as has been the case in the past decade.

\end{itemize}

This chapter reviews the progress made towards achieving the tasks listed above. 
The results presented and discussed reflect the collaborative efforts of the `Software and Computing work package', 
in conjunction with other work packages within the FCC PED feasibility study. 
A key factor in this progress has been the strong software core team at CERN, 
which provides the overall vision, delivers the daily technical coordination, 
and ensures the timely development of essential common tools.

\subsection{The FCC software ecosystem}
\label{sec:fsr_software}

\begin{figure}[ht]
\centering
\includegraphics[width=0.8\textwidth]{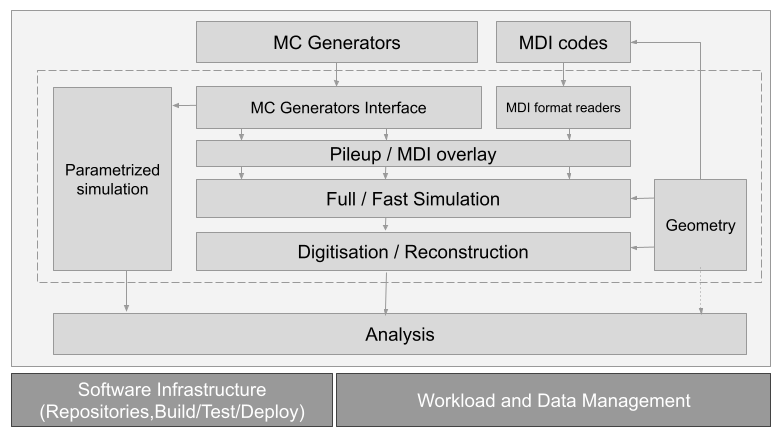}
\caption{Workflows to be supported during the project design and FCC Feasibility Study.}
\label{fig:workflows}
\end{figure}

During the FCC feasibility study, 
the focus has been to deliver the components deemed essential to meet the challenges of FCC-ee 
and to assess the physics potential of the proposed experimental infrastructure: 
a comprehensive suite of $\epem$ event and beam-background generators; 
a proper handling of the complex interaction region and associated backgrounds; 
an infrastructure that facilitated the implementation of detector subsystems and detector concepts
(i.e., their geometries, their simulation, and the development of versatile reconstruction algorithms); 
and tools for event analysis, visualisation, and resource management. 
The corresponding workflows are illustrated in Fig.~\ref{fig:workflows}. 

Because this ambitious plan required human resources far beyond those available during this phase of the project, 
a proposal was made to other $\epem$ particle physicist communities to join forces 
and explore the possibility of developing a common software solution 
that could be adapted to the needs of all collider and detector geometries. 
Based on the initial FCC software framework developed during the Conceptual Design Study, nicknamed \textsc{fccsw}, 
and driven by the belief that the high-energy physics community was ready to 
further expand the number and role of shared components between experiments, 
the \textsc{Key4hep} initiative was launched after two founding workshops held in 2019 and early 2020~\cite{kickBO,kickHK}. 

Like any other large-scale software development, \textsc{Key4hep} is inherently a collaborative effort.
It builds on the experience of the LHC experiments, 
is integrated with community R\&D projects, 
and maximises the reuse of established solutions and packages, 
allowing experiments to benefit from existing community developments.  
Notable examples include the common core framework \textsc{Gaudi}~\cite{Gaudi}), 
developed and used by LHCb and ATLAS following the positive experience of BELLE,
the \textsc{dd4hep}~\cite{dd4hep} R\&D project for detector geometry, 
and other well-known packages like \textsc{Root}~\cite{root}, \textsc{Geant4}~\cite{Geant4}, or \textsc{Podio}~\cite{Podio}. 
By leveraging and contributing to these experiment-independent software packages, 
the \textsc{Key4hep} initiative aims at fostering a broader ecosystem of HEP software,  
enhancing collaboration and innovation across the field.

After recalling the main concepts of \textsc{Key4hep} in Section~\ref{sec:k4h}, the current status in each of the main areas is summarised in the following sections: common event data model (Section~\ref{sec:edm4hep}); event generators (Section~\ref{sec:generators}); parametrised (Section~\ref{sec:delphes}) and full (Section~\ref{sec:geometries}) simulation;  digitisation (Section~\ref{sec:digi}) and reconstruction (Section~\ref{sec:reco});  event analysis (Section~\ref{sec:analyses}); visualisation (Section~\ref{sec:visualisation}); and resources (Section~\ref{sec:resource}).

\subsection{The main components of \textsc{Key4hep}}
\label{sec:k4h}

The \textsc{Key4hep} ecosystem has been designed to address core requirements common to all experiments in a sufficiently robust manner, 
while allowing specific extensions to meet individual demands. 
High-energy physics data processing consists of several basic components that need to be chosen carefully, 
the most important of which are listed below.

\begin{enumerate}

\item A data processing framework provides the structure for integrating and steering all other components. 
For example, past linear collider studies used the \textsc{Marlin}~\cite{Marlin} framework for many years. 
In \textsc{Key4hep}, \textsc{Gaudi}~\cite{Gaudi} was chosen because of 
its widespread adoption by the LHC experiments,
its broader user and developer communities, 
and its recognised potential for future evolution, 
which includes support for accessing heterogeneous resources, 
accommodating different architectures, and enabling task-oriented concurrency.

\item A detector geometry description tool, 
chosen to be \textsc{dd4hep}~\cite{dd4hep} because of its already widespread use in the relevant communities. 
In addition to providing a comprehensive detector description for event simulation, reconstruction, and analysis, 
\textsc{dd4hep} is now also used in production by the CMS and LHCb collaborations. 
This adoption drives a consolidation and development process that is highly advantageous for future experiments, 
in particular in critical areas, such as alignment and calibration.

\item A common event data model, called \textsc{edm4hep} (Section~\ref{sec:edm4hep}).

\item An infrastructure to build, test, and deploy the required software with minimal effort, 
automating the installation and ensuring a consistently linked set of packages with correct dependency resolution. 
The LHC Collaborations addressed this challenge independently, without converging on a unified solution. 
The HEP Software Foundation's packaging working group identified the scientific package manager \textsc{Spack}~\cite{Spack} 
as a potentially promising common solution worth further exploration for use in HEP. 
In line with the \textsc{Key4hep} philosophy, 
\textsc{Spack} was evaluated and successfully used to build \textsc{Key4hep}, 
despite certain limitations~\cite{SPack-CHEP2024-K4H}. 
These challenges include limited support for development workflows and deployment capabilities, 
requiring custom in-house workarounds, 
such as those required to deploy build artifacts on \textsc{CernVM-FS}, 
the widely-adopted technology for software distribution in HEP.

\end{enumerate}

\subsection{The event data model \textsc{edm4hep}}
\label{sec:edm4hep}

The event data model defines the structures needed to describe the required event information in the persistent store. 
While, in principle, this model can be different from what is used in the transient store seen by the data processing algorithms, 
and also from algorithm to algorithm, 
adopting the same event model everywhere reduces the need for conversions, enhances generality, and improves interoperability.

Since the launch of the FCC studies, the choice has been for an event data model managed by \textsc{Podio}~\cite{Podio}, 
a toolkit that generates the data model implementation from templates. 
With this choice, the high-level description in YAML files using Plain-Old-Data (POD) simple types 
are separated from the low-level persistency layer, which can be optimised according to the back-end. 
Once the required classes are defined in the YAML file, the \textsc{Podio} tool creates the source code automatically. 
The same technology has been chosen for \textsc{Key4hep}, with an event data model referred to as \textsc{edm4hep}, 
initially based on the event data models used in past linear collider studies (\textsc{lcio}~\cite{lcio}) 
and in the FCC conceptual studies (\textsc{fccedm}~\cite{fcc-ee-cdr}) classes, and improved from there as needed. 

The underlying assumption is that the same event data model can be used for all types of HEP experiments, 
which is well-suited to the integrated FCC programme, with a lepton collider followed by a hadron collider.
This requirement appears to be fulfilled by \textsc{edm4hep}, 
although a more definitive assessment will become available only when the FCC-hh investigations are extended to more use cases. 

\subsubsection{A specific use case: LEP data in \textsc{edm4hep}}
\label{sec:lepedm4hep}

In the context of the LEP data preservation efforts, an initiative has been launched to migrate those data to \textsc{edm4hep}, 
in a pioneering attempt to analyse these real, non-simulated data in the FCC software ecosystem. 
Beyond the clear advantage of preserving the ability to extract new scientific insights from LEP data for future generations, 
this effort has several key objectives relevant to FCC-ee. 
Most interestingly, LEP and FCC-ee share the same collision type and several centre-of-mass energies. 
This initiative therefore opens a unique opportunity to FCC physicists to gain hands-on experience with real data, 
thereby preparing them for future analyses within the FCC-ee scientific programme. 
It can also provide a platform to validate and improve simulation, reconstruction, and analysis tools, 
thus enhancing their reliability for future collider studies.

A preliminary migration feasibility study is currently underway with data from the ALEPH experiment. 
The project aims at establishing a conversion workflow that integrates both the experiment legacy software, 
executed within dedicated containers with the latest operating systems and validated by the experiment, 
and \textsc{Key4hep}-based applications. 
The initial approach involves extracting data at the analysis 
level\footnote{For ALEPH, the initial focus is on the MINI format~\cite{alephnumbers} 
however, the methodology under development is designed to be adaptable 
and could also be extended to the more comprehensive DST data format.} 
into an operating system-agnostic ASCII format, 
which can then be used to reconstruct the \textsc{edm4hep} structures for FCC analyses. 
Additionally, the project seeks to recreate a bookkeeping database containing relevant ALEPH metadata.

The status of the project was presented in the autumn of 2024~\cite{alephe4hdphep,alephe4hfccfrit}.
The current findings provide encouraging evidence of the overall feasibility of the approach,
despite difficulties in recovering comprehensive documentation, 
retrieving detailed information about the data that are often hardcoded within the legacy software,
and reconstructing detailed metadata about the data samples. 
These findings highlight the importance of generalising this effort as part of a more structured and systematic project.
 
\subsection{Integration of event generators}
\label{sec:generators}

Event generators are a crucial component of the software infrastructure in any HEP experiment, 
to design optimal data analysis strategies, evaluate detector performance requirements, 
and compare them with the response of different detector solutions, 
ultimately maximising the physics potential of the project. 
The richness of the FCC physics programme introduces new challenges for event generators across multiple dimensions. 
The unprecedented statistical precision expected at the \PZ pole for FCC-ee,
and the expanded energy range of FCC-ee and FCC-hh, 
call for substantial theoretical advances (Section~\ref{sec:theory}). 
These very precise calculations need to be translated into similarly accurate, computationally efficient, 
and easily maintainable software packages. 

Many event generators are already included in \textsc{Key4hep}. 
The first set of event generators was derived from the `LCG stacks', 
i.e., software stacks developed and maintained in the EP-SFT group at CERN targeting the cases of ATLAS and LHCb.
These stacks are, however, more and more used by non-LHC projects, 
and have been progressively expanded to include event generators not used at the LHC, 
such as \textsc{Whizard}~\cite{whizard}, a general purpose event generator used for past linear collider studies. 
Today, \textsc{Key4hep} includes an extended set of generic event generators, 
able to cover most of the initial needs of future projects, 
such as \textsc{Pythia8}~\cite{pythia8}. 

The increasing interest for FCC-ee has brought in the need for more specific and more accurate event generators. 
State-of-art generators for high-energy $\epem$ collisions, however, date from the LEP era and have been minimally maintained since. 
One of the first tasks to be addressed in the FCC feasibility study was the recovery of the 
still very relevant LEP event generators, 
such as \textsc{kkmc}~\cite{Jadach:2022mbe}, \textsc{bhlumi}~\cite{bhlumi}, and \textsc{BabaYaga}~\cite{babayaga}. 
The generators and related tools available in \textsc{Key4hep} at the time of writing are listed in Table~\ref{tab:genlist}.

\begin{table}[ht]
\centering
\caption{List of generators and related tools available in \textsc{Key4hep} at the time of writing. 
Most of the generators are available in the upstream \textsc{Spack} repository; 
the ones added by the \textsc{Key4hep-Spack} repository are indicated with an asterisk.}
\label{tab:genlist}
\begin{tabular}{cccccc}
\toprule
\multicolumn{6}{c}{Generators} \\ \midrule
\textsc{BabaYaga}* & \textsc{baurmc} & \textsc{bhlumi}* & \textsc{crmc} & \textsc{EvtGen} & \textsc{Genie} \\
\textsc{Gosam} & \textsc{Guinea-pig}* & \textsc{Herwig3} & \textsc{Herwigpp} & \textsc{kkmc}* & \textsc{MadGraph5amc} \\
\textsc{Photos} & \textsc{Pythia6} & \textsc{Pythia8} & \textsc{Sherpa} & \textsc{Starlight} & \textsc{Superchic} \\
\textsc{Tauola} & \textsc{vbfnlo} & \textsc{Whizard} &  & & \\ \toprule
\multicolumn{6}{c}{Generator tools} \\ \midrule
\textsc{agile} & \textsc{alpgen} & \textsc{ampt} & \textsc{apfel} & \textsc{ccs-qcd} & \textsc{chaplin} \\
\textsc{collier} & \textsc{cuba} & \textsc{dire} & \textsc{feynhiggs} & \textsc{form} & \textsc{HepMC} \\
\textsc{HepMC3} & \textsc{heppdt} & \textsc{hoppet} & \textsc{hztool} & \textsc{lhapdf} & \textsc{lhapdfsets} \\
\textsc{looptools} & \textsc{openloops} & \textsc{professor} & \textsc{prophecy4f} & \textsc{qd} & \textsc{qgraf} \\
\textsc{recola} & \textsc{rivet} & \textsc{syscalc} & \textsc{thepeg} & \textsc{unigen} & \textsc{yoda} \\
\bottomrule
\end{tabular}
\end{table}

\subsubsection{Event generators as packages}

Within the software ecosystem, generators are software packages that come with their own build, test, and deploy recipes. 
The recipes allow \textsc{Key4hep} to build in share installation mode, run the built-in test programs, 
and install the binaries into a distributed shared file system (CernVM-FS).

The generated events need to be injected into the simulation workflow, 
which requires a low level of interoperability between the event generator and the rest of the chain, 
named \emph{Common Data Formats}.
This level ensures that the applications exchange information files that are understood by all relevant processes, 
even on different hardware.\footnote{Passing information through files might impact performance. 
However, reading Monte Carlo events from files is not a limiting factor at the LHC and will likely not be at FCC either.}
The event generators are indeed typically stand-alone applications producing files with the generated events, 
in ASCII or binary data format. 
These event data formats have evolved in time following the requirements set by the community and are briefly reviewed in the next section. 
Some recent event generators offer enhanced interoperability with \emph{Callable Interfaces}, 
i.e., an API of functions streamlining the exchange of information among various components and increasing flexibility. 
These APIs can also be used as interface to a framework, as is described in the next section. 

\subsubsection{Common data formats}

The data formats produced by event generators provide structures to store the lists of particles per event, 
with their momenta and type, information relevant to the event, such as its weight(s), 
and information relevant to the run, such as the configuration and version of the generator. 
Several formats serving similar purposes appeared in the years ahead of the LHC. 
Among these, \textsc{StdHep}~\cite{stdhep} and \textsc{HepEvt}~\cite{hepevt} are first attempts to create standards, 
a challenge that has been assumed by \textsc{HepMC}~\cite{hepMC}, 
the latest version of which (version~3) includes readers for most of the available formats. 
The complexity of the generators developed for LHC brought in new needs, 
essentially to store information about matrix elements and alike. 
They have been included in the `Les Houches Event format' (LHEf). 
This format had initially been designed to transfer the matrix-element information to the parton shower component, 
but has since been used as a generic format at LHC and beyond, including FCC.

To handle the plethora of formats, 
\textsc{Key4hep} ensures that readers for all the formats required by the event generators of interest 
are available for all the simulation chains in use, based either on \textsc{Gaudi} or with standalone applications. 
For \textsc{Gaudi} applications, 
such as those using the framework components provided by \textsc{k4SimDelphes}~\cite{k4simdelphes} 
and/or \textsc{k4SimGeant4}~\cite{k4simgeant4}, 
the readers are implemented as \textsc{Gaudi} algorithms 
and transform the input from the files into \textsc{edm4hep} structures in memory 
that can be parsed by the next simulation algorithm. 
For standalone applications, 
such as those provided by \textsc{k4SimDelphes} (Section~\ref{sec:delphes}) or \textsc{DDsim} (Section~\ref{sec:geometryStrategy}), 
support for the formats is implemented within the applications.  

In the context of the ongoing ECFA activities for electroweak and Higgs factories, 
the \textsc{HepMC3} and \textsc{edm4hep} formats were recommended for the next versions of event generators, 
and the latter is deemed more suitable to meet the needs of the community. 
These recommendations were discussed at a dedicated event in June 2024~\cite{mcsupporttoolsevt}.

\subsubsection{Configuration}

The configuration of generators for Monte Carlo simulations often involves redundant efforts, 
as different generators aim at the same physics processes. 
To streamline this process and improve efficiency, 
the \textsc{k4GeneratorConfig}~\cite{GenConfig} package provides a unified approach to generator inputs. 
The package, which is already included in \textsc{Key4hep}, addresses essentially two challenges: 
lack of standardisation 
(each generator requiring a specific input format despite simulating, in principle, the same physics processes), 
and author involvement 
(relying on authors to adopt a common input format is impractical, 
as they are unlikely to change existing workflows or agree on a unified approach).
The proposed solution consists in the creation of a master input card defining the required parameters and configuration for simulations, 
which is then converted by a dedicated Python module into generator-specific runcards, 
ensuring compatibility across different MC tools. 
This approach will automate the creation of inputs tailored to each generator while maintaining consistency with the master card.

The targetted benefits are reproducibility and error minimisation, 
as a unified configuration process inherently reduces discrepancies and manual errors in input preparation, 
and minimises the potential for mistakes that can arise from manual handling of generator-specific configuration cards. 
The solution also provides a way to improve legacy preservation, 
as the system also supports the older `LEP era generators', 
ensuring continued usability, even when there is no active development by the original authors.

An initial set of generators has already been integrated in the system 
and an infrastructure has been set up to enable running and comparing benchmarks of the tools. 
A status report was presented at the 3$^\text{rd}$ ECFA workshop, in October 2024~\cite{GenConfigECFAParis}.

\subsection{Parametrised simulation}
\label{sec:delphes}

The \textsc{Delphes} parametrised simulation toolkit~\cite{delphes} is integrated into \textsc{Key4hep}. 
By applying resolution and efficiency functions to generator-level objects, \textsc{Delphes} offers a high-level approximation of the detector response in a highly flexible and lightweight tool. Its capabilities encompass the propagation of charged (and neutral) particles in a magnetic field, the modelling of electromagnetic and hadronic calorimeters, and the parametrisation of muon identification efficiencies. 
Basic elements like charged-particle tracks and calorimeter energy clusters are reconstructed from the simulated detector response. A simplistic particle-flow reconstruction combines tracks and calorimeter information to form a list of particle candidates in a global event description. These particles are then used as input to form higher-level objects such as isolated leptons, jets, and total/missing energy and momentum.  Although the physics performance obtained from \textsc{Delphes} represents 
the best that could be achieved with nominal sub-detector resolutions, this tool shows a reasonable agreement with the more detailed full simulation, as discussed in Section~\ref{sec:PhysPerf_FullSim}, and can thus be used to establish detector requirements and to provide rather robust estimates of the FCC-ee physics potential.   

\begin{figure}[ht]
\centering
\includegraphics[width=0.32\columnwidth]{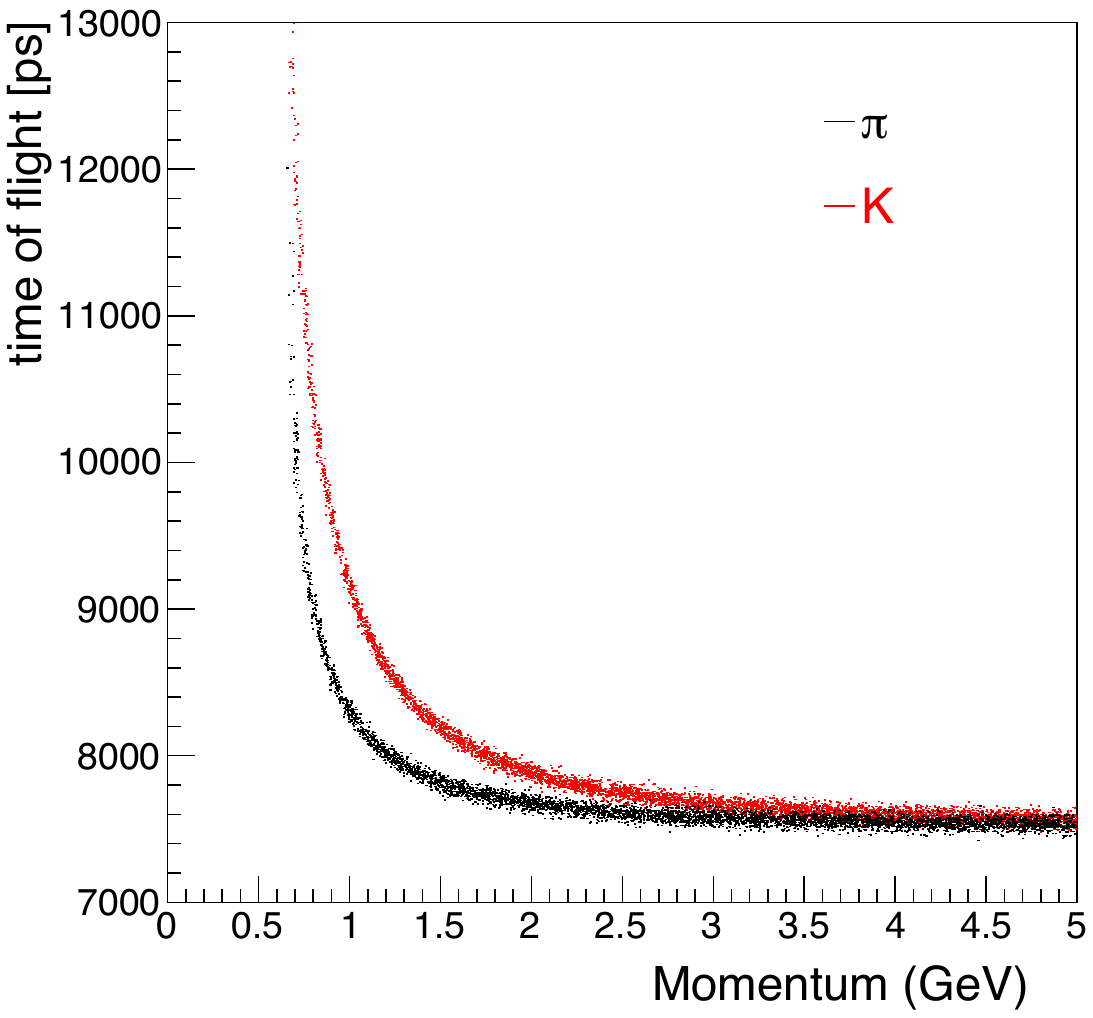}
\includegraphics[width=0.32\columnwidth]{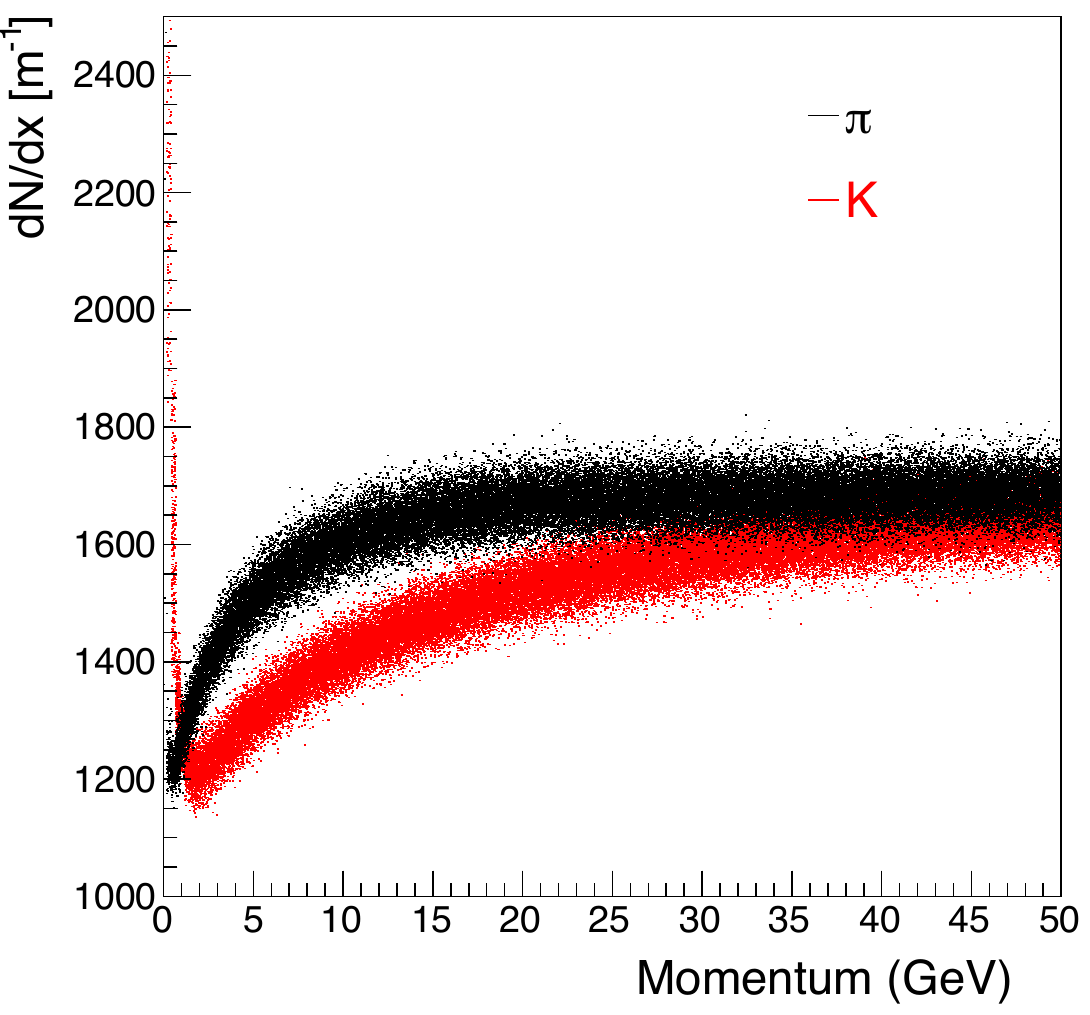}
\includegraphics[width=0.32\columnwidth]{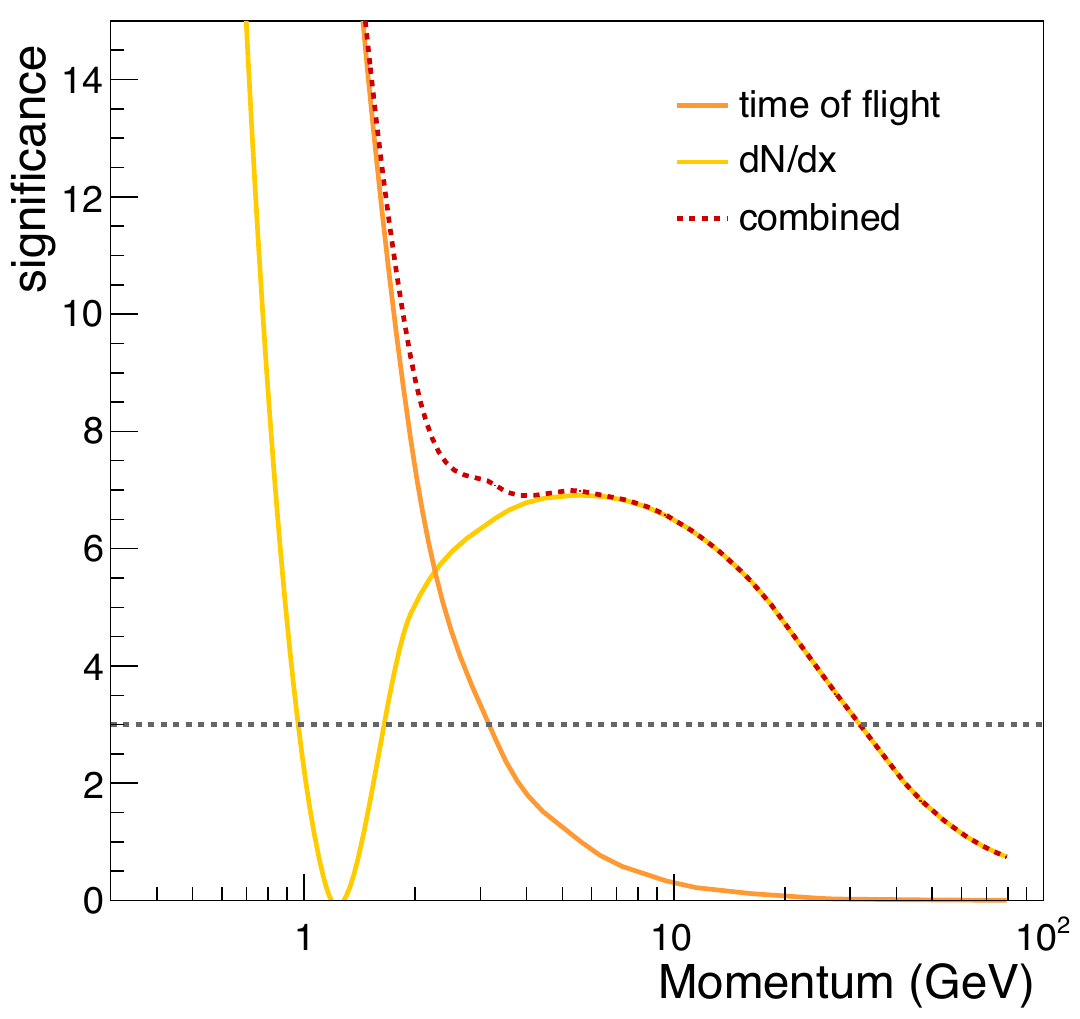}
\caption{Particle identification metrics for charged kaons and pions emitted at $\theta = 90\,^\circ$, 
as derived from the IDEA detector \textsc{Delphes} parametrisation. 
From left to right, the plots display, as a function of the particle momentum,
the time-of-flight in the tracker, 
the number of ionisation clusters per unit length for a gas mixture of 90\% He and 10\% Isobutane, 
and their respective or combined separation power in terms of standard deviations.}
\label{fig:pid_delphes}
\end{figure}

The \textsc{Delphes} toolkit was enhanced with several new features, which include an algorithm reconstructing the charged-particle track parameters and covariance matrix for arbitrary tracking geometries;  a module for deriving time-of-flight measurements; 
and a tool that parametrises the number of ionisation clusters associated with the trajectory of a charged particle in a gaseous detector, based on simulations from \textsc{Garfield}~\cite{garfield}. These tools have been incorporated into the central \textsc{Delphes} package and are employed, in particular, to train a graph neural network for jet flavour identification~\cite{flavorTaggingDelphes}. Illustrating distributions produced with these new modules are shown in Fig.~\ref{fig:pid_delphes}. 

A \textsc{Delphes} configuration card simulating the expected performance of the IDEA detector concept, 
with a detailed layer-by-layer description of its tracking system, 
has been implemented and added to the central repository~\cite{delphes_card_idea}. 
An alternative version of this parametrisation, featuring a full silicon tracking system, 
was also developed to emulate physics performance closer to what is anticipated from the CLD detector concept. 
The integration of CLD calorimeter parametrisations into this card is under development. 
More information on these parametrisations can be found in Ref.~\cite{softwareNote}.

The \textsc{Delphes} integration into \textsc{Key4hep} is achieved through the \textsc{k4SimDelphes} package~\cite{k4simdelphes}, which wraps the core components of \textsc{Delphes} and adds the functionality to convert its output into the \textsc{edm4hep} data model. 
Several command-line tools (\textsc{Delphes*\_edm4hep}) simplify the use of these extensions in stand-alone mode, 
while the \textsc{k4SimDelphesAlg} \textsc{Gaudi} algorithm enables full integration of \textsc{Delphes} simulations 
within a \textsc{Key4hep} workflow, 
allowing users to benefit from features such as the generator functionalities described in Section~\ref{sec:generators}. 

To enhance flavour physics studies in parametrised simulations, an interface between \textsc{Pythia8} and \textsc{EvtGen} is implemented, and is fully integrated with the \textsc{Delphes} applications. This interface enables efficient generation of background contributions in regions of phase space where specific signals are expected (e.g., rare hadron decays or long-lived particles). Users must apply a posteriori normalisations `manually' to ensure consistency between the generated number of events and the reported cross sections.

\subsection{Full simulation}
\label{sec:geometries}

Some applications require simulations with a higher level of detail than \textsc{Delphes}. For example, R\&D efforts and physics analyses that access detector-level information demand a more detailed description of the detector geometry and response. 
To address these needs (and, incidentally, to validate the detector parametrisation outlined in the previous section), a detailed modelling of all the (sub-)detectors considered for FCC is actively being developed. This section provides an overview of the current status of geometry implementation, along with existing examples of full detector configurations using different combinations of sub-detectors.

\subsubsection{Detector description and simulation strategy}
\label{sec:geometryStrategy}

The \textsc{dd4hep} toolkit~\cite{dd4hep} (Section~\ref{sec:k4h}) is used to model FCC detector geometries. 
This modern community standard hides the geometry complexity and allows high-level aspects 
(such as sub-detector dimensions, shape and materials, or the list of sub-detectors to be combined into a complete detector concept) 
to be configured by non-expert users with simple \textsc{xml} (`compact') files in a flexible plug-and-play approach, crucial for detector optimisation efforts.
This approach allows the optimisation of (sub-)detector variants without the need for recompilation and without an in-depth understanding of the technical details of the geometry implementation. 
It is essential, however, that the developer of the C++ detector builder (also called `detector driver') 
incorporates the necessary prescriptions and safeguards to ensure that a `sane' geometry be implemented under all circumstances. 
Several tools are available in \textsc{dd4hep} to validate the resulting geometries, such as checks to ensure that detector volumes do not overlap. 

While implementing a completely new sub-detector geometry is a more complex task, 
numerous examples and a wealth of community expertise greatly facilitate this process. 
Adding the newly created sub-detector to an existing concept, 
or building a new concept using existing sub-detectors, 
is straightforward using the `master' \textsc{xml} files that manage the list of sub-detectors and their envelope dimensions. 
This, again, assumes that the C++ drivers are designed to adapt dynamically to changes in outer envelope dimensions. 
A policy has been set in place to document \textsc{xml} parameters that require additional prescriptions before modification, 
complementing the sanity check tools to reduce the risk of creating ill-defined geometries.

The \textsc{dd4hep} framework is also well-suited for the modelling of individual modules used in test beams. 
Implementing test-beam geometries into this framework for the upcoming R\&D activities would be highly advantageous, 
as it would provide a basis for preparing complete sub-detectors, 
in view of their integration into full detector concepts. 
Occasionally, the availability of complete sub-detector geometries predates the test-beam campaigns. 
In such cases, R\&D teams could simplify and adapt the more complex full geometry to create the individual module.

The \textsc{Geant4} simulations are run on these models through the \textsc{ddSim} tool provided by \textsc{dd4hep}. 
This tool provides many functionalities that help the simulation configuration through command-line arguments or 
Python steering files (or a combination thereof), and can be run with simple particle guns or by providing input files containing the particles from Monte Carlo generators. 
Most common formats are supported: \textsc{HepMC(3)}, \textsc{HepEvt}, \textsc{StdHep}, 
and the so-called `pair' files produced by \textsc{GuineaPig++} 
(in addition to the \textsc{lcio} format). 
The output format can be \textsc{edm4hep}, \textsc{lcio}, or the \textsc{dd4hep} native format.

\subsubsection{Sub-detector models}
\label{sec:subdetectors}

Sub-detector geometries are available in the \textsc{k4geo} \textsc{gitHub} package~\cite{k4geo}.
The currently available \textsc{dd4hep} drivers, relevant for FCC-ee, are listed below. 
A more thorough description of their implementation details can be found in Ref.~\cite{softwareNote}.

\begin{figure}[ht]
\centering
\includegraphics[width=\textwidth]{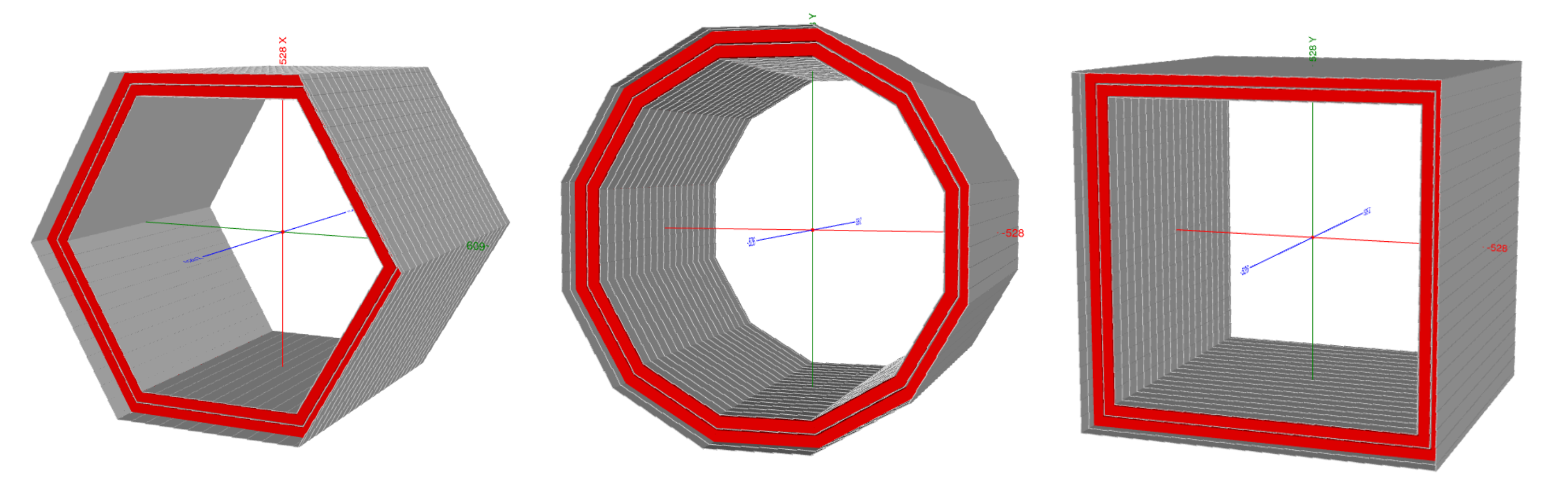}
\caption{Illustration of the flexibility provided by the detector driver that builds $\mu$RWELL-based muon systems:
different barrel implementations with 6, 12, and 4 sides (from left to right). 
The grey layers represent the sensitive detector components, while the return yokes are shown in red.}
\label{fig:muRWELL}
\end{figure}

\begin{itemize}

\item \textbf{Beam-pipe and related MDI components.} 
Two different models exist: one based on `native' \textsc{Geant4} shapes with a simplified geometry 
and the other exploiting the ability of \textsc{dd4hep} to directly import the CAD-based technical drawings, 
thus providing a much higher level of detail at the expense of poorer computing performance.

\item \textbf{Vertex detectors.} 
The driver used to model, for instance, the CLD vertex barrel detector, 
builds an arbitrary number of layers made of staves parallel to the $z$ axis with two different materials 
(one for the sensitive layers and the other for the support). 
The endcaps are constructed with simple octagonal plates perpendicular to the $z$ axis, 
made of an arbitrary number of slices with different thicknesses and materials. 
Another detector builder was recently developed to model the IDEA vertex detector. 
This driver offers a higher level of detail: 
it accounts for non-sensitive edges of the modules and includes staves/petals internal structures, together with support components. 
A switch in this driver allows the modelling of a geometry with almost support-free curved sensors.

\item \textbf{Main Trackers.}
Two versions are available: 
the former modelling a full silicon tracker, following an approach similar to that of the CLD vertex detector builder, described above, and the latter modelling a full stereo gaseous drift chamber. 
The latter features hyperboloid layers and twisted tubes to emulate the cell volume hosting the sense and field wires. 
The full geometry is built with a handful of user-provided parameters:
global dimensions, number of (super)layers, number of cells in the first layer and its increment from one layer to the next, etc. 
A third model, describing a straw-tube tracker, is under preparation.
    
\item \textbf{Dedicated particle identification detector.} 
A first version of the Array of RICH Cells (ARC) concept, described in Section~\ref{sec:concepts}, is implemented. 
The driver builds both the barrel and the endcaps, and omits, for now, the cells in the transition region, given their higher complexity. 
This detector builder includes the radiator gas and aerogel, the mirrors, the sensitive silicon sensors, 
the cooling plates and the outer vessel. 
The mirror positions and inclinations are set according to Ref.~\cite{Tat_ARC}. 
The material budget of the full sub-detector implementation corresponds to approximately 5\% of a radiation length and can be easily tuned.
    
\item \textbf{Calorimeters.} 
The following drivers are available: 
SiW luminosity calorimeter, 
noble liquid calorimeter with inclined absorber/readout planes, 
scintillating tile calorimeter (`tileCal'), 
and two fibre dual readout calorimeters, 
one with a capillary tube-based geometry and one where the fibres are directly inserted into metallic towers.
A detector driver modelling a segmented crystal-based dual readout calorimeter is being integrated in \textsc{Key4hep}. 
The simulation of dual readout calorimeters is especially challenging, 
in particular because of the number of scintillating photons generated, 
and requires special prescriptions~\cite{softwareNote}. 
The high-granularity calorimeters proposed for CLD are built with the generic driver described below.

\item \textbf{Generic detector builders.}
Some drivers have been designed generically to serve multiple purposes. 
For example, \textsc{GenericCalBarrel\_o1\_v01} and \textsc{GenericCalEndcap\_o1\_v01}, 
are used, e.g., to build the ECAL, HCAL, and muon system of CLD. 
These drivers arrange user-defined layer sequences, either radially or along the $z$-axis, 
forming a polyhedron with a customisable number of sides, set in the compact file.
Another example of a generic detector builder is \textsc{muonSystemMuRWell\_o1\_v01},
which defines stacks of layers made of user-defined tiles (e.g., PCB's)
and also provides flexibility on the number of sides, as illustrated in Fig.~\ref{fig:muRWELL}. 
This flexibility is an appealing feature at this stage of the project,
as it allows the software implementation to smoothly and rapidly adapt to detector R\&D choices. 
Unlike the formerly described drivers building the CLD calorimeters and muon system,
\textsc{muonSystemMuRWell\_o1\_v01} allows the user to define some overlap between the tiles 
(e.g., to account for their possibly non-sensitive edges) 
and stagger each side of the polyhedron to ensure a fully hermetic coverage.
This constructor is used to model the IDEA $\mu$RWELL-based muon system and pre-shower sub-dectectors. 

\end{itemize}

\subsubsection{Full detector models}

\begin{figure}[ht]
\centering
\includegraphics[scale=0.3]{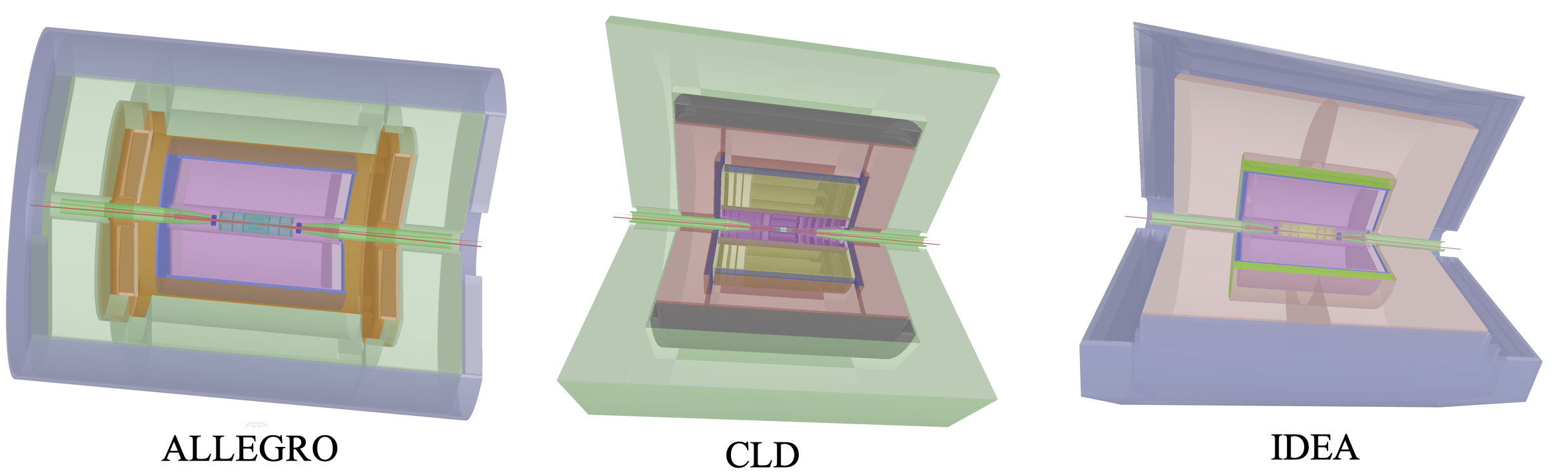}
\caption{Pictures of three detector concepts (ALLEGRO, CLD, and IDEA) 
implemented in \textsc{dd4hep}.}
\label{fig:threeDets}
\end{figure}

The availability of these sub-detector geometries along with the plug-and-play philosophy of \textsc{dd4hep} 
facilitates the creation and implementation of full detector concept alternatives. 
These models are hosted in the \textsc{k4geo} \textsc{gitHub} package~\cite{k4geo} 
and take the MDI components (beam-pipe, compensating solenoids, shields, LumiCal, etc.) 
from a central place, to ease the software maintenance. 
Pictures of three detector concepts, as they are currently implemented in \textsc{dd4hep}, 
are shown on Fig.~\ref{fig:threeDets}.

\begin{figure}[ht]
\centering
\includegraphics[scale=0.3]{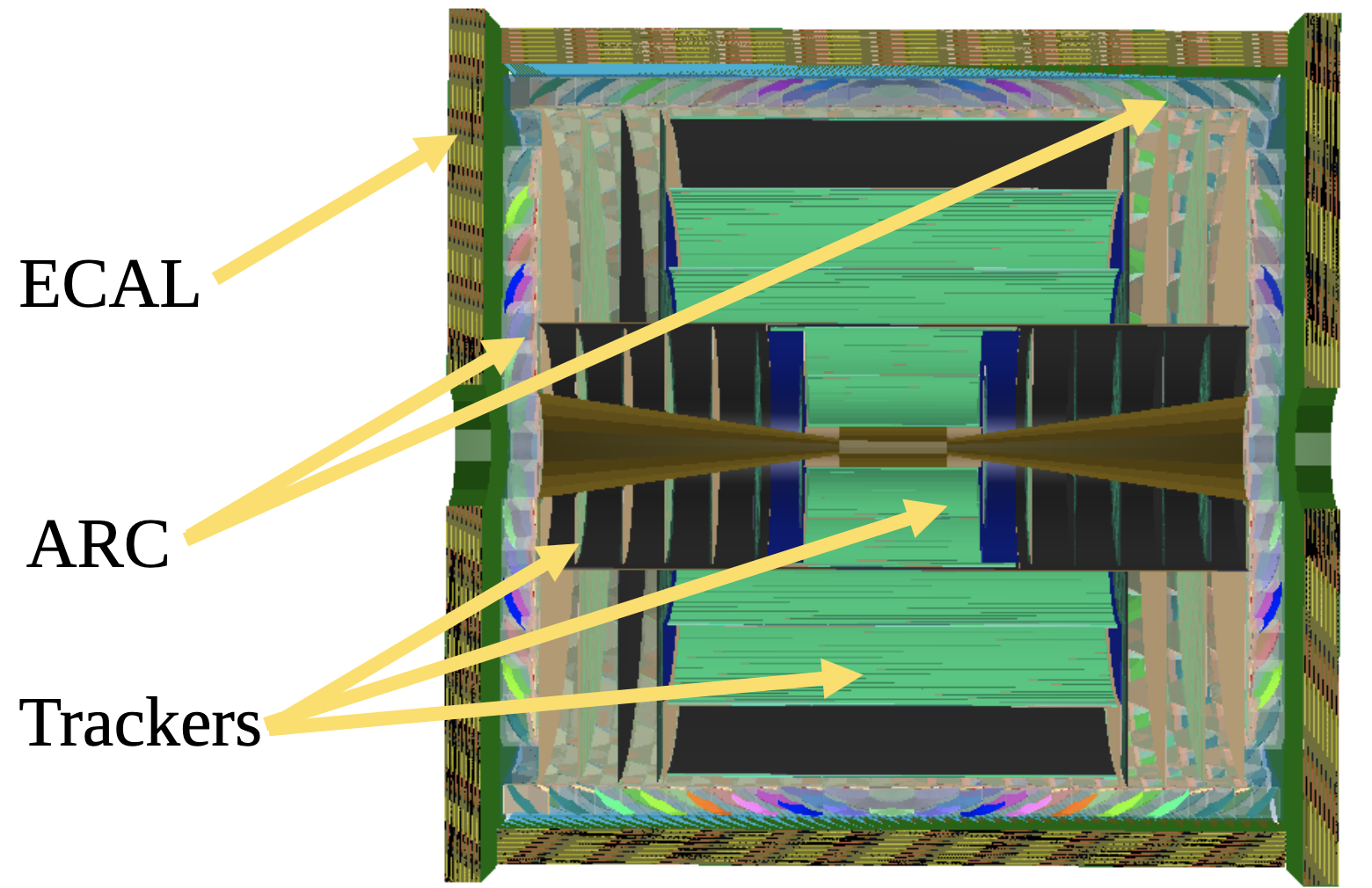}
\caption{Picture of the CLD detector concept model with enhanced particle identification capabilities. 
This variant of CLD includes the ARC sub-detector between the outer tracker and the ECAL. 
The other subsystems (HCAL, muon chambers, etc.) are present in the model but not shown here.}
\label{fig:CLD_ARC}
\end{figure}

A complete model of the CLD detector is available~\cite{CLD}. The main change with respect to the original geometry is the adaptation of the vertex detector to the most recent beam-pipe design (Section~\ref{sec:mdi}). Based on this model, an alternative option including a dedicated PID detector, named ARC, was prepared. This version of CLD features a reduced-size outer tracker in order to fit the ARC between the main tracker and the (unchanged) ECAL, as shown in Fig.~\ref{fig:CLD_ARC}. This enables comprehensive studies to evaluate the benefits of enhanced particle identification capabilities, as well as any potential impact on the performance of the particle-flow reconstruction.

A first complete model of the IDEA detector is also available. It consists of a detailed inner vertex detector with straight staves; a full stereo drift chamber with all wires included; a silicon wrapper constructed with the same driver as that used for the inner vertex detector;\footnote{With the options of changing the definition of the staves in the compact file to have sensitive components on both sides, using strips instead of pixels, and a reduction in the level of details being modelled to reach a computationally affordable description.} a solenoid and an end-plate absorber described as simple cylinders with a thickness $0.75\% X_0$; a pre-sampler; a fibre-based `monolithic' dual-readout calorimeter; and a $\mu$RWELL-based muon system. Based on this implementation, an alternative version of the IDEA detector, featuring an additional crystal-based ECAL, is now being developed.

The ALLEGRO detector concept is implemented as follows: a tracking system (inner vertex detector, drift chamber, and silicon wrapper) directly imported from the IDEA implementation; a noble liquid ECAL inside a lightweight cryostat whose outer wall also accounts for the solenoid material budget;  and a tile HCAL, placed inside a muon system place-holder made of two nested layers of sensitive cylinders.

The \textsc{Geant4} toolkit is used to simulate the passage of particles through matter. 
Its raw simulation output requires further processing to create data objects suitable for analysis. 
The processing workflow includes two key stages: 
digitisation, which transforms simulated detector hits into realistic detector responses, 
and reconstruction, which interprets these responses to identify and characterise the underlying physics objects. 
The next two sections provide an overview of the capabilities currently available in the central FCC software,
with ongoing developments and future directions. 
A more detailed discussion can be found in Ref.~\cite{softwareNote}.

\subsection{Digitisation} 
\label{sec:digi}

For silicon trackers and muon systems, a straightforward and generic digitisation approach is currently implemented. 
The simulated hit positions and time are smeared with Gaussian uncertainties, 
keeping the digitised hit position within the sensitive volume and applying an optional time window cut. 
The spatial smearing is performed in the local coordinate system of the sensitive module and allows for two distinct Gaussian widths. 
For strip-based sensors, the position along the strip direction is set in the middle of the sensor 
and the related uncertainty is defined as the wafer length divided by $\sqrt{12}$. 
This simple method allows a flexible and generic software implementation, making it applicable to various sub-detector types. 
An enhanced digitisation model that includes, in particular, charge-sharing effects, is under development. 
This new model will enable more detailed studies and simulations.

The digitisation of the drift chamber sub-detector follows a similar methodology. 
In this case, the two variables subject to Gaussian smearing are the distance from the simulated hit to the nearest wire 
and the position along the wire. 
In addition to producing digitised hits with smeared positions, 
the algorithm calculates the associated number of interactions (clusters) and the number of electrons emitted per interaction. 
This calculation is based on a parametrised model using the \textsc{Garfield++} simulation framework~\cite{clustercount}. 
A more detailed drift chamber digitisation, which will include the generation and analysis of full waveforms, is under development. 
This evolution will enable more reliable assessments of, e.g., the impact of beam-induced backgrounds on the drift chamber occupancy and on track reconstruction efficiency and purity.

A different approach is required for calorimeters: 
the simulated energy deposits corresponding to the same readout cell are aggregated
and a calibration factor (sampling fraction) is applied alongside other specific processes, as outlined below. 
\begin{itemize}
\item The digitisation of CLD calorimeters is based on the original \textsc{ILCSoft} implementation, interfaced with \textsc{Gaudi} wrappers and data format converters to \textsc{edm4hep}. This implementation supports per-layer sampling fractions, operates in both analogue and digital modes,
and includes zero-suppression emulation.

\item The ALLEGRO calorimeter digitiser is a ‘\textsc{Key4hep}-native’ solution 
(i.e., a \textsc{Gaudi} algorithm with \textsc{edm4hep} input and an output that does not require wrappers or converters) 
that handles radial-layer-dependent calibration, zero suppression, noise addition, and cross-talk emulation. 
The cross-talk emulation is performed in two steps: 
(1)~the derivation of a detector-specific map that encodes which cells ‘communicate’ with each other 
and the associated cross-talk coefficients (sourced from measurements for the ALLEGRO ECAL);
(2)~the detector-agnostic energy distribution across cells, following the guidelines set by the map.

\item The fibre dual-readout calorimeter digitiser accurately emulates the response of the silicon photomultiplier,
generating full waveforms based on photon arrival times using the \textsc{SimSiPM}~\cite{simsipm} package. 
This digitisation depends on parameters that can be obtained from vendors’ data sheets and accounts for a variety of effects, 
including wavelength-dependent efficiency, dark counts, after-pulsing, and cross-talk.

\end{itemize}

\subsection{Reconstruction}
\label{sec:reco}

The \textsc{edm4hep} digitised objects produced by the algorithms mentioned in the previous section 
are subsequently processed through various reconstruction routines. 
Two tracking solutions are currently used in the FCC software: 
one based on a conformal tracking implementation from \textsc{ILCSoft}~\cite{conformaltracking}, 
suitable for silicon tracking systems, 
and another based on a graph neural network~\cite{MLTRACKING_NOTE}, 
successfully applied to both silicon and gaseous trackers, 
as detailed in Sections~\ref{sec:PhysPerf_FullSim} and~\ref{sec:PhysPerf_TrackAngles}. 
The graph neural network solution currently covers only track finding;
ongoing developments will extend its capabilities to track fitting. 
Other `classical' approaches are also investigated, both for gaseous and silicon-based tracking systems.

Calorimeter clustering can be performed using three different \textsc{Key4hep}-native solutions: 
one based on a fixed-size sliding window for finding local maxima, 
another based on ATLAS topological clustering~\cite{topoclustering}, 
and the third using the CMS \textsc{clue} algorithm, 
which has been ported to \textsc{Key4hep} in the \textsc{k4clue} package~\cite{k4clue}. 
A fourth option for calorimeter clustering, used by the CLD detector, 
is available through the particle flow algorithm described below.

A particle-flow-based global event reconstruction is available in the \textsc{PandoraSDK}~\cite{pandorasdk} framework, 
which is integrated into \textsc{ILCSoft} and accessible in \textsc{Key4hep} via wrappers and converters. 
A complete particle-flow-algorithm sequence is in place for the CLD detector, adapted from the ILD solution~\cite{pandorapfa}. 
While this implementation already demonstrates excellent performance~\cite{CLD}, further optimisation is underway. 
The particle flow approach is expected to deliver the highest performance, in particular for jet energy and angular resolutions, 
making it the preferred tool for detector optimisation. 
Consequently, dedicated efforts are focused on several key areas:
\begin{itemize}
\item re-implementing the interface to \textsc{PandoraSDK} as a \textsc{Gaudi} algorithm with \textsc{edm4hep} input and output, 
eliminating the reliance on data converters and wrappers;
\item developing \textsc{PandoraSDK} sequences for the ALLEGRO and IDEA detectors, 
adapted to and exploiting the specificities of their trackers and calorimeters;
\item implementing machine learning-based particle flow algorithms using graph neural networks, 
as described in Ref.~\cite{MLPF_NOTE}.
\end{itemize}

The clustering of \textsc{edm4hep::ReconstructedParticleCollection} particle-flow particle candidates into jets 
is managed through a standard interface to the \textsc{FastJet}~\cite{fastjet} package. 
An additional \textsc{FastJet}-based interface is available for the clustering of calorimeter-only objects 
and can be used for detector setups that do not yet provide reconstructed particle candidates.

Several particle identification algorithms are, or will be, developed to further extend particle-flow capabilities with additional particle identification algorithms. 
These algorithms include a RICH pattern recognition~\cite{Forty1998} for the ARC detector (Section~\ref{sec:ARC}), 
a cluster-counting technique for gaseous trackers with d$N$/d$x$ measurement (Section~\ref{sec:Detectors_MainTracking}), 
and a jet flavour tagging using the deep-learning-based ParticleNet~\cite{particlenet} model. 
The latter is already implemented within the analysis framework described in the next section 
and is being migrated to a \textsc{Gaudi} algorithm for integration into the central reconstruction chain.

\subsection{Analysis tools}
\label{sec:analyses}

To maximise synergies and enhance the productivity of the FCC particle physicist teams, the analysis tools are centrally developed and maintained in \textsc{Key4hep}. 
The samples generated for FCC studies are stored using the \textsc{Root}-based \textsc{edm4hep} data model, 
both for the parametrised and full simulation workflows. 
Since this data model is built with the \textsc{Podio} library, 
\textsc{edm4hep} files can be analysed using the automatically generated helper methods (Section~\ref{sec:edm4hep}) 
in C++ or through the associated \textsc{Python} bindings. 
Although this is not the intended usage, those files can also be processed using `plain' (\textsc{Py})\textsc{Root}, 
i.e., without going through the \textsc{Podio} layer. 
A third solution, relying on \textsc{RDataFrames}~\cite{rdataframe}, is available through \textsc{Key4hep}. 
This package, called \textsc{FCCAnalyses}, has been central to most of the physics analyses presented in this report and is described below.

The \textsc{FCCAnalyses} package equips analysers with a comprehensive and efficient toolkit built around \textsc{RDataFrame}. 
By leveraging the \textsc{RDataFrame} framework, it provides automatic multi-threading, 
with object relations from \textsc{edm4hep} maintained in a structured tabular format 
through an \textsc{RDataSource} instance implemented in \textsc{Podio}. 
Event selection and high-level computations follow the standard \textsc{RDataFrame} syntax, 
while \textsc{FCCAnalyses} includes a set of general-purpose functions for added convenience. 
Samples and their metadata (cross sections, event counts, normalisation factors, etc.) are managed via \textsc{json} files, 
with centrally-produced samples browsable through a web interface~\cite{samplesWebPage} and available on \textsc{eos}. 
A plotting utility facilitates process normalisation, stacked histograms, and background vs.\ signal comparisons. 
In \textsc{FCCAnalyses}, histograms can be produced directly from input samples in a single step, but a `staged' workflow, 
allowing resource-intensive steps to be isolated for efficiency, is also supported. 
To enhance analysis capabilities, \textsc{FCCAnalyses} provides additional features, including the following ones.

\begin{itemize}
\item \textbf{Job submission} through \textsc{HTCondor}~\cite{htcondor} to CERN batch system.
\item \textbf{Jet clustering} via \textsc{FastJet}~\cite{fastjet} and functionalities to perform jet-parton matching.
\item \textbf{Machine learning integration} using \textsc{onnx~\cite{onnx}}, \textsc{tmva}~\cite{tmva} or \textsc{XGBoost}~\cite{xgboost}.
\item \textbf{Flavour tagging} with the deep-learning based \textsc{ParticleNet}~\cite{particlenet} model.
\item Automated \textbf{yield table production} for arbitrary cut-flows. 
\item Direct creation of \textsc{Root} files and associated data cards 
in a format suitable for the \textsc{Combine}~\cite{combine} \textbf{statistical analysis and combination tool}.
\item \textbf{Smearing tools} to generate new object collections from existing \textsc{Delphes} samples 
under different detector performance assumptions.
\end{itemize}
A more detailed overview is given in Ref.~\cite{softwareNote}.

In addition to this well-established toolkit, new initiatives aim at expanding the technology options offered to the analysers. 
For instance, the \textsc{Coffea}~\cite{coffea} \textsc{Python}-based framework, 
known for scalable, efficient processing of large HEP datasets, now supports \textsc{edm4hep}. 
The FCC software team is also exploring future-proof solutions, 
acknowledging shifts in paradigms akin to the HEP community’s transition from Fortran to C++. 
In this regard, the Julia~\cite{Julia} language, celebrated for its speed, ease of use, and expressive syntax, 
could address the `two-language problem' 
(i.e., the reliance on C++ for performance-intensive tasks and Python for user interface). 
To test Julia's suitability, an \textsc{edm4hep} format reader was developed as part of the \textsc{JuliaHEP}~\cite{juliahep} project, 
which aims to consolidate Julia-based tools for HEP.

\subsection{Visualisation}
\label{sec:visualisation}

Visual renderings of the software processes and outputs are crucial for purposes 
such as debugging, validation, physics interpretation, and outreach. 
A variety of related tools have been developed or integrated into the FCC software ecosystem, as described below.

\begin{figure}[ht]
\centering
\includegraphics[width=0.9\textwidth]{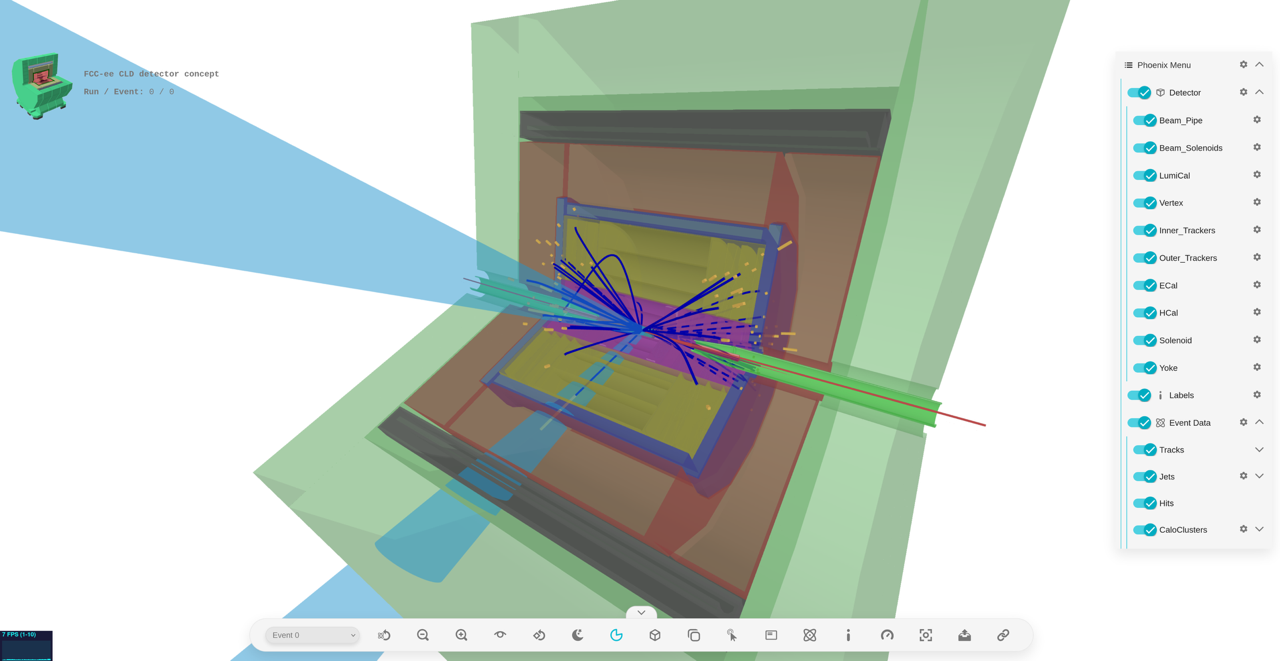}
\caption{Visualisation of a $\ttbar$ event within the CLD detector, 
displayed using the \textsc{Phoenix@FCC} web-based event display.}
\label{fig:phoenix}
\end{figure}

To enable visualisation of FCC events and detectors, 
an experiment-independent web-based event display tool, Phoenix~\cite{phoenix}, 
was extended to support the \textsc{edm4hep} data model. 
Leveraging this technology, a dedicated web interface, \textsc{Phoenix@FCC}~\cite{phoenixatfcc}, 
was created to host the central versions of FCC detector geometries. 
This interface allows users to overlay event data onto a 3D view of the detector, 
with the flexibility to easily incorporate custom detectors. 
An example of an event display generated with this tool is shown in Fig.~\ref{fig:phoenix}.

Beyond the FCC-specific implementation, several other tools are available for displaying detector geometries and event content. 
These include the \textsc{JSRoot} web interface~\cite{jsroot}, 
the \textsc{Qt} plugin integrated with \textsc{Geant4} (accessible through \textsc{ddsim}), 
and the \textsc{CED} event display from \textsc{ILCSoft}~\cite{ced}.

\begin{figure}[ht]
\centering
\includegraphics[width=0.8\textwidth]{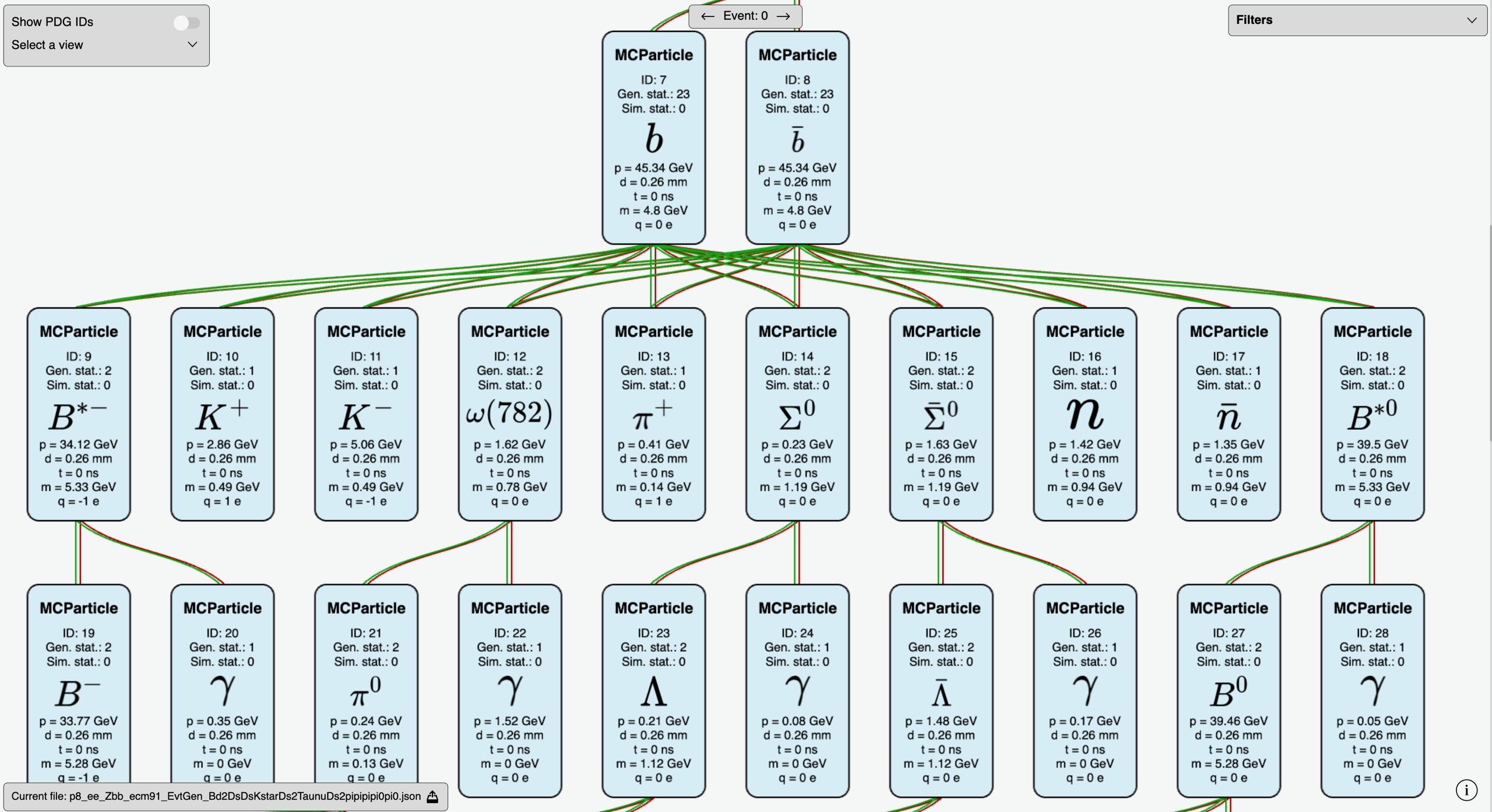}
\caption{Visualisation of the Monte Carlo particle tree for an $\epem \to \PZ \to \PQb\PAQb$ event,
generated using the \textsc{edm4hep} Event Data Explorer~\cite{eede}.}
\label{fig:eede}
\end{figure}

Another visualisation tool was developed as part of the FCC studies, 
to render the content and relationships between objects in \textsc{edm4hep}-based events. 
This tool, called the \textsc{edm4hep} Event Data Explorer (\textsc{eede})~\cite{eede}, 
visualises \textsc{edm4hep} objects as boxes that display the values of their various fields and are connected to related objects 
(e.g., a calorimeter cluster will be linked to its associated individual cells). 
The displayed content can be tuned by applying filters on the objects. 
Figure~\ref{fig:eede} shows a screenshot of this tool in use, with an excerpt of a Monte Carlo generator particle tree.

\subsection{Computing resources}
\label{sec:resource}

The computing resource requirements of a project include the processing power necessary to manage the data generated by its activities 
and the storage capacity needed to maintain those data samples at various stages of processing. 
During the Feasibility Study, the work of particle physicists has primarily focused on `offline' activities, 
which correspond to the workflows in Fig.~\ref{fig:workflows}. 
These activities are expected to remain central during the next phase of the study.
An initial analysis of the FCC offline computing needs is presented in Ref.~\cite{ganhelfccres}. 
This section builds upon and expands that analysis, with further insights. 

The FCC PED activities to date have primarily relied on resources provided by CERN, as displayed in Table~\ref{tab:cernres}. 
These resources represent approximately 0.1\% of those currently available to the LHC experiments, and will need to substantially increase.
The main purpose of this section is to give enough elements to provide reasonable projections of the needs, 
as a function of parameters that will be better defined as the project progresses. 
Current ideas to ensure that an adequate amount of resources is available for the studies are also presented.
The target timescale is that of the next phase of the study (2025--2027), 
possibly extending to the first years of the proto-collaboration forming phase (2028--2030). 
A projection to the needs of the actual \PZ-pole operations, in two decades from now, is also presented.

\begin{table}[t]
\centering
\caption{Resources available at CERN for FCC physics, experiments and detectors studies as of October 2024. 
The (old) benchmark HepSpec06 (HS06) is still used to facilitate comparison with LHC numbers; 
for the purpose of this analysis, HS06 and HS23 are numerically equivalent.}
\label{tab:cernres}
\begin{tabular}{c l}
\toprule
\multicolumn{2}{c}{Storage on \textsc{eos}} \\
\midrule
600\,TB & for central productions \\
100\,TB & for analysis \\ 
\midrule
\multicolumn{2}{c}{Processing power} \\ 
\midrule
9000 HS06 / HS23 & CPU on lxbatch \\
3 GPU & nodes on CERN Openstack \\
Some GPU & nodes on EuroHPC (via CERN OpenLab) \\ 
\bottomrule
\end{tabular}
\end{table}

\subsubsection{Resource modelling} 
\label{sec:resourceshllhc}

Reliable projections require a model for the resource needs. 
When discussing the computing resource needs, it is necessary to treat the two main phases of the project separately: 
design and preparation on the one hand, and data taking on the other.
During the design and preparation phase, the studies are performed with simulated and test-beam data, with needs increasing with time, based on expected integrated luminosities that define the statistical framework, and with many moving targets and unstructured activities (such as the number and type of detector concepts, data formats, algorithms, and levels of software optimisation). During the collider operation, both collected and simulated data need to be processed continuously, with more stable versions of detector implementations, data formats, and core-software, 
albeit with quasi-stable algorithms that will continue to be refined during the operations phase. 

\begin{figure}[ht]
\centering
\includegraphics[width=\textwidth]{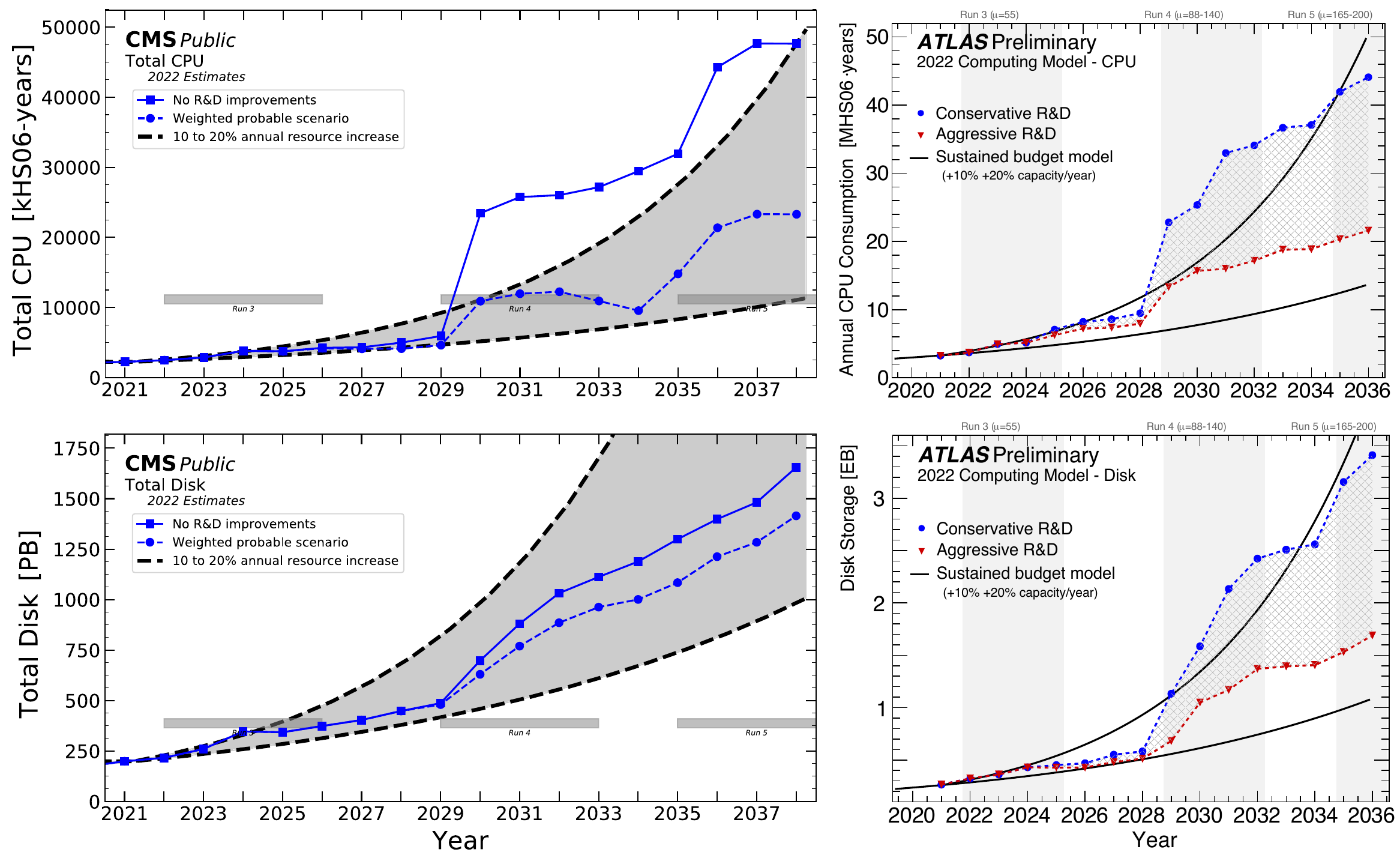}
\caption{Projections of CMS (left) and ATLAS (right) computing (top) and storage (bottom) resource needs for HL-LHC. 
In both cases, the demands in case of no or conservative R\&D and planned or aggressive R\&D 
are compared with projections of future pledged resources. 
From Refs.~\cite{CMSreshllhc} and~\cite{ATLASreshllhc}. 
For ATLAS and CMS, the data and Monte Carlo samples have roughly the same size~\cite{PiparoPriv}.}
\label{fig:resourceshllhc}
\end{figure}

The significant and sustained effort dedicated to modelling the computing resource needs in ATLAS and CMS has been thoroughly documented, and has quantified both the short- and long-term projections for the high-luminosity LHC programme. An example can be found in Ref.~\cite{langecmsres}. The most recent update is shown in Fig.~\ref{fig:resourceshllhc}, from the latest CMS~\cite{CMSreshllhc} and ATLAS~\cite{ATLASreshllhc} approved figures, with similar real and simulated data samples. Not surprisingly, the projected needs are shown to decrease with more aggressive R\&D efforts. In particular, past investments in R\&D, software quality/performance improvement, and optimal use of storage in the past decade, have proven to be critical in reducing the HL-LHC computing needs of the experiments. The projected needs are compared to the extrapolated available computing capacities under a sustainable flat budget model. 
From theses numbers, it is possible to derive that both ATLAS and CMS will have storage needs of the order of a few EB and computing needs of the order of a few MHS06.\footnote{An HS06 = HEPSpec06 is a benchmark to measure the computing capacity of a Computing Element. 
The rule of thumb `core = 10 HS06' is used for the hardware available (at the time of writing) at, for example, the CERN facilities.}

The basic concepts of resource modelling employed by the LHC experiments can be applied to the FCC case, which 
boils down to identifying the activities and the corresponding set of objectives and required workflows. 
The detailed activities and objectives depend on the phase of the project, although macro-activities can be present at all stages.
Today, and in the near future, the identified macro-activities include data analysis, sub-detector development and optimisation, and algorithm design and optimisation.

\paragraph{Data analysis}

This activity will always be present, in all phases of the project. 
During the design phase, the goal is to study the potential of the project with a level of accuracy that depends on the aspect being investigated. 
A first feasibility study of a measurement is typically done using a parametrised simulation of the response of a given detector concept, 
to identify critical points that would require a more accurate simulation. 
Parametrised (or fast) simulation is also required for an experiment during data taking, e.g., for a rapid exploration of large portions of parameter space. 
As the studies progress, the need for accuracy increases, which calls for  fully-simulated and properly-reconstructed events. 
Therefore, the analysis activity will always bring requirements for adequate samples of parametrised and fully simulated-reconstructed events, 
their relative importance depending on the phase of the project. 
An example of a presently ongoing analysis macro-activity with full simulation is discussed in Section~\ref{sec:PhysPerf_FullSim}. 

\paragraph{Sub-detector development and optimisation}

This activity is specific to the design phase, during which the detector response is simulated,
primarily using particle guns to generate particles of a specific type, within a given region of phase space. 
Additionally, samples of simulated events may prove useful and serve as benchmarks for the design process.
While the size of these samples is usually not enormous, 
they may require a higher level of detail than is typical for standard physics analyses, 
potentially resulting in large data volumes. 
Examples of such activities are discussed in Chapter~\ref{sec:concepts}.

\paragraph{Algorithm design and optimisation}

This activity is present during all phases of a project. 
During the design phase, a minimal set of algorithms is identified to evaluate the performance of the proposed (sub)detectors, and to subsequently improve and tune their design. 
During later stages, when the detector design is frozen (or when the detector is built), the algorithms are continuously developed for improved performance. 
This activity makes use, in particular, of particle guns and simulated events enriched in certain topologies. 
Examples of such activities are presented in Chapter~\ref{sec:requirements}.
Advanced AI approaches, such as ML-based tracking and particle flow algorithms, are complementary to more classical approaches in both the design and production phases. During the design phase, they offer a rapid re-optimisation of varying designs and concepts; during the production phase, they are expected to deliver better performance.

\subsubsection{Modelling resources for FCC-ee}

At this stage, quantitative predictions are only possible for the analysis macro-activity. The necessary basic ingredients are the event sizes and the event processing time. The current activities mostly use parametrised simulations, hereafter labelled \textsc{Delphes}, 
for which rather complete sets of events have been produced. 
The numbers presented in the following are taken from the \emph{Winter 2023} production campaign. 
For the full simulation, some measurements have been performed with the simulation of selected samples with CLD, taken as a reference. 
It is assumed that the event sizes and the event processing times scale in the same way as for \textsc{Delphes}. 
This assumption is based on the fact that these numbers depend mostly on the number and type of particles, 
and that they affect the sizes and processing time in the same way in \textsc{Delphes} and full simulation.
The resulting event sizes and processing times are reported in Table~\ref{tab:baselinesizes}. 

\begin{table}[ht]
\centering
\caption{Baseline event sizes and processing times. Parametrised simulation and full simulation with \textsc{Geant4} are denoted  
\textsc{Delphes} and \textsc{Full}, respectively. For \textsc{Full}, the values are extrapolated from measurements performed with $\PZ \to \PQq\PAQq$ events, under the assumptions described in the text. }
\label{tab:baselinesizes}
\begin{tabular}{c c c c c c }
\toprule
Process & $\sqrt{s}$ & \multicolumn{2}{c}{Size / event } & \multicolumn{2}{c}{Processing time / event }  \\ \cline{3-6}
$\epem \to$ & (GeV) & \textsc{Delphes} (kB) & \textsc{Full} (MB) & \textsc{Delphes} (ms) & \textsc{Full} (s) \\ 
\midrule
$\PZ \to \PQq\PAQq$, $\ell^+\ell^-$ & 91.18 & 8.3, 1.2 & 1.1, 0.16      & 14, 0.5 & 11, 1.6 \\
$\PWp\PWm \to \text{all}$, $\PGn\PAGn \ell^+\ell^-$ & 157--163 & 9.5, 1.2 & 1.3, 0.16 & 16, 0.5 & 13, 1.6 \\
$\PH\PZ \to \PGn\PAGn \PQb\PAQb$, $\PQb\PAQb \PQb\PAQb$ & 240 & 8.9, 13 & 1.2, 1.8    & 15, 23 & 12, 18 \\
$\PZ\PZ \to \text{all}$ & 240 & 10 & 1.4                              & 17 & 13 \\
$\ttbar \to \text{all}$ & 365 & 18 & 2.3                           & 30 & 23 \\ 
\bottomrule
\end{tabular}
\end{table}

To project the needs in the years to come, 
an assumption on the needs in terms of sizes of the simulated samples has been made. 
A rule of thumb often used to get approximate values is to assume simulated samples corresponding to the expected integrated luminosity at each centre-of-mass energy, 
as displayed in Table~\ref{tab:seqbaseline} of this report. 
Table~\ref{tab:baselineneedsfull} shows the corresponding needs, for one experiment, in terms of computing resources.

Reconstruction and analysis requirements are not included in Table~\ref{tab:baselineneedsfull} 
because of the lack of sufficient quantitative information. 
However, the size of the \textsc{Delphes} output, designed to emulate the reconstruction process, 
can serve as a basis for estimating the reconstruction needs. 
While the actual reconstruction may require more detailed information than what is currently provided by \textsc{Delphes}, 
it is unlikely to significantly impact the overall resource demands. 
This conclusion is even clearer for the final analysis ntuples, which represent a condensed version of the \textsc{Delphes} output.
The processing power needs are more difficult to quantify. 
For reference, at LEP and Belle~II, the reconstruction step took about 30\% of the full simulation CPU time~\cite{ganhelfccres}, 
which can be taken as a conservative estimate, given that (opportunistic) heterogeneous resources can play a relevant role in here.
The numbers reported \emph{for one experimennt} in Table~\ref{tab:baselineneedsfull} 
give therefore a reasonable estimate of the scale of the 
problem.\footnote{It should be noted that, for \textsc{Delphes}, 
in principle the sample can be reused for a different \emph{detector concept} on the fly 
with minimal use of additional processing resources.}
These numbers show that the computing needs for the \PZ run are of the same order of magnitude as those 
of ATLAS and CMS  for HL-LHC. For the other runs, instead, the table shows that full nominal integrated luminosity samples are achievable today.

\begin{table}[t]
\centering
\caption{Projected needs for the nominal integrated luminosity, for one experiment. 
The amount of HS06 is shown for a reference period of three years, i.e., roughly the duration of the next phase of the study.}
\label{tab:baselineneedsfull}
\begin{tabular}{c c c c c c c}
\toprule
    &         &  & \multicolumn{2}{c}{\textsc{Delphes}}  & \multicolumn{2}{c}{\textsc{Full}} \\ \cline{4-7} 
\multirow{2}{*}{Run} & \multirow{2}{*}{Process} &   Number & Storage & CPU & Storage & CPU \\
    &         &   of events & (PB)          & (HS06) & (PB)        & (HS06) \\ \midrule
\multirow{2}{*}{\PZ} & $\PQq\PAQq$     &  1500 G  & 12.5          & 2.2 k  & 1650         & 2 M \\ 
   & $\ell^+\ell^-$    &  225 G   & 0.275         & 12    &   40         & 40 k \\ \midrule
\PW & $\PWp\PWm$      &  60 M    & $\sim 10^{-3}$&       & 0.075        & 72 \\ \midrule
\multirow{2}{*}{$\PH\PZ$} & $\PH\PZ$     &  500 k   & $\sim 10^{-5}$&       & $\sim 10^{-3}$ & $\sim 1$ \\ 
   & $\rm VBF\-H$ &  16 k    & $\sim 10^{-6}$&       & $\lesssim 10^{-3}$ &  \\ \midrule
\multirow{3}{*}{top} & $\ttbar$ &  500 k   & $\sim 10^{-5}$&       & $\sim 10^{-2}$ & $\sim 1$ \\ 
   & $\hz{}$      &  90 k    & $\sim 10^{-6}$&       & $\lesssim 10^{-3}$ &  \\ 
   & VBF\,H &  23 k    & $\sim 10^{-6}$&       & $\lesssim 10^{-3}$ &  \\ \midrule
\multicolumn{2}{c}{Total} &  1725 G    & {\bf 13}           & 2.2  & {\bf 1690} & {\bf 2 M} \\ 
\bottomrule
\end{tabular}
\end{table}

\subsubsection{Minimal baseline working scenario}

These considerations enable the design of a minimal working scenario to serve as a baseline target. 
This scenario includes, as a minimum, full nominal integrated luminosity samples for the \PW, $\PH\PZ$, and $\ttbar$ datasets, 
as well as `sufficiently large' samples for the \PZ run, 
where the precise definition of `sufficiently large' remains to be determined.
Table~\ref{tab:baselineneedsplan} reports the values, assuming that `large enough' means 100 times the LEP event samples.
The total values per detector and for all four detector concepts are also provided, assuming that they have similar average demands. 
These numbers are not far from the resources currently available and are well within the range of what can be realistically achieved, 
as discussed later. 
Consequently, the baseline `100$\times$LEP' scenario appears to be a realistic and achievable target.

\begin{table}[ht]
\centering
\caption{Projected needs for one experiment, 
for the scenario with nominal integrated luminosity samples for the \PW, $\PH\PZ$ and $\ttbar$ runs, 
and event samples 100 times larger than the LEP samples for the \PZ run. 
The total corresponds to four experiments requiring the same resources. 
The amount of HS06 is shown for a reference period of three years, i.e., roughly the duration of the next phase of the study. 
The totals in bold are beyond today’s availability.}
\label{tab:baselineneedsplan}
\begin{tabular}{c c c c c c c}
\toprule
    &         &  & \multicolumn{2}{c}{\textsc{Delphes}}  & \multicolumn{2}{c}{\textsc{Full}} \\ \cline{4-7} 
\multirow{2}{*}{Run} & \multirow{2}{*}{Process} &   Number & Storage & CPU & Storage & CPU \\
    &         &   of events & (TB)          & (HS06) & (TB)        & (HS06) \\ \midrule
        \PZ & $\PQq\PAQq$    &  400 M    & 3.25        & $\sim 1$  & 440     & 475 \\ 
        (100$\times$LEP) & $\ell^+\ell^-$ &  42.5 M   & 0.05        &           & 6.5     & 7 \\ \midrule
        \PW & $\PWp\PWm$     &  60 M     & 0.6         &           & 75      & 72\\ \midrule
\multirow{2}{*}{$\PH\PZ$} & $\PH\PZ$    &  500 k     & 0.0065      &           & 1       & $\sim 1$ \\ 
                & VBF\,H &  16 k  & $\sim 0.001$ &           & 0.25    &  \\ \midrule
\multirow{3}{*}{top} & $\ttbar$ &  500 k     & 0.009       &           & 9       & $\sim 1$ \\ 
                & $\PH\PZ$     &  90 k  & $\sim 0.001$ &           & 0.2     &  \\ 
                & $\rm VBF\-H$&  23 k   & $\sim 0.001$ &           & 0.25    &  \\ \midrule
\multicolumn{2}{c}{Total}        &  500 M    & 4           & $\sim 1$  & {\bf 530} & {\bf 550} \\ 
\multicolumn{2}{c}{4 experiments}        & 2000 M    & 16          & $\sim 4$  & {\bf 2100} & {\bf 2200} \\ 
\bottomrule
\end{tabular}
\end{table}

Efficiently managing limited computing resources requires a three-fold optimisation approach for the software, the analysis techniques, and the workload and data management. A proactive approach to resource management is essential for securing ongoing support from official bodies, 
especially as computing needs continue to grow. The following paragraphs discuss the current (or planned) pertinent investigations. 

\subsubsection{Optimising resources: Software}

\paragraph{Software quality and efficiency improvements}

Upgrading the code quality and efficiency is fundamental, though it often involves a considerable investment in time and expertise. 
By continually using the latest software versions, projects can leverage optimisations and new capabilities. 
For instance, recent updates to \textsc{Geant4} have nearly doubled the speed of simulation for the ATLAS experiment, 
illustrating the potential performance gains achievable through modernised codebases. 
Moreover, new, faster simulation techniques are emerging and being integrated in the software frameworks 
for use in production\footnote{Parts of the electromagnetic calorimeter of ATLAS are simulated with Machine Learning techniques.}, 
and may offer further efficiency improvements. 
End-to-end fast simulation based on Machine Learning techniques are also being developed, 
which could provide interesting ways to test very large event samples~\cite{flashsim-chep24}. 

\paragraph{Selective and filtered simulation approaches}

Another opportunity for optimisation lies in selective or filtered simulations. 
By applying filters at the generation stage, 
simulations can focus only on event components relevant to specific analyses, 
reducing unnecessary computation. 
For example, simulating only sub-detectors relevant for a given study could save substantial processing power. 
This strategy requires a careful balance between computational savings and scientific accuracy, but could significantly improve throughput.

\paragraph{Leveraging heterogeneous resources for reconstruction and analysis}

Heterogeneous computing environments, combining CPUs, GPUs, and other accelerators, 
offer significant opportunities for resource optimisation, particularly for computationally intensive tasks such as reconstruction and analysis. 
This approach has already proven highly valuable at the LHC, where the use of GPUs and specialised hardware for suitable tasks has improved efficiency, 
freeing traditional CPU resources for other activities. 
Extensive R\&D programmes are underway to develop the potential of these technologies in view of 
HL-LHC.\footnote{An example is the Next Generation Trigger project~\cite{kickNGT}.} 
Integrating these advances into the FCC software ecosystem to ensure seamless compatibility and optimal performance 
across diverse processing architectures is of critical importance for the FCC project.  
    
\subsubsection{Optimising resources: Analysis techniques}

\paragraph{Enhanced interplay between full and parametrised simulation}

In scenarios where full simulation is unnecessary, parametrised simulations, such as those provided by \textsc{Delphes}, 
can produce sufficiently accurate results with far lower resource demands. 
Facilitating interaction between full and parametrised simulations allows researchers to switch between these approaches as needed, 
optimising resource use for each stage of an analysis. 
Developing automated or optimised configurations for \textsc{Delphes} would further support this flexibility, 
making it easier to adapt simulations based on the specific needs of each analysis.

\paragraph{Advancing beyond traditional statistical methods}

By default, HEP analyses rely on a straightforward rule of thumb: 
generating MC samples at least as large as the expected data sample. 
While simple, this approach is resource-intensive, particularly in scenarios with high data volumes and complex simulations. 
Alternative statistical methodologies that allow similar statistical power with fewer simulated events 
are already explored in some fast simulation domains with promising results~\cite{varred}; 
these variance reduction techniques should be further investigated to optimise computational resources.    

\subsubsection{Optimising resources: Workload and data management}

A home-made solution effectively supported the production on the local CERN \textsc{HTCondor} and \textsc{eos} resources until the time of writing. 
The increasing complexity and scale of FCC projects necessitate a more robust and distributed infrastructure.
The primary requirements for the updated system include: 
a centralised file catalogue to organise and track data;
data replication to ensure redundancy and accessibility across sites; 
and remote access capabilities to facilitate seamless interaction with distributed resources.

To address these FCC needs, \textsc{iLCDirac}, the instance of the \textsc{Dirac} framework used in past linear collider studies, has been adopted. 
This framework is widely recognised for its workload management and file catalogue capabilities, and is already employed by several high-energy physics experiments, such as LHCb, Belle~II, BES~III, JUNO, etc. 
Through this integration, the FCC Virtual Organisation (FCC~VO) has become part of \textsc{iLCDirac}.
The CERN \textsc{HTCondor} and \textsc{eos} resources have been associated with the FCC~VO within \textsc{iLCDirac}. 
In addition, steering applications tailored to FCC workflows have been integrated into the framework 
to support specialised production needs.

Production with \textsc{iLCDirac} has begun, marking significant progress in adopting the distributed model. 
Efforts to include external storage elements (SE) have been initiated with sites at CNAF, Bari, and Glasgow. 
While these integrations are still in the testing phase, 
they represent an essential step towards expanding the infrastructure capacity and flexibility.
This approach ensures that FCC projects benefit from a scalable and distributed workload management system, 
ready to handle the challenges of future production and analysis tasks.

\subsubsection{Increasing available resources: pledged resources}

At CERN, FCC currently obtains computing resources from quotas reserved for the `Small and Medium Experiments' (SME) projects. 
These resources, while sufficient for initial activities, are limited and cannot realistically be increased by a large factor in the SME context. 

To address future needs, 
discussions are ongoing with the WLCG and CERN resource management teams to integrate FCC into the WLCG computational infrastructure. 
This integration, which should begin as early as 2025, 
would establish a reliable framework for resource allocation and enable a steady increase in computing and storage capacity over time. 
Current projections suggest that, by the end of the next phase of the study, in 2027, 
a tenfold increase in resources should be achievable, equivalent to approximately 5\,PB of storage and 100\,kHS06 of CPU power. 
These resources would be sufficient to support the minimal baseline `100$\times$LEP' scenario discussed earlier, 
while also providing contingency for targeted studies requiring larger data samples.

\subsubsection{Increasing available resources: opportunistic resources}

To enhance resource availability for FCC projects, leveraging opportunistic resources is a key strategy, which includes high-performance computing (HPC) facilities, national resources, and advanced development initiatives.

On the HPC front, discussions are ongoing with CERN OpenLab to exploit EuroHPC resources for FCC needs. 
The focus is on providing transparent access through the CERN OpenLab interface, 
which is being developed to support diverse hardware architectures, such as AMD, Intel, and ARM CPUs, as well as NVIDIA and AMD GPUs.
Initial estimates suggest access to substantial computing resources over periods ranging from six months to one year, 
which would support specific, time-limited studies requiring intensive computational power.
Investigations are underway to integrate these HPC resources into \textsc{Dirac}, ensuring streamlined access for FCC workflows.
The HPC resources are particularly suitable for specialised use cases that demand intense but temporary computational efforts, 
such as AI and machine learning (ML) developments, and
physics-specific simulations or statistical studies requiring short-term bursts of computing power.

Resources available at individual institutes or through national departments could significantly contribute to FCC workloads. 
These resources can potentially be included within the FCC~VO through integration with the \textsc{Dirac} framework, 
enhancing the overall pool of accessible computational power.

\subsubsection{Projecting to the \texorpdfstring{\PZ}{Z} run}

\begin{figure}[t]
\centering
\includegraphics[scale=0.7]{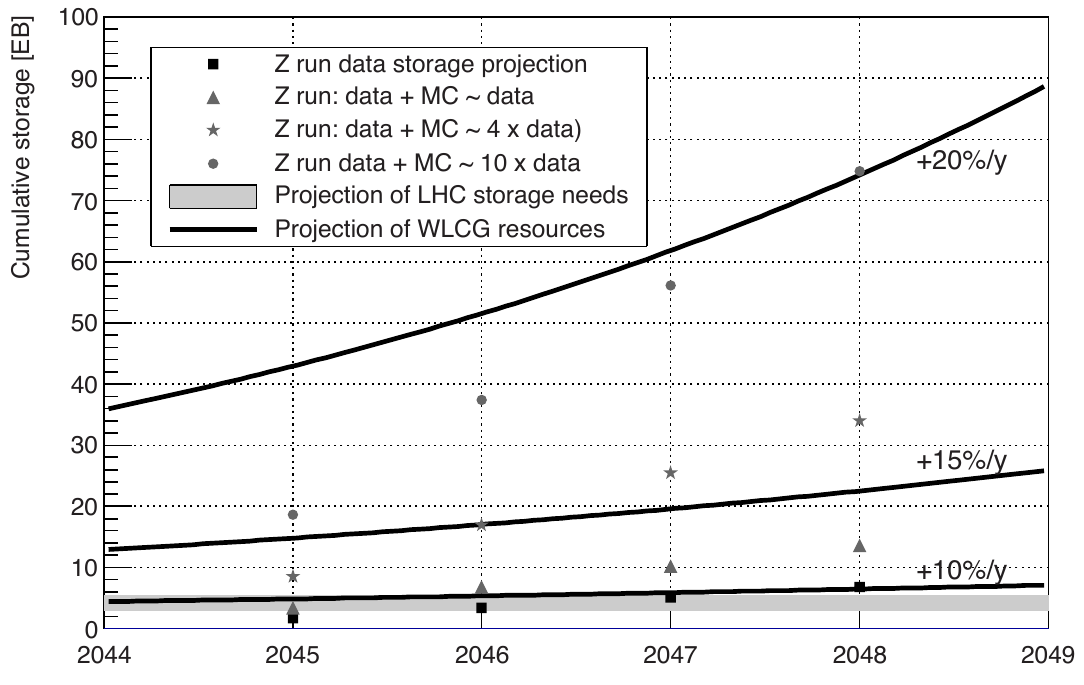}
\caption{Projection of the current storage resources to the FCC-ee \PZ-pole run with four experiments collecting equal amounts of data in four successive years (squares) 
and varying amounts of simulated data (triangles, stars, circles). 
The figure also includes the projected resource needs of the LHC and a hypothetical evolution of WLCG resources, under different scenarios for sustained annual budget increases.}
\label{fig:StorageProjZRun}
\end{figure}

All the above provides a basis for understanding the computing needs during the \PZ run. 
The analysis focuses on storage requirements, likely to be the most critical aspect.
The squares in Fig.~\ref{fig:StorageProjZRun} show the projected FCC storage requirements for the real data collected during the \PZ-pole
run,\footnote{A similar situation applies to CPU needs, 
though it is important to consider the previously mentioned challenge of adapting to heterogeneous resources. 
Nonetheless, it is anticipated that by 2045 this issue will likely no longer be a significant obstacle.} with four experiments and the assumption of four identical runs in four years. Also shown are 
the projections for simulated data with size identical to the real data (triangles), four times larger than the real data (stars) and ten times larger than the real data (circles). 
The horizontal band indicates the HL-LHC's projected storage requirements, 
as derived from Fig.~\ref{fig:resourceshllhc} assuming that ATLAS and CMS account for approximately 90\% of the total.

The storage resources required for FCC-ee data during the \PZ run are comparable to the projected requirements of the LHC. 
The inclusion of simulated samples, however, will substantially increase the storage demands, even in the minimal scenario, 
in which the simulated samples have a size similar to that of the real data samples. A dedicated computing infrastructure, 
similar to the World LHC Computing Grid (WLCG) for the LHC, is therefore required to meet the computing demands of the FCC. 
As previously mentioned, this could possibly be a natural extension of the WLCG. 
The projected storage capacity of the WLCG during the years of the FCC-ee \PZ-pole run, 
under various assumptions regarding sustained annual budget increases, is also shown in Fig.~\ref{fig:StorageProjZRun}. 
While provided for illustrative purposes only, 
this projection suggests that a natural evolution of the current model could effectively supply the required resources.

\subsection{Human resources: status and needs}
\label{sec:humanresource}

The substantial progress achieved in the software ecosystem during the FCC Feasibility Study 
has demonstrated that a dedicated software core team at CERN is pivotal in driving and coordinating the majority of FCC activities. 
This team has evolved over time, initially operating within the CERN EP-SFT group and then, in September 2024, 
moved to the newly-established CERN EP-FCC group. 
The current workforce at CERN includes (in FTEs)
1.4 staff members, providing leadership and continuity, 3 fellows, and 3 students (technical and doctoral).
Additionally, the CERN EP R\&D program has been instrumental in providing fellow support, 
particularly for the development and advancement of \textsc{Key4hep}, a critical component for the FCC activities.
Contributions from external institutes have increased, in particular from collaborators in France, Italy, and the United States. 
This expanding network of external contributors strengthens the FCC effort by bringing diverse expertise, 
resources, and perspectives, complementing the core team’s efforts at CERN.

Consolidating and expanding the team is essential to sustain progress, address emerging challenges, 
and capitalise on synergies with other initiatives. 
To strengthen this team, efforts are being made to secure two additional staff positions for long-term stability and expertise;  
two continued fellows to retain knowledge and maintain operational continuity; 
two technical students, providing critical support for development tasks and creating a pipeline of future talent.
In addition, a dedicated IT contact would be beneficial to foster strong relationships with the CERN~IT service groups, 
and ensure streamlined access to IT services and infrastructure critical to the FCC activities.

The software ecosystem central to FCC workflows, \textsc{Key4hep}, requires permanent development, consolidation, and maintenance. 
Establishing long-term support mechanisms is crucial to ensure the continued relevance and effectiveness of \textsc{Key4hep} . 
Discussions with CERN EP management are underway, 
to secure a sustainable future for the project as it evolves from its current R\&D phase into a baseline activity.

\subsection{Outlook}

Before and during the Feasibility Study, a constant concern has been to overcome the challenges of limited resources, 
while being able at all times to evaluate the FCC detector requirements and physics potential, 
with a robust software and computing infrastructure. 
It was therefore strategically decided to maximize the use of already existing solutions in a common software framework, 
while broadening the applicability of common tools and frameworks to meet the evolving needs of the project. 
The resulting \textsc{Key4hep} ecosystem is now routinely used by FCC particle physicists in their daily work. 
To fully reap the rewards of such a transformative approach, it will be essential to further generalise its use, 
in particular in the scope of the ongoing DRD efforts and, later, within the experimental Collaborations.

The long-term viability and adaptability of the framework will require continuous consolidation and harmonisation, 
to efficiently support the required workflows. 
To maximise the effectiveness of the available computing resources, 
robust support for multi-threading must be guaranteed, 
and mechanisms to handle heterogeneous resources should be implemented wherever applicable. 
These efforts are closely aligned with HL-LHC developments, 
making it imperative to ensure the seamless integration of these advances into \textsc{Key4hep}. 
As \textsc{Key4hep} transitions out of the R\&D phase, 
it will be crucial to establish a sustainable support model with contributions from collaborating institutes. 

Significant progress has been made in the development of full simulation tools, 
including the essential plug-and-play capability, 
but much remains to be accomplished in this domain. 
A critical task will be the implementation of detailed signal digitisers, 
where the DRD teams are expected to play an important role. 
To consolidate the conclusions of the Feasibility Study, 
from beam-induced background simulations to detector requirements and physics potential, 
the complete SIM-DIGI-RECO chain must be established for all present and future detector subsystems and concepts.   
As new sub-detector technologies emerge, new detector configurations, distinct from those studied during the Feasibility Study, will be explored. 
More flexible geometry implementations and improved sub-detector interoperability will be required. 
The generalisation of common reconstruction algorithms, e.g., for tracking, particle identification, and particle-flow reconstruction, 
will empower the community to identify optimal detector designs with a minimal investment of human resources. 
Advanced AI approaches will play a pivotal role in this effort.

Future full simulation studies will demand a significant increase of both computing and human resources. 
Meeting these growing computational needs involves pursuing multiple avenues inspired by the LHC model, 
with integration into the WLCG resource pool, leveraging HPC calls, and exploring opportunistic use of available resources. 
Should the acquisition of additional resources prove challenging, mitigation techniques have been identified and will need to be implemented. 
Improving the interface between centrally-produced samples and analysis frameworks, 
including Python-based frameworks, is essential. 

Closer ties with the LHC community will need to be developed, as they can unlock mutual benefits, 
by leveraging shared technologies that advance both the FCC and LHC objectives.  
In this respect, promoting positions shared between LHC and FCC might foster joint development efforts and enhance expertise exchange. 
Finally, to broaden the project resource base, continued efforts are required to attract more external contributors from institutes worldwide, 
to raise awareness and interest in FCC among the broader scientific community, 
and to encourage partnerships and resource sharing.

\cleardoublepage
\section{Energy calibration, polarisation, monochromatisation}
\label{sec:epol}
\subsection{Overview}

Excellent knowledge of the collision energy, $\sqrts$, is vital for many of the most important measurements that will be performed at FCC-ee, in particular the determination of the \PZ-resonance parameters, and the mass and width of the \PW boson.
To achieve this goal requires calibrating the mean energy of each beam  around the ring, $E_\mathrm{b}$, in principle not identical for electrons and positrons but here designated with a single symbol for simplicity. 
The collision energy $\sqrts$ can then be calculated, provided there is sufficiently good knowledge of the crossing angle of the two beams and all effects that give rise to local shifts of the energy at each interaction point. 

Circular colliders have the unique attribute that transverse polarisation naturally accumulates through the Sokolov--Ternov effect, and the spin tune, which is the ratio of the precession frequency to the revolution frequency, is directly proportional to $E_\mathrm{b}$. 
The spin tune can be directly measured by the procedure of resonant depolarisation (RDP), in which the frequency of a depolarising kicker magnet is adjusted until the polarisation is found to vanish. 
This technique, which has an intrinsic relative precision of $10^{-6}$ or better, has been exploited at many facilities, 
most notably at LEP in scans of the \PZ resonance~\cite{Assmann:1998qb} and more recently at VEPP4 in the determination of the J/$\psi$ and $\psi$(2S) masses~\cite{ANASHIN201550}. 
Alternatively, in a free spin precession (FSP) measurement the depolariser may be used to rotate the spin vector into the horizontal plane, and the precession frequency measured directly. 
The self-polarisation of the beams can be used for these measurements, but this will only be possible for \PZ-pole operation and at energies up to and including the $\PW^+\PW^-$ threshold. 
At energies higher than these, the polarisation level will be too small for RDP and FSP measurements to be practical and the energy scale will have to be determined from physics processes at the experiments, such as $\Pep\Pem \to \PZ (\PGg)$, $\PZ\PZ$ and $\PW^+\PW^-$ production. 
Here, the \PZ and \PW masses measured with high precision at lower energies with RDP
provide a normalisation that can then be applied at higher energies. 

The calculation of $\sqrts$ at each interaction point requires good knowledge of the crossing angle of the two beams, which must be measured by the experiments in real time.  
In addition, it is necessary to account for local energy variations from synchrotron radiation, the RF system and impedance, and to consider the effects of opposite-sign vertical dispersion.

The knowledge of $E_\mathrm{b}$ at LEP was ultimately limited by the sampling rate of RDP measurements, which were performed outside physics operation with a periodicity of around once per week. 
The energy was found to vary significantly between measurements due to several effects, for example earth tides and stray ground electric currents~\cite{Assmann:1998qb}. 
In order to enable the much greater degree of systematic control that the vastly larger event samples at FCC-ee warrant, the operational strategy will be very different to LEP. 
Measurements of $E_\mathrm{b}$ will be performed several times per hour on non-colliding pilot bunches. 
In \PZ running, around 160 pilot bunches will be injected at start of fill, and wiggler magnets will be activated to speed up the polarisation time. 
One to two hours will be required for the polarisation to build, after which the wigglers will be turned off and physics (colliding) bunches injected. 
The RF frequency will be continually adjusted to keep the beams centred in the quadrupoles, thus suppressing tide-driven energy changes, 
which would otherwise be $\cal{O}$(100\,MeV). 
A model will be developed to track residual energy variations between measurements.

A more detailed discussion on the machine aspects concerning the $\sqrts$ calibration can be found in the Volume~2 of this Report~\cite{FCC-PED-FSR-Vol2}. 
The contribution of the experiments to those studies and the current level of understanding of the expected performance are summarised here.
Also included below is a brief discussion of the studies that are underway for monochromatisation of the collision energy when operating at $\sqrts \approx 125$\,GeV, 
corresponding to the mass of the Higgs boson.  
Monochromatisation is motivated by the need to reduce the spread of $\sqrts$ to a value similar to the Higgs width (around 4\,MeV), 
thereby improving the sensitivity to direct Higgs production and allowing tight constraints to be placed on the electron-Yukawa coupling.

\subsection{Input from the experiments}
\label{sec:EPOL_input_from_the_experiments}

The experiments operating at FCC-ee will themselves provide measurements that are essential input to the calibration of the collision energy and related quantities.  
A full discussion of these measurements can be found in Ref.~\cite{Blondel:2019jmp}.  
Here, a brief summary is given, together with some recent updates.
The principal data set for performing these measurements is the very large sample of dimuon events that each experiment will collect, 
arising from the process $\Pep\Pem \to \PGmp\PGmm (\PGg)$, where $\PGg$ indicates the possible presence of initial-state radiation (ISR).  
Analysis of the topology of these events, 
constrained by the total energy and momentum conservation in the final state, 
allows several important quantities to be determined. 
This analysis is, in general, based on the knowledge of the muon directions, 
which in turn imposes demands on the performance of the tracking system (see Section~\ref{sec:PhysPerf_DetectorConcepts}).

\subsubsection{The crossing angle \texorpdfstring{$\alpha$}{alpha}}

The nominal value of the crossing angle is $\alpha = 30$\,mrad, but the true value must be determined throughout data-taking so that the collision energy can be calculated to the required precision.   
At the \PZ~pole, more than $10^6$ dimuon events will be collected every 10~minutes in each detector, which will allow this parameter to be measured with a statistical uncertainty of 0.3\,$\mu$rad, which is sufficient for the physics goals, since a precision of 15\,$\mu$rad leads to an uncertainty of 10\,keV on $\sqrts$.   
The statistical precision will be worse at higher energies, where the production rate is lower, but will not compromise the physics measurements that are targeted in these regimes.

There is an important subtlety in the crossing-angle determination that must be accounted for.  
The electron and positron bunches experience mutual electric and magnetic fields that accelerate (decelerate) the bunches before (after) the collision and  also increase (decrease) the crossing angle.  
The collision energy is invariant, but the change in crossing angle from this effect (estimated to be a relative 0.6\% modification) must be known so that the measured crossing angle can be corrected back to the unaffected quantity and used together with beam energies as determined from RDP to calculate $\sqrts$.

\begin{figure}[ht]
\centering
\includegraphics[width=0.75\linewidth]{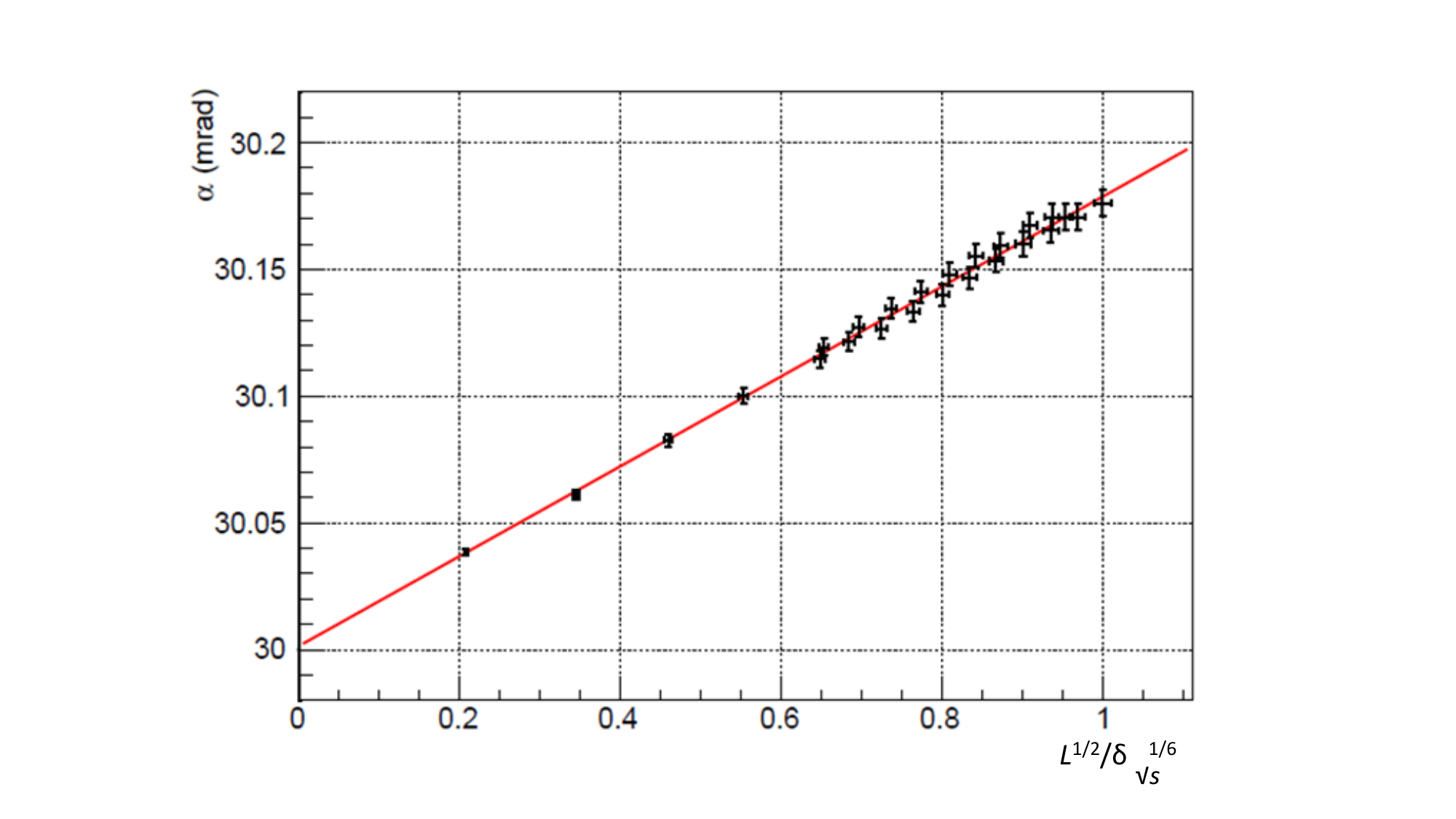}
\caption{Change in the measured crossing-angle $\alpha$ vs.\ $\mathcal{L}^{1/2} \,/\, {\delta_{\sqrts}}^{1/6}$, at various points during the top-up injection.
Extrapolation down to $\mathcal{L}^{1/2} \,/\, {\delta_{\sqrts}}^{1/6}=0$ allows the crossing-angle to be determined in the absence of bunch-bunch effects~\cite{Blondel:2019jmp}.} 
\label{fig:alphakick}
\end{figure}

The magnitude of the variation in $\alpha$ depends on parameters such as the bunch population and the spread in collision energy  $\delta_{\sqrts}$.  
It is found empirically, from simulation studies, that $\alpha$ is proportional to $\mathcal{L}^{1/2} \,/\, {\delta_{\sqrts}}^{1/6}$.  
By measuring $\mathcal{L}$, $\alpha$, and $\delta_{\sqrts}$ from dimuon events for different bunch intensities, it will be possible to extrapolate to zero intensity and determine the value of $\alpha$ in the absence of these effects.  
A good opportunity to perform these measurements would be in the period that top-up injection is taking place.
It is therefore important that the detector can operate during this period and that the beams are stable.  
A simulated study of the measurement of $\alpha$ against $\mathcal{L}^{1/2} \,/\, {\delta_{\sqrts}}^{1/6}$ is presented in Fig.~\ref{fig:alphakick}.

\subsubsection{The longitudinal boost and the collision-energy spread}

\begin{figure}[ht]
\centering
\includegraphics[width=0.7\linewidth]{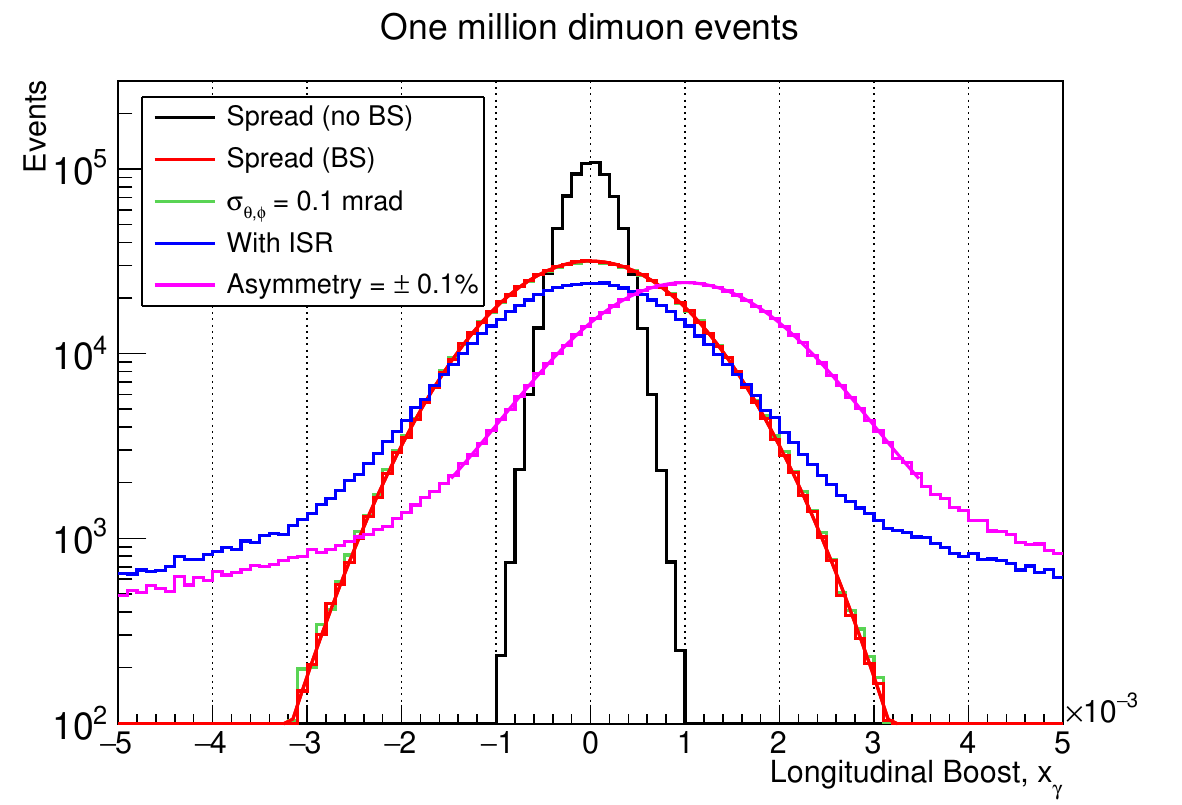} 
\caption{Fitted value of longitudinal boost from one million dimuon events at one of the FCC-ee IPs~\cite{Blondel:2019jmp}. 
Once the ISR is unfolded, this distribution can be used to measure the energy spread. 
The magenta line shows the impact of a centre-of-mass boost on the distribution. 
The shift can be measured with a statistical precision of 40\,keV.  
The other curves indicate the impact of beamstrahlung, angular resolution on the track directions, and ISR.}
\label{fig:ecmspread}
\end{figure}

The dimuon topology allows the longitudinal boost to be determined on an event-by-event basis.  
When averaged over a suitable sample size, this provides invaluable information to constrain the model of the energy loss around the ring and to calculate the local collision energy at each interaction point.  
The width of this distribution (Fig.~\ref{fig:ecmspread}) is a measure of $\delta_{\sqrts}$, which is an essential input to the measurement of certain observables, such as the \PZ and \PW widths. 
Again, the foreseen statistical precision on these quantities is excellent. 
For example, the energy spread can be measured to one part in a thousand with one million dimuon events.   
Recent work~\cite{delta-sqrts-ISR-corrections}
has investigated how sensitive the determination of $\delta_{\sqrts}$ is to the knowledge of the ISR corrections in dimuon production.  
The conclusion is that the measurement is robust; even if it is assumed that the second-order corrections from ISR are unknown (which is not the case), the resulting bias on the extraction of $\delta_{\sqrts}$ is far smaller than the statistical uncertainty.

\subsubsection{Relative \texorpdfstring{$\sqrts$}{ECM} determination in the \texorpdfstring{\PZ}{Z}-resonance scan}

\begin{figure}[t]
\centering
\includegraphics[width=0.49\linewidth]{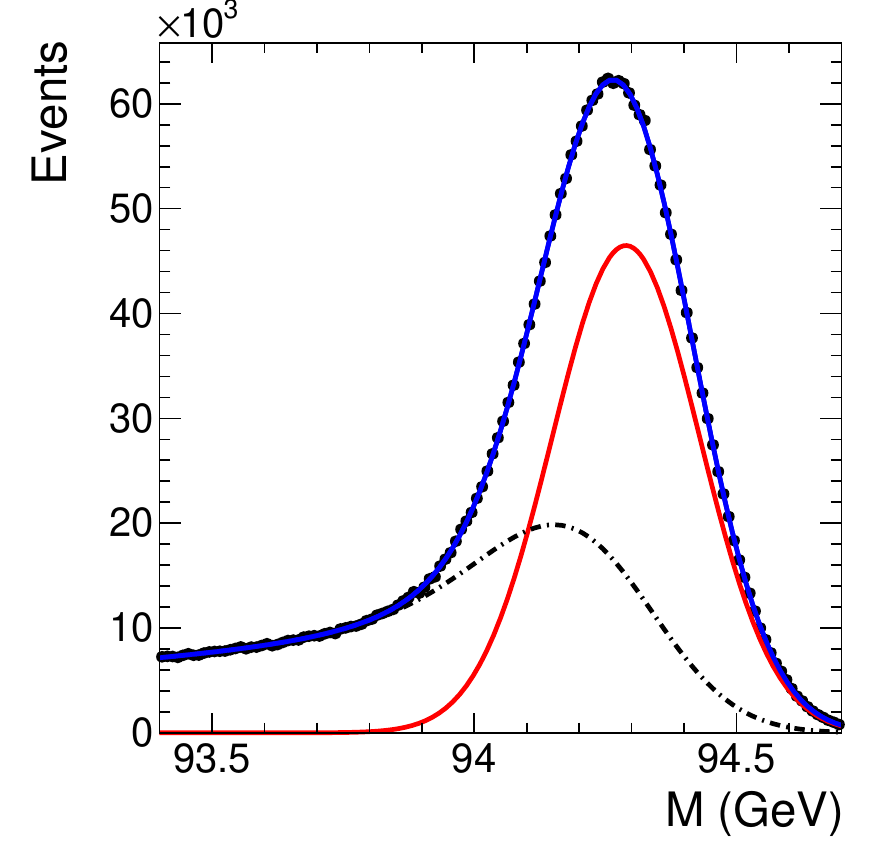} 
\includegraphics[width=0.49\linewidth]{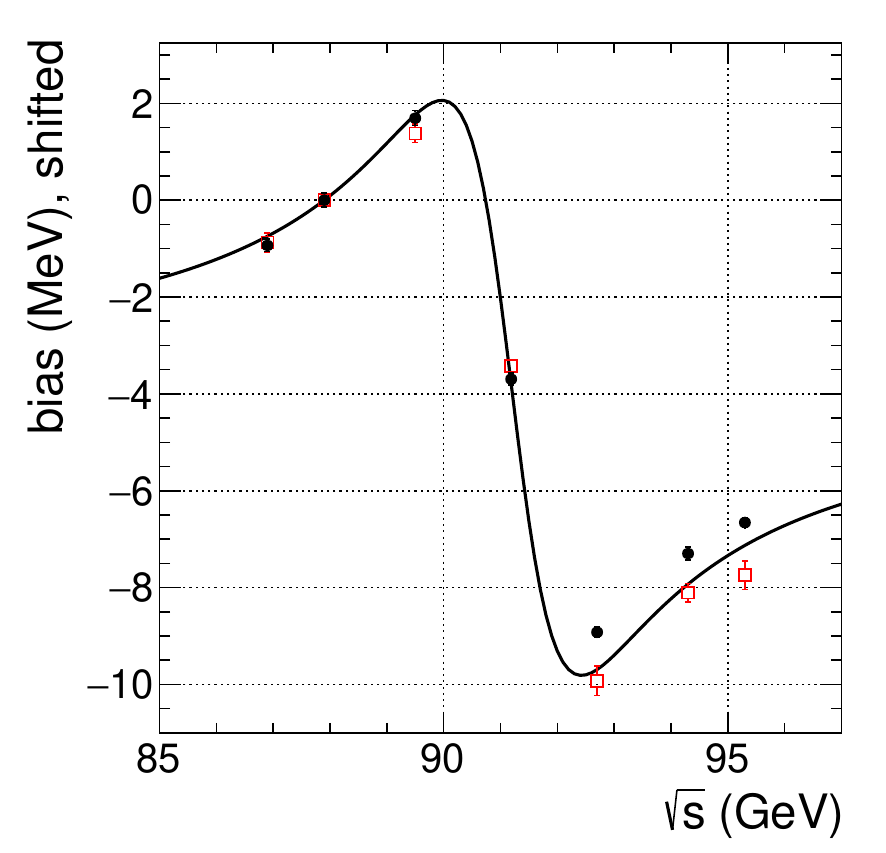} 
\caption{Left: Example fit to the dimuon invariant-mass distribution at $\sqrts = 94.3$\,GeV, where the peak is modelled with the superposition (blue) of a Gaussian function (red) and two exponential functions (black).  
Right: The difference (bias) between the $\sqrts$ extracted from the dimuon invariant-mass fit and the true value.  
Results are shown for the simulated performance of the IDEA detector (red points) and for the expected dependence, including (black points) or not (black curve) ISR/FSR effects.  
Each set of results also has an overall overset of a few MeV, which is energy independent and, hence, not relevant for the determination of the \PZ width. 
A single shift has been applied to the bias, to correct for these offsets, so that the bias is zero at 87.9\,GeV, for all sets of results.}
\label{fig:ecmproxy}
\end{figure}

The reconstructed peak position of the dimuon invariant-mass distribution provides an excellent proxy for the collision energy. 
The difference in this reconstructed position between the points of the \PZ-resonance scan provides a measure of the change in collision energy, 
which is a critical input for several analyses, in particular the measurement of the \PZ boson width. 
The distribution is fitted in bins of the polar angle for back-to-back events.  
An example fit is shown in Fig.~\ref{fig:ecmproxy}~(left).
The statistical precision on this pseudo-$\sqrts$ measurement, when summing the samples from four experiments, is around 20\,keV for each of the two off-peak running points, assuming the momentum resolution of the IDEA detector.
So that the detector does not introduce a bias in the measurement larger than the statistical precision, the momentum scale stability must be controlled at this level. 

The field stability can be tracked with NMR probes and the momentum scale can be directly monitored through the reconstruction of low-mass resonances. 
However, even with a perfect detector, there is a bias in the pseudo-$\sqrts$ measurement in the \PZ resonance scan arising from ISR/FSR effects, 
and from the product of the Breit--Wigner shape of the resonance and the Gaussian distribution of the energy spread of the colliding beams. 
The value of this bias varies by about 8\,MeV when going from $\sqrts = 87.9$ to 94.3\,GeV, as can be seen in Fig.~\ref{fig:ecmproxy}~(right).  
This difference must be corrected for in the measurement, which requires a good understanding of the ISR/FSR effects.   
In a generator-level study, disabling ISR/FSR changes the difference in the bias between the two off-peak points by around 500\,keV.  
Therefore, the theoretical prediction of these ISR/FSR effects to the 1\% level would be sufficient to render their impact negligible for the \PZ-width measurement. 

\subsubsection{Absolute \texorpdfstring{$\sqrts$}{ECM} determination}

At collision energies above the \PZ boson mass, the dimuon events may be used to provide an absolute measurement of $\sqrts$. 
Radiative returns, in which the emission of an initial-state photon means that the dimuon has the \PZ mass, allows for the calibration of events unaffected by ISR. 
The method can be extended to also include multihadron final states.  
This method can be calibrated at the \PW-pair production threshold with RDP, 
and is of great value for physics studies in the regime where no RDP is possible, 
i.e., collision energies above 200\,GeV. 
This approach also provides a useful complementary measurement of $\sqrts$ in the intermediate energies where RDP is possible but challenging. 
The foreseen statistical uncertainty is 
around 280\,keV for 6\,ab$^{-1}$ of integrated luminosity at $\sqrts = 125$\,GeV 
and 260\,keV for 20\,ab$^{-1}$ at $\sqrts = 160$\,GeV.
The performance of the tracking system must be sufficiently good that the precision is not compromised.

\subsection{Expected precision on EW observables from the collision energy and its spread}

Several of the most important electroweak observables are expected to have a dominant or significant systematic uncertainty associated with the knowledge of the collision energy and collision-energy spread.   
The collision-energy uncertainties can be classed in three distinct categories, itemised below.  
These uncertainties propagate to the physics results in an observable-dependent manner, as discussed in Ref.~\cite{Blondel:2019jmp}.

\begin{itemize}
    \item  Uncertainties that are fully correlated between measurements propagate to the knowledge of the absolute energy scale.  
    Examples include the values of the anomalous magnetic moment and the mass of the electron, the frequency of the RF system, and any other systematic bias that occurs at all times and all energies.  
    At this stage in the studies, it is estimated that this uncertainty will be around 100\,keV on the collision energy at the \PZ pole and 300\,keV at the $\PW^+\PW^-$ threshold.  
    This contribution is expected to be the dominant systematic uncertainty in the measurements of the \PZ and \PW boson masses.
    \item A point-to-point contribution comprises biases that occur at all times, or lead to an average shift, but are different for each energy setting.  
    The principal method of determining this uncertainty will be based on the dimuon invariant-mass distribution, as reconstructed by the experiments.  
    The estimated magnitude of this uncorrelated uncertainty is 20\,keV for each off-peak point of the \mbox{\PZ-resonance} scan.  
    The understanding gained at the \PZ pole and complementary measurements will lead to a corresponding uncertainty of around 100\,keV at the $\PW^+\PW^-$ threshold.  
    The point-to-point uncertainty is expected to be the dominant contribution in the measurement of the \PZ width.
   \item The uncertainty on each individual RDP measurement is dominated by an uncertainty that is set by the frequency of the polarimeter sampling or the size of the energy bins where the depolarisation can be located.  
   A reasonable estimate of this uncertainty is 200\,keV at the \PZ pole and 300\,keV  at the $\PW^+\PW^-$ threshold.   
   As this component is statistical in nature, its impact decreases with the square-root of the number of events. 
   As it is planned to collect $\sim$\,$10^4$ measurements at each energy point, the final uncertainty from this source will be essentially negligible, compared to other contributions.  
   However, the importance of making each measurement as precise as possible, and of collecting the largest possible number of measurements, will become more evident when the data set is split into smaller samples to perform systematic checks.
\end{itemize}

\begin{table}[h]
\centering
\caption{Current projected $\sqrts$-related uncertainties on selected electroweak observables.}  
\label{tab:DS-Ecal-errors-final}
\begin{tabular}{cccccc} \toprule
    & \multicolumn{5}{c}{Observable} \\
    Uncertainty 
    & $m_{\PZ}$ (keV) & $\Gamma_{\PZ}$ (keV) 
    &  $\sin^2 \theta^\text{eff}_{\PW}$ ($\times 10^{-6}$) 
    & $\frac{\Delta \alpha_\text{QED}(m^2_{\PZ})}{\alpha_\text{QED}(m^2_{\PZ})}$ ($\times 10^{-5}$) 
    & $m_{\PW}$ (keV) \\ 
\midrule
    Absolute & 100 & 2.5 & -- & 0.1 & 150 \\
    Point-to-point & 14 & 11 & 1.2 & 0.5 & 50 \\
    Sample size & 1 & 1 & 0.1 & -- & 3 \\
    Energy spread & -- & 5 & -- & 0.1 & -- \\ 
\midrule
    Total $\sqrts$-related & 101 & 12 & 1.2 & 0.5 & 158 \\ 
\midrule
    FCC-ee statistical & 4 & 4 & 1.2 & 3.9 & 180 \\  
\bottomrule
\end{tabular}
\end{table}

The contributions from each uncertainty category, and their quadratic sum, are listed in Table~\ref{tab:DS-Ecal-errors-final} for several key electroweak observables.
This table also shows the contribution from the uncertainty in the knowledge of the energy spread, which affects quantities with a strong quadratic dependence on the collision energy.  
Observables that are most susceptible to this uncertainty include the \PZ cross section and the \PZ width.  
A collision-energy spread of 70\,MeV, determined with a precision of $\pm$\,0.05\,MeV, leads to a sub-dominant systematic uncertainty in the measurement of these observables.

With the current expectations, it will be possible to reduce the uncertainty from energy-related quantities by an order of magnitude or more with respect to what was achieved at LEP, such that they will be smaller than, or similar to, the statistical uncertainty for all observables apart from $m_{\PZ}$.  
Indeed, the entries in Table~\ref{tab:DS-Ecal-errors-final} for the $\sqrts$-related systematic uncertainties can be compared to the corresponding LEP values of 
1.7\,MeV for $m_{\PZ}$, 1.2\,MeV for $\Gamma_{\PZ}$, and 9\,MeV for $m_{\PW}$. 

\subsection{Prospects for monochromatisation and the measurement of the electron Yukawa coupling}
\label{sec:eYukawa}

This section describes the tantalising possibility, unique to FCC-ee, 
to observe the $s$-channel process $\Pep\Pem \to \PH$, and thereby determine the electron Yukawa coupling.  
The challenges are very demanding and progress is required from both the accelerator and the experimental analyses, but the current studies, described below, are encouraging.

\subsubsection*{The electron Yukawa via resonant Higgs production at 125\,GeV}

Confirming the mechanism of mass generation for the stable visible elementary particles of the universe, 
composed of $\PQu$ and $\PQd$ quarks plus the electron (and neutrinos), 
is experimentally very challenging because of the low masses of the first-generation fermions and, thereby, 
their small Yukawa couplings to the Higgs field. (The neutrino mass generation remains a BSM problem in itself.) 
In the SM, the Yukawa coupling of the electron is $y_{\Pe} = \sqrt{2} \, m_{\Pe}/v = 2.8 \cdot 10^{-6}$, 
for $m_{\Pe}(m_{\PH}) = 486$\,keV and Higgs vacuum expectation value $v = ( \sqrt{2} \, G_\mathrm{F} )^{-1/2} = 246.22$\,GeV.
Measuring it via $\PH \to \Pep\Pem$ appears hopeless at hadron colliders because the decay has a tiny partial width, 
proportional to the electron mass squared,
\begin{equation}
\Gamma(\PH\to\Pep\Pem) = 
\frac{{G_\mathrm{F}} \, m_{\PH} \, m_{\Pe}^2} {4\,\sqrt{2}\,\pi}
\left(1-\frac{4\,m^2_{\Pe}}{m^2_{\PH}}\right)^{3/2} = 2.14 \times 10^{-11} \, \text{GeV} \, ,
\label{eq:Gamma_H_ee}
\end{equation}
which corresponds to a branching fraction $\mathcal{B}(\PH\to\Pep\Pem) \approx 5 \times 10^{-9}$ 
for the SM Higgs boson. 
At LHC and FCC-hh, such a final state is completely swamped by the Drell--Yan $\Pep\Pem$ continuum. 
The current LHC searches~\cite{ATLAS:2019old,CMS:2022urr}, 
exploiting about 140\,fb$^{-1}$ of pp data at $\sqrts = 13$\,TeV,
reach an observed upper limit on the branching fraction 
of $\mathcal{B}(\PH\to\Pep\Pem) < 3.0 \times 10^{-4}$ at 95\% CL. 
This value translates into the current upper bound on the Higgs boson effective coupling modifier to electrons of $\lvert \kappa_{\Pe} \rvert  < 240$. 
Assuming that the sensitivity to the $\PH \to \Pep\Pem$ decay scales simply with the square root of the integrated luminosity, 
the HL-LHC phase with a $\LumiInt = 2\times 3$\,ab$^{-1}$ data sample 
(combining the ATLAS and CMS results) 
will result in $y_{\Pe} \lesssim 100 ~ y^\mathrm{\textsc{sm}}_{\Pe}$. 
Based on searches for the similar $\PH \to \PGmp\PGmm$ channel, 
one can expect upper limits on $\mathcal{B}(\PH\to\Pep\Pem)$ to be further improved by factors of about four, 
by adding more Higgs production categories and using advanced multivariate analysis methods, 
eventually reaching $y_{\Pe} \lesssim 50 ~ y^\mathrm{\textsc{sm}}_{\Pe}$ at the end of the HL-LHC.

About ten years ago, it was first noticed that the unparalleled integrated luminosities 
of $\LumiInt \approx 10$\,ab$^{-1}$/year expected at $\sqrts = 125$\,GeV at the FCC-ee 
would make it possible to attempt an observation of the direct production of the scalar boson and, thereby, 
a direct measurement of the electron Yukawa coupling~\cite{dEnterria:2014,dEnterria:2017dac}. 
Subsequently, several 
theoretical studies~\cite{Jadach:2015cwa,Greco:2016izi,Dery:2017axi,Davoudiasl:2023huk,Boughezal:2024yjk}, 
simulated data analysis~\cite{dEnterria:2021xij}, 
and accelerator studies~\cite{Zimmermann:2017tjv,telnov2020monochromatizationeecolliderslarge,Faus-Golfe:2021udx} 
have explored the various aspects connected to the $\Pep\Pem\to\PH$ measurement. 

The Feynman diagrams for $s$-channel Higgs production 
(and its statistically most significant decay at FCC-ee, $\PH \to \Pg\Pg$, see below) 
and dominant backgrounds are shown in Fig.~\ref{fig:eeH_diags}~(left). 
The resonant Higgs cross section in $\Pep\Pem$ collisions as a function of $\sqrts$ 
is theoretically given at Born level by the relativistic Breit--Wigner (BW) expression:
\begin{equation}
\sigma_{\Pe\Pe \to \PH} = \frac{4\, \pi\, \Gamma_{\PH}\, \Gamma(\PH\to\Pep\Pem)} {(s-m_{\PH}^2)^2 + m_{\PH}^2\Gamma_{\PH}^2} \, .
\label{eq:sigma_H_ee}
\end{equation}
An accurate knowledge of $m_{\PH}$ is, therefore, critical to maximise the resonant cross section. 
An $\cal{O}$(4\,MeV) precision on the Higgs boson mass can be achieved (Section~\ref{sec:ch_part_tracking}) before any dedicated $\Pep\Pem\to\PH$ run.
In addition,  the FCC-ee beam energies will be monitored with a relative precision of $10^{-6}$, providing a sub-MeV accuracy on the exact point in the Higgs lineshape being probed at any moment.

\begin{figure}[t]
\centering
\includegraphics[width=0.27\columnwidth]{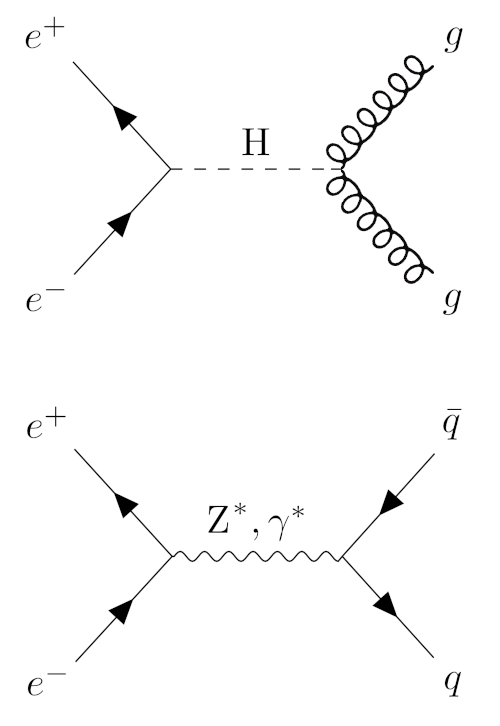}
\includegraphics[width=0.53\columnwidth]{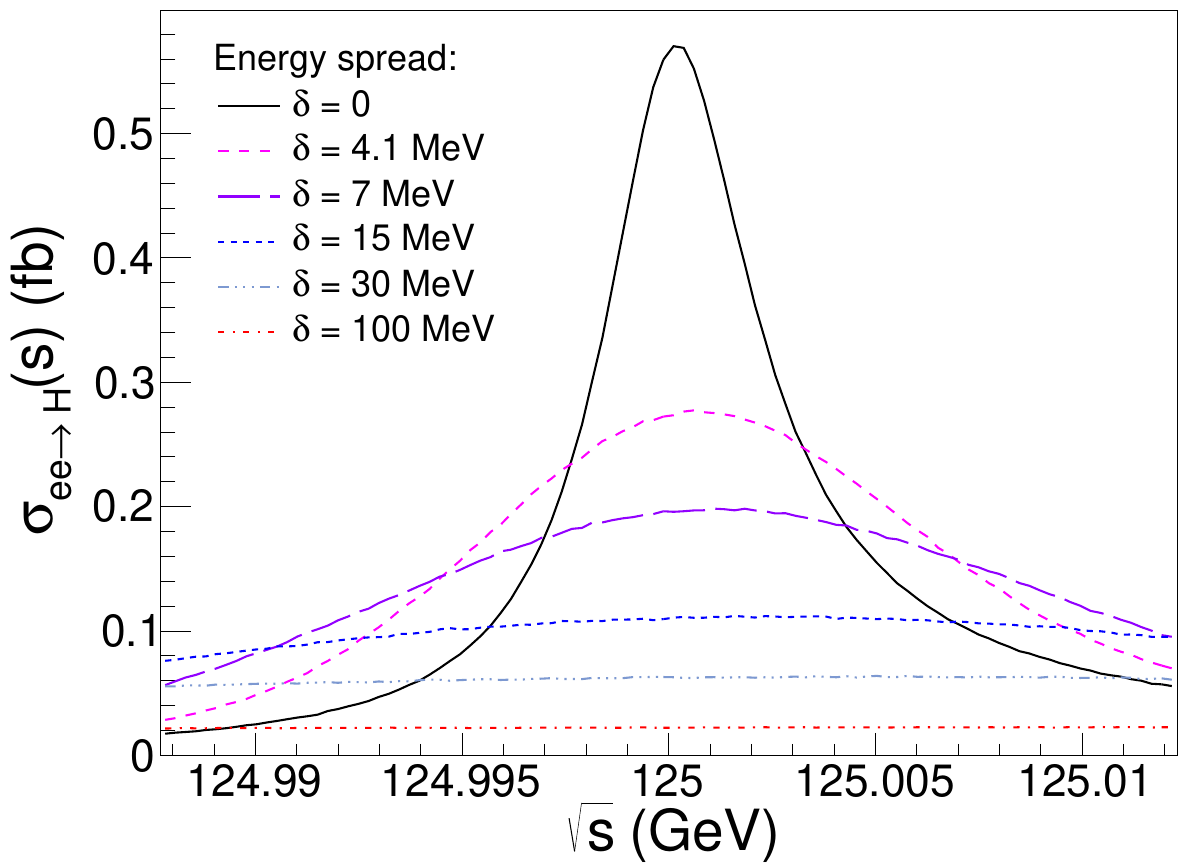}
\caption{Left: Diagrams for the $s$-channel production of the Higgs boson decaying into two gluon jets (top) 
and reducible $\PZ^*$ quark dijet backgrounds (bottom) in $\Pep\Pem$ at $\sqrts = 125$\,GeV. 
Right: Resonant Higgs production cross section at $\sqrts = 125$\,GeV, including ISR effects, 
for several $\Pep\Pem$ collision energy spread values: $\delta_{\sqrts} = 0$, 4.1, 7, 15, 30, and 100\,MeV~\cite{Jadach:2015cwa}.
\label{fig:eeH_diags}}
\end{figure}

For $m_{\PH} = 125$\,GeV, Eq.~\ref{eq:sigma_H_ee} gives 
$\sigma_{\Pe\Pe \to \PH} = 4\,\pi\,\mathcal{B}(\PH\to\Pep\Pem) / m_{\PH}^2 = 1.64$\,fb as peak cross section. 
Two effects, however, lead to a significant reduction of the Born-level prediction: 
(i)~ISR depletes the cross section and generates an asymmetry of the Higgs lineshape; and 
(ii)~the actual beams are never perfectly monoenergetic, i.e., 
the collision $\sqrts$ has a spread $\delta_{\sqrts}$ around its central value,
further leading to a smearing of the BW peak. 
For FCC-ee operating at 125\,GeV, 
the natural spread in collision energy due to synchrotron radiation will be around 70\,MeV. 
Monochromatisation aims at reducing $\delta_{\sqrts}$ to the few MeV scale, while still delivering moderately large (few ab$^{-1}$) integrated luminosities, $\LumiInt$~\cite{Zimmermann:2017tjv,telnov2020monochromatizationeecolliderslarge,Faus-Golfe:2021udx}. 
The reduction of the BW cross section due to  initial-state photon emission(s) alone 
is of a factor of 0.35 and leads to $\sigma_{\Pe\Pe \to \PH} = 0.57$\,fb~\cite{Jadach:2015cwa}. 

The additional impact of a given energy spread on the Higgs BW shape can be quantified through the convolution of BW and Gaussian distributions, i.e., a relativistic Voigtian function. 
Figure~\ref{fig:eeH_diags}~(right) shows the Higgs lineshape for various $\delta_{\sqrts}$ values. 
The combination of ISR plus $\delta_{\sqrts} = \Gamma_{\PH} = 4.1$\,MeV reduces the peak Higgs cross section by a total factor of 0.17, 
down to $\sigma_{\Pe\Pe \to \PH} = 0.28$\,fb. 
Though tiny, the cross section for any other $\Pep\Pem \to \PH$ production process, through intermediate \PW or \PZ bosons, 
is much further suppressed by the electron mass for on-shell external fermions (chirality flip)~\cite{Altmannshofer:2015qra} 
and is negligible as well as any other loop-induced Higgs production mechanisms at $\sqrts = 125$\,GeV 
that is not sensitive to $y_{\Pe}$~\cite{DdE_Dung2025}.

The strategy to observe the resonant production of the Higgs boson~\cite{dEnterria:2021xij} 
is based on identifying final states consistent with any of the \PH decay modes 
that lead to small (but statistically significant when combined together) excesses over the expected backgrounds  
(orders-of-magnitude more abundant than the signal). 
In Ref.~\cite{dEnterria:2021xij}, a detailed study was performed with large simulated event samples of signal 
and associated backgrounds generated with \pythia~8~\cite{pythia8} for eleven Higgs boson decay channels. 
A benchmark monochromatisation point of $(\delta_{\sqrts},\LumiInt)=(4.1\,\text{MeV},10\,\text{ab}^{-1})$, 
corresponding to 2800 Higgs bosons produced, was assumed for the signal. 
A simplified description of the expected experimental performance was considered for the reconstruction and (mis)tagging 
of heavy-quark ($\PQc$, $\PQb$), light-quark ($\PQu\PQd\PQs$), and gluon ($\Pg$) jets, as well as photons, electrons, and hadronically decaying tau leptons. 
Generic preselection criteria were defined to suppress reducible backgrounds while keeping the largest fraction of the signal events. 
A subsequent multivariate analysis of $\cal{O}$(50) kinematic and global topological variables, defined for each event, was carried out. 
Boosted Decision Tree (BDT) classifiers were trained on signal and background events to maximise the signal significance for each individual channel. 
The most significant Higgs decay channels are found to be $\PH \to \Pg\Pg$ 
(for a gluon efficiency of 70\% and a $\PQu\PQd\PQs$-for-$\Pg$ jet mistagging rate of 1\%), 
and $\PH\to\PW\PW^* \to \ell\,\PGn\,jj$. 
The digluon final state is the most sensitive channel to search for the resonant Higgs boson production (Fig.~\ref{fig:eeH_diags}~left, upper diagram) because it has a moderately large branching fraction ($\mathcal{B}\approx 8\%$) 
while the $\PZ^*\to\Pg\Pg$ decay is forbidden by the Landau--Yang theorem. 
The most important experimental challenge is to reduce the light-quark for gluon mistagging rate to the 1\% level 
(while maintaining the efficiency for the $\PH\to\Pg\Pg$ channel at 70\%) 
to keep the overwhelming $\PZ^*\to \uubar$, $\ddbar$, $\ssbar$ backgrounds (Fig.~\ref{fig:eeH_diags}~left, lower diagram) under control. 
Such a mistagging rate is a factor of about seven times better than the current state-of-the-art for jet-flavour tagging algorithms~\cite{Bedeschi:2022rnj}, 
but it is a realistic goal given all the experimental and theoretical improvements in the understanding of parton radiation and hadronisation expected at the FCC-ee~\cite{Proceedings:2017ocd}.

Combining all results for an accelerator operating at $(\delta_{\sqrts},\LumiInt)=(4.1\,\text{MeV},10\,\text{ab}^{-1})$, 
a 1.3\,$\sigma$ signal significance can be reached for the direct production of the Higgs boson, 
corresponding to an upper limit on the electron Yukawa coupling at 1.6 times the SM value: 
$\lvert y_{\Pe} \rvert < 1.6 \, \lvert y^\mathrm{\textsc{sm}}_{\Pe} \rvert $ at 95\% CL for each detector in one  year. 
Based on this benchmark result and the dependence of the resonant Higgs cross section on $\delta_{\sqrts}$ 
(Fig.~\ref{fig:eeH_diags}, right), 
bidimensional maps of $\Pep\Pem\to\PH$ significance and electron-Yukawa sensitivities have been determined in the $(\delta_{\sqrts},\LumiInt)$ plane. 

\begin{figure}[ht]
\centering
\includegraphics[width=0.65\textwidth]{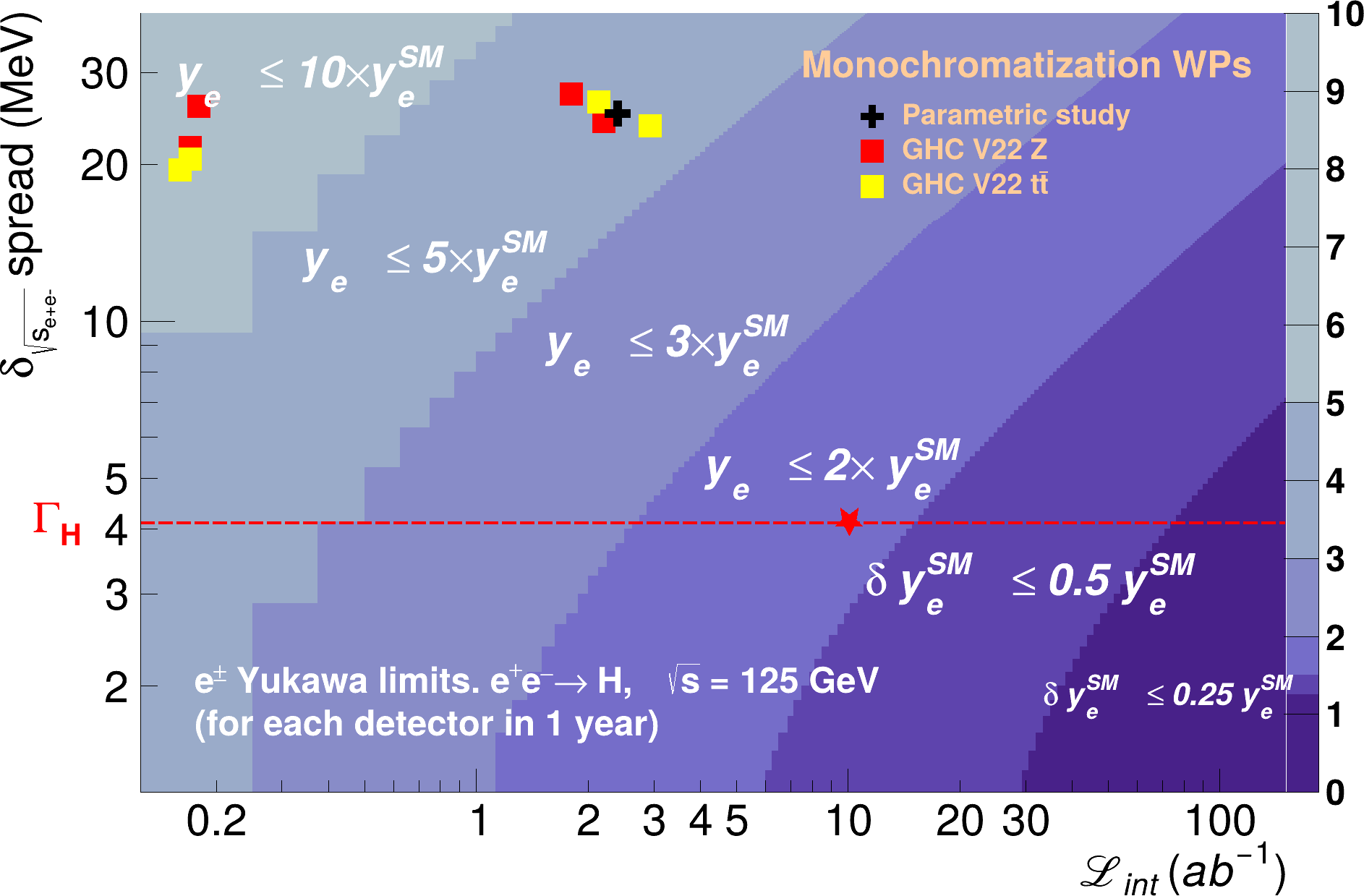}
\caption{Upper limit contours (at 95\% CL) on the electron Yukawa $y_{\Pe}$ (coloured bands) 
in the collision-energy spread $\delta_{\sqrts}$ vs.\ integrated luminosity $\LumiInt$ plane.
The red star over the $\delta_{\sqrts} = \Gamma_{\PH} =4.1$\,MeV red-dashed line indicates the reference point assumed in the physics simulation analysis~\cite{dEnterria:2021xij}. 
The black cross indicates the previously achieved working point with self-consistent parametric monochromatisation~\cite{ValdiviaGarcia:2022nks,Faus-Golfe:2021udx}.
The red and yellow squares indicate the monochromatisation points based on simulations of the `GHC V22 Z' and `GHC V22 $\PQt\PAQt$' optics, respectively 
(see text for details)~\cite{Zhang:2024sao}.
\label{fig:eeH_signif}}
\end{figure}

Figure~\ref{fig:eeH_signif} shows the 95\% CL upper limit contours on the electron Yukawa coupling strength as a function of the energy spread and integrated luminosity with the red star 
(on the red-dashed line corresponding to a reference monochromatised collision-energy spread equal to the Higgs boson width), 
indicating the result of this benchmark study. 
The next section discusses the results of the current monochromatisation simulation studies, shown with red and yellow squares in this figure.

\subsubsection*{FCC-ee monochromatisation}
\begin{figure}[ht]
\centering
\includegraphics[width=0.9\textwidth]{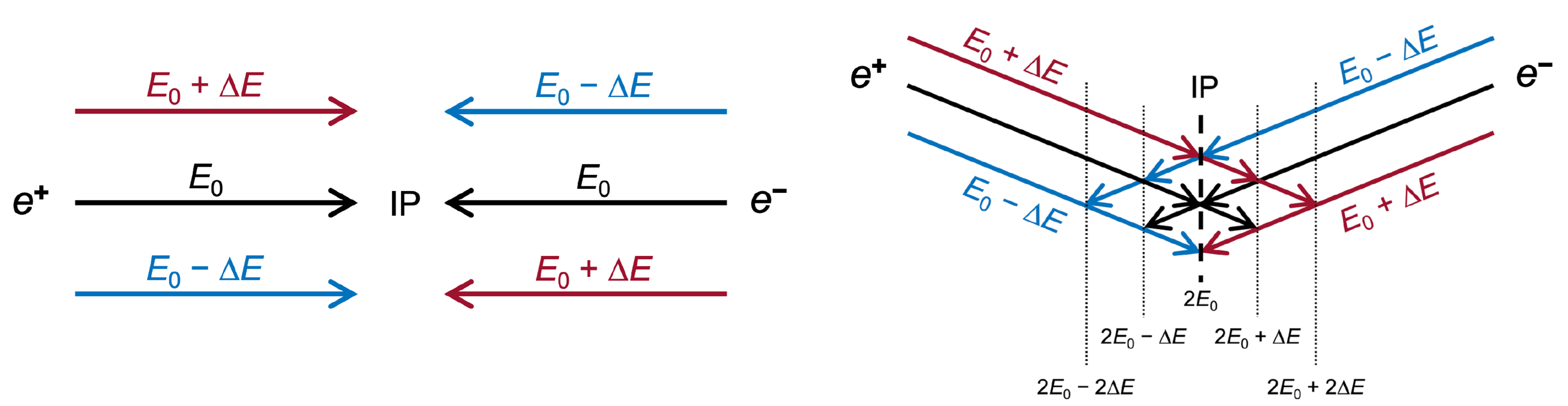}
\caption{Schematic of the principle of monochromatisation for head-on collisions (left) and collisions with a crossing angle (right). 
In both cases, opposite-sign correlations between the transverse position in the beam and the energy lead to a reduction in the collision-energy spread, compared with the uncorrelated case.}
\label{fig:monochromschem}
\end{figure}

Monochromatisation is necessary to reduce 
$\delta_{\sqrts}$ to the few MeV level of the natural SM Higgs width and thereby increase the sensitivity of the electron-Yukawa measurement via $\Pep\Pem \to \PH$. 
This strategy was first proposed 50 years ago~\cite{Renieri:1975wt} and relies on creating opposite correlations between spatial position and energy deviations within the colliding beams with nominal beam energy $E_0$. 
Figure~\ref{fig:monochromschem} shows a schematic of the principle of monochromatisation for beams that collide head on (left) and with a crossing angle $\alpha$ (right).  
The current baseline design of FCC-ee corresponds to the latter configuration. 
In both configurations, the correlations between transverse (either horizontal or vertical) position in the beam and energy lead to a lower spread in collision energy than in the uncorrelated case.

Monochromatisation can be achieved by adding dedicated components at the interaction region (IR) to generate a non-zero dispersion function with opposite signs for the two beams at the IP. 
A non-zero dispersion function at the IP in the horizontal and/or vertical directions ($D_{x,y}^{\ast} \neq 0$) enlarges the IP transverse beam size ($\sigma_{x,y}^{\ast}$), which in turn affects the luminosity, $\mathcal{L} \propto 1/(\sigma_{x}^{\ast} \sigma_{y}^{\ast})$. 
The monochromatisation factor is defined as
\begin{equation}
    \lambda = \sqrt{1 + \sigma_{\delta}^2 
    \left( \frac{D_x^{\ast 2}}{\varepsilon_{x} \, \beta_{x}^{\ast}} + \frac{D_y^{\ast 2}}{\varepsilon_{y} \, \beta_{y}^{\ast}} \right) } \, ,
\label{eq:lambda_monochrom}
\end{equation}
where $\sigma_{\delta}$ is the relative energy spread, $\varepsilon_{x,y}$ the transverse emittances, and $\beta_{x,y}^{\ast}$ the betatron functions at the IP. 
For any value of $\lambda$ achieved, $\delta_{\sqrts}$ and $\mathcal{L}$ in the monochromatisation operation mode are
\begin{equation}
    \delta_{\sqrts} = \frac{\sqrt{2} \, E_0 \, \sigma_\delta}{\lambda} \quad \text{and} \quad 
    \mathcal{L} = \frac{\mathcal{L}_0}{\lambda} \, ,
\label{eq:delta_sqrts}
\end{equation}
where $\mathcal{L}_0$ represents the luminosity for the same values of $\beta_{x,y}^{\ast}$ but without $D_{x,y}^{\ast}$. 
Consequently, the design of a monochromatisation scheme requires considering both the IR beam optics and the optimisation of other collider parameters to maintain the highest possible luminosity. 
Possible approaches to monochromatisation for FCC-ee have been studied for several years, starting from self-consistent parametric studies~\cite{ValdiviaGarcia:2016rdg,Zimmermann:2017tjv,Faus-Golfe:2021udx,ValdiviaGarcia:2022nks}. 
Recent  developments~\cite{Zhang:2024sao,Zhang:2024phd} comprise a detailed study of the IP-region optics required for monochromatisation, exploring different potential configurations and their implementation in the FCC-ee global lattice, along with beam-dynamics simulations and performance evaluations including the impact of beamstrahlung. 

The baseline FCC-ee standard lattice design is the so-called `Global Hybrid Correction' (GHC) optics~\cite{Oide:2016mkm,Keintzel:2023myn,vanRiesen-Haupt:2024ore}.
It allows for four IRs, where the $\Pep$ and $\Pem$ beams are brought to collision 
with an $\alpha= 30$\,mrad angle in the horizontal plane, as well as a potential vertical crab-waist scheme. 
The results of the monochromatisation studies shown in Fig.~\ref{fig:eeH_signif} are based on two versions of this optics: 
`GHC V22 Z', where the lattice is optimised for operation at the \PZ pole, and 
`GHC V22 $\PQt\PAQt$', which is optimised for operation above the $\PQt\PAQt$ threshold 
(in both, `V22' designates the 2022 configuration).
Three approaches to monochromatisation have been investigated.  
In the first, the horizontal dipoles used for the local-chromaticity-correction system are reconfigured to generate a non-zero $D^\ast_x$ of size ${\sim}\, 10$\,cm, 
while maintaining the same $\alpha$ value.  
Given the values of the other parameters in Eq.~(\ref{eq:lambda_monochrom})~\cite{Keintzel:2023myn}, it follows that monochromatisation factors in the range $\lambda = 5$--8 are achievable. 

This study was performed both to provide monochromatisation in all four IRs and then repeated to give monochromatisation in only two IRs.
The second method introduces a non-zero value of $D^\ast_y$ by adjusting the strengths of the skew quadrupoles in the interaction region.  
The very low vertical emittance in FCC-ee implies that similar monochromatisation factors as in the horizontal case 
can be achieved with $D^\ast_y \approx 1$\,mm.  
Finally, schemes involving non-zero values of both $D^\ast_x$ and $D^\ast_y$ have been explored.   
In all cases, the layout of the components around the IR and the parameter values were adjusted to satisfy the boundary conditions in the machine and deliver optimum performance.

Simulations were performed with \textsc{Guinea-Pig}~\cite{Schulte:1998au} to determine  the performance of the different monochromatisation schemes, taking into account the impact of beamstrahlung. 
The particle distribution at the IP was simulated as an ideal Gaussian distribution, comprising 40\,000 particles and defined by the following global optical performance parameters: 
$E_0$, $\sigma_{\delta}$, $\varepsilon_{x,y}$, $\beta_{x,y}^{\ast}$, $D_{x,y}^{\ast}$, $\sigma_{z}$, and $\alpha$. 
For each configuration, the $\delta_{\sqrts}$ (from the distribution of the collision energy) and $\mathcal{L}$ were calculated. 
The results are presented in Tables~\ref{tab:monochromZ} and~\ref{tab:monochromtt} for the `GHC V22 Z' and `GHC V22' $\PQt\PAQt$ optics, respectively.
 
\begin{table}[ht]
\caption{Values of $\delta_{\sqrts}$, $\mathcal{L}$, and $\LumiInt$ for five setups of the `GHC V22 Z' monochromatisation IR optics~\cite{Zhang:2024sao}:
without monochromatisation (`Std.\ ZES'), 
with $D^*_x \ne 0$ in four and two IPs (`ZH4IP' and `ZH2IP'),
with $D^*_y \ne 0$ (`ZV'), 
and with $D^*_{x,y} \ne 0$ (`ZHV').}
\label{tab:monochromZ}
\centering
\begin{tabular}{lccccc}\hline
    & Std.\ ZES & ZH4IP & ZH2IP & ZV & ZHV \\\hline
    CM energy spread $\delta_{\sqrts}$ (MeV) & 69.52 & 26.80 & 24.40 & 25.25 & 20.58 \\
    Luminosity / IP $\mathcal{L}$ ($10^{34}$ cm$^{-2}$s$^{-1}$) & 44.8 & 15.0 & 18.4 & 1.46 & 1.42 \\
    Integrated luminosity / IP / year $\LumiInt$ (ab$^{-1}$) & 5.38 & 1.80 & 2.21 & 0.18 & 0.17 \\\hline
\end{tabular}
\end{table}

\begin{table}[ht]
\caption{Same as previous table, but for the `GHC V22 $\PQt\PAQt$' monochromatisation IR optics~\cite{Zhang:2024sao}.}
\label{tab:monochromtt}
\centering
\begin{tabular}{lccccc}\hline
    & Std.\ TES & TH4IP & TH2IP & TV & THV \\\hline
    CM energy spread $\delta_{\sqrts}$ (MeV) & 67.20 & 27.10 & 23.16 & 20.23 & 21.24 \\
    Luminosity / IP $\mathcal{L}$ ($10^{34}$\,cm$^{-2}$s$^{-1}$) & 71.2 & 17.9 & 24.5 & 1.37 & 1.42 \\
    Integrated luminosity / IP / year $\LumiInt$ (ab$^{-1}$) & 8.54 & 2.15 & 2.94 & 0.16 & 0.17 \\
    \hline
\end{tabular}
\end{table}

All the investigated monochromatisation schemes are successful in reducing $\delta_{\sqrts}$ 
by a factor of two or more with respect to the value without monochromatisation.  
As expected, this reduction in energy spread is accompanied by a reduction in luminosity, which is more marked for the configurations with $D^\ast_y \ne 0$ and combined $D^\ast_{x,y} \ne 0$, where the beamstrahlung leads to a blow up in $\epsilon_y$.   
The corresponding physics performance is plotted as red (yellow) squares for the `GHC V22 Z' (`GHC V22 $\PQt\PAQt$') setups in the $(\delta_{\sqrts},\LumiInt)$ plane in Fig.~\ref{fig:eeH_signif}, from which the corresponding 95\% CL upper limit contours for the $y_{\Pe}$ coupling can be read off. 
The physics performance of all designed monochromatisation IR optics with non-zero $D_{x}^{\ast}$ are comparable to, or even exceed, those of the previous FCC-ee self-consistent parameters (black cross). 
The `MonochroM TH2IP' optics achieves the best $\delta_{\sqrts}$ vs.\ $\LumiInt$ benchmark, with $\delta_{\sqrts} = 23.16$\,MeV and $\LumiInt = 2.94$\,ab$^{-1}$. 
This corresponds to an upper limit (at 95\% CL) of 
$\lvert y_{\Pe} \rvert < 3.2\,\lvert y^\mathrm{\textsc{sm}}_{\Pe} \rvert$ for the Higgs-electron coupling, for each detector in one year. 
With four experiments running at the same luminosity with the `TH4IP' scheme in the `GHC V22 $\PQt\PAQt$' optics with crossing angle, one should be able to set an upper limit (at 95\% CL) of about 2.5 times the SM value in one year of operation. 
This is to be  compared with about 4 times the SM value when operating without monochromatisation. 
Prospects for improvements in the monochromatisation performance are briefly alluded to in Section~\ref{sec:epol_next}.

\subsection{Outlook}
\label{sec:epol_next}

The studies performed before and during the Feasibility Study established a baseline scheme 
for calibration of the collision energy, ensuring that the physics goals of FCC-ee can be met.  
Nevertheless, these studies must be refined in certain areas
and alternative schemes should be considered to further improve performance and operational efficiency.

The absolute uncertainties in $\sqrt{s}$ arising from RDP 
are the dominant systematic uncertainties in the measurements of the \PZ and \PW masses.  
Future studies will investigate whether these uncertainties can be reduced 
beyond the currently assumed values: 100\,keV at the \PZ pole and 300\,keV at the $\PWp\PWm$ threshold.

The measurements of energy-related quantities made by the experiments using dimuon events 
are a critical ingredient in the $\sqrt{s}$ calibration, at all centre-of-mass energies. 
Recently, several of these studies have been deepened to validate their robustness 
with respect to the uncertainties in the knowledge of higher-order ISR/FSR effects;
this work will be extended further.  
It is also important to consolidate the strategy for understanding 
how the measurement of the crossing angle is affected by changes in the bunch intensity.
The impact of detector performance and the interplay with alignment studies will be another focus of attention. 
Finally, the use of other categories of physics events, beyond dimuons, will be investigated.

It is important to have a reliable procedure to accurately translate the mean beam energy, measured by RDP,
to the local collision energy, relevant for the physics measurements.
Full simulations of this procedure will be conducted, at each interaction point,
incorporating the in-situ measurement of the longitudinal boosts of the collisions and
the knowledge of the machine impedances.
Attention will also be paid to the control of energy shifts from possible dispersion effects at each interaction point, 
as well as to the related requirements on the precision of the system of beam-position monitors.

More detailed simulations of the level and lifetime of transverse polarisation will be performed, 
in parallel with changes to account for any evolution in the proposed optics of the accelerator. 
A deeper understanding will be sought of any effects that might bias 
the assumed proportionality between the spin tune and the mean beam energy.  
It will be particularly important to monitor the expected level of polarisation at the $\PWp\PWm$ threshold 
and the RDP strategy in this challenging regime. 
Detailed technical designs will be made of the polarimeter and depolariser systems.

So far, the baseline strategy to get transversally polarised pilot bunches 
has been to inject unpolarised beams and to stimulate the growth of polarisation by activating wigglers at the start of each fill. 
This is a robust approach but has the disadvantage of introducing dead-time, during which no collisions are possible.  
To overcome this inconvenient, 
studies will investigate the possibility of injecting already-polarised pilot bunches,
for which the design of the injection system must be modified. 
Simulations will be required to validate that the bunches retain their polarisation throughout the injection step 
and while travelling in the booster ring.

Further investigations will also take place regarding the feasibility of the electron Yukawa measurement.
In particular, new and refined schemes will be investigated with the aim of improving the monochromatisation of the collision energy.
It will also be necessary to develop and simulate a procedure to monitor and adjust the collision energy in real time, 
to ensure that it remains centred at the Higgs pole.   
Further physics studies will be performed to improve the signal yield and the signal-to-background discrimination.  
Investigations will be performed to evaluate if the significance of the signal can be improved with differential measurements, 
accounting for the expected correlations between the monochromatisation and the longitudinal coordinate of the $\epem$ collision.

\cleardoublepage
\section{Community building}
\label{sec:community}
Community building is critical for the success of the FCC project, because it fosters political, public, and financial support, encourages scientific collaboration and interdisciplinary innovation, promotes edu\-ca\-tion and outreach, and helps ensure the project’s long-term impact and sustainability. 
By creating an engaged, well-informed, and enthusiastic community, FCC can position itself not only as a groundbreaking scientific endeavour but also as a global, inclusive project with far-reaching benefits for humanity. 
This aspect has been given high priority ever since the beginning of the Conceptual Design Study, in 2014, in several directions:

\begin{itemize}
\item {\bf Global scientific networks}: 
The FCC project will be one of the largest scientific endeavours ever attempted and its success crucially depends on the cooperation of a broad international scientific community. 
Engaging researchers, engineers, and institutions worldwide in discussions and collaborations, early on, will ensure that 
FCC leverages the best scientific expertise, knowledge, training, and innovation from across the globe,
from the preparation of the theory and experimental tools to the actual operation of the collider and the detectors, culminating in the data analysis and interpretation.
\item {\bf Transdisciplinary collaboration}: 
The FCC project encompasses a wide range of disciplines, from particle physics theory to detector design and R\&D, from computing to engineering, from accelerator physics to machine-detector interface, etc. 
A well-established community always encourages cross-disciplinary collaboration and cross-fertilisation, enabling experts from various fields to contribute their knowledge and solve complex problems together. 
It also provides the resilience necessary to overcome setbacks, adapt to new circumstances, keep pushing the project forward to its implementation.
\item {\bf Inspiring the next generation of scientists and engineers}: 
Community engagement will help inspire the next generation to pursue careers in high-energy physics, and to build and operate the collider and the experiments. 
The FCC project must be a focal point for educational programmes, workshops, and professional training that generate interest and excitement in scientific discovery. The work done today will likely benefit future generations in ways that cannot yet be predicted. 
By creating a strong, multi-generational community, the project ensures that these long-term benefits are recognised, advocated for, and maximised over time.
\item {\bf Funding:}
Last but not least, the FCC project will require funding from multiple sources, including governments, private donors, and international organisations. 
A strong, active community can advocate for the project, raising awareness about its scientific and societal benefits. 
Demonstrating broad support and interests from the community and advocacy of the long-term scientific benefits is required to establish credibility and increase the likelihood of securing the necessary resources for the project.
\end{itemize}

As a matter of fact, the FCC feasibility study midterm review committees issued a number of recommendations to this effect. 
In particular, they recommended:

\begin{itemize}
\item to work with the scientific community, institutes, laboratories, and funding agencies to ensure support and resources for four experiments, facilitating the exploitation of the full scientific potential offered by the large investment in the FCC facility;
\item to dedicate additional human and financial resources to the project, with a resource-loaded schedule of work and clear priorities;
\item to develop the coordination and structure required to enable the theoretical progress needed to match the anticipated experimental precision of the FCC data, both at CERN (fellows, scientific associates, visitors) and by engaging collaborating institutes (including, for instance, the creation of European networks); and
\item to establish a dedicated FCC team in the research sector at CERN, with specific new positions associated, and to quantify its size and makeup in terms of seniority so that the resources required can be estimated.
\end{itemize}

Such actions have been (at least partially) anticipated a decade ago during the Conceptual Design Study (2014-2019), and intensified during the Feasibility Study (2021--2025). 
The three volumes of the FCC Conceptual Design Report~\cite{fcc-phys-cdr,fcc-ee-cdr,FCC-hhCDR}, released in January 2019, have been signed by 1364 authors, distributed in about 140 institutions, which had become members of the FCC Collaboration with the signature of Memoranda of Understanding (MoU) during this first phase. 
To comply with the recommendations of the 2021 European Strategy Update, the FCC Collaboration was reinforced by inviting more institutions to join and carry out the FCC Feasibility Study. 
One of the goals was to gather a majority of particle physicists behind the project, able to build a consensus in the community at the next European Particle Physics Strategy update that FCC is, indeed, the post-LHC collider option with the broadest scientific impact, at the intensity and the energy frontiers. 

During the five years of the Feasibility Study, the expansion of the Collaboration was pursued in two different ways.

\begin{itemize}

\item[A.] At the Collaboration level, the FCC Global Collaboration (FGC) Working Group continuously engages with current and new participants ---national institutes, laboratories and universities, as well as industry--- to encourage an expanded membership through Memoranda of Understanding. 
The FGC explores opportunities for future prospective participants, supports new participants in the application process, and assists the new participants in defining areas of collaboration, in particular for the accelerator. 
It then concludes relevant agreements to facilitate the integration process. 
It also prepares the foundations for R\&D and contributions by industry, 
as well as fostering interest in geology, geodesy, logistics, materials science, and other areas that, 
while not being at the core of CERN activities, 
are critical for the success of the FCC feasibility study and its implementation on the ground.

\item[B.] From the side of the FCC Physics, Experiments, and Detectors (PED) group, one of the six pillars of the FCC Feasibility Study, contacts are established with all high-energy physics groups in Europe and around the world, to invite them to join and contribute to the FCC project, in parallel to their current activities.
To this effect, an International Forum of National Contacts (IFNC) was created, chaired by two national contacts, with the mandate of attracting the different HEP groups, country by country. 
The quasi-complete list of European states (around 30) are represented in the IFNC, which meets regularly to reinforce the collaboration. 
The IFNC also reaches out to the rest of the world.  
The United States of America are now strongly involved in the Collaboration, following the recent Snowmass process and the positive recommendation of the U.S.\ Particle Physics Project Prioritization Panel (P5) to engage in an off-shore Higgs Factory. 
Most of the other large non-European countries are also active and represented in the IFNC (including Argentina, Brazil, Canada, Chile, India, Mexico, Pakistan, South Korea, Thailand, and Turkey) and the list continues expanding.  
\vskip .05cm
The cases of Japan and China, currently considering projects for colliders on their soil, are treated separately. 
With the goal to integrate teams from these countries in the near future, scientific exchanges with the Japanese and Chinese communities already occur in the FCC/CEPC/LC/ECFA and other national workshops.
Informal discussions with early-career Japanese physicists have also been organised.
\vskip .05cm
To build an extensive community of HEP physicists, the IFNC is also developing a finer structure, identifying institutional contacts for the participating institutes inside a country (up to 20 or 30 individuals for certain countries, such as France, Germany, Italy or the UK, and even more for the US), 
the goal being to have a vast majority of HEP institutes contributing, or on the verge of contributing, to the project by the end of the next European Strategy Update (2026).
\end{itemize}

In parallel, the PED group has also launched a call for Expressions of Interests (EoI's), in October 2024, to encourage the different institutes to get together and collaborate on innovative FCC sub-detector R\&D within the DRD Collaborations, and on detector concept studies, with the FCC integration as main objective. 
A first version of these sub-detector EoI's will be submitted as input to the European Strategy Group by March 31$^\text{st}$, 2025, and will be the basis for continuing development and consolidation during the next phase of the FCC study, until the end of 2027, with more R\&D, prototype construction, detailed simulation, test beams, etc. 
Detector concepts EoI’s will follow a similar development. 
Different combinations of sub-detectors will be integrated and tested in the common software to reach optimal performance, possibly as a function of the many physics objectives. 
Further steps are also envisioned, such as documenting, as an additional input to the European Strategy, the potential contribution of the different countries to FCC in the coming years, should the project be recommended and then approved.

The work with the scientific community progresses on a daily basis in the working groups of the PED pillar and gets a broader exposure twice a year during 
the `FCC Collaboration Weeks' in Spring (2021 at CERN; 2022 in Paris; 2023 in London; 2024 in San Francisco; 2025 in Vienna) 
and the `FCC Physics Workshops' in Winter (2022 in Liverpool; 2023 in Krakow; 2024 in Annecy; 2025 at CERN), 
allowing the community to be reinforced and new scientific collaborations to emerge. 
A strong FCC participation in the US Community Study on the Future of Particle Physics (`Snowmass 2021') acted as a decisive seed for the FCC effort in the US and was followed by three US-FCC workshops in 2023, 2024, and 2025. 
Besides, many national FCC Collaboration workshops have taken place, in essentially all large countries or regions of Europe and in the US, with active participation from the PED coordination group, yielding a significant growth of the PED part of the Collaboration in the past five years.

Several funding agencies have already reacted very positively to this evolution, decisively supported by substantive diplomatic work from the CERN management. 
The most resounding three examples are listed below.

\begin{enumerate}
\item In April 2024, the White House and CERN signed a joint statement of intent~\cite{WHCERN} saying, in particular, \emph{``Should the member states determine that FCC-ee is likely to be CERN’s next world-leading research facility, the US intends to collaborate on its construction and physics exploitation, subject to appropriate domestic approvals''.}
\item In June 2024, the CERN Council approved the Medium Term Plan (MTP) for the period 2025--2029~\cite{CERNmtp2}, including the funding of a bottom-up resource request for FCC-specific new positions.
\item In September 2024, the EU Competitiveness Report~\cite{EUCompetitivenessReport} was publicly released, with the following statements: 
\emph{``One of CERN’s most promising current projects, with significant scientific potential, is the construction of the Future Circular Collider (FCC): a 90-km ring designed initially for an electron collider and later for a hadron collider. [\ldots] 
Refinancing CERN and ensuring its continued global leadership in frontier research should be regarded as a top EU priority''.}
\end{enumerate}

Following the approval of the CERN MTP in June 2024, CERN created a dedicated FCC group in the EP department, effective since September 2024, with dedicated new positions (fellows, students, scientific associates, visitors) over the next three years. 
This sends another strong signal to the HEP community worldwide, regarding the host-lab commitment to FCC. 
Past experience has shown that even a moderately-sized host-lab group provides a significant leverage on contributions from the particle-physics community at other institutes. 
It is now anticipated that the creation of this group will be a strong incentive for CERN contract holders to reassign part of their time to the FCC project, starting with the supervision of the new recruits.
As the project moves to the next phase, many more engineers and detector physicists will be needed soon. 
The current situation is being carefully monitored and it is expected that a significant pool of these engineers and physicists will transition to FCC detector development as the  construction for the HL-LHC upgrades begins to wind down.

Assuming a positive recommendation of the CERN Council at around 2028, the FCC Collaboration expects that an FCC Committee (equivalent to the LHCC for the LHC experiments) will be formed by the CERN management and that a call for FCC detector Conceptual Design Reports will be issued soon after. 
Proto-collaborations (in particular, subsets of the FCC Collaboration, but not only) could then answer this call shortly after 2030 with full detector concept proposals. 

Finally, the coordination and structuring of the theoretical work needed to match the anticipated experimental precision of the FCC data has started during the Feasibility Study. 
Several successful mini-workshops were organised (Targets and Tools, Flavour Physics Programme, BSM Physics Programme, Higgs/Top/EW Physics Programme, Parton Shower, Phenomenology) with between 100 and 350 participants. 
The CERN/TH FCC team currently consists of three physicists, with the occasional participation of up to eight staff members and fellows, reflecting a genuine academic interest in working on the FCC physics among theorists. 
Needless to say, more dedicated resources will be needed for a worldwide organisation during the next phases of the FCC study.

\cleardoublepage

\section{Outlook}
\label{sec:conclusion}
The alignment of stars that led, in 2011/2012, to the concept of a $\sim$\,100\,km electron-positron collider in the same tunnel as a future 100\,TeV proton-proton collider in the Geneva basin and, in 2020, to the update of the European Strategy for Particle Physics (ESPPU) endorsing the FCC feasibility study as a top priority for CERN and its international partners, has presented the global HEP community with a unique and exceptional opportunity to advance the field. 
After almost five years of Feasibility Study performed by a worldwide consortium of scientists and engineers, the particle physicists ---including the early-career contingent--- are now steadily coalescing around the initial-stage machine (FCC-ee) as the first priority for a post-LHC collider at CERN. 

The FCC-ee offers ideal conditions (high luminosity, centre-of-mass energy calibration, possibly mono\-chromatisation, and moderate beam-induced backgrounds) for the intensity-frontier study of the four heaviest particles of the Standard Model, accumulating enormous samples ($6\times 10^{12}$ \PZ bosons at 88--94\,GeV, $5 \times 10^8$ \PW bosons at and above 157\,GeV, $2.8 \times 10^6$ Higgs bosons at and above 240\,GeV, and $4 \times 10^6$ top quarks at and above 340\,GeV) in only a few years of operation at each energy point. With a wealth of opportunities for precision electroweak, QCD, flavour, and Higgs measurements, searches for rare or forbidden processes, and the possible discovery of feebly coupled particles, \mbox{FCC-ee} is sensitive to essentially any kind of new physics from a few GeV up to scales of 10 to 100\,TeV, over an extremely broad range of couplings. These studies may help address the most profound questions of particle physics today. What is the nature of dark matter? How did the antimatter disappear? What is responsible for the non-zero neutrino masses? Even if \mbox{FCC-ee} were to `only' confirm the Standard Model with a precision up to three orders of magnitude better than today, theories that propose solutions to these fundamental questions would become very tightly constrained, thus guiding the development of new models and improving significantly the understanding of the creation of the Universe. It is instructive to recall that one of the foundation stones of modern physics is a null experiment: the absence of discovery of the Ether by the Michelson--Morley experiment in 1887.

The FCC-ee is also the perfect springboard for a 100\,TeV hadron collider (FCC-hh), for which it provides a major fraction of the infrastructure. The complementary and synergistic exploratory physics programmes of these two machines offer a uniquely powerful long-term vision. In a way not too dissimilar from the discovery of Neptune in 1846, which was predicted by the precise measurement of the orbit of Uranus in 1843 (inconsistent with Newton's laws applied to the solar system of the seven planets observed at that time), any deviation with respect to the Standard Model predictions observed at FCC-ee is testable directly with FCC-hh, up to a scale of 40\,TeV. When a new physics signal is eventually observed at FCC-hh or elsewhere, the FCC-ee precision measurements ---whether they agree or not with the Standard Model--- will be an invaluable resource in establishing the nature of the underlying physics. 

The FCC integrated project is also able to measure the Higgs boson interactions with fermions and gauge bosons with high precision (down to a part in a thousand) without theoretical hypotheses. 
The Higgs potential, modelled by the Higgs boson self interactions, plays a fascinating and central role in the cosmological history of the universe. A first determination of the trilinear Higgs boson self-coupling with a precision of the order of 25\% will be provided by HL-LHC through the measurement of the Higgs-pair production cross section. A qualitatively different and complementary determination with a similar precision will exploit the per-mil-level 
measurement of the single-Higgs production cross section at FCC-ee, yielding a combined precision better than 20\%. A unique per-cent level measurement of the trilinear Higgs self-coupling will only become available with FCC-hh.

To date, there is no other collider project that can even remotely compete with such an exploratory breadth and depth: the most complete understanding of the Higgs sector; the unique interplay between the electroweak, flavour, and Higgs measurements at the intensity frontier; the multiple synergies between FCC-ee and FCC-hh; the access to the smallest couplings and the highest energy scales; in all aspects, the FCC integrated programme offers outstanding physics prospects. The FCC project, with its four interaction points,  therefore suits to perfection the scientific ambitions of CERN and of its worldwide community of 15\,000 users. It is also remarkable that the overall duration; the electricity consumption; the total cost; and the life-cycle carbon footprint of the FCC-ee construction and operation are all very significantly smaller~\cite{Blondel:2024mry} than those of other (lower luminosity) Higgs factory options at CERN once normalised to their physics output (i.e., once the running time of these alternative projects is extended to match the same Higgs-coupling precision as FCC-ee, and yet, disregarding the richness of the remainder of the FCC-ee physics programme, which lies beyond these other options). Furthermore, the FCC integrated project minimises the carbon emissions on the road to the highest energies, as the FCC-ee infrastructure will be fully recycled for FCC-hh. The FCC infrastructure could also be repurposed to house a fast accelerator and injector for a very high energy muon collider in the LEP/LHC tunnel.

\begin{figure}[htbp]
\centering
\begin{sideways}
\begin{minipage}{165mm} 
\caption{\label{fig:roadmap} 
Possible timeline of the FCC-ee project between the end of the feasibility study and the start of the first FCC-ee physics run, with key dates for the project as a whole (pink), and separately for the accelerator (left side, green) and for the detectors (right side, blue). (The acronym FC$^3$ stands for `FCC Committee'.) }\vspace{0.5cm}
\includegraphics[width=\columnwidth,angle=0]{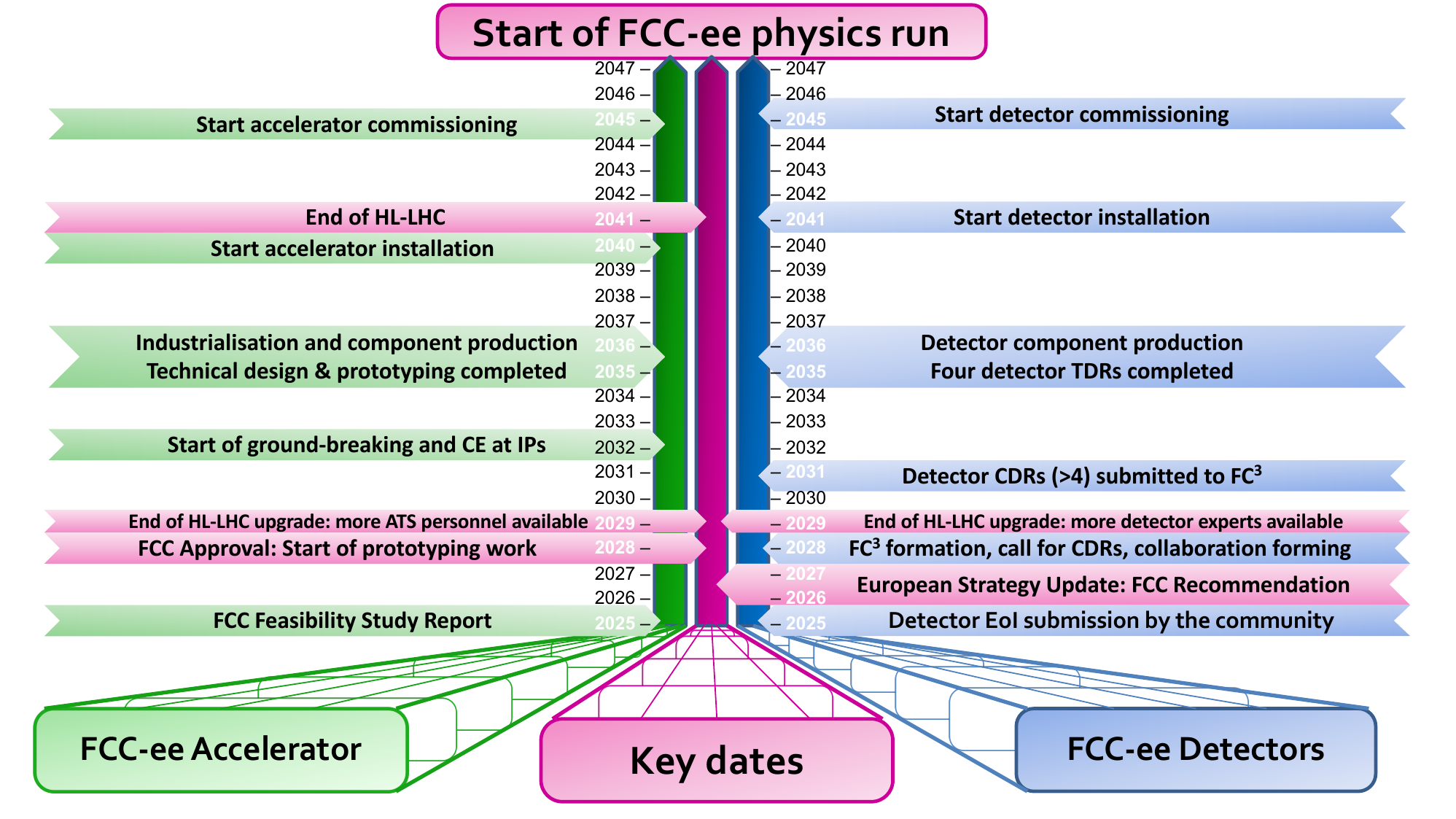}
\end{minipage}
\end{sideways}
\end{figure}

The delivery of this Feasibility Study marks the completion of the first phase of an extended and intensive programme of work.
Even though the $\epem$ collider is currently scheduled to deliver its first collisions in the second half of the 2040's, the timeline of the project, displayed in Fig.~\ref{fig:roadmap} for both the accelerator and the experiments, is tightly filled with important deadlines that must be met and critical milestones that cannot be missed.

The first of these milestones has been for the particle physics community to prepare expressions of interest (EoIs) for critical detector components towards the realisation of up to four experiments, in time for the forthcoming European Strategy Update. More EoIs are foreseen if FCC-ee is recommended as the preferred post-LHC project at CERN by the European Strategy Group. Should the project be approved by CERN Council on the basis of this recommendation, a formal call for the creation of experiment proto-collaborations will be launched, towards the submission of Detector Conceptual Design Reports (CDRs) a few years later, followed by detector Technical Design Reports (TDRs) in the mid-2030's. Achieving these ambitious goals will require an appropriate and timely re-organisation of activities over the coming couple of years, and a significant injection of (human and financial) resources in the project.

In parallel with detector design, construction, installation, and commissioning, the road towards the first collisions is paved with many other exigencies to respond to, a partial list of which follows.  
\begin{itemize}
\item First and foremost, the effort towards the goal of matching theoretical calculation uncertainties to the expected statistical power of the collider (and propagating these calculations to technically accurate Monte Carlo generators), so as to optimally exploit the data from FCC-ee and, later, from FCC-hh, needs to be immediately structured and sustained with appropriate resources, from the clear roadmap established during the Feasibility Study. 

\item In the coming few years, the requirements from detailed physics case studies will continue to be the driving factor in efforts to design the most technologically-advanced (but also the most realistic) FCC-ee detector concepts, to ensure the full coverage of the wide and challenging physics programme. These efforts will have to be intensified, in particular at the \PZ pole, where the requirements from physics are expected to be the most demanding. 

\item These physics studies will need to be backed up with a versatile and reliable software ecosystem, from accurate Monte Carlo event generators to modern event reconstruction and analysis software, passing through detailed detector simulation. Achieving this will demand a significant increase of both computing and human resources, which were a limiting factor during the Feasibility Study. To broaden the project resource base, continued efforts are required to attract (many) more external contributors from institutes worldwide. Promoting positions shared between LHC and FCC will also foster joint development efforts and enhance expertise exchange. 

\item Regarding the interaction region, the integration strategy of the inner sub-detectors (e.g., vertex detectors, luminometers, superconducting pipes, supporting structures) with the accelerator components will be established, as their designs are finalised.   
Methods for opening and maintenance of the detector elements, together with their alignment with respect to the beams will be further developed, and possible consequences on the cavern dimensions will be evaluated. Most importantly, the studies of the impact of all beam-induced backgrounds on the detector performance and mitigation solutions will need to be consolidated for all sub-detectors.

\item The measurement and monitoring of the centre-of-mass energy will be one of the cornerstones of the whole physics programme. The baseline scheme presented in this Volume must be refined and made fully reliable, and alternative schemes will be considered. Specifically, the possibility of injecting already polarised electron and positron bunches will be thoroughly investigated. Improved schemes for monochromatisation will be systematically studied in view of proposing a sustainable method for the electron Yukawa coupling measurement. 

\item Finally, detailed studies of the physics programme of FCC-ee on the one hand, 
and groundwork studies of physics and detectors at FCC-hh on the other, 
will continue to further extend the exploration of the standalone potential of each collider, 
and the synergies between them.
These efforts will also reflect 
the guidelines that will emerge from the review of the Feasibility Study Report and from the ESPPU discussions.

\end{itemize}
While it will be years before all these goals are attained, there is no doubt that the recommendation of the FCC project by the forthcoming ESU will catalyse the engagement of the particle physics community at large on FCC-ee. The growing interest from young scientists in the project must be recognised with appropriate career opportunities.

\begin{figure}[htbp]
\centering
\begin{sideways}
\begin{minipage}{175mm} 
\caption{\label{fig:cdrplus} 
Potential physics programme for FCC-ee, ordered by increasing centre-of-mass energy, without indication of a specific chronological sequence. The events highlighted in red indicate the minimal programme with fifteen years of running, and with the corresponding integrated luminosities and physics outcome. The numbers of \PZ, $\PW\PW$, $\PZ\PH$, and $\ttbar$ events delivered to four interaction points are indicated.  A possible wider physics programme, with additional centre-of-mass energies, is highlighted in blue.}
\includegraphics[width=\columnwidth,angle=0]{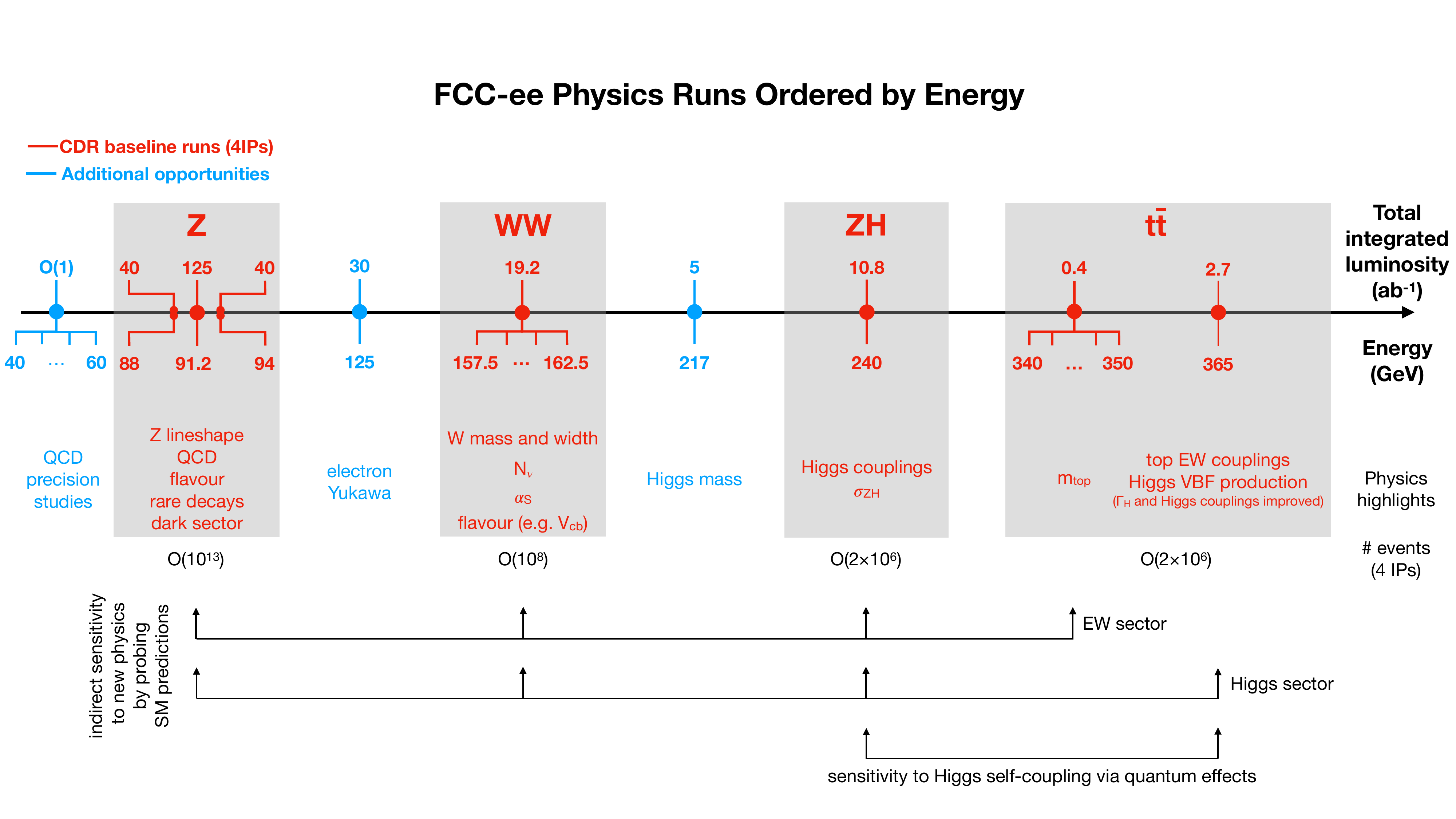}
\end{minipage}
\end{sideways}
\end{figure}

The baseline choice of collision energies and running sequence (\PZ pole, $\PW\PW$ threshold, $\PZ\PH$ maximum, and $\ttbar$ threshold, as shown in Fig.~\ref{fig:cdrplus}) with four interaction points is sufficient to demonstrate that FCC-ee offers an extraordinary list of opportunities (and associated challenges) for milestone measurements with a real chance of discovery. Looking ahead, this list will only become richer as the investigations deepen, and it is already clear that more integrated luminosity and the extension of the programme to encompass additional energy points would increase the FCC science value still further~\cite{note-FCCeeSequence}. In parallel, the FCC-ee machine parameters are now  different from  those that prevailed at the time of the CDR, and will continue to evolve in the next phase of the study, while the  understanding of the  physics performance of the experiments is progressing. This perspective may significantly modify, or enrich, the desirable scientific programme, as pictured in Fig.~\ref{fig:cdrplus}.

For example, the possibility to run at $\sqrt{s} = 125$\,GeV, with a centre-of-mass energy spread of the order of the Higgs boson width, is under consideration, towards a 3\,$\sigma$ evidence for the electron Yukawa coupling. Precision QCD studies with a few ab$^{-1}$ at centre-of-mass energies of 20, 30, 40, \dots, 80\,GeV are very appealing to a significant fraction of the HEP community. Because the study of the  monochromatisation schemes required for the electron Yukawa measurement are not yet at a mature stage; and because $\PZ$-pole data with energetic initial-state radiation can in principle also explore lower centre-of-mass energies; no proposal is made yet to include these energy points in the baseline programme. 

These additional possibilities and corresponding science value further demonstrates the flexibility and breadth of the FCC-ee physics potential. An increase of the FCC-ee specific luminosity would enable the physics programme to be extended to encompass at least some of these possibilities within the currently envisaged 15-year timescale. It will ultimately be up to the experimental collaborations and the relevant scientific committees to optimise the time flexibility and to tailor the FCC-ee operation scenario according to a number of factors that cannot be controlled today. For example, external events such as the FCC-hh magnet readiness or the CERN financial situation may call for a change in the overall duration of the FCC-ee running time. Furthermore, intriguing early results may motivate additional running at a given working  point. Meanwhile, new avenues will be explored towards increasing the luminosities at all energies.

At the end of the HL-LHC operation, 
the LEP-LHC integrated programme will have delivered sustained scientific excellence for over 50 years, and greatly advanced our understanding of the fundamental interactions.   The succession of FCC-ee and FCC-hh in a common tunnel will replicate and magnify this success, with vastly better precision and increased energies. The FCC integrated programme offers outstanding prospects for progress in a wealth of topics, including Higgs, electroweak, and flavour physics, as well as remarkable sensitivity in direct searches for both feebly coupled low-mass particles and those that may exist up to many tens of TeV. The upcoming ESPPU provides an opportunity for the global HEP community to endorse this vision, and then to work together so as to enable the first stage of this project, FCC-ee, to be realised in a timely fashion. Seizing this opportunity will be an important step forward in the journey towards a more complete understanding of the laws of nature.   

\clearpage

\clearpage

\bibliographystyle{cms_unsrt}

\bibliography{biblio}%

\addcontentsline{toc}{section}{References}

\end{document}